\documentclass[12pt]{amsbook}
\usepackage{amssymb}
\usepackage[all]{xy}

\usepackage[colorlinks=true,linkcolor=magenta,citecolor=magenta]{hyperref}
\makeindex

\newtheorem{theorem}{Theorem}[chapter]

\newtheorem{answer}[theorem]{Answer}
\newtheorem{answers}[theorem]{Answers}
\newtheorem{cat}[theorem]{Cat}
\newtheorem{claim}[theorem]{Claim}
\newtheorem{conclusion}[theorem]{Conclusion}
\newtheorem{conjecture}[theorem]{Conjecture}
\newtheorem{definition}[theorem]{Definition}
\newtheorem{dillema}[theorem]{Dillema}
\newtheorem{exercise}[theorem]{Exercise}
\newtheorem{fact}[theorem]{Fact}
\newtheorem{homework}[theorem]{Homework}
\newtheorem{method}[theorem]{Method}
\newtheorem{notations}[theorem]{Notations}
\newtheorem{plan}[theorem]{Plan}
\newtheorem{principle}[theorem]{Principle}
\newtheorem{problem}[theorem]{Problem}
\newtheorem{proposition}[theorem]{Proposition}
\newtheorem{puzzle}[theorem]{Puzzle}
\newtheorem{question}[theorem]{Question}
\newtheorem{questions}[theorem]{Questions}
\newtheorem{warning}[theorem]{Warning}

\begin{document}

\title{Advanced linear algebra}

\author{Teo Banica}
\address{Department of Mathematics, University of Cergy-Pontoise, F-95000 Cergy-Pontoise, France. {\tt teo.banica@gmail.com}}


\begin{abstract}
This is an introduction to advanced linear algebra, with emphasis on geometric aspects, and with some applications included too. We first review basic linear algebra, notably with the spectral theorem in its general form, and with the theory of the resultant and discriminant. Then we discuss the Jordan form and its basic applications to physics, and other advanced decomposition results for the matrices. We then go into positivity topics, involving matrices and bilinear forms, and with a look into curved space-time, and discrete Laplacians. Finally, we discuss the various groups of matrices, with a look at reflection groups, Lie groups, spin matrices and random matrices.
\end{abstract}

\maketitle

\chapter*{Preface}

This is an introduction to advanced linear algebra, with emphasis on geometric aspects, and with some applications included too. The book is organized itself with emphasis on algebra and symmetry, in 4 parts, having 4 chapters each, with each chapter having 24 pages, and consisting of 4 sections, plus an informal exercise section at the end. 

\bigskip

Add to this 8 pages of front matter and 8 pages of back matter, and you have exactly 400 pages, according to the equation $8+16\times 24+8=400$, that took me years and years to find, after countless organization attempts, with some previous books that I wrote. With this being something nice, making among others the text printer-friendly.

\bigskip

Getting now to the contents of the book, this will be certainly algebraic, but less maniac on algebraic aspects, and often insisting on the underlying geometry. Personally I tend to regard any vector as being something alive and dynamic, and any linear map as being something alive and dynamic too, and any matrix as consisting of course of numbers, which themselves are alive and dynamic objects too. With the ``dynamism'' of everything coming from the underlying physics, I mean, after all these years spent doing or teaching mathematics, I have yet to meet an interesting vector, or linear map, or matrix, not having something to do with physics. So, this will be for the general philosophy, geometry and physics, in order to understand linear algebra, and vice versa.

\bigskip

As already mentioned, the book is organized in 4 parts, which are as follows:

\bigskip

I - We review here the basic linear algebra, namely vectors, linear maps, matrices, matrix inversion, determinant, eigenvectors, eigenvalues and diagonalization. Then we review more advanced theory, such as the spectral theorem in its various forms, and the theory of the resultant and discriminant. We include as well all sorts of tricks, as for instance the fact that the diagonalizable matrices over $\mathbb C$ are dense.

\bigskip

II - Unfortunately not all matrices are diagonalizable, and in this second part, we discuss what can be done when they aren't. We first explain the Jordan decomposition theorem, and the exponentiation of matrices. Then we discuss some other advanced decomposition results, and notably the singular value decomposition. Finally, we have a look at dynamical systems, and at infinite dimensions and compact operators too. 

\bigskip

III - Here we go into positivity and negativity topics, motivated by the positivity and negativity of the matrix eigenvalues, notably for the Hessian matrices. We first discuss the classification of the bilinear forms, in terms of their signature, and do not miss the occasion of talking a bit about curved spacetime, and Lorentz geometry. Then we discuss bistochastic matrices, discrete Fourier analysis, designs and discrete Laplacians.

\bigskip

IV - Finally, for ending in beauty, we discuss what matrices can do when operating together, in groups. Nothing or almost can resist to these frightening formations, and we first discuss the finite groups, with emphasis on the reflection groups, and then the compact Lie groups and their representations, with a look into spin matrices and related quantum physics topics too. We end with an introduction to the random matrices.

\bigskip

In the hope that you will find this book useful. As a complement to it, for more applications to classical analysis you have my calculus book \cite{ba1}, for more about groups, of all types, you have my group theory book \cite{ba2}, and for more about what happens to linear algebra in infinite dimensions you have my operator algebra book \cite{ba3}. 

\bigskip

Many thanks go to my students, and especially the undergraduate ones, that I often take into $SU_2$ and $SO_3$, no matter what the course is about, and who invariably like this stuff. Thanks as well to my colleagues, for countless coffee room discussions about linear algebra, and for some joint research on linear algebra too. Finally, many thanks go to my cats, for their teachings on both linearity and non-linearity.

\bigskip

\

{\em Cergy, May 2026}

\smallskip

{\em Teo Banica}

\baselineskip=15.95pt
\tableofcontents
\baselineskip=14pt

\part{Linear algebra}

\ \vskip50mm

\begin{center}
{\em Hey, where did we go

Days when the rains came

Down in the hollow

Playing a new game}
\end{center}

\chapter{Linear maps}

\section*{1a. Spaces, vectors}

As you can see, we live in $\mathbb R^3$, and this is where most of the questions in our mathematics take place. However, you also know from calculus, or from physics, that dealing with the mathematics of $\mathbb R^3$ is no easy matter. So, for this purpose, doing mathematics in $\mathbb R^3$, the best is to regard our space $\mathbb R^3$ as being part of a hierarchy of spaces $\mathbb R^N$, where you can do mathematics, at varying levels of difficulty, as follows:

\bigskip

(1) First comes $\mathbb R$. This has little to no interest in connection with real-life problems, but as you know well from calculus, everything mathematics comes from here, with this meaning sequences, series, functions, continuity, derivatives, integrals and so on. In this book we will assume the basic theory of $\mathbb R$ known. If you need from time to time to revise that, you have here Rudin \cite{ru1}, or Lax-Terrell \cite{lt1}, or the first part of my book \cite{ba1}.

\bigskip

(2) Then comes $\mathbb R^2$. This is the entry point to advanced mathematics, because most of the $\mathbb R^3$ phenomenona have interesting 2D analogues, quite often capturing the whole point. Sometimes, $\mathbb R^2$ can be even your final destination, because many interesting $\mathbb R^3$ questions take place in fact in a plane $\mathbb R^2\subset\mathbb R^3$. And finally, as another key feature of $\mathbb R^2$, we have an isomorphism $\mathbb R^2\simeq\mathbb C$, transforming by some kind of magic your 2-variable questions in $\mathbb R^2$ into routine 1-variable problems, over the complex numbers $\mathbb C$. In this book we will assume $\mathbb R^2$, $\mathbb C$ reasonably known, and in case you need from time to time a reference here, go with the books of Rudin \cite{ru2}, or Lax-Terrell \cite{lt2}, or mine \cite{ba1}.

\bigskip

(3) Then comes $\mathbb R^3$. Here there are no tricks of type $\mathbb R^2\simeq\mathbb C$, so we are definitely into several variables, whose functioning we must understand well. But intuition, helped by the $\mathbb R^3$ surrounding us, can help a lot. As an interesting feature of $\mathbb R^3$, of rather engineering type, and contradicting all the mathematics that you learned, volumes of bodies $V\subset\mathbb R^3$ are easier to compute than areas $A\subset\mathbb R^2$, simply by plunging them into water, and measuring the water displacement. And isn't this genius. Also, mathematically, on $\mathbb R^3$ we have available the vector product $x\times y$, which can be useful for many things.

\bigskip

(4) Then comes $\mathbb R^4$. You would say why bothering with it, but the point is that, according to Einstein's relativity theory, our usual $\mathbb R^3$ does not really exist, in practice, as strange as this might seem, because the space variables $(x,y,z)$ are in fact connected to the time variable $t$. Thus, the correct variable for any physics problem, involving at least a bit or relativity, and there are so many of them, including everything having to do with light, electromagnetism, or quantum physics, is in fact $(x,y,z,t)$, which lives in $\mathbb R^4$, or rather in a technical, curved version of $\mathbb R^4$. Note also that we have $\mathbb R^4\simeq\mathbb C^2$.

\bigskip

(5) Then comes $\mathbb R^N$. This is actually simpler than both $\mathbb R^3$ and $\mathbb R^4$, for most matters, and when we said in (3) above, in relation with $\mathbb R^3$, that ``we are definitely into several variables, whose functioning we must understand well'', we meant by this ``time to learn several variables, first in $\mathbb R^N$, and then in $\mathbb R^3$''. Also, as another interesting feature of $\mathbb R^N$, the vector product $x\times y$ from $\mathbb R^3$ has no analogue in $\mathbb R^N$, and with this being a good thing, forcing us to rewrite many things that we know from $\mathbb R^3$, obtained via $x\times y$, in a more straightforward way in $\mathbb R^N$, by using the rock-solid scalar product $<x,y>$.

\bigskip

(6) Finally, we have $\mathbb R^\infty$. This is normally reserved for quantum mechanics matters, which live there, in infinite dimensions, and to be more precise in $\mathbb C^\infty$, to be fully correct, and in what regards the level of difficulty, with respect to $\mathbb R^3,\mathbb R^4,\mathbb R^N$, this can wildly vary, depending on the type of quantum mechanics questions that you have in mind. That is, for easy questions $\mathbb R^\infty$ can be simpler than $\mathbb R^N$, for the simple reason that there are less tools available, so less mathematics to be done. However, for difficult questions, $\mathbb R^\infty$ can be at the same level of difficulty with $\mathbb R^3,\mathbb R^4$, or even harder. Finally, forgetting about quantum, knowing a bit about $\mathbb R^\infty,\mathbb C^\infty$ can be useful for $\mathbb R,\mathbb R^2,\mathbb R^3,\ldots\,$, and this because the real or complex functions on $\mathbb R^N$ form spaces which are isomorphic to $\mathbb R^\infty,\mathbb C^\infty$.

\bigskip

So, this was the general story with mathematics, and more specifically geometry and analysis, motivated by physics questions, the conclusion being as follows:

\begin{conclusion}
Mathematics inside $\mathbb R^3$ is a tricky business, best learned:
\begin{enumerate}
\item By studying $\mathbb R,\mathbb R^2,\mathbb R^3,\mathbb R^4,\mathbb R^N,\mathbb R^\infty$, which are all useful,

\item One at the time, switching dimensions when needed,

\item And by having an eye on $\mathbb C,\mathbb C^2,\mathbb C^N,\mathbb C^\infty$ too.
\end{enumerate}
\end{conclusion}

What about linear algebra, in relation with all this? Well, linear algebra is the key to geometry and analysis, and therefore to physics too, dealing with the most basic phenomena that can appear, the ``linear'' ones, which are at the core of everything.

\bigskip

There are many things to be learned here, even at the basic level, and I am pretty much sure that you know a bit of this, foundations of linear algebra. However, it won't hurt to review a bit all this, at a more advanced level, and we will do this now. 

\bigskip

To start with, we certainly have some business to do with $\mathbb R^2,\mathbb R^3$, basic things there that you might know or not. Then, as a main objective, we would like to properly understand what the determinant is, and how the diagonalization procedure works. And finally, in view of the few occurrences of $\mathbb C$ instead of $\mathbb R$ in the above, it is better to talk as well, at least from time to time, about linear algebra over arbitrary fields.

\bigskip

In view of this, here will be our plan for the first 2 chapters of the present book:

\begin{plan}
We must review the linear algebra that we know, by learning:
\begin{enumerate}
\item More geometry in $\mathbb R^2,\mathbb R^3$, with focus on the linear maps there.

\item The precise and true meaning of the determinant, in $\mathbb R^N$.

\item The diagonalization procedure, done geometrically, also in $\mathbb R^N$.

\item What happens as well over $\mathbb C$, or over an arbitrary field $F$.
\end{enumerate}
\end{plan}

So this will be our plan, and afterwards in chapters 3-16 we will of course further build on all this, with a number of results that should be new to you, I hope.

\bigskip

Before starting, a few references too. Normally what we will be doing here will be quite self-contained, but quite often coming with very compact proofs, for the linear algebra basics that you are supposed to know. As standard references here, you have Lang \cite{la1} if you are more into algebra, and Lax \cite{lx1} if you are more into analysis.

\bigskip

Getting started for good now, at the beginning of everything, we have:

\begin{definition}
The points $x\in\mathbb R^N$ can be represented as vectors
$$x=\begin{pmatrix}
x_1\\
\vdots\\
x_N
\end{pmatrix}$$
and are subject to the addition and multiplication by scalars operations
$$x+y=\begin{pmatrix}
x_1+y_1\\
\vdots\\
x_N+y_N
\end{pmatrix}\quad,\quad
\lambda x=\begin{pmatrix}
\lambda x_1\\
\vdots\\
\lambda x_N
\end{pmatrix}$$
geometrically corresponding to forming a parallelogram, and dilating by $\lambda$.
\end{definition}

To be more precise here, having a point $x\in\mathbb R^N$ is the same as having its coordinates $x_1,\ldots,x_N\in\mathbb R$, and we agree to designate $x$ by the vertical vector formed by these coordinates $x_1,\ldots,x_N$. In case you wonder why we chose the vertical notation over the horizontal notation, the point here is that a bit later, when we will do linear algebra, we will really need this convention, as for the linear maps $f:\mathbb R^N\to\mathbb R^N$ to take the form $f(x)=Ax$, with $A$ being a $N\times N$ matrix. More on this in a moment.

\bigskip

Regarding now the addition operation, in 2 dimensions this can be understood indeed geometrically, with $x+y$ completing the parallelogram based at $O,x,y$:
$$\xymatrix@R=10pt@C=15pt{
&&&\\
x_2+y_2&&&&\bullet^{x+y}\\
y_2&&\bullet^y\ar@{-}[urr]&&\\
x_2&&&\bullet_x\ar@{-}[uur]&\\
&\bullet\ar@{-}[urr]\ar[rrrr]\ar[uuuu]\ar@{-}[uur]&&&&\\
&&\ y_1\ &\ x_1\ &x_1+y_1}$$

In fact, this interpretation still holds in arbitrary $N$ dimensions, with everything here coming from basic geometry and the Thales theorem, as you surely know. As for the multiplication by scalars operation, this is something very intuitive.

\bigskip

Still talking vectors, at a more advanced level we can talk about scalar products and lengths of vectors, with the basic theory here being summarized as follows:

\begin{theorem}
We can talk about scalar products and lengths, according to
$$<x,y>=\sum_ix_iy_i\quad,\quad ||x||=\sqrt{\sum_ix_i^2}$$
which are related by the following conversion formulae,
$$||x||=\sqrt{<x,x>}\quad,\quad<x,y>=\frac{||x+y||^2-||x-y||^2}{4}$$
and the following happen:
\begin{enumerate}
\item $<\lambda x,y>=<x,\lambda y>=\lambda<x,y>$.

\item $<x+y,z>=<x,z>+<y,z>$.

\item $<x,y+z>=<x,z>+<y,z>$. 

\item $||\lambda x||=|\lambda|\cdot||x||$.

\item $|<x,y>|\leq||x||\cdot||y||$.

\item $||x+y||\leq||x||+||y||$.

\item $x\perp y\iff<x,y>=0$, by definition.

\item $<x,y>=||x||\cdot||y||\cdot\cos t$, with $t$ being the angle between $x,y$, by definition.

\item $<x,y>=<x',y>=<x,y'>$, prime being the projection on the other vector.
\end{enumerate}
\end{theorem}

\begin{proof}
We can certainly talk about scalar products and lengths, as above, and with the second conversion formula, called polarization identity, coming from:
\begin{eqnarray*}
||x+y||^2-||x-y||^2
&=&<x+y,x+y>-<x-y,x-y>\\
&=&||x||^2+||y||^2+2<x,y>-||x||^2-||y||^2+2<x,y>\\
&=&4<x,y>
\end{eqnarray*}

By the way, talking useful identities, we have as well a parallelogram rule, that I forgot to mention in the above statement, which is as follows:
\begin{eqnarray*}
||x+y||^2+||x-y||^2
&=&<x+y,x+y>+<x-y,x-y>\\
&=&||x||^2+||y||^2+2<x,y>+||x||^2+||y||^2-2<x,y>\\
&=&2(||x||^2+||y||^2)
\end{eqnarray*}

As for the various claims in the statement, these are all elementary, as follows:

\medskip

(1-4) All the verifications here are indeed trivial.

\medskip

(5) Given two vectors $x,y\in\mathbb R^N$, consider the following function $f:\mathbb R\to\mathbb R$:
\begin{eqnarray*}
f(t)
&=&||x+ty||^2\\
&=&<x+ty,x+ty>\\
&=&||x||^2+2t<x,y>+t^2||y||^2
\end{eqnarray*}

Thus $f$ is a degree 2 polynomial, and since this polynomial is positive, its discriminant must be negative, $\Delta\leq0$. But the discriminant is given by the following formula:
$$\Delta=4<x,y>^2-4||x||^2||y||^2$$

Thus we have the Cauchy-Schwarz inequality, $|<x,y>|\leq||x||\cdot||y||$, as claimed. 

\medskip

(6) This can be proved by raising to the square and simplifying, as follows:
\begin{eqnarray*}
||x+y||\leq||x||+||y||
&\iff&||x+y||^2\leq||x||^2+||y||^2+2||x||\cdot||y||\\
&\iff&||x||^2+||y||^2+2<x,y>\leq||x||^2+||y||^2+2||x||\cdot||y||\\
&\iff&<x,y>\leq||x||\cdot||y||
\end{eqnarray*}

Indeed, the last inequality holds by (5), so we have our triangle inequality.

\medskip

(7-9) These assertions are something more subtle, because in the lack of good intuition, do we really know what orthogonality and angles really are, in $\mathbb R^N$. So, the best is to proceed as indicated, with orthogonality and angles being defined as in the statement, and with the last formula being an instructive exercise for you, enjoy.
\end{proof}

\section*{1b. Linear maps}

Getting now to linear algebra, we need a definition for the linear maps, our main objects of study. Leaving the arbitrary fields for later, here that definition is:

\index{linear map}
\index{matrix}
\index{rectangular matrix}

\begin{definition}
A map $f:\mathbb R^N\to\mathbb R^M$ is called linear when it satisfies the following equivalent conditions:
\begin{enumerate}
\item Algebraic conditions: $f(x+y)=f(x)+f(y)$ and $f(\lambda x)=\lambda f(x)$.

\item $f$ maps lines to lines: $f(tx+(1-t)y)=tf(x)+(1-t)f(y)$.

\item Each component of $f(x)$ appears as a linear combination $\sum_i\lambda_ix_i$.

\item $f(x)=Ax$, for some rectangular matrix $A\in M_{M\times N}(\mathbb R)$.
\end{enumerate}
\end{definition}

To be more precise, these conditions are something very familiar, with each having its own advantages and disadvantages, and the equivalence between them, that you know well too, is not difficult to establish, the idea with all this being as follows:

\bigskip

(1) The algebraic conditions, although a bit abstract, prove to be in practice very useful, and they will be often our main workhorses here, in order to understand the linear maps. The name ``algebraic conditions'' comes from abstract algebra, because we can talk more generally about linear maps between abstract vector spaces $f:V\to W$, and the operations on such vector spaces being the vector sum $x+y$ and the multiplication by scalars $\lambda x$, being linear simply means ``preserving the algebraic structure''.

\bigskip

(2) This is something certainly more intuitive, among other justifying the same ``linear'' for our maps, I mean that linear must certainly come from ``line'', but go see a line in the axioms (1), personally I don't see any. In practice, while the equivalence with (1) is something clear, this condition, while intuitive and beautiful, is not very useful in practice. Also, as a technical remark, when saying in the above ``maps lines to lines'', by line we mean, as above, a dynamic object, with a parameter $t\in\mathbb R$ involved. When dropping this convention, and regarding the lines as sets, the equivalence with (1) no longer holds, and we will leave finding a counterexample here as an instructive exercise.

\bigskip

(3) This is also something nice, of old-style flavor, which helps understanding what is going on, and with the equivalence with (1) being clear from definitions. However, as we will see in a moment, this condition is clearly equivalent to (4) too, which is more powerful, and so in practice, our condition is somehow stuck between (1) and (4), which are both more powerful, each in its own way, and so, hard life for this condition.

\bigskip

(4) This is something very powerful, and a true rival to (1), usually surpassing it in power, for nearly all concrete applications. The equivalence comes via (3), because according to that condition we can write each component $f(x)_i$ as a linear combination $\sum_jA_{ij}x_j$, which according to the rules of usual matrix multiplication means $(Ax)_i$. Thus, we have our matrix $A\in M_{M\times N}(\mathbb R)$ making the formula $f(x)=Ax$ work, as desired.

\bigskip

Before going further, let us record the following result, focusing on the condition (4) in Definition 1.5, and building a bit more on the equivalence with (1):

\index{scalar product}

\begin{theorem}
The linear maps $f:\mathbb R^N\to\mathbb R^M$ are the maps of the form
$$f(x)=Ax$$
with $A\in M_{M\times N}(\mathbb R)$, and $A$ can be recaptured via $A_{ij}=<f(e_j),e_i>$.
\end{theorem}

\begin{proof}
This is something very standard, the idea being as follows:

\medskip

(1) The first assertion follows the above discussion, or even from Definition 1.5 based on the above discussion, if you prefer. In fact, in what follows, $f(x)=Ax$ will be more or less our definition for the linear maps, for most questions that we will investigate.

\medskip

(2) The second assertion is something clear too, coming as follows:
$$<f(e_j),e_i>
=<Ae_j,e_i>
=(Ae_j)_i
=A_{ij}$$

Thus, we are led to the conclusions in the statement.
\end{proof}

In order to understand now how the above correspondence $f\leftrightarrow A$ works, let us discuss some examples. We have here the following statement, which is a must-know:

\index{rotation}
\index{symmetry}
\index{projection}
\index{rank 1 projection}

\begin{proposition}
The following happen:
\begin{enumerate}
\item The rotation of angle $t\in\mathbb R$ is given by the following matrix:
$$R_t=\begin{pmatrix}\cos t&-\sin t\\ \sin t&\cos t\end{pmatrix}$$

\item The symmetry with respect to the $Ox$ axis rotated by $t/2\in\mathbb R$ is given by:
$$S_t=\begin{pmatrix}\cos t&\sin t\\ \sin t&-\cos t\end{pmatrix}$$

\item The projection on the $Ox$ axis rotated by $t/2\in\mathbb R$ is given by:
$$P_t=\frac{1}{2}\begin{pmatrix}1+\cos t&\sin t\\ \sin t&1-\cos t\end{pmatrix}$$

\item The projection on the all-one vector $\xi\in\mathbb R^N$ is given by:
$$P=\frac{1}{N}
\begin{pmatrix}
1&\ldots&1\\
\vdots&&\vdots\\
1&\ldots&1
\end{pmatrix}$$

\item In fact, the projection on $\mathbb Rx$ is given by $P=||x||^{-2}(x_ix_j)_{ij}$.
\end{enumerate}
\end{proposition}

\begin{proof}
All this is well-known and elementary, the idea being as follows:

\medskip

(1) The rotation being linear, it must correspond to a certain matrix:
$$R_t=\begin{pmatrix}a&b\\ c&d\end{pmatrix}$$

But we can guess this matrix, via its action on the standard coordinate vectors $\binom{1}{0}$ and $\binom{0}{1}$. Indeed, a quick picture shows that we must have:
$$\begin{pmatrix}a&b\\ c&d\end{pmatrix}\begin{pmatrix}1\\ 0\end{pmatrix}=
\begin{pmatrix}\cos t\\ \sin t\end{pmatrix}$$

Also, by paying attention to positives and negatives, we must have:
$$\begin{pmatrix}a&b\\ c&d\end{pmatrix}\begin{pmatrix}0\\ 1\end{pmatrix}=
\begin{pmatrix}-\sin t\\ \cos t\end{pmatrix}$$

Guessing now the matrix is not complicated, because the first equation gives us the first column, and the second equation gives us the second column:
$$\binom{a}{c}=\begin{pmatrix}\cos t\\ \sin t\end{pmatrix}\quad,\quad 
\binom{b}{d}=\begin{pmatrix}-\sin t\\ \cos t\end{pmatrix}$$

Thus, we can just put together these two vectors, and we obtain our matrix.

\medskip

(2) This is again clear on a picture, drawn with $t/2$ instead of $t$, as indicated.

\medskip

(3) Again, picture drawn with $t/2$, as indicated, plus some easy trigonometry.

\medskip

(4) This comes from $Px=A(x)\xi$, with $A(x)$ being the average of the entries of $x$. 

\medskip

(5) Consider a vector $y\in\mathbb R^N$. Its projection $Py$ on the space $\mathbb Rx$ must be a certain multiple of $x$, and we are led in this way to the following formula:
$$Py
=\frac{<y,x>}{<x,x>}\,x
=\frac{1}{||x||^2}<y,x>x$$

With this in hand, we can now compute the entries of $P$, as follows:
\begin{eqnarray*}
P_{ij}
&=&<Pe_j,e_i>\\
&=&\frac{1}{||x||^2}<e_j,x><x,e_i>\\
&=&\frac{x_jx_i}{||x||^2}
\end{eqnarray*}

Thus, we are led to the formula in the statement.
\end{proof}

As another piece of general theory, that you surely know well too, we will need:

\index{eigenvalues}
\index{eigenvectors}
\index{diagonalization}

\begin{definition}
A linear map $f:\mathbb R^N\to\mathbb R^N$ is called diagonalizable if there is a basis  $v_1,\ldots,v_N\in\mathbb R^N$ such that $f$ multiplies by $\lambda_i$ in the direction $v_i$:
$$f(v_i)=\lambda_iv_i$$
In terms of the writing $f(x)=Ax$, we say that the corresponding matrix $A\in M_N(\mathbb R)$ is diagonalizable, with eingenvectors $v_i$ and eigenvalues $\lambda_i$.
\end{definition}

Here the system of vectors $v_1,\ldots,v_N\in\mathbb R^N$ being a basis means as usual that each vector $x\in\mathbb R^N$ must decompose uniquely as $x=\sum_ic_iv_i$, and we can see that the diagonalizability assumption uniquely determines $f$, which follows to be given by:
$$f\left(\sum_ic_iv_i\right)=\sum_i\lambda_ic_iv_i$$

Obviously, being diagonalizable means to be ``good'', and being not diagonalizable means to be ``bad''. In order to understand this, what good and bad mean in linear algebra, let us work out some examples. We have here the following result:

\begin{proposition}
The following happen:
\begin{enumerate}
\item The rotation $R_t$ is not diagonalizable, unless at $t=0$ where it is the identity, $R_0=1$, and at $t=\pi$ where it is minus the identity, $R_\pi=-1$.

\item The symmetry $S_t$ is diagonalizable, with eigenvectors on the symmetry axis, and on its orthogonal, with respective eigenvalues $1,-1$.

\item The projection $P_t$ is diagonalizable, with the eigenvectors exactly as for the symmetry $S_t$, this time with respective eigenvalues $1,0$.

\item In fact, any projection is diagonalizable, with eigenvectors on it image, and on the orthogonal of its image, with respective eigenvalues $1,0$.
\end{enumerate}
\end{proposition}

\begin{proof}
All this is self-explanatory and, we insist, with no need for any computation, and exercise for you, to figure out how all this works, just by drawing pictures, and thinking. Of course, if eager for computations, do not worry, we will have some, later.
\end{proof}

Still in relation with diagonalization, at the general level, we have:

\index{diagonal form}
\index{passage matrix}

\begin{theorem}
Assuming that a matrix $A\in M_N(\mathbb R)$ is diagonalizable, with eigenvectors $v_1,\ldots,v_N$ and corresponding eigenvalues $\lambda_1,\ldots,\lambda_N$, we have
$$A=PDP^{-1}$$
with the matrices $P,D\in M_N(\mathbb R)$ being given by the formulae
$$P=[v_1,\ldots,v_N]\quad,\quad 
D=diag(\lambda_1,\ldots,\lambda_N)$$
and respectively called passage matrix, and diagonal form of $A$.
\end{theorem}

\begin{proof}
We have $Pe_i=v_i$, where $\{e_i\}$ is the standard basis of $\mathbb R^N$, and so:
$$APe_i
=Av_i
=\lambda_iv_i$$

On the other hand, once again by using $Pe_i=v_i$, we have as well:
$$PDe_i
=P\lambda_ie_i
=\lambda_iPe_i
=\lambda_iv_i$$

Thus we have $AP=PD$, and so $A=PDP^{-1}$, as claimed.
\end{proof}

As an illustration for this, you can have some fun with the various matrices from Proposition 1.7, with in each case the corresponding diagonalization formula $A=PDP^{-1}$ coming without much pain, because Proposition 1.9 tells us what both $P,D$ are, in each case, and the only piece of work remaining is that of figuring out what $P^{-1}$ is. Enjoy.

\bigskip

Summarizing, the diagonalizable matrices are the ``good'' ones, and their diagonalization is quite often a matter of doing some geometry. Regarding the non-diagonalizable matrices, these actually fall into two classes, ``bad'' and ``evil''. The bad ones are those which diagonalize over $\mathbb C$, with a main example here being the rotation $R_t$, and more on this later. As for the evil ones, these are evil, a basic example being as follows:

\index{non-diagonalizable}
\index{Jordan block}

\begin{theorem}
The following matrix is not diagonalizable,
$$J=\begin{pmatrix}0&1\\0&0\end{pmatrix}$$
because it has only $1$ eigenvector.
\end{theorem}

\begin{proof}
The above matrix, called $J$ en hommage to Jordan, acts as follows:
$$\begin{pmatrix}0&1\\0&0\end{pmatrix}\binom{x}{y}=\binom{y}{0}$$

Thus the eigenvector/eigenvalue equation $Jv=\lambda v$ reads:
$$\binom{y}{0}=\binom{\lambda x}{\lambda y}$$

We have then two cases, depending on $\lambda$, as follows, which give the result:

\medskip

(1) For $\lambda\neq0$ we must have $y=0$, coming from the second row, and so $x=0$ as well, coming from the first row, so we have no nontrivial eigenvectors. 

\medskip

(2) As for the case $\lambda=0$, here we must have $y=0$, coming from the first row, and so the eigenvectors here are the vectors of the form $\binom{x}{0}$. 
\end{proof}

So long for the foundations of vector calculus and linear algebra. We will be back to all this, and especially to the diagonalization problem, which is something quite subtle, on several occasions, in what follows, and notably in chapter 2 below.

\section*{1c. Geometry, analysis}

Let us see now what we can do with our linear maps. Perhaps the simplest application, which is something of key importance for both mathematics and physics, is the classification of conics. These are the algebraic curves of degree $2$ in the plane:
$$C=\left\{\binom{x}{y}\in\mathbb R^2\Big|P(x,y)=0\right\}\quad,\quad\deg P\leq 2$$

You certainly know from physics that conics appear from gravity, and more specifically, describe the trajectory of one object with respect to another. For instance, in our Solar system, all planets move on ellipses, which are of course conics, around the Sun.

\bigskip

But, are there any other possible trajectories, besides ellipses? And here, with physics and astronomy things are a bit complicated, because with the planets ruled out, we must carefully observe all sorts of small and capricious objects, such as comets, and asteroids. Fortunately math, and linear algebra, come to the rescue, and we first have:

\index{conic}
\index{ellipsis}
\index{parabola}
\index{hyperbola}
\index{degenerate conic}
\index{gravity}

\begin{proposition}
Up to non-degenerate linear transformations of the plane, which are by definition transformations as follows, assumed to be invertible,
$$\binom{x}{y}\to A\binom{x}{y}$$
the conics are the circles, parabolas, hyperbolas, along with some degenerate solutions, namely $\emptyset$, points, lines, pairs of lines, $\mathbb R^2$.
\end{proposition}

\begin{proof}
This is something very classical, the idea being as follows:

\medskip

(1) As a first remark, it looks like we forgot the ellipses, but via linear transformations these become circles, so things fine. As a second remark, all our claimed solutions can appear. Indeed, the circles, parabolas, hyperbolas can appear as follows:
$$x^2+y^2=1\quad,\quad x^2=y\quad,\quad xy=1$$

As for $\emptyset$, points, lines, pairs of lines, $\mathbb R^2$, these can appear too, as follows, and with our polynomial $P$ chosen, whenever possible, to be of degree exactly 2:
$$x^2=-1\quad,\quad x^2+y^2=0\quad,\quad x^2=0\quad,\quad xy=0\quad,\quad 0=0$$

Observe here that, when dealing with these degenerate cases, assuming $\deg P=2$ instead of $\deg P\leq 2$ would only rule out $\mathbb R^2$ itself, which is not worth it. 

\medskip

(2) Getting now to the proof of our result, classification up to linear transformations, consider an arbitrary conic, written as follows, with $a,b,c,d,e,f\in\mathbb R$:
$$ax^2+by^2+cxy+dx+ey+f=0$$

Assume first $a\neq0$. By making a square out of $ax^2$, up to a linear transformation in $(x,y)$, we can get rid of the term $cxy$, and we are left with:
$$ax^2+by^2+dx+ey+f=0$$

In the case $b\neq0$ we can make two obvious squares, and again up to a linear transformation in $(x,y)$, we are left with an equation as follows:
$$x^2\pm y^2=k$$

In the case of positive sign, $x^2+y^2=k$, the solutions are the circle, when $k\geq0$, the point, when $k=0$, and $\emptyset$, when $k<0$. As for the case of negative sign, $x^2-y^2=k$, which reads $(x-y)(x+y)=k$, here once again by linearity our equation becomes $xy=l$, which is a hyperbola when $l\neq0$, and two lines when $l=0$.

\medskip

(3) In the case $b\neq0$ the study is similar, with the same solutions, so we are left with the case $a=b=0$. Here our conic is as follows, with $c,d,e,f\in\mathbb R$:
$$cxy+dx+ey+f=0$$

If $c\neq 0$, by linearity our equation becomes $xy=l$, which produces a hyperbola or two lines, as explained before. As for the remaining case, $c=0$, here our equation is:
$$dx+ey+f=0$$

But this is generically the equation of a line, unless we are in the case $d=e=0$, where our equation is $f=0$, having as solutions $\emptyset$ when $f\neq0$, and $\mathbb R^2$ when $f=0$.
\end{proof}

As a continuation of the above, we can now formulate a final result, as follows:

\index{conic}
\index{degenerate conic}
\index{non-degenerate conic}

\begin{theorem}
The conics, which are the algebraic curves of degree $2$ in the plane,
$$C=\left\{\binom{x}{y}\in\mathbb R^2\Big|P(x,y)=0\right\}$$
with $\deg P\leq 2$, are up to degeneration the ellipses, parabolas and hyperbolas.
\end{theorem}

\begin{proof}
We already have the classification up to linear transformations, and the point is that this classification leads to the classification in general too, by applying linear transformations to the solutions that we found, with the conclusions in the statement.
\end{proof}

Getting back now to physics, our result predicts that there are certain objects in our Solar system moving around the Sun on parabolas, or hyperbolas. And this is indeed true, with certain asteroids doing so, and with the technical remark that these asteroids do not belong in fact to our Solar system, precisely because they are able to escape from the gravitational attraction of the Sun, on parabolic or hyperbolic trajectories.

\bigskip

As another physical remark, there is an interesting discussion to be made here, in relation with degeneracy, because the degenerate conics appearing from mathematics do not exactly coincide with the degenerate conics appearing from physics. For instance in mathematics we have $\emptyset$, the lines, the pairs of lines, and $\mathbb R^2$, which cannot appear as gravitational trajectories, while in physics we have the segment, which is the trajectory of a centered free fall, which is obviously not a conic, in a mathematical sense.

\bigskip

This sounds quite interesting, and as a homework for you, reader, we have:

\index{mathematical conic}
\index{physical conic}
\index{foundations}

\begin{homework}
Fix the foundations of mathematics and physics, as for the mathematical conics to coincide with the physical conics, in the degenerate cases too.
\end{homework}

So long for applications of linear algebra to basic geometry and physics. We can of course use similar methods for many other geometric problems, and we will be back to this, later in this book, when discussing more in detail manifolds and geometry.

\bigskip

Switching topics now, no discussion about matrices and linear algebra would be complete without a word on multivariable calculus, and all the matrices appearing there. In fact, and you might already know this, this is more or less how matrices and linear algebra appeared, in our human mathematics, according to the following scheme:
$${\rm physics}\implies{\rm calculus}\implies{\rm matrices}$$

But probably enough talking, let us get to work, study the functions of several variables, and see where this study gets us into. As a first result here, which is something fundamental, getting us precisely into matrices and linear algebra, we have:

\index{derivative}
\index{partial derivative}

\begin{theorem}
A function $f:\mathbb R^N\to\mathbb R^M$ is continuously differentiable,
$$f(x+t)\simeq f(x)+f'(x)t$$
with $f'(x)$ linear, and $x\to f'(x)$ continuous, precisely when it has partial derivatives,
$$\frac{df_i}{dx_j}(x)=\lim_{t\to 0}\frac{f_i(x+te_j)-f_i(x)}{t}$$
which depend continuously on $x$. In this case the derivative is
$$f'(x)=\left(\frac{df_i}{dx_j}(x)\right)_{ij}\in M_{M\times N}(\mathbb R)$$ 
acting on the vectors $t\in\mathbb R^N$ by usual multiplication.
\end{theorem}

\begin{proof}
As a first observation, the formula in the statement makes sense indeed, as an equality, or rather approximation, of vectors in $\mathbb R^M$, as follows:
$$f\begin{pmatrix}x_1+t_1\\ \vdots\\ x_N+t_N\end{pmatrix}
\simeq f\begin{pmatrix}x_1\\ \vdots\\ x_N\end{pmatrix}
+\begin{pmatrix}
\frac{df_1}{dx_1}(x)&\ldots&\frac{df_1}{dx_N}(x)\\
\vdots&&\vdots\\
\frac{df_M}{dx_1}(x)&\ldots&\frac{df_M}{dx_N}(x)
\end{pmatrix}\begin{pmatrix}t_1\\ \vdots\\ t_N\end{pmatrix}$$

In order to prove now this formula, we can proceed by recurrence, as follows:

\medskip

(1) First of all, at $N=M=1$ what we have is a usual 1-variable function $f:\mathbb R\to\mathbb R$, and the formula in the statement is something that we know well, namely:
$$f(x+t)\simeq f(x)+f'(x)t$$

(2) Next, at $N=2,M=1$ the result holds as well, by using (1) twice, as follows:
\begin{eqnarray*}
f\binom{x_1+t_1}{x_2+t_2}
&\simeq&f\binom{x_1+t_1}{x_2}+\frac{df}{dx_2}(x)t_2\\
&\simeq&f\binom{x_1}{x_2}+\frac{df}{dx_1}(x)t_1+\frac{df}{dx_2}(x)t_2\\
&=&f\binom{x_1}{x_2}+\begin{pmatrix}\frac{df}{dx_1}(x)&\frac{df}{dx_2}(x)\end{pmatrix}\binom{t_1}{t_2}
\end{eqnarray*}

(3) More generally, we can deal in this way with the general case $M=1$, with the formula here, obtained via a straightforward recurrence, being as follows:
\begin{eqnarray*}
f\begin{pmatrix}x_1+t_1\\ \vdots\\ x_N+t_N\end{pmatrix}
&\simeq&f\begin{pmatrix}x_1\\ \vdots\\ x_N\end{pmatrix}+\frac{df}{dx_1}(x)t_1+\ldots+\frac{df}{dx_N}(x)t_N\\
&=&f\begin{pmatrix}x_1\\ \vdots\\ x_N\end{pmatrix}+
\begin{pmatrix}\frac{df}{dx_1}(x)&\ldots&\frac{df}{dx_N}(x)\end{pmatrix}
\begin{pmatrix}t_1\\ \vdots\\ t_N\end{pmatrix}
\end{eqnarray*}

(4) But this gives the result in the case where both $N,M\in\mathbb N$ are arbitrary too. Indeed, consider a function $f:\mathbb R^N\to\mathbb R^M$, and let us write it as follows:
$$f=\begin{pmatrix}f_1\\ \vdots\\ f_M\end{pmatrix}$$

We can apply (3) to each of the components $f_i:\mathbb R^N\to\mathbb R$, and we get:
$$f_i\begin{pmatrix}x_1+t_1\\ \vdots\\ x_N+t_N\end{pmatrix}
\simeq f_i\begin{pmatrix}x_1\\ \vdots\\ x_N\end{pmatrix}+
\begin{pmatrix}\frac{df_i}{dx_1}(x)&\ldots&\frac{df_i}{dx_N}(x)\end{pmatrix}
\begin{pmatrix}t_1\\ \vdots\\ t_N\end{pmatrix}$$

(5) But this collection of $M$ formulae tells us precisely that the following happens, as an equality, or rather approximation, of vectors in $\mathbb R^M$:
$$f\begin{pmatrix}x_1+t_1\\ \vdots\\ x_N+t_N\end{pmatrix}
\simeq f\begin{pmatrix}x_1\\ \vdots\\ x_N\end{pmatrix}
+\begin{pmatrix}
\frac{df_1}{dx_1}(x)&\ldots&\frac{df_1}{dx_N}(x)\\
\vdots&&\vdots\\
\frac{df_M}{dx_1}(x)&\ldots&\frac{df_M}{dx_N}(x)
\end{pmatrix}\begin{pmatrix}t_1\\ \vdots\\ t_N\end{pmatrix}$$

Thus, we are led to the conclusion in the statement.
\end{proof}

Generally speaking, Theorem 1.15 is all you need to know, for extending to several variables the basic results from one-variable calculus. As a basic illustration, we have:

\index{chain rule}

\begin{theorem}
We have the chain derivative formula
$$(f\circ g)'(x)=f'(g(x))\cdot g'(x)$$
as an equality of matrices.
\end{theorem}

\begin{proof}
This is something standard in one variable, and in several variables the proof is similar, by using the abstract notion of derivative coming from Theorem 1.15. To be more precise, consider a composition of functions, as follows:
$$f:\mathbb R^N\to\mathbb R^M\quad,\quad 
g:\mathbb R^K\to\mathbb R^N\quad,\quad 
f\circ g:\mathbb R^K\to\mathbb R^M$$

According to Theorem 1.15, the derivatives of these functions are certain linear maps, corresponding to certain rectangular matrices, as follows:
$$f'(g(x))\in M_{M\times N}(\mathbb R)\quad,\quad 
g'(x)\in M_{N\times K}(\mathbb R)\quad\quad
(f\circ g)'(x)\in M_{M\times K}(\mathbb R)$$

Thus, our formula makes sense indeed. As for proof, this comes from:
\begin{eqnarray*}
(f\circ g)(x+t)
&=&f(g(x+t))\\
&\simeq&f(g(x)+g'(x)t)\\
&\simeq&f(g(x))+f'(g(x))g'(x)t
\end{eqnarray*}

Thus, we are led to the conclusion in the statement.
\end{proof}

Moving on, as further good news, for us linear algebraists, we have as well an important square matrix, which appears for the scalar functions, at order 2, as follows:

\index{second derivative}
\index{Taylor formula}
\index{Hessian matrix}

\begin{theorem}
Given a twice differentiable function $f:\mathbb R^N\to\mathbb R$, we have
$$f(x+t)\simeq f(x)+f'(x)t+\frac{<f''(x)t,t>}{2}$$
with $f'(x)\in M_{1\times N}(\mathbb R)$ being a row vector, and with $f''(x)\in M_N(\mathbb R)$, given by 
$$f''(x)=\left(\frac{d^2f}{dx_idx_j}\right)_{ij}(x)$$
being the Hessian matrix of $f$, at the point $x\in\mathbb R^N$.
\end{theorem}

\begin{proof}
This is something quite tricky, the idea being as follows:

\medskip

(1) As a first remark, at $N=1$ the Hessian matrix is the $1\times1$ matrix having as entry the usual second derivative $f''(x)\in\mathbb R$, and the formula in the statement is something that we know well from one-variable calculus, namely the Taylor formula at order 2:
$$f(x+t)\simeq f(x)+f'(x)t+\frac{f''(x)t^2}{2}$$

(2) In general now, this is in fact something which does not need a whole new proof, because it follows from the one-variable formula above, applied to the restriction of $f$ to the following segment in $\mathbb R^N$, which can be regarded as being a one-variable interval:
$$I=[x,x+t]$$

To be more precise, let $y\in\mathbb R^N$, and consider the following function, with $r\in\mathbb R$:
$$g(r)=f(x+ry)$$

We know from (1) that the Taylor formula for $g$, at the point $r=0$, reads:
$$g(r)\simeq g(0)+g'(0)r+\frac{g''(0)r^2}{2}$$

And our claim is that, with $t=ry$, this is precisely the formula in the statement.

\medskip

(3) So, let us see if our claim is correct. By using the chain rule, we have the following formula, with on the right, as usual, a row vector multiplied by a column vector:
$$g'(r)=f'(x+ry)\cdot y$$

By using again the chain rule, we can compute the second derivative as well:
\begin{eqnarray*}
g''(r)
&=&(f'(x+ry)\cdot y)'\\
&=&\left(\sum_i\frac{df}{dx_i}(x+ry)\cdot y_i\right)'\\
&=&\sum_i\sum_j\frac{d^2f}{dx_idx_j}(x+ry)\cdot\frac{d(x+ry)_j}{dr}\cdot y_i\\
&=&\sum_i\sum_j\frac{d^2f}{dx_idx_j}(x+ry)\cdot y_iy_j\\
&=&<f''(x+ry)y,y>
\end{eqnarray*}

(4) Time now to conclude. We know that we have $g(r)=f(x+ry)$, and according to our various computations above, we have the following formulae:
$$g(0)=f(x)\quad,\quad 
g'(0)=f'(x)\quad,\quad 
g''(0)=<f''(x)y,y>$$

Buit with this data in hand, the usual Taylor formula for our one variable function $g$, at order 2, at the point $r=0$, takes the following form, with $t=ry$:
\begin{eqnarray*}
f(x+ry)
&\simeq&f(x)+f'(x)ry+\frac{<f''(x)y,y>r^2}{2}\\
&=&f(x)+f'(x)t+\frac{<f''(x)t,t>}{2}
\end{eqnarray*}

Thus, we have obtained the formula in the statement.
\end{proof}

As a conclusion to all this, geometry and physics and analysis are all about matrices and linear algebra, or perhaps vice versa, and with the above being most likely just the tip of the iceberg. Which is good to know, and this will be our philosophy in what follows, develop the theory of matrices and linear algebra, as to get to know about the iceberg.

\section*{1d. Complex numbers} 

Let us discuss now what happens over the complex numbers. You certainly know about these numbers, and hopefully even love them, as any mathematician or physicist should do. Before that, however, a bit of general theory. Let us start with:

\index{field}
\index{field axioms}

\begin{definition}
A field is a set $F$ with a sum operation $+$ and a product operation $\times$, subject to the following conditions:
\begin{enumerate}
\item $a+b=b+a$, $a+(b+c)=(a+b)+c$, there exists $0\in F$ such that $a+0=0$, and any $a\in F$ has an inverse $-a\in F$, satisfying $a+(-a)=0$.

\item $ab=ba$, $a(bc)=(ab)c$, there exists $1\in F$ such that $a1=a$, and any $a\neq0$ has a multiplicative inverse $a^{-1}\in F$, satisfying $aa^{-1}=1$.

\item The sum and product are compatible via $a(b+c)=ab+ac$.
\end{enumerate}
\end{definition}

In other words, a field satisfies what we can normally expect from ``numbers'', and as basic examples, we have of course $\mathbb Q$ and $\mathbb R$. There are many other interesting examples of fields, quite often coming from arithmetic, and more on this, later in this book.

\bigskip

Getting now to the complex numbers, their basic theory can be summarized as follows:

\index{complex number}
\index{root of unity}
\index{degree 2 equation}

\begin{theorem}
The complex numbers, $z=a+ib$ with $a,b\in\mathbb R$ and with $i$ being a formal number satisying $i^2=-1$, form a field $\mathbb C$. Moreover:
\begin{enumerate}
\item We have a field embedding $\mathbb R\subset\mathbb C$, given by $a\to a+0\cdot i$.

\item Additively, we have $\mathbb C\simeq\mathbb R^2$, with $z=a+ib$ corresponding to $(a,b)$.

\item The length of vectors $r=|z|$, with $z=a+ib$, is given by $r=\sqrt{a^2+b^2}$.

\item With $z=r(\cos t+i\sin t)$, the products $z=z'z''$ are given by $r=r'r''$, $t=t'+t''$.

\item We have the formula $e^{it}=\cos t+i\sin t$, so we can write $z=re^{it}$.

\item There are $N$ solutions to the equation $z^N=1$, called $N$-th roots of unity.

\item Any degree $2$ equation with complex coefficients has both roots in $\mathbb C$.
\end{enumerate}
\end{theorem}

\begin{proof}
We have indeed a field, with the inversion formula being as follows:
$$(a+ib)^{-1}=\frac{a-ib}{a^2+b^2}$$

As for the rest, this is a mixture of trivial and non-trivial results, as follows:

\medskip

(1-3) These assertions are clear. Observe also that we have $r^2=z\bar{z}$, with $\bar{z}=a-ib$.

\medskip

(4) This is something more subtle. To start with, by dealing with triangles in the plane, and computing lengths and areas, and in the hope that you still remember that, we are led to the following remarkable formulae, for the sines and cosines of sums:
$$\sin(s+t)=\sin s\cos t+\cos s\sin t$$
$$\cos(s+t)=\cos s\cos t-\sin s\sin t$$

But, with these formulae in hand, we have the following computation, as desired:
\begin{eqnarray*}
&&(\cos s+i\sin s)(\cos t+i\sin t)\\
&=&(\cos s\cos t+i^2\sin s\sin t)+i(\sin s\cos t+\cos s\sin t)\\
&=&(\cos s\cos t-\sin s\sin t)+i(\sin s\cos t+\cos s\sin t)\\
&=&\cos(s+t)+i\sin(s+t)
\end{eqnarray*}

(5) Again, this is something subtle. Our first claim is that we can exponentiate the arbitrary complex numbers $z\in\mathbb C$, according to the following formula:
$$e^z=\sum_{k=0}^\infty\frac{z^k}{k!}$$

Indeed, the series converges indeed, as shown by the following computation:
$$|e^z|
=\left|\sum_{k=0}^\infty\frac{z^k}{k!}\right|
\leq\sum_{k=0}^\infty\left|\frac{z^k}{k!}\right|
=\sum_{k=0}^\infty\frac{|z|^k}{k!}
=e^{|z|}<\infty$$

Moreover, $z\to e^z$ is continuous, and $e^{z+z'}=e^ze^{z'}$. Next, observe that we have:
$$e^{\bar{z}}=\sum_{k=0}^\infty\frac{\bar{z}^k}{k!}=\overline{\sum_{k=0}^\infty\frac{z^k}{k!}}=\overline{e^z}$$

Also, we have as well the following computation, again valid for any $z\in\mathbb C$:
$$e^ze^{-z}=e^{z-z}=e^0=1\implies (e^z)^{-1}=e^{-z}$$

Now by using the above two formulae, applied with $z=it$, with $t\in\mathbb R$, we get:
$$e^{-it}=\overline{e^{it}}\quad,\quad (e^{it})^{-1}=e^{-it}$$

Thus $z=e^{it}$ satisfies $z^{-1}=\bar{z}$, and so belongs to the unit circle $\mathbb T\subset\mathbb C$. Moreover, we know that $t\to e^{it}$ is a group morphism, meaning that it maps sums in $\mathbb R$ into products in $\mathbb T$. But in view of this, barring any pathologies, this operation can only appear by ``wrapping". That is, we must have a formula as follows, for a certain $\alpha\in\mathbb R$:
$$e^{it}=\cos(\alpha t)+i\sin(\alpha t)$$

In order now to find the parameter $\alpha\in\mathbb R$, let us look at what happens around $t=0$. And here, we have the following basic estimate, obtained by truncating $\exp$:
$$e^{it}\simeq 1+it$$

On the other hand, according to the basic trigonometry estimates for $\sin$ and $\cos$, from plane geometry, we have as well the following estimate, again around $t=0$:
$$\cos(\alpha t)+i\sin(\alpha t)\simeq 1+i\alpha t$$

Thus $\alpha=1$, as desired. Alternatively, and perhaps a bit more rigorously, in order to prove $e^{it}=\cos t+i\sin t$, we can use the following function $f:\mathbb R\to\mathbb C$:
$$f(t)=\frac{\cos t+i\sin t}{e^{it}}$$

Indeed, by using $\sin'=\cos$, $\cos'=-\sin$, coming from the formulae in (4), we have:
\begin{eqnarray*}
f'(t)
&=&(e^{-it}(\cos t+i\sin t))'\\
&=&-ie^{-it}(\cos t+i\sin t)+e^{-it}(-\sin t+i\cos t)\\
&=&e^{-it}(-i\cos t+\sin t)+e^{-it}(-\sin t+i\cos t)\\
&=&0
\end{eqnarray*}

We conclude that $f:\mathbb R\to\mathbb C$ is constant, equal to $f(0)=1$, as desired.

\medskip

(6-7) These assertions both follow from (5), with $z=w^k$, with $w=e^{2\pi i/N}$ and $k=0,1,\ldots,N-1$ for (6), and with $\sqrt{re^{it}}=\pm\sqrt{r}e^{it/2}$ needed for (7). 
\end{proof}

Many interesting things can be done with the complex numbers. As a first magic result, going well beyond what we can do with the real numbers, we have:

\index{diagonalization of rotation}
\index{complex eigenvalues}

\begin{theorem}
The rotation of angle $t\in\mathbb R$ in the plane diagonalizes as:
$$R_t=\frac{1}{2}\begin{pmatrix}1&1\\i&-i\end{pmatrix}
\begin{pmatrix}e^{-it}&0\\0&e^{it}\end{pmatrix}
\begin{pmatrix}1&-i\\1&i\end{pmatrix}$$
Over the real numbers this is impossible, unless $t=0,\pi$.
\end{theorem}

\begin{proof}
The last assertion is something clear, that we already know, coming from the fact that at $t\neq0,\pi$ our rotation is a ``true'' rotation, having no eigenvectors in the plane. Regarding the first assertion, the point is that we have the following computation:
$$R_t\binom{1}{i}
=\begin{pmatrix}\cos t&-\sin t\\ \sin t&\cos t\end{pmatrix}\binom{1}{i}
=\binom{\cos t-i\sin t}{i\cos t+\sin t}
=e^{-it}\binom{1}{i}$$

We have as well a second eigenvector, as follows:
$$R_t\binom{1}{-i}
=\begin{pmatrix}\cos t&-\sin t\\ \sin t&\cos t\end{pmatrix}\binom{1}{-i}
=\binom{\cos t+i\sin t}{-i\cos t+\sin t}
=e^{it}\binom{1}{-i}$$

Thus our rotation matrix $R_t$ is indeed diagonalizable over $\mathbb C$, with the passage matrix and diagonal form being, according to the above formulae, as follows:
$$P=\begin{pmatrix}1&1\\i&-i\end{pmatrix}\quad,\quad 
D=\begin{pmatrix}e^{-it}&0\\0&e^{it}\end{pmatrix}$$

Now by inverting $P$, we are led to the conclusion in the statement.
\end{proof}

Another thing that we can do with complex numbers is to nicely diagonalize the all-one matrix, that we met in Proposition 1.7. Indeed, over the reals this matrix is certainly diagonalizable, but not in a nice way, due to troubles in finding ``canonical'' solutions of the following equation, which is the eigenvector equation for $\lambda=0$:
$$x_1+\ldots+x_N=0$$

In the complex setting, however, the roots of unity come to the rescue, via:

\index{regular polygon}
\index{sum of roots}
\index{roots of unity}
\index{barycenter}

\begin{proposition}
The roots of unity, $\{w^k\}$ with $w=e^{2\pi i/N}$, have the property
$$\sum_{k=0}^{N-1}(w^k)^s=N\delta_{N|s}$$
for any exponent $s\in\mathbb N$, where on the right we have a Kronecker symbol.
\end{proposition}

\begin{proof}
The numbers in the statement, when written more conveniently as $(w^s)^k$ with $k=0,\ldots,N-1$, form a certain regular polygon in the plane $P_s$. Thus, if we denote by $C_s$ the barycenter of this polygon, we have the following formula:
$$\frac{1}{N}\sum_{k=0}^{N-1}w^{ks}=C_s$$

Now observe that in the case $N\slash\hskip-1.6mm|\,s$ our polygon $P_s$ is non-degenerate, circling around the unit circle, and having center $C_s=0$. As for the case $N|s$, here the polygon is degenerate, lying at 1, and having center $C_s=1$. Thus, we have the following formula:
$$C_s=\delta_{N|s}$$

Thus, we obtain the formula in the statement. Alternatively, the formula in the statement follows of course too by algebrically summing the sum there.
\end{proof}

We have the following definition, inspired by what happens in Proposition 1.21:

\index{Fourier matrix}

\begin{definition}
The Fourier matrix $F_N$ is the following matrix, with $w=e^{2\pi i/N}$:
$$F_N=
\begin{pmatrix}
1&1&1&\ldots&1\\
1&w&w^2&\ldots&w^{N-1}\\
1&w^2&w^4&\ldots&w^{2(N-1)}\\
\vdots&\vdots&\vdots&&\vdots\\
1&w^{N-1}&w^{2(N-1)}&\ldots&w^{(N-1)^2}
\end{pmatrix}$$
That is, $F_N=(w^{ij})_{ij}$, with indices $i,j\in\{0,1,\ldots,N-1\}$, taken modulo $N$.
\end{definition}

Here the terminology comes from the fact that $F_N$ is the matrix of the Fourier transform over the cyclic group $\mathbb Z_N$, and more on this later in this book, when systematically discussing the discrete Fourier transform, in its various versions.

\bigskip

As a first example, at $N=2$ the root of unity is $w=-1$, and with indices as above, namely $i,j\in\{0,1\}$, taken modulo 2, our Fourier matrix is as follows:
$$F_2=\begin{pmatrix}1&1\\1&-1\end{pmatrix}$$

At $N=3$ now, the root of unity is $w=e^{2\pi i/3}$, and the Fourier matrix is:
$$F_3=\begin{pmatrix}1&1&1\\ 1&w&w^2\\ 1&w^2&w\end{pmatrix}$$

At $N=4$ now, the root of unit is $w=i$, and the Fourier matrix is:
$$F_4=\begin{pmatrix}
1&1&1&1\\
1&i&-1&-i\\
1&-1&1&-1\\
1&-i&-1&i
\end{pmatrix}$$

And so on, you get the point, with how this matrix works. Getting back now to the diagonalization problem for the all-one matrix, this can be solved as follows:

\index{flat matrix}
\index{Fourier matrix}

\begin{theorem}
The all-one matrix diagonalizes as follows,
$$\begin{pmatrix}
1&\ldots&\ldots&1\\
\vdots&&&\vdots\\
\vdots&&&\vdots\\
1&\ldots&\ldots&1\end{pmatrix}
=F_N
\begin{pmatrix}
N&&&0\\
&0\\
&&\ddots\\
0&&&0\end{pmatrix}F_N^{-1}$$
with $F_N=(w^{ij})_{ij}$ being the Fourier matrix, and $F_N^{-1}=\frac{1}{N}(w^{-ij})_{ij}$ being its inverse.
\end{theorem}

\begin{proof}
We know that the all-one matrix is $N$ times the projection on the all-one vector, so we are left with finding the 0-eigenvectors, which amounts in solving:
$$x_0+\ldots+x_{N-1}=0$$

But for this purpose, we can use the root of unity $w=e^{2\pi i/N}$, and more specifically, the following standard formula, coming from Proposition 1.21:
$$\sum_{i=0}^{N-1}w^{ij}=N\delta_{j0}$$

Indeed, this formula shows that for $j=1,\ldots,N-1$, the vector $v_j=(w^{ij})_i$ is a 0-eigenvector. Moreover, these vectors are pairwise orthogonal, because we have:
$$<v_j,v_k>
=\sum_iw^{ij-ik}
=N\delta_{jk}$$

Thus, we have our basis $\{v_1,\ldots,v_{N-1}\}$ of 0-eigenvectors, and since the $N$-eigenvector is $\xi=v_0$, the passage matrix $P$ that we are looking is given by:
$$P=\begin{bmatrix}v_0&v_1&\ldots&v_{N-1}\end{bmatrix}$$

But this is precisely the Fourier matrix, $P=F_N$. Finally, the formula of the inverse, namely $F_N^{-1}=\frac{1}{N}(w^{-ij})_{ij}$, comes once again by using Proposition 1.21.
\end{proof}

Many other things can be done with complex numbers in linear algebra. We will be back to this, on numerous occasions, in the remainder of this book.

\section*{1e. Exercises}

This was an elementary chapter, for the most concerned with things that you are supposed to know, and as exercises, for making sure that you know indeed, we have:

\begin{exercise}
Check the formulae given above of the plane rotations $R_t$, plane symmetries $S_t$, and plane projections $P_t$.
\end{exercise}

\begin{exercise}
Diagonalize the plane symmetries $S_t$ and the plane projections $P_t$, geometrically, without any algebraic computations.
\end{exercise}

\begin{exercise}
Prove that the conics are, modulo some degenerate cases, exactly the curves which appear by cutting a two-sided cone with a plane.
\end{exercise}

\begin{exercise}
Compute a few first derivatives, and a few Hessian matrices too, for some multivariable functions of your choice.
\end{exercise}

\begin{exercise}
Learn a bit about finite fields, about the Fermat polynomial, about splitting fields for polynomials, and about the fields $\mathbb F_q$.
\end{exercise}

\begin{exercise}
Learn also about the $p$-adic numbers, including full details on their construction, and what can be done with them.
\end{exercise}

\begin{exercise}
Further meditate on the complex numbers, on the various ways of introducing and denoting them, and on the formula $e^{it}=\cos t+i\sin t$.
\end{exercise}

\begin{exercise}
Learn more about the Fourier matrices and their properties, and learn as well about the complex Hadamard matrices, which generalize them.
\end{exercise}

As bonus exercise, of genuine linear algebra type, find some other matrices that you can diagonalize geometrically, besides the $2\times 2$ matrices $R_t,S_t,P_t$ discussed above.

\chapter{Matrix theory}

\section*{2a. Matrix inversion}

We have seen that most of the interesting maps $f:\mathbb R^N\to\mathbb R^N$ that we know, such as the rotations, symmetries and projections, are linear, and can be written as $f(v)=Av$ with $A\in M_N(\mathbb R)$ being a square matrix. We would like to develop now some general theory for such linear maps. As a first result here, which is motivational, we have:

\index{linear equations}
\index{matrix inversion}

\begin{theorem}
Any linear system of equations 
$$\begin{cases}
a_{11}x_1+a_{12}x_2+\ldots+a_{1N}x_N\!\!\!&=\ v_1\\
a_{21}x_1+a_{22}x_2+\ldots+a_{2N}x_N\!\!\!&=\ v_2\\
\ \ \vdots\\
a_{N1}x_1+a_{N2}x_2+\ldots+a_{NN}x_N\!\!\!&=\ v_N
\end{cases}$$
can be written as $Ax=v$, and when $A$ is invertible, its solution is $x=A^{-1}v$.
\end{theorem}

\begin{proof}
With linear algebra conventions, our system reads:
$$\begin{pmatrix}
a_{11}&a_{12}&\ldots&a_{1N}\\
a_{21}&a_{22}&\ldots&a_{2N}\\
\vdots&&&\vdots\\
a_{N1}&a_{N2}&\ldots&a_{NN}
\end{pmatrix}
\begin{pmatrix}
x_1\\
x_2\\
\vdots\\
x_N
\end{pmatrix}
=\begin{pmatrix}
v_1\\
v_2\\
\vdots\\
v_N
\end{pmatrix}$$

Thus, we are led to the conclusions in the statement.
\end{proof}

In practice, we are led to the question of inverting the matrices $A\in M_N(\mathbb R)$. And this is the same question as inverting the linear maps $f:\mathbb R^N\to\mathbb R^N$, due to:

\index{invertible matrix}
\index{invertible linear map}

\begin{theorem}
A linear map $f:\mathbb R^N\to\mathbb R^N$, written as
$$f(v)=Av$$
is invertible precisely when $A$ is invertible, and in this case we have $f^{-1}(v)=A^{-1}v$.
\end{theorem}

\begin{proof}
This is something that we basically know, coming from the fact that, with the notation $f_A(v)=Av$, we have the following formula:
$$f_Af_B=f_{AB}$$

Thus, we are led to the conclusion in the statement.
\end{proof}

In order to study now invertibility questions, for matrices or linear maps, let us begin with some examples. In the simplest case, in 2 dimensions, the result is as follows:

\index{inversion formula}

\begin{theorem}
We have the following inversion formula, for the $2\times2$ matrices:
$$\begin{pmatrix}a&b\\ c&d\end{pmatrix}^{-1}
=\frac{1}{ad-bc}\begin{pmatrix}d&-b\\ -c&a\end{pmatrix}$$
When $ad-bc=0$, the matrix is not invertible.
\end{theorem}

\begin{proof}
We have two assertions to be proved, the idea being as follows:

\medskip

(1) As a first observation, when $ad-bc=0$ we must have, for some $\lambda\in\mathbb R$:
$$b=\lambda a\quad,\quad 
d=\lambda c$$

Thus our matrix must be of the following special type:
$$\begin{pmatrix}a&b\\ c&d\end{pmatrix}=\begin{pmatrix}a&\lambda a\\ a&\lambda c\end{pmatrix}$$

But in this case the columns are proportional, so the linear map associated to the matrix is not invertible, and so the matrix itself is not invertible either.

\medskip

(2) When $ad-bc\neq 0$, let us look for an inversion formula of the following type:
$$\begin{pmatrix}a&b\\ c&d\end{pmatrix}^{-1}
=\frac{1}{ad-bc}\begin{pmatrix}*&*\\ *&*\end{pmatrix}$$

We must therefore solve the following equations:
$$\begin{pmatrix}a&b\\ c&d\end{pmatrix}
\begin{pmatrix}*&*\\ *&*\end{pmatrix}=
\begin{pmatrix}ad-bc&0\\ 0&ad-bc\end{pmatrix}$$

But the obvious solution here is as follows:
$$\begin{pmatrix}a&b\\ c&d\end{pmatrix}
\begin{pmatrix}d&-b\\ -c&a\end{pmatrix}=
\begin{pmatrix}ad-bc&0\\ 0&ad-bc\end{pmatrix}$$

Thus, we are led to the formula in the statement.
\end{proof}

In order to deal now with the inversion problem in general, for the arbitrary matrices $A\in M_N(\mathbb R)$, we will use the same method as the one above, at $N=2$. Let us write indeed our matrix as follows, with $v_1,\ldots,v_N\in\mathbb R^N$ being its column vectors:
$$A=[v_1,\ldots,v_N]$$

We know that, in order for our matrix $A$ to be invertible, its column vectors $v_1,\ldots,v_N$ must be linearly independent. Thus, we are led into the question of understanding when a family of vectors $v_1,\ldots,v_N\in\mathbb R^N$ are linearly independent, which leads us to:

\index{volume of parallelepiped}
\index{positive determinant}

\begin{definition}
Associated to any vectors $v_1,\ldots,v_N\in\mathbb R^N$ is the volume
$${\rm det}^+(v_1\ldots v_N)=vol<v_1,\ldots,v_N>$$
of the parallelepiped made by these vectors.
\end{definition}

Here the volume is taken in the standard $N$-dimensional sense. At $N=1$ this volume is a length, at $N=2$ this volume is an area, at $N=3$ this is the usual 3D volume, and so on. In general, the volume of a body $X\subset\mathbb R^N$ is by definition the number $vol(X)\in[0,\infty]$ of copies of the unit cube $C\subset\mathbb R^N$ which are needed for filling $X$. Now with this notion in hand, in relation with our inversion problem, we have the following statement:

\index{invertible matrix}

\begin{proposition}
The quantity ${\rm det}^+$ that we constructed, regarded as a function of the corresponding square matrices, formed by column vectors,
$${\rm det}^+:M_N(\mathbb R)\to\mathbb R_+$$
has the property that a matrix $A\in M_N(\mathbb R)$ is invertible precisely when ${\rm det}^+(A)>0$.
\end{proposition}

\begin{proof}
This follows from the fact that a matrix $A\in M_N(\mathbb R)$ is invertible precisely when its column vectors $v_1,\ldots,v_N\in\mathbb R^N$ are linearly independent. But this latter condition is equivalent to the fact that we must have the following strict inequality: 
$$vol<v_1,\ldots,v_N>>0$$

Thus, we are led to the conclusion in the statement.
\end{proof}

Summarizing, all this leads us into the explicit computation of ${\rm det}^+$. As a first observation, in 1 dimension we obtain the absolute value of the real numbers:
$${\rm det}^+(a)=|a|$$

In 2 dimensions now, the computation is non-trivial, and we have the following result, making the link with our main result so far, namely Theorem 2.3:

\begin{theorem}
In $2$ dimensions we have the following formula,
$${\rm det}^+\begin{pmatrix}a&b\\ c&d\end{pmatrix}=|ad-bc|$$
with ${\rm det}^+:M_2(\mathbb R)\to\mathbb R_+$ being the function constructed above.
\end{theorem}

\begin{proof}
We must show that the area of the parallelogram formed by $\binom{a}{c},\binom{b}{d}$ equals $|ad-bc|$. We can assume $a,b,c,d>0$ for simplifying, the proof in general being similar. Moreover, by switching if needed the vectors $\binom{a}{c},\binom{b}{d}$, we can assume that we have:
$$\frac{a}{c}>\frac{b}{d}$$

Now let us slide the upper side of the parallelogram downwards left, until we reach $Oy$. Our parallelogram, which has not changed its area in this process, becomes:
$$
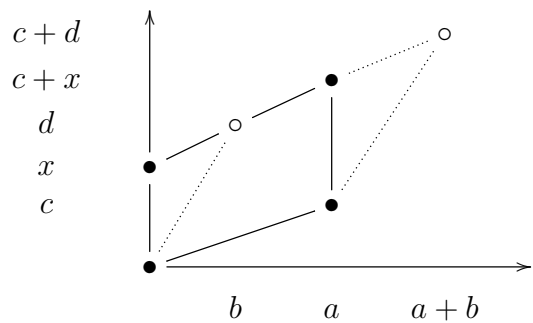
\xymatrix@R=1pt@C=16pt{
&&&\\
c+d&&&&\circ\\
c+x&&&\bullet\ar@{.}[ur]&\\
d&&\circ\ar@{-}[ur]&&&&&\\
x&\bullet\ar@{-}[ur]\ar[uuuu]&&&\\
c&&&\bullet\ar@{-}[uuu]\ar@{.}[uuuur]&\\
&&&&&&\\
&\bullet\ar@{-}[uurr]\ar@{-}[uuu]\ar[rrrr]\ar@{.}[uuuur]&&&&\\
&&\ b\ &\ a\ &a+b}$$

Moreover, we can further modify this parallelogram, once again by not altering its area, by sliding the right side downwards, until we reach the $Ox$ axis:
$$
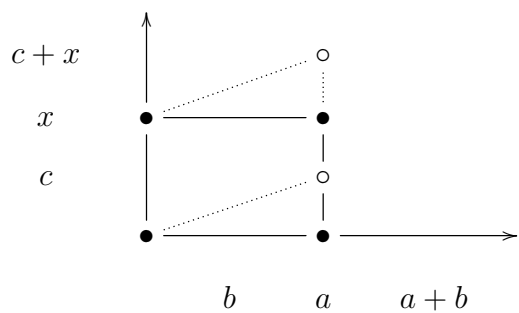
\xymatrix@R=10pt@C=15pt{
&&&\\
c+x&&&\circ&\\
x&\bullet\ar@{.}[urr]\ar[uu]\ar@{-}[rr]&&\bullet\ar@{.}[u]&\\
c&&&\circ\ar@{-}[u]&\\
&\bullet\ar@{.}[urr]\ar@{-}[uu]\ar@{-}[rr]&&\bullet\ar@{-}[u]\ar[rr]&&\\
&&\ b\ &\ a\ &a+b}$$

Let us compute now the area. Since our two sliding operations have not changed the area of the original parallelogram, this area is given by the following formula:
$$A=ax$$

In order to compute the quantity $x$, observe that in the context of the first move, we have two similar triangles, according to the following picture:
$$
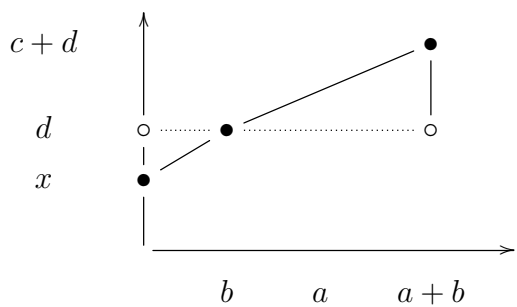
\xymatrix@R=4pt@C=15pt{
&&&&\\
c+d&&&&\bullet\\
&&&&&&\\
d&\circ\ar@{.}[r]\ar[uuu]&\bullet\ar@{.}[rr]\ar@{-}[uurr]&&\circ\ar@{-}[uu]\\
x&\bullet\ar@{-}[u]\ar@{-}[ur]&&&\\
&&&&\\
&\ar@{-}[uu]\ar[rrrr]&&&&\\
&&\ b\ &\ a\ &a+b}$$

Thus, we are led to the following equation for the number $x$:
$$\frac{d-x}{b}=\frac{c}{a}$$

But this gives $x=d-bc/a$, so the area of our parallelogram, or rather of the final rectangle obtained from it, which has the same area as the original parallelogram, is:
$$A=ax=ad-bc$$

We are therefore led to the conclusion in the statement.
\end{proof}

All this is very nice, and obviously we have a beginning of theory here. However, when looking carefully, we can see that our theory has a weakness, because:

\medskip

-- In 1 dimension the number $a$, which is the simplest function of $a$ itself, is certainly a better quantity than the number $|a|$.

\medskip

-- In 2 dimensions the number $ad-bc$, which is linear in $a,b,c,d$, is certainly a better quantity than the number $|ad-bc|$.

\medskip

So, let us upgrade now our theory, by constructing a better function, which takes signed values. In order to do this, we must come up with a way of splitting the systems of vectors $v_1,\ldots,v_N\in\mathbb R^N$ into two classes, say positive and negative. And here, the answer is quite clear, because a bit of thinking leads to the following definition:

\index{oriented system of vectors}
\index{unoriented system of vectors}
\index{sign of system of vectors}

\begin{definition}
A system of vectors $v_1,\ldots,v_N\in\mathbb R^N$ is called:
\begin{enumerate}
\item Oriented, if one can continuously pass from the standard basis to it.

\item Unoriented, otherwise.
\end{enumerate}
The associated sign is $+$ in the oriented case, and $-$ in the unoriented case. 
\end{definition}

As a first example, in 1 dimension the basis consists of the single vector $e=1$, which can be continuously deformed into any vector $a>0$. Thus, the sign is the usual one:
$$sgn(a)=
\begin{cases}
+&{\rm if}\ a>0\\
-&{\rm if}\ a<0
\end{cases}$$

Thus, in connection with our original question, we are definitely on the good track, because when multiplying $|a|$ by this sign we obtain $a$ itself, as desired:
$$a=sgn(a)|a|$$

In 2 dimensions now, the explicit formula of the sign is as follows:

\begin{proposition}
We have the following formula, valid for any $2$ vectors in $\mathbb R^2$,
$$sgn\left[\binom{a}{c},\binom{b}{d}\right]=sgn(ad-bc)$$
with the sign function on the right being the usual one, in $1$ dimension.
\end{proposition}

\begin{proof}
According to our conventions, the sign of $\binom{a}{c},\binom{b}{d}$ is as follows:

\medskip

(1) The sign is $+$ when these vectors come in this order with respect to the counterclockwise rotation in the plane, around 0.

\medskip

(2) The sign is $-$ otherwise, meaning when these vectors come in this order with respect to the clockwise rotation in the plane, around 0. 

\medskip

If we assume now $a,b,c,d>0$ for simplifying, we are left with comparing the angles having the numbers $c/a$ and $d/b$ as tangents, and we obtain in this way:
$$sgn\left[\binom{a}{c},\binom{b}{d}\right]=
\begin{cases}
+&{\rm if}\ \frac{c}{a}<\frac{d}{b}\\
-&{\rm if}\ \frac{c}{a}>\frac{d}{b}
\end{cases}$$

But this gives the formula in the statement. The proof in general is similar.
\end{proof}

Once again, in connection with our original question, we are on the good track, because when multiplying $|ad-bc|$ by this sign we obtain $ad-bc$ itself, as desired:
$$ad-bc=sgn(ad-bc)|ad-bc|$$

At the level of the general results now, we have:

\begin{proposition}
The orientation of a system of vectors changes as follows:
\begin{enumerate}
\item If we switch the sign of a vector, the associated sign switches.

\item If we permute two vectors, the associated sign switches as well.
\end{enumerate}
\end{proposition}

\begin{proof}
Both these assertions are clear from the definition of the sign, because the two operations in question change the orientation of the system of vectors.
\end{proof}

With the above notion in hand, we can now formulate:

\index{determinant}
\index{signed volume}

\begin{definition}
The determinant of $v_1,\ldots,v_N\in\mathbb R^N$ is the signed volume
$$\det(v_1\ldots v_N)=\pm vol<v_1,\ldots,v_N>$$
of the parallelepiped made by these vectors.
\end{definition}

In other words, we are upgrading here Definition 2.4, by adding a sign to the quantity ${\rm det}^+$ constructed there, as to potentially reach to good additivity properties:
$$\det(v_1\ldots v_N)=\pm {\rm det}^+(v_1\ldots v_N)$$

In relation with our original inversion problem for the square matrices, this upgrade does not change what we have so far, and we have the following statement:

\index{invertible matrix}

\begin{theorem}
The quantity $\det$ that we constructed, regarded as a function of the corresponding square matrices, formed by column vectors,
$$\det:M_N(\mathbb R)\to\mathbb R$$
has the property that a matrix $A\in M_N(\mathbb R)$ is invertible precisely when $\det(A)\neq 0$.
\end{theorem}

\begin{proof}
We know from the above that a matrix $A\in M_N(\mathbb R)$ is invertible precisely when ${\rm det}^+(A)=|\det A|$ is strictly positive, and this gives the result.
\end{proof}

Let us try now to compute the determinant. In 1 dimension we have of course the formula $\det(a)=a$, because the absolute value fits, and so does the sign:
$$\det(a)
=sgn(a)\times|a|
=a$$

In 2 dimensions now, we have the following result:

\begin{theorem}
In $2$ dimensions we have the following formula,
$$\begin{vmatrix}a&b\\ c&d\end{vmatrix}=ad-bc$$
with $|\,.\,|=\det$ being the determinant function constructed above.
\end{theorem}

\begin{proof}
According to our definition, to the computation in Theorem 2.6, and to the sign formula from Proposition 2.8, the determinant of a $2\times2$ matrix is given by:
\begin{eqnarray*}
\det\begin{pmatrix}a&b\\ c&d\end{pmatrix}
&=&sgn\left[\binom{a}{c},\binom{b}{d}\right]\times {\rm det}^+\begin{pmatrix}a&b\\ c&d\end{pmatrix}\\
&=&sgn\left[\binom{a}{c},\binom{b}{d}\right]\times|ad-bc|\\
&=&sgn(ad-bc)\times|ad-bc|\\
&=&ad-bc
\end{eqnarray*}

Thus, we have obtained the formula in the statement.
\end{proof}

\section*{2b. The determinant}

In order to discuss now arbitrary dimensions, we will need a number of theoretical results. Here is a first series of formulae, coming straight from the definitions:

\index{permuting columns}

\begin{theorem}
The determinant has the following properties:
\begin{enumerate}
\item When multiplying by scalars, the determinant gets multiplied as well:
$$\det(\lambda_1v_1,\ldots,\lambda_Nv_N)=\lambda_1\ldots\lambda_N\det(v_1,\ldots,v_N)$$

\item When permuting two columns, the determinant changes the sign:
$$\det(\ldots,u,\ldots,v,\ldots)=-\det(\ldots,v,\ldots,u,\ldots)$$

\item The determinant $\det(e_1,\ldots,e_N)$ of the standard basis of $\mathbb R^N$ is $1$.
\end{enumerate}
\end{theorem}

\begin{proof}
All this is clear from definitions, as follows:

\medskip

(1) This follows from definitions, and from Proposition 2.9 (1).

\medskip

(2) This follows as well from definitions, and from Proposition 2.9 (2).

\medskip

(3) This is clear from our definition of the determinant.
\end{proof}

In order to reach to a more advanced theory, let us adopt now the linear map point of view. In this setting, the definition of the determinant reformulates as follows: 

\index{inflation coefficient}

\begin{theorem}
Given a linear map, written as $f(v)=Av$, its ``inflation coefficient'', obtained as the signed volume of the image of the unit cube, is given by:
$$I_f=\det A$$
More generally, $I_f$ is the inflation ratio of any parallelepiped in $\mathbb R^N$, via the transformation $f$. In particular $f$ is invertible precisely when $\det A\neq0$.
\end{theorem}

\begin{proof}
The only non-trivial thing in all this is the fact that the inflation coefficient $I_f$, as defined above, is independent of the choice of the parallelepiped. But this is a generalization of the Thales theorem, which follows from the Thales theorem itself.
\end{proof}

As a first application of the above linear map viewpoint, we have:

\index{determinant of products}

\begin{theorem}
We have the following formula, valid for any matrices $A,B$:
$$\det(AB)=\det A\cdot\det B$$
In particular, we have $\det(AB)=\det(BA)$.
\end{theorem}

\begin{proof}
The first formula follows from the formula $f_{AB}=f_Af_B$ for the associated linear maps. As for $\det(AB)=\det(BA)$, this is clear from the first formula.
\end{proof}

Getting back now to explicit computations, we have the following key result:

\index{product of eigenvalues}
\index{diagonalizable matrix}

\begin{theorem}
The determinant of a diagonalizable matrix 
$$A\sim\begin{pmatrix}
\lambda_1\\ 
&\ddots\\
&&\lambda_N\end{pmatrix}$$
is the product of its eigenvalues, $\det A=\lambda_1\ldots\lambda_N$.
\end{theorem}

\begin{proof}
When the matrix is diagonal we can use Theorem 2.13, which gives:
$$\begin{vmatrix}
\lambda_1\\ 
&\ddots\\
&&\lambda_N\end{vmatrix}
=\lambda_1\ldots\lambda_N
\begin{vmatrix}
1\\ 
&\ddots\\
&&1\end{vmatrix}
=\lambda_1\ldots\lambda_N$$

In general, we know that a diagonalizable matrix can be written in the form $A=PDP^{-1}$, with $D=diag(\lambda_1,\ldots,\lambda_N)$. Now by using Theorem 2.15, we obtain:
\begin{eqnarray*}
\det A
&=&\det(PDP^{-1})\\
&=&\det(DP^{-1}P)\\
&=&\det D\\
&=&\lambda_1\ldots\lambda_N
\end{eqnarray*}

Thus, we are led to the formula in the statement.
\end{proof}

In general now, at the theoretical level, we have the following key result:

\begin{theorem}
The determinant has the additivity property
$$\det(\ldots,u+v,\ldots)
=\det(\ldots,u,\ldots)
+\det(\ldots,v,\ldots)$$
valid for any choice of the vectors involved.
\end{theorem}

\begin{proof}
This follows by doing some elementary geometry, in the spirit of the computations in the proof of Theorem 2.6, by using the Thales theorem. Exercise for you.
\end{proof}

As a basic application of the above result, we have:

\index{upper triangular matrix}
\index{lower triangular matrix}

\begin{theorem}
We have the following results:
\begin{enumerate}
\item The determinant of a diagonal matrix is the product of diagonal entries.

\item The same is true for the upper triangular matrices.

\item The same is true for the lower triangular matrices.
\end{enumerate}
\end{theorem}

\begin{proof}
All this can be deduced by using our various general formulae, as follows:

\medskip

(1) This is something that we already know, from Theorem 2.16.

\medskip

(2) This follows by using our various formulae, then (1), as follows:
\begin{eqnarray*}
\begin{vmatrix}
\lambda_1&&&*\\ 
&\lambda_2\\
&&\ddots\\
0&&&\lambda_N\end{vmatrix}
&=&\begin{vmatrix}
\lambda_1&0&&*\\ 
&\lambda_2\\
&&\ddots\\
0&&&\lambda_N\end{vmatrix}\\
&&\vdots\\
&&\vdots\\
&=&\begin{vmatrix}
\lambda_1&&&0\\ 
&\lambda_2\\
&&\ddots\\
0&&&\lambda_N\end{vmatrix}\\
&=&\lambda_1\ldots\lambda_N
\end{eqnarray*}

(3) This follows as well from our various formulae, then (1), by proceeding this time from right to left, from the last column towards the first column.
\end{proof}

As an important theoretical result now, we have:

\index{abstract determinant}

\begin{theorem}
The determinant of square matrices is the unique map
$$\det:M_N(\mathbb R)\to\mathbb R$$
satisfying the conditions found above.
\end{theorem}

\begin{proof}
Any map $\det':M_N(\mathbb R)\to\mathbb R$ satisfying our conditions must indeed coincide with $\det$ on the upper triangular matrices, and then on all the matrices.
\end{proof}

Here is now another important theoretical result:

\index{row expansion}

\begin{theorem}
The determinant is subject to the row expansion formula
\begin{eqnarray*}
\begin{vmatrix}a_{11}&\ldots&a_{1N}\\
\vdots&&\vdots\\
a_{N1}&\ldots&a_{NN}\end{vmatrix}
&=&a_{11}\begin{vmatrix}a_{22}&\ldots&a_{2N}\\
\vdots&&\vdots\\
a_{N2}&\ldots&a_{NN}\end{vmatrix}
-a_{12}\begin{vmatrix}a_{21}&a_{23}&\ldots&a_{2N}\\
\vdots&\vdots&&\vdots\\
a_{N1}&a_{N3}&\ldots&a_{NN}\end{vmatrix}\\
&&+\ldots\ldots
+(-1)^{N+1}a_{1N}\begin{vmatrix}a_{21}&\ldots&a_{2,N-1}\\
\vdots&&\vdots\\
a_{N1}&\ldots&a_{N,N-1}\end{vmatrix}
\end{eqnarray*}
and this method fully computes it, by recurrence.
\end{theorem}

\begin{proof}
This follows indeed from the fact that the above formula produces a certain function $\det:M_N(\mathbb R)\to\mathbb R$, which has the properties required by Theorem 2.19.
\end{proof}

We can expand as well over the columns, as follows:

\index{column expansion}

\begin{theorem}
The determinant is subject to the column expansion formula
\begin{eqnarray*}
\begin{vmatrix}a_{11}&\ldots&a_{1N}\\
\vdots&&\vdots\\
a_{N1}&\ldots&a_{NN}\end{vmatrix}
&=&a_{11}\begin{vmatrix}a_{22}&\ldots&a_{2N}\\
\vdots&&\vdots\\
a_{N2}&\ldots&a_{NN}\end{vmatrix}
-a_{21}\begin{vmatrix}a_{12}&\ldots&a_{1N}\\
a_{32}&\ldots&a_{3N}\\
\vdots&&\vdots\\
a_{N2}&\ldots&a_{NN}\end{vmatrix}\\
&&+\ldots\ldots
+(-1)^{N+1}a_{N1}\begin{vmatrix}a_{12}&\ldots&a_{1N}\\
\vdots&&\vdots\\
a_{N-1,2}&\ldots&a_{N-1,N}\end{vmatrix}
\end{eqnarray*}
and this method fully computes it, by recurrence.
\end{theorem}

\begin{proof}
This follows indeed by using the same argument as for the rows.
\end{proof}

As a first application of the above methods, we can now prove:

\index{Sarrus formula}

\begin{theorem}
The determinant of the $3\times3$ matrices is given by
$$\begin{vmatrix}a&b&c\\ d&e&f\\ g&h&i\end{vmatrix}=aei+bfg+cdh-ceg-bdi-afh$$
which can be memorized by using Sarrus' triangle method, ``triangles parallel to the diagonal, minus triangles parallel to the antidiagonal".
\end{theorem}

\begin{proof}
Here is the computation, by using the above results:
\begin{eqnarray*}
\begin{vmatrix}a&b&c\\ d&e&f\\ g&h&i\end{vmatrix}
&=&a\begin{vmatrix}e&f\\h&i\end{vmatrix}
-b\begin{vmatrix}d&f\\g&i\end{vmatrix}
+c\begin{vmatrix}d&e\\g&h\end{vmatrix}\\
&=&a(ei-fh)-b(di-fg)+c(dh-eg)\\
&=&aei-afh-bdi+bfg+cdh-ceg\\
&=&aei+bfg+cdh-ceg-bdi-afh
\end{eqnarray*}

Thus, we obtain the formula in the statement.
\end{proof}

Let us discuss now the general formula of the determinant, at arbitrary values $N\in\mathbb N$ of the matrix size, generalizing the formulae that we have at $N=2,3$. We will need:

\index{permutation}
\index{symmetric group}

\begin{definition}
A permutation of $\{1,\ldots,N\}$ is a bijection, as follows:
$$\sigma:\{1,\ldots,N\}\to\{1,\ldots,N\}$$
The set of such permutations is denoted $S_N$.
\end{definition}

There are many possible notations for the permutations, the simplest one consisting in writing the numbers $1,\ldots,N$, and below them, their permuted versions:
$$\sigma=\begin{pmatrix}
1&2&3&4&5\\
2&1&4&5&3
\end{pmatrix}$$

Another method, which is better for most purposes, and faster too, remember that time is money, is by denoting permutations as diagrams, going from top to bottom:
$$\xymatrix@R=3mm@C=3.5mm{
&\ar@{-}[ddr]&\ar@{-}[ddl]&\ar@{-}[ddrr]&\ar@{-}[ddl]&\ar@{-}[ddl]\\
\sigma=\\
&&&&&}$$

There are many interesting things that can be said about permutations. In what concerns us, we will need the following key result:

\index{permutation}
\index{signature}
\index{number of inversions}
\index{transposition}

\begin{theorem}
The permutations have a signature function
$$\varepsilon:S_N\to\{\pm1\}$$
which can be defined in the following equivalent ways:
\begin{enumerate}
\item As $(-1)^c$, where $c$ is the number of inversions.

\item As $(-1)^t$, where $t$ is the number of transpositions.

\item As $(-1)^o$, where $o$ is the number of odd cycles.

\item As $(-1)^x$, where $x$ is the number of crossings.

\item As the sign of the corresponding permuted basis of $\mathbb R^N$.
\end{enumerate}
\end{theorem}

\begin{proof}
Let us begin with the precise definition of $c,t,o,x$, as numbers modulo 2:

\medskip

(1) The idea here is that given any two numbers $i<j$ among $1,\ldots,N$, the permutation  can either keep them in the same order, $\sigma(i)<\sigma(j)$, or invert them:
$$\sigma(j)>\sigma(i)$$

Now by making $i<j$ vary over all pairs of numbers in $1,\ldots,N$, we can count the number of inversions, and call it $c$. This is an integer, $c\in\mathbb N$, which is well-defined.

\medskip

(2) Here the idea, which is something quite intuitive, is that any permutation appears as a product of switches, also called transpositions: 
$$i\leftrightarrow j$$

The decomposition as a product of transpositions is not unique, but the number $t$ of the needed transpositions is unique, when considered modulo 2. This follows for instance from the equivalence of (2) with (1,3,4,5), explained below.

\medskip

(3) Here the point is that any permutation decomposes, in a unique way, as a product of cycles, which are by definition permutations of the following type:
$$i_1\to i_2\to i_3\to\ldots\ldots\to i_k\to i_1$$

Some of these cycles have even length, and some others have odd length. By counting those having odd length, we obtain a well-defined number $o\in\mathbb N$.

\medskip

(4) Here the method is that of drawing the permutation, as we usually do, and by avoiding triple crossings, and then counting the number of crossings. This number $x$ depends on the way we draw the permutations, but modulo 2, we always get the same number. Indeed, this follows from the fact that we can continuously pass from a drawing to each other, and that when doing so, the number of crossings can only jump by $\pm2$.

\medskip

Summarizing, we have 4 different definitions for the signature of the permutations, which all make sense, constructed according to (1-4) above. Regarding now the fact that we always obtain the same number, this can be established as follows:

\medskip

(1)=(2) This is clear, because any transposition inverts once, modulo 2.

\medskip

(1)=(3) This is clear as well, because the odd cycles invert once, modulo 2.

\medskip

(1)=(4) This comes from the fact that the crossings correspond to inversions.

\medskip

(2)=(3) This follows by decomposing the cycles into transpositions.

\medskip

(2)=(4) This comes from the fact that the crossings correspond to transpositions.

\medskip

(3)=(4) This follows by drawing a product of cycles, and counting the crossings.

\medskip

Finally, in what regards the equivalence of all these constructions with (5), here simplest is to use (2). Indeed, we already know that the sign of a system of vectors switches when interchanging two vectors, and so the equivalence between (2,5) is clear. 
\end{proof}

Now back to linear algebra, we can formulate a key result, as follows:

\index{determinant formula}
\index{Sarrus formula}

\begin{theorem}
We have the following formula for the determinant,
$$\det A=\sum_{\sigma\in S_N}\varepsilon(\sigma)A_{1\sigma(1)}\ldots A_{N\sigma(N)}$$
with the signature function being the one introduced above.
\end{theorem}

\begin{proof}
This follows by recurrence over $N\in\mathbb N$, as follows:

\medskip

(1) When developing the determinant over the first column, we obtain a signed sum of $N$ determinants of size $(N-1)\times(N-1)$. But each of these determinants can be computed by developing over the first column too, and so on, and we are led to the conclusion that we have a formula as in the statement, with $\varepsilon(\sigma)\in\{-1,1\}$ being certain coefficients.

\medskip

(2) But these latter coefficients $\varepsilon(\sigma)\in\{-1,1\}$ can only be the signatures of the corresponding permutations $\sigma\in S_N$, with this being something that can be viewed again by recurrence, with either of the definitions (1-5) in Theorem 2.24 for the signature.
\end{proof}

The above result is something quite tricky. As a first, basic illustration, in 2 dimensions we recover the usual formula of the determinant, the details being as follows:
\begin{eqnarray*}
\begin{vmatrix}a&b\\ c&d\end{vmatrix}
&=&\varepsilon(|\,|)\cdot ad+\varepsilon(\slash\hskip-2mm\backslash)\cdot cb\\
&=&1\cdot ad+(-1)\cdot cb\\
&=&ad-bc
\end{eqnarray*}

In 3 dimensions, we recover the Sarrus formula, that we know from Theorem 2.22:
$$\begin{vmatrix}a&b&c\\ d&e&f\\ g&h&i\end{vmatrix}=aei+bfg+cdh-ceg-bdi-afh$$

Observe that the triangles in the Sarrus formula correspond to the permutations of $\{1,2,3\}$, and their signs correspond to the signatures of these permutations:
\begin{eqnarray*}
\det
&=&\begin{pmatrix}*&&\\ &*&\\ &&*\end{pmatrix}
+\begin{pmatrix}&*&\\ &&*\\ *&&\end{pmatrix}
+\begin{pmatrix}&&*\\ *&&\\ &*&\end{pmatrix}\\
&-&\begin{pmatrix}&&*\\ &*&\\ *&&\end{pmatrix}
+\begin{pmatrix}&*&\\ *&&\\ &&*\end{pmatrix}
+\begin{pmatrix}*&&\\ &&*\\ &*&\end{pmatrix}
\end{eqnarray*}

In 4 dimensions now, by using our technology, we can formulate:

\begin{theorem}
The determinant of the $4\times4$ matrices is given by
\begin{eqnarray*}
&&\begin{vmatrix}a_1&a_2&a_3&a_4\\ b_1&b_2&b_3&b_4\\ c_1&c_2&c_3&c_4\\ d_1&d_2&d_3&d_4\end{vmatrix}\\
&=&a_1b_2c_3d_4-a_1b_2c_4d_3-a_1b_3c_2d_4+a_1b_3c_4d_2+a_1b_4c_2d_3-a_1b_4c_3d_2\\
&-&a_2b_1c_3d_4+a_2b_1c_4d_3+a_2b_3c_1d_4-a_2b_3c_4d_1-a_2b_4c_1d_3+a_2b_4c_3d_1\\
&+&a_3b_1c_2d_4+a_3b_1c_4d_2-a_3b_2c_1d_4+a_3b_2c_4d_1+a_3b_4c_1d_2-a_3b_4c_2d_1\\
&-&a_4b_1c_2d_3+a_4b_1c_3d_2-a_4b_2c_1d_3-a_4b_2c_3d_1-a_4b_3c_1d_2+a_4b_3c_2d_1
\end{eqnarray*}
with the generic term being of the following form, with $\sigma\in S_4$,
$$\pm a_{\sigma(1)}b_{\sigma(2)}c_{\sigma(3)}d_{\sigma(4)}$$
and with the sign being $\varepsilon(\sigma)$, computable by using Theorem 2.24.
\end{theorem}

\begin{proof}
This follows indeed from Theorem 2.25, with the various permutations appearing in the statement being listed according to the lexicographic order.
\end{proof}

As yet another application, we have the following key result:

\index{transpose matrix}

\begin{theorem}
We have the formula
$$\det A=\det A^t$$
valid for any square matrix $A$.
\end{theorem}

\begin{proof}
This follows from the formula in Theorem 2.25. Indeed, we have:
\begin{eqnarray*}
\det A^t
&=&\sum_{\sigma\in S_N}\varepsilon(\sigma)(A^t)_{1\sigma(1)}\ldots(A^t)_{N\sigma(N)}\\
&=&\sum_{\sigma\in S_N}\varepsilon(\sigma)A_{\sigma(1)1}\ldots A_{\sigma(N)N}\\
&=&\sum_{\sigma\in S_N}\varepsilon(\sigma)A_{1\sigma^{-1}(1)}\ldots A_{N\sigma^{-1}(N)}\\
&=&\sum_{\sigma\in S_N}\varepsilon(\sigma^{-1})A_{1\sigma^{-1}(1)}\ldots A_{N\sigma^{-1}(N)}\\
&=&\sum_{\sigma\in S_N}\varepsilon(\sigma)A_{1\sigma(1)}\ldots A_{N\sigma(N)}\\
&=&\det A
\end{eqnarray*}

Thus, we are led to the formula in the statement.
\end{proof}

There are countless other applications of the formula in Theorem 2.25. Importantly, that formula allows us to deal now with the complex matrices too, as follows:

\index{complex matrix}
\index{deterrminant}

\begin{theorem}
If we define the determinant of a complex matrix $A\in M_N(\mathbb C)$ to be
$$\det A=\sum_{\sigma\in S_N}\varepsilon(\sigma)A_{1\sigma(1)}\ldots A_{N\sigma(N)}$$
then this determinant has the same properties as the determinant of the real matrices.
\end{theorem}

\begin{proof}
This follows by doing some sort of reverse engineering, with respect to what we have been doing in this section, and we reach to the conclusion that $\det$ has indeed all the good properties that we are familiar with. Except of course for those at the very beginning of this chapter, in relation with volumes, which don't extend well to $\mathbb C^N$.
\end{proof}

\section*{2c. Some applications}

Good news, we have now in our bag all the needed techniques for computing the determinant. As a first application, we can now invert the $3\times3$ matrices, as follows:

\index{matrix inversion}

\begin{theorem}
The inverses of the $3\times3$ matrices are given by
$$\begin{pmatrix}a&b&c\\ d&e&f\\ g&h&i\end{pmatrix}^{-1}
=\frac{1}{D}\begin{pmatrix}ei-fh&ch-bi&bf-ce\\ fg-di&ai-cg&cd-af\\ dh-eg&bg-ah&ae-bd\end{pmatrix}$$
with $D$ being the determimant. When $D=0$, the matrix is not invertible.
\end{theorem}

\begin{proof}
As before for the $2\times 2$ matrices, in order for our matrix to be invertible, we must have $D\neq0$. The trick now is to look for solutions of the following problem:
$$\begin{pmatrix}a&b&c\\ d&e&f\\ g&h&i\end{pmatrix}
\begin{pmatrix}*&*&*\\ *&*&*\\ *&*&*\end{pmatrix}
=\begin{pmatrix}D&0&0\\ 0&D&0\\ 0&0&D\end{pmatrix}$$

We know from Theorem 2.22 that the determinant is given by:
$$D=aei+bfg+cdh-ceg-bdi-afh$$

But this leads, via some obvious choices, to the following solution:
$$\begin{pmatrix}*&*&*\\ *&*&*\\ *&*&*\end{pmatrix}
=\begin{pmatrix}ei-fh&ch-bi&bf-ce\\ fg-di&ai-cg&cd-af\\ dh-eg&bg-ah&ae-bd\end{pmatrix}$$

Thus, by rescaling, we obtain the formula in the statement.
\end{proof}

In fact, we can now fully solve the inversion problem, as follows:

\index{matrix inversion}

\begin{theorem}
The inverse of a square matrix having nonzero determinant is
$$A^{-1}=\frac{1}{\det A}
\begin{pmatrix}
\det A^{(11)}&-\det A^{(21)}&\det A^{(31)}&\ldots\\
-\det A^{(12)}&\det A^{(22)}&-\det A^{(32)}&\ldots\\
\det A^{(13)}&-\det A^{(23)}&\det A^{(33)}&\ldots\\
\vdots&\vdots&\vdots&
\end{pmatrix}$$
where $A^{(ij)}$ is the matrix $A$, with the $i$-th row and $j$-th column removed.
\end{theorem}

\begin{proof}
This follows indeed by using the row expansion formula from Theorem 2.20, which in terms of the matrix $A^{-1}$ in the statement reads $AA^{-1}=1$.
\end{proof}

Switching topics, as a first true application of the determinants, let us discuss now the Gram-Schmidt procedure. We have the following key result:

\begin{theorem}
Any system of linearly independent vectors $\{f_1,\ldots,f_n\}$ can be turned into an orthogonal system $\{e_1,\ldots,e_n\}$ by using the Gram-Schmidt procedure,
$$e_1=f_1$$
$$e_2=f_2+\alpha_1f_1$$
$$e_3=f_3+\beta_1f_1+\beta_2f_2$$
$$e_4=f_4+\gamma_1f_1+\gamma_2f_2+\gamma_3f_3$$
$$\vdots$$
with the needed scalars $\alpha_i,\beta_i,\gamma_i,\ldots$ being uniquely determined.
\end{theorem}

\begin{proof}
Many things can be said here, depending on how sharp you want to be, with the essentials of what is to be known being as follows:

\medskip

(1) Let us first study the case $n=2$. With $e_1=f_1$ and $e_2=f_2+\alpha_1f_1$ as in the statement, the needed orthogonality condition can be processed as follows:
\begin{eqnarray*}
e_1\perp e_2
&\iff&<f_1,f_2+\alpha_1f_1>=0\\
&\iff&\alpha_1<f_1,f_1>=-<f_1,f_2>\\
&\iff&\alpha_1=-\frac{<f_1,f_2>}{<f_1,f_1>}
\end{eqnarray*}

Thus, we get our result, and with the remark that, alternatively, we can set:
$$e_2=f_2-Proj_{e_1}(f_2)$$

Indeed, with the above formula of $\alpha_1$ in hand, the vector $e_2=f_2+\alpha_1f_1$ that we get is precisely this one. Or, we can simply argue that this latter vector $e_2$ does the job, and with some basic linear algebra telling us that this vector $e_2$ is indeed unique.

\medskip 

(2) At $n=3$ now, with $e_1,e_2$ already constructed, and with $e_3=f_3+\beta_1f_1+\beta_2f_2$ as in the statement, the first orthogonality condition can be processed as follows:
\begin{eqnarray*}
e_1\perp e_3
&\iff&<f_1,f_3+\beta_1f_1+\beta_2f_2>=0\\
&\iff&\beta_1<f_1,f_1>+\beta_2<f_1,f_2>=-<f_1,f_3>
\end{eqnarray*}

As for the second orthogonality condition, this can be now processed as follows:
\begin{eqnarray*}
e_2\perp e_3
&\iff&<f_2,f_3+\beta_1f_1+\beta_2f_2>=0\\
&\iff&\beta_1<f_2,f_1>+\beta_2<f_2,f_2>=-<f_2,f_3>
\end{eqnarray*}

Thus, we are led to the following system, for the parameters $\beta_1,\beta_2$:
$$\beta_1<f_1,f_1>+\beta_2<f_1,f_2>=-<f_1,f_3>$$
$$\beta_1<f_2,f_1>+\beta_2<f_2,f_2>=-<f_2,f_3>$$

Now let us compute the determinant of this system. This is given by:
\begin{eqnarray*}
D
&=&\begin{vmatrix}
<f_1,f_1>&<f_1,f_2>\\
<f_2,f_1>&<f_2,f_2>
\end{vmatrix}\\
&=&<f_1,f_1><f_2,f_2>-<f_1,f_2><f_2,f_1>\\
&=&||f_1||^2||f_2||^2-<f_1,f_2>^2
\end{eqnarray*}

But this is exactly the quantity from the Cauchy-Schwarz inequality, so we have $D\geq0$, with equality when $f_1,f_2$ are proportional. Now since $f_1,f_2$ were assumed to be linearly independent, we conclude that we have $D>0$, so our system has indeed solutions.

\medskip

(3) Alternatively, we can say at $n=3$ that with the vectors $e_1,e_2$ being already constructed, we can construct the vector $e_3$ as follows, obviously doing the orthogonality job, and with its uniqueness coming from standard linear algebra:
$$e_3=f_3-Proj_{e_1}(f_3)-Proj_{e_2}(f_3)$$

(4) Summarizing, we have two possible proofs for our result. Getting now to the general case, as a first proof, which is perhaps the most straightforward, we can set:
$$e_1=f_1$$
$$e_2=f_2-Proj_{e_1}(f_2)$$
$$e_3=f_3-Proj_{e_1}(f_3)-Proj_{e_2}(f_3)$$
$$e_4=f_4-Proj_{e_1}(f_4)-Proj_{e_2}(f_4)-Proj_{e_3}(f_4)$$
$$\vdots$$

Indeed, these vectors do indeed the needed orthogonality job, and their uniqueness is clear too, via some basic linear algebra, that we will leave here as an exercise.

\medskip

(5) Alternatively, by doing some explicit computations, as in (1) and (2), we must prove that a certain determinant is nonzero. To be more precise, at step $k+1$ of the orthogonalization algorithm, the system to be solved is as follows:
$$x_1<f_1,f_1>+x_2<f_1,f_2>+\ldots+x_k<f_1,f_k>=-<f_1,f_{k+1}>$$
$$x_1<f_2,f_1>+x_2<f_2,f_2>+\ldots+x_k<f_2,f_k>=-<f_2,f_{k+1}>$$
$$\vdots$$
$$x_1<f_k,f_1>+x_2<f_k,f_2>+\ldots+x_k<f_k,f_k>=-<f_k,f_{k+1}>$$

Thus, the determinant to be studied, in order to prove that our system has indeed solutions, is the Gram determinant of $f_1,\ldots,f_k$, given by the following formula:
$$D_k=\begin{vmatrix}
<f_1,f_1>&<f_1,f_2>&\ldots&<f_1,f_k>\\
<f_2,f_1>&<f_2,f_2>&\ldots&<f_2,f_k>\\
&&\vdots\\
<f_k,f_1>&<f_k,f_2>&\ldots&<f_k,f_k>
\end{vmatrix}$$

(6) Now in relation with this latter question, we have already seen in (2) that we have $D_2>0$, but with this being something quite complicated, coming from Cauchy-Schwarz. So, not very good news, but fortunately, linear algebra comes to the rescue. Consider the square matrix formed by our vectors $f_1,\ldots,f_k$, arranged horizontally, as follows:
$$F=\begin{pmatrix}
(f_1)_1&\ldots&(f_1)_k\\
&\vdots\\
(f_k)_1&\ldots&(f_k)_k
\end{pmatrix}$$

We have then the following computation, for any two indices $i,j$:
\begin{eqnarray*}
(FF^t)_{ij}
&=&\sum_lF_{il}(F^t)_{lj}\\
&=&\sum_lF_{il}F_{jl}\\
&=&\sum_l(f_i)_l(f_j)_l\\
&=&<f_i,f_j>
\end{eqnarray*}

We conclude that at the matrix level, we have the following formula:
$$FF^t=\begin{pmatrix}
<f_1,f_1>&<f_1,f_2>&\ldots&<f_1,f_k>\\
<f_2,f_1>&<f_2,f_2>&\ldots&<f_2,f_k>\\
&&\vdots\\
<f_k,f_1>&<f_k,f_2>&\ldots&<f_k,f_k>
\end{pmatrix}$$

Thus, at the level of the corresponding determinants we obtain, as desired:
$$D_k=\det(FF^t)=(\det F)^2>0$$

(7) Finally, and getting back now to the system, we can work out some explicit formulae for $e_i$, alternative to those in (4), based on this. To be more precise, we have:
$$e_k=\frac{1}{D_{k-1}}\begin{vmatrix}
<f_1,f_1>&<f_1,f_2>&\ldots&<f_1,f_k>\\
<f_2,f_1>&<f_2,f_2>&\ldots&<f_2,f_k>\\
&&\vdots\\
<f_{k-1},f_1>&<f_{k-1},f_2>&\ldots&<f_{k-1},f_k>\\
f_1&f_2&\ldots&f_k
\end{vmatrix}$$

And we will leave some illustrations here as an instructive exercise, and please do better than my students, who usually stop after 2-3 steps.
\end{proof}

Finally, no discussion about determinants would be complete without:

\index{Vandermonde formula}

\begin{theorem}
We have the Vandermonde determinant formula
$$\begin{vmatrix}
1&1&1&\ldots\ldots&1\\
x_1&x_2&x_3&\ldots\ldots&x_N\\
x_1^2&x_2^2&x_3^2&\ldots\ldots&x_N\\
\vdots&\vdots&\vdots&&\vdots\\
\vdots&\vdots&\vdots&&\vdots\\
x_1^{N-1}&x_2^{N-1}&x_3^{N-1}&\ldots\ldots&x_N^{N-1}
\end{vmatrix}
=\prod_{i>j}(x_i-x_j)$$
valid for any $x_1,\ldots,x_N\in\mathbb R$.
\end{theorem}

\begin{proof}
By expanding over the columns, we can see that the determinant in question, say $D$, is a polynomial in the variables $x_1,\ldots,x_N$, having degree $N-1$ in each variable. Now observe that when setting $x_i=x_j$, for some indices $i\neq j$, our matrix will have two identical columns, and so its determinant $D$ will vanish:
$$x_i=x_j\implies D=0$$

But this gives us the key to the computation of $D$. Indeed, $D$ must be divisible by $x_i-x_j$ for any $i\neq j$, and so we must have a formula of the following type:
$$D=c\prod_{i>j}(x_i-x_j)$$

Moreover, since the product on the right is, exactly as $D$ itself, a polynomial in the variables $x_1,\ldots,x_N$, having degree $N-1$ in each variable, we conclude that $c$ must be a constant, not depending on any of the variables $x_1,\ldots,x_N$. It remains to find the value of $c$, and this can be done for instance by recurrence, and we obtain $c=1$, as desired.
\end{proof}

\section*{2d. Diagonalization} 

Let us discuss now the diagonalization question for linear maps and matrices. The basic diagonalization theory, formulated in terms of matrices, is as follows:

\index{eigenvalue}
\index{eigenvector}
\index{diagonalization}
\index{passage matrix}

\begin{theorem}
A vector $v\in\mathbb R^N$ is called eigenvector of $A\in M_N(\mathbb R)$, with corresponding eigenvalue $\lambda$, when $A$ multiplies by $\lambda$ in the direction of $v$:
$$Av=\lambda v$$
In the case where $\mathbb R^N$ has a basis $v_1,\ldots,v_N$ formed by eigenvectors of $A$, with corresponding eigenvalues $\lambda_1,\ldots,\lambda_N$, in this new basis $A$ becomes diagonal, as follows:
$$A\sim\begin{pmatrix}\lambda_1\\&\ddots\\&&\lambda_N\end{pmatrix}$$
Equivalently, if we denote by $D=diag(\lambda_1,\ldots,\lambda_N)$ the above diagonal matrix, and by $P=[v_1\ldots v_N]$ the square matrix formed by the eigenvectors of $A$, we have:
$$A=PDP^{-1}$$
In this case we say that the matrix $A$ is diagonalizable.
\end{theorem}

\begin{proof}
This is something that we know from chapter 1, the idea being as follows:

\medskip

(1) The first assertion is clear, because the matrix which multiplies each basis element $v_i$ by a number $\lambda_i$ is precisely the diagonal matrix $D=diag(\lambda_1,\ldots,\lambda_N)$.

\medskip

(2) The second assertion follows from the first one, by changing the basis. We can prove this by a direct computation as well, because we have $Pe_i=v_i$, and so:
$$PDP^{-1}v_i
=PDe_i
=P\lambda_ie_i
=\lambda_iPe_i
=\lambda_iv_i$$

Thus, the matrices $A$ and $PDP^{-1}$ coincide, as stated.
\end{proof}

In order to study the diagonalization problem, the idea is that the eigenvectors can be grouped into linear spaces, called eigenspaces, as follows:

\index{eigenspace}

\begin{theorem}
Let $A\in M_N(\mathbb R)$, and for any eigenvalue $\lambda\in\mathbb R$ define the corresponding eigenspace as being the vector space formed by the corresponding eigenvectors:
$$E_\lambda=\left\{v\in\mathbb R^N\Big|Av=\lambda v\right\}$$
These eigenspaces $E_\lambda$ are then in a direct sum position, in the sense that given vectors $v_1\in E_{\lambda_1},\ldots,v_k\in E_{\lambda_k}$ corresponding to different eigenvalues $\lambda_1,\ldots,\lambda_k$, we have:
$$\sum_ic_iv_i=0\implies c_i=0$$
In particular, we have $\sum_\lambda\dim(E_\lambda)\leq N$, with the sum being over all the eigenvalues, and our matrix is diagonalizable precisely when we have equality.
\end{theorem}

\begin{proof}
We prove the first assertion by recurrence on $k\in\mathbb N$. Assume by contradiction that we have a formula as follows, with the scalars $c_1,\ldots,c_k$ being not all zero:
$$c_1v_1+\ldots+c_kv_k=0$$

By dividing by one of these scalars, we can assume that our formula is:
$$v_k=c_1v_1+\ldots+c_{k-1}v_{k-1}$$

Now let us apply $A$ to this vector. On the left we obtain:
$$Av_k
=\lambda_kv_k
=\lambda_kc_1v_1+\ldots+\lambda_kc_{k-1}v_{k-1}$$

On the right we obtain something different, as follows:
\begin{eqnarray*}
A(c_1v_1+\ldots+c_{k-1}v_{k-1})
&=&c_1Av_1+\ldots+c_{k-1}Av_{k-1}\\
&=&c_1\lambda_1v_1+\ldots+c_{k-1}\lambda_{k-1}v_{k-1}
\end{eqnarray*}

We conclude from this that the following equality must hold:
$$\lambda_kc_1v_1+\ldots+\lambda_kc_{k-1}v_{k-1}=c_1\lambda_1v_1+\ldots+c_{k-1}\lambda_{k-1}v_{k-1}$$

On the other hand, we know by recurrence that the vectors $v_1,\ldots,v_{k-1}$ must be linearly independent. Thus, the coefficients must be equal, at right and at left:
$$\lambda_kc_1=c_1\lambda_1$$
$$\vdots$$
$$\lambda_kc_{k-1}=c_{k-1}\lambda_{k-1}$$

Now since at least one of the numbers $c_i$ must be nonzero, from $\lambda_kc_i=c_i\lambda_i$ we obtain $\lambda_k=\lambda_i$, which is a contradiction. Thus our proof by recurrence of the first assertion is complete. As for the second assertion, this follows from the first one.
\end{proof}

In order to reach now to more advanced results, we can use the following key fact:

\index{characteristic polynomial}
\index{roots of polynomials}
\index{multiplicity of root}

\begin{theorem}
Given a matrix $A\in M_N(\mathbb R)$, consider its characteristic polynomial:
$$P(x)=\det(A-x1_N)$$
The eigenvalues of $A$ are then the roots of $P$. Also, we have the inequality
$$\dim(E_\lambda)\leq m_\lambda$$
where $m_\lambda$ is the multiplicity of $\lambda$, as root of $P$.
\end{theorem}

\begin{proof}
The first assertion follows from the following computation, using the fact that a linear map is bijective when the determinant of the associated matrix is nonzero:
\begin{eqnarray*}
\exists v,Av=\lambda v
&\iff&\exists v,(A-\lambda 1_N)v=0\\
&\iff&\det(A-\lambda 1_N)=0
\end{eqnarray*}

Regarding now the second assertion, given an eigenvalue $\lambda$ of our matrix $A$, consider the dimension $d_\lambda=\dim(E_\lambda)$ of the corresponding eigenspace. By changing the basis of $\mathbb R^N$, as for the eigenspace $E_\lambda$ to be spanned by the first $d_\lambda$ basis elements, our matrix becomes as follows, with $B$ being a certain smaller matrix:
$$A\sim\begin{pmatrix}\lambda 1_{d_\lambda}&0\\0&B\end{pmatrix}$$

We conclude that the characteristic polynomial of $A$ is of the following form:
$$P_A
=P_{\lambda 1_{d_\lambda}}P_B
=(\lambda-x)^{d_\lambda}P_B$$

Thus the multiplicity $m_\lambda$ of our eigenvalue $\lambda$, viewed as a root of $P$, is subject to the estimate $m_\lambda\geq d_\lambda$, and this leads to the conclusion in the statement.
\end{proof}

With this discussed, what is next? In answer, trading $\mathbb R$ for the complex numbers $\mathbb C$, due to the following key result, called fundamental theorem of algebra:

\index{roots of polynomial}
\index{complex roots}

\begin{theorem}
Any polynomial $P\in\mathbb C[X]$ decomposes as
$$P=c(X-a_1)\ldots (X-a_N)$$
with $c\in\mathbb C$ and with $a_1,\ldots,a_N\in\mathbb C$.
\end{theorem}

\begin{proof}
The problem is that of proving that our polynomial has at least one root, because afterwards we can proceed by recurrence. We prove this by contradiction. So, assume that $P$ has no roots, and pick a number $z\in\mathbb C$ where $|P|$ attains its minimum:
$$|P(z)|=\min_{x\in\mathbb C}|P(x)|>0$$ 

Since $Q(t)=P(z+t)-P(z)$ is a polynomial which vanishes at $t=0$, this polynomial must be of the form $ct^k$ + higher terms, with $c\neq0$, and with $k\geq1$ being an integer. We obtain from this that, with $t\in\mathbb C$ small, we have the following estimate:
$$P(z+t)\simeq P(z)+ct^k$$

Now let us write $t=rw$, with $r>0$ small, and with $|w|=1$. Our estimate becomes:
$$P(z+rw)\simeq P(z)+cr^kw^k$$

Now recall that we have assumed $P(z)\neq0$. We can therefore choose $w\in\mathbb T$ such that $cw^k$ points in the opposite direction to that of $P(z)$, and we obtain in this way:
\begin{eqnarray*}
|P(z+rw)|
&\simeq&|P(z)+cr^kw^k|\\
&=&|P(z)|(1-|c|r^k)
\end{eqnarray*}

Now by choosing $r>0$ small enough, as for the error in the first estimate to be small, and overcame by the negative quantity $-|c|r^k$, we obtain from this:
$$|P(z+rw)|<|P(z)|$$

But this contradicts our definition of $z\in\mathbb C$, as a point where $|P|$ attains its minimum. Thus $P$ has a root, and by recurrence it has $N$ roots, as stated.
\end{proof}

Getting back now to linear algebra, in view of Theorem 2.36, it makes sense to upgrade our discussion, by changing the ground field from $\mathbb R$ to $\mathbb C$, where the matrices have more chances to have eigenvalues. The main results that we have so far, namely Theorems 2.33, 2.34, 2.35, extend in a straightforward way to the complex number setting, and when coupled with Theorem 2.36, they lead to the following remarkable result:

\index{eigenvalue}
\index{eigenvector}
\index{characteristic polynomial}
\index{diagonalization}

\begin{theorem}
Given a matrix $A\in M_N(\mathbb C)$, consider its characteristic polynomial
$$P(X)=\det(A-X1_N)$$ 
then factorize this polynomial, by computing the complex roots, with multiplicities,
$$P(X)=(-1)^N(X-\lambda_1)^{n_1}\ldots(X-\lambda_k)^{n_k}$$
and finally compute the corresponding eigenspaces, for each eigenvalue found:
$$E_i=\left\{v\in\mathbb C^N\Big|Av=\lambda_iv\right\}$$
The dimensions of these eigenspaces satisfy then the following inequalities,
$$\dim(E_i)\leq n_i$$
and $A$ is diagonalizable precisely when we have equality for any $i$.
\end{theorem}

\begin{proof}
This follows by combining the above results. Indeed, by summing the inequalities $\dim(E_\lambda)\leq m_\lambda$ from Theorem 2.35, we obtain an inequality as follows:
$$\sum_\lambda\dim(E_\lambda)\leq\sum_\lambda m_\lambda\leq N$$

On the other hand, we know from Theorem 2.34 that our matrix is diagonalizable when we have global equality. Thus, we are led to the conclusion in the statement.
\end{proof}

This was for the main result of linear algebra. There are countless applications of this, and generally speaking, advanced linear algebra consists in building on Theorem 2.37. 

\bigskip

In practice now, the use of Theorem 2.37 requires some practice, and skill. In relation with this, let us record as well a useful algorithmic version of the above result:

\index{diagonalization algorithm}

\begin{theorem}
The square matrices $A\in M_N(\mathbb C)$ can be diagonalized as follows:
\begin{enumerate}
\item Compute the characteristic polynomial.

\item Factorize the characteristic polynomial.

\item Compute the eigenvectors, for each eigenvalue found.

\item If there are no $N$ eigenvectors, $A$ is not diagonalizable.

\item Otherwise, $A$ is diagonalizable, $A=PDP^{-1}$.
\end{enumerate}
\end{theorem}

\begin{proof}
This is an informal reformulation of Theorem 2.37, with (4) referring to the total number of linearly independent eigenvectors found in (3), and with $A=PDP^{-1}$ in (5) being the usual diagonalization formula, with $P,D$ being as before.
\end{proof}

As a remark here, in step (3) it is always better to start with the eigenvalues having big multiplicity. Indeed, a multiplicity 1 eigenvalue, for instance, can never lead to the end of the computation, via (4), simply because the eigenvectors always exist.

\section*{2e. Exercises}

This was another basic chapter, all starting level linear algebra material, that you are basically supposed to know, and as exercises on all this, we have:

\begin{exercise}
Have some fun with computing the orientation of various systems of vectors in space, of your choice.
\end{exercise}

\begin{exercise}
Forget if needed the definition and theory of the determinant that you learned in school, and relearn it as indicated above, as a signed volume.
\end{exercise}

\begin{exercise}
Further meditate on the uniqueness statement for the determinant given above, and learn about some other such uniqueness statements as well.
\end{exercise}

\begin{exercise}
Review the theory of permutations and their signatures, and learn also about the alternating group $A_N\subset S_N$.
\end{exercise}

\begin{exercise}
In relation with Gram-Schmidt, learn about the Legendre, Chebycheff, Jacobi, Laguerre and Hermite polynomials.
\end{exercise}

\begin{exercise}
Learn about the Hadamard determinant bound, the Hadamard matrices, their various properties, and the Hadamard Conjecture, and its status.
\end{exercise}

\begin{exercise}
Learn the various standard algebraic and analytic tricks for factorizing polynomials.
\end{exercise}

\begin{exercise}
Try to understand what happens to our diagonalization theory in the case of basic non-diagonalizable matrices, such as the basic Jordan block $J$.
\end{exercise}

As bonus exercise, diagonalize some matrices, as many as you can. Normally you cannot call yourself a scientist until you do $3\times3$ matrices in 15 minutes, or less.

\chapter{Spectral theorems}

\section*{3a. Self-adjoints}

Let us go back to the diagonalization question, discussed in the previous chapter. We would like to present here some diagonalization results which are far more powerful, concerning various types of special matrices, collectively known as ``spectral theorems''. 

\bigskip

As explained in the previous chapter, the advanced theory takes place over $\mathbb C$, so as a first task, we have to review the vector basics in this context. Let us start with:

\begin{theorem}
We can talk about scalar products and lengths in $\mathbb C^N$, according to
$$<x,y>=\sum_ix_i\bar{y}_i\quad,\quad ||x||=\sqrt{\sum_i|x_i|^2}$$
and the following happen:
\begin{enumerate}
\item Linearity: $<x,y>$ is linear in $x$, and antilinear in $y$.

\item Norm dilation formula: $||\lambda x||=|\lambda|\cdot||x||$.

\item Main norm formula: $||x||=\sqrt{<x,x>}$.

\item Polarization: $<x,y>
=(||x+y||^2-||x-y||^2
+i||x+iy||^2-i||x-iy||^2)/4$.

\item Parallelogram rule: $||x+y||^2+||x-y||^2=2(||x||^2+||y||^2)$.

\item Cauchy-Schwarz: $|<x,y>|\leq||x||\cdot||y||$.

\item Triangle inequality: $||x+y||\leq||x||+||y||$.

\item Orthogonality: $x\perp y\iff<x,y>=0$, by definition.

\item $<x,y>=<x',y>=<x,y'>$, prime being the projection on the other vector.
\end{enumerate}
\end{theorem}

\begin{proof}
We follow the proof from the real case, from chapter 1. We can certainly talk about scalar products and lengths, as above, and (1,2,3) are clear. Regarding now (4), which is more complicated in the present complex setting, this comes from:
\begin{eqnarray*}
&&||x+y||^2-||x-y||^2+i||x+iy||^2-i||x-iy||^2\\
&=&||x||^2+||y||^2-||x||^2-||y||^2+i||x||^2+i||y||^2-i||x||^2-i||y||^2\\
&&+2Re(<x,y>)+2Re(<x,y>)+2iIm(<x,y>)+2iIm(<x,y>)\\
&=&4<x,y>
\end{eqnarray*}

Next, (5) follows as in the real case. Regarding now (6), again inspired by the real case, consider the following quantity, depending on $t\in\mathbb R$, and on $w\in\mathbb T$:
\begin{eqnarray*}
f(t)
&=&||wx+ty||^2\\
&=&<wx+ty,wx+ty>\\
&=&||x||^2+2tRe(w<x,y>)+t^2||y||^2
\end{eqnarray*}

Since $f$ is obviously positive, its discriminant must be negative, which gives:
$$|Re(w<x,y>)|\leq||x||\cdot||y||$$

Now the point is that we can choose $w\in\mathbb T$ such that $w<x,y>\in\mathbb R$, so we obtain the Cauchy-Schwarz inequality, as desired. Finally, (7) follows from (6), exactly as in the real case, then (8) is just a definition, and (9) follows again as in the real case.
\end{proof}
 
Still talking complex scalar products, in relation with linear algebra, we have:

\begin{theorem}
Given a linear map $T:\mathbb C^N\to\mathbb C^N$, the following happen:
\begin{enumerate}
\item We can construct its adjoint $T^*:\mathbb C^N\to\mathbb C^N$, given by $<Tx,y>=<x,T^*y>$.

\item This corresponds to the adjoint matrix $A^*\in M_N(\mathbb C)$, given by $(A^*)_{ij}=\bar{A}_{ji}$.

\item $T(x)=Ux$ with $U\in M_N(\mathbb C)$ is an isometry precisely when $U^*=U^{-1}$.

\item $T(x)=Px$ with $P\in M_N(\mathbb C)$ is a projection precisely when $P^2=P^*=P$.
\end{enumerate}
\end{theorem}

\begin{proof}
This is something very standard, the idea being as follows:

\medskip

(1) Given a linear map $T:\mathbb C^N\to\mathbb C^N$, fix $y\in\mathbb C^N$, and consider the linear form $\varphi(x)=<Tx,y>$. This form must be as follows, for a certain vector $T^*y\in\mathbb C^N$:
$$\varphi(x)=<x,T^*y>$$

Thus, we have constructed a map $y\to T^*y$ as in the statement, which is obviously linear, and we can denote this linear map $T^*:\mathbb C^N\to\mathbb C^N$, and call it adjoint of $T$. 

\medskip

(2) We use the formula from (1), defining the adjoint. By taking the vectors $x,y\in\mathbb C^N$ to be elements of the standard basis of $\mathbb C^N$, our defining formula for $T^*$ reads:
$$<Te_i,e_j>=<e_i,T^*e_j>$$

By reversing the scalar product on the right, this formula can be written as:
$$<T^*e_j,e_i>=\overline{<Te_i,e_j>}$$

But this means that the matrix of $T^*$ is given by $(A^*)_{ij}=\bar{A}_{ji}$, as desired.

\medskip

(3) Given a matrix $U\in M_N(\mathbb C)$, we have indeed the following equivalences, with the first one coming from the polarization identity, and the other ones being clear:
\begin{eqnarray*}
||Ux||=||x||
&\iff&<Ux,Uy>=<x,y>\\
&\iff&<x,U^*Uy>=<x,y>\\
&\iff&U^*Uy=y\\
&\iff&U^*U=1\\
&\iff&U^*=U^{-1}
\end{eqnarray*}

(4) Given a matrix $P\in M_N(\mathbb C)$, in order for $x\to Px$ to be an oblique projection, we must have $P^2=P$. Now observe that this projection is orthogonal when:
\begin{eqnarray*}
<Px-x,Py>=0
&\iff&<P^*Px-P^*x,y>=0\\
&\iff&P^*Px-P^*x=0\\
&\iff&P^*P-P^*=0\\
&\iff&P^*P=P^*
\end{eqnarray*}

The point now is that by conjugating the last formula, we obtain $P^*P=P$. Thus we must have $P=P^*$, and this gives the result. 
\end{proof}

Summarizing, the linear maps come in pairs $T,T^*$, and the associated matrices come as well in pairs $A,A^*$. Getting now to our diagonalization business, we first have:

\index{self-adjoint matrix}
\index{adjoint matrix}
\index{spectral theorem}

\begin{theorem}
Any matrix $A\in M_N(\mathbb C)$ which is self-adjoint, $A=A^*$, is diagonalizable, with the diagonalization being of the following type,
$$A=UDU^*$$
with $U\in U_N$, meaning $U^*=U^{-1}$, and $D\in M_N(\mathbb R)$ diagonal. The converse holds too.
\end{theorem}

\begin{proof}
As a first remark, the converse trivially holds, because if we take a matrix of the form $A=UDU^*$, with $U$ unitary and $D$ diagonal and real, then we have:
$$A^*
=(UDU^*)^*
=UD^*U^*
=UDU^*
=A$$

In the other sense now, assume that $A$ is self-adjoint, $A=A^*$.  Our first claim is that the eigenvalues are real. Indeed, assuming $Av=\lambda v$, we have:
\begin{eqnarray*}
\lambda<v,v>
&=&<Av,v>\\
&=&<v,Av>\\
&=&\bar{\lambda}<v,v>
\end{eqnarray*}

Thus we obtain $\lambda\in\mathbb R$, as claimed. Our next claim now is that the eigenspaces corresponding to different eigenvalues are pairwise orthogonal. Assume indeed that:
$$Av=\lambda v\quad,\quad 
Aw=\mu w$$

We have then the following computation, using the fact that we have $\mu\in\mathbb R$:
\begin{eqnarray*}
\lambda<v,w>
&=&<Av,w>\\
&=&<v,Aw>\\
&=&\mu<v,w>
\end{eqnarray*}

Thus $\lambda\neq\mu$ implies $v\perp w$, as claimed. In order now to finish the proof, it remains to prove that the eigenspaces of $A$ span the whole space $\mathbb C^N$. For this purpose, we will use a recurrence method. Let us pick an eigenvector of our matrix:
$$Av=\lambda v$$

Assuming now that we have a vector $w$ orthogonal to it, $v\perp w$, we have:
\begin{eqnarray*}
<Aw,v>
&=&<w,Av>\\
&=&<w,\lambda v>\\
&=&\lambda<w,v>\\
&=&0
\end{eqnarray*}

Thus, if $v$ is an eigenvector, then the vector space $v^\perp$ is invariant under $A$. Moreover, since a matrix $A$ is self-adjoint precisely when $<Av,v>\in\mathbb R$ for any vector $v\in\mathbb C^N$, as one can see by expanding the scalar product, the restriction of $A$ to the subspace $v^\perp$ is self-adjoint. Thus, we can proceed by recurrence, and we obtain the result.
\end{proof}

Observe that, as a consequence of the above result, that you certainly might have heard of, any symmetric matrix $A\in M_N(\mathbb R)$ is diagonalizable. In fact, we have:

\index{symmetric matrix}

\begin{theorem}
Any matrix $A\in M_N(\mathbb R)$ which is symmetric, $A=A^t$, is diagonalizable over $\mathbb R$, with the diagonalization being of the following type,
$$A=UDU^t$$
with $U\in O_N$, meaning $U^t=U^{-1}$, and with $D$ diagonal. The converse holds too.
\end{theorem}

\begin{proof}
As before, the converse trivially holds, because if we take a matrix of the form $A=UDU^t$, with $U$ orthogonal and $D$ diagonal and real, then we have $A^t=A$. In the other sense now, this follows from Theorem 3.3, and its proof.
\end{proof}

As basic examples of self-adjoint matrices, we have the orthogonal projections:

\index{projection}

\begin{proposition}
The matrices $P\in M_N(\mathbb C)$ which are projections, $P^2=P^*=P$, 
are precisely those which diagonalize as follows,
$$P=UDU^*$$
with $U\in U_N$, and with $D\in M_N(0,1)$ being diagonal.
\end{proposition}

\begin{proof}
Since we have $P^*=P$, by using Theorem 3.3, the eigenvalues must be real. Then, by using $P^2=P$, assuming that we have $Pv=\lambda v$, we obtain:
\begin{eqnarray*}
\lambda<v,v>
&=&<P^2v,v>\\
&=&<Pv,Pv>\\
&=&\lambda^2<v,v>
\end{eqnarray*}

Thus $\lambda\in\{0,1\}$, and the diagonalization must be as follows, with $e_i\in \{0,1\}$:
$$P\sim\begin{pmatrix}
e_1\\
&\ddots\\
&&e_N
\end{pmatrix}$$

To be more precise, the number of 1 values is the dimension of the image of $P$.
\end{proof}

In the real case, the result regarding the projections is as follows:

\begin{proposition}
The matrices $P\in M_N(\mathbb R)$ which are projections, $P^2=P^t=P$, are precisely those which diagonalize as follows,
$$P=UDU^t$$
with $U\in O_N$, and with $D\in M_N(0,1)$ being diagonal.
\end{proposition}

\begin{proof}
This follows indeed from Proposition 3.5, and its proof. 
\end{proof}

An important class of self-adjoint matrices, which includes for instance all the projections, are the positive matrices. The theory here is as follows:

\index{positive matrix}
\index{square root}
\index{positive eigenvalues}

\begin{theorem}
For a matrix $A\in M_N(\mathbb C)$ the following conditions are equivalent, and if they are satisfied, we say that $A$ is positive:
\begin{enumerate}
\item $A=B^2$, with $B=B^*$.

\item $A=CC^*$, for some $C\in M_N(\mathbb C)$.

\item $<Ax,x>\geq0$, for any vector $x\in\mathbb C^N$.

\item $A=A^*$, and the eigenvalues are positive, $\lambda_i\geq0$.

\item $A=UDU^*$, with $U\in U_N$ and with $D\in M_N(\mathbb R_+)$ diagonal.
\end{enumerate}
\end{theorem}

\begin{proof}
The idea is that the equivalences in the statement basically follow from some elementary computations, with only Theorem 3.3 needed, at some point:

\medskip

$(1)\implies(2)$ This is clear, because we can take $C=B$.

\medskip

$(2)\implies(3)$ This follows from the following computation:
$$<Ax,x>
=<CC^*x,x>
=<C^*x,C^*x>
\geq0$$

$(3)\implies(4)$ By using the fact that $<Ax,x>$ is real, we have:
$$<Ax,x>
=<x,A^*x>
=<A^*x,x>$$

Thus we have $A=A^*$, and the remaining assertion, regarding the eigenvalues, follows from the following computation, assuming $Ax=\lambda x$:
$$<Ax,x>
=<\lambda x,x>
=\lambda<x,x>
\geq0$$

$(4)\implies(5)$ This follows indeed by using Theorem 3.3.

\medskip

$(5)\implies(1)$ Assuming $A=UDU^*$ with $U\in U_N$, and with $D\in M_N(\mathbb R_+)$ diagonal, we can set $B=U\sqrt{D}U^*$. Then $B$ is self-adjoint, and $B^2=A$, which gives the result.
\end{proof}

Let us record as well the following technical version of the above result:

\index{strictly positive matrix}

\begin{theorem}
For a matrix $A\in M_N(\mathbb C)$ the following conditions are equivalent, and if they are satisfied, we say that $A$ is strictly positive:
\begin{enumerate}
\item $A=B^2$, with $B=B^*$, invertible.

\item $A=CC^*$, for some $C\in M_N(\mathbb C)$ invertible.

\item $<Ax,x>>0$, for any nonzero vector $x\in\mathbb C^N$.

\item $A=A^*$, and the eigenvalues are strictly positive, $\lambda_i>0$.

\item $A=UDU^*$, with $U\in U_N$ and with $D\in M_N(\mathbb R_+^*)$ diagonal.
\end{enumerate}
\end{theorem}

\begin{proof}
This follows either from Theorem 3.7, by adding the above various extra assumptions, or from the proof of Theorem 3.7, by modifying where needed.
\end{proof}

Let us discuss now some applications of the above. Some very basic examples of symmetric matrices are the adjacency matrices of graphs, and we first have here:

\index{eigenvectors}
\index{eigenvalues}
\index{harmonic function}
\index{sum over neighbors}
\index{average over neighbors}

\begin{proposition}
Given a graph $X$, with adjacency matrix $d\in M_N(0,1)$, the eigenvalues of $d$, with eigenvalue $\lambda$, can be identified with the functions $f$ satisfying:
$$\lambda f(i)=\sum_{i-j}f(j)$$ 
That is, the value of $f$ at each vertex must be the rescaled average, over the neighbors. 
\end{proposition}

\begin{proof}
We have indeed the following computation, valid for any vector $f$:
\begin{eqnarray*}
(df)_i
&=&\sum_jd_{ij}f_j\\
&=&\sum_{i-j}d_{ij}f_j+\sum_{i\not-j}d_{ij}f_j\\
&=&\sum_{i-j}1\cdot f_j+\sum_{i\not-j}0\cdot f_j\\
&=&\sum_{i-j}f_j
\end{eqnarray*}

Thus, we are led to the conclusion in the statement.
\end{proof}

The above result is quite interesting, and as an illustration, when assuming that our graph is $k$-regular, for the particular value $\lambda=k$, the eigenvalue condition reads:
$$f(i)=\frac{1}{k}\sum_{i-j}f(j)$$ 

Thus, we can see here a relation with harmonic functions. There are many things that can be said here, and we will be back to this later in this book. Getting now to the point, we can apply to $d\in M_N(0,1)$ our spectral theorems, and we obtain:

\index{trace of matrix}

\begin{theorem}
The adjacency matrix $d\in M_N(0,1)$ of any graph is diagonalizable, with the diagonalization being of the following type,
$$d=UDU^t$$
with $U\in O_N$, and with $D\in M_N(\mathbb R)$ diagonal. Moreover, we have $Tr(D)=0$.
\end{theorem}

\begin{proof}
Here the first assertion follows from Theorem 3.4, because $d$ is by definition real and symmetric. As for the last assertion, this deserves some explanations:

\medskip

(1) Generally speaking, in analogy with the last assertions in Theorem 3.3 and Theorem 3.4, which are something extremely useful, we would like to know under which assumptions on a rotation $U\in O_N$, and on a diagonal matrix $D\in M_N(\mathbb R)$, the real symmetric matrix $d=UDU^t$ has 0-1 entries, and 0 on the diagonal.

\medskip

(2) Unfortunately, both these questions are obviously difficult, there is no simple answer to them, and things are like that. So, gone the possibility of a converse. However, as a small consolation, we can make the remark that, with $d=UDU^t$, we have:
$$Tr(d)=Tr(UDU^t)=Tr(D)$$

Thus we have at least $Tr(D)=0$, as a necessary condition on $(U,D)$, as stated.
\end{proof}

In view of the above difficulties with the bijectivity, it is perhaps wise to formulate as well the graph particular case of Theorem 3.3. The statement here is as follows:

\begin{theorem}
The adjacency matrix $d\in M_N(0,1)$ of any graph is diagonalizable, with the diagonalization being of the following type,
$$d=UDU^*$$
with $U\in U_N$, and with $D\in M_N(\mathbb R)$ diagonal. Moreover, we have $Tr(D)=0$.
\end{theorem}

\begin{proof}
This follows from Theorem 3.3, via the various remarks from the proof of Proposition 3.9 and Theorem 3.10. But the simplest is to say that the statement itself is just a copy of Theorem 3.10, with $U\in O_N$ replaced by the weaker condition $U\in U_N$.
\end{proof}

As a concrete illustration now for all this, let us look at the simplest graph of them all, namely the simplex, or complete graph. And for this graph, not only the above results successfully apply, but we can also see why Theorem 3.11 is something wise:

\index{flat matrix}
\index{Fourier matrix}

\begin{theorem}
The adjacency matrix of the simplex diagonalizes as follows,
$$\begin{pmatrix}
0&1&\ldots&1&1\\
1&0&\ldots&1&1\\
\vdots&\vdots&&\vdots&\vdots\\
1&1&\ldots&0&1\\
1&1&\ldots&1&0
\end{pmatrix}
=\frac{1}{N}\,F_N
\begin{pmatrix}
N-1&&&&0\\
&-1\\
&&\ddots&\\
&&&-1\\
0&&&&-1
\end{pmatrix}F_N^*$$
with $F_N=(w^{ij})_{ij}$ with $w=e^{2\pi i/N}$ being as usual the Fourier matrix.
\end{theorem}

\begin{proof}
The adjacency matrix of the simplex is $d=\mathbb I_N-1_N$, with $\mathbb I_N$ being the all-one, or flat matrix. So, let us first attempt to diagonalize the flat matrix:
$$\mathbb I_N=\begin{pmatrix}
1&\ldots&\ldots&1\\
\vdots&&&\vdots\\
\vdots&&&\vdots\\
1&\ldots&\ldots&1\end{pmatrix}$$

But here, we already know, since chapter 1, that the simplest way to diagonalize this matrix is as follows, with $F_N=(w^{ij})_{ij}$ being the Fourier matrix:
$$\begin{pmatrix}
1&\ldots&\ldots&1\\
\vdots&&&\vdots\\
\vdots&&&\vdots\\
1&\ldots&\ldots&1\end{pmatrix}
=\frac{1}{N}\,F_N
\begin{pmatrix}
N&&&&0\\
&0\\
&&\ddots\\
&&&0\\
0&&&&0\end{pmatrix}F_N^*$$

But this gives the result, by substracting $-1$ from everything.
\end{proof}

All the above is quite useful, and we will use these results on a regular basis, in what follows. There are also some positivity considerations that can be made, in relation with graphs, and we will be back to this later, when talking Laplacians of graphs.

\section*{3b. Rotations, unitaries}

Let us discuss now the case of the unitary matrices. We have here:

\index{unitary}
\index{spectral theorem}

\begin{theorem}
Any matrix $U\in M_N(\mathbb C)$ which is unitary, $U^*=U^{-1}$, is diagonalizable, with the eigenvalues on $\mathbb T$. More precisely we have
$$U=VDV^*$$
with $V\in U_N$, and with $D\in M_N(\mathbb T)$ diagonal. The converse holds too.
\end{theorem}

\begin{proof}
As a first remark, the converse trivially holds, because given a matrix of type $U=VDV^*$, with $V\in U_N$, and with $D\in M_N(\mathbb T)$ being diagonal, we have:
$$U^*
=VD^*V^*
=(VDV^*)^{-1}
=U^{-1}$$

Let us prove now the first assertion, stating that the eigenvalues of a unitary matrix $U\in U_N$ belong to $\mathbb T$. Indeed, assuming $Uv=\lambda v$, we have:
\begin{eqnarray*}
<v,v>
&=&<U^*Uv,v>\\
&=&<Uv,Uv>\\
&=&<\lambda v,\lambda v>\\
&=&|\lambda|^2<v,v>
\end{eqnarray*}

Thus $\lambda\in\mathbb T$, as claimed. Our next claim now is that the eigenspaces corresponding to different eigenvalues are pairwise orthogonal. Assume indeed that $Uv=\lambda v$, $Uw=\mu w$. We have then the following computation, using $U^*=U^{-1}$ and $\lambda,\mu\in\mathbb T$:
\begin{eqnarray*}
\lambda<v,w>
&=&<Uv,w>\\
&=&<v,U^*w>\\
&=&<v,U^{-1}w>\\
&=&<v,\mu^{-1}w>\\
&=&\mu<v,w>
\end{eqnarray*}

Thus $\lambda\neq\mu$ implies $v\perp w$, as claimed. In order now to finish the proof, it remains to prove that the eigenspaces of $U$ span the whole space $\mathbb C^N$. For this purpose, we will use a recurrence method. Let us pick an eigenvector of our matrix:
$$Uv=\lambda v$$

Assuming that we have a vector $w$ orthogonal to it, $v\perp w$, we have:
\begin{eqnarray*}
<Uw,v>
&=&<w,U^*v>\\
&=&<w,U^{-1}v>\\
&=&<w,\lambda^{-1}v>\\
&=&\lambda<w,v>\\
&=&0
\end{eqnarray*}

Thus, if $v$ is an eigenvector, then the vector space $v^\perp$ is invariant under $U$. Now since $U$ is an isometry, so is its restriction to this space $v^\perp$. Thus this restriction is a unitary, and so we can proceed by recurrence, and we obtain the result.
\end{proof}

Let us record as well the real version of the above result, in a weak form:

\index{orthogonal matrix}

\begin{proposition}
Any matrix $U\in M_N(\mathbb R)$ which is orthogonal, $U^t=U^{-1}$,
is diagonalizable, with the eigenvalues on $\mathbb T$. More precisely we have
$$U=VDV^*$$
with $V\in U_N$, and with $D\in M_N(\mathbb T)$ being diagonal.
\end{proposition}

\begin{proof}
This follows indeed from Theorem 3.13.
\end{proof}

Observe that the above result does not provide us with a complete characterization of the matrices $U\in M_N(\mathbb R)$ which are orthogonal. To be more precise, the question left is that of understanding when the matrices of type $U=VDV^*$, with $V\in U_N$, and with $D\in M_N(\mathbb T)$ being diagonal, are real, and this is something non-trivial.

\bigskip

As an illustration for the above, for the simplest unitaries that we know, namely the rotations in the real plane, we have the following result:

\begin{theorem}
The rotation of angle $t\in\mathbb R$ in the real plane, namely
$$R_t=\begin{pmatrix}\cos t&-\sin t\\ \sin t&\cos t\end{pmatrix}$$
can be diagonalized over the complex numbers, as follows:
$$R_t=\frac{1}{2}\begin{pmatrix}1&1\\i&-i\end{pmatrix}
\begin{pmatrix}e^{-it}&0\\0&e^{it}\end{pmatrix}
\begin{pmatrix}1&-i\\1&i\end{pmatrix}$$
Over the real numbers this is impossible, unless $t=0,\pi$.
\end{theorem}

\begin{proof}
This is indeed something that we know since chapter 1, and we refer to the discussion there for the above assertions, both over $\mathbb R$ and $\mathbb C$.
\end{proof}

In two complex dimensions now, it is convenient to restrict the attention to the unitaries $U\in SU_2$ of determinant 1, which are subject to the following key result:

\begin{theorem}
We have the following formula,
$$SU_2=\left\{\begin{pmatrix}a&b\\ -\bar{b}&\bar{a}\end{pmatrix}\ \Big|\ |a|^2+|b|^2=1\right\}$$
which makes $SU_2$ isomorphic to the unit sphere $S^1_\mathbb C\subset\mathbb C^2$.
\end{theorem}

\begin{proof}
Consider indeed an arbitrary $2\times 2$ matrix, written as follows:
$$U=\begin{pmatrix}a&b\\ c&d\end{pmatrix}$$

Assuming that we have $\det U=1$, the equation $U^{-1}=U^*$ reads:
$$\begin{pmatrix}d&-b\\ -c&a\end{pmatrix}
=\begin{pmatrix}\bar{a}&\bar{c}\\ \bar{b}&\bar{d}\end{pmatrix}$$

We are therefore led to the following equations, for the matrix entries:
$$d=\bar{a}\quad,\quad 
c=-\bar{b}$$

Thus our matrix must be of the following special form in the statement. Moreover, since the determinant of this matrix is 1, we must have, as stated:
$$|a|^2+|b|^2=1$$

Thus, done with one inclusion. As for the reverse inclusion, this is clear.
\end{proof}

According to Theorem 3.13 the matrices in Theorem 3.16 are diagonalizable, and we will leave their diagonalization as an instructive exercise. Moving forward now, in 3 dimensions things are more complicated, and in order to discuss this, we will need:

\begin{theorem}
We have the following formula,
$$SU_2=\left\{pc_1+qc_2+rc_3+sc_4\ \Big|\ p^2+q^2+r^2+s^2=1\right\}$$
where $c_1,c_2,c_3,c_4$ are the following matrices,
$$c_1=\begin{pmatrix}1&0\\ 0&1\end{pmatrix}\quad,\quad
c_2=\begin{pmatrix}i&0\\ 0&-i\end{pmatrix}\quad,\quad
c_3=\begin{pmatrix}0&1\\ -1&0\end{pmatrix}\quad,\quad 
c_4=\begin{pmatrix}0&i\\ i&0\end{pmatrix}$$
called Pauli spin matrices.
\end{theorem}

\begin{proof}
We know from Theorem 3.16 that the group $SU_2$ can be parametrized by the points of the real sphere $S^3_\mathbb R\subset\mathbb R^4$, in the following way:
$$SU_2=\left\{\begin{pmatrix}p+iq&r+is\\ -r+is&p-iq\end{pmatrix}\ \Big|\ p^2+q^2+r^2+s^2=1\right\}$$

But this gives the formula in the statement, with the Pauli matrices $c_1,c_2,c_3,c_4$ being the coefficients of $p,q,r,s$, in this parametrization.
\end{proof}

The Pauli matrices have a number of remarkable properties, as follows:

\begin{proposition}
The Pauli matrices multiply according to the following formulae,
$$c_2^2=c_3^2=c_4^2=-1$$
$$c_2c_3=-c_3c_2=c_4$$
$$c_3c_4=-c_4c_3=c_2$$
$$c_4c_2=-c_2c_4=c_3$$
they conjugate according to the following rules,
$$c_1^*=c_1,\ c_2^*=-c_2,\ c_3^*=-c_3,\ c_4^*=-c_4$$
and they form an orthonormal basis of $M_2(\mathbb C)$, with respect to the scalar product
$$<x,y>=tr(xy^*)$$
with $tr:M_2(\mathbb C)\to\mathbb C$ being the normalized trace of $2\times 2$ matrices, $tr=Tr/2$.
\end{proposition}

\begin{proof}
The various formulae are all clear. Next, the fact that the Pauli matrices are pairwise orthogonal follows from computations of the following type, for $i\neq j$:
$$<c_i,c_j>
=tr(c_ic_j^*)
=tr(\pm c_ic_j)
=tr(\pm c_k)
=0$$

As for the fact that the Pauli matrices have norm 1, this follows from:
$$<c_i,c_i>
=tr(c_ic_i^*)
=tr(\pm c_i^2)
=tr(c_1)
=1$$

Thus, we are led to the conclusion in the statement.
\end{proof}

Getting now towards $SO_3$, we first have the following result:

\begin{proposition}
The adjoint action $SU_2\curvearrowright M_2(\mathbb C)$, given by $T_U(A)=UAU^*$, leaves invariant the following real vector subspace of $M_2(\mathbb C)$,
$$\mathbb R^4=span(c_1,c_2,c_3,c_4)$$
and we obtain in this way a group morphism $SU_2\to GL_4(\mathbb R)$.
\end{proposition}

\begin{proof}
We have two assertions to be proved, as follows:

\medskip

(1) We must first prove that, with $E\subset M_2(\mathbb C)$ being the real vector space in the statement, we have the following implication:
$$U\in SU_2,A\in E\implies UAU^*\in E$$

But this is clear from the multiplication rules for the Pauli matrices, from Proposition 3.18. Indeed, let us write our matrices $U,A$ as follows:
$$U=xc_1+yc_2+zc_3+tc_4$$
$$A=ac_1+bc_2+cc_3+dc_4$$

We know that the coefficients $x,y,z,t$ and $a,b,c,d$ are all real, due to $U\in SU_2$ and $A\in E$. The point now is that when computing $UAU^*$, by using the various rules from Proposition 3.18, we obtain a matrix of the same type, namely a combination of $c_1,c_2,c_3,c_4$, with real coefficients. Thus, we have $UAU^*\in E$, as desired.

\medskip

(2) In order to conclude, let us identify $E\simeq\mathbb R^4$, by using the basis $c_1,c_2,c_3,c_4$. The result found in (1) shows that we have a correspondence as follows:
$$SU_2\to M_4(\mathbb R)\quad,\quad U\to (T_U)_{|E}$$

Now observe that for any $U\in SU_2$ and any $A\in M_2(\mathbb C)$ we have:
$$T_{U^*}T_U(A)=U^*UAU^*U=A$$

Thus $T_{U^*}=T_U^{-1}$, and so the correspondence that we found can be written as:
$$SU_2\to GL_4(\mathbb R)\quad,\quad U\to (T_U)_{|E}$$

But this a group morphism, due to the following computation:
$$T_UT_V(A)=UVAV^*U^*=T_{UV}(A)$$

Thus, we are led to the conclusion in the statement.
\end{proof}

The above result is quite interesting, and as a continuation of it, we have:

\begin{proposition}
With respect to the standard basis $c_1,c_2,c_3,c_4$ of the vector space $\mathbb R^4=span(c_1,c_2,c_3,c_4)$, the morphism $T:SU_2\to GL_4(\mathbb R)$ is given by:
$$T_U=\begin{pmatrix}
1&0&0&0\\
0&p^2+q^2-r^2-s^2&2(qr-ps)&2(pr+qs)\\
0&2(ps+qr)&p^2+r^2-q^2-s^2&2(rs-pq)\\
0&2(qs-pr)&2(pq+rs)&p^2+s^2-q^2-r^2
\end{pmatrix}$$
Thus, when looking at $T$ as a group morphism $SU_2\to O_4$, what we have in fact is a group morphism $SU_2\to O_3$, and even $SU_2\to SO_3$.
\end{proposition}

\begin{proof}
With notations from Proposition 3.19 and its proof, let us first look at the action $L:SU_2\curvearrowright\mathbb R^4$ by left multiplication, $L_U(A)=UA$. We have:
$$L_U=\begin{pmatrix}
p&-q&-r&-s\\
q&p&-s&r\\
r&s&p&-q\\
s&-r&q&p
\end{pmatrix}$$

Similarly, in what regards now the action $R:SU_2\curvearrowright\mathbb R^4$ by right multiplication, $R_U(A)=AU^*$, the corresponding matrix is given by:
$$R_U=\begin{pmatrix}
p&q&r&s\\
-q&p&-s&r\\
-r&s&p&-q\\
-s&-r&q&p
\end{pmatrix}$$

Now since we have $T_U=R_UL_U$, we obtain the formula in the statement.
\end{proof}

We can now formulate a famous result, due to Euler-Rodrigues, as follows:

\index{double cover map}
\index{Euler-Rodrigues formula}
\index{rotation}

\begin{theorem}
We have the Euler-Rodrigues formula
$$U=\begin{pmatrix}
p^2+q^2-r^2-s^2&2(qr-ps)&2(pr+qs)\\
2(ps+qr)&p^2+r^2-q^2-s^2&2(rs-pq)\\
2(qs-pr)&2(pq+rs)&p^2+s^2-q^2-r^2
\end{pmatrix}$$
with $p^2+q^2+r^2+s^2=1$, for the generic elements of $SO_3$.
\end{theorem}

\begin{proof}
We know from the above that we have a group morphism $SU_2\to SO_3$, given by the formula in the statement, and the problem now is that of proving that this is a double cover map, in the sense that it is surjective, and with kernel $\{\pm1\}$.

\medskip

(1) Regarding the kernel, this is elementary to compute, as follows:
\begin{eqnarray*}
\ker(SU_2\to SO_3)
&=&\left\{U\in SU_2\Big|UA=AU,\forall A\in E\right\}\\
&=&\left\{U\in SU_2\Big|Uc_i=c_iU,\forall i\right\}\\
&=&\{\pm1\}
\end{eqnarray*}

(2) Thus, we are left with proving that our group morphism $SU_2\to SO_3$ is surjective. For this purpose, we will use the well-known fact, due to Euler, that any rotation in 3D space $U\in SO_3$ has a rotation axis. Indeed, we have the following computation:
\begin{eqnarray*}
\det(U-1)
&=&\det(U^t-1)\\
&=&\det(U^t(1-U))\\
&=&\det(U^t)\det(1-U)\\
&=&\det(1-U)
\end{eqnarray*}

Thus $\det(U-1)=0$, which tells us that $U$ must have a $1$-eigenvector:
$$U\xi=\xi$$

Thus, we got our rotation axis for our abstract rotation $U\in SO_3$, as desired. 

\medskip

(3) More concretely now, in relation with our problem, we must look for fixed points of rotations of type $U\in Im(SU_2\to SO_3)$. By linearity it is enough to look for fixed points belonging to the sphere $S^2_\mathbb R\subset\mathbb R^3$. Now recall that in our picture for the quotient map $SU_2\to SO_3$, the space $\mathbb R^3$ appears as $F=span_\mathbb R(c_2,c_3,c_4)$, naturally embedded into the space $\mathbb R^4$ appearing as $E=span_\mathbb R(c_1,c_2,c_3,c_4)$. Thus, we must look for fixed points belonging to the sphere $S^3_\mathbb R\subset\mathbb R^4$ whose first coordinate vanishes. But, in our $\mathbb R^4=E$ picture, this sphere $S^3_\mathbb R$ is the group $SU_2$. Thus, we must look for fixed points $V\in SU_2$ whose first coordinate with respect to $c_1,c_2,c_3,c_4$ vanishes, which amounts in saying that the diagonal entries of $V$ must be purely imaginary numbers.

\medskip

(4) Long story short, via our various identifications, we are led into solving the equation $UV=VU$ with $U,V\in SU_2$, and with $V$ having a purely imaginary diagonal. So, with standard notations for $SU_2$, we must solve the following equation, with $p\in i\mathbb R$:
$$\begin{pmatrix}a&b\\-\bar{b}&\bar{a}\end{pmatrix}
\begin{pmatrix}p&q\\-\bar{q}&\bar{p}\end{pmatrix}
=\begin{pmatrix}p&q\\-\bar{q}&\bar{p}\end{pmatrix}
\begin{pmatrix}a&b\\-\bar{b}&\bar{a}\end{pmatrix}$$

But this is something which is routine. Indeed, by identifying coefficients we obtain the following equations, each appearing twice:
$$b\bar{q}=\bar{b}q\quad,\quad b(p-\bar{p})=(a-\bar{a})q$$

In the case $b=0$ the only equation which is left is $q=0$, and reminding that we must have $p\in i\mathbb R$, we do have solutions, namely two of them, as follows:
$$V=\pm\begin{pmatrix}i&0\\0&i\end{pmatrix}$$

(5) In the remaining case $b\neq0$, the first equation reads $b\bar{q}\in\mathbb R$, so we must have $q=\lambda b$ with $\lambda\in\mathbb R$. Now with this substitution made, the second equation reads $p-\bar{p}=\lambda(a-\bar{a})$, and since we must have $p\in i\mathbb R$, this gives $2p=\lambda(a-\bar{a})$. Thus, our equations are:
$$q=\lambda b\quad,\quad p=\lambda\cdot\frac{a-\bar{a}}{2}$$

Getting back now to our problem about finding fixed points, assuming $|a|^2+|b|^2=1$ we must find $\lambda\in\mathbb R$ such that the above numbers $p,q$ satisfy $|p|^2+|q|^2=1$. But:
\begin{eqnarray*}
|p|^2+|q|^2
&=&|\lambda b|^2+\left|\lambda\cdot\frac{a-\bar{a}}{2}\right|^2\\
&=&\lambda^2(|b|^2+Im(a)^2)\\
&=&\lambda^2(1-Re(a)^2)
\end{eqnarray*}

Thus, we have again two solutions to our fixed point problem, given by:
$$\lambda=\pm\frac{1}{\sqrt{1-Re(a)^2}}$$

Summarizing, we have proved that any rotation $U\in Im(SU_2\to SO_3)$ has an axis, and with the direction of this axis, corresponding to a pair of opposite points on the sphere $S^2_\mathbb R\subset\mathbb R^3$, being given by the above formulae, via $S^2_\mathbb R\subset S^3_\mathbb R=SU_2$.

\medskip

(6) Now since $U\in SO_3$ is uniquely determined by its rotation axis, which can be regarded as a point of $S^2_\mathbb R/\{\pm1\}$, plus its rotation angle $t\in[0,2\pi)$, by using $S^2_\mathbb R\subset S^3_\mathbb R=SU_2$ we are led to the conclusion that $U$ is uniquely determined by an element of $SU_2/\{\pm 1\}$, and so appears indeed via the Euler-Rodrigues formula, as desired.
\end{proof}

\section*{3c. Normal matrices} 

Back to generalities, the self-adjoint matrices and the unitary matrices are particular cases of the general notion of a ``normal matrix'', and we have here:

\index{normal matrix}
\index{diagonalizable matrix}
\index{spectral theorem}

\begin{theorem}
Any matrix $A\in M_N(\mathbb C)$ which is normal, $AA^*=A^*A$, is diagonalizable, with the diagonalization being of the following type,
$$A=UDU^*$$
with $U\in U_N$, and with $D\in M_N(\mathbb C)$ diagonal. The converse holds too.
\end{theorem}

\begin{proof}
As a first remark, the converse trivially holds, because if we take a matrix of the form $A=UDU^*$, with $U$ unitary and $D$ diagonal, then we have:
$$AA^*
=UDD^*U^*
=UD^*DU^*
=A^*A$$

In the other sense now, this is something more technical. Our first claim is that a matrix $A$ is normal precisely when the following happens, for any vector $v$:
$$||Av||=||A^*v||$$

Indeed, the above equality can be written as follows:
$$<AA^*v,v>=<A^*Av,v>$$

But this is equivalent to $AA^*=A^*A$, by expanding the scalar products. Our claim now is that $A,A^*$ have the same eigenvectors, with conjugate eigenvalues:
$$Av=\lambda v\implies A^*v=\bar{\lambda}v$$

Indeed, this follows from the following computation, and from the trivial fact that if $A$ is normal, then so is any matrix of type $A-\lambda 1_N$:
\begin{eqnarray*}
||(A^*-\bar{\lambda}1_N)v||
&=&||(A-\lambda 1_N)^*v||\\
&=&||(A-\lambda 1_N)v||\\
&=&0
\end{eqnarray*}

Let us prove now, by using this, that the eigenspaces of $A$ are pairwise orthogonal. Assume that we have two eigenvectors, corresponding to different eigenvalues, $\lambda\neq\mu$:
$$Av=\lambda v\quad,\quad 
Aw=\mu w$$

We have the following computation, which shows that $\lambda\neq\mu$ implies $v\perp w$:
\begin{eqnarray*}
\lambda<v,w>
&=&<\lambda v,w>\\
&=&<Av,w>\\
&=&<v,A^*w>\\
&=&<v,\bar{\mu}w>\\
&=&\mu<v,w>
\end{eqnarray*}

In order to finish, it remains to prove that the eigenspaces of $A$ span the whole $\mathbb C^N$. This is something that we have already seen for the self-adjoint matrices, and for unitaries, and we will use here these results, in order to deal with the general normal case. As a first observation, given an arbitrary matrix $A$, the matrix $AA^*$ is self-adjoint:
$$(AA^*)^*=AA^*$$

Thus, we can diagonalize this matrix $AA^*$, as follows, with the passage matrix being a unitary, $V\in U_N$, and with the diagonal form being real, $E\in M_N(\mathbb R)$:
$$AA^*=VEV^*$$

Now observe that, for matrices of type $A=UDU^*$, which are those that we supposed to deal with, we have the following formulae:
$$V=U\quad,\quad 
E=D\bar{D}$$

In particular, the matrices $A$ and $AA^*$ have the same eigenspaces. So, this will be our idea, proving that the eigenspaces of $AA^*$ are eigenspaces of $A$. In order to do so, let us pick two eigenvectors $v,w$ of the matrix $AA^*$, corresponding to different eigenvalues, $\lambda\neq\mu$. The eigenvalue equations are then as follows:
$$AA^*v=\lambda v\quad,\quad 
AA^*w=\mu w$$

We have the following computation, using the normality condition $AA^*=A^*A$, and the fact that the eigenvalues of $AA^*$, and in particular $\mu$, are real:
\begin{eqnarray*}
\lambda<Av,w>
&=&<\lambda Av,w>\\
&=&<A\lambda v,w>\\
&=&<AAA^*v,w>\\
&=&<AA^*Av,w>\\
&=&<Av,AA^*w>\\
&=&<Av,\mu w>\\
&=&\mu<Av,w>
\end{eqnarray*}

We conclude that we have $<Av,w>=0$. But this reformulates as follows:
$$\lambda\neq\mu\implies A(E_\lambda)\perp E_\mu$$

Now since the eigenspaces of $AA^*$ are pairwise orthogonal, and span the whole $\mathbb C^N$, we deduce from this that these eigenspaces are invariant under $A$:
$$A(E_\lambda)\subset E_\lambda$$

But with this result in hand, we can now finish. Indeed, we can decompose the problem, and the matrix $A$ itself, following these eigenspaces of the matrix $AA^*$, which in practice amounts in saying that we can assume that we only have 1 eigenspace. By rescaling, this is the same as assuming that we have $AA^*=1$, and so we are now into the unitary case, that we know how to solve, as explained in Theorem 3.13.
\end{proof}

So long for spectral theorems. As a basic application, we have the following result:

\index{absolute value}
\index{modulus of matrix}

\begin{theorem}
Given a matrix $A\in M_N(\mathbb C)$, the following happen:
\begin{enumerate}
\item We can construct its modulus $|A|\geq0$ via the formula $|A|=\sqrt{A^*A}$.

\item When $A$ is normal, the modulus is equally given by $|A|=\sqrt{AA^*}$.

\item When $A$ is invertible, we have $A=U|A|$, with $U$ being a unitary.

\item In general, we still have $A=U|A|$, with $U$ being a partial isometry.
\end{enumerate}
\end{theorem}

\begin{proof}
This is something very standard, the idea being as follows:

\medskip

(1) We have $A^*A\geq0$, so according to Theorem 3.7, we can diagonalize this matrix in the following way, with $V\in U_N$, and with $D\in M_N(\mathbb R_+)$ being diagonal:
$$A^*A=VDV^*$$

But with this done, we can construct the modulus as $|A|=V\sqrt{D}V^*$, as desired.

\medskip

(2) This is indeed something which is clear from definitions.

\medskip

(3) According to our definition of the modulus, $|A|=\sqrt{A^*A}$, we have:
\begin{eqnarray*}
<|A|x,|A|y>
&=&<x,|A|^2y>\\
&=&<x,A^*Ay>\\
&=&<Ax,Ay>
\end{eqnarray*}

We conclude that the following linear map is well-defined, and isometric:
$$U:Im|A|\to Im(A)\quad,\quad 
|A|x\to Ax$$

Now when $A$ is invertible, this map is a unitary $U\in U_N$, and $A=U|A|$, as desired.

\medskip

(4) In the general case, we can further extend the above linear isometric map $U$ into a partial isometry $U:\mathbb C^N\to\mathbb C^N$, in a straightforward way, by setting:
$$Ux=0\quad,\quad\forall x\in Im|A|^\perp$$

And the point is that, with this convention, the result follows. All this was of course quite quick, but we will be back to this, with further details, in chapter 6.
\end{proof}

\section*{3d. Spectral measures}

We would like to discuss now some interesting applications of our various spectral theorems to probability theory. Let us start with something basic, as follows:

\index{probability space}
\index{random variable}
\index{moments}
\index{law}
\index{distribution}

\begin{definition}
Let $X$ be a probability space, that is, a space with a probability measure, and with the corresponding integration denoted $E$, and called expectation.
\begin{enumerate}
\item The random variables are the real functions $f\in L^\infty(X)$.

\item The moments of such a variable are the numbers $M_k(f)=E(f^k)$.

\item The law of such a variable is the measure given by $M_k(f)=\int_\mathbb Rx^kd\mu_f(x)$.
\end{enumerate}
\end{definition}

Here, and in what follows, we use the term ``law'' for ``probability distribution'', which means exactly the same thing, and is more convenient. Regarding now the fact that the law $\mu_f$ exists indeed, this is true, but not exactly trivial. By linearity, we would like to have a probability measure making hold the following formula, for any $P\in\mathbb C[X]$:
$$E(P(f))=\int_\mathbb RP(x)d\mu_f(x)$$

By using a standard continuity argument, it is enough to have this formula for the characteristic functions $\chi_I$ of the arbitrary measurable sets of real numbers $I\subset\mathbb R$:
$$E(\chi_I(f))=\int_\mathbb R\chi_I(x)d\mu_f(x)$$

But this latter formula, which reads $P(f\in I)=\mu_f(I)$, can serve as a definition for $\mu_f$, and we are done. Alternatively, assuming some familiarity with measure theory, $\mu_f$ is the push-forward of the probability measure on $X$, via the function $f:X\to\mathbb R$. 

\bigskip

Let us summarize this discussion in the form of a theorem, as follows:

\begin{theorem}
The law $\mu_f$ of a random variable $f$ exists indeed, and we have
$$E(\varphi(f))=\int_\mathbb R\varphi(x)d\mu_f(x)$$
for any integrable function $\varphi:\mathbb R\to\mathbb C$.
\end{theorem}

\begin{proof}
This follows from the above discussion, and with the precise assumption on $\varphi:\mathbb R\to\mathbb C$, which is its integrability, in the abstract mathematical sense, being in fact something that we will not really need, in what follows. In fact, for most purposes we will get away with polynomials $\varphi\in\mathbb C[X]$, and by linearity this means that we can get away with monomials $\varphi(x)=x^k$, which brings us back to Definition 3.24 (3), as stated.
\end{proof}

Getting now to the case of the matrices $A\in M_N(\mathbb C)$, here it is quite tricky to figure out what the law of $A$ should mean, based on intuition only. So, in the lack of a bright idea, let us just reproduce Definition 3.24, with a few modifications, as follows:

\begin{definition}
Let $N\in\mathbb N$, and consider the algebra $M_N(\mathbb C)$ of complex $N\times N$ matrices, with its normalized trace $tr:M_N(\mathbb C)\to\mathbb C$, given by $tr(A)=Tr(A)/N$.
\begin{enumerate}
\item We call random variables the self-adjoint matrices $A\in M_N(\mathbb C)$.

\item The moments of such a variable are the numbers $M_k(A)=tr(A^k)$.

\item The law of such a variable is the measure given by $M_k(A)=\int_\mathbb Rx^kd\mu_A(x)$.
\end{enumerate}
\end{definition}

What we have here looks quite reasonable, but as before with the usual random variables $f\in L^\infty(X)$, some discussion is needed, in order to understand if the law $\mu_A$ exists indeed, and by which mechanism. And, good news here, in the case of the simplest self-adjoint matrices, namely the real diagonal ones, we have:

\begin{theorem}
For any diagonal matrix $A\in M_N(\mathbb R)$ we have the formula
$$tr(P(A))=\frac{1}{N}(P(\lambda_1)+\ldots+P(\lambda_N))$$
where $\lambda_1,\ldots,\lambda_N\in\mathbb R$ are the diagonal entries of $A$. Thus the measure
$$\mu_A=\frac{1}{N}(\delta_{\lambda_1}+\ldots+\delta_{\lambda_N})$$
can be regarded as being the law of $A$, in the sense of Definition 3.26.
\end{theorem}

\begin{proof}
Assume indeed that we have a real diagonal matrix, as follows, with the convention that the matrix entries which are missing are by definition 0 entries:
$$A=\begin{pmatrix}
\lambda_1\\
&\ddots\\
&&\lambda_N\end{pmatrix}$$

Given a polynomial $P\in\mathbb C[X]$, we have then the following formula:
$$P(A)=\begin{pmatrix}
P(\lambda_1)\\
&\ddots\\
&&P(\lambda_N)\end{pmatrix}$$

Thus, the first formula in the statement holds indeed. In particular, we conclude that the moments of $A$ are given by the following formula:
$$M_k(A)=tr(A^k)=\frac{1}{N}\sum_i\lambda_i^k$$

On the other hand, with $\mu_A=\frac{1}{N}(\delta_{\lambda_1}+\ldots+\delta_{\lambda_N})$ as in the statement, we have:
$$\int_\mathbb Rx^kd\mu_A(x)
=\frac{1}{N}\sum_i\int_\mathbb Rx^kd\delta_{\lambda_i}(x)
=\frac{1}{N}\sum_i\lambda_i^k$$

Thus that the law of $A$ exists indeed, and is the measure $\mu_A$, as claimed.
\end{proof}

The point now is that, by using the spectral theorem for self-adjoint matrices, we have the following generalization of Theorem 3.27, dealing with the general case:

\begin{theorem}
For a self-adjoint matrix $A\in M_N(\mathbb C)$ we have the formula
$$tr(P(A))=\frac{1}{N}(P(\lambda_1)+\ldots+P(\lambda_N))$$
where $\lambda_1,\ldots,\lambda_N\in\mathbb R$ are the eigenvalues of $A$. Thus the measure
$$\mu_A=\frac{1}{N}(\delta_{\lambda_1}+\ldots+\delta_{\lambda_N})$$
can be regarded as being the law of $A$, in the sense of Definition 3.26.
\end{theorem}

\begin{proof}
We already know, from Theorem 3.27, that the result holds indeed for the diagonal matrices. In the general case now, that of an arbitrary self-adjoint matrix, we know from Theorem 3.3 that our matrix is diagonalizable, as follows:
$$A=UDU^*$$

Now observe that the moments of $A$ are given by the following formula:
\begin{eqnarray*}
tr(A^k)
&=&tr(UDU^*\cdot UDU^*\ldots UDU^*)\\
&=&tr(UD^kU^*)\\
&=&tr(D^k)
\end{eqnarray*}

We conclude from this, by reasoning by linearity, that the matrices $A,D$ have the same law, $\mu_A=\mu_D$, and this gives all the assertions in the statement.
\end{proof}

The above theory is not the end of the story, because we can talk about complex random variables, $f:X\to\mathbb C$, and about non-self-adjoint matrices too, $A\neq A^*$. We will see that, with a bit of know-how, we can have some law technology going on, for both.

\bigskip

Let us start with the complex variables $f\in L^\infty(X)$. The main difference with respect to the real case comes from the fact that we have now a pair of variables instead of one, namely $f:X\to\mathbb C$ itself, and its conjugate $\bar{f}:X\to\mathbb C$. Thus, we are led to:

\index{colored integers}
\index{colored moments}

\begin{definition}
The moments a complex variable $f\in L^\infty(X)$ are the numbers
$$M_k(f)=E(f^k)$$
depending on colored integers $k=\circ\bullet\bullet\circ\ldots\,$, with the conventions
$$f^\emptyset=1\quad,\quad f^\circ=f\quad,\quad f^\bullet=\bar{f}$$
and multiplicativity, in order to define the colored powers $f^k$.
\end{definition}

Observe that, since $f,\bar{f}$ commute, we can permute terms, and restrict the attention to exponents of type $k=\ldots\circ\circ\circ\bullet\bullet\bullet\bullet\ldots\,$, if we want to. However, our various results below will look better without doing this, so we will use Definition 3.29 as stated.

\bigskip

Regarding now the notion of law, this extends too, the result being as follows:

\begin{theorem}
Each complex variable $f\in L^\infty(X)$ has a law, which is by definition a complex probability measure $\mu_f$ making the following formula hold,
$$M_k(f)=\int_\mathbb Cz^kd\mu_f(z)$$
for any colored integer $k$. Moreover, we have in fact the formula
$$E(\varphi(f))=\int_\mathbb C\varphi(x)d\mu_f(x)$$
valid for any integrable function $\varphi:\mathbb C\to\mathbb C$.
\end{theorem}

\begin{proof}
The first assertion follows exactly as in the real case, and with $z^k$ being defined exactly as $f^k$, namely by the following formulae, and multiplicativity: 
$$z^\emptyset=1\quad,\quad z^\circ=z\quad,\quad z^\bullet=\bar{z}$$

As for the second assertion, this basically follows from this by linearity and continuity, by using standard measure theory, again as in the real case.
\end{proof}

Moving ahead towards matrices, all this leads to a mixture of easy and complicated problems. First, Definition 3.29 has the following straightforward analogue:

\index{colored integers}
\index{colored moments}

\begin{definition}
The moments a matrix $A\in M_N(\mathbb C)$ are the numbers
$$M_k(A)=tr(A^k)$$
depending on colored integers $k=\circ\bullet\bullet\circ\ldots\,$, with the usual conventions
$$A^\emptyset=1\quad,\quad A^\circ=A\quad,\quad A^\bullet=A^*$$
and multiplicativity, in order to define the colored powers $A^k$.
\end{definition}

As a first observation about this, unless the matrix is normal, $AA^*=A^*A$, we cannot switch to exponents of type $k=\ldots\circ\circ\circ\bullet\bullet\bullet\bullet\ldots\,$, as it was theoretically possible for the complex variables $f\in L^\infty(X)$. Here is an explicit counterexample for this:

\begin{proposition}
The following matrix, which is not normal,
$$J=\begin{pmatrix}0&1\\0&0\end{pmatrix}$$
has the property $tr(JJ^*JJ^*)\neq tr(JJJ^*J^*)$.
\end{proposition}

\begin{proof}
We have the following formulae, which show that $J$ is not normal:
$$JJ^*=\begin{pmatrix}1&0\\0&0\end{pmatrix}\quad,\quad 
J^*J=\begin{pmatrix}0&0\\0&1\end{pmatrix}$$

Let us compute now the quantities in the statement. We first have:
$$tr(JJ^*JJ^*)
=tr\left(\begin{pmatrix}1&0\\0&0\end{pmatrix}\begin{pmatrix}1&0\\0&0\end{pmatrix}\right)
=tr\begin{pmatrix}1&0\\0&0\end{pmatrix}
=\frac{1}{2}$$

On the other hand, we have as well the following formula:
$$tr(JJJ^*J^*)
=tr\left(\begin{pmatrix}0&0\\0&0\end{pmatrix}
\begin{pmatrix}0&0\\0&0\end{pmatrix}
\right)
=tr\begin{pmatrix}0&0\\0&0\end{pmatrix}
=0$$

Thus, we are led to the conclusion in the statement.
\end{proof}

The above counterexample makes it quite clear that things will be complicated, when attempting to talk about the law of an arbitrary matrix $A\in M_N(\mathbb C)$. But, there is solution to everything. By being a bit smart, we can formulate things as follows:

\index{distribution}
\index{law}

\begin{definition}
The law of a complex matrix $A\in M_N(\mathbb C)$ is the following functional, on the algebra of polynomials in two noncommuting variables $X,X^*$:
$$\mu_A:\mathbb C<X,X^*>\to\mathbb C\quad,\quad P\to tr(P(A))$$
In the case where we have a complex probability measure $\mu_A\in\mathcal P(\mathbb C)$ such that
$$tr(P(A))=\int_\mathbb CP(x)\,d\mu_A(x)$$
we identify this complex measure with the law of $A$.
\end{definition}

As mentioned above, this is something smart, that will take us some time to understand. As a first observation, knowing the law is the same as knowing the moments, because if we write our polynomial as $P=\sum_kc_kX^k$, then we have:
$$tr(P(A))
=tr\left(\sum_kc_kA^k\right)
=\sum_kc_kM_k(A)$$

Let us try now to compute some matrix laws, and see what we get. We already did some computations in the real case, and then for the basic $2\times2$ Jordan block $J$ too, and based on all this, we can formulate the following result, with mixed conclusions:

\begin{theorem}
The following happen:
\begin{enumerate}
\item If $A=A^*$ then $\mu_A=\frac{1}{N}(\lambda_1+\ldots+\lambda_N)$, with $\lambda_i\in\mathbb R$ being the eigenvalues.

\item If $A$ is diagonal, $\mu_A=\frac{1}{N}(\lambda_1+\ldots+\lambda_N)$, with $\lambda_i\in\mathbb C$ being the eigenvalues.

\item For the basic Jordan block $J$, the law $\mu_J$ is not a complex measure.

\item In fact, assuming $AA^*\neq A^*A$, the law $\mu_A$ is not a complex measure.
\end{enumerate}
\end{theorem}

\begin{proof}
This follows from the above, with only (4) being new. Assuming $AA^*\neq A^*A$, in order to show that $\mu_A$ is not a measure, we can use a positivity trick, as follows:
\begin{eqnarray*}
AA^*-A^*A\neq0
&\implies&(AA^*-A^*A)^2>0\\
&\implies&AA^*AA^*-AA^*A^*A-A^*AAA^*+A^*AA^*A>0\\
&\implies&tr(AA^*AA^*-AA^*A^*A-A^*AAA^*+A^*AA^*A)>0\\
&\implies&tr(AA^*AA^*+A^*AA^*A)>tr(AA^*A^*A+A^*AAA^*)\\
&\implies&tr(AA^*AA^*)>tr(AAA^*A^*)
\end{eqnarray*}

Thus, we can conclude as in the proof for $J$, the point being that we cannot obtain both the above numbers by integrating $|z|^2$ with respect to a measure $\mu_A\in\mathcal P(\mathbb C)$.
\end{proof}

Fortunately, by using the spectral theorem for normal matrices, we have:

\index{normal matrix}
\index{law}
\index{distribution}

\begin{theorem}
Given a matrix $A\in M_N(\mathbb C)$ which is normal, $AA^*=A^*A$, we have the following formula, valid for any polynomial $P\in\mathbb C<X,X^*>$,
$$tr(P(A))=\frac{1}{N}(P(\lambda_1)+\ldots+P(\lambda_N))$$
where $\lambda_1,\ldots,\lambda_N\in\mathbb C$ are the eigenvalues of $A$. Thus the complex measure
$$\mu_A=\frac{1}{N}(\delta_{\lambda_1}+\ldots+\delta_{\lambda_N})$$
is the law of $A$. In the non-normal case, the law $\mu_A$ is not a measure.
\end{theorem}

\begin{proof}
As before in the diagonal case, since our matrix is normal, $AA^*=A^*A$, knowing its law in the abstract sense of generalized probability is the same as knowing the restriction of this abstract distribution to the usual polynomials in two variables:
$$\mu_A:\mathbb C[X,X^*]\to\mathbb C\quad,\quad 
P\to tr(P(A))$$

In order now to compute this functional, we can write $A=UDU^*$, as in Theorem 3.22, and then change the basis via $U$, which in practice means that we can simply assume $U=1$. Thus if we denote by $\lambda_1,\ldots,\lambda_N$ the diagonal entries of $D$, which are the eigenvalues of $A$, the law that we are looking for is the following functional:
$$\mu_A:\mathbb C[X,X^*]\to\mathbb C\quad,\quad 
P\to\frac{1}{N}(P(\lambda_1)+\ldots+P(\lambda_N))$$

But this functional corresponds to integrating $P$ with respect to the following complex measure, that we agree to still denote by $\mu_A$, and call distribution of $A$:
$$\mu_A=\frac{1}{N}(\delta_{\lambda_1}+\ldots+\delta_{\lambda_N})$$

Thus, we are led to the conclusion in the statement.
\end{proof}

We will be back to such things later, when discussing the random matrices.

\section*{3e. Exercises}

This was a quite tricky chapter, and as exercises on this, we have:

\begin{exercise}
Learn more about graphs, and related Laplace operators.
\end{exercise}

\begin{exercise}
Work out diagonalization results for various products of graphs.
\end{exercise}

\begin{exercise}
Find a geometric proof for the diagonalization of the rotation $R_t$.
\end{exercise}

\begin{exercise}
Learn more about the Pauli spin matrices, and their properties.
\end{exercise}

\begin{exercise}
Learn alternative proofs for the spectral theorem for normal matrices.
\end{exercise}

\begin{exercise}
Learn about the polar decomposition of complex matrices.
\end{exercise}

\begin{exercise}
Learn some probability theory, as to be at ease with the above.
\end{exercise}

\begin{exercise}
Find formulae for the colored moments of the Jordan blocks.
\end{exercise}

As bonus exercise, learn some operator theory as well, which is related to all this.

\chapter{Polynomials, roots}

\section*{4a. Resultant}

We have seen that the diagonalization of a matrix $A\in M_N(\mathbb C)$ normally requires the computation and factorization of its characteristic polynomial $P\in\mathbb C[X]$. Also, of particular interest is the question of deciding whether $P$ has simple roots or not. 

\bigskip

We will investigate here such questions, computing the roots of the polynomials $P\in\mathbb C[X]$, and their multiplicities. Besides being related to linear algebra via the above-mentioned characteristic polynomial considerations, we will see that these questions actually require linear algebra, and more specifically determinants, in order to be solved. In short, expect some subtle material to come, related in many ways to linear algebra.  

\bigskip

Let us start with something that we know well, but is always good to remember:

\index{degree 2 equation}

\begin{theorem}
The solutions of $ax^2+bx+c=0$ with $a,b,c\in\mathbb C$ are
$$x_{1,2}=\frac{-b\pm\sqrt{b^2-4ac}}{2a}$$
with the square root of complex numbers being defined as $\sqrt{re^{it}}=\sqrt{r}e^{it/2}$.
\end{theorem}

\begin{proof}
We can indeed write our equation in the following way:
\begin{eqnarray*}
ax^2+bx+c=0
&\iff&x^2+\frac{b}{a}x+\frac{c}{a}=0\\
&\iff&\left(x+\frac{b}{2a}\right)^2=\frac{b^2-4ac}{4a^2}\\
&\iff&x+\frac{b}{2a}=\pm\frac{\sqrt{b^2-4ac}}{2a}
\end{eqnarray*}

Here we have used the fact, mentioned in the statement, that any complex number $z=re^{it}$ has indeed a square root, given by $\sqrt{z}=\sqrt{r}e^{it/2}$, plus in fact a second square root as well, namely $-\sqrt{z}$. Thus, we are led to the conclusion in the statement.
\end{proof}

Moving now to degree 3 and higher, things here are far more complicated, and as a first objective, we would like to understand what the analogue of the discriminant $\Delta=b^2-4ac$ is. But even this is something quite tricky, because we would like to have $\Delta=0$ precisely when $(P,P')\neq1$, which leads us into the question of deciding, given two polynomials $P,Q\in\mathbb C[X]$, if these polynomials have a common root, $(P,Q)\neq1$, or not. 

\bigskip

Fortunately this latter question has a nice answer. We will need:

\index{symmetric functions}

\begin{theorem}
Given a monic polynomial $P\in\mathbb C[X]$, factorized as
$$P=(X-a_1)\ldots(X-a_k)$$
the following happen:
\begin{enumerate}
\item The coefficients of $P$ are symmetric functions in $a_1,\ldots,a_k$.

\item The symmetric functions in $a_1,\ldots,a_k$ are polynomials in the coefficients of $P$.
\end{enumerate}
\end{theorem}

\begin{proof}
This is something standard, the idea being as follows:

\medskip

(1) By expanding our polynomial, we have the following formula:
$$P=\sum_{r=0}^k(-1)^r\sum_{i_1<\ldots<i_r}a_{i_1}\ldots a_{i_r}\cdot X^{k-r}$$

Thus the coefficients of $P$ are, up to some signs, the following functions:
$$f_r=\sum_{i_1<\ldots<i_r}a_{i_1}\ldots a_{i_r}$$

But these are indeed symmetric functions in $a_1,\ldots,a_k$, as claimed. 

\medskip

(2) Conversely now, let us look at the symmetric functions in the roots $a_1,\ldots,a_k$. These appear as linear combinations of the basic symmetric functions, given by:
$$S_r=\sum_ia_i^r$$

Moreover, when allowing polynomials instead of linear combinations, we need in fact only the first $k$ such sums, namely $S_1,\ldots,S_k$. That is, the symmetric functions $\mathcal F$ in our variables $a_1,\ldots,a_k$, with integer coefficients, appear as follows:
$$\mathcal F=\mathbb Z[S_1,\ldots,S_k]$$ 

(3) The point now is that, alternatively, the symmetric functions in our variables $a_1,\ldots,a_k$ appear as well as linear combinations of the functions $f_r$ that we found in (1), and that when allowing polynomials instead of linear combinations, we need in fact only the first $k$ functions, namely $f_1,\ldots,f_k$. That is, we have as well:
$$\mathcal F=\mathbb Z[f_1,\ldots,f_k]$$

But this gives the result, because we can pass from $\{S_r\}$ to $\{f_r\}$, and vice versa.
\end{proof}

Getting back now to our original question, namely that of deciding whether two polynomials $P,Q\in\mathbb C[X]$ have a common root or not, this has the following nice answer:

\index{resultant}
\index{common roots}

\begin{theorem}
Given two polynomials $P,Q\in\mathbb C[X]$, written as
$$P=c(X-a_1)\ldots(X-a_k)\quad,\quad 
Q=d(X-b_1)\ldots(X-b_l)$$
the following quantity, which is called resultant of $P,Q$,
$$R(P,Q)=c^ld^k\prod_{ij}(a_i-b_j)$$
is a certain polynomial in the coefficients of $P,Q$, with integer coefficients, and we have $R(P,Q)=0$ precisely when $P,Q$ have a common root.
\end{theorem}

\begin{proof}
This is something quite tricky, the idea being as follows:

\medskip

(1) Given two polynomials $P,Q\in\mathbb C[X]$, we can certainly construct the quantity $R(P,Q)$ in the statement, with the role of the normalization factor $c^ld^k$ to become clear later on, and then we have $R(P,Q)=0$ precisely when $P,Q$ have a common root:
$$R(P,Q)=0\iff \exists i,j,a_i=b_j$$

(2) As bad news, however, this quantity $R(P,Q)$, defined in this way, is a priori not very useful in practice, because it depends on the roots $a_i,b_j$ of our polynomials $P,Q$, that we cannot compute in general. However, and here comes our point, as we will prove below, it turns out that $R(P,Q)$ is in fact a polynomial in the coefficients of $P,Q$, with integer coefficients, and this is where the power of $R(P,Q)$ comes from.

\medskip

(3) You might perhaps say, nice, but why not doing things the other way around, that is, formulating our theorem with the explicit formula of $R(P,Q)$, in terms of the coefficients of $P,Q$, and then proving that we have $R(P,Q)=0$, via roots and everything. Good point, but this is not exactly obvious, the formula of $R(P,Q)$ in terms of the coefficients of $P,Q$ being something terribly complicated. In short, trust me, let us prove our theorem as stated, and for alternative formulae of $R(P,Q)$, we will see later. 

\medskip

(4) Getting started now, let us expand the formula of $R(P,Q)$, by making all the multiplications there, abstractly, in our head. Everything being symmetric in $a_1,\ldots,a_k$, we obtain in this way certain symmetric functions in these variables, which will be therefore certain polynomials in the coefficients of $P$. Moreover, due to our normalization factor $c^l$, these polynomials in the coefficients of $P$ will have integer coefficients.

\medskip

(5) With this done, let us look now what happens with respect to the remaining variables $b_1,\ldots,b_l$, which are the roots of $Q$. Once again what we have here are certain symmetric functions in these variables $b_1,\ldots,b_l$, and these symmetric functions must be certain polynomials in the coefficients of $Q$. Moreover, due to our normalization factor $d^k$, these polynomials in the coefficients of $Q$ will have integer coefficients.

\medskip

(6) Thus, we are led to the conclusion in the statement, that $R(P,Q)$ is a polynomial in the coefficients of $P,Q$, with integer coefficients, and with the remark that the $c^ld^k$ factor is there for these latter coefficients to be indeed integers, instead of rationals.
\end{proof}

All the above might seem a bit complicated, so as an illustration, let us work out an example. Consider the case of a polynomial of degree 2, and a polynomial of degree 1:
$$P=ax^2+bx+c\quad,\quad 
Q=dx+e$$

In order to compute the resultant, let us factorize our polynomials:
$$P=a(x-p)(x-q)\quad,\quad 
Q=d(x-r)$$

The resultant can be then computed as follows, by using the method above:
\begin{eqnarray*}
R(P,Q)
&=&ad^2(p-r)(q-r)\\
&=&ad^2(pq-(p+q)r+r^2)\\
&=&cd^2+bd^2r+ad^2r^2\\
&=&cd^2-bde+ae^2
\end{eqnarray*}

Finally, observe that $R(P,Q)=0$ corresponds indeed to the fact that $P,Q$ have a common root. Indeed, the root of $Q$ is $r=-e/d$, and we have:
$$P(r)
=\frac{ae^2}{d^2}-\frac{be}{d}+c
=\frac{R(P,Q)}{d^2}$$

Regarding now the explicit formula of the resultant $R(P,Q)$, this is something quite complicated, but fortunately, linear algebra comes to the rescue. We have indeed:

\index{resultant}

\begin{theorem}
The resultant of two polynomials, written as 
$$P=p_kX^k+\ldots+p_1X+p_0\quad,\quad 
Q=q_lX^l+\ldots+q_1X+q_0$$
appears as the determinant of an associated matrix, as follows,
$$R(P,Q)=
\begin{vmatrix}
p_k&&&q_l\\
\vdots&\ddots&&\vdots&\ddots\\
p_0&&p_k&q_0&&q_l\\
&\ddots&\vdots&&\ddots&\vdots\\
&&p_0&&&q_0
\end{vmatrix}
$$
with the matrix having size $k+l$, and having $0$ coefficients at the blank spaces.
\end{theorem}

\begin{proof}
This is something quite clever, due to Sylvester, as follows:

\medskip

(1) Consider the vector space $\mathbb C_k[X]$ formed by the polynomials of degree $<k$:
$$\mathbb C_k[X]=\left\{P\in\mathbb C[X]\Big|\deg P<k\right\}$$

This is a vector space of dimension $k$, having as basis the monomials $1,X,\ldots,X^{k-1}$. Now given polynomials $P,Q$ as in the statement, consider the following linear map:
$$\Phi:\mathbb C_l[X]\times\mathbb C_k[X]\to\mathbb C_{k+l}[X]\quad,\quad (A,B)\to AP+BQ$$

\medskip

(2) Our first claim is that with respect to the standard bases for all the vector spaces involved, namely those consisting of the monomials $1,X,X^2,\ldots$, the matrix of $\Phi$ is the matrix in the statement. But this is something which is clear from definitions.

\medskip

(3) Our second claim is that $\det\Phi=0$ happens precisely when $P,Q$ have a common root. Indeed, our polynomials $P,Q$ having a common root means that we can find $A,B$ such that $AP+BQ=0$, and so that $(A,B)\in\ker\Phi$, which reads $\det\Phi=0$.

\medskip

(4) Finally, our claim is that we have $\det\Phi=R(P,Q)$. But this follows from the uniqueness of the resultant, up to a scalar, and with this uniqueness property being elementary to establish, along the lines of the proofs of Theorems 4.2 and 4.3.
\end{proof}

As an illustration for this, consider our favorite polynomials, as before:
$$P=ax^2+bx+c\quad,\quad 
Q=dx+e$$

According to the above result, the resultant should be then, as it should:
$$R(P,Q)
=\begin{vmatrix}
a&d&0\\
b&e&d\\
c&0&e
\end{vmatrix}
=ae^2-bde+cd^2$$

We will be back to more computations of resultants later.

\section*{4b. Discriminant}

We can go back now to our original question regarding discriminants, and we have:

\index{discriminant}
\index{double root}
\index{single roots}

\begin{theorem}
Given a polynomial $P\in\mathbb C[X]$, written as
$$P(X)=aX^N+bX^{N-1}+cX^{N-2}+\ldots$$
its discriminant, defined as being the following quantity,
$$\Delta(P)=\frac{(-1)^{\binom{N}{2}}}{a}\,R(P,P')$$
is a polynomial in the coefficients of $P$, with integer coefficients, and $\Delta(P)=0$ happens precisely when $P$ has a double root.
\end{theorem}

\begin{proof}
The fact that the discriminant $\Delta(P)$ is a polynomial in the coefficients of $P$, with integer coefficients, comes from Theorem 4.3, coupled with the fact that the division by the leading coefficient $a$ is indeed possible, under $\mathbb Z$, as being shown by the following formula, which is written of course a bit informally, coming from Theorem 4.4:
$$R(P,P')=
\begin{vmatrix}
a&&&Na\\
\vdots&\ddots&&\vdots&\ddots\\
z&&a&y&&Na\\
&\ddots&\vdots&&\ddots&\vdots\\
&&z&&&y
\end{vmatrix}
$$

Also, the fact that we have $\Delta(P)=0$ precisely when $P$ has a double root is clear from Theorem 4.3. Finally, let us mention that the sign $(-1)^{\binom{N}{2}}$ is there for various reasons, including the compatibility with some well-known formulae, at small values of $N\in\mathbb N$, such as $\Delta(P)=b^2-4ac$ in degree 2, that we will discuss in a moment.
\end{proof}

As already mentioned, by using Theorem 4.4, we have an explicit formula for the discriminant, as the determinant of a certain matrix. There is a lot of theory here, and in order to get into this, let us first see what happens in degree 2. Here we have:
$$P=aX^2+bX+c\quad,\quad 
P'=2aX+b$$

Thus, the resultant is given by the following formula:
\begin{eqnarray*}
R(P,P')
&=&ab^2-b(2a)b+c(2a)^2\\
&=&4a^2c-ab^2\\
&=&-a(b^2-4ac)
\end{eqnarray*}

It follows that the discriminant of our polynomial is, as it should:
$$\Delta(P)=b^2-4ac$$

Alternatively, we can use the formula in Theorem 4.4, and we obtain:
\begin{eqnarray*}
\Delta(P)=
&=&-\frac{1}{a}\begin{vmatrix}
a&2a&\\
b&b&2a\\
c&&b
\end{vmatrix}\\
&=&-\begin{vmatrix}
1&2&\\
b&b&2a\\
c&&b
\end{vmatrix}\\
&=&-b^2+2(b^2-2ac)\\
&=&b^2-4ac
\end{eqnarray*}

We will be back later to such formulae, in degree 3, and in degree 4 as well, with the comment however, coming in advance, that these formulae are not very beautiful.

\bigskip

At the theoretical level now, we have the following result, which is not trivial:

\begin{theorem}
The discriminant of a polynomial $P$ is given by the formula
$$\Delta(P)=a^{2N-2}\prod_{i<j}(r_i-r_j)^2$$
where $a$ is the leading coefficient, and $r_1,\ldots,r_N$ are the roots.
\end{theorem}

\begin{proof}
This is something quite tricky, the idea being as follows:

\medskip

(1) The first thought goes to the formula in Theorem 4.3, so let us see what that formula teaches us, in the case $Q=P'$. Let us write $P,P'$ as follows:
$$P=a(x-r_1)\ldots(x-r_N)$$ 
$$P'=Na(x-p_1)\ldots(x-p_{N-1})$$

According to Theorem 4.3, the resultant of $P,P'$ is then given by:
$$R(P,P')=a^{N-1}(Na)^N\prod_{ij}(r_i-p_j)$$

And bad news, this is not exactly what we wished for, namely the formula in the statement. That is, we are on the good way, but certainly have to work some more.

\medskip

(2) Obviously, we must get rid of the roots $p_1,\ldots,p_{N-1}$ of the polynomial $P'$. In order to do this, let us rewrite the formula that we found in (1) in the following way:
\begin{eqnarray*}
R(P,P')
&=&N^Na^{2N-1}\prod_i\left(\prod_j(r_i-p_j)\right)\\
&=&N^Na^{2N-1}\prod_i\frac{P'(r_i)}{Na}\\
&=&a^{N-1}\prod_iP'(r_i)
\end{eqnarray*}

(3) In order to compute now $P'$, and more specifically the values $P'(r_i)$ that we are interested in, we can use the Leibnitz rule. So, consider our polynomial:
$$P(x)=a(x-r_1)\ldots(x-r_N)$$ 

The Leibnitz rule for derivatives tells us that $(fg)'=f'g+fg'$, but then also that $(fgh)'=f'gh+fg'h+fgh'$, and so on. Thus, for our polynomial, we obtain:
$$P'(x)=a\sum_i(x-r_1)\ldots\underbrace{(x-r_i)}_{missing}\ldots(x-r_N)$$ 

Now when applying this formula to one of the roots $r_i$, we obtain:
$$P'(r_i)=a(r_i-r_1)\ldots\underbrace{(r_i-r_i)}_{missing}\ldots(r_i-r_N)$$ 

By making now the product over all indices $i$, this gives the following formula:
$$\prod_iP'(r_i)=a^N\prod_{i\neq j}(r_i-r_j)$$

(4) Time now to put everything together. By taking the formula in (2), making the normalizations in Theorem 4.5, and then using the formula found in (3), we obtain:
\begin{eqnarray*}
\Delta(P)
&=&(-1)^{\binom{N}{2}}a^{N-2}\prod_iP'(r_i)\\
&=&(-1)^{\binom{N}{2}}a^{2N-2}\prod_{i\neq j}(r_i-r_j)\\
&=&a^{2N-2}\prod_{i<j}(r_i-r_j)^2
\end{eqnarray*}

Thus, we are led to the conclusion in the statement.
\end{proof}

As applications now, the formula in Theorem 4.6 is quite useful for the real polynomials $P\in\mathbb R[X]$ in small degree, because it allows to say when the roots are real, or complex, or at least have some partial information about this. For instance, we have:

\index{degree 2 polynomial}
\index{degree 3 polynomial}
\index{real roots}
\index{complex roots}

\begin{proposition}
Consider a polynomial with real coefficients, $P\in\mathbb R[X]$, assumed for simplicity to have nonzero discriminant, $\Delta\neq0$.
\begin{enumerate}
\item In degree $2$, the roots are real when $\Delta>0$, and complex when $\Delta<0$.

\item In degree $3$, all roots are real precisely when $\Delta>0$.
\end{enumerate}
\end{proposition}

\begin{proof}
This is very standard, the idea being as follows:

\medskip

(1) The first assertion is something that you certainly know well, since ages, formally coming from Theorem 4.1, but let us see how this comes via the formula in Theorem 4.6. In degree $N=2$, this formula looks as follows, with $r_1,r_2$ being the roots:
$$\Delta(P)=a^2(r_1-r_2)^2$$

Thus $\Delta>0$ amounts in saying that we have $(r_1-r_2)^2>0$. Now since $r_1,r_2$ are conjugate, and with this being something trivial, meaning no need here for the computations in Theorem 4.1, we conclude that $\Delta>0$ means that $r_1,r_2$ are real, as stated.

\medskip

(2) In degree $N=3$ now, we know from analysis that $P$ has at least one real root, and the problem is whether the remaining 2 roots are real, or complex conjugate. For this purpose, we can use the formula in Theorem 4.6, which in degree 3 reads:
$$\Delta(P)=a^4(r_1-r_2)^2(r_1-r_3)^2(r_2-r_3)^2$$

We can see that in the case $r_1,r_2,r_3\in\mathbb R$, we have $\Delta(P)>0$. Conversely now, assume that $r_1=r$ is the real root, coming from analysis, and that the other roots are $r_2=z$ and $r_3=\bar{z}$, with $z$ being a complex number, which is not real. We have then:
\begin{eqnarray*}
\Delta(P)
&=&a^4(r-z)^2(r-\bar{z})^2(z-\bar{z})^2\\
&=&a^4|r-z|^4(2i Im(z))^2\\
&=&-4a^4|r-z|^4Im(z)^2\\
&<&0
\end{eqnarray*}

Thus, we are led to the conclusion in the statement.
\end{proof}

Let us work out now in detail what happens in degree 3, with the explicit computation of the discriminant, in terms of the coefficients. Here the result is as follows:

\begin{theorem}
The discriminant of a degree $3$ polynomial,
$$P=aX^3+bX^2+cX+d$$
is given by $\Delta(P)=b^2c^2-4ac^3-4b^3d-27a^2d^2+18abcd$.
\end{theorem}

\begin{proof}
We have two methods available, based on Theorem 4.3 and Theorem 4.4, and both being instructive, we will try them both. The computations are as follows:

\medskip

(1) Let us first go the pedestrian way, based on the definition of the resultant, from Theorem 4.3. Consider two polynomials, of degree 3 and degree 2, written as follows:
$$P=aX^3+bX^2+cX+d$$
$$Q=eX^2+fX+g=e(X-s)(X-t)$$

The resultant of these two polynomials is then given by:
\begin{eqnarray*}
R(P,Q)
&=&a^2e^3(p-s)(p-t)(q-s)(q-t)(r-s)(r-t)\\
&=&a^2\cdot e(p-s)(p-t)\cdot e(q-s)(q-t)\cdot e(r-s)(r-t)\\
&=&a^2Q(p)Q(q)Q(r)\\
&=&a^2(ep^2+fp+g)(eq^2+fq+g)(er^2+fr+g)
\end{eqnarray*}

By expanding, we obtain the following formula for this resultant:
\begin{eqnarray*}
\frac{R(P,Q)}{a^2}
&=&e^3p^2q^2r^2+e^2f(p^2q^2r+p^2qr^2+pq^2r^2)\\
&+&e^2g(p^2q^2+p^2r^2+q^2r^2)+ef^2(p^2qr+pq^2r+pqr^2)\\
&+&efg(p^2q+pq^2+p^2r+pr^2+q^2r+qr^2)+f^3pqr\\
&+&eg^2(p^2+q^2+r^2)+f^2g(pq+pr+qr)\\
&+&fg^2(p+q+r)+g^3
\end{eqnarray*}

Note in passing that we have 27 terms on the right, as we should, and with this kind of check being mandatory, when doing such computations. Next, we have:
$$p+q+r=-\frac{b}{a}\quad,\quad
pq+pr+qr=\frac{c}{a}\quad,\quad 
pqr=-\frac{d}{a}$$

By using these formulae, we can produce some more, as follows:
$$p^2+q^2+r^2=(p+q+r)^2-2(pq+pr+qr)=\frac{b^2}{a^2}-\frac{2c}{a}$$
$$p^2q+pq^2+p^2r+pr^2+q^2r+qr^2=(p+q+r)(pq+pr+qr)-3pqr=-\frac{bc}{a^2}+\frac{3d}{a}$$
$$p^2q^2+p^2r^2+q^2r^2=(pq+pr+qr)^2-2pqr(p+q+r)=\frac{c^2}{a^2}-\frac{2bd}{a^2}$$

By plugging now this data into the formula of $R(P,Q)$, we obtain:
\begin{eqnarray*}
R(P,Q)
&=&a^2e^3\cdot\frac{d^2}{a^2}-a^2e^2f\cdot\frac{cd}{a^2}+a^2e^2g\left(\frac{c^2}{a^2}-\frac{2bd}{a^2}\right)+a^2ef^2\cdot\frac{bd}{a^2}\\
&+&a^2efg\left(-\frac{bc}{a^2}+\frac{3d}{a}\right)-a^2f^3\cdot\frac{d}{a}\\
&+&a^2eg^2\left(\frac{b^2}{a^2}-\frac{2c}{a}\right)+a^2f^2g\cdot\frac{c}{a}-a^2fg^2\cdot\frac{b}{a}+a^2g^3
\end{eqnarray*}

Thus, we have the following formula for the resultant:
\begin{eqnarray*}
R(P,Q)
&=&d^2e^3-cde^2f+c^2e^2g-2bde^2g+bdef^2-bcefg+3adefg\\
&-&adf^3+b^2eg^2-2aceg^2+acf^2g-abfg^2+a^2g^3
\end{eqnarray*}

Getting back now to our discriminant problem, with $Q=P'$, which corresponds to $e=3a$, $f=2b$, $g=c$, we obtain the following formula:
\begin{eqnarray*}
R(P,P')
&=&27a^3d^2-18a^2bcd+9a^2c^3-18a^2bcd+12ab^3d-6ab^2c^2+18a^2bcd\\
&-&8ab^3d+3ab^2c^2-6a^2c^3+4ab^2c^2-2ab^2c^2+a^2c^3
\end{eqnarray*}

By simplifying terms, and dividing by $a$, we obtain the following formula:
$$-\Delta(P)=27a^2d^2-18abcd+4ac^3+4b^3d-b^2c^2$$

But this gives the formula in the statement, namely:
$$\Delta(P)=b^2c^2-4ac^3-4b^3d-27a^2d^2+18abcd$$

(2) Let us see as well how the computation does, by using Theorem 4.4, which is our most advanced tool, so far. Consider a polynomial of degree 3, and its derivative:
$$P=aX^3+bX^2+cX+d$$
$$P'=3aX^2+2bX+c$$

By using now Theorem 4.4 and computing the determinant, we obtain:
\begin{eqnarray*}
R(P,P')
&=&\begin{vmatrix}
a&&3a\\
b&a&2b&3a\\
c&b&c&2b&3a\\
d&c&&c&2b\\
&d&&&c
\end{vmatrix}\\
&=&\begin{vmatrix}
a&&\\
b&a&-b&3a\\
c&b&-2c&2b&3a\\
d&c&-3d&c&2b\\
&d&&&c
\end{vmatrix}\\
&=&a\begin{vmatrix}
a&-b&3a\\
b&-2c&2b&3a\\
c&-3d&c&2b\\
d&&&c
\end{vmatrix}\\
&=&-ad\begin{vmatrix}
-b&3a\\
-2c&2b&3a\\
-3d&c&2b
\end{vmatrix}
+ac\begin{vmatrix}
a&-b&3a\\
b&-2c&2b\\
c&-3d&c
\end{vmatrix}\\
&=&-ad(-4b^3-27a^2d+12abc+3abc)\\
&&+ac(-2ac^2-2b^2c-9abd+6ac^2+b^2c+6abd)\\
&=&a(4b^3d+27a^2d^2-15abcd+4ac^3-b^2c^2-3abcd)\\
&=&a(4b^3d+27a^2d^2-18abcd+4ac^3-b^2c^2)
\end{eqnarray*}

Now according to Theorem 4.5, the discriminant of our polynomial is given by:
\begin{eqnarray*}
\Delta(P)
&=&-\frac{R(P,P')}{a}\\
&=&-4b^3d-27a^2d^2+18abcd-4ac^3+b^2c^2\\
&=&b^2c^2-4ac^3-4b^3d-27a^2d^2+18abcd
\end{eqnarray*}

Thus, we have again obtained the formula in the statement.
\end{proof}

Still talking degree 3 equations, let us try now to solve such an equation $P=0$, with $P=aX^3+bX^2+cX+d$ as above. By linear transformations we can assume $a=1,b=0$, and then it is convenient to write $c=3p,d=2q$. Thus, our equation becomes:
$$x^3+3px+2q=0$$

Regarding such equations, many things can be said, and to start with, we have the following famous result, dealing with real roots, due to Cardano:

\index{Cardano formula}
\index{degree 3 polynomial}
\index{degree 3 equation}

\begin{theorem}
For a normalized degree $3$ equation, namely 
$$x^3+3px+2q=0$$
the discriminant is $\Delta=-108(p^3+q^2)$. Assuming $p,q\in\mathbb R$ and $\Delta<0$, the number
$$x=\sqrt[3]{-q+\sqrt{p^3+q^2}}+\sqrt[3]{-q-\sqrt{p^3+q^2}}$$
is a real solution of our equation.
\end{theorem}

\begin{proof}
The formula of $\Delta$ is clear from definitions, and with $108=4\times 27$. Now with $x$ as in the statement, by using $(a+b)^3=a^3+b^3+3ab(a+b)$, we have:
\begin{eqnarray*}
x^3
&=&\left(\sqrt[3]{-q+\sqrt{p^3+q^2}}+\sqrt[3]{-q-\sqrt{p^3+q^2}}\right)^3\\
&=&-2q+3\sqrt[3]{-q+\sqrt{p^3+q^2}}\cdot\sqrt[3]{-q-\sqrt{p^3+q^2}}\cdot x\\
&=&-2q+3\sqrt[3]{q^2-p^3-q^2}\cdot x\\
&=&-2q-3px
\end{eqnarray*}

Thus, we are led to the conclusion in the statement.
\end{proof}

Regarding the other roots, we know from Proposition 4.7 that these are both real when $\Delta<0$, and complex conjugate when $\Delta<0$. Thus, in the context of Theorem 4.9, the other two roots are complex conjugate, the formula for them being as follows:

\index{root of unity}

\begin{proposition}
For a normalized degree $3$ equation, namely 
$$x^3+3px+2q=0$$
with $p,q\in\mathbb R$ and discriminant $\Delta=-108(p^3+q^2)$ negative, $\Delta<0$, the numbers
$$z=w\sqrt[3]{-q+\sqrt{p^3+q^2}}+w^2\sqrt[3]{-q-\sqrt{p^3+q^2}}$$
$$\bar{z}=w^2\sqrt[3]{-q+\sqrt{p^3+q^2}}+w\sqrt[3]{-q-\sqrt{p^3+q^2}}$$
with $w=e^{2\pi i/3}$ are the complex conjugate solutions of our equation.
\end{proposition}

\begin{proof}
As before, by using $(a+b)^3=a^3+b^3+3ab(a+b)$, we have:
\begin{eqnarray*}
z^3
&=&\left(w\sqrt[3]{-q+\sqrt{p^3+q^2}}+w^2\sqrt[3]{-q-\sqrt{p^3+q^2}}\right)^3\\
&=&-2q+3\sqrt[3]{-q+\sqrt{p^3+q^2}}\cdot\sqrt[3]{-q-\sqrt{p^3+q^2}}\cdot z\\
&=&-2q+3\sqrt[3]{q^2-p^3-q^2}\cdot z\\
&=&-2q-3pz
\end{eqnarray*}

Thus, we are led to the conclusion in the statement.
\end{proof}

As a conclusion, we have the following statement, unifying the above:

\index{Cardano formula}
\index{degree 3 equation}

\begin{theorem}
For a normalized degree $3$ equation, namely 
$$x^3+3px+2q=0$$
the discriminant is $\Delta=-108(p^3+q^2)$. Assuming $p,q\in\mathbb R$ and $\Delta<0$, the numbers
$$x=w\sqrt[3]{-q+\sqrt{p^3+q^2}}+w^2\sqrt[3]{-q-\sqrt{p^3+q^2}}$$
with $w=1,e^{2\pi i/3},e^{4\pi i/3}$ are the solutions of our equation.
\end{theorem}

\begin{proof}
This follows indeed from Theorem 4.9 and Proposition 4.10. Alternatively, we can redo the computation in their proof, which was nearly identical anyway, in the present setting, with $x$ being given by the above formula, by using $w^3=1$.
\end{proof}

As a comment here, the formula in Theorem 4.11 holds of course in the case $\Delta>0$ too, and also when the coefficients are complex numbers, $p,q\in\mathbb C$, and this due to the fact that the proof rests on the nearly trivial computation from the proof of Theorem 4.9, or of Proposition 4.10. However, these extensions are quite often not very useful, because when it comes to extract all the above square and cubic roots, for complex numbers, you can well end up with the initial question, the one that you started with. 

\section*{4c. Higher degree}

In higher degree things become quite complicated. In degree 4, to start with, we first have the following result, dealing with the discriminant and its applications:

\index{degree 4 polynomial}
\index{degree 4 equation}

\begin{theorem}
The discriminant of $P=ax^4+bx^3+cx^2+dx+e$ is given by:
\begin{eqnarray*}
\Delta
&=&256a^3e^3-192a^2bde^2-128a^2c^2e^2+144a^2cd^2e-27a^2d^4\\
&&+144ab^2ce^2-6ab^2d^2e-80abc^2de+18abcd^3+16ac^4e\\
&&-4ac^3d^2-27b^4e^2+18b^3cde-4b^3d^3-4b^2c^3e+b^2c^2d^2
\end{eqnarray*}
In the case $\Delta<0$ we  have $2$ real roots and $2$ complex conjugate roots, and in the case $\Delta>0$ the roots are either all real or all complex.
\end{theorem}

\begin{proof}
The formula of $\Delta$ follows from the definition of the discriminant, from Theorem 4.5, with the resultant computed via Theorem 4.4, as follows:
$$\Delta
=\frac{1}{a}\begin{vmatrix}
a&&&4a\\
b&a&&3b&4a\\
c&b&a&2c&3b&4a\\
d&c&b&d&2c&3b&4a\\
e&d&c&&d&2c&3b\\
&e&d&&&d&2c\\
&&e&&&&d
\end{vmatrix}$$

As for the last assertion, the study here is routine, a bit as in degree 3.
\end{proof}

In practice, as in degree 3, we can do first some manipulations on our polynomials, as to have them in simpler form, and we have the following version of Theorem 4.12:

\begin{proposition}
The discriminant of $P=x^4+cx^2+dx+e$, normalized degree $4$ polynomial, is given by the following formula:
$$\Delta=16c^4e-4c^3d^2-128c^2e^2+144cd^2e-27d^4+256e^3$$
As before, if $\Delta<0$ we  have $2$ real roots and $2$ complex conjugate roots, and if $\Delta>0$ the roots are either all real or all complex.
\end{proposition}

\begin{proof}
This is a consequence of Theorem 4.10, with $a=1,b=0$, but we can deduce this as well directly. Indeed, the formula of $\Delta$ follows, quite easily, from:
$$\Delta
=\frac{1}{a}\begin{vmatrix}
1&&&4\\
&1&&&4\\
c&&1&2c&&4\\
d&c&&d&2c&&4\\
e&d&c&&d&2c&\\
&e&d&&&d&2c\\
&&e&&&&d
\end{vmatrix}$$

As for the last assertion, this is something that we know, from Theorem 4.12.
\end{proof}

We still have some work to do. Indeed, looking back at what we did in degree 3, the passage there from Theorem 4.8 to Theorem 4.9 was made of two operations, namely ``depressing'' the equation, that is, getting rid of the next-to-highest term, and then rescaling the coefficients, as for the formula of $\Delta$ to become as simple as possible. 

\bigskip

In our present setting now, degree 4, with the depressing done as above, in Proposition 4.13, it remains to rescale the coefficients, as for the formula of $\Delta$ to become as simple as possible. And here, a bit of formula hunting, in relation with $2,3$ powers, leads to:

\begin{theorem}
The discriminant of a normalized degree $4$ polynomial, written as
$$P=x^4+6px^2+4qx+3r$$ 
is given by the following formula:
$$\Delta=256\times27\times\big(9p^4r-2p^3q^2-6p^2r^2+6pq^2r-q^4+r^3\big)$$
In the case $\Delta<0$ we have $2$ real roots and $2$ complex conjugate roots, and in the case $\Delta>0$ the roots are either all real or all complex.
\end{theorem}

\begin{proof}
This follows from Proposition 4.13, with $c=6p,d=4q,e=3r$, but we can deduce this as well directly. Indeed, the formula of $\Delta$ follows, quite easily, from:
$$\Delta
=\begin{vmatrix}
1&&&4\\
&1&&&4\\
6p&&1&12p&&4\\
4q&6p&&4q&12p&&4\\
3r&4q&6p&&4q&12p&\\
&3r&4q&&&4q&12p\\
&&3r&&&&4q
\end{vmatrix}$$

As for the last assertion, this is something that we know from Theorem 4.12.
\end{proof}

Time now to get to the real thing, solving the equation. We have here:

\index{degree 4 equation}
\index{Cardano formula}

\begin{theorem}
The roots of a normalized degree $4$ equation, written as
$$x^4+6px^2+4qx+3r=0$$ 
are as follows, with $y$ satisfying the equation $(y^2-3r)(y-3p)=2q^2$,
$$x_1=\frac{1}{\sqrt{2}}\left(-\sqrt{y-3p}+\sqrt{-y-3p+\frac{4q}{\sqrt{2y-6p}}}\right)$$
$$x_2=\frac{1}{\sqrt{2}}\left(-\sqrt{y-3p}-\sqrt{-y-3p+\frac{4q}{\sqrt{2y-6p}}}\right)$$
$$x_3=\frac{1}{\sqrt{2}}\left(\sqrt{y-3p}+\sqrt{-y-3p-\frac{4q}{\sqrt{2y-6p}}}\right)$$
$$x_4=\frac{1}{\sqrt{2}}\left(\sqrt{y-3p}-\sqrt{-y-3p-\frac{4q}{\sqrt{2y-6p}}}\right)$$
and with $y$ being computable via the Cardano formula.
\end{theorem}

\begin{proof}
This is something quite tricky, the idea being as follows:

\medskip

(1) To start with, let us write our equation in the following form:
$$x^4=-6px^2-4qx-3r$$

The idea will be that of adding a suitable common term, to both sides, as to make square on both sides, as to eventually end with a sort of double quadratic equation. For this purpose, our claim is that what we need is a number $y$ satisfying:
$$(y^2-3r)(y-3p)=2q^2$$

Indeed, assuming that we have this number $y$, our equation becomes:
\begin{eqnarray*}
(x^2+y)^2
&=&x^4+2x^2y+y^2\\
&=&-6px^2-4qx-3r+2x^2y+y^2\\
&=&(2y-6p)x^2-4qx+y^2-3r\\
&=&(2y-6p)x^2-4qx+\frac{2q^2}{y-3p}\\
&=&\left(\sqrt{2y-6p}\cdot x-\frac{2q}{\sqrt{2y-6p}}\right)^2
\end{eqnarray*}

(2) Which looks very good, leading us to the following degree 2 equations:
$$x^2+y+\sqrt{2y-6p}\cdot x-\frac{2q}{\sqrt{2y-6p}}=0$$
$$x^2+y-\sqrt{2y-6p}\cdot x+\frac{2q}{\sqrt{2y-6p}}=0$$

Now let us write these two degree 2 equations in standard form, as follows:
$$x^2+\sqrt{2y-6p}\cdot x+\left(y-\frac{2q}{\sqrt{2y-6p}}\right)=0$$
$$x^2-\sqrt{2y-6p}\cdot x+\left(y+\frac{2q}{\sqrt{2y-6p}}\right)=0$$

(3) Regarding the first equation, the solutions there are as follows:
$$x_1=\frac{1}{2}\left(-\sqrt{2y-6p}+\sqrt{-2y-6p+\frac{8q}{\sqrt{2y-6p}}}\right)$$
$$x_2=\frac{1}{2}\left(-\sqrt{2y-6p}-\sqrt{-2y-6p+\frac{8q}{\sqrt{2y-6p}}}\right)$$

As for the second equation, the solutions there are as follows:
$$x_3=\frac{1}{2}\left(\sqrt{2y-6p}+\sqrt{-2y-6p-\frac{8q}{\sqrt{2y-6p}}}\right)$$
$$x_4=\frac{1}{2}\left(\sqrt{2y-6p}-\sqrt{-2y-6p-\frac{8q}{\sqrt{2y-6p}}}\right)$$

(4) Now by cutting a $\sqrt{2}$ factor from everything, this gives the formulae in the statement. As for the last claim, regarding the nature of $y$, this comes from Cardano.
\end{proof}

We still have to compute the number $y$ appearing in the above via Cardano, and the result here, adding to what we already have in Theorem 4.15, is as follows:

\begin{theorem}[continuation]
The value of $y$ in the previous theorem is
$$y=t+p+\frac{a}{t}$$
where the number $t$ is given by the formula
$$t=\sqrt[3]{b+\sqrt{b^2-a^3}}$$
with $a=p^2+r$ and $b=2p^2-3pr+q^2$.
\end{theorem}

\begin{proof}
The legend has it that this is what comes from Cardano, but depressing and normalizing and solving $(y^2-3r)(y-3p)=2q^2$ makes it for too many operations, so the most pragmatic is to simply check this equation. With $y$ as above, we have:
\begin{eqnarray*}
y^2-3r
&=&t^2+2pt+(p^2+2a)+\frac{2pa}{t}+\frac{a^2}{t^2}-3r\\
&=&t^2+2pt+(3p^2-r)+\frac{2pa}{t}+\frac{a^2}{t^2}
\end{eqnarray*}

With this in hand, we have the following computation:
\begin{eqnarray*}
(y^2-3r)(y-3p)
&=&\left(t^2+2pt+(3p^2-r)+\frac{2pa}{t}+\frac{a^2}{t^2}\right)\left(t-2p+\frac{a}{t}\right)\\
&=&t^3+(a-4p^2+3p^2-r)t+(2pa-6p^3+2pr+2pa)\\
&&+(3p^2a-ra-4p^2a+a^2)\frac{1}{t}+\frac{a^3}{t^3}\\
&=&t^3+(a-p^2-r)t+2p(2a-3p^2+r)+a(a-p^2-r)\frac{1}{t}+\frac{a^3}{t^3}\\
&=&t^3+2p(-p^2+3r)+\frac{a^3}{t^3}
\end{eqnarray*}

Now by using the formula of $t$ in the statement, this gives:
\begin{eqnarray*}
(y^2-3r)(y-3p)
&=&b+\sqrt{b^2-a^3}-4p^2+6pr+\frac{a^3}{b+\sqrt{b^2-a^3}}\\
&=&b+\sqrt{b^2-a^3}-4p^2+6pr+b-\sqrt{b^2-a^3}\\
&=&2b-4p^2+6pr\\
&=&2(2p^2-3pr+q^2)-4p^2+6pr\\
&=&2q^2
\end{eqnarray*}

Thus, we are led to the conclusion in the statement.
\end{proof}

In higher degree, things become more complicated. In order to discuss this, following Galois and others, we will need some field theory basics. Let us start with:

\index{field}
\index{arbitrary field}
\index{characteristic of field}
\index{prime field}
\index{Fermat theorem}
\index{Fermat polynomial}
\index{finite field}
\index{multiplicative group}

\begin{theorem}
Given a field $F$, define its characteristic $p=char(F)$ as being the smallest $p\in\mathbb N$ such that the following happens, and as $p=0$, if this never happens:
$$\underbrace{1+\ldots+1}_{p\ times}=0$$
Then, assuming $p>0$, this number $p$ must be prime, we have a field embedding $\mathbb F_p\subset F$, and $q=|F|$ must be of the form $q=p^k$, with $k\in\mathbb N$. Also, we have the formulae
$$(a+b)^p=a^p+b^p\quad,\quad a^q=a$$
valid for any $a,b\in F$, and the Fermat poynomial $X^q-X$ factorizes as:
$$X^q-X=\prod_{a\in F}(X-a)$$
Also, regardless of $p$, any finite multiplicative subgroup $G\subset F-\{0\}$ must be cyclic.
\end{theorem}

\begin{proof}
This is a quite crowded statement, the idea with all this being as follows:

\medskip

(1) The fact that $p>0$ must be prime comes by contradiction, by using:
$$(\underbrace{1+\ldots+1}_{a\ times})\times(\underbrace{1+\ldots+1}_{b\ times})=\underbrace{1+\ldots+1}_{ab\ times}$$

Indeed, assuming that we have $p=ab$ with $a,b>1$, the above formula corresponds to an equality of type $AB=0$ with $A,B\neq0$ inside $F$, which is impossible.

\medskip

(2) Back to the general case, $F$ has a smallest subfield $E\subset F$, called prime field, consisting of the various sums $1+\ldots+1$, and their quotients. In the case $p=0$ we obviously have $E=\mathbb Q$. In the case $p>0$ now, the multiplication formula in (1) shows that the set $S=\{1+\ldots+1\}$ is stable under taking quotients, and so $E=S$.

\medskip

(3) Now with $E=S$ in hand, we obviously have $(E,+)=\mathbb Z_p$, and since the multiplication is given by the formula in (1), we conclude that we have $E=\mathbb F_p$, as a field. Thus, in the case $p>0$, we have constructed an embedding $\mathbb F_p\subset F$, as claimed.

\medskip

(4) In the context of the above embedding $\mathbb F_p\subset F$, we can say that $F$ is a vector space over $\mathbb F_p$, and so we have $|F|=p^k$, with $k\in\mathbb N$ being the dimension of this space.

\medskip

(5) The baby Fermat formula $(a+b)^p=a^p+b^p$ can be established as follows:
$$(a+b)^p=\sum_{k=0}^p\binom{p}{k}a^kb^{p-k}=a^p+b^p$$

(6) As for the Fermat formula $a^q=a$ itself, which implies the assertion about $X^q-X$, this follows from the last assertion, which can be proved via some basic arithmetic inside $F$, and which for $G=F-\{0\}$ itself, with $|F|=q$, gives $a^{q-1}=1$, for any $a\neq0$.
\end{proof}

At a more advanced level, following Galois, we have the following key result:

\index{field extension}
\index{separable extension}
\index{Galois theorem}
\index{root of polynomial}

\begin{theorem} 
Given a field extension $E\subset F$, we can talk about its Galois group $G$, as the group of automorphisms of $F$ fixing $E$. The intermediate fields
$$E\subset K\subset F$$
are then in correspondence with the subgroups $H\subset G$, with such a field $K$ corresponding to the subgroup $H$ consisting of automorphisms $g\in G$ fixing $K$.
\end{theorem}

\begin{proof}
This is something self-explanatory, and follows indeed from some algebra, under suitable assumptions, in order for that algebra to properly apply.
\end{proof}

Getting now towards polynomials and their roots, we have here:

\index{splitting field}
\index{algebraic closure}
\index{algebrically closed}

\begin{theorem}
Given a field $F$ and a polynomial $P\in F[X]$, we can talk about the abstract splitting field of $P$, where this polynomial decomposes as:
$$P(X)=c\prod_i(X-a_i)$$
In particular, any field $F$ has a certain algebraic closure $\bar{F}$, where all the polynomials $P\in F[X]$, and in fact all polynomials $P\in\bar{F}[X]$ too, have roots.
\end{theorem}

\begin{proof}
This is again something which is quite self-explanatory, which follows from Theorem 4.18 and from some extra algebra, under suitable assumptions.
\end{proof}

Good news, with this in hand, we can now elucidate the structure of finite fields:

\index{finite field}
\index{uniqueness of finite fields}

\begin{theorem}
For any prime power $q=p^k$ there is a unique field $\mathbb F_q$ having $q$ elements. At $k=1$ this is the usual $\mathbb F_p$. In general, this is the splitting field of:
$$P=X^q-X$$
Moreover, we can construct an explicit model for $\mathbb F_q$, at $q=p^2$ or higher, as
$$\mathbb F_q=\mathbb F_p[X]/(Q)$$
with $Q\in\mathbb F_p[X]$ being a suitable irreducible polynomial, of degree $k$. 
\end{theorem}

\begin{proof}
There are several assertions here, the idea being as follows:

\medskip

(1) The first assertion, regarding the existence and uniqueness of $\mathbb F_q$, follows from Theorem 4.19. Indeed, we know from Theorem 4.17 that given a finite field, $|F|=q$ with $k\in\mathbb N$, the Fermat polynomial $P=X^q-X$ factorizes as follows:
$$X^q-X=\prod_{a\in F}(X-a)$$

Thus $F$ must be the splitting field of $P$, and so is unique. As for the existence, this follows also from Theorem 4.19, telling us that this splitting field always exists.

\medskip

(2) In what regards now the modeling of $\mathbb F_q$, at $q=p$ there is nothing to do, because we have our usual $\mathbb F_p$ here. At $q=p^2$ and higher, we know from commutative algebra that we have an isomorphism as follows, whenever $Q\in\mathbb F_p[X]$ is taken irreducible:
$$\mathbb F_q=\mathbb F_p[X]/(Q)$$

(3) Regarding now the best choice of the irreducible polynomial $Q\in\mathbb F_p[X]$, providing us with a good model for the finite field $\mathbb F_q$, that we can use in practice, this question depends on the value of $q=p^k$, and many things can be said here. All in all, the models are quite similar to $\mathbb C=\mathbb R[i]$, with $i$ being a formal number satisfying $i^2=-1$.
\end{proof}

As another application of the above, which motivated Galois, we have:

\index{degree 5 polynomial}
\index{roots}
\index{Galois theory}
\index{solvable group}
\index{tower of extensions}
\index{field extension}
\index{separable extension}

\begin{theorem}
Unlike in degree $N\leq4$, there is no formula for the roots of polynomials of degree $N=5$ and higher, with the reason for this, coming from Galois theory, being that $S_5$ is not solvable. The simplest numeric example is $P=X^5-X-1$.
\end{theorem}

\begin{proof}
This is something quite tricky, the idea being as follows:

\medskip

(1) The first assertion, for generic polynomials, is due to Abel-Ruffini, but Galois theory helps in better understanding this, and comes with a number of bonus points too, namely the possibility of formulating a finer result, with Abel-Ruffini's original ``generic'', which was something algebraic, being now replaced by an analytic ``generic'', and also with the possibility of dealing with concrete polynomials, such as:
$$P=X^5-X-1$$

(2) Regarding now the details of the Galois proof of the Abel-Ruffini theorem, assume that the roots of a polynomial $P\in F[X]$ can be computed by using iterated roots, a bit as for the degree 2 equation, or for the degree 3 and 4 equations, via Cardano. Then, algebrically speaking, this gives rise to a tower of fields as folows, with $F_0=F$, and each $F_{i+1}$ being obtained from $F_i$ by adding a root, $F_{i+1}=F_i(x_i)$, with $x_i^{n_i}\in F_i$:
$$F_0\subset F_1\subset\ldots\subset F_k$$

(3) In order for Galois theory to apply well to this situation, we must make all the extensions normal, which amounts in replacing each $F_{i+1}=F_i(x_i)$ by its extension $K_i(x_i)$, with $K_i$ extending $F_i$ by adding a $n_i$-th root of unity. Thus, with this replacement, we can assume that the tower in (2) in normal, meaning that all Galois groups are cyclic.

\medskip

(4) Now by Galois theory, at the level of the corresponding Galois groups we obtain a tower of groups as follows as follows, which is a resolution of the last group $G_k$, the Galois group of $P$, in the sense of group theory, in the sense that all quotients are cyclic:
$$G_1\subset G_2\subset\ldots\subset G_k$$

As a conclusion, Galois theory tells us that if the roots of a polynomial $P\in F[X]$ can be computed by using iterated roots, then its Galois group $G=G_k$ must be solvable.

\medskip

(5) In the generic case, the conclusion is that Galois theory tells us that, in order for all polynomials of degree 5 to be solvable, via square roots, the group $S_5$, which appears there as Galois group, must be solvable, in the sense of group theory. But this is wrong, because the alternating subgroup $A_5\subset S_5$ is simple, and therefore not solvable.

\medskip

(6) Finally, regarding the polynomial $P=X^5-X-1$, some elementary computations here, based on arithmetic over $\mathbb F_2,\mathbb F_3$, and involving various cycles of length $2,3,5$, show that its Galois group is $S_5$. Thus, we have our counterexample.

\medskip

(7) To be more precise, our polynomial factorizes over $\mathbb F_2$ as follows:
$$X^5-X-1=(X^2+X+1)(X^3+X^2+1)$$

We deduce from this the existence of an element $\tau\sigma\in G\subset S_5$, with $\tau\in S_5$ being a transposition, and with $\sigma\in S_5$ being a 3-cycle, disjoint from it. Thus, we have:
$$\tau=(\tau\sigma)^3\in G$$

(8) On the other hand since $P=X^5-X-1$ is irreducible over $\mathbb F_5$, we have as well available a certain 5-cycle $\rho\in G$. Now since $<\tau,\rho>=S_5$, we conclude that the Galois group of $P$ is full, $G=S_5$, and by (4) and (5) we have our counterexample.

\medskip

(9) Finally, as mentioned in (1), all this shows as well that a random polynomial of degree 5 or higher is not solvable by square roots, and with this being an elementary consequence of the main result from (5), via some standard analysis arguments.
\end{proof}

\section*{4d. Density results}

Getting back now to Earth, and to basic linear algebra and mathematics, we can use the resultant and discriminant technology developed above, in relation with our diagonalization questions for the usual matrices, as formulated in chapter 2, as follows:

\begin{theorem}
For a matrix $A\in M_N(\mathbb C)$ the following conditions are equivalent, and in this case, the matrix is diagonalizable:
\begin{enumerate}
\item The eigenvalues are different, $\lambda_i\neq\lambda_j$.

\item The characteristic polynomial $P$ has simple roots.

\item The characteristic polynomial satisfies $(P,P')=1$.

\item The resultant of $P,P'$ is nonzero, $R(P,P')\neq0$.

\item The discriminant of $P$ is nonzero, $\Delta(P)\neq0$.
\end{enumerate}
\end{theorem}

\begin{proof}
This follows from the general theory that we have, as follows:

\medskip

(1) To start with, the fact that a matrix is diagonalizable when the eigenvalues are different is something elementary, that we know well from chapter 2.

\medskip

(2) The equivalence $(1)\iff(2)$ is something that we know from chapter 2 too, coming from the basic theory of the characteristic polynomial.

\medskip

(3) As for the equivalences $(2)\iff(3)\iff(4)\iff(5)$, which are valid for any poynomial $P$, these follow from the above theory of the resultant and discriminant.
\end{proof}

The above result is quite interesting, and as a continuation to it, we can now formulate a quite tricky and powerful result, having countless potential applications, as follows:

\index{invertible matrix}
\index{distinct eigenvalues}
\index{density}

\begin{theorem}
The following happen, inside $M_N(\mathbb C)$:
\begin{enumerate}
\item The invertible matrices are dense.

\item The matrices having distinct eigenvalues are dense.

\item The diagonalizable matrices are dense.
\end{enumerate}
\end{theorem}

\begin{proof}
These are quite advanced linear algebra results, which can be proved as follows, with the technology that we have so far:

\medskip

(1) This is clear, intuitively speaking, because the invertible matrices are given by the condition $\det A\neq 0$. Thus, the set formed by these matrices appears as the complement of the hypersurface $\det A=0$, and so must be dense inside $M_N(\mathbb C)$, as claimed. 

\medskip

(2) Here we can use a similar argument, this time by saying that the set formed by the matrices having distinct eigenvalues appears as the complement of the hypersurface given by $\Delta(P_A)=0$, and so must be dense inside $M_N(\mathbb C)$, as claimed. 

\medskip

(3) This follows from (2), via the standard fact that the matrices having distinct eigenvalues are diagonalizable, that we know from Theorem 4.22. There are of course some other proofs as well, for instance by putting the matrix in Jordan form, and we will discuss this later in this book, after working out the Jordan form.
\end{proof}

As a first observation, the above result is something extremely useful, more or less allowing you in practice to assume that any matrix $A\in M_N(\mathbb C)$ is diagonalizable. But of course do not try this at home, unless you know what you're doing. 

\bigskip

As an application of the above results, and of our methods in general, we can now establish a number of useful and interesting linear algebra results, as follows:

\index{products of matrices}
\index{functions of matrices}
\index{functional calculus}

\begin{theorem}
The following happen:
\begin{enumerate}
\item We have $P_{AB}=P_{BA}$, for any two matrices $A,B\in M_N(\mathbb C)$.

\item $AB,BA$ have the same eigenvalues, with the same multiplicities.

\item If $A$ has eigenvalues $\lambda_1,\ldots,\lambda_N$, then $f(A)$ has eigenvalues $f(\lambda_1),\ldots,f(\lambda_N)$.
\end{enumerate}
\end{theorem}

\begin{proof}
These results, which are quite non-trivial to prove with bare hands, can be all deduced by using the density tricks from Theorem 4.23, as follows:

\medskip

(1) To start with, it follows from definitions that the characteristic polynomial of a matrix is invariant under conjugation, in the sense that we have:
$$P_C=P_{ACA^{-1}}$$

Now observe that, when assuming that $A$ is invertible, we have:
$$AB=A(BA)A^{-1}$$

Thus, we have the result when $A$ is invertible. By using now Theorem 4.23 (1), we conclude that this formula holds for any matrix $A$, by continuity. 

\medskip

(2) This is a reformulation of (1) above, via the fact that $P$ encodes the eigenvalues, with multiplicities, which is hard to prove with bare hands. Let us also mention here that such things are well-known to fail for the infinite matrices, a basic counterexample here being provided by the shift $A=S$ and its adjoint $B=S^*$, which are given by:
$$S=\begin{pmatrix}
0&0&0&\ldots\\
1&0&0&\ldots\\
0&1&0&\ldots\\
\vdots&\vdots&\vdots
\end{pmatrix}
\qquad,\qquad
S^*=\begin{pmatrix}
0&1&0&\ldots\\
0&0&1&\ldots\\
0&0&0&\ldots\\
\vdots&\vdots&\vdots
\end{pmatrix}$$

Indeed, we have the following two product formulae, for these infinite matrices:
$$SS^*=\begin{pmatrix}
0&0&0&\ldots\\
0&1&0&\ldots\\
0&0&1&\ldots\\
\vdots&\vdots&\vdots
\end{pmatrix}
\qquad,\qquad
S^*S=\begin{pmatrix}
1&0&0&\ldots\\
0&1&0&\ldots\\
0&0&1&\ldots\\
\vdots&\vdots&\vdots
\end{pmatrix}$$

Thus $SS^*$ is a projection, having $0$ as eigenvalue, while $S^*S$ is the identity, having only $1$ eigenvalues. More on this later in this book, when doing infinite dimensions.

\medskip

(3) This is something more informal, the idea being that this is clear for the diagonal matrices $D$, then for the diagonalizable matrices $PDP^{-1}$, and finally for all the matrices, by using Theorem 4.23 (3), provided that $f$ has suitable regularity properties. We will be back to all this later in this book, with details, when doing spectral theory.
\end{proof}

As a conclusion to all this, there is a nice and fruitful relationship between linear algebra on one hand, and the theory of the resultant and discriminant on the other hand, with applications in both senses. We will be back to this, later in this book.

\bigskip

Let us also mention that many of the above results extend to the case where we are dealing with linear algebra and polynomials over an arbitrary field $F$. We will be back to this later in this book, towards the end, when discussing arithmetic aspects.

\bigskip

And good news, that is all. As a matter, however, of making sure that we have not forgotten anything basic, in this introductory Part I, it is probably wise to ask the cat. And cat, who as usual, is more of a physicist than a mathematician, declares:

\begin{cat}
Matrices being generically diagonalizable, if you assume that they are diagonalizable, you might fail sometimes, but can certainly catch some mice.
\end{cat}

Thanks cat, it is all about hunting matters, indeed. This being said, we will have a look right next, in chapter 5 below, at the non-diagonalizable case too, following Jordan and others. This is certainly interesting, mathematically speaking, and with a bit of luck, we might even catch something, perhaps not mice, but some form of human food.

\section*{4e. Exercises}

This was our first truly advanced chapter, and as exercises on this, we have:

\begin{exercise}
Clarify what has been said above, about symmetric functions.
\end{exercise}

\begin{exercise}
Clarify as well all the details in relation with the resultant.
\end{exercise}

\begin{exercise}
Learn the other formulations of the Cardano formula in degree $3$.
\end{exercise}

\begin{exercise}
Complete the Cardano proof in degree $4$ by using the degree $3$ formula.
\end{exercise}

\begin{exercise}
Learn the other formulations of the Cardano formula in degree $4$.
\end{exercise}

\begin{exercise}
Learn more about field extensions, and Galois theory.
\end{exercise}

\begin{exercise}
Work out all the Galois theory details for $P=X^5-X-1$.
\end{exercise}

\begin{exercise}
Find some other proofs for the density of diagonalizable matrices.
\end{exercise}

As bonus exercise, learn some basic algebraic geometry. All good old stuff.

\part{Advanced results}

\ \vskip50mm

\begin{center}
{\em Oh, my life is changing everyday

In every possible way

And oh, my dreams

It's never quite as it seems}
\end{center}

\chapter{Jordan form}

\section*{5a. Linear equations}

Welcome to advanced linear algebra. We know from Part I that the generic matrices $A\in M_N(\mathbb C)$ are diagonalizable, but this is of course not the end of the story. The question being, what if the particular matrix $A\in M_N(\mathbb C)$ appearing on our path, in relation with this or that problem, was chosen by the Devil to be non-diagonalizable?

\bigskip

In answer, work, more work, and even more work. We will see in the present Part II all sorts of tricks, invented by the mathematicians, in order to deal with the non-diagonalizable matrices. And with these consisting of the Jordan form, which is the main theorem around, plus a myriad of other useful decomposition techniques.

\bigskip

Before that, however, some motivational talk. Here is a good, concrete question, which appears in mathematics, physics, and science in general, that we would like to solve:

\begin{question}
How to solve differential equations?
\end{question}

Obviously, this question is quite broad, and as a first concrete example, let us examine the case of a falling object. If we denote by $x=x(t):\mathbb R\to\mathbb R^3$ the position of our falling object, then its speed $v=v(t):\mathbb R\to\mathbb R^3$ and acceleration $a=a(t):\mathbb R\to\mathbb R^3$ are given by the following formulae, with the dots standing for derivatives with respect to time $t$:
$$v=\dot{x}\quad,\quad a=\dot{v}=\ddot{x}$$

Regarding now the equation of motion, this is as follows, coming from Newton, with $m$ being the mass of our object, and with $F$ being the gravitational force:
$$m\cdot a(t)=F(x(t))$$

Thus, in terms of derivatives as above, in order to have as only unknown the position vector $x=x(t):\mathbb R\to\mathbb R^3$, the equation of motion is as follows:
$$m\cdot\ddot{x}(t)=F(x(t))$$

Which looks nice, but since we have here a degree 2 equation, instead of a degree 1 one, which would be better, was it really a good idea to get rid of speed $v:\mathbb R\to\mathbb R^3$ and acceleration $a:\mathbb R\to\mathbb R^3$, and reformulate everything in terms of position $x:\mathbb R\to\mathbb R^3$.

\bigskip

So, going all over again, with the aim this time of reaching to a degree 1 equation, let us replace our 3-dimensional unknown $x:\mathbb R\to\mathbb R^3$ with the 6-dimensional unknown $(x,v):\mathbb R\to\mathbb R^6$. And with this done, good news, we have our degree 1 system:
$$\begin{cases}
\dot{x}(t)=v(t)\\
\dot{v}(t)=\frac{1}{m}F(x(t))
\end{cases}$$

Which was a nice trick, wasn't it. So, before going further, let us record the following conclusion, that we will come back to in a moment, after done with gravity:

\begin{conclusion}
We can convert differential equations of higher order into differential equations of first order, by suitably enlarging the size of our unknown vectors.
\end{conclusion}

Now back to gravity and free falls, and to the degree 1 system found above, let us assume for simplicity that our object is subject to a free fall under a uniform gravitational field. In practice, this means that $F$ is constant, given by the following formula, with $m>0$ being as usual the mass of our object, and with $g>0$ being a certain constant:
$$F(x)=-mg\begin{pmatrix}0\\ 0\\1\end{pmatrix}$$

With this data, the system that we found takes the following form:
$$\begin{cases}
\dot{x}(t)=v(t)\\
\dot{v}(t)=-g\begin{pmatrix}0\\ 0\\1\end{pmatrix}
\end{cases}$$

But this latter system is very easy to solve. Indeed, the second equation gives:
$$v(t)=v(0)-g\begin{pmatrix}0\\ 0\\1\end{pmatrix}t$$

Now by integrating once again, we can recover as well the formula of $x$, as follows:
$$x(t)=x(0)+v(0)t-\frac{g}{2}\begin{pmatrix}0\\ 0\\1\end{pmatrix}t^2$$

Which is very nice, good work that we did here, so let us record our findings, along with a bit more, in the form of a complete statement, as follows:

\begin{theorem}
For a free fall in a uniform gravitational field, with gravitational acceleration constant $g>0$, the equation of motion is
$$x(t)=x(0)+v(0)t-\frac{g}{2}\begin{pmatrix}0\\ 0\\1\end{pmatrix}t^2$$
and the trajectory is a parabola, unless in the case where the free fall is straight downwards, where the trajectory is a line.
\end{theorem}

\begin{proof}
This is a conclusion to what we found above, namely equation of motion, and its obvious implications, at the level of the corresponding trajectory.
\end{proof}

Now back to theory, let us go back to Conclusion 5.2, which was our main theoretical finding so far, and further comment on that. Of course in the case of extremely simple equations, like the above uniform gravity one, there is no really need to use this trick, because you can directly integrate twice. However, in general, this remains a very useful trick, worth some discussion, and we will discuss this now.

\bigskip

Let us start with some generalities in one variable. We have here:

\begin{definition}
A general ordinary differential equation (ODE) is an equation as follows, with a function $x=x(t):\mathbb R\to\mathbb R$ as unknown, 
$$F(t,x,\dot{x},\ldots,x^{(k)})=0$$
depending on a given function $F:U\to\mathbb R$, with $U\subset\mathbb R^{k+2}$ being an open set.
\end{definition}

As a first observation, under suitable assumptions on our function $F:U\to\mathbb R$, and more specifically non-vanishing of its partial derivatives, in all directions, we can use the implicit function theorem, in order to reformulate our equation as follows, for a certain function $f:V\to\mathbb R$, with $V\subset\mathbb R^{k+1}$ being a certain open set:
$$x^{(k)}=f(t,x,\dot{x},\ldots,x^{(k-1)})$$

In practice, we will make this change, which often comes by default, when investigating questions coming from physics, and these will be the ODE that we will be interested in. 

\bigskip

Now moving to several variables, more generally, let us formulate:

\begin{definition}
A standard system of ODE is a system as follows,
$$x_1^{(k)}=f_1(t,x,\dot{x},\ldots,x^{(k-1)})$$
$$\vdots$$
$$x_N^{(k)}=f_N(t,x,\dot{x},\ldots,x^{(k-1)})$$
with the unknown being a vector function $x=x(t):\mathbb R\to\mathbb R^N$.
\end{definition}

Here the adjective ``standard'' refers to the implicit function theorem manipulation made above, which can be of course made in the context of several variables too. Now with these abstract definitions in hand, we can go back to Conclusion 5.2, and formulate a more precise version of that observation, in the form of a theorem, as follows:

\begin{theorem}
We can convert any standard system of ODE into a standard order $1$ system of ODE, by suitably enlarging the size of the unknown vector.
\end{theorem}

\begin{proof}
This is indeed clear from definitions, because with $y=(x,\dot{x},\ldots,x^{(k-1)})$, in the context of Definition 5.5, the system there takes the following form, as desired:
$$\dot{y}_1=y_2$$
$$\dot{y}_2=y_3$$
$$\vdots$$
$$\dot{y}_{k-1}=y_k$$
$$\dot{y}_k=f(t,y)$$

Thus, we are led to the conclusion in the statement. There are of course many explicit applications of this method, and further comments that can be made. More later.
\end{proof}

Getting now to the point where we wanted to get, in order to get truly started with all this, with some mathematics going on, let us have a look at the systems of ODE which are linear. That is, we would like to solve equations as follows, with $f_i$ being linear:
$$x_1^{(k)}=f_1(t,x,\dot{x},\ldots,x^{(k-1)})$$
$$\vdots$$
$$x_N^{(k)}=f_N(t,x,\dot{x},\ldots,x^{(k-1)})$$

By doing the manipulation in Theorem 5.6, and assuming that we are in the ``autonomous'' case, where there is no time $t$ in the linear function $f$ producing the system, we are led to a vector equation as follows, with $A\in M_N(\mathbb R)$ being a certain matrix:
$$x'=Ax$$

But here, we are in familiar territory, namely standard calculus, because in the 1D case, the solution simply appears by exponentiating, as follows: 
$$x=e^{tA}x_0$$

Which is something very nice, and with this understood, we can go back now to our original Question 5.1, from the beginning of this chapter. As already mentioned, that question was something very broad, and as something more concrete now, we have:

\begin{question}
The solution of a system of linear differential equations,
$$x'=Ax\quad,\quad x(0)=x_0$$
with $A\in M_N(\mathbb R)$, is normally given by $x=e^{tA}x_0$, and this because we should have:
$$(e^{tA})'=Ae^{tA}$$
But, what exactly is $e^{tA}$, and then, importantly, how to explicitly compute $e^{tA}$? 
\end{question}

To be more precise, again as with Question 5.1, this question appears indeed in a myriad contexts, all across physics and science, and with all this needing no further presentation. Observe also that, due to Theorem 5.6, this question allows us to deal with differential equations of higher order too, by enlarging the size of our vectors.

\section*{5b. Matrix exponential} 

So, let us attempt to solve Question 5.7. We would first like to talk about exponentials of matrices. But here, the answer can only be given by the following formula:
$$e^A=\sum_{k=0}^\infty\frac{A^k}{k!}$$

Now since what we have here is a series, whose convergence must be justified, all this leads us into analysis over $M_N(\mathbb R)$, or over $M_N(\mathbb C)$, if we want to deal directly with the complex case. So, getting started with our study, let us begin with:

\begin{theorem}
The following quantity, with sup over the norm $1$ vectors,
$$||A||=\sup_{||x||=1}||Ax||$$
where $||x||=\sqrt{\sum|x_i|^2}$ as usual, is a norm on $M_N(\mathbb C)$. We have
$$||AB||\leq||A||\cdot||B||$$
for any two matrices $A,B\in M_N(\mathbb C)$. Also, we have the estimate
$$||A||\geq\sup_i|\lambda_i|$$
where $\lambda_1,\ldots\lambda_N\in\mathbb C$ are the eigenvalues, which is an equality when $A$ is diagonal.
\end{theorem}

\begin{proof}
All this is elementary, coming from definitions, the idea being as follows:

\medskip

(1) Regarding the norm conditions, $||A||\geq0$ with equality precisely when $A=0$ is clear, $||\lambda A||=|\lambda|\cdot||A||$ is clear too, and finally $||A+B||\leq||A||+||B||$ is clear too.

\medskip

(2) Regarding the second assertion, $||AB||\leq||A||\cdot||B||$, this follows from:
\begin{eqnarray*}
||AB||
&=&\sup_{||x||=1}||ABx||\\
&\leq&||A||\sup_{||x||=1}||Bx||\\
&=&||A||\cdot||B||
\end{eqnarray*}

(3) In order to prove the last assertion, observe first that by linearity we have:
$$||A||=\sup_{x\neq0}\frac{||Ax||}{||x||}$$

Now assuming that we have an eigenvector of our matrix, $Ax=\lambda x$, we obtain:
$$||A||\geq\frac{||Ax||}{||x||}=\frac{||\lambda x||}{||x||}=|\lambda|$$

Thus we have indeed $||A||\geq\sup_i|\lambda_i|$, with $\lambda_1,\ldots\lambda_N\in\mathbb C$ being the eigenvalues. 

\medskip

(4) Next, and finishing the proof of our theorem, as stated, observe that when our matrix is diagonal, $A=diag(\lambda_1,\ldots,\lambda_N)$, we obtain an equality, due to:
\begin{eqnarray*}
||Ax||
&=&\sqrt{\sum_i|\lambda_ix_i|^2}\\
&\leq&\sup_i|\lambda_i|\cdot\sqrt{\sum_i|x_i|^2}\\
&=&\sup_i|\lambda_i|\cdot||x||
\end{eqnarray*}

(5) We should mention that for non-diagonal matrices, the inequality $||A||\geq\sup_i|\lambda_i|$ might be strict. In fact, the correct formula, valid for any matrix, is as follows, with $\mu_1,\ldots\mu_N\geq0$ being the eigenvalues of the modulus $|A|=\sqrt{A^*A}$, from chapter 3:
$$||A||=\sup_i\mu_i$$

In other words, the norm of a complex matrix is its largest singular value. And, more on this in chapter 6 below, when discussing the singular value theorem.

\medskip

(6) Finally, as a general comment, we already saw in fact such things in Part I, in an indirect form, when talking about density results inside $M_N(\mathbb C)$. Note also that the space $M_N(\mathbb C)$ being finite dimensional, all the possible norms on it are equivalent.
\end{proof}

Now with the above result in hand, we can do analysis over $M_N(\mathbb C)$, and in particular we can investigate our exponentiation problem, with the following conclusions:

\begin{theorem}
We can talk about the exponentials of matrices $A\in M_N(\mathbb C)$, given by
$$e^A=\sum_{k=0}^\infty\frac{A^k}{k!}$$
and these exponentials have the following basic properties:
\begin{enumerate}
\item $||e^A||\leq e^{||A||}$.

\item If $D=diag(\lambda_i)$ then $e^D=diag(e^{\lambda_i})$.

\item If $P$ is invertible, $e^{PDP^{-1}}=Pe^DP^{-1}$.

\item If $A=PDP^{-1}$ with $D=diag(\lambda_i)$, then $e^A=Pdiag(e^{\lambda_i})P^{-1}$.
\end{enumerate}
\end{theorem}

\begin{proof}
The fact that our exponential series converges indeed follows from (1), so we are left with proving (1-4), and this can be done as follows:

\medskip

(1) We have indeed the following computation, using the various properties of the norm, and notably the formula $||AB||\leq||A||\cdot||B||$, from Theorem 5.8:
$$||e^A||
\leq\sum_{k=0}^\infty\left|\left|\frac{A^k}{k!}\right|\right|
\leq\sum_{k=0}^\infty\frac{||A||^k}{k!}
=e^{||A||}$$

(2) This is clear from definitions, with the computation being as follows:
$$\exp\begin{pmatrix}
\lambda_1\\
&\ddots\\
&&\lambda_N\end{pmatrix}
=\sum_{k=0}^\infty\begin{pmatrix}
\lambda_1^k\\
&\ddots\\
&&\lambda_N^k\end{pmatrix}\Big/k!
=\begin{pmatrix}
e^{\lambda_1}\\
&\ddots\\
&&e^{\lambda_N}\end{pmatrix}$$

(3) Again, this is clear from definitions, the computation being as follows:
\begin{eqnarray*}
e^{PDP^{-1}}
&=&\sum_{k=0}^\infty\frac{(PDP^{-1})^k}{k!}\\
&=&\sum_{k=0}^\infty\frac{PDP^{-1}\cdot PDP^{-1}\ldots PDP^{-1}}{k!}\\
&=&\sum_{k=0}^\infty\frac{PD^kP^{-1}}{k!}\\
&=&P\left(\sum_{k=0}^\infty\frac{D^k}{k!}\right)P^{-1}\\
&=&Pe^DP^{-1}
\end{eqnarray*}

(4) This follows indeed by combining (2) and (3).
\end{proof}

As a consequence of our theory, we can now state, in relation with Question 5.7:

\begin{theorem}
Given a matrix $A\in M_N(\mathbb C)$, the solution of the system 
$$x'=Ax\quad,\quad x(0)=x_0$$
is given by $x=e^{tA}x_0$.
\end{theorem}

\begin{proof}
We have two things to be proved, the idea being as follows:

\medskip

(1) Regarding the existence, $x=e^{tA}x_0$ satisfies the equation $x'=Ax$, due to:
\begin{eqnarray*}
x'
&=&(e^{tA}x_0)'\\
&=&\left(\sum_{k=0}^\infty\frac{(tA)^kx_0}{k!}\right)'\\
&=&\sum_{k=0}^\infty\frac{kt^{k-1}A^kx_0}{k!}\\
&=&A\sum_{k=1}^\infty\frac{t^{k-1}A^{k-1}x_0}{(k-1)!}\\
&=&A\sum_{l=0}^\infty\frac{t^lA^lx_0}{l!}\\
&=&Ae^{tA}x_0\\
&=&Ax
\end{eqnarray*}

Also, the formula $x(0)=x_0$ is clear from $e^{0_N}=1_N$, that is, from the fact that the exponential of the null $N\times N$ matrix is the identity $N\times N$ matrix.

\medskip

(2) Regarding the uniqueness, if $x$ is a solution of $x'=Ax$, then by using the trivial fact that $A$ commutes with the exponentials $e^{-tA}$, we conclude that $y=e^{-tA}x$ satisfies:
\begin{eqnarray*}
y'
&=&(e^{-tA}x)'\\
&=&-Ae^{-tA}x+e^{-tA}x'\\
&=&-Ae^{-tA}x+e^{-tA}Ax\\
&=&-Ae^{-tA}x+Ae^{-tA}x\\
&=&0
\end{eqnarray*}

Thus $y$ is constant, and more specifically $y=x_0$, and so $x=e^{tA}x_0$, as desired.
\end{proof}

As a key result now, which shows that things are certainly more complicated with matrices than with real numbers, when computing exponentials, we have:

\begin{theorem}
We have the following formula, when $A,B$ commute:
$$e^{A+B}=e^Ae^B$$
However, when the matrices $A,B$ do not commute, this formula might fail.
\end{theorem}

\begin{proof}
We have two assertions here, the idea being as follows:

\medskip

(1) As a first observation, when two matrices $A,B$ commute we can compute the powers $(A+B)^k$ as for the usual numbers, and we have a binomial formula, namely:
\begin{eqnarray*}
(A+B)^k
&=&(A+B)(A+B)\ldots(A+B)\\
&=&A^k+kA^{k-1}B+\ldots+kAB^{k-1}+B^k\\
&=&\sum_{r=0}^k\binom{k}{r}A^rB^{k-r}
\end{eqnarray*}

Now by using this binomial formula for $A,B$ we obtain, as for the usual numbers:
\begin{eqnarray*}
e^{A+B}
&=&\sum_{k=0}^\infty\frac{(A+B)^k}{k!}\\
&=&\sum_{k=0}^\infty\sum_{r=0}^k\binom{k}{r}\frac{A^rB^{k-r}}{k!}\\
&=&\sum_{k=0}^\infty\sum_{r=0}^k\frac{A^rB^{k-r}}{r!(k-r)!}\\
&=&\sum_{r=0}^\infty\sum_{s=0}^\infty\frac{A^rB^s}{r!s!}\\
&=&\sum_{r=0}^\infty\frac{A^r}{r!}\sum_{s=0}^\infty\frac{B^s}{s!}\\
&=&e^Ae^B
\end{eqnarray*}

(2) In order to find now a counterexample to $e^{A+B}=e^Ae^B$, we need some matrices which do not commute, $AB\neq BA$, and the simplest such matrices are as follows:
$$J=\begin{pmatrix}0&1\\0&0\end{pmatrix}\quad,\quad J^*=\begin{pmatrix}0&0\\1&0\end{pmatrix}$$

Indeed, the products of these two matrices are given by the following formulae:
$$JJ^*=\begin{pmatrix}1&0\\0&0\end{pmatrix}\quad,\quad J^*J=\begin{pmatrix}0&0\\0&1\end{pmatrix}$$

Now observe that, since these two products are both diagonal, we can compute right away their exponentials, and we are led to the following conclusion:
$$e^{JJ^*}=\begin{pmatrix}e&0\\0&0\end{pmatrix}\neq 
\begin{pmatrix}0&0\\0&e\end{pmatrix}=e^{J^*J}$$

Thus, we have a counterexample to $e^{AB}=e^{BA}$, but bad luck, this being not exactly the counterexample we were looking for, there is still some work to do. So, let us exponentiate our matrices. Regarding $J$, by using the formula $J^2=0$, we obtain:
\begin{eqnarray*}
e^J
&=&\sum_{k=0}^\infty\begin{pmatrix}0&1\\0&0\end{pmatrix}^k\Big/k!\\
&=&\begin{pmatrix}1&0\\0&1\end{pmatrix}+\begin{pmatrix}0&1\\0&0\end{pmatrix}+\begin{pmatrix}0&0\\0&0\end{pmatrix}+\ldots\\
&=&\begin{pmatrix}1&1\\0&1\end{pmatrix}
\end{eqnarray*}

Similarly, regarding $J^*$, by using the formula $(J^*)^2=0$, we obtain:
\begin{eqnarray*}
e^{J^*}
&=&\sum_{k=0}^\infty\begin{pmatrix}0&0\\1&0\end{pmatrix}^k\Big/k!\\
&=&\begin{pmatrix}1&0\\0&1\end{pmatrix}+\begin{pmatrix}0&0\\1&0\end{pmatrix}+\begin{pmatrix}0&0\\0&0\end{pmatrix}+\ldots\\
&=&\begin{pmatrix}1&0\\1&1\end{pmatrix}
\end{eqnarray*}

Now by making products, we obtain the following formulae:
$$e^Je^{J^*}=\begin{pmatrix}1&1\\0&1\end{pmatrix}
\begin{pmatrix}1&0\\1&1\end{pmatrix}
=\begin{pmatrix}2&1\\1&1\end{pmatrix}$$
$$e^{J^*}e^J=\begin{pmatrix}1&0\\1&1\end{pmatrix}
\begin{pmatrix}1&1\\0&1\end{pmatrix}
=\begin{pmatrix}1&1\\1&2\end{pmatrix}$$

But these two formulae give, at least in theory, our counterexample to the multiplication formula $e^{A+B}=e^Ae^B$, due to the following logical implication:
$$e^Je^{J^*}\neq e^{J^*}e^J\implies e^{J+J^*}\neq e^Je^{J^*}\ {\rm or}\ e^{J^*+J}\neq e^{J^*}e^J$$

This being said, let us do a clean work, and find out the explicit counterexample. For this purpose, we must compute $e^{J+J^*}$. The matrix to be exponentiated is:
$$J+J^*=\begin{pmatrix}0&1\\1&0\end{pmatrix}$$

Now this matrix being a symmetry, $(J+J^*)^2=1$, we are led to the following formula:
\begin{eqnarray*}
e^{J+J^*}
&=&\sum_{k=0}^\infty\begin{pmatrix}0&1\\1&0\end{pmatrix}^k\Big/k!\\
&=&\sum_{l=0}^\infty\begin{pmatrix}1&0\\0&1\end{pmatrix}\Big/(2l)!
+\sum_{l=0}^\infty\begin{pmatrix}0&1\\1&0\end{pmatrix}\Big/(2l+1)!\\
&=&\begin{pmatrix}\cosh 1&\sinh 1\\ \sinh 1&\cosh 1\end{pmatrix}
\end{eqnarray*}

Which looks quite exciting. In any case, this matrix being clearly different from $e^Je^{J^*}$, and from $e^{J^*}e^J$ too, we have now our counterexample to $e^{A+B}=e^Ae^B$, as desired.
\end{proof}

Moving forward, in order to compute exponentials, the main tool remains the formula from Theorem 5.9 (4). So, let us see how that formula works. We can actually use here as input the symmetry $J+J^*$ from the previous proof, and we get, as we should:
\begin{eqnarray*}
\exp\begin{pmatrix}0&1\\1&0\end{pmatrix}
&=&\exp\left[\frac{1}{2}
\begin{pmatrix}1&1\\1&-1\end{pmatrix}
\begin{pmatrix}1&0\\0&-1\end{pmatrix}
\begin{pmatrix}1&1\\1&-1\end{pmatrix}\right]\\
&=&\frac{1}{2}
\begin{pmatrix}1&1\\1&-1\end{pmatrix}
\begin{pmatrix}e&0\\0&e^{-1}\end{pmatrix}
\begin{pmatrix}1&1\\1&-1\end{pmatrix}\\
&=&\frac{1}{2}
\begin{pmatrix}1&1\\1&-1\end{pmatrix}
\begin{pmatrix}e&e\\e^{-1}&-e^{-1}\end{pmatrix}\\
&=&\frac{1}{2}
\begin{pmatrix}e+e^{-1}&e-e^{-1}\\e-e^{-1}&e+e^{-1}\end{pmatrix}\\
&=&\begin{pmatrix}\cosh 1&\sinh 1\\ \sinh 1&\cosh 1\end{pmatrix}
\end{eqnarray*}

Beyond the diagonalizable case, the only computations that we have so far are those for the matrices $J,J^*$, from the above proof. But these computations, crucially based on the fact that $J,J^*$ are nilpotent, suggest formulating a general result, as follows:

\begin{theorem}
Assuming that $A\in M_N(\mathbb C)$ is nilpotent, $A^s=0$, we have:
$$e^A=\sum_{k=0}^{s-1}\frac{A^k}{k!}$$
More generally, assuming $A^s=0$, we have the following formula,
$$e^{\lambda+A}=e^{\lambda}\sum_{k=0}^{s-1}\frac{A^k}{k!}$$
valid for any parameter $\lambda\in\mathbb C$.
\end{theorem}

\begin{proof}
The first formula is clear from definitions, and the second one follows from it, by using the fact that the matrices $\lambda I$ and $A$ commute, as follows:
\begin{eqnarray*}
e^{\lambda I+A}
&=&e^{\lambda I}e^A\\
&=&(e^\lambda I)\sum_{k=0}^{s-1}\frac{A^k}{k!}\\
&=&e^{\lambda}\sum_{k=0}^{s-1}\frac{A^k}{k!}
\end{eqnarray*}

Thus, we are led to the conclusions in the statement.
\end{proof}

Before going further with our study, which normally means going head-first into the non-diagonalizable case, let us have a listen to cat, who's meowing something, as usual since I started this book, about the diagonalizable matrices being dense. Good point, cat, and double meal for you tonight, because thinking well, by using that density result we can indeed say something nice about matrix exponentials, as follows:

\begin{theorem}
We have the following formula,
$$\det(e^A)=e^{Tr(A)}$$
valid for any matrix $A\in M_N(\mathbb C)$.
\end{theorem}

\begin{proof}
This is something quite tricky, because according to the definition of the exponential, the computation that we have to do looks of extreme difficulty, as follows:
$$\det\left(\sum_{k=0}^\infty\frac{A^k}{k!}\right)=?$$

But we won't be discouraged by this. For the diagonal matrices, we have:
\begin{eqnarray*}
\det\left[\exp\begin{pmatrix}
\lambda_1\\
&\ddots\\
&&\lambda_N
\end{pmatrix}\right]
&=&\det\begin{pmatrix}
e^{\lambda_1}\\
&\ddots\\
&&e^{\lambda_N}
\end{pmatrix}\\
&=&e^{\lambda_1+\ldots+\lambda_N}\\
&=&\exp\left[Tr\begin{pmatrix}
\lambda_1\\
&\ddots\\
&&\lambda_N
\end{pmatrix}\right]
\end{eqnarray*}

Next, by using this, for the diagonalizable matrices, $A=PDP^{-1}$, we have:
\begin{eqnarray*}
\det(e^A)
&=&\det(e^{PDP^{-1}})\\
&=&\det(Pe^DP^{-1})\\
&=&\det(e^D)\\
&=&e^{Tr(D)}\\
&=&e^{Tr(PDP^{-1})}\\
&=&e^{Tr(A)}
\end{eqnarray*}

And finally, since the diagonalizable matrices are dense, as we know well since chapter 4, we get by continuity our result in general. As simple as that.
\end{proof}

So long for the matrix exponential, using beautiful mathematics and tricks. But, everything has to come to an end, and time now to get into some dirty work.

\section*{5c. The Jordan form}

In order to advance, let us go back to the general diagonalization material from chapter 2. We know from there how to diagonalize the real or complex matrices, in case these are indeed diagonalizable. Also, we have seen several general results, known as ``spectral theorems'', guaranteeing that a matrix is diagonalizable, in chapter 3.

\bigskip

In order to deal with the general case, let us start with the following definition:

\begin{definition}
Given a matrix $A\in M_N(\mathbb C)$, and a vector $x\in\mathbb C^N$, we set
$$C_x=span(x,Ax,A^2x,\ldots)$$
and call it cyclic subspace of $A$, associated to $x$.
\end{definition}

Here the terminology comes from the fact that $A$ acts in a somewhat cyclic way on $C_x$, or at least on the above spanning vectors, according to the following formula:
$$A(A^ix)=A^{i+1}x$$

In order to have some mathematics going, out of this observation, the first remark is that the cyclic space $C_x\subset\mathbb C^N$ is of course finite dimensional. Thus, in the above definition, we can say that $C_x$ appears as follows, with $k\in\mathbb N$ being chosen minimal, such that the space on the right coincides indeed with $C_x$, as constructed above:
$$C_x=span(x,Ax,A^2x,\ldots,A^{k-1}x)$$

In practice, the number $k\in\mathbb N$ must be minimal, as to have a formula as follows:
$$A^kx=a_0x+a_1Ax+a_2A^2x+\ldots+a_{k-1}A^{k-1}x$$

And with this, we are now ready to state our first theorem about the arbitrary matrices, going beyond the general diagonalization material from chapter 2, as follows:

\begin{theorem}
Given a matrix $A\in M_N(\mathbb C)$ and a vector $x\in\mathbb C^N$, the restriction of $A$ to the cyclic subspace $C_x$, with respect to the basis $x,Ax,A^2x,\ldots,A^{k-1}x$, is
$$C=\begin{pmatrix}
0&0&\ldots&0&a_0\\
1&0&\ldots&0&a_1\\
0&1&\ldots&0&a_2\\
&&\ddots&&\vdots\\
0&0&\ldots&1&a_{k-1}
\end{pmatrix}$$
where $a_i\in\mathbb C$ are such that $A^kx=a_0x+a_1Ax+a_2A^2x+\ldots+a_{k-1}A^{k-1}x$.
\end{theorem}

\begin{proof}
This follows from the above discussion. Indeed, in what regards the first $k-1$ columns of the restriction, these are indeed those above, with this coming from:
$$A(A^ix)=A^{i+1}x$$

As for the last column, this is again the one above, with this coming from:
$$A(A^{k-1}x)=a_0x+a_1Ax+a_2A^2x+\ldots+a_{k-1}A^{k-1}x$$

Thus, we are led to the conclusion in the statement.
\end{proof}

In order to further advance, our next observation is that, in the context of Theorem 5.15, what only matters is the following polynomial:
$$P(t)=t^k-a_{k-1}t^{k-1}-\ldots-a_2t^2-a_1t-a_0$$

So, let us reformulate everything in terms of such polynomials. We are led in this way into the following notion, which is something independent of the above discussion:

\begin{definition}
Given an arbitrary monic polynomial, written as
$$P(t)=t^k+b_{k-1}t^{k-1}+\ldots+b_2t^2+b_1t+b_0$$
the following matrix,
$$C_P=\begin{pmatrix}
0&0&\ldots&0&-b_0\\
1&0&\ldots&0&-b_1\\
0&1&\ldots&0&-b_2\\
&&\ddots&&\vdots\\
0&0&\ldots&1&-b_{k-1}
\end{pmatrix}$$
is called its companion matrix.
\end{definition}

Which looks quite good, so our plan now will be to study such companion matrices, and come back afterwards to Theorem 5.15. In what regards the first task, we have:

\begin{theorem}
The companion matrix $C_P$ of a polynomial $P$,
$$C_P=\begin{pmatrix}
0&0&\ldots&0&-b_0\\
1&0&\ldots&0&-b_1\\
0&1&\ldots&0&-b_2\\
&&\ddots&&\vdots\\
0&0&\ldots&1&-b_{k-1}
\end{pmatrix}$$
has the following properties:
\begin{enumerate}
\item Its characteristic polynomial is $P$.

\item Its minimal polynomial is $P$, too.

\item All the eigenspaces are $1$-dimensional.

\item $C_P$ is diagonalizable when the roots of $P$ are distinct.
\end{enumerate}
\end{theorem}

\begin{proof}
This is something straightforward, the idea being as follows:

\medskip

(1) In order to compute the characteristic polynomial, we switch the first two rows, and we eliminate $t$ from the first column. This leads to the following formula:
\begin{eqnarray*}
\det(t-C_P)
&=&\begin{vmatrix}
t&0&0&\ldots&0&0&b_0\\
-1&t&0&\ldots&0&0&b_1\\
0&-1&t&\ldots&0&0&b_2\\
&&\ddots&\ddots&&\vdots\\
0&0&0&\ldots&t&0&b_{k-3}\\
0&0&0&\ldots&-1&t&b_{k-2}\\
0&0&0&\ldots&0&-1&t+b_{k-1}
\end{vmatrix}\\
&=&-\begin{vmatrix}
-1&t&0&\ldots&0&0&b_1\\
t&0&0&\ldots&0&0&b_0\\
0&-1&t&\ldots&0&0&b_2\\
&&\ddots&\ddots&&\vdots\\
0&0&0&\ldots&t&0&b_{k-3}\\
0&0&0&\ldots&-1&t&b_{k-2}\\
0&0&0&\ldots&0&-1&t+b_{k-1}
\end{vmatrix}\\
&=&-\begin{vmatrix}
-1&t&0&\ldots&0&0&b_1\\
0&t^2&0&\ldots&0&0&b_0+b_1t\\
0&-1&t&\ldots&0&0&b_2\\
&&\ddots&\ddots&&\vdots\\
0&0&0&\ldots&t&0&b_{k-3}\\
0&0&0&\ldots&-1&t&b_{k-2}\\
0&0&0&\ldots&0&-1&t+b_{k-1}
\end{vmatrix}
\end{eqnarray*}

Next, we switch the second and third rows, and we eliminate $t^2$ from the second column, again with the help of the $-1$ on the diagonal. We obtain in this way:
\begin{eqnarray*}
\det(t-C_P)
&=&\begin{vmatrix}
-1&t&0&\ldots&0&0&b_1\\
0&-1&t&\ldots&0&0&b_2\\
0&t^2&0&\ldots&0&0&b_0+b_1t\\
&&\ddots&\ddots&&\vdots\\
0&0&0&\ldots&t&0&b_{k-3}\\
0&0&0&\ldots&-1&t&b_{k-2}\\
0&0&0&\ldots&0&-1&t+b_{k-1}
\end{vmatrix}\\
&=&\begin{vmatrix}
-1&t&0&\ldots&0&0&b_1\\
0&-1&t&\ldots&0&0&b_2\\
0&0&0&\ldots&0&0&b_0+b_1t+b_2t^2\\
&&\ddots&\ddots&&\vdots\\
0&0&0&\ldots&t&0&b_{k-3}\\
0&0&0&\ldots&-1&t&b_{k-2}\\
0&0&0&\ldots&0&-1&t+b_{k-1}
\end{vmatrix}
\end{eqnarray*}

And so on by reccurence, and in the end we obtain, as desired:
\begin{eqnarray*}
\det(t-C_P)
&=&(-1)^{k-1}\begin{vmatrix}
-1&t&0&\ldots&0&0&b_1\\
0&-1&t&\ldots&0&0&b_2\\
0&0&-1&\ldots&0&0&b_3\\
&&&\ddots&&\vdots\\
0&0&0&\ldots&-1&0&b_{k-2}\\
0&0&0&\ldots&0&-1&t+b_{k-1}\\
0&0&0&\ldots&0&0&P(t)
\end{vmatrix}\\
&=&P(t)
\end{eqnarray*}

(2) Regarding now the minimal polynomial, this is clearly $P$ too.

\medskip

(3) In order to discuss now the eigenspaces, assume $\det(\lambda-C_P)=0$, which means $P(\lambda)=0$. We know from the above that we have a row equivalence, as follows:
$$\lambda-C_P\sim\begin{pmatrix}
-1&t&0&\ldots&0&0&b_1\\
0&-1&t&\ldots&0&0&b_2\\
0&0&-1&\ldots&0&0&b_3\\
&&&\ddots&&\vdots\\
0&0&0&\ldots&-1&0&b_{k-2}\\
0&0&0&\ldots&0&-1&t+b_{k-1}\\
0&0&0&\ldots&0&0&P(t)
\end{pmatrix}$$

Thus, the eigenspaces are indeed 1-dimensional, as stated.

\medskip

(4) Finally, the fact that our companion matrix $C_P$ is diagonalizable precisely when the roots of the polynomial $P$ are distinct is clear too, from the above formulae.
\end{proof}

Time now to go back to cyclic subspaces. We are first led to the following result:

\begin{theorem}[Cayley-Hamilton]
Any matrix $A\in M_N(\mathbb C)$ satisfies:
$$P_A(A)=0$$
In particular, the minimal polynomial divides the characteristic polynomial.
\end{theorem}

\begin{proof}
In order to prove this, pick a nonzero vector $x\in\mathbb C^N$, construct the associated cyclic subspace $C_x$, and then pick a complement for $C_x$, as to have:
$$\mathbb C^N=C_x\oplus V$$

With respect to this decomposition, our matrix $A$ becomes block-diagonal:
$$A=\begin{pmatrix}C_P&B\\0&D\end{pmatrix}$$

At the level of the characteristic polynomial, this gives a formula as follows:
$$P_A(t)=P_{C_P}(t)P_D(t)=P(t)P_D(t)$$

We know from Theorem 5.17 that we have $P(C_p)=0$, and it follows that we have $P_A(A)x=0$. But since $x\neq0$ was arbitrary, this gives $P_A(A)=0$, as desired.
\end{proof}

As a second result now, which truly advances us in our study, we have:

\begin{theorem}
Any matrix $A\in M_N(\mathbb C)$ can be written, up to a base change, as
$$A=\begin{pmatrix}
C_{P_1}\\
&\ddots\\
&&C_{P_k}
\end{pmatrix}$$
with each $C_{P_i}$ being a companion matrix. In this picture we have
$$P_A=P_1\ldots P_k$$
and $A$ is diagonalizable precisely when each $P_i$ has distinct roots.
\end{theorem}

\begin{proof}
This follows indeed from what we have in the above, via a straightforward recurrence, and we will leave the details here as an instructive exercise.
\end{proof}

Many other things can be said about Theorem 5.19, which is called cyclic subspace decomposition, and for more on this, we refer to the literature on the subject.

\bigskip

Next, we are led in this way to the Jordan form, which applies too to any matrix:

\index{eigenspaces}
\index{Jordan form}
\index{Jordan blocks}

\begin{theorem}
Any matrix $A\in M_N(\mathbb C)$ can be written, up to a base change, as
$$A=\begin{pmatrix}
J_1\\
&\ddots\\
&&J_k
\end{pmatrix}$$
with each $J_i$ being a Jordan block, meaning a matrix as follows,
$$J_i=\begin{pmatrix}
\lambda_i&1\\
&\lambda_i&1\\
&&\ddots&\ddots\\
&&&\lambda_i&1\\
&&&&\lambda_i
\end{pmatrix}$$
with our usual convention that blank spaces stand for $0$ entries.
\end{theorem}

\begin{proof}
We can deduce this from the cyclic subspace decomposition, as follows:

\medskip

(1) As a first ingredient, due to Jordan-Chevalley, we can decompose any matrix $A\in M_N(\mathbb C)$ as $A=B+C$, with $B$ diagonalizable, $C^N=0$, and $BC=CB$. Indeed, let us factor the minimal polynomial of our matrix $A\in M_N(\mathbb C)$, as follows:
$$R_A(t)=(t-\lambda_1)^{m_1}\ldots(t-\lambda_k)^{m_k}$$

Now if we set $L_i=\ker(A-\lambda_i)^{m_i}$, we have a direct sum decomposition, as follows:
$$\mathbb C^N=L_1\oplus\ldots\oplus L_k$$

But with this done, we can define matrices $B,C\in M_N(\mathbb C)$ block-diagonally, by:
$$B_{|L_i}=\lambda_i1_{L_i}\quad,\quad C_{|L_i}=A_{|L_i}-\lambda_i1_{L_i}$$

Now observe that we have indeed $A=B+C$, with $B$ being diagonalizable, and with $BC=CB$. Finally, since $R_A(A)=0$, we have as well $C^N=0$, as desired.

\medskip

(2) Next, let us apply the Jordan-Chevalley decomposition, as performed above, to a companion matrix $C_P$ coming from a polynomial of type $P(t)=(t-\lambda)^k$. We conclude that such a companion matrix must be similar to a Jordan block, as follows:
$$C_P\sim\begin{pmatrix}
\lambda&1\\
&\lambda&1\\
&&\ddots&\ddots\\
&&&\lambda&1\\
&&&&\lambda
\end{pmatrix}$$

(3) Let us turn now to the proof of the theorem. By using the Jordan-Chevalley decomposition, it is enough to prove the theorem for matrices of type $A=\lambda 1_N+C$, with $C^N=0$. But here, the result follows from the cyclic subspace decomposition.
\end{proof}

\section*{5d. Basic applications} 

As a first application now, we can go back to exponentials, and compute $e^A$ for any matrix, decomposed in Jordan form. In fact, we have already seen such computations, in the proof of Theorem 5.11, and the computations in general are quite similar. 

\bigskip

To be more precise, let us write the matrix to be exponentiated in Jordan form, as in Theorem 5.20, as follows, with $P$ denoting the passage matrix used there:
$$A=P\begin{pmatrix}
J_1\\
&\ddots\\
&&J_k
\end{pmatrix}P^{-1}$$

According to Theorem 5.9, the exponential is then given by the following formula:
$$e^A=P\begin{pmatrix}
e^{J_1}\\
&\ddots\\
&&e^{J_k}
\end{pmatrix}P^{-1}$$

Thus, it is enough to know how to exponentiate Jordan blocks. So, consider a Jordan block, as follows, with our usual convention that blank spaces stand for $0$ entries:
$$J=\begin{pmatrix}
\lambda&1\\
&\lambda&1\\
&&\ddots&\ddots\\
&&&\lambda&1\\
&&&&\lambda
\end{pmatrix}$$

In order to exponentiate this matrix, the best is to use Theorem 5.12. Indeed, what we have here is a multiple of the identity, summed with a nilpotent matrix:
$$J=\lambda+\begin{pmatrix}
0&1\\
&0&1\\
&&\ddots&\ddots\\
&&&0&1\\
&&&&0
\end{pmatrix}$$

Thus, we have the following formula for the exponential of our Jordan block:
$$e^J=e^{\lambda}\exp\begin{pmatrix}
0&1\\
&0&1\\
&&\ddots&\ddots\\
&&&0&1\\
&&&&0
\end{pmatrix}$$

So, we are led to the question of exponentiating the matrix on the right, namely:
$$N=\begin{pmatrix}
0&1\\
&0&1\\
&&\ddots&\ddots\\
&&&0&1\\
&&&&0
\end{pmatrix}$$

Now in order to exponentiate this latter matrix, we can use the fact that this matrix is nilpotent. Indeed, the square of this matrix is given by the following formula: 
$$N^2=\begin{pmatrix}
0&0&1\\
&0&0&1\\
&&\ddots&\ddots&\ddots\\
&&&0&0&1\\
&&&&0&0\\
&&&&&0
\end{pmatrix}$$

Then, the third power of this matrix is given by the following formula:
$$N^3=\begin{pmatrix}
0&0&0&1\\
&0&0&0&1\\
&&\ddots&\ddots&\ddots&\ddots\\
&&&0&0&0&1\\
&&&&0&0&0\\
&&&&&0&0\\
&&&&&&0
\end{pmatrix}$$

And so on up to the $(s-1)$-th power, with $s$ being the size of our matrix, which is given by the following formula, with our usual convention for blank spaces:
$$N^s=\begin{pmatrix}
0&&\ldots&\ldots&0&1\\
&0&&&&0\\
&&\ddots&&&\vdots\\
&&&\ddots&&\vdots\\
&&&&0&\\
&&&&&0
\end{pmatrix}$$

Now by using the exponentiating formula in Theorem 5.12, for this nilpotent matrix $N$, we obtain the following formula, for its exponential:
$$e^N=\begin{pmatrix}
1&1&\frac{1}{2}&\frac{1}{6}&&\ldots&&\frac{1}{(s-1)!}\\
&1&1&\frac{1}{2}&\frac{1}{6}\\
&&\ddots&\ddots&\ddots&\ddots&&\vdots\\
&&&\ddots&\ddots&\ddots&\ddots\\
&&&&1&1&\frac{1}{2}&\frac{1}{6}\\
&&&&&1&1&\frac{1}{2}\\
&&&&&&1&1\\
&&&&&&&1
\end{pmatrix}$$

Summarizing, done with our computation, and we can now formulate:

\begin{theorem}
For a matrix written in Jordan form, as follows,
$$A=P\begin{pmatrix}
J_1\\
&\ddots\\
&&J_k
\end{pmatrix}P^{-1}$$
the corresponding exponential is given by the following formula,
$$e^A=P\begin{pmatrix}
e^{J_1}\\
&\ddots\\
&&e^{J_k}
\end{pmatrix}P^{-1}$$
with the exponential of each Jordan block being computed by the formula
$$\exp\begin{pmatrix}
\lambda&1\\
&\lambda&1\\
&&\ddots&\ddots\\
&&&\lambda&1\\
&&&&\lambda
\end{pmatrix}
=e^{\lambda}\begin{pmatrix}
1&1&\frac{1}{2}&\frac{1}{6}&&\ldots&&\frac{1}{(s-1)!}\\
&1&1&\frac{1}{2}&\frac{1}{6}\\
&&\ddots&\ddots&\ddots&\ddots&&\vdots\\
&&&\ddots&\ddots&\ddots&\ddots\\
&&&&1&1&\frac{1}{2}&\frac{1}{6}\\
&&&&&1&1&\frac{1}{2}\\
&&&&&&1&1\\
&&&&&&&1
\end{pmatrix}$$
with $s$ being the size of our Jordan block.
\end{theorem}

\begin{proof}
This follows indeed from the above discussion.
\end{proof}

And good news, this is all we need to know, being obviously something very powerful, closing any further mathematical discussion about exponentiation. We will be back to this later, in chapter 7, with a systematic discussion of the differential equations.

\bigskip

As another application of our theory, we can recover the density of the diagonalizable matrices, that we can get as well via the Jordan form, by perturbing the diagonal.

\bigskip

As yet another related topic, getting back to our spectral measure considerations from chapter 3, recall from there that any normal matrix has a spectral measure, formed by the Dirac masses at the eigenvalues. In the non-normal case, things can be quite complicated, and there are some interesting computations here for the Jordan blocks.

\section*{5e. Exercises}

This was a quite fundamental chapter, at the origin of all possible advaced and modern linear algebra topics, and as exercises on all this, we have:

\begin{exercise}
In relation with equations, have a look at the non-uniform gravitational falls too, first in $1$ dimension, and then in $2$ dimensions.
\end{exercise}

\begin{exercise}
Also in relation with equations, and with the general theory developed above, learn if needed the implicit function theorem.
\end{exercise}

\begin{exercise}
Learn a bit about normed spaces, generalities about them, as needed in the above, in order to talk about the exponential of matrices.
\end{exercise}

\begin{exercise}
Try remembering, and then finding matrix analogues, of some other formulae involving $\exp$, that you know from calculus.
\end{exercise}

\begin{exercise}
Work out all the details of the proof of the Cayley-Hamilton theorem, and find some applications of this theorem too.
\end{exercise}

\begin{exercise}
Fill in all the details for the proof of the block decomposition into companion matrices, and work out some applications of this, too.
\end{exercise}

\begin{exercise}
Work out all the details for the Jordan decomposition theorem, along the lines explained in the above.
\end{exercise}

\begin{exercise}
Learn as well some explicit algorithms for finding the Jordan form, based on the above material, or on some alternative approaches too.
\end{exercise}

As bonus exercise for this chapter, and no surprise here, put various matrices of your choice in Jordan form, the more the better, and the bigger the better, too.

\chapter{Singular values}

\section*{6a. Singular values}

We have seen in the previous chapters some powerful spectral theorems, allowing us to diagonalize many classes of interesting matrices, and the Jordan decomposition theorem too, which applies to any matrix. However, the story is not over here, because the complex matrices $A\in M_N(\mathbb C)$ are subject to far more useful decomposition results.

\bigskip

Here is a concrete, fundamental question, that we would like to discuss now:

\begin{question}
The complex numbers $a\in\mathbb C$, and so the $1\times1$ matrices $A\in M_1(\mathbb C)$ too, are subject to the following decomposition results:
\begin{enumerate}
\item $a=b+ic$, with $b,c\in\mathbb R$.

\item $a=(d-e)+i(f-g)$, with $d,e,f,g\geq0$.

\item $a=rw$ with $r=|a|$ and with $|w|=1$.

\item $a=re^{it}$ with $r=|a|$ and with $t\in\mathbb R$.
\end{enumerate}
What are the analogues of these decomposition results for $A\in M_N(\mathbb C)$, with $N\geq2$?
\end{question}

In answer, since we already talked a bit about polar decomposition in chapter 3, let us discuss the multiplicative questions (3,4) first, as a continuation of that material, and leave the additive questions (1,2), which require more preparations, for later.

\bigskip

So, getting back to the material from chapter 3, here is an account of what we know from there about the modulus of the matrices $A\in M_N(\mathbb C)$, along with a bit more:

\index{modulus of operator}
\index{absolute value}
\index{square root}

\begin{theorem}
Given $A\in M_N(\mathbb C)$, we can construct its modulus $|A|\in M_N(\mathbb C)$ as
$$|A|=\sqrt{A^*A}$$
by using the fact that $A^*A$ is diagonalizable, with positive eigenvalues, and:
\begin{enumerate}
\item The modulus is positive, $|A|\geq0$, and its square is $|A|^2=A^*A$.

\item The modulus is the unique matrix having these two properties.

\item At $N=1$, we obtain the usual absolute value of the complex numbers.

\item When $A$ is normal, the modulus is equally given by $|A|=\sqrt{AA^*}$.
\end{enumerate}
\end{theorem}

\begin{proof}
This is something very standard, the idea being as follows:

\medskip

(1) We already know this, from chapter 3. Indeed, the matrix $A^*A$ is positive, and according to our spectral theorems there, we can diagonalize this positive matrix $A^*A\geq0$ as follows, with $U\in U_N$, and with $D\in M_N(\mathbb R_+)$ being diagonal:
$$A^*A=UDU^*$$

Next, we can extract the square root $\sqrt{D}$, in the obvious way, as follows:
$$D=\begin{pmatrix}
\lambda_1\\
&\ddots\\
&&\lambda_N
\end{pmatrix}\quad\implies\quad 
\sqrt{D}=\begin{pmatrix}
\sqrt{\lambda_1}\\
&\ddots\\
&&\sqrt{\lambda_N}
\end{pmatrix}$$
 
But with this done, we can construct the modulus $|A|$, by setting:
$$|A|=U\sqrt{D}U^*$$

Indeed, this matrix does the job, being positive, and with its square being, as needed:
$$|A|^2=U\sqrt{D}U^*\cdot U\sqrt{D}U^*=UDU^*=A^*A$$

(2) Regarding the uniqueness, this follows from the fact that we can uniquely apply the continuous function $f(x)=\sqrt{x}$ to any positive matrix, and in particular to $A^*A\geq0$, and more on this later in this chapter, when discussing functional calculus. In the meantime, here is an ad-hoc proof. Our uniqueness claim amounts in proving that:
$$B^2=C^2,\ B,C\geq0\implies B=C$$

As a first observation, $B,C$ must have the same eigenvalues, appearing as the square roots of the eigenvalues of $B^2=C^2$. Next, by using the spectral theorem for both $B,C$, we can assume that $B=D$ is diagonal, and that $C=UDU^*$, with $U\in U_N$. Thus, we would like to prove the following, for $U\in U_N$, and $D\in M_N(\mathbb R_+)$ diagonal:
$$(UDU^*)^2=D^2\implies UDU^*=D$$

With the usual convention $[A,B]=0$ when two matrices commute, this reads:
$$[U,D^2]=0\implies[U,D]=0$$

But this follows from the spectral theorem applied to $U$. Indeed, our unitary matrix $U\in U_N$ can be diagonalized as follows, with $V\in U_N$, and $E\in M_N(\mathbb T)$ diagonal:
$$U=VEV^*$$

Thus the commutant of $U$ is given by $U'=VE'V^*$, and since the commutant $E'$ is easily seen to be closed under taking square roots, so must be $U'$, as desired.

\medskip

(3) This is indeed something clear, coming from definitions.

\medskip

(4) The point here is that it is possible to talk as well about $\sqrt{AA^*}$, which is in general different from $\sqrt{A^*A}$. But when $A$ is normal, there is no issue, because we have:
$$AA^*=A^*A\implies\sqrt{AA^*}=\sqrt{A^*A}$$

Thus, last assertion proved, and in relation with the complex numbers and (3), the good way of remembering this is by saying that ``for the $1\times1$ matrices the left and right moduli coincide, because the $1\times1$ matrices are automatically normal''.
\end{proof}

Still talking modulus, coming as a useful complement to Theorem 6.2, and answering some questions that appeared in chapter 5, we have the following key result:

\begin{theorem}
For any complex matrix $A\in M_N(\mathbb C)$, the following happen:
\begin{enumerate}
\item $||A||=||A^*||$.

\item $||A^*A||=||A||^2$.

\item $A$ and $|A|$ have the same norm.

\item $||A||$ is the largest eigenvalue of its modulus $|A|\geq0$.
\end{enumerate}
\end{theorem}

\begin{proof}
This is something quite tricky, the idea being as follows:

\medskip

(1) We recall that the adjoint matrix $A^*$ is subject to the following key formula:
$$<Ax,y>=<x,A^*y>$$

But with this, we can prove that $A$ and $A^*$ have the same norm, as follows:
\begin{eqnarray*}
||A||
&=&\sup_{||x||=1}\sup_{||y||=1}<Ax,y>\\
&=&\sup_{||y||=1}\sup_{||x||=1}<x,A^*y>\\
&=&||A^*||
\end{eqnarray*}

(2) In order to prove this, which is the key assertion, observe first that we have:
$$||A^*A||
\leq||A^*||\cdot||A||
=||A||^2$$

On the other hand, we have as well the reverse estimate, obtained as follows:
\begin{eqnarray*}
||A||^2
&=&\sup_{||x||=1}|<Ax,Ax>|\\
&=&\sup_{||x||=1}|<x,A^*Ax>|\\
&\leq&||A^*A||
\end{eqnarray*}

(3) This comes indeed from (2), via the following computation, using the standard fact that for a positive matrix $B\geq0$ we have $||B^2||=||B||^2$, which itself is an elementary consequence of the spectral theorem for the positive matrices, from chapter 3:
$$\Big|\Big||A|\Big|\Big|=||\sqrt{A^*A}||=\sqrt{||A^*A||}=||A||$$

(4) This comes as a consequence of (3), because the norm of a positive matrix $B\geq0$ is its largest eigenvalue, as shown for instance by the spectral theorem, from chapter 3.
\end{proof}

Regarding now the polar decomposition formula, for the complex matrices, this is again something that we know from chapter 3, the result here being as follows:

\index{polar decomposition}
\index{partial isometry}

\begin{theorem}
Given a matrix $A\in M_N(\mathbb C)$, the following happen:
\begin{enumerate}
\item When $A$ is invertible, we have $A=U|A|$, with $U$ being a unitary.

\item In general, we still have $A=U|A|$, with $U$ being a partial isometry.
\end{enumerate}
\end{theorem}

\begin{proof}
This is something that we know since chapter 3, but always good to talk about it again. According to our definition of the modulus, $|A|=\sqrt{A^*A}$, we have:
\begin{eqnarray*}
<|A|x,|A|y>
&=&<x,|A|^2y>\\
&=&<x,A^*Ay>\\
&=&<Ax,Ay>
\end{eqnarray*}

We conclude that the following linear application is well-defined, and isometric:
$$U:Im|A|\to Im(A)\quad,\quad 
|A|x\to Ax$$

But now we can further extend this linear isometric map $U$ into a partial isometry $U:\mathbb C^N\to\mathbb C^N$, in a straightforward way, by setting:
$$Ux=0\quad,\quad\forall x\in Im|A|^\perp$$

And the point is that, with this convention, the result follows. 
\end{proof}

As a continuation of this, let us discuss now the singular value theorem, which is a key result in linear algebra. The result can be formulated, a bit abstractly, as follows:

\begin{theorem}
We can write the action of any matrix $A\in M_N(\mathbb C)$ in the following form, with $\{e_n\}$, $\{f_n\}$ being orthonormal families, and with $\lambda_n\geq 0$:
$$A(x)=\sum_n\lambda_n<x,e_n>f_n$$
The numbers $\lambda_n$, called singular values of $A$, are the eigenvalues of the modulus $|A|$. In fact, the polar decomposition of $A$ is given by $A=U|A|$, with
$$|A|(x)=\sum_n\lambda_n<x,e_n>e_n$$
and with $U$ being given by $Ue_n=f_n$, and $U=0$ on the complement of $span(e_i)$.
\end{theorem}

\begin{proof}
This basically comes from what we already have, as follows:

\medskip

(1) Given two orthonormal families $\{e_n\}$, $\{f_n\}$, and a sequence of real numbers $\lambda_n\geq0$, consider the linear map given by the formula in the statement, namely:
$$A(x)=\sum_n\lambda_n<x,e_n>f_n$$

The adjoint of this linear map is then given by the following formula: 
$$A^*(x)=\sum_n\lambda_n<x,f_n>e_n$$

Thus, when composing $A^*$ with $A$, we obtain the following linear map:
$$A^*A(x)=\sum_n\lambda_n^2<x,e_n>e_n$$

Now by extracting the square root, we obtain the formula in the statement, namely:
$$|A|(x)=\sum_n\lambda_n<x,e_n>e_n$$

(2) Conversely, consider a matrix $A\in M_N(\mathbb C)$. Then $A^*A$ is self-adjoint, so we have a formula as follows, with $\{e_n\}$ being a certain orthonormal family, and with $\lambda_n\geq0$:
$$A^*A(x)=\sum_n\lambda_n^2<x,e_n>e_n$$

By extracting the square root we obtain the formula of $|A|$ in the statement, namely:
$$|A|(x)=\sum_n\lambda_n<x,e_n>e_n$$

Moreover, by setting $U(e_n)=f_n$, we obtain a second orthonormal family, $\{f_n\}$, such that the following formula holds:
$$A(x)
=U|A|
=\sum_n\lambda_n<x,e_n>f_n$$

Thus, our matrix $A\in M_N(\mathbb C)$ appears indeed as in the statement.
\end{proof}

As before with the polar decomposition, there are many possible applications of the singular value theorem. We will be back to this, on several occasions, in what follows. 

\bigskip

As a technical remark now, it is possible to slightly improve a part of the above statement. Consider indeed a linear map of the following form, with $\{e_n\}$, $\{f_n\}$ being orthonormal families as before, and with $\lambda_n$ being now complex numbers:
$$A(x)=\sum_n\lambda_n<x,e_n>f_n$$

The adjoint of this linear map is then given by the following formula: 
$$A^*(x)=\sum_n\bar{\lambda}_n<x,f_n>e_n$$

Thus, when composing $A^*$ with $A$, we obtain the following linear map:
$$A^*A(x)=\sum_n|\lambda_n|^2<x,e_n>e_n$$

Now by extracting the square root, we conclude that the polar decomposition of $A$ is given by $A=U|A|$, with the modulus $|A|$ being as follows:
$$|A|(x)=\sum_n|\lambda_n|<x,e_n>e_n$$

As for the partial isometry $U$, this is given by $Ue_n=w_nf_n$, and $U=0$ on the complement of $span(e_i)$, where $w_n\in\mathbb T$ are such that $\lambda_n=|\lambda_n|w_n$.

\bigskip

As already mentioned in the above, there are many possible applications of the singular value theorem. We will be back to this, on several occasions, in what follows. Also, we will discuss in chapter 8 below a remarkable generalization of the above results, to the case of certain special infinite matrices, whose associated linear operators have suitable compactness properties, making them quite similar to the usual matrices.

\bigskip

Finally, in relation with Question 6.1, (3) there was certainly answered, but (4) still needs some discussion. In relation with this, the problem is how to write the unitaries $U\in U_N$ as exponentials. We will be back to this question later in this book.

\section*{6b. Functional calculus}

Getting now to the additive problems in Question 6.1, namely (1) and (2) there, in order to solve them, we are in need of more systematic functional calculus. So, recall from chapter 5 that we know how to exponentiate any matrix $A\in M_N(\mathbb C)$, as follows:
$$e^A=\sum_{k=0}^\infty\frac{A^k}{k!}$$

As our next topic for this chapter, which is something very useful to know, we would like to systematically discuss the functional calculus for the complex matrices:
$$A\to f(A)$$

The general principle here, that we already met at the end of chapter 4, and used in chapter 5, is something quite intuitive, and which is very simple to state, as follows:

\begin{principle}
Under suitable regularity assumptions, on both functions and matrices, we can apply complex functions $f:\mathbb C\to\mathbb C$ to the complex matrices $A\in M_N(\mathbb C)$,
$$A\to f(A)$$
and at the level of the corresponding eigenvalues, if $\lambda_1,\ldots,\lambda_N\in\mathbb C$ are the eigenvalues of $A$, then $f(\lambda_1),\ldots,f(\lambda_N)\in\mathbb C$ should be the eigenvalues of $A$.
\end{principle}

This is obviously something quite general, and potentially something very useful too, for all sorts of purposes. And, as explained in chapter 4, one good reason for which we can expect this principle to hold comes from the density of the diagonalizable matrices. 

\bigskip

Indeed, assume first that our matrix is diagonalizable, as follows:
$$A=P\begin{pmatrix}
\lambda_1\\
&\ddots\\
&&\lambda_N
\end{pmatrix}
P^{-1}$$

In this case, we can in principle define a matrix $f(A)\in M_N(\mathbb C)$ by the following formula, and with its eigenvalues being indeed those predicted by Principle 6.6:
\begin{eqnarray*}
f(A)
&=&f\left[P\begin{pmatrix}
\lambda_1\\
&\ddots\\
&&\lambda_N
\end{pmatrix}
P^{-1}\right]\\
&=&Pf\begin{pmatrix}
\lambda_1\\
&\ddots\\
&&\lambda_N
\end{pmatrix}
P^{-1}\\
&=&P\begin{pmatrix}
f(\lambda_1)\\
&\ddots\\
&&f(\lambda_N)
\end{pmatrix}
P^{-1}
\end{eqnarray*}

As for the general case, where our matrix $A\in M_N(\mathbb C)$ is no longer assumed to be diagonalizable, this should normally follow from what we have above, by using the fact, that we know well from chapter 4, that the diagonalizable matrices are dense.

\bigskip

So, this was for the story, and in addition to this, we have already seen in chapter 5 how this method effectively works for the exponential function, $f(x)=e^x$.

\bigskip

In practice now, things are more complicated that this, because both our computation for the diagonal matrices, and the extension to the general case via density, normally require some regularity assumptions, on both $f$ and $A$, in order to truly work.

\bigskip

Summarizing, things to do for us, and for dealing with this problem, we will use:

\begin{method}
In order to establish Principle 6.6, we can use:
\begin{enumerate}
\item Various direct methods.

\item Density of the diagonalizable matrices.

\item The Jordan form, and other more specialized results.
\end{enumerate}
\end{method}

Getting started now, let us first discuss in detail the case of the polynomials. As a warm-up target function here, we have $f(x)=x^2$, with the result being as follows:

\begin{theorem}
We can apply $f(x)=x^2$ to any matrix $A\in M_N(\mathbb C)$, 
$$A\to A^2$$
and if $\lambda_1,\ldots,\lambda_N$ are the eigenvalues of $A$, then $\lambda_1^2,\ldots,\lambda_N^2$ are the eigenvalues of $A^2$.
\end{theorem}

\begin{proof}
This does not look difficult to establish, but let us have a detailed discussion about this, which will be quite instructive, by following Method 6.7: 

\medskip

(1) Starting with bare hands, let us see if we can solve the problem at $N=2$, via a direct computation, without using any trick. So, consider a $2\times2$ matrix, as follows:
$$A=\begin{pmatrix}
a&b\\ 
c&d
\end{pmatrix}$$

The square of this matrix is then given by the following formula:
$$A^2=\begin{pmatrix}
a&b\\ 
c&d
\end{pmatrix}\begin{pmatrix}
a&b\\ 
c&d
\end{pmatrix}
=\begin{pmatrix}
a^2+bc&ab+bd\\ 
ac+cd&bc+d^2
\end{pmatrix}$$

Now let us compare the eigenvalues of the matrices $A,A^2$. Those of the initial matrix $A$, say $r,s\in\mathbb C$, are subject to the following two equations:
$$r+s=Tr(A)=a+d$$
$$rs=\det(A)=ad-bc$$

As for the eigenvalues of the squared matrix $A^2$, say $R,S\in\mathbb C$, these are subject to some similar equations, which are as follows:
$$R+S=Tr(A^2)=a^2+d^2+2bc$$
$$RS=\det(A^2)=\det(A)^2$$

The second equality suggests that we should have $R=r^2,S=s^2$, so let us prove now that it is indeed so. For this purpose, we just have to check that the numbers $R=r^2,S=s^2$ sum up to the quantity computed above, and this is done as follows:
\begin{eqnarray*}
R+S
&=&r^2+s^2\\
&=&(r+s)^2-2rs\\
&=&(a+d)^2-2(ad-bc)\\
&=&a^2+d^2+2bc
\end{eqnarray*}

Summarizing, result proved with bare hands at $N=2$. However, when thinking a bit, such methods will become quite complicated at $N\geq3$, so we will stop here, with this.

\medskip

(2) Still with bare hands, but allowing us some tricks, namely the use of square roots of complex numbers, here is how the result can be established, at any $N\in\mathbb N$. Consider our matrix $A\in M_N(\mathbb C)$, and let us factorize its characteristic polynomial, as follows:
$$\det(A-x)=\prod_i(\lambda_i-x)$$

We have then the following computation, for the characteristic polynomial of $A^2$:
\begin{eqnarray*}
\det(A^2-x)
&=&\det((A-\sqrt{x})(A+\sqrt{x}))\\
&=&\det(A-\sqrt{x})\det(A+\sqrt{x})\\
&=&\prod_i(\lambda_i-\sqrt{x})\prod_i(\lambda_i+\sqrt{x})\\
&=&\prod_i(\lambda_i-\sqrt{x})(\lambda_i+\sqrt{x})\\
&=&\prod_i(\lambda_i^2-x)
\end{eqnarray*}

Thus, claimed proved, eventually, with bare hands, or almost.

\medskip

(3) Getting now to more advanced tricks, bringing heavy simplifications, we can use here density arguments. Indeed, assume first that our matrix is diagonalizable:
$$A=P\begin{pmatrix}
\lambda_1\\
&\ddots\\
&&\lambda_N
\end{pmatrix}
P^{-1}$$

In this case, we have the following computation, for the squared matrix:
\begin{eqnarray*}
A^2
&=&P\begin{pmatrix}
\lambda_1\\
&\ddots\\
&&\lambda_N
\end{pmatrix}
P^{-1}\cdot P\begin{pmatrix}
\lambda_1\\
&\ddots\\
&&\lambda_N
\end{pmatrix}
P^{-1}\\
&=&P\begin{pmatrix}
\lambda_1\\
&\ddots\\
&&\lambda_N
\end{pmatrix}^2
P^{-1}\\
&=&P\begin{pmatrix}
\lambda_1^2\\
&\ddots\\
&&\lambda_N^2
\end{pmatrix}
P^{-1}
\end{eqnarray*}

Thus, claim proved in this case, and the general case follows now by using the fact, that we know well from chapter 4, that the diagonalizable matrices are dense.

\medskip

(4) Finally, for our discussion to be complete, let us see as well what we get, by using a nuclear bomb, that is, the Jordan form. So, let us write the matrix to be squared in Jordan form, as in chapter 5, as follows, with $P$ denoting the passage matrix:
$$A=P\begin{pmatrix}
J_1\\
&\ddots\\
&&J_k
\end{pmatrix}P^{-1}$$

The square of this matrix is then given by the following formula:
$$e^A=P\begin{pmatrix}
J_1^2\\
&\ddots\\
&&J_k^2
\end{pmatrix}P^{-1}$$

Thus, it is enough to know how to square the Jordan blocks. So, consider a Jordan block, as follows, with our usual convention that blank spaces stand for $0$ entries:
$$J=\begin{pmatrix}
\lambda&1\\
&\lambda&1\\
&&\ddots&\ddots\\
&&&\lambda&1\\
&&&&\lambda
\end{pmatrix}$$

The square of this Jordan block is then given by the following formula:
$$J=\begin{pmatrix}
\lambda^2&2\lambda&1\\
&\lambda^2&2\lambda&1\\
&&\ddots&\ddots&\ddots\\
&&&\lambda^2&2\lambda&1\\
&&&&\lambda^2&2\lambda\\
&&&&&\lambda^2
\end{pmatrix}$$

Thus, we have the square $\lambda^2$ of the eigenvalue $\lambda$ on the diagonal, and by putting everything together, we are again led to the conclusion in the statement.
\end{proof}

As a next target, we have $f(x)=x^k$, with $k\in\mathbb N$. Here the result is as follows:

\begin{theorem}
We can apply $f(x)=x^k$ to any matrix $A\in M_N(\mathbb C)$, 
$$A\to A^k$$
and if $\lambda_1,\ldots,\lambda_N$ are the eigenvalues of $A$, then $\lambda_1^k,\ldots,\lambda_N^k$ are the eigenvalues of $A^k$.
\end{theorem}

\begin{proof}
We must extend the proof of Theorem 6.8, and skipping the discussion there at the totally bare hand level, a simple way is by using the roots of unity. Consider indeed our matrix $A\in M_N(\mathbb C)$, and let us factorize its characteristic polynomial, as follows:
$$\det(A-x)=\prod_i(\lambda_i-x)$$

We have then the following computation for the matrix $A^k$, with $w=e^{2\pi i/k}$:
\begin{eqnarray*}
\det(A^k-x)
&=&\det\left(\prod_j\left(A-w^j\sqrt[k]{x}\,\right)\right)\\
&=&\prod_j\det\left(A-w^j\sqrt[k]{x}\,\right)\\
&=&\prod_j\prod_i\left(\lambda_i-w^j\sqrt[k]{x}\,\right)\\
&=&\prod_i\prod_j\left(\lambda_i-w^j\sqrt[k]{x}\,\right)\\
&=&\prod_i(\lambda_i^k-x)
\end{eqnarray*}

Thus, we are led to the conclusion in the statement. Alternatively, we can of course use our usual, and elegant, density argument, or the Jordan form.
\end{proof}

With a bit more work, we can have the result for arbitrary polynomials, as follows:

\begin{theorem}
We can apply any $f\in\mathbb C[X]$ to any matrix $A\in M_N(\mathbb C)$, 
$$A\to f(A)$$
and if $\lambda_1,\ldots,\lambda_N$ are the eigenvalues of $A$, then $f(\lambda_1),\ldots,f(\lambda_N)$ are those of $f(A)$.
\end{theorem}

\begin{proof}
We must extend here the proof of Theorem 6.9. So, consider our matrix $A\in M_N(\mathbb C)$, and let us factorize its characteristic polynomial, as follows:
$$\det(A-x)=\prod_i(\lambda_i-x)$$

Now fix $x\in\mathbb C$, and let us factorize the polynomial $f(z)-x$, as follows:
$$f(z)-x=c\prod_j(z-x_j)$$

We have then the following computation, for the matrix $f(A)$:
\begin{eqnarray*}
\det(f(A)-x)
&=&\det\left(c\prod_j(A-x_j)\right)\\
&=&c^k\prod_j\det(A-x_j)\\
&=&c^k\prod_j\prod_i(\lambda_i-x_j)\\
&=&\prod_i\left(c\prod_j(\lambda_i-x_j)\right)\\
&=&\prod_i(f(\lambda_i)-x)
\end{eqnarray*}

Thus, we are led to the conclusion in the statement. Alternatively, we can of course use our usual, and elegant, density argument, or the Jordan form.
\end{proof}

Getting now to more complicated functions, we first have the inverse function:
$$x\to x^{-1}$$

This function can only be applied to the invertible matrices, and we have here:

\begin{theorem}
We can apply $x\to x^{-1}$ to any invertible matrix $A\in M_N(\mathbb C)$, 
$$A\to A^{-1}$$
and if $\lambda_1,\ldots,\lambda_N$ are the eigenvalues of $A$, then $\lambda_1^{-1},\ldots,\lambda_N^{-1}$ are those of $A^{-1}$.
\end{theorem}

\begin{proof}
There are many possible arguments that we can use here, as follows:

\medskip

(1) To start with, we can say that, given an invertible matrix $A\in M_N(\mathbb C)$, the matrix $A-\lambda$ is invertible precisely when $A^{-1}-\lambda^{-1}$ is, and this due to:
\begin{eqnarray*}
A^{-1}-\lambda^{-1}
&=&A^{-1}(1-\lambda^{-1}A)\\
&=&\lambda^{-1}A^{-1}(\lambda-A)
\end{eqnarray*}

(2) Alternatively, we can say that we have the following formula:
\begin{eqnarray*}
\det(A^{-1}-x)
&=&\det(A^{-1}(1-Ax))\\
&=&\det(-xA^{-1}(A-x^{-1}))\\
&=&(-x)^N\det(A^{-1})\det(A-x^{-1})\\
&=&\frac{(-x)^N}{\det A}\cdot\det(A-x^{-1}) 
\end{eqnarray*}

Thus, we are once again led to the conclusion in the statement.  Alternatively, we can of course use our usual density argument, or the Jordan form.
\end{proof}

Time now to have a break, and look at what we have in the above. In answer, what we have are Theorem 6.10 and Theorem 6.11, which in addition come with quite similar proofs. So, let us unify now these two statements. For this purpose, we will need:

\begin{definition}
A rational function $f\in\mathbb C(X)$ is a quotient of polynomials:
$$f=\frac{P}{Q}$$
Assuming that $P,Q$ are prime to each other, we can regard $f$ as a usual function,
$$f:\mathbb C-X\to\mathbb C$$
with $X$ being the set of zeroes of $Q$, also called poles of $f$.
\end{definition}

Here the term ``poles'' comes from the fact that, if you want to imagine the graph of such a rational function $f$, in two complex dimensions, what you get is some sort of tent, supported by poles of infinite height, situated at the zeroes of $Q$. For more on all this, and on complex analysis in general, we refer as usual to Rudin \cite{ru2}. Although a look at an abstract algebra book, such as Lang \cite{la1}, can be interesting as well.

\bigskip

Now that we have our class of functions, the next step consists in applying them to matrices. Here we cannot expect $f(A)$ to make sense for any $f$ and any $A$, for instance because $A^{-1}$ is defined only when $A$ is invertible. We are led in this way to:

\begin{definition}
Given a matrix $A\in M_N(\mathbb C)$, and a rational function $f=P/Q$ having poles outside the eigenvalues of $A$, we can construct the following matrix,
$$f(A)=P(A)Q(A)^{-1}$$
that we can denote as a usual fraction, as follows,
$$f(A)=\frac{P(A)}{Q(A)}$$
due to the fact that $P(A),Q(A)$ commute, so that the order is irrelevant.
\end{definition}

To be more precise, $f(A)$ is indeed well-defined, and the fraction notation is justified too. In more formal terms, we can say that we have a morphism of algebras as follows, with $\mathbb C(X)^A$ being the rational functions having poles outside the eigenvalues of $A$:
$$\mathbb C(X)^A\to M_N(\mathbb C)\quad,\quad f\to f(A)$$

Summarizing, we have now a good class of functions, generalizing both the polynomials and the inverse map $x\to x^{-1}$. We can now unify Theorems 6.10 and 6.11, as follows:

\index{rational calculus}

\begin{theorem}
Given a matrix $A\in M_N(\mathbb C)$, we can apply to it any rational function $f\in\mathbb C(X)$ having its poles outside the eigenvalues of $A$,
$$A\to f(A)$$
and if $\lambda_1,\ldots,\lambda_N$ are the eigenvalues of $A$, then $f(\lambda_1),\ldots,f(\lambda_N)$ are those of $f(A)$.
\end{theorem}

\begin{proof}
We pick a scalar $\lambda\in\mathbb C$, we write $f=P/Q$, and we set:
$$F=P-\lambda Q$$

By using what we found in Theorem 6.10, for this polynomial $F\in\mathbb C[X]$, we have the following equivalence, with $\sigma(.)$ standing by definition for the set of eigenvalues:
\begin{eqnarray*}
\lambda\in\sigma(f(A))
&\iff&F(A)\notin M_N(\mathbb C)^{-1}\\
&\iff&0\in\sigma(F(A))\\
&\iff&0\in F(\sigma(A))\\
&\iff&\exists\mu\in\sigma(A),F(\mu)=0\\
&\iff&\lambda\in f(\sigma(A))
\end{eqnarray*}

Thus, the eigenvalue set is the good one, and with a bit more work, say by using characteristic polynomials, as in the proofs of Theorem 6.10 and Theorem 6.11, the multiplicities match too. Thus, we are led to the conclusion in the statement.
\end{proof}

\section*{6c. Complex functions}

Getting now to more general functions, our next objective will be that of unifying Theorem 6.14, dealing with the rational functions $f\in\mathbb C(X)$, with what we know from chapter 5 regarding the exponential, $f(x)=e^x$. Hang on, this will take some time.

\bigskip

To start with, we need to talk about holomorphic functions. Let us start with:

\index{differentiable function}
\index{complex function}
\index{holomorphic function}

\begin{definition}
We say that a function $f:X\to\mathbb C$ is differentiable in the complex sense when the following limit is defined for any $z\in X$:
$$f'(z)=\lim_{t\to0}\frac{f(z+t)-f(z)}{t}$$
In this case, we also say that $f$ is holomorphic, and we write $f\in Hol(X)$.
\end{definition}

As a basic example, we have $f(z)=z^n$. In fact, any poynomial $P\in\mathbb C[X]$ is differentiable, with its derivative being given by the same formula as in the real case, namely:
$$P(z)=\sum_{k=0}^nc_kz^k\implies P'(z)=\sum_{k=1}^nkc_kz^{k-1}$$

More generally, any rational function $f\in\mathbb C(X)$ is differentiable on its domain, that is, outsides its poles, because if we write $f=P/Q$ with $P,Q\in\mathbb C[X]$, we have:
$$f'=\left(\frac{P}{Q}\right)'=\frac{P'Q-PQ'}{Q^2}$$

Observe now that all the examples of holomorphic functions that we have are infinitely differentiable. In order to discuss this phenomenon, let us start with:

\begin{proposition}
Each power series $f(z)=\sum_nc_nz^n$ has a radius of convergence 
$$R\in[0,\infty]$$
which is such that $f$ converges for $|z|<R$, and diverges for $|z|>R$. We have:
$$R=\frac{1}{C}\quad,\quad C=\limsup_{n\to\infty}\sqrt[n]{|c_n|}$$
Also, in the case $|z|=R$ the function $f$ can either converge, or diverge.
\end{proposition}

\begin{proof}
This follows from the Cauchy criterion for series, from basic calculus, which says that a series $\sum_nx_n$ converges if $c<1$, and diverges if $c>1$, where:
$$c=\limsup_{n\to\infty}\sqrt[n]{|x_n|}$$

Indeed, with $x_n=|c_nz^n|$ we obtain that the convergence radius $R\in[0,\infty]$ exists, and is given by the formula in the statement. As for the last claim, this is standard.
\end{proof}

Back now to our questions regarding derivatives, we have:

\index{analytic function}

\begin{theorem}
Assuming that a function $f:X\to\mathbb C$ is analytic, in the sense that it is a series, around each point $z\in X$, 
$$f(z+t)=\sum_{n=0}^\infty c_nt^n$$
it follows that $f$ is infinitely differentiable, in the complex sense. In particular, $f'$ exists, and so $f$ is holomorphic in our sense.
\end{theorem}

\begin{proof}
Assuming that $f$ is analytic, as in the statement, we have:
$$f'(z+t)=\sum_{n=1}^\infty nc_nt^{n-1}$$

Moreover, the radius of convergence is the same, as shown by the following computation, using the Cauchy formula for the convergence radius, and $\sqrt[n]{n}\to1$:
$$\frac{1}{R'}
=\limsup_{n\to\infty}\sqrt[n]{|nc_n|}
=\limsup_{n\to\infty}\sqrt[n]{|c_n|}
=\frac{1}{R}$$

Thus $f'$ exists and is analytic, on the same domain, and this gives the result.
\end{proof}

Quite remarkably, the converse of Theorem 6.17 holds. In fact, we have:

\index{holomorphic function}
\index{infinitely differentiable}
\index{analytic function}
\index{Cauchy formula}

\begin{theorem}
The following conditions are equivalent, for a function $f:X\to\mathbb C$:
\begin{enumerate}
\item $f$ is holomorphic.

\item $f$ is infinitely differentiable.

\item $f$ is analytic.

\item $f$ satisfies the Cauchy formula $f(z)=\frac{1}{2\pi i}\int_\gamma\frac{f(y)}{y-z}\,dy$.
\end{enumerate}
\end{theorem}

\begin{proof}
This is something non-trivial, that we will not attempt to explain here in detail, and for the proof of this, and more, you can check for instance Rudin \cite{ru2}.
\end{proof}

Back to matrices, we can extend now what we know about $f(x)=e^x$, as follows:

\begin{theorem}
Given a matrix $A\in M_N(\mathbb C)$, we can apply to it any entire holomorphic function $f:\mathbb C\to\mathbb C$,
$$A\to f(A)$$
and if $\lambda_1,\ldots,\lambda_N$ are the eigenvalues of $A$, then $f(\lambda_1),\ldots,f(\lambda_N)$ are those of $f(A)$.
\end{theorem}

\begin{proof}
For the existence part, this is something that we already know for $f(x)=e^x$, and the proof in general is similar, by using the Taylor series of $f$, as follows:
$$f(x)=\sum_kc_kx^k\quad\implies\quad f(A)=\sum_kc_kA^k$$

 As for the second assertion, again this is something that we know for $f(x)=e^x$, coming via density of the diagonalizable matrices, or via the Jordan form, and the proof in general is similar, as explained earlier in this chapter, for the polynomials.
\end{proof}

Very nice all this, but as a drawback, we are now again in the dark, with Theorem 6.14 with Theorem 6.19 waiting to be unified. So, still lots of work to be done.

\bigskip

In general, the holomorphic functions are not entire, and the above method won't cover the rational functions $f\in\mathbb C(X)^A$ that we want to generalize. Thus, we must use something else. And the answer here comes from the Cauchy formula:
$$f(a)=\frac{1}{2\pi i}\int_\gamma\frac{f(z)}{z-a}\,dz$$

Indeed, given a rational function $f\in\mathbb C(X)^A$, the matrix $f(A)\in M_N(\mathbb C)$, as constructed in Definition 6.13, can be recaptured in an analytic way, as follows:
$$f(A)=\frac{1}{2\pi i}\int_\gamma\frac{f(z)}{z-A}\,dz$$

Now given an arbitrary function $f\in Hol(\sigma(A))$, we can define $f(A)\in M_N(\mathbb C)$ by the exactly same formula, and we obtain in this way the desired correspondence:
$$Hol(\sigma(A))\to M_N(\mathbb C)\quad,\quad 
f\to f(A)$$

This was for the plan. In practice now, all this needs a bit of care, with many verifications needed, and with the technical remark that a winding number must be added to the above Cauchy formulae, for things to be correct. Let us start with:

\begin{definition}
If $\gamma$ is a loop in $\mathbb C$ the number of times $\gamma$ goes around a point $z\in\mathbb C-\gamma$ is computed by the following integral, called winding number:
$$Ind(\gamma,z)=\frac{1}{2\pi i}\int_\gamma\frac{d\xi}{\xi -z}$$
We say that $\gamma$ turns around $z$ if $Ind(\gamma,z)=1$, and that it does not turn if $Ind(\gamma,z)=0$. Otherwise, we say that $\gamma$ turns around $z$ many times, or in the bad sense, or both.
\end{definition}

Let $f:U\to\mathbb C$ be an holomorphic function defined on an open subset of $\mathbb C$, and $\gamma$ be a loop in $U$. If $Ind(\gamma,z)\neq 0$ for $z\in\mathbb C-U$ then $f(z)$ is given by the Cauchy formula:
$$Ind(\gamma,z)f(z)=\frac{1}{2\pi i}\int_\gamma\frac{f(\xi )}{\xi -z}\,d\xi$$

Also, if $Ind(\gamma,z)=0$ for $z\in\mathbb C-U$ then the integral of $f$ on $\gamma$ is zero:
$$\int_\gamma f(\xi)\,d\xi=0$$

It is convenient to use formal combinations of loops, called cycles:
$$\Sigma=n_1\gamma_1+\ldots +n_r\gamma_r$$

The winding number for $\Sigma$ is by definition the corresponding linear combination of winding numbers of its loop components, and the Cauchy formula holds for arbitrary cycles. Now by getting back to our questions regarding matrices, we can formulate:

\begin{definition}
Let $A\in M_N(\mathbb C)$, and let $f:U\to\mathbb C$ be an holomorphic function defined on an open set containing $\sigma(A)$. We can define a matrix $f(A)$ by the formula
$$f(A)=\frac{1}{2\pi i}\int_\Sigma\frac{f(\xi)}{\xi-A}\,d\xi$$
where $\Sigma$ is a cycle in $U-\sigma(A)$ which turns around $\sigma(A)$ and doesn't turn around $\mathbb C-U$.
\end{definition}

The formula makes sense because $\Sigma$ is in $U-\sigma(A)$. Also, $f(A)$ is independent of the choice of $\Sigma$. Indeed, let $\Sigma_1$ and $\Sigma_2$ be two cycles. Their difference $\Sigma_1-\Sigma_2$ is a cycle which doesn't turn around $\sigma(a)$, neither around $\mathbb C-U$. The function $z\to f(z)/(z-A)$ being holomorphic $U-\sigma(A)\to M_N(\mathbb C)$, its integral on $\Sigma_1-\Sigma_2$ must be zero:
$$\int_{\Sigma_1-\Sigma_2}\frac{f(\xi)}{\xi-A}\,d\xi =0$$

Thus $f(A)$ is the same with respect to $\Sigma_1$ and to $\Sigma_2$, and so Definition 6.21 is fully justified. Now with this definition in hand, we first have the following result:

\begin{proposition}
We have the formula
$$f(A)g(A)=(fg)(A)$$
whenever the equality makes sense.
\end{proposition}

\begin{proof}
Let $\Sigma_1$ be a cycle in $U-\sigma(A)$ around $\sigma(A)$ and consider the following set:
$$Int(\Sigma_1)=\left\{z\in\mathbb C-\Sigma_1\Big|Ind(\Sigma_1,z)\neq 0\right\}\cup\Sigma_1$$

This is a compact set, included in $U$ and containing the spectrum of $A$:
$$\sigma(T)\subset Int(\Sigma_1)\subset U$$

Let $\Sigma_2$ be a cycle in $U-Int(\Sigma_1)$ turning around $Int(\Sigma_1)$. Consider two holomorphic functions $f,g$ defined around $\sigma(A)$, so that the statement make sense. We have:
\begin{eqnarray*}
f(A)g(A)
&=&\left(\frac{1}{2\pi i}\right)^2\left(\int_{\Sigma_1}\frac{f(\xi)}{\xi -A}\,d\xi\right) 
\left(\int_{\Sigma_2}\frac{g(\eta)}{\eta-A}\,d\eta\right)\\
&=&\left(\frac{1}{2\pi i}\right)^2\int_{\Sigma_1}\int_{\Sigma_2}\frac{f(\xi)g(\eta )}{(\xi-A)(\eta-A)}\,d\eta d\xi
\end{eqnarray*}

In order to integrate, we can use the following identity:
$$\frac{1}{(\xi-A)(\eta-A)}=\frac{1}{(\eta-\xi)(\xi-A)}+\frac{1}{(\xi-\eta)(\eta-A)}$$

Thus our integral, and so our formula for $f(A)g(A)$, splits into two terms. The first term can be computed by integrating first over $\Sigma_2$, and we obtain:
$$\frac{1}{2\pi i}\int_{\Sigma_1}\frac{f(\xi)g(\xi)}{\xi-A}\,d\xi =(fg)(A)$$

As for the second term, here we can integrate first over $\Sigma_1$, and we get:
$$\frac{1}{2\pi i}\int_{\Sigma_2}\frac{g(\eta)}{\eta-A}
\left( \frac{1}{2\pi i}\int_{\Sigma_1}\frac{f(\xi )}{\xi -\eta}d\xi\right)d\eta =0$$

It follows that $f(A)g(A)$ is equal to $(fg)(A)$, as claimed.
\end{proof}

We can now formulate our theorem regarding holomorphic calculus, as follows:

\begin{theorem}
Given $A\in M_N(\mathbb C)$, we have a morphism of algebras as follows, where $Hol(\sigma(A))$ is the algebra of functions which are holomorphic around $\sigma(A)$,
$$Hol(\sigma(A))\to M_N(\mathbb C)\quad,\quad f\to f(A)$$
which extends the previous rational functional calculus $f\to f(A)$. We have:
$$\sigma(f(A))=f(\sigma(A))$$
Moreover, if $\sigma(A)$ is contained in an open set $U$ and $f_n,f:U\to\mathbb C$ are holomorphic functions such that $f_n\to f$ uniformly on compact subsets of $U$ then $f_n(A)\to f(A)$.
\end{theorem}

\begin{proof}
There are several things to be proved here, as follows:

\medskip

(1) Consider indeed the algebra $Hol(\sigma(A))$, with the convention that two functions are identified if they coincide on an open set containing $\sigma(A)$. We have then a construction $f\to f(A)$ as in the statement, provided by Definition 6.21 and Proposition 6.22.

\medskip

(2) Let us prove now that our construction extends the one for rational functions. Since $1,z$ generate $\mathbb C(X)$, it is enough to show that $f(z)=1$ implies $f(A)=1$, and that $f(z)=z$ implies $f(A)=A$. For this purpose, we prove that $f(z)=z^n$ implies $f(A)=A^n$ for any $n$. But this follows by integrating over a circle $\gamma$ of big radius, as follows:
\begin{eqnarray*}
f(A)
&=&\frac{1}{2\pi i}\int_\gamma\frac{\xi^n}{\xi-A}\,d\xi\\
&=&\frac{1}{2\pi i}\int_\gamma \xi^{n-1}\left(1-\frac{A}{\xi}\right)^{-1}d\xi\\
&=&\frac{1}{2\pi i}\int_\gamma \xi^{n-1}\left(\sum_{k=0}^\infty \xi^{-k}A^k\right) d\xi\\
&=&\sum_{k=0}^\infty\left(\frac{1}{2\pi i}\int_\gamma\xi^{n-k-1}d\xi\right)A^k\\
&=&A^n
\end{eqnarray*}

(3) Regarding $\sigma(f(A))=f(\sigma(A))$, it is enough to prove that this equality holds on the point $0$, and we can do this by double inclusion, as follows:

\medskip

``$\supset$''. Assume that $f(\sigma(A))$ contains $0$, and let $z_0\in\sigma(A)$ be such that $f(z_0)=0$. Consider the function $g(z)=f(z)/(z-z_0)$.  We have $g(A)(A-z_0)=f(A)$ by using the morphism property. Since $A-z_0$ is not invertible, $f(A)$ is not invertible either.

\medskip

``$\subset$''. Assume now that $f(\sigma(A))$ does not contain $0$. With the holomorphic function $g(z)=1/f(z)$ we get $g(A)=f(A)^{-1}$, so $f(A)$ is invertible, and we are done.

\medskip

(4) Finally, regarding the last assertion, this is clear from definitions. And with the remark that this can be applied to holomorphic functions written as series:
$$f(z)=\sum_{k=0}^\infty c_k(z-z_0)^k$$

Indeed, if this is the expansion of $f$ around $z_0$, with convergence radius $r$, and if $\sigma(A)$ is contained in the disc centered at $z_0$ of radius $r$, then $f(A)$ is given by:
$$f(A)=\sum_{k=0}^\infty c_k(A-z_0)^k$$

Summarizing, we have proved the result, and fully extended Theorem 6.14.
\end{proof}

\section*{6d. Normal matrices}

Good work that we did, with Theorem 6.23 looking like a ultimate result. However, this is not exactly the case, with many more functions, such as the continuous ones, or even the measurable ones, still waiting to be investigated. In order to discuss this, let us start with the following result, based on the spectral theorems from chapter 3:

\begin{theorem}
Given a normal matrix $A\in M_N(\mathbb C)$, we have a morphism of algebras as follows, extending the previous holomorphic functional calculus,
$$C(\sigma(A))\to M_N(\mathbb C)\quad,\quad f\to f(A)$$
and if $\lambda_1,\ldots,\lambda_N$ are the eigenvalues of $A$, then $f(\lambda_1),\ldots,f(\lambda_N)$ are those of $f(A)$.
\end{theorem}

\begin{proof}
The idea here is to ``complete'' our previous functional calculus results. In practice, the simplest is to start with the polynomial calculus morphism, namely:
$$\mathbb C[X]\to M_N(\mathbb C)\quad,\quad 
P\to P(A)$$

We know from the above that this morphism is continuous, and is in fact isometric, when regarding the polynomials $P\in\mathbb C[X]$ as functions on $\sigma(A)$:
$$||P(A)||=||P_{|\sigma(A)}||$$

We conclude from this that we have a unique isometric extension, as follows:
$$C(\sigma(A))\to M_N(\mathbb C)\quad,\quad  
f\to f(A)$$

It remains to prove $\sigma(f(A))=f(\sigma(A))$, and we can do this by double inclusion:

\medskip

``$\subset$'' Given a continuous function $f\in C(\sigma(A))$, we must prove that we have:
$$\lambda\notin f(\sigma(A))\implies\lambda\notin\sigma(f(A))$$

For this purpose, consider the following function, which is well-defined:
$$\frac{1}{f-\lambda}\in C(\sigma(A))$$

We can therefore apply this function to $A$, and we obtain:
$$\left(\frac{1}{f-\lambda}\right)A=\frac{1}{f(A)-\lambda}$$

In particular $f(A)-\lambda$ is invertible, so  $\lambda\notin\sigma(f(A))$, as desired.

\medskip

``$\supset$'' Given a continuous function $f\in C(\sigma(A))$, we must prove that we have: 
$$\lambda\in f(\sigma(A))\implies\lambda\in\sigma(f(A))$$

But this is the same as proving that we have:
$$\mu\in\sigma(A)\implies f(\mu)\in\sigma(f(A))$$

For this purpose, we approximate our function by polynomials, $P_n\to f$, and we examine the following convergence, which follows from $P_n\to f$:
$$P_n(A)-P_n(\mu)\to f(A)-f(\mu)$$

We know from polynomial functional calculus that we have:
$$P_n(\mu)
\in P_n(\sigma(A))
=\sigma(P_n(T))$$

Thus, the matrices $P_n(A)-P_n(\mu)$ are not invertible. On the other hand, we know that the set formed by the invertible matrices is open, so its complement is closed. Thus the limit $f(A)-f(\mu)$ is not invertible either, and so $f(\mu)\in\sigma(f(A))$, as desired.
\end{proof}

At a more advanced level, we have as well the following result:

\begin{theorem}
Given a normal matrix $A\in M_N(\mathbb C)$, we have a morphism of algebras as follows, with $L^\infty$ standing for the abstract measurable functions
$$L^\infty(\sigma(A))\to M_N(\mathbb C)\quad,\quad 
f\to f(A)$$
and if $\lambda_1,\ldots,\lambda_N$ are the eigenvalues of $A$, then $f(\lambda_1),\ldots,f(\lambda_N)$ are those of $f(A)$.
\end{theorem}

\begin{proof}
As before, the idea will be that of ``completing'' what we have. To be more precise, we can use the Riesz theorem and a polarization trick, as follows:

\medskip

(1) Given a vector $x\in\mathbb C^N$, consider the following functional:
$$C(\sigma(A))\to\mathbb C\quad,\quad 
g\to<g(A)x,x>$$

By the Riesz theorem, this functional must be the integration with respect to a certain measure $\mu$ on the space $\sigma(A)$. Thus, we have a formula as follows:
$$<g(A)x,x>=\int_{\sigma(A)}g(z)d\mu(z)$$

Now given an arbitrary Borel function $f\in L^\infty(\sigma(A))$, as in the statement, we can define a number $<f(A)x,x>\in\mathbb C$, by using exactly the same formula, namely:
$$<f(A)x,x>=\int_{\sigma(A)}f(z)d\mu(z)$$

Thus, we have managed to define numbers $<f(A)x,x>\in\mathbb C$, for all vectors $x\in\mathbb C^N$, and in addition we can recover these numbers as follows, with $g_n\in C(\sigma(A))$:
$$<f(A)x,x>=\lim_{g_n\to f}<g_n(A)x,x>$$ 

(2) In order to define now numbers $<f(A)x,y>\in\mathbb C$, for all vectors $x,y\in\mathbb C^N$, we can use a polarization trick. Indeed, for any matrix $B\in M_N(\mathbb C)$ we have:
\begin{eqnarray*}
<B(x+y),x+y>
&=&<Bx,x>+<By,y>\\
&&+<Bx,y>+<By,x>
\end{eqnarray*}

By replacing $y\to iy$, we have as well the following formula:
\begin{eqnarray*}
<B(x+iy),x+iy>
&=&<Bx,x>+<By,y>\\
&&-i<Bx,y>+i<By,x>
\end{eqnarray*}

By multiplying this latter formula by $i$, we obtain the following formula:
\begin{eqnarray*}
i<B(x+iy),x+iy>
&=&i<Bx,x>+i<By,y>\\
&&+<Bx,y>-<By,x>
\end{eqnarray*}

Now by summing this latter formula with the first one, we obtain:
\begin{eqnarray*}
<B(x+y),x+y>+i<B(x+iy),x+iy>
&=&(1+i)[<Bx,x>+<By,y>]\\
&&+2<Bx,y>
\end{eqnarray*}

(3) But with this, we can now finish. Indeed, by combining (1,2), given a Borel function $f\in L^\infty(\sigma(A))$, we can define numbers $<f(A)x,y>\in\mathbb C$ for any $x,y\in\mathbb C^N$, and it is routine to check, by using approximation by continuous functions $g_n\to f$ as in (1), that we obtain in this way a matrix $f(A)\in M_N(\mathbb C)$, having all the desired properties.
\end{proof}

Good work that we did, and time now for some applications, in relation with Question 6.1. We know that any number $a\in\mathbb C$ can be written as follows, with $b,c\in\mathbb R$:
$$a=b+ic$$

Also, we know that both the real and imaginary parts $b,c\in\mathbb R$, and more generally any real number $d\in\mathbb R$, can be written as follows, with $e,f\geq0$: 
$$d=e-f$$

Here is the matrix-theoretical generalization of these results:

\begin{theorem}
Given a matrix $A\in M_N(\mathbb C)$, the following happen:
\begin{enumerate}
\item We can write $A=B+iC$, with $B,C\in M_N(\mathbb C)$ being self-adjoint.

\item When $A=A^*$, we can write $A=E-F$, with $E,F\in M_N(\mathbb C)$ being positive.

\item Thus, we can write any $A$ as a linear combination of $4$ positive matrices.
\end{enumerate}
\end{theorem}

\begin{proof}
All this follows from basic spectral theory, as follows:

\medskip

(1) This is something which comes from the following decomposition formula:
$$A=\frac{A+A^*}{2}+i\cdot\frac{A-A^*}{2i}$$

(2) This follows from the spectral theorem for self-adjoint matrices, by applying some suitable functions. Indeed, we can use the following decomposition formula on $\mathbb R$:
$$1=\chi_{[0,\infty)}+\chi_{(-\infty,0)}$$

To be more precise, let us multiply by $z$, and rewrite this formula as follows:
$$z=\chi_{[0,\infty)}z-\chi_{(-\infty,0)}(-z)$$

Now by applying these measurable functions to $A$, we obtain as formula as follows, with both the matrices $A_+,A_-\in M_N(\mathbb C)$ being positive, as desired:
$$A=A_+-A_-$$

(3) This follows indeed by combining the results in (1) and (2) above.
\end{proof}

Going ahead with our decomposition results, another basic thing that we know about complex numbers is that any $a\in\mathbb C$ appears as a real multiple of a unitary:
$$a=re^{it}$$

In the case of the arbitrary matrices, finding the correct analogue of this key decomposition result is something quite tricky. As a basic result here, we have:

\begin{theorem}
Given a matrix $A\in M_N(\mathbb C)$, the following happen:
\begin{enumerate}
\item When $A=A^*$ and $||A||\leq1$, we can write $A$ as an average of $2$ unitaries:
$$A=\frac{U+V}{2}$$

\item In the general $A=A^*$ case, we can write $A$ as a rescaled sum of unitaries:
$$A=\lambda(U+V)$$

\item Thus, in general, we can write $A$ as a rescaled sum of $4$ unitaries.
\end{enumerate}
\end{theorem}

\begin{proof}
This follows from the results that we have, as follows:

\medskip

(1) Assuming $A=A^*$ and $||A||\leq1$, it follows that we have:
$$1-A^2\geq0$$

Our claim is that the decomposition result that we are looking can be taken as follows, with both the components on the right being unitaries:
$$A=\frac{A+i\sqrt{1-A^2}}{2}+\frac{A-i\sqrt{1-A^2}}{2}$$

To be more precise, the square root can be extracted in the usual way, and the check of the unitarity of the components goes as follows:
\begin{eqnarray*}
(A+i\sqrt{1-A^2})(A-i\sqrt{1-A^2})
&=&A^2-i^2\left(\sqrt{1-A^2}\right)^2\\
&=&A^2+(1-A^2)\\
&=&1
\end{eqnarray*}

(2) This simply follows by applying (1) to the following matrix:
$$A'=\frac{A}{||A||}$$

(3) Assuming first $||A||\leq1$, we know from Theorem 6.26 (1) that we can write $A$ as follows, with $B,C$ being self-adjoint, and satisfying $||B||,||C||\leq1$:
$$A=B+iC$$

Now by applying (1) to both $B$ and $C$, we obtain a decomposition of $A$ as follows:
$$A=\frac{U+V+W+X}{2}$$

In general, we can apply this to the matrix $A/||A||$, and we obtain the result.
\end{proof}

As before with other such results, we will be back to this in chapter 8 below, with some further details, directly in the infinite matrix setting, to be investigated there.

\section*{6e. Exercises}

This was a quite advanced chapter, and as exercises on all this, we have:

\begin{exercise}
Learn more about the polar decomposition, and its applications.
\end{exercise}

\begin{exercise}
Learn more about singular values, and their applications.
\end{exercise}

\begin{exercise}
Clarify all details, for the polynomial functional calculus.
\end{exercise}

\begin{exercise}
Clarify all details, for the rational functional calculus.
\end{exercise}

\begin{exercise}
Clarify all details, for the holomorphic functional calculus.
\end{exercise}

\begin{exercise}
Clarify all details, for the continuous functional calculus.
\end{exercise}

\begin{exercise}
Clarify all details, for the measurable functional calculus.
\end{exercise}

\begin{exercise}
Learn about some other decomposition results for matrices.
\end{exercise}

As bonus exercise, learn some systematic complex analysis, as much as you can.

\chapter{Dynamical systems}

\section*{7a. Differential equations}

Time for some applications to analysis, along the lines discussed in chapter 5. Let us recall from there that a standard system of ordinary differential equations (ODE) is a system as follows, with the unknown being a vector function $x=x(t):\mathbb R\to\mathbb R^N$:
$$x_1^{(k)}=f_1(t,x,\dot{x},\ldots,x^{(k-1)})$$
$$\vdots$$
$$x_N^{(k)}=f_N(t,x,\dot{x},\ldots,x^{(k-1)})$$

The point now is that, up to suitably enlarging the size of the unknown vector, we can convert this standard system of ODE into a standard order $1$ system of ODE. Indeed, with $y=(x,\dot{x},\ldots,x^{(k-1)})$, the system takes the following form, as desired:
$$\dot{y}_1=y_2\quad,\quad 
\dot{y}_2=y_3\quad,\quad 
\ldots\quad,\quad 
\dot{y}_{k-1}=y_k\quad,\quad 
\dot{y}_k=f(t,y)$$

Moreover, in the autonomous case, that where the function $f$ does not depend on time $t$, we can further set $z=(t,y)$, and we are led in this way to a system as follows:
$$\dot{z}_1=1\quad,\quad 
\dot{z}_2=z_3\quad,\quad
\ldots\quad,\quad
\dot{z}_k=z_{k+1}\quad,\quad
\dot{z}_{k+1}=f(z)$$

Our first goal in this chapter will be that of finding existence and uniqueness results for the solutions of such systems of ODE. But let us begin with some examples, in 1D. More specifically, we will be interested in the following type of equations:

\begin{definition}
An autonomous order 1 ODE is an equation of type
$$\dot{x}=f(x)\quad,\quad x(0)=x_0$$
with $f\in C(\mathbb R)$ being a certain function.
\end{definition}

In order to solve now our equation, assume that we are in the case $f(x_0)\neq 0$. Then, around $t=0$, we can divide our equation by $f(x(s))$, and then integrate:
$$\int_0^t\frac{\dot{x}(s)}{f(x(s))}\,ds=t$$

In view of this observation, consider the following function:
$$F(x)=\int_{x_0}^x\frac{1}{f(y)}\,dy$$

We have then the following computation, taking into account our equation:
$$F(x(t))
=\int_{x_0}^{x(t)}\frac{1}{f(y)}\,dy
=\int_0^t\frac{\dot{x}(s)}{f(x(s))}\,ds
=t$$

Obviously, the converse holds too, so our original equation is equivalent to:
$$F(x(t))=t$$

Now recall that we assumed $f(x_0)\neq 0$. But this means that $F(x)$ is monotone around $x_0$, and so invertible, so we have a unique solution to our equation, given by:
$$\varphi(t)=F^{-1}(t)$$

Note also that we have, as we should, as required by Definition 7.1:
$$\varphi(0)=F^{-1}(0)=x_0$$

With this discussion made, which was something local, let us turn now to global problems. We have here the following question, that we would like to solve:

\begin{question}
In the context of the above autonomous order 1 ODE, and discussion, what is the interval where the solution is defined? And, when is this interval $\mathbb R$ itself?
\end{question}

In order to discuss this latter question, in view of $f(x_0)\neq0$, assume that we are in the case $f(x_0)>0$, with the other case, $f(x_0)<0$, being similar. We have then $f>0$ on a certain interval $(x_1,x_2)$ around $x_0$. Now consider the following two limits:
$$T_-=\lim_{x\searrow x_1}F(x)\in[-\infty,0)\quad,\quad T_+=\lim_{x\nearrow x_2}F(x)\in(0,\infty]$$

Since $\lim_{t\searrow T_-}\varphi(t)=x_1$, the solution $\varphi$ exists for any $t<0$ precisely when:
$$T_-=\int_{x_1}^{x_0}\frac{1}{f(y)}\,dy=-\infty$$

Similarly, since $\lim_{t\nearrow T_+}\varphi(t)=x_2$, the solution $\varphi$ exists for any $t>0$ when:
$$T_+=\int_{x_0}^{x_2}\frac{1}{f(y)}\,dy=\infty$$

Summarizing, we are led to the following answer to Question 7.2:

\begin{answer}
In the context of the above autonomous order 1 ODE, and discussion involving the interval $(x_1,x_2)$ around $x_0$, the solution $\varphi$ is as follows:
\begin{enumerate}
\item $\varphi$ exists for any $t<0$ when $1/f$ is not integrable around $x_1$.

\item $\varphi$ exists for any $t>0$ when $1/f$ is not integrable around $x_2$.
\end{enumerate}
\end{answer}

All this was quite theoretical, so let us work out now some examples. For $f(x)=x$, and with $x_0>0$, we have $(x_1,x_2)=(0,\infty)$, and the function $F$ is given by:
$$F(x)=\log\left(\frac{x}{x_0}\right)$$

Also, we have $T_\pm=\pm\infty$, and the solution is as follows, defined on the whole $\mathbb R$:
$$\varphi(t)=x_0e^t$$

As a second example now, let us take $f(x)=x^2$, and $x_0>0$. In this case we have $(x_1,x_2)=(0,\infty)$, and the function $F$ is given by:
$$F(x)=\frac{1}{x_0}-\frac{1}{x}$$

Also, in this case we have $T_-=-\infty$ and $T_+=1/x_0$, and the solution of our equation is as follows, defined on the interval $(-\infty,1/x_0)$:
$$\varphi(t)=\frac{x_0}{1-x_0t}$$

We will see some other examples for all this, in what follows.

\bigskip

As a continuation of the above discussion, dealing with the case $f(x_0)\neq0$, it remains now to discuss the case $f(x_0)=0$. Here we have the trivial solution $\varphi(t)=x_0$, and we can have as well non-trivial solutions. Assume for instance that we have:
$$\left|\int_{x_0}^{x_0+\varepsilon}\frac{1}{f(y)}\,dy\right|<\infty$$

Then, we have the following non-trivial solution to our equation:
$$\varphi(t)=F^{-1}(t)\quad,\quad F(x)=\int_{x_0}^x\frac{1}{f(y)}\,dy$$

Again, in order to understand this, nothing better than an explicit example. Let us take $f(x)=\sqrt{|x|}$. In the case $x_0>0$, studied before, we have $(x_1,x_2)=(0,\infty)$, then $F(x)=2(\sqrt{x}-\sqrt{x_0})$, and the solution is as follows, with $t\in(-2\sqrt{x_0},\infty)$:
$$\varphi(t)=\left(\sqrt{x_0}+\frac{t}{2}\right)^2$$

In the case $x_0=0$, however, we have several solutions, that can be obtained by gluing the trivial solution, and the generic solution. Indeed, we can take:
$$\varphi(t)=\begin{cases}
-\frac{(t-t_0)^2}{4}&{\rm for}\ t\leq t_0\\
0&{\rm for}\ t_0\leq t\leq t_1\\
\frac{(t-t_1)^2}{4}&{\rm for}\ t_1\leq t
\end{cases}$$

Based on the above study, and on our various examples, let us formulate:

\begin{conclusion}
In the context of the above autonomous order 1 ODE:
\begin{enumerate}
\item Even when the function $f$ is $C^\infty$, we can only have local solutions.

\item Also, in general, we do not have the uniqueness of the solution. 
\end{enumerate}
\end{conclusion}

Before getting into a heavier theoretical study of the existence and uniqueness of solutions, let us discuss as well a few tricks for the ODE, sometimes leading to explicit solutions. A useful method is that of using a change of variables, as follows:
$$(t,x)\to (s,y)$$

To be more precise, we are looking for suitable functions $\sigma,\eta$, as follows:
$$s=\sigma(t,x)\quad,\quad y=\eta(t,x)$$

In order to have a change of variables, our transformation must be of course invertible. However, this assumption is not enough, at the level of solutions, because by rotating the graph of a function, we do not necessarily obtain the graph of a function.

\bigskip

In view of this, a reasonable assumption is that our transformations must preserve the fibers, with ``fiber'' meaning here corresponding to constant time. That is, we are looking for changes of variables, suitably adapted to our ODE, of the following special type:
$$s=\sigma(t)\quad,\quad y=\eta(t,x)$$ 

Now assume that we have such a transformation, which is invertible, as any change of variables should be, with inverse given by formulae as follows:
$$t=\tau(s)\quad,\quad x=\xi(s,y)$$

Then $\varphi(t)$ is a solution of $\dot{x}=f(t,x)$ precisely when $\psi(s)=\eta(\tau(s),\varphi(\tau(s)))$ is a solution of the following equation, where $\tau=\tau(s)$ and $\xi=\xi(s,y)$:
$$\dot{y}=\dot{\tau}\left(\frac{d\eta}{dt}(\tau,\xi)+\frac{d\eta}{dx}(\tau,\xi)f(t,\xi)\right)$$

Which is quite nice, becase we can get some concrete results in this way, that is, explicit solutions for explicit ODE, by doing some reverse engineering, based on this.

\bigskip

Finally, for ending this preliminary section on general ODE theory, let us discuss some well-known equations. First we have the Bernoulli equations, which are as follows:
$$\dot{x}=f(t)x+g(t)x^n$$

Assuming $n\neq1$, we can set $y=x^{1-n}$, and our equation takes the following form:
$$\dot{y}=(1-n)f(t)y+(1-n)g(t)$$

But this is a linear equation, that we can solve by using the linear algebra methods from chapter 5. We will leave the computations here as an exercise.

\bigskip

As a second class of well-known equations, again coming from a variety of questions from physics, we have the Riccati equations, which are as follows:
$$\dot{x}=f(t)x+g(t)x^2+h(t)$$

Now assuming that we have found a particular solution $x_p(t)$, we can set:
$$y=\frac{1}{x-x_p(t)}$$

With this change of variables, our equation takes the following form:
$$\dot{y}=-(f(t)+2x_p(t)g(t))y-g(t)$$

But this is again a linear equation, that we can solve by using the linear algebra methods from chapter 5. Again, we will leave the computations here as an exercise.

\section*{7b. Functional analysis}

With the above discussed, which remains something a bit ad-hoc, let us try now to develop some general theory. We would like to solve the following problem:

\begin{problem}
Do we have the local existence and uniqueness of the solutions of
$$\dot{x}=f(t,x)\quad,\quad x(t_0)=x_0$$
under suitable assumptions on the function $f\in C(U,\mathbb R^N)$, with $U\subset\mathbb R^{N+1}$ open?
\end{problem}

In order to solve this latter question, we have a strategy which is quite straightforward. Indeed, we can integrate our equation, which takes the following form:
$$x(t)=x_0+\int_{t_0}^tf(s,x(s))ds$$

Based on this observation, consider the following function:
$$K(x)(t)=x_0+\int_{t_0}^tf(s,x(s))ds$$

In terms of this function, our original equation reads:
$$K(x)=x$$

So, all in all, we are into a fixed point problem. But, as you certainly know from basic calculus, such questions can be solved simply by iterating. Thus, we are led to:

\begin{questions}
In relation with the above strategy, for solving Problem 7.5:
\begin{enumerate}
\item Can we develop a theory of infinite dimensional complete normed spaces?

\item Do we have fixed point theorems, inside such complete normed spaces?

\item Can we apply these fixed point theorems, as to solve our ODE problem?
\end{enumerate}
\end{questions}

We will see that the answers to these latter questions are yes, yes, yes. However, this is something quite technical, which will take some time. Let us start with:

\begin{definition}
A normed space is a complex vector space $V$ with a map
$$||.||:V\to\mathbb R_+$$
called norm, subject to the following conditions:
\begin{enumerate}
\item $||x||=0$ implies $x=0$.

\item $||\lambda x||=|\lambda|\cdot||x||$, for any $x\in V$, and $\lambda\in\mathbb C$.

\item $||x+y||\leq||x||+||y||$, for any $x,y\in V$.
\end{enumerate}
When $V$ is complete with respect to $d(x,y)=||x-y||$, we say that $V$ is a Banach space.
\end{definition}

In relation with this, observe that the function $d(x,y)=||x-y||$ is indeed a distance, with the key distance axiom, which is the triangle inequality $d(x,y)\leq d(x,z)+d(y,z)$, coming from our third norm axiom above, namely $||x+y||\leq||x||+||y||$.

\bigskip

As a basic example now, which is finite dimensional, we have the space $V=\mathbb C^N$, with the norm on it being the usual length of the vectors, namely:
$$||x||=\sqrt{\sum_i|x_i|^2}$$

Indeed, for this space (1) is clear, (2) is clear too, and (3), which is equivalent to the triangle inequality in $\mathbb C^N$, can be deduced from the Cauchy-Schwarz inequality, as explained in chapter 3. More on this, with some generalizations, in a moment.

\bigskip

In order to construct further examples, let us start with a basic result, as follows:

\begin{theorem}[Jensen]
Given a convex function $f:\mathbb R\to\mathbb R$, we have the following inequality, for any $x_1,\ldots,x_N\in\mathbb R$, and any $\lambda_1,\ldots,\lambda_N>0$ summing up to $1$,
$$f(\lambda_1x_1+\ldots+\lambda_Nx_N)\leq\lambda_1f(x_1)+\ldots+\lambda_Nx_N$$
with equality when $x_1=\ldots=x_N$. In particular, by taking the weights $\lambda_i$ to be all equal, we obtain the following inequality, valid for any $x_1,\ldots,x_N\in\mathbb R$,
$$f\left(\frac{x_1+\ldots+x_N}{N}\right)\leq\frac{f(x_1)+\ldots+f(x_N)}{N}$$
and once again with equality when $x_1=\ldots=x_N$. We have a similar statement holds for the concave functions, with all the inequalities being reversed.
\end{theorem}

\begin{proof}
This is indeed something very standard, the idea being as follows:

\medskip

(1) First, we can talk about convex functions in a usual, intuitive way, with this meaning by definition that the following inequality must be satisfied:
$$f\left(\frac{x+y}{2}\right)\leq\frac{f(x)+f(y)}{2}$$

(2) But this means, via a simple argument, by approximating numbers $t\in[0,1]$ by sums of powers $2^{-k}$, that for any $t\in[0,1]$ we must have:
$$f(tx+(1-t)y)\leq tf(x)+(1-t)f(y)$$

Alternatively, via yet another simple argument, this time by doing some geometry with triangles, this means that we must have:
$$f\left(\frac{x_1+\ldots+x_N}{N}\right)\leq\frac{f(x_1)+\ldots+f(x_N)}{N}$$

But then, again alternatively, by combining the above two simple arguments, the following must happen, for any $\lambda_1,\ldots,\lambda_N>0$ summing up to $1$:
$$f(\lambda_1x_1+\ldots+\lambda_Nx_N)\leq\lambda_1f(x_1)+\ldots+\lambda_Nx_N$$

(3) Summarizing, all our Jensen inequalities, at $N=2$ and at $N\in\mathbb N$ arbitrary, are equivalent. The point now is that, if we look at what the first Jensen inequality, that we took as definition for the convexity, means, this is simply equivalent to:
$$f''(x)\geq0$$

(4) Thus, we are led to the conclusions in the statement, regarding the convex functions. As for the concave functions, the proof here is similar. Alternatively, we can say that $f$ is concave precisely when $-f$ is convex, and get the results from what we have.
\end{proof}

As a basic application of the Jensen inequality, we have:

\begin{proposition}
For $p\in(1,\infty)$ we have the following inequality,
$$\left|\frac{x_1+\ldots+x_N}{N}\right|^p\leq\frac{|x_1|^p+\ldots+|x_N|^p}{N}$$
and for $p\in(0,1)$ we have the following reverse inequality,
$$\left|\frac{x_1+\ldots+x_N}{N}\right|^p\geq\frac{|x_1|^p+\ldots+|x_N|^p}{N}$$
with in both cases equality precisely when $|x_1|=\ldots=|x_N|$.
\end{proposition}

\begin{proof}
This follows indeed from Theorem 7.8, because we have:
$$(x^p)''=p(p-1)x^{p-2}$$

Thus $x^p$ is convex for $p>1$ and concave for $p<1$, which gives the results.
\end{proof}

As another basic application of the Jensen inequality, we have:

\begin{theorem}[Young]
We have the following inequality,
$$ab\leq \frac{a^p}{p}+\frac{b^q}{q}$$
valid for any $a,b\geq0$, and any exponents $p,q>1$ satisfying $\frac{1}{p}+\frac{1}{q}=1$. 
\end{theorem}

\begin{proof}
We use the logarithm function, which is concave on $(0,\infty)$, due to:
$$(\log x)''=\left(-\frac{1}{x}\right)'=-\frac{1}{x^2}$$

Thus we can apply the Jensen inequality, and we obtain in this way:
\begin{eqnarray*}
\log\left(\frac{a^p}{p}+\frac{b^q}{q}\right)
&\geq&\frac{\log(a^p)}{p}+\frac{\log(b^q)}{q}\\
&=&\log(a)+\log(b)\\
&=&\log(ab)
\end{eqnarray*}

Now by exponentiating, we obtain the Young inequality.
\end{proof}

Moving forward now, as a consequence of the Young inequality, we have:

\begin{theorem}[H\"older]
Assuming that $p,q\geq 1$ are conjugate, in the sense that 
$$\frac{1}{p}+\frac{1}{q}=1$$
we have the following inequality, valid for any two vectors $x,y\in\mathbb C^N$,
$$\sum_i|x_iy_i|\leq\left(\sum_i|x_i|^p\right)^{1/p}\left(\sum_i|y_i|^q\right)^{1/q}$$
with the convention that an $\infty$ exponent produces a $\max|x_i|$ quantity.
\end{theorem}

\begin{proof}
This is something very standard, the idea being as follows:

\medskip

(1) Assume first that we are dealing with finite exponents, $p,q\in(1,\infty)$. By linearity we can assume that $x,y$ are normalized, in the following way:
$$\sum_i|x_i|^p=\sum_i|y_i|^q=1$$

In this case, we want to prove that the following inequality holds:
$$\sum_i|x_iy_i|\leq1$$

For this purpose, we use the Young inequality, which gives, for any $i$:
$$|x_iy_i|\leq\frac{|x_i|^p}{p}+\frac{|y_i|^q}{q}$$

By summing now over $i=1,\ldots,N$, we obtain from this, as desired:
\begin{eqnarray*}
\sum_i|x_iy_i|
&\leq&\sum_i\frac{|x_i|^p}{p}+\sum_i\frac{|y_i|^q}{q}\\
&=&\frac{1}{p}+\frac{1}{q}\\
&=&1
\end{eqnarray*}

(2) In the case $p=1$ and $q=\infty$, or vice versa, the inequality holds too, trivially, with the convention that an $\infty$ exponent produces a max quantity, according to:
$$\lim_{p\to\infty}\left(\sum_i|x_i|^p\right)^{1/p}=\max|x_i|$$

Thus, we are led to the conclusion in the statement.
\end{proof}

As a consequence now of the H\"older inequality, we have:

\begin{theorem}[Minkowski]
Assuming $p\in[1,\infty]$, we have the inequality
$$\left(\sum_i|x_i+y_i|^p\right)^{1/p}\leq \left(\sum_i|x_i|^p\right)^{1/p}+\left(\sum_i|y_i|^p\right)^{1/p}$$
for any two vectors $x,y\in\mathbb C^N$, with our usual conventions at $p=\infty$.
\end{theorem}

\begin{proof}
We have indeed the following estimate, using the H\"older inequality, and the conjugate exponent $q\in[1,\infty]$, given by $1/p+1/q=1$:
\begin{eqnarray*}
&&\sum_i|x_i+y_i|^p\\
&=&\sum_i|x_i+y_i|\cdot|x_i+y_i|^{p-1}\\
&\leq&\sum_i|x_i|\cdot|x_i+y_i|^{p-1}+\sum_i|y_i|\cdot|x_i+y_i|^{p-1}\\
&\leq&\left(\sum_i|x_i|^p\right)^{1/p}\left(\sum_i|x_i+y_i|^{(p-1)q}\right)^{1/q}
+\left(\sum_i|y_i|^p\right)^{1/p}\left(\sum_i|x_i+y_i|^{(p-1)q}\right)^{1/q}\\
&=&\left[\left(\sum_i|x_i|^p\right)^{1/p}+\left(\sum_i|y_i|^p\right)^{1/p}\right]\left(\sum_i|x_i+y_i|^p\right)^{1-1/p}
\end{eqnarray*}

Here we have used $(p-1)q=p$ at the end, coming from $1/p+1/q=1$. Now by dividing both sides by the last quantity at the end, we obtain the result.
\end{proof}

Good news, done with inequalities, and as a consequence of the above results, and more specifically of the Minkowski inequality obtained above, we can formulate:

\begin{theorem}
Given an exponent $p\in[1,\infty]$, the formula
$$||x||_p=\left(\sum_i|x_i|^p\right)^{1/p}$$
with usual conventions at $p=\infty$, defines a norm on $\mathbb C^N$, making it a Banach space.
\end{theorem}

\begin{proof}
Here the normed space assertion follows from the Minkowski inequality, and the Banach space assertion is trivial, our space being finite dimensional.
\end{proof}

Very nice all this, but you might wonder at this point, what is the relation of all this with functions. In answer, Theorem 7.13 can be reformulated as follows:

\begin{theorem}
Given an exponent $p\in[1,\infty]$, the formula
$$||f||_p=\left(\int|f(x)|^p\right)^{1/p}$$
with usual conventions at $p=\infty$, defines a norm on the space of functions 
$$f:\{1,\ldots,N\}\to\mathbb C$$
making it a Banach space.
\end{theorem}

\begin{proof}
This is a just fancy reformulation of Theorem 7.13, by using the fact that the space formed by the functions $f:\{1,\ldots,N\}\to\mathbb C$ is canonically isomorphic to $\mathbb C^N$.
\end{proof}

In order to further extend the above result, let us start with:

\begin{theorem}
Given two functions $f,g:X\to\mathbb C$ and an exponent $p\geq1$, we have
$$\left(\int_X|f+g|^p\right)^{1/p}\leq\left(\int_X|f|^p\right)^{1/p}+\left(\int_X|g|^p\right)^{1/p}$$
called Minkowski inequality. Also, assuming that $p,q\geq1$ satisfy $1/p+1/q=1$, we have
$$\int_X|fg|\leq \left(\int_X|f|^p\right)^{1/p}\left(\int_X|g|^q\right)^{1/q}$$
called H\"older inequality. These inequalities hold as well for $\infty$ values of the exponents.
\end{theorem}

\begin{proof}
This is very standard, exactly as in the case of sequences, as follows:

\medskip

(1) Let us first prove H\"older, in the case of finite exponents, $p,q\in(1,\infty)$. By linearity we can assume that $f,g$ are normalized, in the following way:
$$\int_X|f|^p=\int_X|g|^q=1$$

We can use as before the Young inequality, which gives, for any $x\in X$:
$$|f(x)g(x)|\leq\frac{|f(x)|^p}{p}+\frac{|g(x)|^q}{q}$$

By integrating now over $x\in X$, we obtain from this, as desired:
$$\int_X|fg|
\leq\int_X\frac{|f(x)|^p}{p}+\int_X\frac{|g(x)|^q}{q}
=1$$

(2) Regarding now Minkowski, again in case $p\in(1,\infty)$, this follows from:
\begin{eqnarray*}
&&\int_X|f+g|^p\\
&=&\int_X|f+g|\cdot|f+g|^{p-1}\\
&\leq&\int_X|f|\cdot|f+g|^{p-1}+\int_X|g|\cdot|f+g|^{p-1}\\
&\leq&\left(\int_X|f|^p\right)^{1/p}\left(\int_X|f+g|^{(p-1)q}\right)^{1/q}
+\left(\int_X|g|^p\right)^{1/p}\left(\int_X|f+g|^{(p-1)q}\right)^{1/q}\\
&=&\left[\left(\int|f|^p\right)^{1/p}+\left(\int_X|g|^p\right)^{1/p}\right]\left(\int_X|f+g|^p\right)^{1-1/p}
\end{eqnarray*}

(3) Finally, in the infinite exponent cases we have similar results, which are trivial this time, with the convention that an $\infty$ exponent produces an essential supremum.
\end{proof}

We can now extend Theorem 7.14, into something very general, as follows:

\begin{theorem}
Given a measured space $X$, and $p\in[1,\infty]$, the following space, with the convention that functions are identified up to equality almost everywhere,
$$L^p(X)=\left\{f:X\to\mathbb C\Big|\int_I|f(x)|^pdx<\infty\right\}$$
is a vector space, and the following quantity
$$||f||_p=\left(\int_X|f(x)|^p\right)^{1/p}$$
is a norm on it, making it a Banach space.
\end{theorem}

\begin{proof}
This follows indeed from Theorem 7.15, with due attention to the null sets, and this because of the first normed space axiom, namely:
$$||x||=0\implies x=0$$

To be more precise, in order for this axiom to hold, we must identify the functions up to equality almost everywhere, as indicated in the statement.
\end{proof}

\section*{7c. Existence, uniqueness}

Getting now towards our ODE business, existence and uniqueness results, as explained before, we would like to use some fixed point technology. So, let us formulate:

\begin{definition}
Let $V$ be a Banach space, and $K:C\subset V\to C$ be a linear map, with $C$ being closed. We say that $K$ is a contraction if 
$$||K(x)-K(y)||\leq\theta||x-y||$$
for some $\theta\in[0,1)$. Also, we call fixed point of $K$ any $x\in C$ such that $K(x)=x$.
\end{definition}

Observe that the fixed point of a contraction, if it exists, is unique, due to our assumption $\theta<1$. Now with these notions in hand, we have the following result:

\begin{theorem}
Any contraction $K:C\subset V\to C$ has a unique fixed point $\bar{x}\in C$, which can be obtained by starting with any point $x\in C$, and iterating $K$:
$$\bar{x}=\lim_{n\to\infty}K^n(x)$$
In addition, we have the following estimate,
$$||K^n(x)-\bar{x}||\leq\frac{\theta^n}{1-\theta}||K(x)-x||$$
valid for any $x\in C$, regarding the convergence $K^{(n)}(x)\to\bar{x}$.
\end{theorem}

\begin{proof}
As explained in the above, the uniqueness of the fixed point is clear, coming from our assumption $\theta<1$. Regarding now the existence part, and the precise estimate in the statement too, pick $x=x_0\in C$, and set $x_n=K^n(x_0)$. We have then:
\begin{eqnarray*}
||x_{n+1}-x_n||
&\leq&\theta||x_n-x_{n-1}||\\
&\leq&\theta^2||x_{n-1}-x_{n-2}||\\
&&\vdots\\
&\leq&\theta^n||x_1-x_0||
\end{eqnarray*}

Now by using the triangle inequality, we obtain from this, for $n>m$:
\begin{eqnarray*}
||x_n-x_m||
&\leq&\sum_{j=m+1}^n||x_j-x_{j-1}||\\
&\leq&\theta^m\sum_{j=0}^{n-m-1}\theta^j||x_1-x_0||\\
&\leq&\frac{\theta^m}{1-\theta}||x_1-x_0||
\end{eqnarray*}

Thus the sequence $\{x_n\}$ is Cauchy, and since we are in a Banach space, this sequence converges. Moreover, since $C\subset V$ was chosen closed, the limit belongs to $C$:
$$x_n\to\bar{x}\in C$$

Now since our map $K$ was assumed to be a contraction, it is continuous, and by continuity we obtain, as desired, that we have indeed a fixed point, due to:
$$||K(\bar{x})-\bar{x}||=\lim_{n\to\infty}||x_{n+1}-x_n||=0$$

Finally, in what regards the estimate at the end, in the statement, let us go back to the main estimate obtained before, which was as follows, for any $n>m$:
$$||x_n-x_m||\leq\frac{\theta^m}{1-\theta}||x_1-x_0||$$ 

But this gives, with $m\to\infty$, the estimate in the statement, as desired.
\end{proof}

Now by getting back to our ODE questions, recall from our discussion before that the map which was needing fixed points was as follows:
$$K(x)(t)=x_0+\int_{t_0}^tf(s,x(s))ds$$

Thus, we are led into the question on whether such a map $K$ is a contraction or not. In order to discuss this, let us introduce the following technical definition:

\begin{definition}
A map $f\in C(U,\mathbb R^N)$, with $U\subset\mathbb R^{N+1}$ open, is called locally Lipschitz with respect to $x$, uniformly with respect to $t$, if for any $V\subset U$ compact we have
$$\frac{|f(t,x)-f(t,y)|}{||x-y||}\leq L$$
for any $(t,x)\neq(t,y)\in V$, for a certain number $L\in(0,\infty)$.
\end{definition}

Observe that in the case $L\leq 1$, our map is a contraction, at any $t$. Now with this notion in hand, we can formulate, following Cauchy-Lipschitz and Picard-Lindel\"of:

\begin{theorem}
An equation as follows, with $f\in C(U,\mathbb R^N)$, with $U\subset\mathbb R^{N+1}$ open, has a unique local solution, 
$$\dot{x}=f(t,x)\quad,\quad x(t_0)=x_0$$
provided that $f$ is locally Lipschitz with respect to $x$, uniformly with respect to $t$.
\end{theorem}

\begin{proof}
Consider, as already indicated above, the following map:
$$K(x)(t)=x_0+\int_{t_0}^tf(s,x(s))ds$$

We assume for simplifying $t_0=0$. In order to verify that $K$ is a contraction, for $t>0$ small, consider the following Banach space, with $T>0$ to be determined later:
$$V=C(I,\mathbb R^N)\quad,\quad I=[0,T]$$

Let also $\delta>0$, and consider the following closed ball, inside this space $V$:
$$C=\bar{B}_\delta(x_0)$$

We would like to apply Theorem 7.18, and in order to do so, we need to check two things, namely that we have indeed $K:C\to C$, and that $K$ is a contraction.

\medskip

(1) Let us first check that we have $K:C\to C$. For this purpose, let us set:
$$W=[0,T]\times C\subset U$$

We have then the following estimate, coming from definitions:
\begin{eqnarray*}
|K(x)(t)-x_0|
&\leq&\int_0^t|f(s,x(s))|ds\\
&\leq&t\max_{(t,v)\in W}|f(t,x)|
\end{eqnarray*}

In view of this, consider the number appearing on the right, namely:
$$M=\max_{(t,v)\in W}|f(t,x)|$$

With this notation, we conclude from our estimate above that we have:
$$TM\leq\delta\implies |K(x)(t)-x_0|\leq\delta,\ \forall t\in[0,T]$$

On the other hand, inside the Banach space $C([0,T],\mathbb R^N)$, we have:
$$||K(x)-x_0||=\sup_{t\in[0,T]}|K(x)(t)-x_0|$$

Thus, under the above assumption $TM\leq\delta$, the following happens:
$$||K(x)-x_0||\leq\delta$$

But this shows that we have $K(x)\in \bar{B}_\delta(x_0)=C$, and so that we have, as desired:
$$K:C\to C$$

(2) With this done, let us turn now to the second check, that of the fact that our linear map $K$ is indeed a contraction. For this purpose, we use the Lipschitz property of $f$ from the statement, or rather from Definition 7.19, namely:
$$\frac{|f(t,x)-f(t,y)|}{||x-y||}\leq L$$

By using this, and integrating, we obtain the following estimate:
\begin{eqnarray*}
\int_0^t|f(s,x(s))-f(s,y(s))|ds
&\leq&L\int_0^t|x(s)-y(s)|ds\\
&\leq&Lt\sup_{0\leq s\leq t}|x(s)-y(s)|
\end{eqnarray*}

Thus, in terms of our linear map $K$, we have the following estimate:
$$||K(x)-K(y)||\leq LT||x-y||$$

But this shows that, with $T\leq 1/L$, we have indeed a contraction, as desired.

\medskip

(3) Summarizing, we have shown that we have $K:C\to C$, and that this map is a contraction. Thus Theorem 7.18 applies, and gives the result.
\end{proof}

Before getting into further theory, let us discuss a simple application of the above. Consider the following linear equation, that we certainly know how to solve:
$$\dot{x}=x\quad,\quad x(0)=1$$

Observe that $f(t,x)=x$ is indeed Lipschitz as in Definition 7.19, with $L=1$. Regarding now the linear map $K$, this is given by the following formula:
\begin{eqnarray*}
K(x)(t)
&=&x_0+\int_{t_0}^tf(s,x(s))ds\\
&=&1+\int_0^tx(s)ds
\end{eqnarray*}

By choosing now $y=1$ as starting point, the iteration goes as follows:
$$K(y)=1+\int_0^t1ds=1+t$$
$$K^2(y)=1+\int_0^t(1+s)ds=1+t+\frac{t^2}{2}$$
$$K^3(y)=1+\int_0^t\left(1+s+\frac{s^2}{2}\right)ds=1+t+\frac{t^2}{2}+\frac{t^3}{6}$$
$$\vdots$$

Thus we obtain in the limit, as we should, the following solution:
$$K^\infty(y)=\sum_{n=0}^\infty\frac{t^n}{n!}=e^t$$

Getting now to technical comments, in relation with Theorem 7.20, many things can be said here, and here are two of them, which are of particular importance:

\bigskip

(1) In the context of Theorem 7.20, it is possible to prove that if $f\in C^k(U,\mathbb R^N)$ with $k\geq1$, then the solution is $C^{k+1}$. This is indeed elementary, by recurrence on $k$.

\bigskip

(2) Also in the context of Theorem 7.20, assume that $[t_0,T]\times\mathbb R^N\subset U$ is such that:
$$\int_{t_0}^TL(t)dt<\infty\quad,\quad L(t)=\sup_{x\neq y\in\mathbb R^N}\frac{|f(t,x)-f(t,y)|}{||x-y||}$$

Then, by suitably changing the Banach space norm, and suitably modifying the contraction principle too, it is possible to prove that the solution is defined on $[t_0,T]$.

\bigskip

We refer to the ODE literature for more on the above, which is something quite standard. As a main question now that we would like to solve, we have:

\begin{question}
How does the solution depend on the initial data?
\end{question}

Obviously, this question is of key importance, in relation with our general order vs chaos problematics. However, this question is non-trivial, and our tools so far, which are quite abstract, do not provide a direct answer to it. So, we have to work some more.

\bigskip

In order to solve our question, let us begin with a key technical statement, of classical analysis type, not obviously related to equations, due to Gronwall, as follows:

\begin{proposition}
Assume that a function $\psi$ satisfies the estimate
$$\psi(t)\leq\alpha(t)+\int_0^t\beta(s)\psi(s)ds$$
for any $t\in[0,T]$, with $\alpha(t)\in\mathbb R$, and $\beta(t)>0$. We have then
$$\psi(t)\leq\alpha(t)+\int_0^t\alpha(s)\beta(s)\exp\left(\int_s^t\beta(r)dr\right)ds$$
for any $t\in[0,T]$. Moreover, assuming that $\alpha$ is increasing, we have
$$\psi(t)\leq\alpha(t)\exp\left(\int_0^t\beta(s)ds\right)$$
for any $t\in[0,T]$. 
\end{proposition}

\begin{proof}
This is something quite tough, and for the story, it happened to me more than once, when teaching this to our graduate math students in Cergy, for one student to leave the class during or after the proof, in protest, never to be seen again. Well, in the hope that these guys  eventually found some friends, spouses and jobs, not quite sure about that, and here is the proof of the result, that I personally find quite cute:

\medskip

(1) Let us first prove the first assertion, which is the main one. For this purpose, we use a trick. Consider the following function:
$$\phi(t)=\exp\left(-\int_0^t\beta(s)ds\right)$$

We have then the following computation, using the Leibnitz rule for derivatives, and also using at the end our assumption on $\psi$ from the statement:
\begin{eqnarray*}
&&\frac{d}{dt}\left[\phi(t)\int_0^t\beta(s)\psi(s)ds\right]\\
&=&\left[\frac{d}{dt}\,\phi(t)\right]\int_0^t\beta(s)\psi(s)ds+\phi(t)\left[\frac{d}{dt}\int_0^t\beta(s)\psi(s)ds\right]\\
&=&-\beta(t)\phi(t)\int_0^t\beta(s)\psi(s)ds+\phi(t)\beta(t)\psi(t)\\
&=&\beta(t)\psi(t)\left(\psi(t)-\int_0^t\beta(s)\psi(s)ds\right)\\
&\leq&\alpha(t)\beta(t)\phi(t)
\end{eqnarray*}

Now by integrating with respect to $t$, we obtain from this:
$$\phi(t)\int_0^t\beta(s)\psi(s)ds\leq\int_0^t\alpha(s)\beta(s)\phi(s)ds$$

We conclude that we have the following estimate:
$$\int_0^t\beta(s)\psi(s)ds\leq\int_0^t\alpha(s)\beta(s)\frac{\phi(s)}{\phi(t)}ds$$

By adding now $\alpha(t)$ to both sides, we obtain the following estimate:
$$\alpha(t)+\int_0^t\beta(s)\psi(s)ds\leq\alpha(t)+\int_0^t\alpha(s)\beta(s)\frac{\phi(s)}{\phi(t)}ds$$

But in this situation, we can use once again our assumption on $\psi$ from the statement, and we obtain the following estimate:
$$\psi(t)\leq\alpha(t)+\int_0^t\alpha(s)\beta(s)\frac{\phi(s)}{\phi(t)}ds$$

Now let us look at the fraction on the right. This is given by:
\begin{eqnarray*}
\frac{\phi(s)}{\phi(t)}
&=&\frac{\exp\left(-\int_0^s\beta(r)dr\right)}{\exp\left(-\int_0^t\beta(r)dr\right)}\\
&=&\exp\left(\int_0^t\beta(r)dr-\int_0^s\beta(r)dr\right)\\
&=&\exp\left(\int_s^t\beta(r)dr\right)
\end{eqnarray*}

We conclude that the estimate that we found above reads:
$$\psi(t)\leq\alpha(t)+\int_0^t\alpha(s)\beta(s)\exp\left(\int_s^t\beta(r)dr\right)ds$$

But this is precisely what we wanted to prove, the first estimate in the statement.

\medskip

(2) With this done, let us turn now to the second assertion in the statement. So, assume that the function $\alpha$ there is increasing. We have then:
\begin{eqnarray*}
\psi(t)
&\leq&\alpha(t)+\int_0^t\alpha(s)\beta(s)\exp\left(\int_s^t\beta(r)dr\right)ds\\
&\leq&\alpha(t)+\int_0^t\alpha(t)\beta(s)\exp\left(\int_s^t\beta(r)dr\right)ds\\
&=&\alpha(t)\left[1+\int_0^t\beta(s)\exp\left(\int_s^t\beta(r)dr\right)ds\right]\\
&=&\alpha(t)\left[1+\int_0^t\beta(s)\exp\left(\int_0^t\beta(r)dr-\int_0^s\beta(r)dr\right)ds\right]\\
&=&\alpha(t)\left[1+\exp\left(\int_0^t\beta(r)dr\right)\int_0^t\beta(s)\exp\left(-\int_0^s\beta(r)dr\right)ds\right]
\end{eqnarray*}

Now recall that we can consider, as in (1), the following function:
$$\phi(t)=\exp\left(-\int_0^t\beta(s)ds\right)$$

The derivative of this function satisfies then the following formula:
$$\phi'(t)=-\beta(t)\phi(t)$$

Thus, we have the following formula, for this derivative:
$$\phi'(s)=-\beta(s)\exp\left(-\int_0^s\beta(r)dr\right)$$

We conclude that the estimate found before reformulates as:
\begin{eqnarray*}
\psi(t)
&\leq&\alpha(t)\left[1+\exp\left(\int_0^t\beta(r)dr\right)\int_0^t\beta(s)\exp\left(-\int_0^s\beta(r)dr\right)ds\right]\\
&=&\alpha(t)\left[1+\exp\left(\int_0^t\beta(r)dr\right)\left(-\phi'\right)\Big|_0^t\right]\\
&=&\alpha(t)\left[1+\exp\left(\int_0^t\beta(r)dr\right)(1-\phi(t))\right]
\end{eqnarray*}

In order to finish, consider the following number, depending on $t$:
$$K=\int_0^t\beta(r)dr$$

In terms of this number, the estimate that we found above reads:
\begin{eqnarray*}
\psi(t)
&\leq&\alpha(t)(1+e^K(1-e^{-K}))\\
&=&\alpha(t)(1+e^K-1)\\
&=&\alpha(t)e^K
\end{eqnarray*}

Thus, as a conclusion, we have reached to the following estimate:
$$\psi(t)\leq\alpha(t)\exp\left(\int_0^t\beta(s)ds\right)$$

But this is exactly what we wanted to prove, namely second estimate in the statement, and so, eventually, done. So, very good all this, and still with me, I hope. 
\end{proof}

As a continuation of the above, we won't leave such beautiful things like this, we would definitely love to spend more time with them, we have:

\begin{proposition}
Assume that a function $\psi$ satisfies the estimate
$$\psi(t)\leq\alpha(t)+\int_0^t(\beta\psi(s)+\gamma)ds$$
for any $t\in[0,T]$, with $\alpha\in\mathbb R$, $\beta\geq 0$ and $\gamma\in\mathbb R$. We have then
$$\psi(t)\leq\alpha\exp(\beta t)+\frac{\gamma}{\beta}(\exp(\beta t)-1)$$
for any $t\in[0,T]$. 
\end{proposition}

\begin{proof}
In order to prove this result, consider the following function:
$$\widetilde{\psi}(t)=\psi(t)+\frac{\gamma}{\beta}$$

In terms of this function $\widetilde{\psi}$, our assumption on $\psi$ in the statement reads: 
$$\widetilde{\psi}-\frac{\gamma}{\beta}\leq\alpha+\beta\int_0^t\widetilde{\psi}(s)ds$$

Thus, our modified function $\widetilde{\psi}$ satisfies the following estimate: 
$$\widetilde{\psi}\leq\left(\alpha+\frac{\gamma}{\beta}\right)+\beta\int_0^t\widetilde{\psi}(s)ds$$

Thus, we can apply the second assertion in Proposition 7.22, with $\alpha(t)=\alpha+\gamma/\beta$, and $\beta(t)=\beta$, both constant functions, and we obtain in this way:
$$\widetilde{\psi}\leq\left(\alpha+\frac{\gamma}{\beta}\right)\exp(\beta t)$$

But this gives, in terms of the original function $\psi$, the following estimate:
\begin{eqnarray*}
\psi(t)
&\leq&\left(\alpha+\frac{\gamma}{\beta}\right)\exp(\beta t)-\frac{\gamma}{\beta}\\
&=&\alpha\exp(\beta t)+\frac{\gamma}{\beta}(\exp(\beta t)-1)
\end{eqnarray*}

Thus, we have reached to the conclusion in the statement.
\end{proof}

Now back to the ODE, the above results apply, and we can answer Question 7.21. To be more precise, in the general context of Theorem 7.20, we have the following result:

\begin{theorem}
Assume that $f,g\in C(U,\mathbb R^N)$, with $U\subset\mathbb R^{N+1}$ open, are locally Lipschitz with respect to $x$, uniformly with respect to $t$. If $x,y$ are solutions of
$$\dot{x}=f(t,x)\quad,\quad x(t_0)=x_0$$
$$\dot{y}=g(t,y)\quad,\quad y(t_0)=y_0$$
then we have the following estimate, for any $t$ in the interval of definition of $x,y$,
$$||x(t)-y(t)||\leq||x_0-y_0||e^{L|t-t_0|}+\frac{M}{L}\left(e^{L|t-t_0|}-1\right)$$
with the constant $M$ on the right being given by the following formula,
$$M=\sup_{(t,x)\in U}|f(t,x)-g(t,x)|$$
and with $L>0$ being a common Lipschitz constant for both $f,g$.
\end{theorem}

\begin{proof}
We know from Theorem 7.20 that the above equations have indeed solutions. We can assume for simplifying that we have $t_0=0$. Now observe that we have:
\begin{eqnarray*}
&&||x(t)-y(t)||\\
&\leq&||x_0-y_0||+\int_0^t|f(s,x(s))-g(s,y(s))|ds\\
&\leq&||x_0-y_0||+\int_0^t\Big(|f(s,x(s))-f(s,y(s))|+|f(s,y(s))-g(s,y(s))|\Big)ds\\
&\leq&||x_0-y_0||+\int_0^t\Big(L||x(s)-y(s)||+M\Big)ds
\end{eqnarray*}

In view of this estimate, consider the following function:
$$\psi(t)=||x(t)-y(t)||$$

In terms of this function, the estimate that we found above reads:
$$\psi(t)\leq||x_0-y_0||+\int_0^t(L\psi(s)+M)ds$$

But this shows that the Gronwall estimate from Proposition 7.23 applies, with the following choices for the constants $\alpha\in\mathbb R$, $\beta\geq 0$ and $\gamma\in\mathbb R$ appearing there:
$$\alpha=||x_0-y_0||\quad,\quad \beta=L\quad,\quad \gamma=M$$

So, let us apply Proposition 7.23, with these values of $\alpha,\beta,\gamma$. We obtain:
\begin{eqnarray*}
\psi(t)
&\leq&\alpha\exp(\beta t)+\frac{\gamma}{\beta}(\exp(\beta t)-1)\\
&=&||x_0-y_0||e^{L|t-t_0|}+\frac{M}{L}\left(e^{L|t-t_0|}-1\right)
\end{eqnarray*}

But this is exactly the estimate in the statement, as desired.
\end{proof}

\section*{7d. Dynamical systems}

We discuss in the remainder of this chapter a very fruitful linearization idea, in the context of the dynamical systems, based on the following principle:

\begin{principle}[Linearization]
In order to deal with an arbitrary, non-linear system
$$\dot{x}=f(x)\quad,\quad x_0=0$$
we can write the function $f$ as follows, with $A=f'(x)\in M_N(\mathbb R)$ being its derivative,
$$f(x)=Ax+o(||x||)$$
and then use, by perturbing, the results regarding the linear system $\dot{x}=Ax$.
\end{principle}

Which sounds very good. In practice now, let us first go back to the general theory of the linear equations $\dot{x}=Ax$, as developed in chapter 5. We will need:

\begin{notations}
Given $A\in M_N(\mathbb R)$, we write its characteristic polynomial as
$$P(z)=\prod_i(z-\alpha_i)^{a_i}$$
so that we have a direct sum decomposition of the ambient space, as follows:
$$\mathbb C^N=\bigoplus_i\ker\left[(A-\alpha_i)^{a_i}\right]$$
We also consider the corresponding geometric multiplicities, given by
$$g_i=\dim\ker(A-a_i)$$
and satisfying $g_i\leq a_i$, with equalities when $A$ is diagonalizable.
\end{notations}

We refer to chapter 5 for more on all this, theory and applications. Now back to the linear systems, $\dot{x}=Ax$, let us formulate the following key definition:

\begin{definition}
We say that a linear system $\dot{x}=Ax$ is hyperbolic when
$$Re(\alpha)\neq0$$
for any eigenvalue $\alpha$. In this case, we consider the linear spaces
$$E^\pm=\bigoplus_{\pm Re(\alpha_i)<0}\ker\left[(A-\alpha_i)^{a_i}\right]$$
which are therefore in direct sum position, $\mathbb C^N=E^+\oplus E^-$.
\end{definition}

So, studying these hyperbolic linear systems, and then extending our results to the hyperbolic non-linear systems, by using the derivative, will be our job, in what follows.

\bigskip

In what regards the study in the linear case, this is quickly done, by using the Jordan form for the matrix $A\in M_N(\mathbb R)$, with the result being as follows:

\begin{theorem}
For a hyperbolic linear system $\dot{x}=Ax$, the following happen:
\begin{enumerate}
\item The spaces $E^\pm$ are both invariant by the flow of the system.

\item Any integral curve departing from $E^\pm$ converges to $0$, with $t\to\pm\infty$.

\item In fact, we have the following explicit estimate for the decay,
$$|e^{tA}x_\pm|\leq Ce^{\pm t\alpha}|x_\pm|$$
for any $\pm t>0$ and any $x_\pm\in E^\pm$, with $\alpha>0$ subject to
$$\alpha<\min\Big\{|Re(\alpha_i)|:\pm Re(\alpha_i)<0\Big\}$$
and with $C>0$ depending on $\alpha$.
\end{enumerate}
\end{theorem}

\begin{proof}
This is something quite straightforward, the idea being as follows:

\medskip

(1) This is something which is obvious.

\medskip

(2) This is something that we already know, as a consequence of our general results from before, and which follows also from (3), that we will prove next.

\medskip

(3) We just discuss here the proof of the ``$+$'' result, with the proof of the ``$-$'' result being similar. We put $A$ in Jordan form, and we consider the following quantity:
$$m=\min\Big\{|Re(\alpha_i)|:Re(\alpha_i)<0\Big\}$$

Now let $\alpha<m$ as in the statement, and set $\varepsilon=m-\alpha$. Then, for any eigenvalue satisfying $Re(\alpha_i)<0$, the entry of maximal absolute value, say $M_i$, of the corresponding component $e^{tJ_{a_i}}$ of the matrix $e^{tA}$, can be estimated as follows:
\begin{eqnarray*}
M_i
&=&\frac{|t^ne^{a_it}|}{F}\\
&\leq&\frac{|t^ne^{-\varepsilon t}|e^{-\alpha t}}{F}\\
&\leq&Ce^{-\alpha t}
\end{eqnarray*}

To be more precise, here $F$ is a certain factorial, namely $F=(s-1)!$, with $s$ being the size of the Jordan block, and $C>0$ at the end is a certain constant, depending on this nunber $F$, and on $\alpha$. Thus, we are led to the conclusion in the statement. 
\end{proof}

Now consider a non-linear equation $\dot{x}=f(x)$, and denote by $\Phi(t,x)$ its flow, describing the solution in time $t$, with initial data $x(0)=x$. We have:

\begin{definition}
We associate to the equation $\dot{x}=f(x)$ the following sets, 
$$W^\pm(x_0)=\left\{x\Big|\lim_{t\to\pm\infty}\Phi(t,x)=x_0\right\}$$
gathering the initial data $x$ such that the solution converges to $x_0$, with $t\to\pm\infty$.
\end{definition}

Observe that both the above sets $W^\pm(x_0)$ are stable under the flow. In order now to compute these sets, we use our linearization idea. So, let us introduce as well:

\begin{definition}
We associate to the equation $\dot{x}=f(x)$ the sets
$$M^{\pm,\alpha}=\left\{x\Big|\gamma_\pm(x)\subset U(x_0),\ \sup_{\pm t\geq0}e^{\pm\alpha t}|\Phi(t,x)-x_0|<\infty\right\}$$
and then we consider the intersection of these sets, over eigenvalues,
$$M^\pm(x_0)=\bigcup_{\alpha>0}M^{\pm,\alpha}$$
which in the linear case, $\dot{x}=Ax$, are the spaces $E^\pm$ that we knew from before.
\end{definition}

To be more precise here, in the linear case, $\dot{x}=Ax$, the spaces $M^{\pm,\alpha}$ constructed above correspond to the spaces $E^{\pm,\alpha}$ spanned by the eigenvectors of $A$ corresponding to the eingenvalues satisfying $Re(\lambda)\geq\alpha$ and $Re(\lambda)\leq-\alpha$, and so by intersecting, we obtain indeed the spaces $E^\pm$ that we knew from before, as claimed in the above.

\bigskip

We can now formulate our main linearization result, as follows:

\begin{theorem}
For a hyperbolic point $x_0$, the following happen:
\begin{enumerate}
\item $M^\pm(x_0)$ is a $C^1$ manifold.

\item $M^\pm(x_0)$ is tangent to $E^\pm$ at $0$.

\item $M^\pm(x_0)=W^\pm(x_0)$.
\end{enumerate}
\end{theorem}

\begin{proof}
The idea here is that of using the standard direct sum decomposition $\mathbb R^N=E^+\oplus E^-$, in order to decompose everything, and prove the various assertions. For background and details, we refer to any geometry and dynamical systems books.
\end{proof}

\section*{7e. Exercises}

Tough analytic chapter that we had here, and as exercises, we have:

\begin{exercise}
Learn more about the Bernoulli equations, and their solutions.
\end{exercise}

\begin{exercise}
Learn more about the Riccati equations, and their solutions.
\end{exercise}

\begin{exercise}
Clarify what has been said above, about convex functions and Jensen.
\end{exercise}

\begin{exercise}
Learn more about $L^p$ spaces, and about Banach spaces, in general.
\end{exercise}

\begin{exercise}
In case you skipped the proof of Gronwall, go back and read it.
\end{exercise}

\begin{exercise}
Learn more about hyperbolic linear systems, and their properties.
\end{exercise}

\begin{exercise}
Apply the theory developed in this chapter, to some concrete equations.
\end{exercise}

\begin{exercise}
Have a look as well at the non-linear hyperbolic systems.
\end{exercise}

As bonus exercise, and no surprise here, purchase an ODE book, and read it.

\chapter{Infinite matrices}

\section*{8a. Infinite matrices}

We have learned many things about linear algebra and its applications, and we would like to end the present Part II with a discussion about what happens in infinite dimensions. As we will soon discover, some key results, like the spectral theorems, have suitable extensions to this setting, and some more specialized results, like the singular value decomposition, extend too, under a certain compactness assumption on our linear maps.

\bigskip

In order to get started, for doing our mathematics we need vector spaces, which can be infinite dimensional, and which have scalar products. So, let us formulate:

\index{Hilbert space}
\index{scalar product}

\begin{definition}
A Hilbert space is a complex vector space $H$ with a scalar product $<x,y>$, taken linear at left and antilinear at right,
$$<\lambda x,y>=\lambda<x,y>\quad,\quad <x,\lambda y>=\bar{\lambda}<x,y>$$
which is complete with respect to corresponding norm
$$||x||=\sqrt{<x,x>}$$
in the sense that any sequence $\{x_n\}$ which is a Cauchy sequence, having the property $||x_n-x_m||\to0$ with $n,m\to\infty$, has a limit, $x_n\to x$.
\end{definition}

Here the fact that $||x||=\sqrt{<x,x>}$ is indeed a norm, satisfying $||x+y||\leq||x||+||y||$, follows from Cauchy-Schwarz, $|<x,y>|\leq||x||\cdot||y||$, which itself follows as in the $H=\mathbb C^N$ case, by looking at $f(t)=||wx+ty||^2$, as explained in chapter 3. At the level of examples of Hilbert spaces, generalizing the basic $H=\mathbb C^N$ example, we have:

\index{square-summable}
\index{Cauchy-Schwarz}

\begin{theorem}
Given an index set $I$, which can be finite or not, the space of square-summable vectors having indices in $I$, namely
$$l^2(I)=\left\{(x_i)_{i\in I}\Big|\sum_i|x_i|^2<\infty\right\}$$
is a Hilbert space, with scalar product as follows:
$$<x,y>=\sum_ix_i\bar{y}_i$$
When $I$ is finite, $I=\{1,\ldots,N\}$, we obtain in this way the usual space $H=\mathbb C^N$.
\end{theorem}

\begin{proof}
We have already met such things before, but let us recall all this:

\medskip

(1) We know that $l^2(I)\subset\mathbb C^I$ is the space of vectors satisfying $||x||<\infty$. We want to prove that $l^2(I)$ is a vector space, that $<x,y>$ is a scalar product on it, that $l^2(I)$ is complete with respect to $||.||$, and finally that for $|I|<\infty$ we have $l^2(I)=\mathbb C^{|I|}$.

\medskip

(2) The last assertion, $l^2(I)=\mathbb C^{|I|}$ for $|I|<\infty$, is clear, because in this case the sums are finite, so the condition $||x||<\infty$ is automatic. So, we know at least one thing.

\medskip

(3) Regarding the rest, we have Cauchy-Schwarz, $|<x,y>|\leq||x||\cdot||y||$, proved as usual, by looking at $f(t)=||wx+ty||^2$, and we deduce that we have $||x+y||\leq||x||+||y||$. Thus $l^2(I)$ is indeed a vector space, the other vector space conditions being trivial.

\medskip

(4) Also, the quantity $<x,y>$ is surely a scalar product on this vector space, because all the conditions for a scalar product are trivially satisfied.

\medskip

(5) Finally, the fact that our space $l^2(I)$ is indeed complete with respect to its norm $||.||$ follows in the obvious way, the limit of a Cauchy sequence $\{x_n\}$ being the vector $y=(y_i)$ given by $y_i=\lim_{n\to\infty}x_{ni}$, with all the verifications here being trivial.
\end{proof}

Going now a bit more abstract, we have, more generally, the following result:

\index{square-summable}
\index{square-integrable}

\begin{theorem}
Given an arbitrary space $X$ with a positive measure $\mu$ on it, the space of square-summable complex functions on it, namely
$$L^2(X)=\left\{f:X\to\mathbb C\Big|\int_X|f(x)|^2\,d\mu(x)<\infty\right\}$$
is a Hilbert space, with scalar product as follows:
$$<f,g>=\int_Xf(x)\overline{g(x)}\,d\mu(x)$$
When $X=I$ is discrete, meaning that the measure $\mu$ on it is the counting measure, $\mu(\{x\})=1$ for any $x\in X$, we obtain in this way the previous spaces $l^2(I)$.
\end{theorem}

\begin{proof}
This is something routine, remake of Theorem 8.2, as follows:

\medskip

(1) The proof of the first, and main assertion is something perfectly similar to the proof of Theorem 8.2, by replacing everywhere the sums by integrals. 

\medskip

(2) With the remark that we forgot to say in the statement that the $L^2$ functions are by definition taken up to equality almost everywhere, $f=g$ when $||f-g||=0$.

\medskip

(3) As for the last assertion, when $\mu$ is the counting measure all our integrals here become usual sums, and so we recover in this way Theorem 8.2.
\end{proof}

As a third and last theorem about Hilbert spaces, that we will need, we have:

\index{orthonormal basis}
\index{Gram-Schmidt}
\index{separable space}
\index{Zorn lemma}

\begin{theorem}
Any Hilbert space $H$ has an orthonormal basis $\{e_i\}_{i\in I}$, which is by definition a set of vectors whose span is dense in $H$, and which satisfy
$$<e_i,e_j>=\delta_{ij}$$
with $\delta$ being a Kronecker symbol. The cardinality $|I|$ of the index set, which can be finite, countable, or worse, depends only on $H$, and is called dimension of $H$. We have
$$H\simeq l^2(I)$$
in the obvious way, mapping $\sum\lambda_ie_i\to(\lambda_i)$. The Hilbert spaces with $\dim H=|I|$ being countable, including $l^2(\mathbb N)$ and $L^2(\mathbb R)$, are all isomorphic, and are called separable.
\end{theorem}

\begin{proof}
We have many assertions here, the idea being as follows:

\medskip

(1) In finite dimensions an orthonormal basis $\{e_i\}_{i\in I}$ can be constructed by starting with any vector space basis $\{x_i\}_{i\in I}$, and using the Gram-Schmidt procedure from chapter 2. As for the other assertions, these are all clear, from basic linear algebra.

\medskip

(2) In general, the same method works, namely Gram-Schmidt, with a subtlety coming from the fact that the basis $\{e_i\}_{i\in I}$ will not span in general the whole $H$, but just a dense subspace of it, as it is in fact obvious by looking at the standard basis of $l^2(\mathbb N)$. 

\medskip

(3) And there is a second subtlety as well, coming from the fact that the recurrence procedure needed for Gram-Schmidt must be replaced by some sort of ``transfinite recurrence'', using scary tools from logic, and more specifically the Zorn lemma.

\medskip

(4) Finally, everything at the end is clear from definitions, except perhaps for the fact that $L^2(\mathbb R)$ is separable. But here we can argue that, since functions can be approximated by polynomials, we have a countable algebraic basis, namely $\{x^n\}_{n\in\mathbb N}$, called the Weierstrass basis, that we can orthogonalize afterwards by using Gram-Schmidt.
\end{proof}

Moving ahead, now that we know what our vector spaces are, we can talk about infinite matrices with respect to them. And the situation here is as follows:

\index{infinite matrix}
\index{linear operator}
\index{bounded operator}
\index{continuous operator}

\begin{theorem}
Given a Hilbert space $H$, consider the linear operators $T:H\to H$, and for each such operator define its norm by the following formula:
$$||T||=\sup_{||x||=1}||Tx||$$
The operators which are bounded, $||T||<\infty$, form then a complex algebra $B(H)$, which is complete with respect to $||.||$. When $H$ comes with a basis $\{e_i\}_{i\in I}$, we have
$$B(H)\subset M_I(\mathbb C)$$
with the correspondence $T\to M$ coming via the usual linear algebra formulae, namely:
$$T(x)=Mx\quad,\quad M_{ij}=<Te_j,e_i>$$
In infinite dimensions, the inclusion $B(H)\subset M_I(\mathbb C)$ is not an equality.
\end{theorem}

\begin{proof}
This is something straightforward, the idea being as follows:

\medskip

(1) The fact that we have indeed an algebra, satisfying the product condition in the statement, follows from the following estimates, which are all elementary:
$$||S+T||\leq||S||+||T||\quad,\quad 
||\lambda T||=|\lambda|\cdot||T||\quad,\quad 
||ST||\leq||S||\cdot||T||$$

(2) Regarding now the completness assertion, if $\{T_n\}\subset B(H)$ is Cauchy then $\{T_nx\}$ is Cauchy for any $x\in H$, so we can define the limit $T=\lim_{n\to\infty}T_n$ by setting:
$$Tx=\lim_{n\to\infty}T_nx$$

Let us first check that the application $x\to Tx$ is linear. We have:
\begin{eqnarray*}
T(x+y)
&=&\lim_{n\to\infty}T_n(x+y)\\
&=&\lim_{n\to\infty}T_n(x)+T_n(y)\\
&=&\lim_{n\to\infty}T_n(x)+\lim_{n\to\infty}T_n(y)\\
&=&T(x)+T(y)
\end{eqnarray*}

Similarly, we have $T(\lambda x)=\lambda T(x)$, and we conclude that $x\to Tx$ is linear.

\medskip

(3) With this done, it remains to prove now that we have $T\in B(H)$, and that $T_n\to T$ in norm. For this purpose, observe that we have:
\begin{eqnarray*}
||T_n-T_m||\leq\varepsilon\ ,\ \forall n,m\geq N
&\implies&||T_nx-T_mx||\leq\varepsilon\ ,\ \forall||x||=1\ ,\ \forall n,m\geq N\\
&\implies&||T_nx-Tx||\leq\varepsilon\ ,\ \forall||x||=1\ ,\ \forall n\geq N\\
&\implies&||T_Nx-Tx||\leq\varepsilon\ ,\ \forall||x||=1\\
&\implies&||T_N-T||\leq\varepsilon
\end{eqnarray*}

But this gives both $T\in B(H)$, and $T_N\to T$ in norm, and we are done.

\medskip

(4) Regarding the embedding, the correspondence $T\to M$ in the statement is indeed linear, and its kernel is $\{0\}$, so we have indeed an embedding as follows, as claimed:
$$B(H)\subset M_I(\mathbb C)$$

In finite dimensions we have an isomorphism, because any matrix $M\in M_N(\mathbb C)$ determines a linear operator $T:\mathbb C^N\to\mathbb C^N$, given by the following formula:
$$<Te_j,e_i>=M_{ij}$$

However, in infinite dimensions we have matrices not producing operators, as for instance the all-one matrix, so the embedding $B(H)\subset M_I(\mathbb C)$ is not an isomorphism.
\end{proof}

Finally, as a second and last basic result regarding the operators, we will need:

\index{adjoint operator}
\index{adjoint matrix}

\begin{theorem}
Each operator $T\in B(H)$ has an adjoint $T^*\in B(H)$, given by: 
$$<Tx,y>=<x,T^*y>$$
The operation $T\to T^*$ is antilinear, antimultiplicative, involutive, and satisfies:
$$||T||=||T^*||\quad,\quad ||TT^*||=||T||^2$$
When $H$ comes with a basis $\{e_i\}_{i\in I}$, the operation $T\to T^*$ corresponds to
$$(M^*)_{ij}=\overline{M}_{ji}$$ 
at the level of the associated matrices $M\in M_I(\mathbb C)$.
\end{theorem}

\begin{proof}
This is standard too, and can be proved in 3 steps, as follows:

\medskip

(1) The existence of the adjoint operator $T^*$, given by the formula in the statement, comes from the fact that the function $\varphi(x)=<Tx,y>$ being a linear map $H\to\mathbb C$, we must have a formula as follows, for a certain vector $T^*y\in H$:
$$\varphi(x)=<x,T^*y>$$

Moreover, since this vector is unique, $T^*$ is unique too, and we have as well:
$$(S+T)^*=S^*+T^*\quad,\quad
(\lambda T)^*=\bar{\lambda}T^*\quad,\quad 
(ST)^*=T^*S^*\quad,\quad 
(T^*)^*=T$$

Observe also that we have indeed $T^*\in B(H)$, because:
\begin{eqnarray*}
||T||
&=&\sup_{||x||=1}\sup_{||y||=1}<Tx,y>\\
&=&\sup_{||y||=1}\sup_{||x||=1}<x,T^*y>\\
&=&||T^*||
\end{eqnarray*}

(2) Regarding now $||TT^*||=||T||^2$, which is a key formula, observe that we have:
$$||TT^*||
\leq||T||\cdot||T^*||
=||T||^2$$

On the other hand, we have as well the following estimate:
\begin{eqnarray*}
||T||^2
&=&\sup_{||x||=1}|<Tx,Tx>|\\
&=&\sup_{||x||=1}|<x,T^*Tx>|\\
&\leq&||T^*T||
\end{eqnarray*}

By replacing $T\to T^*$ we obtain from this $||T||^2\leq||TT^*||$, as desired.

\medskip

(3) Finally, when $H$ comes with a basis, the formula $<Tx,y>=<x,T^*y>$ applied with $x=e_i$, $y=e_j$ translates into the formula $(M^*)_{ij}=\overline{M}_{ji}$, as desired.
\end{proof}

\section*{8b. Spectral radius}

Let us discuss now the diagonalization problem for the operators $T\in B(H)$, in analogy with the diagonalization problem for the usual matrices $A\in M_N(\mathbb C)$. We first have:

\begin{definition}
The spectrum of an operator $T\in B(H)$ is the set
$$\sigma(T)=\left\{\lambda\in\mathbb C\Big|T-\lambda\not\in B(H)^{-1}\right\}$$
where $B(H)^{-1}\subset B(H)$ is the set of invertible operators.
\end{definition}

As a basic example, in the finite dimensional case, $H=\mathbb C^N$, the spectrum of a usual matrix $A\in M_N(\mathbb C)$ is the collection of its eigenvalues, taken without multiplicities. We will see many other examples. In general, the spectrum has the following properties:

\index{eigenvalue}

\begin{proposition}
The spectrum of $T\in B(H)$ contains the eigenvalue set
$$\varepsilon(T)=\left\{\lambda\in\mathbb C\Big|\ker(T-\lambda)\neq\{0\}\right\}$$
and $\varepsilon(T)\subset\sigma(T)$ is an equality in finite dimensions, but not in infinite dimensions.
\end{proposition}

\begin{proof}
We have several assertions here, the idea being as follows:

\medskip

(1) First of all, the eigenvalue set is indeed the one in the statement, because $Tx=\lambda x$ tells us precisely that $T-\lambda$ must be not injective. The fact that we have $\varepsilon(T)\subset\sigma(T)$ is clear as well, because if $T-\lambda$ is not injective, it is not bijective.

\medskip

(2) In finite dimensions we have $\varepsilon(T)=\sigma(T)$, because $T-\lambda$ is injective if and only if it is bijective, with the boundedness of the inverse being automatic. 

\medskip

(3) In infinite dimensions we can assume $H=l^2(\mathbb N)$, and the shift operator $S(e_i)=e_{i+1}$ is injective but not surjective. Thus $0\in\sigma(T)-\varepsilon(T)$.
\end{proof}

The above result might look quite surprising, and philosophically, the best way of thinking at all this is as follows: the numbers $\lambda\notin\sigma(T)$ are good, because we can invert $T-\lambda$, the numbers $\lambda\in\sigma(T)-\varepsilon(T)$ are bad, because so they are, and the eigenvalues $\lambda\in\varepsilon(T)$ are evil. Welcome to operator theory, where some things are upside-down.

\bigskip

As a second basic result about the spectrum, which is of great use, we have:

\begin{proposition}
The spectrum of an operator $T\in B(H)$ satisfies:
$$\sigma(T)\subset D_0(||T||))$$
In other words, the spectral radius $\rho(T)=\sup_{\lambda\in\sigma(T)}|\lambda|$ satisfies $\rho(T)\leq||T||$. 
\end{proposition}

\begin{proof}
We use the standard fact that for $||S||<1$, we have the following formula:
$$(1-S)^{-1}=1+S+S^2+\ldots$$

Indeed, by using this inversion formula, we have the following computation:
\begin{eqnarray*}
\lambda>||T||
&\implies&\Big|\Big|\frac{T}{\lambda}\Big|\Big|<1\\
&\implies&1-\frac{T}{\lambda}\in B(H)^{-1}\\
&\implies&\lambda-T\in B(H)^{-1}\\
&\implies&\lambda\notin\sigma(T)
\end{eqnarray*}

Thus, we are led to the conclusion in the statement.
\end{proof}

Let us develop now some more general theory. Here is a key result about products:

\index{spectrum of products}

\begin{theorem}
We have the following formula, valid for any operators $S,T$:
$$\sigma(ST)\cup\{0\}=\sigma(TS)\cup\{0\}$$
In finite dimensions we have $\sigma(ST)=\sigma(TS)$, but this fails in infinite dimensions.
\end{theorem}

\begin{proof}
There are several assertions here, the idea being as follows:

\medskip

(1) Let us first prove the main assertion, stating that the sets $\sigma(ST),\sigma(TS)$ coincide outside 0. We first prove that we have the following implication:
$$1\notin\sigma(ST)\implies1\notin\sigma(TS)$$

Assume indeed that $1-ST$ is invertible, with inverse denoted $R$:
$$R=(1-ST)^{-1}$$

We have then $RST=R-1$, and by using this, we have the following computation:
\begin{eqnarray*}
(1+TRS)(1-TS)
&=&1+TRS-TS-TRSTS\\
&=&1+TRS-TS-TRS+TS\\
&=&1
\end{eqnarray*}

Similarly, $(1-TS)(1+TRS)=1$. Thus $1-TS$ is invertible, proving our claim. Now by multiplying by scalars, we deduce that for any $\lambda\in\mathbb C-\{0\}$ we have, as desired:
$$\lambda\notin\sigma(ST)\implies\lambda\notin\sigma(TS)$$

(2) Regarding now the counterexample to the formula $\sigma(ST)=\sigma(TS)$, in general, let us take $S$ to be the shift on $H=L^2(\mathbb N)$, given by the following formula:
$$S(e_i)=e_{i+1}$$

As for $T$, we can take it to be the adjoint of $S$, which is the following operator:
$$S^*(e_i)=\begin{cases}
e_{i-1}&{\rm if}\ i>0\\
0&{\rm if}\ i=0
\end{cases}$$

Let us compose now these two operators. In one sense, we have:
$$S^*S=1\implies 0\notin\sigma(SS^*)$$

In the other sense, however, the situation is different, as follows:
$$SS^*=Proj(e_0^\perp)\implies 0\in\sigma(SS^*)$$

Thus, the spectra do not match on $0$, and we have our counterexample, as desired.
\end{proof}

As our next result, inspired from the theory from chapter 6, we have:

\index{rational calculus}

\begin{theorem}
We have the ``rational functional calculus'' formula
$$\sigma(f(T))=f(\sigma(T))$$
valid for any rational function $f\in\mathbb C(X)$ having poles outside $\sigma(T)$.
\end{theorem}

\begin{proof}
This can be proved in two steps, as follows:

\medskip

(1) Assume first that our rational function $f\in\mathbb C(X)$ is a usual polynomial $P\in\mathbb C[X]$. We pick a scalar $\lambda\in\mathbb C$, and we decompose the polynomial $P-\lambda$, as follows:
$$P(X)-\lambda=c(X-r_1)\ldots(X-r_n)$$

We have then the following equivalences, which give the result:
\begin{eqnarray*}
\lambda\notin\sigma(P(T))
&\iff&P(T)-\lambda\in B(H)^{-1}\\
&\iff&c(T-r_1)\ldots(T-r_n)\in B(H)^{-1}\\
&\iff&T-r_1,\ldots,T-r_n\in B(H)^{-1}\\
&\iff&r_1,\ldots,r_n\notin\sigma(T)\\
&\iff&\lambda\notin P(\sigma(T))
\end{eqnarray*}

(2) In general now, we pick a scalar $\lambda\in\mathbb C$, we write $f=P/Q$, and we set $F=P-\lambda Q$. By using what we found in (1), for this polynomial $F\in\mathbb C[X]$, we obtain:
\begin{eqnarray*}
\lambda\in\sigma(f(T))
&\iff&F(T)\notin B(H)^{-1}\\
&\iff&0\in\sigma(F(T))\\
&\iff&0\in F(\sigma(T))\\
&\iff&\exists\mu\in\sigma(T),F(\mu)=0\\
&\iff&\lambda\in f(\sigma(T))
\end{eqnarray*}

Thus, we are led to the formula in the statement.
\end{proof}

As a first application of the above methods, we have the following key result:

\begin{theorem}
The following happen:
\begin{enumerate}
\item For a unitary operator, $U^*=U^{-1}$, we have $\sigma(U)\subset\mathbb T$. 

\item For a self-adjoint operator, $T=T^*$, we have $\sigma(T)\subset\mathbb R$.
\end{enumerate}
\end{theorem}

\begin{proof}
This is something quite tricky, based on Theorem 8.11, as follows:

\medskip

(1) Assuming $U^*=U^{-1}$, we have the following norm computation:
$$||U||
=\sqrt{||UU^*||}
=\sqrt{1}
=1$$

Now if we denote by $D$ the unit disk, we obtain from this, using Proposition 8.9:
$$\sigma(U)\subset D$$

On the other hand, once again by using $U^*=U^{-1}$, we have as well:
$$||U^{-1}||
=||U^*||
=||U||
=1$$

Thus, as before with $D$ being the unit disk in the complex plane, we have:
$$\sigma(U^{-1})\subset D$$

Now by using Theorem 8.11, we obtain $\sigma(U)
\subset D\cap D^{-1}
=\mathbb T$, as desired.

\medskip

(2) Consider the following rational function, depending on a parameter $r\in\mathbb R$:
$$f(z)=\frac{z+ir}{z-ir}$$

Then for $r>>0$ the operator $f(T)$ is well-defined, and we have:
$$\left(\frac{T+ir}{T-ir}\right)^*
=\frac{T-ir}{T+ir}
=\left(\frac{T+ir}{T-ir}\right)^{-1}$$

Thus $f(T)$ is unitary, and by (1) we have $\sigma(T)\subset f^{-1}(\mathbb T)=\mathbb R$, as desired.
\end{proof}

Next, we have the following key result, which is something quite far-reaching:

\begin{theorem}
The spectral radius of an operator $T\in B(H)$ is given by
$$\rho(T)=\lim_{n\to\infty}||T^n||^{1/n}$$
and in this formula, we can replace the limit by an inf.
\end{theorem}

\begin{proof}
We have several things to be proved, the idea being as follows:

\medskip

(1) Our first claim is that the numbers $u_n=||T^n||^{1/n}$ satisfy:
$$(n+m)u_{n+m}\leq nu_n+mu_m$$

Indeed, we have the following estimate, using the Young inequality $ab\leq a^p/p+b^q/q$ from chapter 7, with exponents $p=(n+m)/n$ and $q=(n+m)/m$:
\begin{eqnarray*}
u_{n+m}
&=&||T^{n+m}||^{1/(n+m)}\\
&\leq&||T^n||^{1/(n+m)}||T^m||^{1/(n+m)}\\
&\leq&||T^n||^{1/n}\cdot\frac{n}{n+m}+||T^m||^{1/m}\cdot\frac{m}{n+m}\\
&=&\frac{nu_n+mu_m}{n+m}
\end{eqnarray*}

(2) Our second claim is that the second assertion holds, namely:
$$\lim_{n\to\infty}||T^n||^{1/n}=\inf_n||T^n||^{1/n}$$

For this purpose, we just need the inequality found in (1). Indeed, fix $m\geq1$, let $n\geq1$, and write $n=lm+r$ with $0\leq r\leq m-1$. By using twice $u_{ab}\leq u_b$, we get:
$$u_n
\leq\frac{1}{n}( lmu_{lm}+ru_r)
\leq u_{m}+\frac{r}{n}\,u_1$$

It follows that we have $\lim\sup_nu_n\leq u_m$, which proves our claim.

\medskip

(3) Summarizing, we are left with proving the main formula, which is as follows, and with the remark that we already know that the sequence on the right converges:
$$\rho(T)=\lim_{n\to\infty}||T^n||^{1/n}$$

In one sense, we can use the polynomial calculus formula $\sigma(T^n)=\sigma(T)^n$. Indeed, this gives the following estimate, valid for any $n$, as desired:
\begin{eqnarray*}
\rho(T)
&=&\sup_{\lambda\in\sigma(T)}|\lambda|\\
&=&\sup_{\rho\in\sigma(T)^n}|\rho|^{1/n}\\
&=&\sup_{\rho\in\sigma(T^n)}|\rho|^{1/n}\\
&=&\rho(T^n)^{1/n}\\
&\leq&||T^n||^{1/n}
\end{eqnarray*}

(4) For the reverse inequality, we fix a number $\rho>\rho(T)$, and we want to prove that we have $\rho\geq\lim_{n\to\infty}||T^n||^{1/n}$. By using the Cauchy formula, we have:
\begin{eqnarray*}
\frac{1}{2\pi i}\int_{|z|=\rho}\frac{z^n}{z-T}\,dz
&=&\frac{1}{2\pi i}\int_{|z|=\rho}\sum_{k=0}^\infty z^{n-k-1}T^k\,dz\\
&=&\sum_{k=0}^\infty\frac{1}{2\pi i}\left(\int_{|z|=\rho}z^{n-k-1}dz\right)T^k\\
&=&\sum_{k=0}^\infty\delta_{n,k+1}T^k\\
&=&T^{n-1}
\end{eqnarray*}

By applying the norm we obtain from this formula the following estimate:
$$||T^{n-1}||
\leq\frac{1}{2\pi}\int_{|z|=\rho}\left|\left|\frac{z^n}{z-T}\right|\right|\,dz
\leq\rho^n\cdot\sup_{|z|=\rho}\left|\left|\frac{1}{z-T}\right|\right|$$

Since the sup does not depend on $n$, by taking $n$-th roots, we obtain in the limit:
$$\rho\geq\lim_{n\to\infty}||T^n||^{1/n}$$

Now recall that $\rho$ was by definition an arbitrary number satisfying $\rho>\rho(T)$. Thus, we have obtained the following estimate, valid for any $T\in B(H)$:
$$\rho(T)\geq\lim_{n\to\infty}||T^n||^{1/n}$$

Thus, we are led to the conclusion in the statement.
\end{proof}

In the case of the normal elements, we have the following finer result:

\index{normal operator}

\begin{theorem}
The spectral radius of a normal element,
$$TT^*=T^*T$$
is equal to its norm.
\end{theorem}

\begin{proof}
We can proceed in two steps, as follows:

\medskip

\underline{Step 1}. In the case $T=T^*$ we have $||T^n||=||T||^n$ for any exponent of the form $n=2^k$, by using the formula $||TT^*||=||T||^2$, and by taking $n$-th roots we get:
$$\rho(T)\geq||T||$$

Thus, we are done with the self-adjoint case, with the result $\rho(T)=||T||$.

\medskip

\underline{Step 2}. In the general normal case $TT^*=T^*T$ we have $T^n(T^n)^*=(TT^*)^n$, and by using this, along with the result from Step 1, applied to $TT^*$, we obtain:
\begin{eqnarray*}
\rho(T)
&=&\lim_{n\to\infty}||T^n||^{1/n}\\
&=&\sqrt{\lim_{n\to\infty}||T^n(T^n)^*||^{1/n}}\\
&=&\sqrt{\lim_{n\to\infty}||(TT^*)^n||^{1/n}}\\
&=&\sqrt{\rho(TT^*)}\\
&=&\sqrt{||T||^2}\\
&=&||T||
\end{eqnarray*}

Thus, we are led to the conclusion in the statement.
\end{proof}

\section*{8c. Normal operators}

By using Theorem 8.14 we can say a number of non-trivial things about the normal operators, commonly known as ``spectral theorem for normal operators''. We first have:

\begin{theorem}
Given $T\in B(H)$ normal, we have a morphism of algebras
$$\mathbb C[X]\to B(H)\quad,\quad 
P\to P(T)$$
having the properties $||P(T)||=||P_{|\sigma(T)}||$, and $\sigma(P(T))=P(\sigma(T))$.
\end{theorem}

\begin{proof}
This is an improvement of Theorem 8.10 for polynomials, in the normal case, with the extra assertion being the norm estimate. But the element $P(T)$ being normal, we can apply to it the spectral radius formula for normal elements, and we obtain:
\begin{eqnarray*}
||P(T)||
&=&\rho(P(T))\\
&=&\sup_{\lambda\in\sigma(P(T))}|\lambda|\\
&=&\sup_{\lambda\in P(\sigma(T))}|\lambda|\\
&=&||P_{|\sigma(T)}||
\end{eqnarray*}

Thus, we are led to the conclusions in the statement.
\end{proof}

At a more advanced level now, we have the following result:

\begin{theorem}
Given $T\in B(H)$ normal, we have a morphism of algebras
$$C(\sigma(T))\to B(H)\quad,\quad 
f\to f(T)$$
which is isometric, $||f(T)||=||f||$, and has the property $\sigma(f(T))=f(\sigma(T))$.
\end{theorem}

\begin{proof}
The idea here is to ``complete'' the morphism in Theorem 8.15. Indeed, by Stone-Weierstrass, that morphism has a unique isometric extension, as follows:
$$C(\sigma(T))\to B(H)\quad,\quad  
f\to f(T)$$

It remains to prove $\sigma(f(T))=f(\sigma(T))$, and we can do this by double inclusion:

\medskip

``$\subset$'' Given a continuous function $f\in C(\sigma(T))$, we must prove that we have:
$$\lambda\notin f(\sigma(T))\implies\lambda\notin\sigma(f(T))$$

For this purpose, consider the following function, which is well-defined:
$$\frac{1}{f-\lambda}\in C(\sigma(T))$$

We can therefore apply this function to $T$, and we obtain:
$$\left(\frac{1}{f-\lambda}\right)T=\frac{1}{f(T)-\lambda}$$

In particular $f(T)-\lambda$ is invertible, so  $\lambda\notin\sigma(f(T))$, as desired.

\medskip

``$\supset$'' Given a continuous function $f\in C(\sigma(T))$, we must prove that we have: 
$$\lambda\in f(\sigma(T))\implies\lambda\in\sigma(f(T))$$

But this is the same as proving that we have:
$$\mu\in\sigma(T)\implies f(\mu)\in\sigma(f(T))$$

For this purpose, we approximate our function by polynomials, $P_n\to f$, and we examine the following convergence, which follows from $P_n\to f$:
$$P_n(T)-P_n(\mu)\to f(T)-f(\mu)$$

We know from polynomial functional calculus that we have:
$$P_n(\mu)
\in P_n(\sigma(T))
=\sigma(P_n(T))$$

Thus, the operators $P_n(T)-P_n(\mu)$ are not invertible. On the other hand, we know that the set formed by the invertible operators is open, so its complement is closed. Thus the limit $f(T)-f(\mu)$ is not invertible either, and so $f(\mu)\in\sigma(f(T))$, as desired.
\end{proof}

Even more generally now, we have the following result:

\begin{theorem}
Given $T\in B(H)$ normal, we have a morphism of algebras as follows, with $L^\infty$ standing for abstract measurable functions, or Borel functions,
$$L^\infty(\sigma(T))\to B(H)\quad,\quad 
f\to f(T)$$
which is isometric, $||f(T)||=||f||$, and has the property $\sigma(f(T))=f(\sigma(T))$.
\end{theorem}

\begin{proof}
As before, the idea will be that of ``completing'' what we have:

\medskip

(1) Given a vector $x\in H$, consider the following functional:
$$C(\sigma(T))\to\mathbb C\quad,\quad 
g\to<g(T)x,x>$$

By the Riesz theorem, this functional must be the integration with respect to a certain measure $\mu$ on the space $\sigma(T)$. Thus, we have a formula as follows:
$$<g(T)x,x>=\int_{\sigma(T)}g(z)d\mu(z)$$

Now given an arbitrary Borel function $f\in L^\infty(\sigma(T))$, as in the statement, we can define a number $<f(T)x,x>\in\mathbb C$, by using exactly the same formula, namely:
$$<f(T)x,x>=\int_{\sigma(T)}f(z)d\mu(z)$$

Thus, we have managed to define numbers $<f(T)x,x>\in\mathbb C$, for all vectors $x\in H$, and in addition we can recover these numbers as follows, with $g_n\in C(\sigma(T))$:
$$<f(T)x,x>=\lim_{g_n\to f}<g_n(T)x,x>$$ 

(2) In order to define now numbers $<f(T)x,y>\in\mathbb C$, for all vectors $x,y\in H$, we can use a polarization trick. Indeed, for any operator $S\in B(H)$ we have:
$$<S(x+y),x+y>=<Sx,x>+<Sy,y>+<Sx,y>+<Sy,x>$$

By replacing $y\to iy$, we have as well the following formula:
$$<S(x+iy),x+iy>=<Sx,x>+<Sy,y>-i<Sx,y>+i<Sy,x>$$

By multiplying this formula by $i$, and summing with the first one, we obtain:
\begin{eqnarray*}
<S(x+y),x+y>+i<S(x+iy),x+iy>
&=&(1+i)[<Sx,x>+<Sy,y>]\\
&+&2<Sx,y>
\end{eqnarray*}

(3) But with this, we can finish. Indeed, by combining (1,2), given a Borel function $f\in L^\infty(\sigma(T))$, we can define numbers $<f(T)x,y>\in\mathbb C$ for any $x,y\in H$, and we obtain in this way a certain operator $f(T)\in B(H)$, having all the desired properties.
\end{proof}

Good news, we can now diagonalize the normal operators. Let us start with:

\begin{theorem}
Any self-adjoint operator $T\in B(H)$ can be diagonalized,
$$T=U^*M_fU$$
with $U:H\to L^2(X)$ being a unitary operator from $H$ to a certain $L^2$ space associated to $T$, with $f:X\to\mathbb R$ being a certain function, once again associated to $T$, and with
$$M_f(g)=fg$$
being the usual multiplication operator by $f$, on the Hilbert space $L^2(X)$.
\end{theorem}

\begin{proof}
The construction of $U,f$ can be done in several steps, as follows:

\medskip

(1) We first prove the result in the special case where our operator $T$ has a cyclic vector $x\in H$, with this meaning that the following holds:
$$\overline{span\left(T^kx\Big|n\in\mathbb N\right)}=H$$

For this purpose, let us go back to the proof of Theorem 8.17. We will use the following formula from there, with $\mu$ being the measure on $X=\sigma(T)$ associated to $x$:
$$<g(T)x,x>=\int_{\sigma(T)}g(z)d\mu(z)$$

Our claim is that we can define a unitary $U:H\to L^2(X)$, first on the dense part spanned by the vectors $T^kx$, by the following formula, and then by continuity:
$$U[g(T)x]=g$$

Indeed, the following computation shows that $U$ is well-defined, and isometric:
\begin{eqnarray*}
||g(T)x||^2
&=&<g(T)x,g(T)x>\\
&=&<g(T)^*g(T)x,x>\\
&=&<|g|^2(T)x,x>\\
&=&\int_{\sigma(T)}|g(z)|^2d\mu(z)\\
&=&||g||_2^2
\end{eqnarray*}

We can then extend $U$ by continuity into a unitary $U:H\to L^2(X)$, as claimed. Now observe that we have the following formula:
\begin{eqnarray*}
UTU^*g
&=&U[Tg(T)x]\\
&=&U[(zg)(T)x]\\
&=&zg
\end{eqnarray*} 

Thus our result is proved in the present case, with $U$ as above, and with $f(z)=z$.

\medskip

(2) We discuss now the general case. Our first claim is that $H$ has a decomposition as follows, with each $H_i$ being invariant under $T$, and admitting a cyclic vector $x_i$:
$$H=\bigoplus_iH_i$$

Indeed, this is something elementary, the construction being by recurrence in finite dimensions, in the obvious way, and by using the Zorn lemma in general. Now with this decomposition in hand, we can make a direct sum of the diagonalizations obtained in (1), for each of the restrictions $T_{|H_i}$, and we obtain the formula in the statement.
\end{proof}

We have the following technical generalization of the above result:

\begin{theorem}
Any family of commuting self-adjoint operators $T_i\in B(H)$ can be jointly diagonalized,
$$T_i=U^*M_{f_i}U$$
with $U:H\to L^2(X)$ being a unitary operator from $H$ to a certain $L^2$ space associated to $\{T_i\}$, with $f_i:X\to\mathbb R$ being certain functions, once again associated to $T_i$, and with
$$M_{f_i}(g)=f_ig$$
being the usual multiplication operator by $f_i$, on the Hilbert space $L^2(X)$.
\end{theorem}

\begin{proof}
This is similar to the proof of Theorem 8.18, by suitably modifying the measurable calculus formula, and the measure $\mu$ itself, as to have this formula working for all the operators $T_i$. With this modification done, everything extends.
\end{proof}

We can now discuss the case of the arbitrary normal operators, as follows:

\begin{theorem}
Any normal operator $T\in B(H)$ can be diagonalized,
$$T=U^*M_fU$$
with $U:H\to L^2(X)$ being a unitary operator from $H$ to a certain $L^2$ space associated to $T$, with $f:X\to\mathbb C$ being a certain function, once again associated to $T$, and with
$$M_f(g)=fg$$
being the usual multiplication operator by $f$, on the Hilbert space $L^2(X)$.
\end{theorem}

\begin{proof}
Consider the decomposition of $T$ into its real and imaginary parts:
$$T=\frac{T+T^*}{2}+i\cdot\frac{T-T^*}{2i}$$

We know that the real and imaginary parts are self-adjoint operators. Now since $T$ was assumed to be normal, $TT^*=T^*T$, these real and imaginary parts commute:
$$\left[\frac{T+T^*}{2}\,,\,\frac{T-T^*}{2i}\right]=0$$

Thus Theorem 8.19 applies to these real and imaginary parts, and gives the result.
\end{proof}

Getting now to applications, the above results are quite powerful, and many things can be said, in analogy with what we know about usual matrices. Let us record here:

\index{polar decomposition}
\index{partial isometry}

\begin{theorem}
Any bounded operator $T\in B(H)$ can be decomposed as
$$T=U|T|$$
with $U$ being a partial isometry, and with $|T|=\sqrt{T^*T}$.
\end{theorem}

\begin{proof}
The operator $T^*T$ being self-adjoint, and even positive, in the sense that we have $<T^*Tx,x>\geq0$ for any $x\in H$, we can extract its square root $|T|=\sqrt{T^*T}$, by using the spectral theorem. Now observe that we have the following formula:
\begin{eqnarray*}
<|T|x,|T|y>
&=&<x,|T|^2y>\\
&=&<x,T^*Ty>\\
&=&<Tx,Ty>
\end{eqnarray*}

We conclude that the following linear application is well-defined, and isometric:
$$U:Im|T|\to Im(T)\quad,\quad 
|T|x\to Tx$$

Now by continuity we can extend this isometry $U$ into an isometry between certain Hilbert subspaces of $H$, as follows:
$$U:\overline{Im|T|}\to\overline{Im(T)}\quad,\quad 
|T|x\to Tx$$

Moreover, we can further extend $U$ into a partial isometry $U:H\to H$, by setting $Ux=0$, for any $x\in\overline{Im|T|}^\perp$, and with this convention, the result follows. 
\end{proof}

\section*{8d. Compact operators} 

We will restrict now the attention to the compact operators, which share many properties with the usual matrices. Let us start with a basic definition, as follows:

\index{finite rank operator}

\begin{definition}
An operator $T\in B(H)$ is said to be of finite rank if its image 
$$Im(T)\subset H$$ 
is finite dimensional. The set of such operators is denoted $F(H)$.
\end{definition}

There are many interesting examples of finite rank operators, the most basic ones being the finite rank projections, on the finite dimensional subspaces $K\subset H$. We have:

\begin{proposition}
The set of finite rank operators
$$F(H)\subset B(H)$$
is a two-sided $*$-ideal.
\end{proposition}

\begin{proof}
It is clear that $F(H)$ is a vector space. Let us prove now that $F(H)$ is stable under $*$. Given $T\in F(H)$, we can regard it as an invertible operator, as follows:
$$T:(\ker T)^\perp\to Im(T)$$

We conclude that we have the following dimension equality:
$$\dim((\ker T)^\perp)=\dim(Im(T))$$

On the other hand, we have equalities as follows, which give the result:
\begin{eqnarray*}
\dim(Im(T^*))
&=&\dim(\overline{Im(T^*)})\\
&=&\dim((\ker T)^\perp)\\
&=&\dim(Im(T))
\end{eqnarray*}

Regarding now the ideal property, this follows from the following two formulae, valid for any $S,T\in B(H)$, which are once again clear from definitions:
$$\dim(Im(ST))\leq\dim(Im(T))$$
$$\dim(Im(TS))\leq\dim(Im(T))$$

Thus, we are led to the conclusion in the statement.
\end{proof}

Let us discuss now the compact operators. These are introduced as follows:

\index{compact operator}
\index{unit ball}

\begin{definition}
An operator $T\in B(H)$ is said to be compact if the closed set
$$\overline{T(B_1)}\subset H$$
is compact, where $B_1\subset H$ is the unit ball. The set of such operators is denoted $K(H)$.
\end{definition}

In finite dimensions any operator is compact. In general, as a first observation, any finite rank operator is compact. We have in fact the following result:

\begin{proposition}
Any finite rank operator is compact,
$$F(H)\subset K(H)$$
and the finite rank operators are dense inside the compact operators.
\end{proposition}

\begin{proof}
The first assertion is clear, because if $Im(T)$ is finite dimensional, then the following subset is closed and bounded, and so it is compact:
$$\overline{T(B_1)}\subset Im(T)$$

Regarding the second assertion, let us pick a compact operator $T\in K(H)$, and a number $\varepsilon>0$. By compactness of $T$ we can find a finite set $S\subset B_1$ such that:
$$T(B_1)\subset\bigcup_{x\in S}B_\varepsilon(Tx)$$

Consider now the orthogonal projection $P$ onto the following finite dimensional space:
$$E=span\left(Tx\Big|x\in S\right)$$

Since the set $S$ is finite, this space $E$ is finite dimensional, and so $P$ is of finite rank, $P\in F(H)$. Now observe that for any norm one $y\in H$ and any $x\in S$ we have:
\begin{eqnarray*}
||Ty-Tx||^2
&=&||Ty-PTx||^2\\
&=&||Ty-PTy+PTy-PTx||^2\\
&=&||Ty-PTy||^2+||PTx-PTy||^2
\end{eqnarray*}

Now by picking $x\in S$ such that the ball $B_\varepsilon(Tx)$ covers the point $Ty$, we conclude from this that we have the following estimate:
$$||Ty-PTy||\leq||Ty-Tx||\leq\varepsilon$$ 

Thus we have $||T-PT||\leq\varepsilon$, which gives the density result.
\end{proof}

Quite remarkably, the set of compact operators is closed, and we have:

\begin{theorem}
The set of compact operators
$$K(H)\subset B(H)$$
is a closed two-sided $*$-ideal.
\end{theorem}

\begin{proof}
We have several assertions here, the idea being as follows:

\medskip

(1) It is clear that $K(H)$ is a vector space. In order to prove now that $K(H)$ is closed, assume that $T_n\in K(H)$ converges to $T\in B(H)$. Given $\varepsilon>0$, pick $N\in\mathbb N$ such that:
$$||T-T_N||\leq\varepsilon$$

By compactness of $T_N$ we can find a finite set $S\subset B_1$ such that:
$$T_N(B_1)\subset\bigcup_{x\in S}B_\varepsilon(T_Nx)$$

We conclude that for any $y\in B_1$ there exists $x\in S$ such that:
\begin{eqnarray*}
||Ty-Tx||
&\leq&||Ty-T_Ny||+||T_Ny-T_Nx||+||T_Nx-Tx||\\
&\leq&\varepsilon+\varepsilon+\varepsilon\\
&=&3\varepsilon
\end{eqnarray*}

Thus, we have an inclusion as follows, showing that $T$ is indeed compact:
$$T(B_1)\subset\bigcup_{x\in S}B_{3\varepsilon}(Tx)$$

(2) Regarding now the fact that $K(H)$ is stable under involution, this follows from Proposition 8.23, Proposition 8.25 and (1). Indeed, by using Proposition 8.25, given $T\in K(H)$ we can write it as a limit of finite rank operators, as follows:
$$T=\lim_{n\to\infty}T_n$$

Now by applying the adjoint, we have as well $T^*=\lim_{n\to\infty}T_n^*$. We know from Proposition 8.23 that the operators $T_n^*$ are of finite rank, and so compact by Proposition 8.25, and by using (1) we obtain that $T^*$ is compact too, as desired.

\medskip

(3) Finally, regarding the ideal property of $K(H)$, this is clear from definitions.
\end{proof}

Here is now a second key result regarding the compact operators:

\begin{theorem}
A bounded operator $T\in B(H)$ is compact precisely when
$$Te_n\to0$$
for any orthonormal system $\{e_n\}\subset H$.
\end{theorem}

\begin{proof}
We have two implications to be proved, the idea being as follows:

\medskip

``$\implies$'' Assume that $T$ is compact. By contradiction, assume $Te_n\not\to0$. This means that there exists $\varepsilon>0$ and a subsequence satisfying $||Te_{n_k}||>\varepsilon$, and by replacing $\{e_n\}$ with this subsequence, we can assume that the following holds, with $\varepsilon>0$:
$$||Te_n||>\varepsilon$$

Since $T$ is compact, a certain subsequence $\{Te_{n_k}\}$ must converge. Thus, by replacing once again $\{e_n\}$ with a subsequence, we can assume that we have, with $x\neq0$:
$$Te_n\to x$$

But this is a contradiction, as desired, because we obtain in this way:
\begin{eqnarray*}
<x,x>
&=&\lim_{n\to\infty}<Te_n,x>\\
&=&\lim_{n\to\infty}<e_n,T^*x>\\
&=&0
\end{eqnarray*}

``$\Longleftarrow$'' Assume $Te_n\to0$, for any orthonormal system $\{e_n\}\subset H$. In order to prove $T\in K(H)$, we will prove that $T$ is in the closure of the space of finite rank operators:
$$T\in\overline{F(H)}$$

We do this by contradiction. So, assume that there exists $\varepsilon>0$ such that:
$$S\in F(H)\implies||T-S||>\varepsilon$$

As a first observation, by using $S=0$ we obtain $||T||>\varepsilon$. Thus, we can find a norm one vector $e_1\in H$ such that the following holds:
$$||Te_1||>\varepsilon$$

Our claim, which will bring the desired contradiction, is that we can construct by recurrence vectors $e_1,\ldots,e_n$ such that the following holds, for any $i$: 
$$||Te_i||>\varepsilon$$

Indeed, assume that we have constructed $e_1,\ldots,e_n$. Let $E\subset H$ be the linear space spanned by these vectors, and set $P=Proj(E)$. Since $TP$ has finite rank, our assumption above shows that we have $||T-TP||>\varepsilon$. Thus, we can find $x\in H$ such that:
$$||(T-TP)x||>\varepsilon$$

We have then $x\not\in E$, and so we can consider the following nonzero vector:
$$y=(1-P)x$$

With this nonzero vector $y$ constructed, now let us set:
$$e_{n+1}=\frac{y}{||y||}$$

This vector $e_{n+1}$ is then orthogonal to $E$, has norm one, and satisfies:
$$||Te_{n+1}||\geq||y||^{-1}\varepsilon\geq\varepsilon$$

Thus we are done with our construction by recurrence, and this contradicts our assumption that $Te_n\to0$, for any orthonormal system $\{e_n\}\subset H$, as desired.
\end{proof}

Let us discuss now the spectral theory of the compact operators. We first have:

\begin{proposition}
Assuming that $T\in B(H)$, with $\dim H=\infty$, is compact and self-adjoint, the following happen:
\begin{enumerate}
\item The eigenvalues of $T$ form a sequence $\lambda_n\to0$.

\item All eigenvalues $\lambda_n\neq0$ have finite multiplicity.
\end{enumerate}
\end{proposition}

\begin{proof}
We prove both the assertions at the same time. For this purpose, we fix a number $\varepsilon>0$, we consider all the eigenvalues satisfying $|\lambda|\geq\varepsilon$, and for each such eigenvalue we consider the corresponding eigenspace $E_\lambda\subset H$. Let us set:
$$E=span\left(E_\lambda\,\Big|\,|\lambda|\geq\varepsilon\right)$$ 

Our claim, which will prove both (1) and (2), is that this space $E$ is finite dimensional. In now to prove now this claim, we can proceed as follows:

\medskip

(1) We know that we have $E\subset Im(T)$. Our claim is that we have:
$$\bar{E}\subset Im(T)$$

Indeed, assume that we have a sequence $g_n\in E$ which converges, $g_n\to g\in\bar{E}$. Let us write $g_n=Tf_n$, with $f_n\in H$. By definition of $E$, the following condition is satisfied:
$$h\in E\implies||Th||\geq\varepsilon||h||$$

Now since the sequence $\{g_n\}$ is Cauchy we obtain from this that the sequence $\{f_n\}$ is Cauchy as well, and with $f_n\to f$ we have $Tf_n\to Tf$, as desired.

\medskip

(2) Consider now the projection $P\in B(H)$ onto the above space $\bar{E}$. The composition $PT$ is then as follows, surjective on its target:
$$PT:H\to\bar{E}$$

On the other hand since $T$ is compact so must be $PT$, and if follows from this that the space $\bar{E}$ is finite dimensional. Thus $E$ itself must be finite dimensional too, and as explained in the beginning of the proof, this gives (1) and (2), as desired.
\end{proof}

In order to construct now eigenvalues, we will need:

\begin{proposition}
If $T$ is compact and self-adjoint, one of the numbers 
$$||T||\ ,\ -||T||$$
must be an eigenvalue of $T$.
\end{proposition}

\begin{proof}
We know from the spectral theory of the self-adjoint operators that the spectral radius $||T||$ of our operator $T$ is attained, and so one of the numbers $||T||,-||T||$ must be in the spectrum. In order to prove now that one of these numbers must actually appear as an eigenvalue, we must use the compactness of $T$, as follows:

\medskip

(1) First, we can assume $||T||=1$. By functional calculus this implies $||T^3||=1$ too, and so we can find a sequence of norm one vectors $x_n\in H$ such that:
$$|<T^3x_n,x_n>|\to1$$

By using our assumption $T=T^*$, we can rewrite this formula as follows:
$$|<T^2x_n,Tx_n>|\to1$$

Now since $T$ is compact, and $\{x_n\}$ is bounded, we can assume, up to changing the sequence $\{x_n\}$ to one of its subsequences, that the sequence $Tx_n$ converges:
$$Tx_n\to y$$

Thus, the convergence formula found above reformulates as follows, with $y\neq0$:
$$|<Ty,y>|=1$$

(2) Our claim now, which will finish the proof, is that this latter formula implies $Ty=\pm y$. Indeed, by using Cauchy-Schwarz and $||T||=1$, we have:
$$|<Ty,y>|\leq||Ty||\cdot||y||\leq1$$

We know that this must be an equality, so $Ty,y$ must be proportional. But since $T$ is self-adjoint the proportionality factor must be $\pm1$, and so we obtain, as claimed:
$$Ty=\pm y$$

Thus, we have constructed an eigenvector for $\lambda=\pm1$, as desired.
\end{proof}

We can further build on the above results in the following way:

\index{compact and self-adjoint}
\index{self-adjoint and compact}

\begin{proposition}
If $T$ is compact and self-adjoint, there is an orthogonal basis of $H$ made of eigenvectors of $T$.
\end{proposition}

\begin{proof}
We use Proposition 8.28. According to the results there, we can arrange the nonzero eigenvalues of $T$, taken with multiplicities, into a sequence $\lambda_n\to0$. Let $y_n\in H$ be the corresponding eigenvectors, and consider the following space:
$$E=\overline{span(y_n)}$$

The result follows then from the following observations:

\medskip

(1) Since we have $T=T^*$, both $E$ and its orthogonal $E^\perp$ are invariant under $T$. 

\medskip

(2) On the space $E$, our operator $T$ is by definition diagonal.

\medskip

(3) On the space $E^\perp$, our claim is that we have $T=0$. Indeed, assuming that the restriction $S=T_{E^\perp}$ is nonzero, we can apply Proposition 8.29 to this restriction, and we obtain an eigenvalue for $S$, and so for $T$, contradicting the maximality of $E$.
\end{proof}

With the above results in hand, we can now formulate a first theorem, as follows:

\index{basis of eigenvectors}

\begin{theorem}
Assuming that $T\in B(H)$, with $\dim H=\infty$, is compact and self-adjoint, the following happen:
\begin{enumerate}
\item The spectrum $\sigma(T)\subset\mathbb R$ consists of a sequence $\lambda_n\to0$.

\item All spectral values $\lambda\in\sigma(T)-\{0\}$ are eigenvalues.

\item All eigenvalues $\lambda\in\sigma(T)-\{0\}$ have finite multiplicity.

\item There is an orthogonal basis of $H$ made of eigenvectors of $T$.
\end{enumerate}
\end{theorem}

\begin{proof}
In view of our various results above, it is enough to establish (2). So, assume that $\lambda\neq0$ belongs to the spectrum $\sigma(T)$, but is not an eigenvalue. By using Proposition 8.30, pick an orthonormal basis $\{e_n\}$ of $H$ consisting of eigenvectors of $T$, and set:
$$Sx=\sum_n\frac{<x,e_n>}{\lambda_n-\lambda}\,e_n$$

Then $S$ is an inverse for $T-\lambda$, and so we have $\lambda\notin\sigma(T)$, as desired.
\end{proof}

Finally, we have the following result, regarding the general case:

\begin{theorem}
The compact operators $T\in B(H)$, with $\dim H=\infty$, are the operators of the following form, with $\{e_n\}$, $\{f_n\}$ being orthonormal families, and with $\lambda_n\searrow0$:
$$T(x)=\sum_n\lambda_n<x,e_n>f_n$$
The numbers $\lambda_n$, called singular values of $T$, are the eigenvalues of $|T|$. In fact, the polar decomposition of $T$ is given by $T=U|T|$, with
$$|T|(x)=\sum_n\lambda_n<x,e_n>e_n$$
and with $U$ being given by $Ue_n=f_n$, and $U=0$ on the complement of $span(e_i)$.
\end{theorem}

\begin{proof}
This basically follows from Theorem 8.31, as follows:

\medskip

(1) Given two orthonormal families $\{e_n\}$, $\{f_n\}$, and a sequence of real numbers $\lambda_n\searrow0$, consider the linear operator given by the formula in the statement, namely:
$$T(x)=\sum_n\lambda_n<x,e_n>f_n$$

Our first claim is that $T$ is bounded. Indeed, when assuming $|\lambda_n|\leq\varepsilon$ for any $n$, which is something that we can do if we want to prove that $T$ is bounded, we have:
\begin{eqnarray*}
||T(x)||^2
&=&\left|\sum_n\lambda_n<x,e_n>f_n\right|^2\\
&=&\sum_n|\lambda_n|^2|<x,e_n>|^2\\
&\leq&\varepsilon^2\sum_n|<x,e_n>|^2\\
&\leq&\varepsilon^2||x||^2
\end{eqnarray*}

(2) The next observation is that this operator is indeed compact, because it appears as the norm limit, $T_N\to T$, of the following sequence of finite rank operators:
$$T_N=\sum_{n\leq N}\lambda_n<x,e_n>f_n$$

(3) Regarding now the polar decomposition assertion, for the above operator, this follows once again from definitions. Indeed, the adjoint is given by:
$$T^*(x)=\sum_n\lambda_n<x,f_n>e_n$$

Thus, when composing $T^*$ with $T$, we obtain the following operator:
$$T^*T(x)=\sum_n\lambda_n^2<x,e_n>e_n$$

Now by extracting the square root, we obtain the formula in the statement, namely:
$$|T|(x)=\sum_n\lambda_n<x,e_n>e_n$$

(4) Conversely now, assume that $T\in B(H)$ is compact. Then $T^*T$, which is self-adjoint, must be compact as well, and so by Theorem 8.31 we have a formula as follows, with $\{e_n\}$ being a certain orthonormal family, and with $\lambda_n\searrow0$:
$$T^*T(x)=\sum_n\lambda_n^2<x,e_n>e_n$$

By extracting the square root we obtain the formula of $|T|$ in the statement, and then by setting $U(e_n)=f_n$ we obtain a second orthonormal family, $\{f_n\}$, such that:
$$T(x)
=U|T|
=\sum_n\lambda_n<x,e_n>f_n$$

Thus, our compact operator $T\in B(H)$ appears indeed as in the statement.
\end{proof}

\section*{8e. Exercises}

This was a quite straightforward chapter, and as exercises, we have:

\begin{exercise}
In relation with separability, learn about orthogonal polynomials.
\end{exercise}

\begin{exercise}
Clarify the details in the proof of the measurable calculus theorem.
\end{exercise}

\begin{exercise}
Jointly diagonalize the families of commuting normal operators.
\end{exercise}

\begin{exercise}
Check the positivity details, in the proof of the polar decomposition.
\end{exercise}

\begin{exercise}
Learn as well about the strictly positive operators, $T>0$.
\end{exercise}

\begin{exercise}
Work out other decomposition results, for the linear operators.
\end{exercise}

\begin{exercise}
Learn about the trace class operators, and their properties.
\end{exercise}

\begin{exercise}
Learn as well about Hilbert-Schmidt operators, and their properties.
\end{exercise}

As bonus exercise, in addition to this, learn some operator algebras as well.

\part{Positive matrices}

\ \vskip50mm

\begin{center}
{\em She'll carry on through it all

She's a waterfall

She'll carry on through it all

She's a waterfall}
\end{center}

\chapter{Positive matrices}

\section*{9a. Jacobian, Hessian}

Welcome to positivity. We have certainly met positive matrices in the above, but that material was quite theoretical, and time now to have a closer look at positivity, which is what makes this world go round, via various subtle mathematical mechanisms.

\bigskip

Our claim is that positivity is closely related to geometry. Indeed, assume that in your algebra, geometry or analysis calculations you came upon a quantity as follows:
$$f(x)=\sum_{ij}A_{ij}x_ix_j$$

Then, assuming $A=A^*$, we can change the basis as for $A\in M_N(\mathbb C)$ to become diagonal, and the quantity above takes the following form, with $\lambda_i\in\mathbb R$:
$$f(x)=\sum_i\lambda_ix_i^2$$

And, it is quite clear that for each eigenvalue $\lambda_i$, the fact that we have $\lambda_i<0$, or $\lambda_i=0$, or $\lambda_i>0$, matters a lot. Thus, we are into matrix positivity, which moreover is of clear geometric flavor, because in order to understand $f$, nothing better than looking first at the surface $f(x)=0$, whose shape will certainly depend on the signs of the $\lambda_i$.

\bigskip

Quite interesting all this, hope you agree with me. In practice now, many potential things to be done, and as a first objective, forgetting about the above, which might be something a bit specialized, say for later, and going straight to the basics, which are the good old multivariable calculus, let us try to answer the following question:

\begin{question}
What are the matrices $A$ appearing in relation with the functions
$$f:\mathbb R^N\to\mathbb R^M$$
and how do the positivity properties of $A$ translate into geometric properties of $f$? 
\end{question}

So, this will be our plan, reviewing the basic multivariable calculus, with linear algebra and positivity ideas in mind, then developing more linear algebra and positivity theory if needed, in order to reach to applications, and finally, once this done, getting into more specialized questions, say regarding the surfaces of type $\sum_i\lambda_ix_i^2=0$, and their shape. 

\bigskip

Let us start with something that we know well from chapter 1, namely:

\index{partial derivatives}
\index{continuously differentiable}

\begin{theorem}
A function $f:\mathbb R^N\to\mathbb R^M$ is continuously differentiable,
$$f(x+t)\simeq f(x)+f'(x)t$$
with $f'(x)$ linear, and $x\to f'(x)$ continuous, precisely when it has partial derivatives,
$$\frac{df_i}{dx_j}(x)=\lim_{t\to 0}\frac{f_i(x+te_j)-f_i(x)}{t}$$
which depend continuously on $x$. In this case the derivative is
$$f'(x)=\left(\frac{df_i}{dx_j}(x)\right)_{ij}\in M_{M\times N}(\mathbb R)$$ 
acting on the vectors $t\in\mathbb R^N$ by usual multiplication.
\end{theorem}

\begin{proof}
The formula in the statement makes sense indeed, as follows:
$$f\begin{pmatrix}x_1+t_1\\ \vdots\\ x_N+t_N\end{pmatrix}
\simeq f\begin{pmatrix}x_1\\ \vdots\\ x_N\end{pmatrix}
+\begin{pmatrix}
\frac{df_1}{dx_1}(x)&\ldots&\frac{df_1}{dx_N}(x)\\
\vdots&&\vdots\\
\frac{df_M}{dx_1}(x)&\ldots&\frac{df_M}{dx_N}(x)
\end{pmatrix}\begin{pmatrix}t_1\\ \vdots\\ t_N\end{pmatrix}$$

As for the proof, at $N=M=1$ what we have is a usual 1-variable function $f:\mathbb R\to\mathbb R$, and the formula in the statement is something that we know well, namely:
$$f(x+t)\simeq f(x)+f'(x)t$$

But this gives the result at any $N,M\in\mathbb N$, by performing a standard recurrence on $N\in\mathbb N$, and a trivial recurrence on $M\in\mathbb N$, as explained in chapter 1.
\end{proof}

Summarizing, and coming as good news, we do have matrices in multivariable calculus. As a second result now, showing that these matrices are indeed useful, we have:

\index{chain rule}

\begin{theorem}
We have the chain derivative formula
$$(f\circ g)'(x)=f'(g(x))\cdot g'(x)$$
as an equality of matrices.
\end{theorem}

\begin{proof}
This is something that we know too from chapter 1, coming from:
\begin{eqnarray*}
(f\circ g)(x+t)
&=&f(g(x+t))\\
&\simeq&f(g(x)+g'(x)t)\\
&\simeq&f(g(x))+f'(g(x))g'(x)t
\end{eqnarray*}

Thus, we are led to the formula in the statement.
\end{proof}

Still at order 1, as a third result now, which is something new, making appear a subtle linear algebra notion, namely the determinant, we have, in relation with integration:

\index{change of variables}
\index{Jacobian}
\index{matrix determinant}
\index{volume inflation}

\begin{theorem}
Given a transformation $\varphi=(\varphi_1,\ldots,\varphi_N)$, we have
$$\int_Ef(x)dx=\int_{\varphi^{-1}(E)}f(\varphi(t))|J_\varphi(t)|dt$$
with the $J_\varphi$ quantity, called Jacobian, being given by
$$J_\varphi(t)=\det\left[\left(\frac{d\varphi_i}{dx_j}(x)\right)_{ij}\right]$$ 
and with this generalizing the usual formula from one variable calculus.
\end{theorem}

\begin{proof}
This is something quite tricky, the idea being as follows:

\medskip

(1) Observe first that the above formula generalizes indeed the change of variable formula in 1 dimension, the point here being that the absolute value on the derivative appears as to compensate for the lack of explicit bounds for the integral.

\medskip 

(2) As a second observation, we can assume if we want, by linearity, that we are dealing with the constant function $f=1$. For this function, our formula reads:
$$vol(E)=\int_{\varphi^{-1}(E)}|J_\varphi(t)|dt$$

In terms of $D={\varphi^{-1}(E)}$, this amounts in proving that we have:
$$vol(\varphi(D))=\int_D|J_\varphi(t)|dt$$

And here, as a first remark, our formula is clear for the linear maps $\varphi$, by using the definition of the determinant of real matrices, from chapter 2, as a signed volume.

\medskip

(3) However, the extension of this to the case of non-linear maps $\varphi$ is something non-trivial, so we will not follow this path. In order to prove now the result, as stated, our first claim is that the validity of the theorem is stable under taking compositions of transformations $\varphi$. In order to prove this claim, consider a composition, as follows:
$$\varphi:E\to F\quad,\quad 
\psi:D\to E\quad,\quad 
\varphi\circ\psi:D\to F$$

Assuming that the theorem holds for $\varphi,\psi$, we deduce that we have, as desired:
\begin{eqnarray*}
\int_Ff(x)dx
&=&\int_Ef(\varphi(s))|J_\varphi(s)|ds\\
&=&\int_Df(\varphi\circ\psi(t))|J_\varphi(\psi(t))|\cdot|J_\psi(t)|dt\\
&=&\int_Df(\varphi\circ\psi(t))|J_{\varphi\circ\psi}(t)|dt
\end{eqnarray*}

(4) Next, as a key ingredient, let us examine the case where we are in $N=2$ dimensions, and our transformation $\varphi$ has one of the following special forms:
$$\varphi(x,y)=(\psi(x,y),y)\quad,\quad\varphi(x,y)=(x,\psi(x,y))$$

By symmetry, it is enough to deal with the first case. Here the Jacobian is $d\psi/dx$, and by replacing if needed $\psi\to-\psi$, we can assume that this Jacobian is positive, $d\psi/dx>0$. Now by assuming as before that $D=\varphi^{-1}(E)$ is a rectangle, $D=[a,b]\times[c,d]$, we can prove our formula by using the change of variables in 1 dimension, as follows:
\begin{eqnarray*}
\int_Ef(s)ds
&=&\int_{\varphi(D)}f(x,y)dxdy\\
&=&\int_c^d\int_{\psi(a,y)}^{\psi(b,y)}f(x,y)dxdy\\
&=&\int_c^d\int_a^bf(\psi(x,y),y)\frac{d\psi}{dx}\,dxdy\\
&=&\int_Df(\varphi(t))J_\varphi(t)dt
\end{eqnarray*}

(5) But with this, we can now prove the theorem, in $N=2$ dimensions. Indeed, given a transformation $\varphi=(\varphi_1,\varphi_2)$, consider the following two transformations:
$$\phi(x,y)=(\varphi_1(x,y),y)\quad,\quad \psi(x,y)=(x,\varphi_2\circ\phi^{-1}(x,y))$$

We have then $\varphi=\psi\circ\phi$, and by using (4) for $\psi,\phi$, which are of the special form there, and then (3) for composing, we conclude that the theorem holds indeed for $\varphi$, as desired. Thus, theorem proved in $N=2$ dimensions, and the extension of the above proof to arbitrary $N$ dimensions is straightforward, that we will leave here as an exercise.
\end{proof}

At the level of the main applications, in 2 dimensions, we have:

\index{polar coordinates}

\begin{proposition}
We have polar coordinates in $2$ dimensions,
$$\begin{cases}
x\!\!\!&=\ r\cos t\\
y\!\!\!&=\ r\sin t
\end{cases}$$
the corresponding Jacobian being $J=r$.
\end{proposition}

\begin{proof}
This is indeed elementary, with the Jacobian being as follows:
$$J
=\begin{vmatrix}
\cos t&-r\sin t\\
\sin t&r\cos t
\end{vmatrix}=r$$

Thus, we are led to the conclusions in the statement.
\end{proof}

We can now compute the Gauss integral, which is the best calculus formula ever:

\index{Gauss integral}

\begin{theorem}
We have the following formula,
$$\int_\mathbb Re^{-x^2}dx=\sqrt{\pi}$$
called Gauss integral formula.
\end{theorem}

\begin{proof}
Let $I$ be the above integral. By using polar coordinates, we obtain:
\begin{eqnarray*}
I^2
&=&\int_\mathbb R\int_\mathbb Re^{-x^2-y^2}dxdy\\
&=&\int_0^{2\pi}\int_0^\infty e^{-r^2}rdrdt\\
&=&2\pi\int_0^\infty\left(-\frac{e^{-r^2}}{2}\right)'dr\\
&=&\pi
\end{eqnarray*}

Thus, we are led to the formula in the statement.
\end{proof}

Very nice all this, and as a conclusion, for doing some probability theory, we need Theorem 9.6, which needs Theorem 9.4, which needs linear algebra. Good to know.

\bigskip

In relation with Question 9.1, however, we are not yet into matrix positivity. For reaching to that, we will have to upgrade Theorem 9.2 to order 2. Let us start with:

\index{higher derivative}
\index{Clairaut formula}

\begin{theorem}
Given $f:\mathbb R^N\to\mathbb R$, we can talk about its higher derivatives
$$\frac{d^kf}{dx_{i_1}\ldots dx_{i_k}}=\frac{d}{dx_{i_1}}\cdots\frac{d}{dx_{i_k}}(f)$$
provided that these derivatives exist indeed. Moreover, due to the Clairaut formula,
$$\frac{d^2f}{dx_idx_j}=\frac{d^2f}{dx_jdx_i}$$
the order in which these higher derivatives are computed is irrelevant.
\end{theorem}

\begin{proof}
This is self-explanatory, based on the Clairaut formula, which itself is something elementary, coming from the mean value theorem. Observe that, in practice, we can permute the order of our partial derivative computations, and a standard way of doing this is by differentiating first with respect to $x_1$, then with respect to $x_2$, and so on:
$$\frac{d^kf}{dx_1^{k_1}\ldots dx_N^{k_N}}=\frac{d^{k_1}}{dx_1^{k_1}}\cdots\frac{d^{k_N}}{dx_N^{k_N}}(f)$$

To be more precise, here $k\in\mathbb N$ is as usual the global order of our derivatives, the exponents $k_1,\ldots,k_N\in\mathbb N$ are subject to the condition $k_1+\ldots+k_N=k$, and the operations on the right are the familiar one-variable higher derivative operations.
\end{proof}

Regarding now the Taylor formula, in several variables, at order 2, we have:

\index{local minimum}
\index{local maximum}
\index{Hessian matrix}

\begin{theorem}
Given a function $f:\mathbb R^N\to\mathbb R$, construct its Hessian, as being:
$$f''(x)=\left(\frac{d^2f}{dx_idx_j}(x)\right)_{ij}$$ 
We have then the following order $2$ approximation of $f$ around a given $x\in\mathbb R^N$,
$$f(x+t)\simeq f(x)+f'(x)t+\frac{<f''(x)t,t>}{2}$$
relating the positivity properties of $f''$ to the local minima and maxima of $f$.
\end{theorem}

\begin{proof}
This is something that we know from chapter 1, the idea being to look at the restriction of $f$ to the segment $I=[x,x+t]$. To be more precise, let $y\in\mathbb R^N$, and set:
$$g(r)=f(x+ry)$$

As explained in chapter 1, by using the chain rule, we have the following formulae:
$$g(0)=f(x)\quad,\quad 
g'(0)=f'(x)\quad,\quad 
g''(0)=<f''(x)y,y>$$

But with this data in hand, the usual Taylor formula for our one variable function $g$, at order 2, at the point $r=0$, takes the following form, with $t=ry$:
\begin{eqnarray*}
f(x+ry)
&\simeq&f(x)+f'(x)ry+\frac{<f''(x)y,y>r^2}{2}\\
&=&f(x)+f'(x)t+\frac{<f''(x)t,t>}{2}
\end{eqnarray*}

Thus, we have obtained the formula in the statement. Finally, the last assertion, regarding the local extrema, is something standard, as in the one-variable case.
\end{proof}

As a complement to Theorem 9.8, very useful in practice, let us record:

\index{Hessian eigenvalues}

\begin{theorem}
Given a twice differentiable function $f:\mathbb R^N\to\mathbb R$, assume that $f'(x)=0$, and let $\lambda_1,\ldots,\lambda_N$ be the eigenvalues of $f''(x)$. Then:
\begin{enumerate}
\item $\lambda_i\geq0$ is needed for $x$ to be a local minimum.

\item $\lambda_i>0$ guarantees that $x$ is a local minimum.

\item $\lambda_i\leq0$ is needed for $x$ to be a local maximum.

\item $\lambda_i<0$ guarantees that $x$ is a local maximum.
\end{enumerate}
\end{theorem}

\begin{proof}
We know from chapter 3 that the Hessian matrix $f''(x)$, which is symmetric, is diagonalized by a certain matrix $U\in O_N$. But with this in hand, we can change the basis of $\mathbb R^N$, with the help of this matrix $U\in O_N$, and the Taylor formula becomes:
$$f(x+t)\simeq f(x)+\sum_{i=1}^N\lambda_it_i^2$$

And this latter formula, obviously, gives all the assertions in the statement.
\end{proof}

Finally, no discussion about second derivatives would be complete without a word on the Laplace operator. We first have the following result, which is a bit heuristic:

\begin{proposition}
Given a function $f:\mathbb R^N\to\mathbb R$, its Laplacian, given by
$$\Delta f=\sum_{i=1}^N\frac{d^2f}{dx_i^2}$$
measures how much different is $f(x)$, compared to the average of $f(y)$, with $y\simeq x$.
\end{proposition}

\begin{proof}
As mentioned, this is something a bit heuristic, but which is good to know. Let us write the Taylor formula at order 2, as such, and with $t\to-t$ too:
$$f(x+t)\simeq f(x)+f'(x)t+\frac{<f''(x)t,t>}{2}$$
$$f(x-t)\simeq f(x)-f'(x)t+\frac{<f''(x)t,t>}{2}$$

By making the average, we obtain the following formula:
$$\frac{f(x+t)+f(x-t)}{2}\simeq f(x)+\frac{<f''(x)t,t>}{2}$$

Thus, thinking a bit, we are led to the conclusion in the statement, modulo some discussion about integrating all this, that we will not really need, in what follows.
\end{proof}

Summarizing, the quantity $\Delta f$ appears as the correct numeric analogue, in higher dimensions, of the second derivative $f''$. With this understood, the problem is now, what can we say about the mathematics of $\Delta$? Inspired by linear algebra, we have: 

\begin{question}
The Laplace operator being linear,
$$\Delta(af+bg)=a\Delta f+b\Delta g$$
what can we say about it, inspired by usual linear algebra?
\end{question}

In answer now, the space of functions $f:\mathbb R^N\to\mathbb R$, on which $\Delta$ acts, being infinite dimensional, the usual tools from linear algebra do not apply as such, and we must be extremely careful. For instance, we cannot really expect to diagonalize $\Delta$, via some sort of explicit procedure, as we usually do in linear algebra, for the usual matrices.

\bigskip

Thinking some more, there is actually a real bug too with our problem, because at $N=1$ this problem becomes ``what can we say about the second derivatives $f'':\mathbb R\to\mathbb R$ of the functions $f:\mathbb R\to\mathbb R$, inspired by linear algebra'', with answer ``not much''.

\bigskip

And by thinking even more, still at $N=1$, there is a second bug too, because if a function $f:\mathbb R\to\mathbb R$ is twice differentiable, nothing will guarantee that its second derivative $f:\mathbb R\to\mathbb R$ is twice differentiable too. Thus, we have some issues with the domain and range of $\Delta$, regarded as linear operator, and these problems will persist at higher $N$.

\bigskip

So, shall we trash Question 9.11? Not so quick, because, very remarkably, some magic comes at $N=2$ and higher in relation with complex analysis, according to:

\begin{principle}
The functions $f:\mathbb R^N\to\mathbb R$ which are $0$-eigenvectors of $\Delta$,
$$\Delta f=0$$
called harmonic functions, have the following properties:
\begin{enumerate}
\item At $N=1$, nothing spectacular, these are just the linear functions.

\item At $N=2$, these are, locally, the real parts of holomorphic functions.

\item At $N\geq 3$, these still share many properties with the holomorphic functions. 
\end{enumerate}
\end{principle}

In order to understand this principle, which is something quite deep, or at least get introduced to it, let us first look at the case $N=2$. Here, any function $f:\mathbb R^2\to\mathbb R$ can be regarded as function $f:\mathbb C\to\mathbb R$, depending on the following variable:
$$z=x+iy$$

But, in view of this, it is natural to enlarge the attention to the functions $f:\mathbb C\to\mathbb C$, and ask which of these functions are harmonic, $\Delta f=0$. And here, we have the following remarkable result, making the link with complex analysis:

\begin{theorem}
Any holomorphic function $f:\mathbb C\to\mathbb C$, when regarded as function
$$f:\mathbb R^2\to\mathbb C$$
is harmonic. Moreover, the conjugates $\bar{f}$ of holomorphic functions are harmonic too.
\end{theorem}

\begin{proof}
The first assertion comes from the following computation, with $z=x+iy$:
\begin{eqnarray*}
\Delta z^n
&=&\frac{d^2z^n}{dx^2}+\frac{d^2z^n}{dy^2}\\
&=&\frac{d(nz^{n-1})}{dx}+\frac{d(inz^{n-1})}{dy}\\
&=&n(n-1)z^{n-2}-n(n-1)z^{n-2}\\
&=&0
\end{eqnarray*}

As for the second assertion, this follows from $\Delta\bar{f}=\overline{\Delta f}$, which is clear from definitions, and which shows that if $f$ is harmonic, then so is its conjugate $\bar{f}$.
\end{proof}

Many more things can be said, along these lines, notably with a proof of the assertion (2) in Principle 9.12, which is however a quite tough piece of mathematics, and then with a clarification of (3) too, from that same principle, which again requires some substantial mathematics. For more on all this, you can check for instance Rudin \cite{ru2}.

\section*{9b. Positive matrices}

Motivated by the above, and more specifically by Theorems 9.8 and 9.9, which are at the core of multivariable calculus, let us have a closer look at the positivity properties of the matrices $A\in M_N(\mathbb R)$, and more generally, of the matrices $A\in M_N(\mathbb C)$. 

\bigskip

To start with, as explained in chapter 3, the basics of positivity are as follows:

\begin{theorem}
For a matrix $A\in M_N(\mathbb C)$ the following conditions are equivalent, and if they are satisfied, we say that $A$ is positive, and write $A\geq0$:
\begin{enumerate}
\item $A=B^2$, with $B=B^*$.

\item $A=CC^*$, for some $C\in M_N(\mathbb C)$.

\item $<Ax,x>\geq0$, for any vector $x\in\mathbb C^N$.

\item $A=A^*$, and the eigenvalues are positive, $\lambda_i\geq0$.

\item $A=UDU^*$, with $U\in U_N$ and with $D\in M_N(\mathbb R_+)$ diagonal.
\end{enumerate}
\end{theorem}

\begin{proof}
We know this from chapter 3, the  idea being as follows:

\medskip

$(1)\implies(2)$ This is clear, because we can take $C=B$.

\medskip

$(2)\implies(3)$ This comes from $<CC^*x,x>=<C^*x,C^*x>\geq0$.

\medskip

$(3)\implies(4)$ From $<Ax,x>\in\mathbb R$ we get $<Ax,x>=<x,A^*x>=<A^*x,x>$, which gives $A=A^*$. Also, assuming $Ax=\lambda x$ we have $<Ax,x>=\lambda<x,x>$, so $\lambda\geq0$.

\medskip

$(4)\implies(5)$ This comes from the spectral theorem for the self-adjoint matrices.

\medskip

$(5)\implies(1)$ If we set $B=U\sqrt{D}U^*$ then $B=B^*$ and $B^2=A$, as desired.
\end{proof}

Also as explained in chapter 3, we have the following version of the above result:

\index{strictly positive matrix}

\begin{theorem}
For a matrix $A\in M_N(\mathbb C)$ the following conditions are equivalent, and if they are satisfied, we say that $A$ is strictly positive, and write $A>0$:
\begin{enumerate}
\item $A=B^2$, with $B=B^*$, invertible.

\item $A=CC^*$, for some $C\in M_N(\mathbb C)$ invertible.

\item $<Ax,x>>0$, for any nonzero vector $x\in\mathbb C^N$.

\item $A=A^*$, and the eigenvalues are strictly positive, $\lambda_i>0$.

\item $A=UDU^*$, with $U\in U_N$ and with $D\in M_N(\mathbb R_+^*)$ diagonal.
\end{enumerate}
\end{theorem}

\begin{proof}
This follows either from Theorem 9.14, by adding the above various extra assumptions, or from the proof of Theorem 9.14, by modifying where needed.
\end{proof}

Getting now to the linear algebra theory developed afterwards, in chapters 5-6, the Jordan decomposition and the singular value decomposition are not of great use, because the spectral theorem for positive matrices, Theorem 9.14 (5), obviously beats everything. However, we can say a few interesting things in relation with the exponentiation of matrices from chapter 5, and with the functional calculus from chapter 6, as follows:

\begin{theorem}
Given a positive matrix $A\in M_N(\mathbb C)$, the following happen:
\begin{enumerate}
\item We can apply any measurable function $f:\mathbb R_+\to\mathbb C$ to this matrix, and in terms of the diagonalization $A=UDU^*$, with $U\in U_N$, we have $f(A)=Uf(D)U^*$.

\item The above procedure $f\to f(A)$ is unique. In particular, with $f(x)=\sqrt{x}$, our matrix has a unique square root $\sqrt{A}$, which is positive, with square $A$.

\item For the basic examples of positive matrices, for projections we have $\sqrt{P}=P$, and for the positive exponentials, $e^A$ with $A=A^*$, we have $\sqrt{e^A}=e^{A/2}$.
\end{enumerate}
\end{theorem}

\begin{proof}
Many things can be said here, the idea being as follows:

\medskip

(1) This is indeed something that we know from chapter 6, with the explicit construction of $f(A)$ being as follows, $\lambda_1,\ldots,\lambda_N\geq0$ being the eigenvalues of $A$:
$$A=U\begin{pmatrix}
\lambda_1\\ 
&\ddots\\
&&\lambda_N
\end{pmatrix}U^*
\ \implies\ 
f(A)=U\begin{pmatrix}
f(\lambda_1)\\ 
&\ddots\\
&&f(\lambda_N)
\end{pmatrix}U^*$$

(2) This is again something that we know from chapter 6, and with the uniqueness of the square root being discussed in detail in the beginning of chapter 6.

\medskip

(3) For the projections we clearly have $\sqrt{P}=P$. Regarding now the exponentials, from $e^A=\sum_kA^k/k!$ we obtain $(e^A)^*=e^{A^*}$, so when $A=A^*$ we have $e^A=(e^A)^*$. Moreover, when assuming $A=A^*$, the eigenvalues of $e^A$ will be $e^{\lambda_1},\ldots,e^{\lambda_N}>0$, where $\lambda_1,\ldots,\lambda_N\in\mathbb R$ are the eigenvalues of $A$, so we have $e^A>0$ in this case. Finally, still under the assumption $A=A^*$, the formula $\sqrt{e^A}=e^{A/2}$ is clear from definitions.
\end{proof}

What is next? Still motivated by Theorems 9.8 and 9.9, let us have a closer look at low dimensions. And here, at $N=2$ we have the following useful result:

\begin{proposition}
For a $2\times2$ complex matrix
$$A=\begin{pmatrix}
a&b\\ c&d\end{pmatrix}$$
with real eigenvalues $\lambda_1,\lambda_2$, the following conditions are equivalent:
\begin{enumerate}
\item The eigenvalues are positive, $\lambda_1,\lambda_2\geq0$.

\item The trace and determinant are positive, $Tr(A),\det A\geq0$.
\end{enumerate}
\end{proposition}

\begin{proof}
We know that the eigenvalues $\lambda_1,\lambda_2$ are subject to:
$$\lambda_1+\lambda_2=Tr(A)\quad,\quad 
\lambda_1\lambda_2=\det A$$

But this gives the result, by using the following elementary equivalence:
$$\begin{cases}x\geq0\\ y\geq0\end{cases}\iff
\begin{cases}x+y\geq0\\ xy\geq0\end{cases}$$

Thus, we are led to the conclusion in the statement.
\end{proof}

Summarizing, we know how to decide if a matrix $A\in M_2(\mathbb C)$ is positive or not, without explicitly computing its eigenvalues. Quite remarkably, the same happens in 3D:

\begin{theorem}
For a $3\times3$ complex matrix
$$A=\begin{pmatrix}
a&b&c\\ d&e&f\\ g&h&i\end{pmatrix}$$
with real eigenvalues $\lambda_1,\lambda_2,\lambda_3$, the following conditions are equivalent:
\begin{enumerate}
\item The eigenvalues are positive, $\lambda_1,\lambda_2,\lambda_3\geq0$.

\item We have $Tr(A),S(A),\det A\geq0$, where $S(A)=|^{ab}_{de}|+|^{ac}_{gi}|+|^{ef}_{hi}|$.
\end{enumerate}
\end{theorem}

\begin{proof}
There are several things going on here, the idea being as follows:

\medskip

(1) To start with, the characteristic polynomial of the $3\times3$ matrices is given by the following formula, obtained by identifying the coefficients of $1,x,x^2,x^3$:
\begin{eqnarray*}
\begin{vmatrix}
a-x&b&c\\ d&e-x&f\\ g&h&i-x\end{vmatrix}
&=&-x^3+x^2(a+e+i)\\
&&-x\left(\begin{vmatrix}a&b\\ d&e\end{vmatrix}
+\begin{vmatrix}a&c\\ g&i\end{vmatrix}
+\begin{vmatrix}e&f\\ h&i\end{vmatrix}\right)\\
&&+\begin{vmatrix}
a&b&c\\ d&e&f\\ g&h&i\end{vmatrix}
\end{eqnarray*}

In other words, with the notations in the statement, we have:
$$\det(A-x)=-x^3+x^2Tr(A)-xS(A)+\det A$$

(2) On the other hand, in terms of the eigenvalues $\lambda_1,\lambda_2,\lambda_3$, we have:
\begin{eqnarray*}
\det(A-x)
&=&-(x-\lambda_1)(x-\lambda_2)(x-\lambda_3)\\
&=&-x^3+(\lambda_1+\lambda_2+\lambda_3)x^2-(\lambda_1\lambda_2+\lambda_1\lambda_3+\lambda_2\lambda_3)x+\lambda_1\lambda_2\lambda_3
\end{eqnarray*}

Thus, the eigenvalues $\lambda_1,\lambda_2,\lambda_3$ are subject to the following formulae:
$$\lambda_1+\lambda_2+\lambda_3=Tr(A)$$
$$\lambda_1\lambda_2+\lambda_1\lambda_3+\lambda_2\lambda_3=S(A)$$
$$\lambda_1\lambda_2\lambda_3=\det A$$

(3) Our claim now is that we have the following equivalence, valid for any three real numbers $x,y,z$, which will give the result:
$$\begin{cases}x\geq0\\ y\geq0\\ z\geq0\end{cases}\iff
\begin{cases}x+y+z\geq0\\ xy+xz+yz\geq0\\ xyz\geq0\end{cases}$$

Indeed, in one sense this is clear. In the other sense, assume that the conditions on the right are satisfied. We will prove by contradiction that $x,y,z\geq0$. Assuming that this is wrong, in view of $xyz\geq0$, and by symmetry, we can assume that:
$$x\geq 0\quad,\quad y\leq 0\quad,\quad z\leq0$$

Now in view of $x+y+z\geq0$, we can write $x,y,z$ as follows, with $a,b,c\geq0$:
$$x=a+b+c\quad,\quad y=-b\quad,\quad z=-c$$

But with this done, we have the following computation:
\begin{eqnarray*}
xy+xz+yz
&=&-(a+b+c)(b+c)+bc\\
&=&-a(b+c)-(b+c)^2+bc\\
&=&-a(b+c)-b^2-c^2-bc\\
&\leq&0
\end{eqnarray*}

Thus we have our contradiction, so claim proved, which gives the result.
\end{proof}

In higher dimensions things are more complicated, and we no longer have analogues of the above results. Here is however a basic positivity criterion, which can be useful:

\begin{theorem}
The Gram matrix of any system of vectors
$$A_{ij}=<v_i,v_j>$$
is positive, and any positive matrix appears in this way.
\end{theorem}

\begin{proof}
In one sense, let us denote by $B\in M_{N\times M}(\mathbb C)$ the rectangular matrix formed by the entries of the vectors $v_1,\ldots,v_M\in\mathbb C^N$, taken together:
$$B=\begin{bmatrix}v_1&\ldots&v_M\end{bmatrix}$$

We have then the following formula, for the Gram matrix in the statement:
\begin{eqnarray*}
A_{ij}
&=&<v_i,v_j>\\
&=&\sum_k(v_i)_k\overline{(v_j)_k}\\
&=&\sum_kB_{ki}\bar{B}_{kj}\\
&=&(B^t\bar{B})_{ij}
\end{eqnarray*}

Thus with $C=B^t$ we have the following formula, which gives the result:
$$A=CC^*\geq0$$ 

As for the converse, this comes from the standard fact, that we know from Theorem 9.14, that the positive matrices are precisely those of the form $A=CC^*$.
\end{proof}

\section*{9c. Sylvester theorem}

As a continuation of the above material, we would like to discuss now a useful criterion for positivity, which is both subtle and quick, due to Sylvester, as follows:

\begin{claim}[Sylvester]
An arbitrary self-adjoint square matrix
$$A=\begin{pmatrix}
a_{11}&\ldots&a_{1N}\\
\vdots&&\vdots\\
a_{N1}&\ldots&a_{NN}\end{pmatrix}$$
is positive precisely when the following quantities, called leading principal minors,
$$d_k=\begin{vmatrix}
a_{11}&\ldots&a_{1k}\\
\vdots&&\vdots\\
a_{k1}&\ldots&a_{kk}\end{vmatrix}$$
with $k\leq N$, are all positive.
\end{claim}

As an illustration for this, which is something non-trivial, let us work out the case $N=2$. So, consider an arbitrary $2\times2$ matrix, written as follows:
$$A=\begin{pmatrix}
a&b\\ c&d\end{pmatrix}$$

In this case, the Sylvester positivity criterion takes the following form:
$$a\geq 0\quad,\quad ad-bc\geq0$$

Thus, what we have here is of slightly different nature from what we previously had in Proposition 9.17, the point coming from the self-adjointness of $A$.

\bigskip

In order to prove now Claim 9.20, we will use a recurrence on $N\in\mathbb N$. Let us write our matrix as follows, with $B$ being a $(N-1)\times(N-1)$ matrix, $v\in\mathbb C^{N-1}$ and $x\in\mathbb R$:
$$A=\begin{pmatrix}
B&v\\
v^*&x\end{pmatrix}$$

Obviously, in order to do our recurrence work, we are in need of relating the determinants of $A$ and of $B$. And here, we can use the following standard fact:

\begin{theorem}
We have the following determinant formula,
$$\det\begin{pmatrix}A&B\\C&D\end{pmatrix}
=\det A\cdot \det(D-CA^{-1}B)$$
valid for any matrix decomposed into blocks.
\end{theorem}

\begin{proof}
This is something very standard, the idea being as follows:

\medskip

(1) As a first observation, the formula holds indeed for a usual $2\times2$ matrix:
\begin{eqnarray*}
\begin{vmatrix}
a&b\\ c&d\end{vmatrix}
&=&ad-bc\\
&=&a(d-a^{-1}bc)\\
&=&a(d-ca^{-1}b)
\end{eqnarray*}

(2) Next, the formula holds for the $3\times3$ matrices split as $2+1$. Indeed, we have:
$$\begin{vmatrix}
a&b&c\\ d&e&f\\ g&h&i\end{vmatrix}
=aei-bdi-ceg+bfg+cdh-afh$$

On the other hand, we have the following computation, yielding the same number:
\begin{eqnarray*}
&&\begin{vmatrix}
a&b\\ d&e\end{vmatrix}\cdot
\left[i-\begin{pmatrix}g&h\end{pmatrix}
\begin{pmatrix}
a&b\\ d&e\end{pmatrix}^{-1}
\begin{pmatrix}
c\\
f\end{pmatrix}\right]\\
&=&(ae-bd)\cdot
\left[i-\begin{pmatrix}g&h\end{pmatrix}
\frac{1}{ae-bd}
\begin{pmatrix}
e&-b\\ -d&a\end{pmatrix}
\begin{pmatrix}
c\\
f\end{pmatrix}\right]\\
&=&(ae-bd)i-\begin{pmatrix}g&h\end{pmatrix}
\begin{pmatrix}
e&-b\\ -d&a\end{pmatrix}
\begin{pmatrix}
c\\
f\end{pmatrix}\\
&=&(ae-bd)i-\begin{pmatrix}g&h\end{pmatrix}
\begin{pmatrix}
ce-bf\\ -cd+af\end{pmatrix}\\
&=&aei-bdi-ceg+bfg+cdh-afh
\end{eqnarray*}

(3) Next, the formula holds for the $3\times3$ matrices split as $1+2$. Indeed, consider the same matrix as before, with its determinant computed by Sarrus:
$$\begin{vmatrix}
a&b&c\\ d&e&f\\ g&h&i\end{vmatrix}
=aei-bdi-ceg+bfg+cdh-afh$$

On the other hand, we have the following computation, yielding the same number:
\begin{eqnarray*}
&&a\cdot\det\left[\begin{pmatrix}e&f\\h&i\end{pmatrix}
-\begin{pmatrix}d\\ g\end{pmatrix}a^{-1}
\begin{pmatrix}b&c\end{pmatrix}\right]\\
&=&a\cdot\det\left[\begin{pmatrix}e&f\\h&i\end{pmatrix}
-a^{-1}\begin{pmatrix}bd&cd\\
bg&cg\end{pmatrix}\right]\\
&=&a^{-1}\cdot\det\left[a\begin{pmatrix}e&f\\h&i\end{pmatrix}
-\begin{pmatrix}bd&cd\\
bg&cg\end{pmatrix}\right]\\
&=&a^{-1}\cdot\det\left[\begin{pmatrix}ae&af\\ah&ai\end{pmatrix}
-\begin{pmatrix}bd&cd\\
bg&cg\end{pmatrix}\right]\\
&=&a^{-1}\begin{vmatrix}ae-bd&af-cd\\ ah-bg&ai-cg\end{vmatrix}\\
&=&a^{-1}[(ae-bd)(ai-cg)-(ah-bg)(af-cd)]\\
&=&a^{-1}(a^2ei-aceg-abdi+bcdg-a^2fh+acdh+abfg-bcdg)\\
&=&a^{-1}(a^2ei-aceg-abdi-a^2fh+acdh+abfg)\\
&=&aei-ceg-bdi-afh+cdh+bfg
\end{eqnarray*}

(4) In general now, the above verifications show that a direct algebraic approach would be quite difficult. Fortunately, there is a clever solution to this, coming from:
\begin{eqnarray*}
\begin{pmatrix}1&0\\CA^{-1}&1\end{pmatrix}
\begin{pmatrix}A&0\\0&D-CA^{-1}B\end{pmatrix}
\begin{pmatrix}1&A^{-1}B\\0&1\end{pmatrix}
&=&\begin{pmatrix}A&0\\C&D-CA^{-1}B\end{pmatrix}
\begin{pmatrix}1&A^{-1}B\\0&1\end{pmatrix}\\
&=&\begin{pmatrix}A&B\\C&D\end{pmatrix}
\end{eqnarray*}

Now by applying the determinant on both sides, we obtain the result.
\end{proof}

Good news, we can now establish the Sylvester positivity criterion, as follows:

\begin{theorem}[Sylvester]
An arbitrary self-adjoint square matrix
$$A=\begin{pmatrix}
a_{11}&\ldots&a_{1N}\\
\vdots&&\vdots\\
a_{N1}&\ldots&a_{NN}\end{pmatrix}$$
is positive precisely when the following quantities, called leading principal minors,
$$d_k=\begin{vmatrix}
a_{11}&\ldots&a_{1k}\\
\vdots&&\vdots\\
a_{k1}&\ldots&a_{kk}\end{vmatrix}$$
with $k\leq N$, are all positive.
\end{theorem}

\begin{proof}
We can prove this by recurrence on $N\in\mathbb N$, as follows:

\medskip

(1) Let us write our matrix as follows, with $B$ being a certain $(N-1)\times(N-1)$ complex matrix, $v\in\mathbb C^{N-1}$ being a certain vector, and $x\in\mathbb R$:
$$A=\begin{pmatrix}
B&v\\
v^*&x\end{pmatrix}$$

We have then the following computation, for any $w\in\mathbb C^{N-1}$ and any $z\in\mathbb C$:
\begin{eqnarray*}
\left<A\begin{pmatrix}w\\ z\end{pmatrix},\begin{pmatrix}w\\ z\end{pmatrix}\right>
&=&\left<\begin{pmatrix}
B&v\\
v^*&x\end{pmatrix}
\begin{pmatrix}w\\ z\end{pmatrix},\begin{pmatrix}w\\ z\end{pmatrix}\right>\\
&=&\left<\begin{pmatrix}
Bw+zv\\
<w,v>+xz\end{pmatrix},\begin{pmatrix}w\\ z\end{pmatrix}\right>\\
&=&<Bw,w>+z<v,w>+\bar{z}<w,v>+x|z|^2\\
&=&<Bw,w>+2Re(z<v,w>)+x|z|^2
\end{eqnarray*}

(2) Now let us write $z=\alpha y$, with $y\geq 0$ and $|\alpha|=1$. Our formula becomes:
$$\left<A\begin{pmatrix}w\\ z\end{pmatrix},\begin{pmatrix}w\\ z\end{pmatrix}\right>
=<Bw,w>+2yRe(\alpha<v,w>)+xy^2$$

In order to prove that this quantity, which is a degree 2 polynomial in $y\in\mathbb R$, is positive, we can look at its discriminant $\Delta$. Indeed, we must have $\Delta\leq0$, so we must prove that, for any vector $w\in\mathbb C^{N-1}$, and any number $\alpha\in\mathbb T$, we have:
$$x<Bw,w>\ \geq\ Re(\alpha<v,w>)^2$$

Equivalently, we must prove that for any vector $w\in\mathbb C^{N-1}$, we have:
$$x<Bw,w>\ \geq\ |<v,w>|^2$$

(3) For this latter purpose, we can use the block determinant formula from Theorem 9.21. For our matrix $A$, decomposed as above, that block determinant formula reads:
\begin{eqnarray*}
\det A
&=&\det B\cdot (x-v^*B^{-1}v)\\
&=&\det B\cdot (x-<B^{-1}v,v>)
\end{eqnarray*}

Now since both $\det A,\det B$ are positive, we obtain the following inequality:
$$x\ \geq\ <B^{-1}v,v>$$

(4) As a second ingredient, we have as well the following elementary estimate, using the positivity of $B$, and then the Cauchy-Schwarz inequality, at the end:
\begin{eqnarray*}
<B^{-1}v,v><Bw,w>
&=&<\sqrt{B^{-1}}v,\sqrt{B^{-1}}v><\sqrt{B}w,\sqrt{B}w>\\
&=&||\sqrt{B^{-1}}v||^2||\sqrt{B}w||^2\\
&\geq&|<\sqrt{B^{-1}}v,\sqrt{B}w>|^2\\
&=&|<v,w>|^2
\end{eqnarray*}

(5) Now by putting (3) and (4) together, we obtain the following estimate:
\begin{eqnarray*}
x<Bw,w>
&\geq&<B^{-1}v,v><Bw,w>\\
&\geq&|<v,w>|^2
\end{eqnarray*}

Thus we have obtained the estimate needed in (2), as desired.
\end{proof}

As a conclusion to all this, in the context of the Hessian matrices coming from calculus, via Theorems 9.8 and 9.9, the positivity criteria from Proposition 9.17 and Theorem 9.18 apply at $N=2,3$, and the Sylvester criterion from Theorem 9.22 applies in general. 

\bigskip

All this is very useful in practice, when doing various calculus computations.

\section*{9d. Infinite matrices} 

Getting back to generalities, let us see as well what happens in infinite dimensions, as a continuation of our discussion about linear operators from chapter 8. As a main goal, we would like to have a generalization of Theorems 9.14 and 9.15, in this setting.

\bigskip

In order to do so, we need to know more about orthogonality in arbitrary Hilbert spaces. Let us start with something basic and elementary, as follows:

\begin{theorem}
Let $H$ be a Hilbert space, and $E\subset H$ be a closed subspace.
\begin{enumerate}
\item Given $x\in H$, we can find a unique $y\in E$, minimizing $||x-y||$.

\item With $x,y$ as above, we have $x=y+z$, for a certain $z\in E^\perp$.

\item Thus, we have a direct sum decomposition $H=E\oplus E^\perp$.

\item In terms of $H=E\oplus E^\perp$, the projection $x\to y$ is given by $P(x,y)=x$.
\end{enumerate}
\end{theorem}

\begin{proof}
This is something very standard, the idea being as follows:

\medskip

(1) Given $x\in H$ and two vectors $v,w\in E$, we have the following estimate:
\begin{eqnarray*}
||x-v||^2+||x-w||^2
&=&2\left(\left|\left|x-\frac{v+w}{2}\right|\right|^2+\left|\left|\frac{v-w}{2}\right|\right|^2\right)\\
&\geq&2d(x,E)^2+\frac{||v-w||^2}{2}
\end{eqnarray*}

But this shows that any sequence in $E$ realizing the inf in the definition of $d(x,E)$ is Cauchy, so it converges to a vector $y$. Since $E$ is closed we have $y\in E$, so $y$ realizes the inf. Moreover, again from the above inequality, such a $y$ realizing the inf is unique.

\medskip

(2) In order to prove $x-y\in E^\perp$, let $v\in E$ and choose $w\in\mathbb T$ such that $w<x-y,v>$ is a real number. For any $t\in\mathbb R$ we have the following equality:
$$||x-y+twv||=||x-y||^2+2tw<x-y,v>+t^2||v||^2$$

By construction of the vector $y$ we know that this function has a minimum at $t=0$. But this function is a degree 2 polynomial, so the middle term must vanish:
$$2w<x-y,v>=0$$

Now since this must hold for any $v\in E$, we must have $x-y\in E^\perp$, as desired.

\medskip

(3) This is consequence of what we found in (1,2).

\medskip

(4) This is also a consequence of what we found in (1,2).
\end{proof}

Many things can be said, as a continuation of the above, as for instance with:

\begin{theorem}
For a closed subspace $E\subset H$, we have:
$$E^{\perp\perp}=E$$
More generally, for an arbitrary linear subspace $E\subset H$, we have
$$E^{\perp\perp}=\bar{E}$$
and with the closing operation being needed, in infinite dimensions.
\end{theorem}

\begin{proof}
All this comes indeed as an elementary application of our orthogonal projection technology from Theorem 9.23, and we will leave the details here, which are all straightforward, with no extra difficulties involved, as an instructive exercise.
\end{proof}

In relation now with the linear operators, and their adjoints, we have:

\begin{theorem}
Given a bounded operator $T\in B(H)$, the following happen:
\begin{enumerate}
\item $\ker T^*=(Im T)^\perp$.

\item $\overline{Im T^*}=(\ker T)^\perp$.
\end{enumerate}
\end{theorem}

\begin{proof}
Both these assertions are elementary, as follows:

\medskip

(1) Let us first prove ``$\subset$''. Assuming $T^*x=0$, we have indeed $x\perp ImT$, because:
$$<x,Ty>
=<T^*x,y>
=0$$

As for ``$\supset$'', assuming $<x,Ty>=0$ for any $y$, we have $T^*x=0$, because:
$$<T^*x,y>
=<x,Ty>
=0$$

(2) This can be deduced from (1), applied to the operator $T^*$, as follows:
$$(\ker T)^\perp
=(Im T^*)^{\perp\perp}
=\overline{Im T^*}$$

Here we have used the formula $E^{\perp\perp}=\bar{E}$, valid for any linear subspace $E\subset H$ of a Hilbert space, which for $E$ closed reads $E^{\perp\perp}=E$, and comes from $H=E\oplus E^\perp$, and which in general follows from $E^{\perp\perp}\subset\bar{E}^{\perp\perp}=\bar{E}$, as explained in Theorem 9.24.
\end{proof}

Getting now to positivity matters, we have the following result:

\begin{theorem}
For an operator $T\in B(H)$, the following are equivalent:
\begin{enumerate}
\item $<Tx,x>\geq0$, for any $x\in H$.

\item $T$ is normal, and $\sigma(T)\subset[0,\infty)$.

\item $T=S^2$, for some $S\in B(H)$ satisfying $S=S^*$.

\item $T=R^*R$, for some $R\in B(H)$.
\end{enumerate}
If these conditions are satisfied, we call $T$ positive, and write $T\geq0$.
\end{theorem}

\begin{proof}
This is something very standard, the idea being as follows:

\medskip

$(1)\implies(2)$ Assuming $<Tx,x>\geq0$, with $S=T-T^*$ we have:
\begin{eqnarray*}
<Sx,x>
&=&<Tx,x>-<T^*x,x>\\
&=&<Tx,x>-<x,Tx>\\
&=&<Tx,x>-\overline{<Tx,x>}\\
&=&0
\end{eqnarray*}

The next step is to use a polarization trick, as follows:
\begin{eqnarray*}
<Sx,y>
&=&<S(x+y),x+y>-<Sx,x>-<Sy,y>-<Sy,x>\\
&=&-<Sy,x>\\
&=&<y,Sx>\\
&=&\overline{<Sx,y>}
\end{eqnarray*}

Thus we must have $<Sx,y>\in\mathbb R$, and with $y\to iy$ we obtain $<Sx,y>\in i\mathbb R$ too, and so $<Sx,y>=0$. Thus $S=0$, which gives $T=T^*$. Now since $T$ is self-adjoint, it is normal as claimed. Moreover, by self-adjointness, we have:
$$\sigma(T)\subset\mathbb R$$

In order to prove now that we have indeed $\sigma(T)\subset[0,\infty)$, as claimed, we must invert $T+\lambda$, for any $\lambda>0$. For this purpose, observe that we have:
\begin{eqnarray*}
<(T+\lambda)x,x>
&=&<Tx,x>+<\lambda x,x>\\
&\geq&<\lambda x,x>\\
&=&\lambda||x||^2
\end{eqnarray*}

But this shows that $T+\lambda$ is injective. In order to prove now the surjectivity, and the boundedness of the inverse, observe first that we have:
\begin{eqnarray*}
Im(T+\lambda)^\perp
&=&\ker(T+\lambda)^*\\
&=&\ker(T+\lambda)\\
&=&\{0\}
\end{eqnarray*}

Thus $Im(T+\lambda)$ is dense. On the other hand, observe that we have:
\begin{eqnarray*}
||(T+\lambda)x||^2
&=&<Tx+\lambda x,Tx+\lambda x>\\
&=&||Tx||^2+2\lambda<Tx,x>+\lambda^2||x||^2\\
&\geq&\lambda^2||x||^2
\end{eqnarray*}

Thus for any vector in the image $y\in Im(T+\lambda)$ we have:
$$||y||\geq\lambda\big|\big|(T+\lambda)^{-1}y\big|\big|$$

As a conclusion to what we have so far, $T+\lambda$ is bijective and invertible as a bounded operator from $H$ onto its image, with the following norm bound:
$$||(T+\lambda)^{-1}||\leq\lambda^{-1}$$

But this shows that $Im(T+\lambda)$ is complete, hence closed, and since we already knew that $Im(T+\lambda)$ is dense, our operator $T+\lambda$ is surjective, and we are done.

\medskip

$(2)\implies(3)$ Since $T$ is normal, and with spectrum contained in $[0,\infty)$, we can use the continuous functional calculus theory for the normal operators from chapter 8, with the function $f(x)=\sqrt{x}$, as to construct a square root $S=\sqrt{T}$. 

\medskip

$(3)\implies(4)$ This is trivial, because we can set $R=S$. 

\medskip

$(4)\implies(1)$ This is clear, because we have the following computation:
$$<R^*Rx,x>
=<Rx,Rx>
=||Rx||^2$$

Thus, we have the equivalences in the statement.
\end{proof}

It is possible to talk as well about strictly positive operators, and we have here:

\begin{theorem}
For an operator $T\in B(H)$, the following are equivalent:
\begin{enumerate}
\item $T$ is positive and invertible.

\item $T$ is normal, and $\sigma(T)\subset(0,\infty)$.

\item $T=S^2$, for some $S\in B(H)$ invertible, satisfying $S=S^*$.

\item $T=R^*R$, for some $R\in B(H)$ invertible.
\end{enumerate}
If these conditions are satisfied, we call $T$ strictly positive, and write $T>0$.
\end{theorem}

\begin{proof}
Our claim is that the above conditions (1-4) are precisely the conditions (1-4) in Theorem 9.26, with the assumption ``$T$ is invertible'' added. Indeed:

\medskip

(1) This is clear by definition.

\medskip

(2) In the context of Theorem 9.26 (2), namely when $T$ is normal, and $\sigma(T)\subset[0,\infty)$, the invertibility of $T$, which means $0\notin\sigma(T)$, gives $\sigma(T)\subset(0,\infty)$, as desired.

\medskip

(3) In the context of Theorem 9.26 (3), namely when $T=S^2$, with $S=S^*$, by using the basic properties of the functional calculus for normal operators, the invertibility of $T$ is equivalent to the invertibility of its square root $S=\sqrt{T}$, as desired.

\medskip

(4) In the context of Theorem 9.26 (4), namely when $T=RR^*$, the invertibility of $T$ is equivalent to the invertibility of $R$. This can be either checked directly, or deduced via the equivalence $(3)\iff(4)$ from Theorem 9.26, by using the above argument (3).
\end{proof}

As a subtlety now, we have the following complement to the above result:

\begin{theorem}
For a strictly positive operator, $T>0$, we have
$$<Tx,x>>0\quad,\quad\forall x\neq0$$
but the converse of this fact is not true, unless we are in finite dimensions.
\end{theorem}

\begin{proof}
We have several things to be proved, the idea being as follows:

\medskip

(1) Regarding the main assertion, the inequality can be deduced as follows, by using the fact that the operator $S=\sqrt{T}$ is invertible, and in particular injective:
\begin{eqnarray*}
<Tx,x>
&=&<S^2x,x>\\
&=&<Sx,S^*x>\\
&=&<Sx,Sx>\\
&=&||Sx||^2\\
&>&0
\end{eqnarray*}

(2) In finite dimensions, assuming $<Tx,x>>0$ for any $x\neq0$, we know from Theorem 9.26 that we have $T\geq0$. Thus we have $\sigma(T)\subset[0,\infty)$, and assuming by contradiction $0\in\sigma(T)$, we obtain that $T$ has $\lambda=0$ as eigenvalue, and the corresponding eigenvector $x\neq0$ has the property $<Tx,x>=0$, contradiction. Thus $T>0$, as claimed.

\medskip

(3) Regarding now the counterexample, consider the following operator on $l^2(\mathbb N)$:
$$T=\begin{pmatrix}
1\\
&\frac{1}{2}\\
&&\frac{1}{3}\\
&&&\ddots
\end{pmatrix}$$

This operator $T$ is well-defined and bounded, and we have $<Tx,x>>0$ for any $x\neq0$. However $T$ is not invertible, and so the converse does not hold, as stated.
\end{proof}

Summarizing, we have a suitable generalization of Theorems 9.14 and 9.15, in the infinite dimensional setting. This is of course just the beginning of the story, and for more advanced positivity theory in this setting, we recommend any operator theory book.

\section*{9e. Exercises}

This was a quite routine analysis chapter, and as exercises on this, we have:

\begin{exercise}
Check the details in the proof of the change of variable formula.
\end{exercise}

\begin{exercise}
Learn some other proofs of the change of variable formula.
\end{exercise}

\begin{exercise}
Compute the Gauss integral without polar coordinates. Can you?
\end{exercise}

\begin{exercise}
Compute the arbitrary polynomial integrals over the spheres.
\end{exercise}

\begin{exercise}
Work out the multivariable Taylor formula, at order $k\in\mathbb N$.
\end{exercise}

\begin{exercise}
Learn more, from physicists, about the Laplace operator.
\end{exercise}

\begin{exercise}
Learn about the harmonic functions, and their various properties.
\end{exercise}

\begin{exercise}
Make a list, from physics, of interesting positive matrices or operators.
\end{exercise}

As bonus exercise, read if needed Rudin \cite{ru1}, \cite{ru2}, or an equivalent text.

\chapter{Forms, signature}

\section*{10a. Bilinear forms}

It is good time now to talk about geometry. We already know about conics, from chapter 1. However, when getting to $\mathbb R^3$, we are right away into a dillema, because the plane curves have two generalizations. First we have the algebraic curves in $\mathbb R^3$:

\begin{definition}
An algebraic curve in $\mathbb R^3$ is a curve as follows,
$$C=\left\{(x,y,z)\in\mathbb R^3\Big|P(x,y,z)=0,\,Q(x,y,z)=0\right\}$$
appearing as the joint zeroes of two polynomials $P,Q$.
\end{definition}

These curves look of course like the usual plane curves, and at the level of the phenomena that can appear, these are similar to those in the plane, involving singularities and so on, but also knotting, which is a new phenomenon. However, it is hard to say something with bare hands about knots. I mean, just try, and you will understand.

\bigskip

On the other hand, as another natural generalization of the plane curves, and this might sound a bit surprising, we have the surfaces in $\mathbb R^3$, constructed as follows:

\begin{definition}
An algebraic surface in $\mathbb R^3$ is a surface as follows,
$$S=\left\{(x,y,z)\in\mathbb R^3\Big|P(x,y,z)=0\right\}$$
appearing as the zeroes of a polynomial $P$.
\end{definition}

The point indeed is that, as it was the case with the plane curves, what we have here is something defined by a single equation. And with respect to many questions, having a single equation matters a lot, and this is why surfaces in $\mathbb R^3$ are ``simpler'' than curves in $\mathbb R^3$. In fact, believe me, they are even the correct generalization of the curves in $\mathbb R^2$.

\bigskip

As an example of what can be done with surfaces, which is very similar to what we did with the conics $C\subset\mathbb R^2$ in chapter 1, we have the following result:

\begin{theorem}
The degree $2$ surfaces $S\subset\mathbb R^3$, called quadrics, are the ellipsoid
$$\left(\frac{x}{a}\right)^2+\left(\frac{y}{b}\right)^2+\left(\frac{z}{c}\right)^2=1$$
which is the only compact one, plus $16$ more, which can be explicitly listed.
\end{theorem}

\begin{proof}
We will be quite brief here, because we intend to rediscuss all this in a moment, with full details, in arbitrary $N$ dimensions, the idea being as follows:

\medskip

(1) The equations for a quadric $S\subset\mathbb R^2$ are best written as follows, with $A\in M_3(\mathbb R)$ being a matrix, $B\in M_{1\times 3}(\mathbb R)$ being a row vector, and $C\in\mathbb R$ being a constant:
$$<Au,u>+Bu+C=0$$

(2) By doing now the linear algebra, and we will come back to this in a moment, with details, or by invoking the theorem of Sylvester on quadratic forms, we are left, modulo degeneracy and linear transformations, with signed sums of squares, as follows:
$$\pm x^2\pm y^2\pm z^2=0,1$$

(3) Thus the sphere is the only compact quadric, up to linear transformations, and by applying now linear transformations to it, we are led to the ellipsoids in the statement.

\medskip

(4) As for the other quadrics, there are many of them, a bit similar to the parabolas and hyperbolas in 2 dimensions, and some work here leads to a 16 item list.
\end{proof}

With this done, instead of further insisting on the surfaces $S\subset\mathbb R^3$, or getting into their rivals, the curves $C\subset\mathbb R^3$, which appear as intersections of such surfaces, $C=S\cap S'$, let us get instead to arbitrary $N$ dimensions, see what the axiomatics looks like there, with the hope that this will clarify our dimensionality dillema, curves vs surfaces.

\bigskip

So, moving to $N$ dimensions, we have here the following definition, to start with:

\index{hypersurface}
\index{algebraic hypersurface}

\begin{definition}
An algebraic hypersurface in $\mathbb R^N$ is a space of the form
$$S=\left\{(x_1,\ldots,x_N)\in\mathbb R^N\Big|P(x_1,\ldots,x_N)=0,\forall i\right\}$$
appearing as the zeroes of a polynomial $P\in\mathbb R[x_1,\ldots,x_N]$.
\end{definition}

Again, this is a quite general definition, covering both the plane curves $C\subset\mathbb R$ and the surfaces $S\subset\mathbb R^2$, which is certainly worth a systematic exploration. But, no hurry with this, for the moment we are here for talking definitons and axiomatics.

\bigskip

In order to have now a full collection of beasts, in all possible dimensions $N\in\mathbb N$, and of all possible dimensions $k\in\mathbb N$, we must intersect such algebraic hypersurfaces. We are led in this way to the zeroes of families of polynomials, as follows:

\index{algebraic manifold}
\index{zero of polynomials}
\index{intesection of surfaces}

\begin{definition}
An algebraic manifold in $\mathbb R^N$ is a space of the form
$$X=\left\{(x_1,\ldots,x_N)\in\mathbb R^N\Big|P_i(x_1,\ldots,x_N)=0,\forall i\right\}$$
with $P_i\in\mathbb R[x_1,\ldots,x_N]$ being a family of polynomials.
\end{definition}

As a first observation, as already mentioned, such a manifold appears as an intersection of hypersurfaces $S_i$, those associated to the various polynomials $P_i$:
$$X=S_1\cap\ldots\cap S_r$$

There is actually a bit of a discussion needed here, regarding the parameter $r\in\mathbb N$, shall we allow this parameter to be $r=\infty$ too, or not. However, with some abstract commutative algebra helping, the idea is that allowing $r=\infty$ forces in fact $r<\infty$.

\bigskip

As an announcement now, good news, what we have in Definition 10.5 is the good and final notion of algebraic manifold, very general, and with the branch of mathematics studying such manifolds being called algebraic geometry. In what follows we will discuss a bit what can be done with this, as a continuation of our previous work on the plane curves, at the elementary level, by using our accumulated linear algebra knowledge.

\bigskip

Let us first look more in detail at the hypersurfaces. We have here:

\index{degree 2}
\index{quadric}
\index{sphere}
\index{Sylvester theorem}
\index{bilinear form}
\index{symmetric matrix}

\begin{theorem}
The degree $2$ hypersurfaces $S\subset\mathbb R^N$, called quadrics, are up to degeneracy and to linear transformations the hypersurfaces of the following form,
$$\pm x_1^2\pm\ldots\pm x_N^2=0,1$$
and with the sphere being the only compact one.
\end{theorem}

\begin{proof}
We have two statements here, the idea being as follows:

\medskip

(1) The equations for a quadric $S\subset\mathbb R^N$ are best written as follows, with $A\in M_N(\mathbb R)$ being a matrix, $B\in M_{1\times N}(\mathbb R)$ being a row vector, and $C\in\mathbb R$ being a constant:
$$<Ax,x>+Bx+C=0$$

(2) By doing the linear algebra, or by invoking the theorem of Sylvester on quadratic forms, we are left, modulo linear transformations, with signed sums of squares:
$$\pm x_1^2\pm\ldots\pm x_N^2=0,1$$

(3) To be more precise, with linear algebra, by evenly distributing the terms $x_ix_j$ above and below the diagonal, we can assume that our matrix $A\in M_N(\mathbb R)$ is symmetric. Thus $A$ must be diagonalizable, and by changing the basis of $\mathbb R^N$, as to have it diagonal, our equation becomes as follows, with $D\in M_N(\mathbb R)$ being now diagonal:
$$<Dx,x>+Ex+F=0$$

(4) But now, by making squares in the obvious way, which amounts in applying yet another linear transformation to our quadric, the equation takes the following form, with $G\in M_N(-1,0,1)$ being diagonal, and with $H\in\{0,1\}$ being a constant:
$$<Gx,x>=H$$

(5) Now barring the degenerate cases, we can further assume $G\in M_N(-1,1)$, and we are led in this way to the equation claimed in (2) above, namely:
$$\pm x_1^2\pm\ldots\pm x_N^2=0,1$$

(6) In particular we see that, up to some degenerate cases, namely emptyset and point, the only compact quadric, up to linear transformations, is the one given by:
$$x_1^2+\ldots+x_N^2=1$$

(7) But this is the unit sphere, so are led to the conclusions in the statement.
\end{proof}

Many other things can be said, as a continuation of this, and we can only recommend here learning some algebraic geometry, say from Harris \cite{har} or Shafarevich \cite{sha}.

\section*{10b. Smooth manifolds}

We already know about the algebraic curves, then surfaces and other algebraic manifolds, generalizing the conics, from the above. A second idea now, in order to generalize the conics, is to look at the smooth manifolds, in the following sense:

\begin{definition}
A smooth manifold is a space $X$ which is locally isomorphic to $\mathbb R^N$. To be more precise, this space $X$ must be covered by charts, bijectively mapping open pieces of it to open pieces of $\mathbb R^N$, with the changes of charts being $C^\infty$ functions.
\end{definition}

It is of course possible to talk as well about $C^k$ manifolds, with $k<\infty$, but this is rather technical material, that we will not get into, in this book.

\bigskip

As basic examples of smooth manifolds, we have of course $\mathbb R^N$ itself, or any open subset $X\subset\mathbb R^N$, with only 1 chart being needed here. Other basic examples include the circle, or curves like ellipses and so on, for obvious reasons. To be more precise, the unit circle can be covered by 2 charts as above, by using polar coordinates, in the obvious way, and then by applying dilations, translations and other such transformations, namely bijections which are smooth, we obtain a whole menagery of circle-looking manifolds.  

\bigskip

Here is a more precise statement in this sense, covering the conics:

\index{deformation of manifold}

\begin{theorem}
The following are smooth manifolds, in the plane:
\begin{enumerate}
\item The circles.

\item The ellipses.

\item The non-degenerate conics.

\item Smooth deformations of these.
\end{enumerate}
\end{theorem}

\begin{proof}
All this is quite intuitive, the idea being as follows:

\medskip

(1) Consider the unit circle, $x^2+y^2=1$. We can write then $x=\cos t$, $y=\sin t$, with $t\in[0,2\pi)$, and we seem to have here the solution to our problem, just using 1 chart. But this is of course wrong, because $[0,2\pi)$ is not open, and we have a problem at $0$. In practice we need to use 2 such charts, say with the first one being with $t\in(0,3\pi/2)$, and the second one being with $t\in(\pi,5\pi/2)$. As for the fact that the change of charts is indeed smooth, this comes by writing down the formulae, or just thinking a bit, and arguing that this change of chart being actually a translation, it is automatically linear.

\medskip

(2) This follows from (1), by pulling the circle in both the $Ox$ and $Oy$ directions, and the  formulae here, based on those for ellipses from chapter 1, are left to you reader.

\medskip

(3) We already have the ellipses, and the case of the parabolas and hyperbolas is elementary as well, and in fact simpler than the case of the ellipses. Indeed, a parablola is clearly homeomorphic to $\mathbb R$, and a hyperbola, to two copies of $\mathbb R$.

\medskip

(4) This is something which is clear too, depending of course on what exactly we mean by ``smooth deformation'', and by using a bit of multivariable calculus if needed.
\end{proof}

In higher dimensions, as basic examples, we have the spheres, as shown by:

\begin{theorem}
The sphere is a smooth manifold.
\end{theorem}

\begin{proof}
There are several proofs for this, all instructive, as follows:

\medskip

(1) A first idea is to use spherical coordinates, which are as follows:
$$\begin{cases}
x_1\!\!\!&=\ r\cos t_1\\
x_2\!\!\!&=\ r\sin t_1\cos t_2\\
\vdots\\
x_{N-1}\!\!\!&=\ r\sin t_1\sin t_2\ldots\sin t_{N-2}\cos t_{N-1}\\
x_N\!\!\!&=\ r\sin t_1\sin t_2\ldots\sin t_{N-2}\sin t_{N-1}
\end{cases}$$

Indeed, these produce explicit charts for the sphere.

\medskip

(2) A second idea, which makes use of less charts, namely 2 charts only, is to use the stereographic projection, which is given by inverse maps as follows:
$$\Phi:\mathbb R^N\to S^N_\mathbb R-\{\infty\}\quad,\quad 
\Psi:S^N_\mathbb R-\{\infty\}\to\mathbb R^N$$

To be more precise, the formulae of these maps, which are elementary to establish, are as follows, with the convention $\mathbb R^{N+1}=\mathbb R\times\mathbb R^N$, and with the coordinate of $\mathbb R$ denoted $x_0$, and with the coordinates of $\mathbb R^N$ denoted $x_1,\ldots,x_N$:
$$\Phi(v)=(1,0)+\frac{2}{1+||v||^2}\,(-1,v)\quad,\quad 
\Psi(c,x)=\frac{x}{1-c}$$

Indeed, we get in this way explicit charts for the sphere.

\medskip

(3) We have as well cylindrical coordinates, as well as many other types of more specialized coordinates, which can be useful in physics, plus of course, in disciplines like geography, economics and so on. There are many interesting computations that can be done here, and we will be back to these, on a regular basis in what follows, once we will know more about smooth manifolds, and their properties.
\end{proof}

Other basic examples of smooth manifolds include the projective spaces:

\begin{theorem}
The projective space $P^{N-1}_\mathbb R$ is a smooth manifold, with charts
$$(x_1,\ldots,x_N)\to\left(\frac{x_1}{x_i},\ldots,\frac{x_{i-1}}{x_i},\frac{x_{i+1}}{x_i},\ldots,\frac{x_N}{x_i}\right)$$
where $x_i\neq0$. This manifold is compact, and of dimension $N-1$.
\end{theorem}

\begin{proof}
We know that $P^{N-1}_\mathbb R$ appears by definition as the space of lines in $\mathbb R^N$ passing through the origin, so we have the following formula, with $\sim$ being the proportionality of vectors, given as usual by $x\sim y$ when $x=\lambda y$, for some scalar $\lambda\neq0$:
$$P^{N-1}_\mathbb R=\mathbb R^N-\{0\}/\sim$$

Alternatively, we can restrict if we want the attention to the vectors on the unit sphere $S^{N-1}_\mathbb  R\subset\mathbb R^N$, and this because any line in $\mathbb R^N$ passing through the origin will certainly cross this sphere. Moreover, it is clear that our line will cross the sphere in exactly two points $\pm x$, and we conclude that we have the following formula, with $\sim$ being now the proportionality of vectors on the sphere, given by $x\sim y$ when $x=\pm y$:
$$P^{N-1}_\mathbb R=S^{N-1}_\mathbb R/\sim$$

With this discussion made, let us get now to what is to be proved. Obviously, once we fix an index $i\in\{1,\ldots,N\}$, the condition $x_i\neq0$ on the vectors $x\in\mathbb R^N-\{0\}$ defines an open subset $U_i\subset P^{N-1}_\mathbb R$, and the open subsets that we get in this way cover $P^{N-1}_\mathbb R$:
$$P^{N-1}_\mathbb R=U_1\cup\ldots\cup U_N$$

Moreover, the map in the statement is injective $U_i\to\mathbb R^{N-1}$, and it is clear too that the changes of charts are $C^\infty$. Thus, we have our smooth manifold, as claimed.
\end{proof}

As a continuation of this, we can talk about Riemannian manifolds, which are those smooth manifolds where we can talk about length, area, volume, integration and so on. 

\bigskip

The theory of Riemannian manifolds is far more advanced with respect to what can be done with the arbitrary smooth manifolds, for somewhat obvious reasons. At the level of applications, we will see later in this chapter that, save for a slight modification in the axioms, we can use such manifolds for understanding our surrounding space-time.

\bigskip

As a last topic of discussion, in regards with general differential geometry, let us go back now to the optimization questions for functions, discussed in the previous chapter. Thinking well, the functions that we have to minimize or maximize, in the real life, are often defined on a manifold, instead of being defined on the whole $\mathbb R^N$. Fortunately, the good old principle $f'(x)=0$ can be adapted to the manifold case, as follows:

\begin{principle}
In order for a function $f:X\to\mathbb R$ defined on a manifold $X$ to have a local extremum at $x\in X$, we must have, as usual 
$$f'(x)=0$$
but with this taking into account the fact that the equations defining the manifold count as well as ``zero'', and so must be incorporated into the formula $f'(x)=0$. 
\end{principle}

In what follows, we will take this principle as granted. In practice, the idea is that we must have a formula as follows, with $g_i$ being the constraint functions for our manifold $X$, and with $\lambda_i\in\mathbb R$ being certain scalars, called Lagrange multipliers:
$$f'(x)=\sum_i\lambda_ig_i'(x)$$

As a basic illustration for this, our claim is that, by using a suitable manifold, and a suitable function, and Lagrange multipliers, we can prove in this way the H\"older inequality, that we know well of course, but without any computation. Let us start with:

\begin{proposition}
For any exponent $p>1$, the following set
$$S_p=\left\{x\in\mathbb R^N\Big|\sum_i|x_i|^p=1\right\}$$
is a submanifold of $\mathbb R^N$.
\end{proposition}

\begin{proof}
We know from the above that the unit sphere in $\mathbb R^N$ is a manifold. In our terms, this solves our problem at $p=2$, because this unit sphere is:
$$S_2=\left\{x\in\mathbb R^N\Big|\sum_ix_i^2=1\right\}$$

Now observe that we have a bijection $S_p\simeq S_2$, at least on the part where all the coordinates are positive, $x_i>0$, given by the following function:
$$x_i\to x_i^{2/p}$$

Thus we obtain that $S_p$ is indeed a manifold, as claimed.
\end{proof}

We already know that the manifold $S_p$ constructed above is the unit sphere, in the case $p=2$. In order to have a better geometric picture of what is going on, in general, observe that $S_p$ can be constructed as well at $p=1$, as follows:
$$S_1=\left\{x\in\mathbb R^N\Big|\sum_i|x_i|=1\right\}$$

However, this is no longer a manifold, as we can see for instance at $N=2$, where we obtain a square. Now observe that we can talk as well about $p=\infty$, as follows:
$$S_\infty=\left\{x\in\mathbb R^N\Big|\sup_i|x_i|=1\right\}$$

This letter set is no longer a manifold either, as we can see for instance at $N=2$, where we obtain again a square, containing the previous square, the one at $p=1$.

\bigskip

With these limiting constructions in hand, we can have now a better geometric picture of what is going on, in the general context of Proposition 10.12. Indeed, let us draw, at $N=2$ for simplifying, our sets $S_p$ at the values $p=1,2,\infty$ of the exponent:
$$
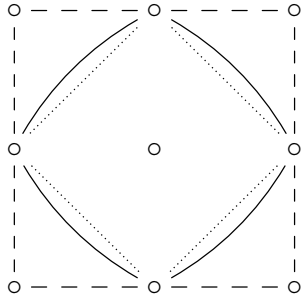
\xymatrix@R=40pt@C=40pt{
\circ\ar@{--}[r]\ar@{--}[d]&\circ\ar@{--}[r]\ar@{.}[dr]\ar@{.}[dl]\ar@{-}@/^/[dr]&\circ\ar@{--}[d]\\
\circ\ar@{--}[d]\ar@{.}[dr]\ar@{-}@/^/[ur]&\circ&\circ\ar@{--}[d]\ar@{.}[dl]\ar@{-}@/^/[dl]\\
\circ\ar@{--}[r]&\circ\ar@{--}[r]\ar@{-}@/^/[ul]&\circ
}$$

We can see that what we have is a small square, at $p=1$, becoming smooth and inflating towards the circle, in the parameter range $p\in(1,2]$, and then further inflating, in the parameter range $p\in[2,\infty)$, towards the big square appearing at $p=\infty$.

\bigskip

With these preliminaries in hand, we can formulate our result, as follows:

\begin{theorem}
The local extrema over $S_p$ of the function
$$f(x)=\sum_ix_iy_i$$ 
can be computed by using Lagrange multipliers, and this gives
$$\left|\sum_ix_iy_i\right|\leq\left(\sum_i|x_i|^p\right)^{1/p}\left(\sum_i|y_i|^q\right)^{1/q}$$
with $1/p+1/q=1$, that is, the H\"older inequality, with a purely geometric proof.
\end{theorem}

\begin{proof}
We can restrict the attention to the case where all the coordinates are positive, $x_i>0$ and  $y_i>0$. The derivative of the function in the statement is:
$$f'(x)=(y_1,\ldots,y_N)$$

On the other hand, we know that the manifold $S_p$ appears by definition as the set of zeroes of the function $\varphi(x)=\sum_ix_i^p-1$, having derivative as follows:
$$\varphi'(x)=p(x_1^{p-1},\ldots,x_N^{p-1})$$

Thus, by using Lagrange multipliers, the critical points of $f$ must satisfy:
$$(y_1,\ldots,y_N)\sim(x_1^{p-1},\ldots,x_N^{p-1})$$

In other words, the critical points must satisfy $x_i=\lambda y_i^{1/(p-1)}$, for some $\lambda>0$, and by using now $\sum_ix_i^p=1$ we can compute the precise value of $\lambda$, and we get:
$$\lambda=\left(\sum_iy_i^{p/(p-1)}\right)^{-1/p}$$ 

Now let us see what this means. Since the critical point is unique, this must be a maximum of our function, and we conclude that for any $x\in S_p$, we have:
$$\sum_ix_iy_i\leq\sum_i\lambda y_i^{1/(p-1)}\cdot y_i=\left(\sum_iy_i^{p/(p-1)}\right)^{1-1/p}=\left(\sum_iy_i^q\right)^{1/q}$$

Thus we have H\"older, and the general case follows from this, by rescaling.
\end{proof}

There are many other possible applications of the Lagrange multipliers, to all sorts of science questions, and for more on all this, including of course a proof of Principle 10.11 too, we refer to any solid differential geometry book, of pure or applied type.

\section*{10c. Relativity theory} 

Let us discuss now some applications of the above to basic theoretical physics. Based on experiments by Fizeau, then Michelson-Morley and others, and some physics by Maxwell and Lorentz too, Einstein came upon the following principles:

\index{Einstein principles}
\index{speed of light}
\index{faster than light}

\begin{fact}[Einstein principles]
The following happen:
\begin{enumerate}
\item Light travels in vacuum at a finite speed, $c<\infty$.

\item This speed $c$ is the same for all inertial observers.

\item In non-vacuum, the light speed is lower, $v<c$.

\item Nothing can travel faster than light, $v\not>c$.
\end{enumerate}
\end{fact} 

The point now is that, obviously, something is wrong here. Indeed, assuming for instance that we have a train, running in vacuum at speed $v>0$, and someone on board lights a flashlight $\ast$ towards the locomotive, then an observer $\circ$ on the ground will see the light traveling at speed $c+v>c$, which is a contradiction:
$$
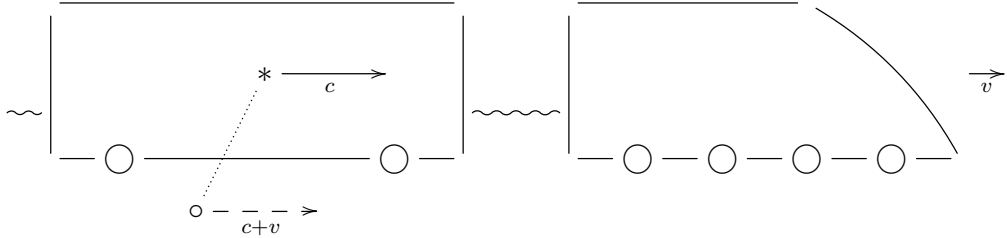
\xymatrix@R=5pt@C=13pt{
&\ar@{-}[rrrrrr]&&&&&&&&\ar@{-}[rrr]&&&\\
&&&&&&&&&\\
&&&&\ast\ar[rr]_c&&&&&&&&&&\ar[r]_v&\\
\ar@{~}[r]&&&&&&&\ar@{~}[rr]&&\\
&\ar@{-}[r]\ar@{-}[uuuu]&\bigcirc\ar@{-}[rrrr]&&&&\bigcirc\ar@{-}[r]&\ar@{-}[uuuu]&&\ar@{-}[r]\ar@{-}[uuuu]&\bigcirc\ar@{-}[r]&\bigcirc\ar@{-}[r]&\bigcirc\ar@{-}[r]&\bigcirc\ar@{-}[r]&\ar@/_/@{-}[uuuull]\\
&&&\circ\ar@{-->}[rr]_{c+v}\ar@{.}[uuur]&&
}$$

Equivalently, with the same train running, in vacuum at speed $v>0$, if the observer on the ground lights a flashlight $\ast$ towards the back of the train, then viewed from the train, that light will travel at speed $c+v>c$, which is a contradiction again:
$$
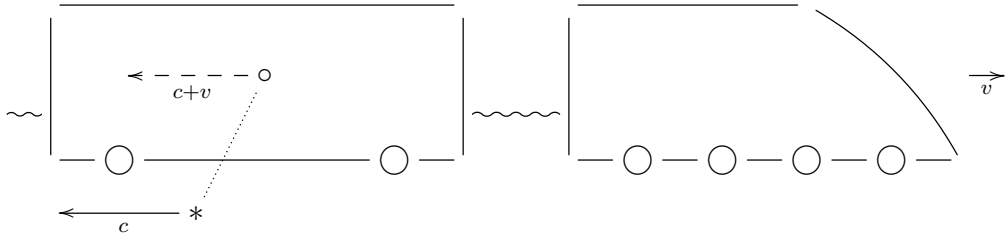
\xymatrix@R=5pt@C=13pt{
&\ar@{-}[rrrrrr]&&&&&&&&\ar@{-}[rrr]&&&\\
&&&&&&&&&\\
&&&&\circ\ar@{-->}[ll]^{c+v}&&&&&&&&&&\ar[r]_v&\\
\ar@{~}[r]&&&&&&&\ar@{~}[rr]&&\\
&\ar@{-}[r]\ar@{-}[uuuu]&\bigcirc\ar@{-}[rrrr]&&&&\bigcirc\ar@{-}[r]&\ar@{-}[uuuu]&&\ar@{-}[r]\ar@{-}[uuuu]&\bigcirc\ar@{-}[r]&\bigcirc\ar@{-}[r]&\bigcirc\ar@{-}[r]&\bigcirc\ar@{-}[r]&\ar@/_/@{-}[uuuull]\\
&&&\ast\ar[ll]^c\ar@{.}[uuur]&&
}$$

Summarizing, Fact 10.14 implies $c+v=c$, so contradicts classical mechanics, which therefore needs a fix. By dividing all speeds by $c$, as to have $c=1$, and by restricting the attention to the 1D case, to start with, we are led to the following puzzle:

\begin{puzzle}
How to define speed addition on the space of $1{\rm D}$ speeds, which is
$$I=[-1,1]$$
with our $c=1$ convention, as to have $1+c=1$, as required by physics?
\end{puzzle}

In view of our basic geometric knowledge, a natural idea here would be that of wrapping $[-1,1]$ into a circle, and then stereographically projecting on $\mathbb R$. Indeed, we can then ``import'' to $[-,1,1]$ the usual addition on $\mathbb R$, via the inverse of this map.

\bigskip

So, let us see where all this leads us. First, the formula of our map is as follows:

\index{stereographic projection}

\begin{proposition}
The map wrapping $[-1,1]$ into the unit circle, and then stereographically projecting on $\mathbb R$ is given by the formula
$$\varphi(u)=\tan\left(\frac{\pi u}{2}\right)$$
with the convention that our wrapping is the most straightforward one, making correspond $\pm 1\to i$, with negatives on the left, and positives on the right.
\end{proposition}

\begin{proof}
Regarding the wrapping, as indicated, this is given by:
$$u\to e^{it}\quad,\quad t=\pi u-\frac{\pi}{2}$$

Indeed, this correspondence wraps $[-1,1]$ as above, the basic instances of our correspondence being as follows, and with everything being fine modulo $2\pi$:
$$-1\to\frac{\pi}{2}\quad,\quad -\frac{1}{2}\to-\pi\quad,\quad 0\to-\frac{\pi}{2}\quad,\quad \frac{1}{2}\to 0\quad,\quad 1\to \frac{\pi}{2}$$

Regarding now the stereographic projection, the picture here is as follows:
$$
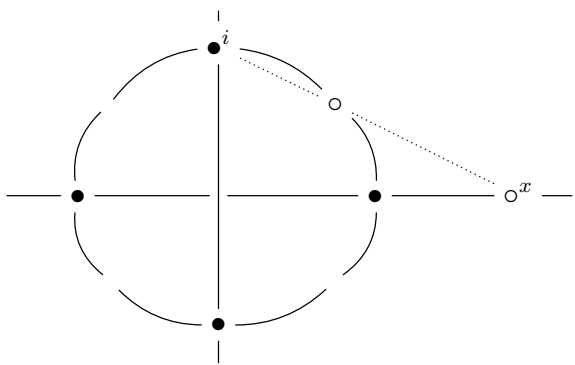
\xymatrix@R=7pt@C=2.1pt{
&&&&&&&&\ar@{-}[d]\\
&&&&&&&&\bullet^i\ar@{-}[dddddd]\ar@{-}@/_/[dllll]\ar@{.}[drrrr]&\\
&&&&\ar@{-}@/_/[ddl]&&&&&&&&\circ\ar@{-}@/_/[ullll]\ar@{.}[ddrrrrrr]\\
&&&&&&&&&&&&\\
\ar@{-}[rrr]&&&\bullet\ar@{-}@/_/[ddr]\ar@{-}[rrrrr]&&&&&\ar@{-}[rrrrr]&&&&&\bullet\ar@{-}@/_/[uul]\ar@{-}[rrrrr]&&&&&\circ^x\ar@{-}[rr]&&\\
&&&&&&&&&&&&\\
&&&&\ar@{-}@/_/[drrrr]&&&&&&&&\ar@{-}@/_/[uur]\\
&&&&&&&&\bullet\ar@{-}@/_/[urrrr]\ar@{-}[d]\\
&&&&&&&&}$$

Thus, by Thales, the formula of the stereographic projection is as follows:
$$\frac{\cos t}{x}=\frac{1-\sin t}{1}\implies x=\frac{\cos t}{1-\sin t}$$

Now if we compose our wrapping operation above with the stereographic projection, what we get is, via the above Thales formula, and some trigonometry:
\begin{eqnarray*}
x
&=&\frac{\cos t}{1-\sin t}\\
&=&\frac{\cos\left(\pi u-\frac{\pi}{2}\right)}{1-\sin\left(\pi u-\frac{\pi}{2}\right)}\\
&=&\frac{\cos\left(\frac{\pi}{2}-\pi u\right)}{1+\sin\left(\frac{\pi}{2}-\pi u\right)}\\
&=&\frac{\sin(\pi u)}{1+\cos(\pi u)}\\
&=&\frac{2\sin\left(\frac{\pi u}{2}\right)\cos\left(\frac{\pi u}{2}\right)}{2\cos^2\left(\frac{\pi u}{2}\right)}\\
&=&\tan\left(\frac{\pi u}{2}\right)
\end{eqnarray*}

Thus, we are led to the conclusion in the statement.
\end{proof}

The above result is very nice, but when it comes to physics, things do not work, for instance because of the wrong slope of the function $\varphi(u)=\tan\left(\frac{\pi u}{2}\right)$ at the origin, which makes our summing on $[-1,1]$ not compatible with the Galileo addition, at low speeds.

\bigskip

So, what to do? Obviously, trash Proposition 10.16, and start all over again. Getting back now to Puzzle 10.15, this has in fact a simpler solution, based this time on algebra, and which in addition is the good, physically correct solution, as follows:

\index{sum of speeds}
\index{Einstein formula}
\index{relativistic sum}
\index{Galileo formula}

\begin{theorem}
If we sum the speeds according to the Einstein formula
$$u+_ev=\frac{u+v}{1+uv}$$
then the Galileo formula still holds, approximately, for low speeds
$$u+_ev\simeq u+v$$
and if we have $u=1$ or $v=1$, the resulting sum is $u+_ev=1$.
\end{theorem}

\begin{proof}
All this is self-explanatory, and clear from definitions, and with the Einstein formula of $u+_ev$ itself being just an obvious solution to Puzzle 10.15, provided that, importantly, we know 0 geometry, and rely on very basic algebra only.
\end{proof}

So, very nice, problem solved, at least in 1D. But, shall we give up with geometry, and the stereographic projection? Certainly not, let us try to recycle that material. In order to do this, let us recall that the usual trigonometric functions are given by:
$$\sin x=\frac{e^{ix}-e^{-ix}}{2i}\quad,\quad \cos x=\frac{e^{ix}+e^{-ix}}{2}\quad,\quad \tan x=\frac{e^{ix}-e^{-ix}}{i(e^{ix}+e^{-ix})}$$

The point now is that, and you might know this from calculus, the above functions have some natural ``hyperbolic'' or ``imaginary'' analogues, constructed as follows:
$$\sinh x=\frac{e^x-e^{-x}}{2}\quad,\quad \cosh x=\frac{e^x+e^{-x}}{2}\quad,\quad \tanh x=\frac{e^x-e^{-x}}{e^x+e^{-x}}$$

But the function on the right, $\tanh$, starts reminding the formula of Einstein addition, from Theorem 10.17. So, we have our idea, and we are led to the following result:

\index{hyperbolic function}
\index{hyperbolic tangent}

\begin{theorem}
The Einstein speed summation in 1D is given by
$$\tanh x+_e\tanh y=\tanh(x+y)$$
with $\tanh:[-\infty,\infty]\to[-1,1]$ being the hyperbolic tangent function.
\end{theorem}

\begin{proof}
This follows by putting together our various formulae above, but it is perhaps better, for clarity, to prove this directly. Our claim is that we have: 
$$\tanh(x+y)=\frac{\tanh x+\tanh y}{1+\tanh x\tanh y}$$

But this can be checked via direct computation, from the definitions, as follows:
\begin{eqnarray*}
&&\frac{\tanh x+\tanh y}{1+\tanh x\tanh y}\\
&=&\left(\frac{e^x-e^{-x}}{e^x+e^{-x}}+\frac{e^y-e^{-y}}{e^y+e^{-y}}\right)
\Big/\left(1+\frac{e^x-e^{-x}}{e^x+e^{-x}}\cdot\frac{e^y-e^{-y}}{e^y+e^{-y}}\right)\\
&=&\frac{(e^x-e^{-x})(e^y+e^{-y})+(e^x+e^{-x})(e^y-e^{-y})}{(e^x+e^{-x})(e^y+e^{-y})+(e^x-e^{-x})(e^y+e^{-y})}\\
&=&\frac{2(e^{x+y}-e^{-x-y})}{2(e^{x+y}+e^{-x-y})}\\
&=&\tanh(x+y)
\end{eqnarray*}

Thus, we are led to the conclusion in the statement.
\end{proof}

Very nice all this, hope you agree. As a conclusion, passing from the Riemann stereographic projection sum to the Einstein summation basically amounts in replacing:
$$\tan\to\tanh$$

Let us formulate as well this finding more philosophically, as follows:

\begin{conclusion}
The Einstein speed summation in 1D is the imaginary analogue of the summation on $[-1,1]$ obtained via Riemann's stereographic projection.
\end{conclusion}

Getting now to several dimensions, we have an analogue of Puzzle 10.15 here, and after doing the math, we are led to the following conclusion:

\index{vector product}
\index{Einstein sum}

\begin{theorem}
When defining the Einstein speed summation in $3{\rm D}$ as 
$$u+_ev=\frac{1}{1+<u,v>}\left(u+v+\frac{u\times(u\times v)}{1+\sqrt{1-||u||^2}}\right)$$
in $c=1$ units, the following happen:
\begin{enumerate}
\item When $u\sim v$, we recover the previous $1{\rm D}$ formula.

\item We have $||u||,||v||<1\implies||u+_ev||<1$.

\item When $||u||=1$, we have $u+_ev=u$.

\item When $||v||=1$, we have $||u+_ev||=1$.

\item However, $||v||=1$ does not imply $u+_ev=v$.

\item Also, the formula $u+_ev=v+_eu$ fails.
\end{enumerate}
In addition, the above formula is physically correct, agreeing with experiments.
\end{theorem}

\begin{proof}
This is something quite tricky, with the key physics claim at the end being indeed true, and with the idea with the mathematical part being as follows:

\medskip

(1) This is something which follows from definitions.

\medskip

(2) In order to simplify notation, let us set $\delta=\sqrt{1-||u||^2}$, which is the inverse of the quantity $\gamma=1/\sqrt{1-||u||^2}$. With this convention, we have:
\begin{eqnarray*}
u+_ev
&=&\frac{1}{1+<u,v>}\left(u+v+\frac{<u,v>u-||u||^2v}{1+\delta}\right)\\
&=&\frac{(1+\delta+<u,v>)u+(1+\delta-||u||^2)v}{(1+<u,v>)(1+\delta)}
\end{eqnarray*}

Taking now the squared norm and computing gives the following formula:
$$||u+_ev||^2
=\frac{(1+\delta)^2||u+v||^2+(||u||^2-2(1+\delta))(||u||^2||v||^2-<u,v>^2)}{(1+<u,v>)^2(1+\delta)^2}$$

But this formula can be further processed by using $\delta=\sqrt{1-||u||^2}$, and by navigating through the various quantities which appear, we obtain, as a final product:
$$||u+_ev||^2=\frac{||u+v||^2-||u||^2||v||^2+<u,v>^2}{(1+<u,v>)^2}$$

But this type of formula is exactly what we need, for what we want to do. Indeed, by assuming $||u||,||v||<1$, we have the following estimate:
\begin{eqnarray*}
||u+_ev||^2<1
&\iff&||u+v||^2-||u||^2||v||^2+<u,v>^2<(1+<u,v>)^2\\
&\iff&||u+v||^2-||u||^2||v||^2<1+2<u,v>\\
&\iff&||u||^2+||v||^2-||u||^2||v||^2<1\\
&\iff&(1-||u||^2)(1-||v||^2)>0
\end{eqnarray*}

Thus, we are led to the conclusion in the statement.

\medskip

(3) This is something elementary, coming from definitions.

\medskip

(4) This comes from the squared norm formula established in the proof of (2) above, because when assuming $||v||=1$, we obtain:
\begin{eqnarray*}
||u+_ev||^2
&=&\frac{||u+v||^2-||u||^2+<u,v>^2}{(1+<u,v>)^2}\\
&=&\frac{||u||^2+1+2<u,v>-||u||^2+<u,v>^2}{(1+<u,v>)^2}\\
&=&\frac{1+2<u,v>+<u,v>^2}{(1+<u,v>)^2}\\
&=&1
\end{eqnarray*}

(5) This is clear, from the obvious lack of symmetry of our formula.

\medskip

(6) This is again clear, from the obvious lack of symmetry of our formula.
\end{proof}

Time now to draw some concrete conclusions, from the above speed computations. Since speed $v=d/t$ is distance over time, we must fine-tune distance $d$, or time $t$, or both. Let us first discuss, following as usual Einstein, what happens to time $t$. Here the result, which might seem quite surprising, at a first glance, is as follows:

\index{Lorentz dilation}
\index{Lorentz factor}
\index{time dilation}
\index{relativistic time}

\begin{theorem}
Relativistic time is subject to Lorentz dilation
$$t\to\gamma t$$
where the number $\gamma\geq1$, called Lorentz factor, is given by the formula
$$\gamma=\frac{1}{\sqrt{1-v^2/c^2}}$$
with $v$ being the moving speed, at which time is measured. 
\end{theorem}

\begin{proof}
Assume indeed that we have a train, moving to the right with speed $v$, through vacuum. In order to compute the height $h$ of the train, the passenger onboard switches on the ceiling light bulb, measures the time $t$ that the light needs to hit the floor, by traveling at speed $c$, and concludes that the train height is $h=ct$:
$$
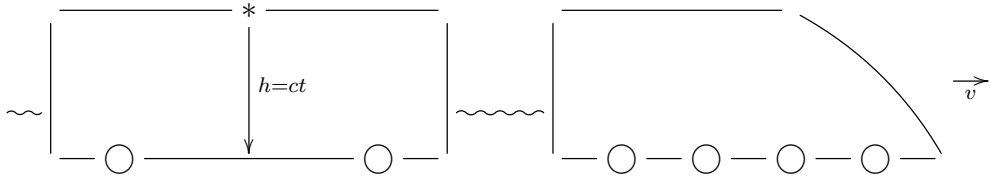
\xymatrix@R=5pt@C=13pt{
&\ar@{-}[rrr]&&&\ast\ar[dddd]^{h=ct}\ar@{-}[rrr]&&&&&\ar@{-}[rrr]&&&\\
&&&&&&&&&\\
&&&&&&&&&&&&&&\ar[r]_v&\\
\ar@{~}[r]&&&&&&&\ar@{~}[rr]&&\\
&\ar@{-}[r]\ar@{-}[uuuu]&\bigcirc\ar@{-}[rrrr]&&&&\bigcirc\ar@{-}[r]&\ar@{-}[uuuu]&&\ar@{-}[r]\ar@{-}[uuuu]&\bigcirc\ar@{-}[r]&\bigcirc\ar@{-}[r]&\bigcirc\ar@{-}[r]&\bigcirc\ar@{-}[r]&\ar@/_/@{-}[uuuull]
}$$

On the other hand, an observer on the ground will see here something different, namely a right triangle, with on the vertical the height of the train $h$, on the horizontal the distance $vT$ that the train has traveled, and on the hypotenuse the distance $cT$ that light has traveled, with $T$ being the duration of the event, according to his watch:
$$
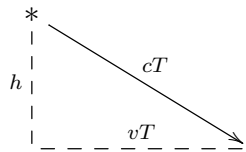
\xymatrix@R=5pt@C=13pt{
\ast\ar@{--}[dddd]_h\ar[ddddrrrr]^{cT}\\
&\\
&\\
&\\
\ar@{--}[rrrr]^{vT}&&&&
}$$

Now by Pythagoras applied to this triangle, we have the following formula:
$$h^2+(vT)^2=(cT)^2$$

Thus, the observer on the ground will reach to the following formula for $h$:
$$h=\sqrt{c^2-v^2}\cdot T$$

But $h$ must be the same for both observers, so we have the following formula:
$$\sqrt{c^2-v^2}\cdot T=ct$$

It follows that the two times $t$ and $T$ are indeed not equal, and are related by:
$$T=\frac{ct}{\sqrt{c^2-v^2}}=\frac{t}{\sqrt{1-v^2/c^2}}=\gamma t$$

Thus, we are led to the formula in the statement.
\end{proof}

Let us discuss now what happens to length. We have here the following result:

\index{Lorentz contraction}
\index{length contraction}
\index{relativistic length}

\begin{theorem}
Relativistic length is subject to Lorentz contraction
$$L\to L/\gamma$$
where the number $\gamma\geq1$, called Lorentz factor, is given by the usual formula
$$\gamma=\frac{1}{\sqrt{1-v^2/c^2}}$$
with $v$ being the moving speed, at which length is measured. 
\end{theorem}

\begin{proof}
As before in the proof of Theorem 10.21, meaning in the same train traveling at speed $v$, in vacuum, imagine now that the passenger wants to measure the length $L$ of the car. For this purpose he switches on the light bulb, now at the rear of the car, and measures the time $t$ needed for the light to reach the front of the car, and get reflected back by a mirror installed there, according to the following scheme:
$$
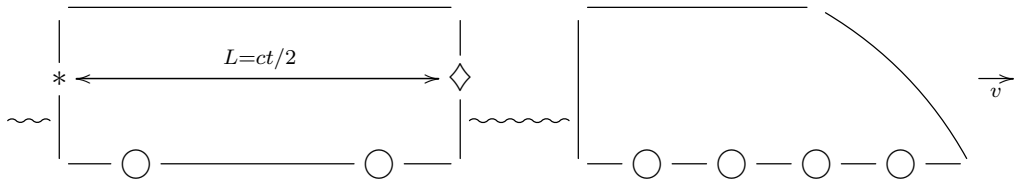
\xymatrix@R=4pt@C=13pt{
&\ar@{-}[rrrrrr]&&&&&&&&\ar@{-}[rrr]&&&\\
&&&&&&&&&\\
&\ast\ar@{-}[uu]\ar[rrrrrr]^{L=ct/2}&&&&&&\diamondsuit\ar[llllll]\ar@{-}[uu]&&&&&&&\ar[r]_v&\\
\ar@{~}[r]&&&&&&&\ar@{~}[rr]&&\\
&\ar@{-}[r]\ar@{-}[uu]&\bigcirc\ar@{-}[rrrr]&&&&\bigcirc\ar@{-}[r]&\ar@{-}[uu]&&\ar@{-}[r]\ar@{-}[uuuu]&\bigcirc\ar@{-}[r]&\bigcirc\ar@{-}[r]&\bigcirc\ar@{-}[r]&\bigcirc\ar@{-}[r]&\ar@/_/@{-}[uuuull]
}$$

He concludes that, as marked above, the length $L$ of the car is given by:
$$L=\frac{ct}{2}$$ 

Now viewed from the ground, the duration of the event is $T=T_1+T_2$, where $T_1>T_2$ are respectively the time needed for the light to travel forward, among others for beating $v$, and the time for the light to travel back, helped this time by $v$. More precisely, if $l$ denotes the length of the train car viewed from the ground, the formula of $T$ is:
$$T
=T_1+T_2
=\frac{l}{c-v}+\frac{l}{c+v}
=\frac{2lc}{c^2-v^2}$$

With this data, the formula $T=\gamma t$ of time dilation established before reads:
$$\frac{2lc}{c^2-v^2}=\gamma t=\frac{2\gamma L}{c}$$

Thus, the two lengths $L$ and $l$ are indeed not equal, and related by:
$$l
=\frac{\gamma L(c^2-v^2)}{c^2}
=\gamma L\left(1-\frac{v^2}{c^2}\right)
=\frac{\gamma L}{\gamma^2}
=\frac{L}{\gamma}$$

Thus, we are led to the conclusion in the statement.
\end{proof}

\section*{10d. Curved spacetime}

With the above discussed, time now to get into the real thing, namely happens to our usual $\mathbb R^4$. The result here, which is something quite tricky, is as follows:

\begin{theorem}
In the context of a relativistic object moving with speed $v$ along the $x$ axis, the frame change is given by the Lorentz transformation
$$x'=\gamma(x-vt)$$
$$y'=y$$
$$z'=z$$
$$t'=\gamma(t-vx/c^2)$$
with $\gamma=1/\sqrt{1-v^2/c^2}$ being as usual the Lorentz factor.
\end{theorem}

\begin{proof}
We know that, with respect to the non-relativistic formulae, $x$ is subject to the Lorentz dilation by $\gamma$, and we obtain as desired:
$$x'=\gamma(x-vt)$$

Regarding $y,z$, these are obviously unchanged, so done with these too. Finally, for $t$ we must use the reverse Lorentz transformation, given by the following formulae:
$$x=\gamma(x'+vt')$$
$$y=y'$$
$$z=z'$$

By using the formula of $x'$ we can compute $t'$, and we obtain the following formula:
\begin{eqnarray*}
t'
&=&\frac{x-\gamma x'}{\gamma v}\\
&=&\frac{x-\gamma^2(x-vt)}{\gamma v}\\
&=&\frac{\gamma^2vt+(1-\gamma^2)x}{\gamma v}
\end{eqnarray*}

On the other hand, we have the following computation:
$$\gamma^2=\frac{c^2}{c^2-v^2}
\implies\gamma^2(c^2-v^2)=c^2
\implies(\gamma^2-1)c^2=\gamma^2v^2$$

Thus we can finish the computation of $t'$ as follows:
\begin{eqnarray*}
t'
&=&\frac{\gamma^2vt+(1-\gamma^2)x}{\gamma v}\\
&=&\frac{\gamma^2vt-\gamma^2v^2x/c^2}{\gamma v}\\
&=&\gamma\left(t-\frac{vx}{c^2}\right)
\end{eqnarray*}

We are therefore led to the conclusion in the statement.
\end{proof}

Now since $y,z$ are irrelevant, we can put them at the end, and put the time $t$ first, as to be close to $x$. By multiplying as well the time equation by $c$, our system becomes:
$$ct'=\gamma(ct-vx/c)$$
$$x'=\gamma(x-vt)$$
$$y'=y$$
$$z'=z$$

In linear algebra terms, the result is as follows:

\index{Lorentz transform}

\begin{theorem}
The Lorentz transformation is given by
$$\begin{pmatrix}
\gamma&-\beta\gamma&0&0\\
-\beta\gamma&\gamma&0&0\\
0&0&1&0\\
0&0&0&1
\end{pmatrix}
\begin{pmatrix}ct\\ x\\ y\\ z\end{pmatrix}
=\begin{pmatrix}ct'\\ x'\\ y'\\ z'\end{pmatrix}$$
where $\gamma=1/\sqrt{1-v^2/c^2}$ as usual, and where $\beta=v/c$.
\end{theorem}

\begin{proof}
In terms of $\beta=v/c$, replacing $v$, the system looks as follows:
$$ct'=\gamma(ct-\beta x)$$
$$x'=\gamma(x-\beta ct)$$
$$y'=y$$
$$z'=z$$

But this gives the formula in the statement.
\end{proof}

As a nice and basic theoretical application of the Lorentz transform, this brings a new viewpoint on the Einstein speed addition formula, the result being follows:

\begin{theorem}
The speed addition formula in $3D$ relativity is
$$u+_ev=\frac{1}{1+<u,v>}\left(u+v+\frac{u\times(u\times v)}{1+\sqrt{1-||u||^2}}\right)$$
in $c=1$ units.
\end{theorem}

\begin{proof}
We already know this, but the point is that we can derive this as well from the formula of the Lorentz transform, by computing some derivatives, as follows:

\medskip

(1) The idea will be that of differentiating $x,y,z,t$ in the formulae for the inverse Lorentz transform, which are as follows:
$$x=\gamma(x'+ut')$$
$$y=y'$$
$$z=z'$$
$$t=\gamma(t'+ux'/c^2)$$

(2) Indeed, by differentiating these equalities, we obtain the following formulae:
$$dx=\gamma(dx'+udt')$$
$$dy=dy'$$
$$dz=dz'$$
$$dt=\gamma(dt'+udx'/c^2)$$

(3) Now by dividing the first three formulae by the fourth one, we obtain:
$$\frac{dx}{dt}=\frac{dx'+udt'}{dt'+udx'/c^2}$$
$$\frac{dy}{dt}=\frac{dy'}{\gamma(dt'+udx'/c^2)}$$
$$\frac{dz}{dt}=\frac{dz'}{\gamma(dt'+udx'/c^2)}$$

(4) We can make these look better by dividing everywhere by $dt'$, and we get:
$$\frac{dx}{dt}=\frac{dx'/dt'+u}{1+u/c^2\cdot dx'/dt'}$$
$$\frac{dy}{dt}=\frac{dy'/dt'}{\gamma(1+u/c^2\cdot dx'/dt')}$$
$$\frac{dz}{dt}=\frac{dz'/dt'}{\gamma(1+u/c^2\cdot dx'/dt')}$$

(5) In terms of speeds now, this means that we have, with $w=u+_ev$:
$$w_x=\frac{v_x+u}{1+u/c^2\cdot v_x}$$
$$w_y=\frac{v_y}{\gamma(1+u/c^2\cdot v_x)}$$
$$w_z=\frac{v_z}{\gamma(1+u/c^2\cdot v_x)}$$

(6) Now in $c=1$ units, these formulae are as follows, with $w=u+_ev$:
$$w_x=\frac{v_x+u}{1+uv_x}$$
$$w_y=\frac{v_y}{\gamma(1+uv_x)}$$
$$w_z=\frac{v_z}{\gamma(1+uv_x)}$$

(7) In vector notation now, the above formulae show that we have:
\begin{eqnarray*}
u+_ev
&=&\frac{1}{1+uv_x}\begin{pmatrix}v_x+u\\ v_y/\gamma\\ v_z/\gamma\end{pmatrix}\\
&=&\frac{1}{1+<u,v>}\left(u+\begin{pmatrix}v_x\\v_y/\gamma\\v_z/\gamma\end{pmatrix}\right)
\end{eqnarray*}

(8) On the other hand, we have the following computation:
\begin{eqnarray*}
u\times(u\times v)
&=&\begin{pmatrix}u\\0\\0\end{pmatrix}\times\left[
\begin{pmatrix}u\\0\\0\end{pmatrix}\times
\begin{pmatrix}v_x\\v_y\\v_z\end{pmatrix}\right]\\
&=&\begin{pmatrix}u\\0\\0\end{pmatrix}\times
\begin{pmatrix}0\\-uv_z\\uv_y\end{pmatrix}\\
&=&\begin{pmatrix}0\\-u^2v_y\\-u^2v_z\end{pmatrix}
\end{eqnarray*}

(9) We deduce from this that we have the following formula:
\begin{eqnarray*}
v+\frac{u\times(u\times v)}{1+\sqrt{1-u^2}}
&=&\begin{pmatrix}v_x\\v_y(1-u^2/(1+\sqrt{1-u^2}))\\v_z(1-u^2/(1+\sqrt{1-u^2}))\end{pmatrix}\\
&=&\begin{pmatrix}v_x\\v_y\sqrt{1-u^2}\\v_z\sqrt{1-u^2}\end{pmatrix}\\
&=&\begin{pmatrix}v_x\\v_y/\gamma\\v_z/\gamma\end{pmatrix}
\end{eqnarray*}

(10) Here we have used the following identity, which is something trivial:
$$1-\frac{u^2}{1+\sqrt{1-u^2}}=\sqrt{1-u^2}$$

(11) The point now is that the formula from (9) shows that the speed addition formula established in (7) can be written in the following way:
$$u+_ev=\frac{1}{1+<u,v>}\left(u+v+\frac{u\times(u\times v)}{1+\sqrt{1-||u||^2}}\right)$$

Summarizing, we have recovered the formula for speed addition in relativity, from before, in our present configuration, with $u$ assumed to be along the $Ox$ axis.

\medskip

(12) Finally, there is a discussion for passing from the standard configuration investigated above, where the movement is along the $Ox$ axis, to the general configuration, where the movement is arbitary. But this can be done either by decomposing one speed vector with respect to the other, or simply by arguing that everything is rotation invariant.
\end{proof}

Many other things can be said, as a continuation of the above. For the moment, what we have will do, with the conclusion being that, when it comes to high speeds, $v\simeq c$, spacetime is curved. So, let us try now to better understand this phenomenon.

\bigskip

To start with, recall that in the non-relativistic setting two events are separated by space $\Delta x$ and time $\Delta t$, with these two separation variables being independent. In relativistic physics this is no longer true, and the correct analogue of this comes from:

\index{spacetime separation}
\index{invariant}
\index{frame change}
\index{relativistic distance}
\index{Lorentz invariance}

\begin{theorem}
The following quantity, called relativistic spacetime separation
$$\Delta s^2=c^2\Delta t^2-(\Delta x^2+\Delta y^2+\Delta z^2)$$
is invariant under relativistic frame changes.
\end{theorem}

\begin{proof}
This is something important, and as before with such things, we will take our time, and carefully understand how this result works:

\medskip

(1) Let us first examine the case of the standard configuration. We must prove that the quantity $K=c^2t^2-x^2-y^2-z^2$ is invariant under Lorentz transformations, in the standard configuration. For this purpose, observe that we have:
$$K=\left<\begin{pmatrix}
1&0&0&0\\
0&-1&0&0\\
0&0&-1&0\\
0&0&0&-1
\end{pmatrix}
\begin{pmatrix}ct\\ x\\ y\\ z\end{pmatrix}
,\begin{pmatrix}ct\\ x\\ y\\ z\end{pmatrix}\right>$$

Now recall that the Lorentz transformation is given in standard configuration by the following formula, where $\gamma=1/\sqrt{1-v^2/c^2}$ as usual, and where $\beta=v/c$:
$$\begin{pmatrix}ct'\\ x'\\ y'\\ z'\end{pmatrix}=
\begin{pmatrix}
\gamma&-\beta\gamma&0&0\\
-\beta\gamma&\gamma&0&0\\
0&0&1&0\\
0&0&0&1
\end{pmatrix}
\begin{pmatrix}ct\\ x\\ y\\ z\end{pmatrix}$$

Thus, if we denote by $L$ the matrix of the Lorentz transformation, and by $E$ the matrix found before, we must prove that for any vector $\xi$ we have:
$$<E\xi,\xi>=<EL\xi,L\xi>$$

Since the matrix $L$ is symmetric, we have the following formula:
$$<EL\xi,L\xi>=<LEL\xi,\xi>$$

Thus, we must prove that the following happens:
$$E=LEL$$

But this is the same as proving that we have the following equality:
$$L^{-1}E=EL$$

Moreover, by using the fact that the passage $L\to L^{-1}$ is given by $\beta\to-\beta$, what we eventually want to prove is that:
$$L_{-\beta}E=EL_\beta$$

So, let us prove this. As usual we can restrict the attention to the upper left corner, call that NW corner, and  here we have the following computation:
\begin{eqnarray*}
(L_{-\beta}E)_{NW}
&=&\begin{pmatrix}\gamma&\beta\gamma\\ \beta\gamma&\gamma\end{pmatrix}
\begin{pmatrix}1&0\\ 0&-1\end{pmatrix}\\
&=&\begin{pmatrix}\gamma&-\beta\gamma\\ \beta\gamma&-\gamma\end{pmatrix}
\end{eqnarray*}

On the other hand, we have as well the following computation:
\begin{eqnarray*}
(EL_\beta)_{NW}
&=&\begin{pmatrix}1&0\\ 0&-1\end{pmatrix}
\begin{pmatrix}\gamma&-\beta\gamma\\ -\beta\gamma&\gamma\end{pmatrix}\\
&=&\begin{pmatrix}\gamma&-\beta\gamma\\ \beta\gamma&-\gamma\end{pmatrix}
\end{eqnarray*}

The matrices on the right being equal, this gives the result.

\medskip

(2) Now let us prove the invariance in general. The matrix of the Lorentz transformation being a bit complicated, in this case, as explained in the above, the best is to use for this the raw formulae of $x',t'$, that we found in the above, namely:
$$x'=x+(\gamma-1)\frac{<v,x>v}{||v||^2}-\gamma tv$$
$$t'=\gamma\left(t-\frac{<v,x>}{c^2}\right)$$

With these formulae in hand, we have the following computation:
\begin{eqnarray*}
&&(ct')^2-||x'||^2\\
&=&\gamma^2\left(ct-\frac{<v,x>}{c}\right)^2\\
&&-||x||^2-(\gamma-1)^2\frac{<v,x>^2}{||v||^2}-\gamma^2t^2||v||^2\\
&&-2(\gamma-1)\frac{<v,x>^2}{||v||^2}+2\gamma t<v,x>+2\gamma(\gamma-1)t<v,x>\\
&=&\gamma^2t^2(c^2-||v||^2)-||x||^2\\
&&+<v,x>(-2\gamma^2 t+2\gamma t+2\gamma(\gamma-1)t)\\
&&+<v,x>^2\left(\frac{\gamma^2}{c^2}-\frac{(\gamma-1)^2}{||v||^2}-\frac{2(\gamma-1)}{||v||^2}\right)\\
&=&c^2t^2-||x||^2+<v,x>^2\left(\frac{\gamma^2}{c^2}-\frac{\gamma^2-1}{||v||^2}\right)\\
&=&c^2t^2-||x||^2
\end{eqnarray*}

Here we have used the following trivial formula, for the coefficient of $t^2$:
$$\gamma^2(c^2-||v||^2)
=\frac{c^2-||v||^2}{1-||v||^2/c^2}
=c^2$$

Also, we have used the following formula, for the coefficient of $<v,x>^2$:
\begin{eqnarray*}
\frac{\gamma^2}{c^2}-\frac{\gamma^2-1}{||v||^2}
&=&\gamma^2\left(\frac{1}{c^2}-\frac{1}{||v||^2}\right)+\frac{1}{||v||^2}\\
&=&\frac{1}{1-||v||^2/c^2}\cdot\frac{||v||^2-c^2}{c^2||v||^2}+\frac{1}{||v||^2}\\
&=&\frac{c^2}{c^2-||v||^2}\cdot\frac{||v||^2-c^2}{c^2||v||^2}+\frac{1}{||v||^2}\\
&=&-\frac{1}{||v||^2}+\frac{1}{||v||^2}\\
&=&0
\end{eqnarray*}

Thus, we are led to the conclusion in the statement.
\end{proof}

In relation with the above, it is possible to do some sort of reverse engineering, by recovering the formula of the Lorentz transform, from the spacetime separation invariance. Thus, we are led in this way to yet another axiomatization of the theory.

\bigskip

Finally, what we have in Theorem 10.26 suggests that curved spacetime is some sort of Riemannian manifold, that is, a smooth manifold whose tangent spaces come with scalar products, but with a $-$ sign appearing in the various formulae, instead of a $+$. 

\bigskip

This is indeed true, with such manifolds being called ``Lorentz manifolds'', and with all this being related to everything algebraic and differential geometry, and linear algebra positivity and negativity too, that we learned in the beginning of this chapter.

\bigskip

Which sounds quite exciting, doesn't it. There are many things that can be learned here, both math and physics, and in the hope that you will go this way, and enjoy it.

\section*{10e. Exercises}

This was an introduction to modern math and physics, and as exercises, we have:

\begin{exercise}
Have some fun with learning about plane algebraic curves.
\end{exercise}

\begin{exercise}
Learn some commutative algebra, notably the Nullstellensatz.
\end{exercise}

\begin{exercise}
Learn how to compute lengths of curves, and areas of surfaces.
\end{exercise}

\begin{exercise}
Learn the formal definition and basic properties of smooth manifolds.
\end{exercise}

\begin{exercise}
Learn also about Riemannian manifolds, as much as you can.
\end{exercise}

\begin{exercise}
Still in relation with geometry, learn about knots and links, too.
\end{exercise}

\begin{exercise}
Recover the Einstein speed summation in $3D$, by yourself.
\end{exercise}

\begin{exercise}
Meditate on the $c\pm v$ speeds used for the Lorentz contraction.
\end{exercise}

As bonus exercise, that we warmly recommend, read Einstein's book \cite{ein}.

\chapter{Special matrices}

\section*{11a. Circulant matrices}

Back now to more traditional linear algebra, we discuss in this chapter various classes of ``special matrices''. As a first and central example here, we have the flat matrix:

\index{flat matrix}
\index{all-one matrix}
\index{all-one vector}

\begin{definition}
The flat matrix $\mathbb I_N$ is the all-one $N\times N$ matrix:
$$\mathbb I_N=\begin{pmatrix}
1&\ldots&1\\
\vdots&&\vdots\\
1&\ldots&1\end{pmatrix}$$
Equivalently, $\mathbb I_N/N$ is the orthogonal projection on the all-one vector $\xi\in\mathbb C^N$.
\end{definition}

A first interesting question regarding $\mathbb I_N$ concerns its diagonalization. As explained in chapter 1, this is best solved by using complex numbers, in the following way:

\index{flat matrix}
\index{Fourier matrix}

\begin{theorem}
The flat matrix $\mathbb I_N$ diagonalizes as follows,
$$\begin{pmatrix}
1&\ldots&\ldots&1\\
\vdots&&&\vdots\\
\vdots&&&\vdots\\
1&\ldots&\ldots&1\end{pmatrix}
=\frac{1}{N}\,F_N
\begin{pmatrix}
N\\
&0\\
&&\ddots\\
&&&0\end{pmatrix}F_N^*$$
where $F_N=(w^{ij})_{ij}$ with $w=e^{2\pi i/N}$ is the Fourier matrix.
\end{theorem}

\begin{proof}
The flat matrix being $N$ times the projection on the all-one vector, we are left with finding the 0-eigenvectors, which amounts in solving the following equation:
$$x_0+\ldots+x_{N-1}=0$$

But for this purpose, we use the root of unity $w=e^{2\pi i/N}$, which is subject to:
$$\sum_{i=0}^{N-1}w^{ij}=N\delta_{j0}$$

Indeed, this formula shows that for $j=1,\ldots,N-1$, the vector $v_j=(w^{ij})_i$ is a 0-eigenvector. Moreover, these vectors are pairwise orthogonal, because we have:
$$<v_j,v_k>
=\sum_iw^{ij-ik}
=N\delta_{jk}$$

Thus, we have our basis $\{v_1,\ldots,v_{N-1}\}$ of 0-eigenvectors, and since the $N$-eigenvector is $\xi=v_0$, the passage matrix $P$ that we are looking is given by:
$$P=\begin{bmatrix}v_0&v_1&\ldots&v_{N-1}\end{bmatrix}$$

But this is precisely the Fourier matrix, $P=F_N$. In order to finish now, observe that the above computation of $<v_i,v_j>$ shows that $F_N/\sqrt{N}$ is unitary, and so:
$$F_N^{-1}=\frac{1}{N}\,F_N^*$$

Thus, we are led to the diagonalization formula in the statement.
\end{proof}

Now observe that the flat matrix $\mathbb I_N$ is circulant, in the following sense:

\index{circulant matrix}

\begin{definition}
A real or complex matrix $M$ is called circulant if
$$M_{ij}=\xi_{j-i}$$
for a certain vector $\xi$, with the indices taken modulo $N$.
\end{definition}

The circulant matrices are beautiful mathematical objects, which appear of course in many serious problems as well. As an example, at $N=4$, we must have:
$$M=\begin{pmatrix}
a&b&c&d\\
d&a&b&c\\
c&d&a&b\\
b&c&d&a
\end{pmatrix}$$

The point now is that, while certainly gently looking, these matrices can be quite diabolic, when it comes to diagonalization, and other problems. For instance, when $M$ is real, the computations with $M$ are usually quite complicated over the real numbers. Fortunately the complex numbers and the Fourier matrices are there, and we have:

\index{Fourier-diagonal matrix}
\index{discrete Fourier transform}
\index{Fourier-diagonal}

\begin{theorem}
For a matrix $M\in M_N(\mathbb C)$, the following are equivalent:
\begin{enumerate}
\item $M$ is circulant, in the sense that we have, for a certain vector $\xi\in\mathbb C^N$:
$$M_{ij}=\xi_{j-i}$$

\item $M$ is Fourier-diagonal, in the sense that, for a certain diagonal matrix $Q$:
$$M=F_NQF_N^*$$
\end{enumerate}
In addition, if these conditions hold, then $\xi,Q$ are related by the formula
$$\xi=F_N^*q$$
where $q\in\mathbb C^N$ is the column vector formed by the diagonal entries of $Q$.
\end{theorem}

\begin{proof}
This follows from some basic computations with roots of unity, as follows:

\medskip

$(1)\implies(2)$ Assuming $M_{ij}=\xi_{j-i}$, the matrix $Q=F_N^*MF_N$ is indeed diagonal, as shown by the following computation:
\begin{eqnarray*}
Q_{ij}
&=&\sum_{kl}w^{-ik}M_{kl}w^{lj}\\
&=&\sum_{kl}w^{jl-ik}\xi_{l-k}\\
&=&\sum_{kr}w^{j(k+r)-ik}\xi_r\\
&=&\sum_rw^{jr}\xi_r\sum_kw^{(j-i)k}\\
&=&N\delta_{ij}\sum_rw^{jr}\xi_r
\end{eqnarray*}

$(2)\implies(1)$ Assuming $Q=diag(q_1,\ldots,q_N)$, the matrix $M=F_NQF_N^*$ is indeed circulant, as shown by the following computation:
$$M_{ij}
=\sum_kw^{ik}Q_{kk}w^{-jk}
=\sum_kw^{(i-j)k}q_k$$

To be more precise, in this formula the last term depends only on $j-i$, and so shows that we have $M_{ij}=\xi_{j-i}$, with $\xi$ being the following vector:
$$\xi_i
=\sum_kw^{-ik}q_k
=(F_N^*q)_i$$

Thus, we are led to the conclusions in the statement.
\end{proof}

As a basic illustration for the above result, for the circulant matrix $M=\mathbb I_N$ we recover in this way the diagonalization result from Theorem 11.1, namely:
$$\begin{pmatrix}
1&\ldots&\ldots&1\\
\vdots&&&\vdots\\
\vdots&&&\vdots\\
1&\ldots&\ldots&1\end{pmatrix}
=\frac{1}{N}\,F_N
\begin{pmatrix}
N\\
&0\\
&&\ddots\\
&&&0\end{pmatrix}F_N^*$$

There are many other examples, as for instance those coming from the adjacency matrices of the circulant graphs, and with the remark that, up to a copy of $1_N$, the flat matrix $\mathbb I_N$ is indeed of this type. We will leave some exploration here as an exercise.

\bigskip

The above result is something quite powerful, and very useful, and suggests doing everything in Fourier, when dealing with circulant matrices. And we can use here:

\index{discrete Fourier transform}
\index{circulant unitary matrix}
\index{circulant orthogonal matrix}

\begin{theorem}
The various basic sets of $N\times N$ circulant matrices are as follows, with the convention that associated to any $q\in\mathbb C^N$ is the matrix  $Q=diag(q_1,\ldots,q_N)$:
\begin{enumerate}
\item The set of all circulant matrices is:
$$M_N(\mathbb C)^{circ}=\left\{F_NQF_N^*\Big|q\in\mathbb C^N\right\}$$

\item The set of all circulant unitary matrices is:
$$U_N^{circ}=\left\{\frac{1}{N}F_NQF_N^*\Big|q\in\mathbb T^N\right\}$$

\item The set of all circulant orthogonal matrices is:
$$O_N^{circ}=\left\{\frac{1}{N}F_NQF_N^*\Big|q\in\mathbb T^N,\bar{q}_i=q_{-i},\forall i\right\}$$
\end{enumerate}
In addition, in this picture, the first row vector of $F_NQF_N^*$ is given by $\xi=F_N^*q$.
\end{theorem}

\begin{proof}
All this follows from Theorem 11.4, as follows:

\medskip

(1) This assertion, along with the last one, is Theorem 11.4 itself.

\medskip

(2) This is clear from (1), and from the fact that the rescaled matrix $F_N/\sqrt{N}$ is unitary, because the eigenvalues of a unitary matrix must be on the unit circle $\mathbb T$.

\medskip

(3) This follows from (2), because the matrix is real when $\xi_i=\bar{\xi}_i$, and in Fourier transform, $\xi=F_N^*q$, this corresponds to the condition $\bar{q}_i=q_{-i}$.
\end{proof}

There are many other interesting things that can be said about the circulant matrices, along the above lines. Importantly, all this can be generalized to the matrices which are $(N_1,\ldots,N_k)$ patterned, with the matrix doing the job here being as follows:
$$F_{N_1,\ldots,N_k}=F_{N_1}\otimes\ldots\otimes F_{N_k}$$

To be more precise here, in order to have some better notations and formalism, consider a finite abelian group $G$, decomposed as $G=\mathbb Z_{N_1}\times\ldots\times\mathbb Z_{N_k}$. We can then talk about the corresponding Fourier matrix $F_G$, which coincides with the above one:
$$F_G=F_{N_1}\otimes\ldots\otimes F_{N_k}$$

We will discuss more in detail such matrices, which are called generalized Fourier matrices, and are basic examples of complex Hadamard matrices, in a moment. In the meantime, we have the following result, which is standard in discrete Fourier analysis, extending what we previously knew from the above, in the circulant case:

\index{Fourier-diagonal}
\index{discrete Fourier transform}

\begin{theorem}
For a matrix $A\in M_N(\mathbb C)$, the following are equivalent,
\begin{enumerate}
\item $A$ is $G$-invariant, $A_{ij}=\xi_{j-i}$, for a certain vector $\xi\in\mathbb C^N$,

\item $A$ is Fourier-diagonal, $A=F_GQF_G^*$, for a certain diagonal matrix $Q$,
\end{enumerate}
and if so, $\xi=F_G^*q$, where $q\in\mathbb C^N$ is the vector formed by the diagonal entries of $Q$.
\end{theorem}

\begin{proof}
This is something that we know from the above in the cyclic case, $G=\mathbb Z_N$, and the proof in general is similar, by using matrix indices as follows:
$$i,j\in G$$

To be more precise, in order to get started, with our generalization, let us decompose our finite abelian group $G$ as a product of cyclic groups, as follows:
$$G=\mathbb Z_{N_1}\times\ldots\times\mathbb Z_{N_s}$$

The corresponding Fourier matrix decomposes then as well, as follows:
$$F_G=F_{N_1}\otimes\ldots\otimes F_{N_s}$$

Now if we set $w_i=e^{2\pi i/N_i}$, this means that we have the following formula:
$$(F_G)_{ij}=w_1^{i_1j_1}\ldots w_s^{i_sj_s}$$

We can now prove the equivalence in the statement, as follows:

\medskip

$(1)\implies(2)$ Assuming $A_{ij}=\xi_{j-i}$, the matrix $Q=F_G^*AF_G$ is diagonal, as shown by the following computation, with all indices being group elements:
\begin{eqnarray*}
Q_{ij}
&=&\sum_{kl}\overline{(F_G)}_{ki}A_{kl}(F_G)_{lj}\\
&=&\sum_{kl}w_1^{-k_1i_1}\ldots w_s^{-k_si_s}\cdot\xi_{l-k}\cdot w_1^{l_1j_1}\ldots w_s^{l_sj_s}\\
&=&\sum_{kl}w_1^{l_1j_1-k_1i_1}\ldots w_s^{l_sj_s-k_si_s}\xi_{l-k}\\
&=&\sum_{kr}w_1^{(k_1+r_1)j_1-k_1i_1}\ldots w_s^{(k_s+r_s)j_s-k_si_s}\xi_r\\
&=&\sum_rw_1^{r_1j_1}\ldots w_s^{r_sj_s}\xi_r\sum_kw_1^{k_1(j_1-i_1)}\ldots w_s^{k_s(j_s-i_s)}\\
&=&\sum_rw_1^{r_1j_1}\ldots w_s^{r_sj_s}\xi_r\cdot N_1\delta_{i_1j_1}\ldots N_s\delta_{i_sj_s}\\
&=&N\delta_{ij}\sum_r(F_G)_{jr}\xi_r
\end{eqnarray*}

$(2)\implies(1)$ Assuming $Q=diag(q_1,\ldots,q_N)$, the matrix $A=F_GQF_G^*$ is $G$-invariant, as shown by the following computation, again with all indices being group elements:
\begin{eqnarray*}
A_{ij}
&=&\sum_{kl}(F_G)_{ik}Q_{kk}\overline{(F_G)}_{kj}\\
&=&\sum_kw_1^{i_1k_1}\ldots w_s^{i_sk_s}\cdot q_k\cdot w_1^{-j_1k_1}\ldots w_s^{-j_sk_s}\\
&=&\sum_kw_1^{(i_1-j_1)k_1}\ldots w_s^{(i_s-j_s)k_s}q_k
\end{eqnarray*}

To be more precise, in this formula the last term depends only on $j-i$, and so shows that we have $A_{ij}=\xi_{j-i}$, with $\xi$ being the following vector:
\begin{eqnarray*}
\xi_i
&=&\sum_kw_1^{-i_1k_1}\ldots w_s^{-i_sk_s}q_k\\
&=&\sum_k(F_G^*)_{ik}q_k\\
&=&(F_G^*q)_i
\end{eqnarray*}

Thus, we are led to the conclusions in the statement.
\end{proof}

As before with the circulant matrices, there are many illustrations for the above result, as for instance those coming from the adjacency matrices of the graphs having a transitive action of an abelian group. We will leave some exploration here as an exercise.

\section*{11b. Hadamard matrices}

Switching topics, but somehow still in relation with the Fourier matrices, taken in a generalized sense, we would like to discuss now the Hadamard matrices. Let us begin with the real Hadamard matrix case, which is very old, and very interesting too:

\begin{definition}
An Hadamard matrix is a square binary matrix
$$H\in M_N(\pm1)$$
whose rows are pairwise orthogonal.
\end{definition}

Here is a basic example of such a matrix, called first Walsh matrix:
$$W_2=\begin{pmatrix}1&1\\ 1&-1\end{pmatrix}$$

This matrix is quite trivial, of size $2\times2$, but by taking tensor powers of it, we have as examples the higher Walsh matrices as well, having size $2^k\times 2^k$, given by:
$$W_{2^k}=W_2^{\otimes k}$$

Observe that the first Walsh matrix $W_2$ is in fact the Fourier matrix of $\mathbb Z_2$, and that, more generally, the Walsh matrix $W_{2^k}$ is the Fourier matrix of $\mathbb Z_2^k$. We will be back to this, group-theoretical aspects of the Hadamard matrices, later in this chapter.

\bigskip

Getting back to Definition 11.7 as stated, we have seen that we have examples of such matrices at any $N=2^k$. But, what happens then in arbitrary size $N\times N$? It is clear that we must have $2|N$, and along the same lines, it is easy to see, by playing around with the first rows, that once your matrix has $N\geq3$ rows, we must have $4|N$:

\begin{proposition}
The size of an Hadamard matrix $H\in M_N(\pm1)$ must satisfy 
$$N\in\{2\}\cup 4\mathbb N$$
with this coming from the orthogonality condition between the first $3$ rows.
\end{proposition}

\begin{proof}
This is something very standard and intuitive, the idea being that by permuting the rows and columns or by multiplying them by $-1$, as to rearrange the first 3 rows, we can always assume that our matrix looks as follows:
$$H=\begin{pmatrix}
1\ldots\ldots 1&1\ldots\ldots 1&1\ldots\ldots 1&1\ldots\ldots 1\\
1\ldots\ldots 1&1\ldots\ldots 1&-1\ldots -1&-1\ldots -1\\
1\ldots\ldots 1&-1\ldots -1&1\ldots\ldots 1&-1\ldots -1\\
\underbrace{\ldots\ldots\ldots}_x&\underbrace{\ldots\ldots\ldots}_y&\underbrace{\ldots\ldots\ldots}_z&\underbrace{\ldots\ldots\ldots}_t
\end{pmatrix}$$

Now if we denote by $x,y,z,t$ the sizes of the block columns, as indicated above, the orthogonality conditions between the first 3 rows give $x=y=z=t$. We conclude that the matrix size $N=x+y+z+t$ must be a multiple of 4, as claimed.
\end{proof}

The above result is something quite interesting, and the point is that a similar analysis with 4 rows or more does not give any further restriction on the possible values of the size $N\in\mathbb N$. In fact, we are led in this way to the following famous conjecture:

\begin{conjecture}[Hadamard]
There is an Hadamard matrix of order $N$,
$$H\in M_N(\pm1)$$
for any $N\in4\mathbb N$.
\end{conjecture}

Normally this is an analytic question, because in practice the number of Hadamard matrices grows exponentially with $N$, and so in order to prove the conjecture, you just need a modest lower estimate on this number. But, no one knows how to do this, and this despite the Hadamard conjecture being open for more than 100 years.

\bigskip

This being said, what we can do with our algebraic methods is to verify at least the Hadamard conjecture at small values of $N\in4\mathbb N$. And here, with $N=4,8$ being solved by the Walsh matrices, we are faced with constructing a matrix at $N=12$.

\bigskip

In order to solve this question, let $q=p^k$ be an odd prime power, and set: 
$$\chi(a)=\begin{cases}
0&{\rm if}\ a=0\\
1&{\rm if}\ a=b^2,b\neq0\\
-1&{\rm otherwise}
\end{cases}$$

Then set $Q_{ab}=\chi(b-a)$, with indices in $\mathbb F_q$. With these conventions, the Paley construction of Hadamard matrices, which works well at $N=12$, is as follows:

\begin{theorem}
Given an odd prime power $q=p^k$, construct $Q_{ab}=\chi(b-a)$ as above. We have then constructions of Hadamard matrices, as follows:
\begin{enumerate}
\item Paley $1$: if $q=3(4)$ we have a matrix of size $N=q+1$, as follows:
$$P_N^1=1+\begin{pmatrix}
0&1&\ldots&1\\
-1\\
\vdots&&Q\\
-1
\end{pmatrix}$$

\item Paley $2$: if $q=1(4)$ we have a matrix of size $N=2q+2$, as follows:
$$P_N^2=\begin{pmatrix}
0&1&\ldots&1\\
1\\
\vdots&&Q\\
1
\end{pmatrix}\quad:\quad 0\to\begin{pmatrix}1&-1\\ -1&-1\end{pmatrix}\quad,\quad\pm1\to\pm\begin{pmatrix}1&1\\1&-1\end{pmatrix}$$
\end{enumerate}
These matrices are skew-symmetric $(H+H^t=2)$, respectively symmetric $(H=H^t)$.
\end{theorem}

\begin{proof}
In order to simplify the presentation, we denote by $1$ all the identity matrices, of any size, and by $\mathbb I$ all the rectangular all-one matrices, of any size as well. It is elementary to check that the matrix $Q_{ab}=\chi(a-b)$ has the following properties:
$$QQ^t=q1-\mathbb I\quad,\quad 
Q\mathbb I=\mathbb IQ=0$$

In addition, we have the following formulae, which are elementary as well, coming from the fact that $-1$ is a square in $\mathbb F_q$ precisely when $q=1(4)$:
$$q=1(4)\implies Q=Q^t\quad,\quad
q=3(4)\implies Q=-Q^t$$

With these observations in hand, the proof goes as follows:

\medskip

(1) With our above conventions for $1$ and $\mathbb I$, the matrix in the statement is:
$$P_N^1=\begin{pmatrix}1&\mathbb I\\ -\mathbb I&1+Q\end{pmatrix}$$
 
With this formula in hand, the Hadamard matrix condition follows from:
\begin{eqnarray*}
P_N^1(P_N^1)^t
&=&\begin{pmatrix}1&\mathbb I\\ -\mathbb I&1+Q\end{pmatrix}\begin{pmatrix}1&-\mathbb I\\ \mathbb I&1-Q\end{pmatrix}\\
&=&\begin{pmatrix}N&0\\ 0&\mathbb I+1-Q^2\end{pmatrix}\\
&=&\begin{pmatrix}N&0\\ 0&N\end{pmatrix}
\end{eqnarray*}

(2) If we denote by $G,F$ the $2\times2$ matrices in the statement, which replace respectively the $0,1$ entries, then we have the following formula for our matrix:
$$P_N^2=\begin{pmatrix}0&\mathbb I\\ \mathbb I&Q\end{pmatrix}\otimes F+1\otimes G$$

With this formula in hand, the Hadamard matrix condition follows from:
\begin{eqnarray*}
(P_N^2)^2
&=&\begin{pmatrix}0&\mathbb I\\ \mathbb I&Q\end{pmatrix}^2\otimes F^2+\begin{pmatrix}1&0\\ 0&1\end{pmatrix}\otimes G^2+\begin{pmatrix}0&\mathbb I\\ \mathbb I&Q\end{pmatrix}\otimes(FG+GF)\\
&=&\begin{pmatrix}q&0\\ 0&q\end{pmatrix}\otimes 2+\begin{pmatrix}1&0\\ 0&1\end{pmatrix}\otimes 2+\begin{pmatrix}0&\mathbb I\\ \mathbb I&Q\end{pmatrix}\otimes0\\
&=&\begin{pmatrix}N&0\\ 0&N\end{pmatrix}
\end{eqnarray*}

Finally, the last assertion is clear, from the above formulae relating $Q,Q^t$.
\end{proof}

In practice now, by using the Walsh and Paley constructions, the next problem when verifying the Hadamard Conjecture appears at $N=92$. But here, we have:

\begin{theorem}
Assuming that $A,B,C,D\in M_K(\pm1)$ are circulant, symmetric, pairwise commute and satisfy the condition
$$A^2+B^2+C^2+D^2=4K$$
the following $4K\times4K$ matrix is Hadamard, called of Williamson type:
$$H=\begin{pmatrix}
A&B&C&D\\
-B&A&-D&C\\
-C&D&A&-B\\
-D&-C&B&A
\end{pmatrix}$$
Moreover, matrices $A,B,C,D$ as above exist at $K=23$, where $4K=92$.
\end{theorem}

\begin{proof}
Consider the standard quaternion units $1,i,j,k\in M_4(0,1)$. These matrices are by definition those describing the positions of the $A,B,C,D$ entries in the matrix $H$ from the statement, and so this matrix can be written as follows:
$$H=A\otimes 1+B\otimes i+C\otimes j+D\otimes k$$

Assuming now that $A,B,C,D$ are symmetric, we have:
\begin{eqnarray*}
HH^t
&=&(A\otimes 1+B\otimes i+C\otimes j+D\otimes k)\\
&&(A\otimes 1-B\otimes i-C\otimes j-D\otimes k)\\
&=&(A^2+B^2+C^2+D^2)\otimes 1-([A,B]-[C,D])\otimes i\\
&&-([A,C]-[B,D])\otimes j-([A,D]-[B,C])\otimes k
\end{eqnarray*}

Now assume that our matrices $A,B,C,D$ pairwise commute, and satisfy the condition in the statement. In this case, it follows from the above formula that we have:
$$HH^t=4K$$

Thus, we obtain indeed an Hadamard matrix, as claimed. However, finding such matrices is in general a difficult task, and this is where Williamson's extra assumption in the statement, that $A,B,C,D$ should be taken circulant, comes from. Finally, regarding the $K=23$ and $N=92$ example, this comes via a computer search. 
\end{proof}

At higher $N$ things become more technical, and more complicated constructions, along the lines of those of Paley and Williamson, are needed. Quite curiously, as of now, early 21th century, the human knowledge stops at the number of the beast, namely:
$$\mathfrak N=666$$

That is, explicit examples of Hadamard matrices have been constructed for all multiples of four $N\leq 664$, but no such matrix is known so far at $N=668$. 

\bigskip

Switching topics now, another well-known open question concerns the circulant case. Given a binary vector $\gamma\in(\pm 1)^N$, one can ask whether the matrix $H\in M_N(\pm 1)$ defined by $H_{ij}=\gamma_{j-i}$ is Hadamard or not. Here is a solution to the problem:
$$K_4=\begin{pmatrix}-1&1&1&1\\ 1&-1&1&1\\ 1&1&-1&1\\ 1&1&1&-1\end{pmatrix}$$

More generally, any vector $\gamma\in(\pm 1)^4$ satisfying $\sum\gamma_i=\pm 1$ is a solution to the problem. The following conjecture, from the 50s, states that there are no other solutions:

\index{Circulant Hadamard conjecture}
\index{CHC}

\begin{conjecture}[Circulant Hadamard Conjecture (CHC)]
The only Hadamard matrices which are circulant are
$$K_4=\begin{pmatrix}-1&1&1&1\\ 1&-1&1&1\\ 1&1&-1&1\\ 1&1&1&-1\end{pmatrix}$$
and its conjugates, regardless of the value of $N\in\mathbb N$.
\end{conjecture}

The fact that such a simple-looking problem is still open might seem quite surprising. Indeed, if we denote by $S\subset\{1,\ldots,N\}$ the set of positions of the $-1$ entries of $\gamma$, the Hadamard matrix condition is simply, for any $k\neq 0$, taken modulo $N$:
$$|S\cap(S+k)|=|S|-N/4$$ 

Thus, the above conjecture simply states that at $N\neq 4$, such a set $S$ cannot exist. This is a well-known problem in combinatorics, raised by Ryser a long time ago.

\section*{11c. Complex Hadamard}

We have seen that the theory of real Hadamard matrices leads to many difficult questions. As a further twist to the plot, bringing some sort of solution to this, we have:

\index{Fourier matrix}
\index{complex Hadamard matrix}
\index{Hadamard conjecture}

\begin{theorem}
When enlarging the attention to the complex Hadamard matrices, $H\in M_N(\mathbb T)$ having the rows pairwise orthogonal, the Fourier matrix,
$$F_N=
\begin{pmatrix}
1&1&1&\ldots&1\\
1&w&w^2&\ldots&w^{N-1}\\
1&w^2&w^4&\ldots&w^{2(N-1)}\\
\vdots&\vdots&\vdots&&\vdots\\
1&w^{N-1}&w^{2(N-1)}&\ldots&w^{(N-1)^2}
\end{pmatrix}$$
with $w=e^{2\pi i/N}$, provides an example of such a matrix, at any $N\in\mathbb N$. Thus, the Hadamard Conjecture problematics dissapears, in the complex setting.
\end{theorem}

\begin{proof}
We have seen before that the rescaling $U=F_N/\sqrt{N}$ is unitary. Thus the rows of $U$ are pairwise orthogonal, and so follow to be the rows of $F_N$.
\end{proof}

In view of the above result, let us study more in detail the complex Hadamard matrices. Many examples can be constructed, quite often by using the combinatorics of roots of unity, and as a basic example here, we have the tensor products of Fourier matrices:
$$F_{N_1,\ldots,N_k}=F_{N_1}\otimes\ldots\otimes F_{N_k}$$ 

Moreover, we have as well deformations, and we are led in this way to:

\begin{theorem}
The only Hadamard matrices at $N=4$ are, up to equivalence
$$F_4^q=\begin{pmatrix}
1&1&1&1\\
1&-1&1&-1\\
1&q&-1&-q\\ 
1&-q&-1&q
\end{pmatrix}$$
with $q\in\mathbb T$, which appear as suitable deformations of $W_4=F_2\otimes F_2$.
\end{theorem}

\begin{proof}
This is something quite self-explanatory, and we will leave working out all this, namely finding the correct meaning of the equivalence relation, and of the deformation notion used, along of course with the proof, as an instructive exercise.
\end{proof}

Moving on, at $N=5$ the situation is considerably more complicated, with $F_5$ being the only matrix. The key technical result here, due to Haagerup, is as follows:

\begin{proposition}
Given an Hadamard matrix $H\in M_5(\mathbb T)$, chosen dephased,
$$H=\begin{pmatrix}
1&1&1&1&1\\
1&a&x&*&*\\
1&y&b&*&*\\
1&*&*&*&*\\
1&*&*&*&*
\end{pmatrix}$$
the numbers $a,b,x,y$ must satisfy the equation $(x-y)(x-ab)(y-ab)=0$.
\end{proposition}

\begin{proof}
Consider the upper 3-row truncation of $H$, which looks as follows:
$$H'=\begin{pmatrix}
1&1&1&1&1\\
1&a&x&p&q\\
1&y&b&r&s
\end{pmatrix}$$

By using the orthogonality of the rows, we have:
$$(1+a+x)(1+\bar{b}+\bar{y})(1+\bar{a}y+b\bar{x})
=-(p+q)(r+s)(\bar{p}r+\bar{q}s)$$

On the other hand, by using $p,q,r,s\in\mathbb T$, we have:
\begin{eqnarray*}
(p+q)(r+s)(\bar{p}r+\bar{q}s)
&=&(r+p\bar{q}s+\bar{p}qr+s)(\bar{r}+\bar{s})\\
&=&1+p\bar{q}\bar{r}s+\bar{p}q+\bar{r}s+r\bar{s}+p\bar{q}+\bar{p}qr\bar{s}+1\\
&=&2Re(1+p\bar{q}+r\bar{s}+p\bar{q}r\bar{s})\\
&=&2Re[(1+p\bar{q})(1+r\bar{s})]
\end{eqnarray*}

We conclude that we have the following formula, involving $a,b,x,y$ only:
$$(1+a+x)(1+\bar{b}+\bar{y})(1+\bar{a}y+b\bar{x})\in\mathbb R$$

Now this is a product of type $(1+\alpha)(1+\beta)(1+\gamma)$, with the first summand being 1, and with the last summand, namely $\alpha\beta\gamma$, being real as well, as shown by the above general $p,q,r,s\in\mathbb T$ computation. Thus, when expanding, and we are left with:
\begin{eqnarray*}
&&(a+x)+(\bar{b}+\bar{y})+(\bar{a}y+b\bar{x})+(a+x)(\bar{b}+\bar{y})\\
&+&(a+x)(\bar{a}y+b\bar{x})+(\bar{b}+\bar{y})(\bar{a}y+b\bar{x})\in\mathbb R
\end{eqnarray*}

By expanding all the products, our formula looks as follows:
\begin{eqnarray*}
&&a+x+\bar{b}+\bar{y}+\bar{a}y+b\bar{x}+a\bar{b}+a\bar{y}+\bar{b}x+x\bar{y}\\
&+&1+ab\bar{x}+\bar{a}xy+b+\bar{a}\bar{b}y+\bar{x}+\bar{a}+b\bar{x}\bar{y}\in\mathbb R
\end{eqnarray*}

By removing from this all terms of type $z+\bar{z}$, we are left with:
$$a\bar{b}+x\bar{y}+ab\bar{x}+\bar{a}\bar{b}y+\bar{a}xy+b\bar{x}\bar{y}\in\mathbb R$$

Now by getting back to our Hadamard matrix, all this remains true when transposing it, which amounts in interchanging $x\leftrightarrow y$. Thus, we have as well:
$$a\bar{b}+\bar{x}y+ab\bar{y}+\bar{a}\bar{b}x+\bar{a}xy+b\bar{x}\bar{y}\in\mathbb R$$

By substracting now the two equations that we have, we obtain:
$$x\bar{y}-\bar{x}y+ab(\bar{x}-\bar{y})+\bar{a}\bar{b}(y-x)\in\mathbb R$$

Now observe that this number, say $Z$, is purely imaginary, because $\bar{Z}=-Z$. Thus our equation reads $Z=0$. On the other hand, we have the following formula:
\begin{eqnarray*}
abxyZ
&=&abx^2-aby^2+a^2b^2(y-x)+xy(y-x)\\
&=&(y-x)(a^2b^2+xy-ab(x+y))\\
&=&(y-x)(ab-x)(ab-y)
\end{eqnarray*}

Thus, our equation $Z=0$ corresponds to the formula in the statement.
\end{proof}

We are led in this way to the following theorem, due to Haagerup:

\begin{theorem}
The only Hadamard matrix at $N=5$ is the Fourier matrix,
$$F_5=\begin{pmatrix}
1&1&1&1&1\\
1&w&w^2&w^3&w^4\\
1&w^2&w^4&w&w^3\\
1&w^3&w&w^4&w^2\\
1&w^4&w^3&w^2&w
\end{pmatrix}$$
with $w=e^{2\pi i/5}$, up to the standard equivalence relation for such matrices. 
\end{theorem}

\begin{proof}
Assume that have an Hadamard matrix $H\in M_5(\mathbb T)$, chosen dephased, and written as in Proposition 11.15, with emphasis on the upper left $2\times2$ subcorner.

\medskip

(1) We know from Proposition 11.15, applied to $H$ itself, and to its transpose $H^t$ as well, that the entries $a,b,x,y$ must satisfy the following equations:
$$(a-b)(a-xy)(b-xy)=0$$
$$(x-y)(x-ab)(y-ab)=0$$

Our first claim is that, by doing some combinatorics, we can actually obtain from this $a=b$ and $x=y$, up to the equivalence relation for the Hadamard matrices: 
$$H\sim\begin{pmatrix}
1&1&1&1&1\\
1&a&x&*&*\\
1&x&a&*&*\\
1&*&*&*&*\\
1&*&*&*&*
\end{pmatrix}$$

Indeed, the above two equations lead to 9 possible cases, the first of which is, as desired, $a=b$ and $x=y$. As for the remaining 8 cases, here again things are determined by 2 parameters, and in practice, we can always permute the first 3 rows and 3 columns, and then dephase our matrix, as for our matrix to take the above special form.

\medskip

(2) With this result in hand, the combinatorics of the scalar products between the first 3 rows, and between the first 3 columns as well, becomes something which is quite simple to investigate. By doing a routine study here, and then completing it with a study of the lower right $2\times2$ corner as well, we are led to 2 possible cases, as follows:
$$H\sim\begin{pmatrix}
1&1&1&1&1\\
1&a&b&c&d\\
1&b&a&d&c\\
1&c&d&a&b\\
1&d&c&b&a
\end{pmatrix}\quad,\quad 
H\sim\begin{pmatrix}
1&1&1&1&1\\
1&a&b&c&d\\
1&b&a&d&c\\
1&c&d&b&a\\
1&d&c&a&b
\end{pmatrix}$$

(3) Our claim now is that the first case is in fact not possible. Indeed, we must have:
\begin{eqnarray*}
a+b+c+d&=&-1\\
2Re(a\bar{b})+2Re(c\bar{d})&=&-1\\
2Re(a\bar{c})+2Re(b\bar{d})&=&-1\\
2Re(a\bar{d})+2Re(b\bar{c})&=&-1
\end{eqnarray*}

Now since $|Re(x)|\leq1$ for any $x\in\mathbb T$, we deduce from the second equation that:
$$Re(a\bar{b})\leq 1/2$$

In other words, the arc length between $a,b$ satisfies:
$$\theta(a,b)\geq\pi/3$$

The same argument applies to $c,d$, and to the other pairs of numbers in the last 2 equations. Now since our equations are invariant under permutations of $a,b,c,d$, we can assume that $a,b,c,d$ are ordered in this way on the unit circle, and by the above, separated by $\geq\pi/3$ arc lengths. But this tells us that we have the following inequalities:
$$\theta(a,c)\geq 2\pi/3\quad,\quad 
\theta(b,d)\geq 2\pi/3$$

These two inequalities give the following estimates:
$$Re(a\bar{c})\leq-1/2\quad,\quad 
Re(b\bar{d})\leq-1/2$$

But these estimates contradict the third equation. Thus, our claim is proved.

\medskip

(4) Summarizing, we have proved so far that our matrix must be as follows:
$$H\sim\begin{pmatrix}
1&1&1&1&1\\
1&a&b&c&d\\
1&b&a&d&c\\
1&c&d&b&a\\
1&d&c&a&b
\end{pmatrix}$$

We are now in position of finishing. The orthogonality equations are as follows:
\begin{eqnarray*}
a+b+c+d&=&-1\\
2Re(a\bar{b})+2Re(c\bar{d})&=&-1\\
a\bar{c}+c\bar{b}+b\bar{d}+d\bar{a}&=&-1
\end{eqnarray*}

The third equation can be written in the following equivalent form:
\begin{eqnarray*}
Re[(a+b)(\bar{c}+\bar{d})]&=&-1\\
Im[(a-b)(\bar{c}-\bar{d})]&=&0
\end{eqnarray*}

By using now $a,b,c,d\in\mathbb T$, we obtain from this:
$$\frac{a+b}{a-b}\in i\mathbb R\quad,\quad 
\frac{c+d}{c-d}\in i\mathbb R$$

Thus we can find $s,t\in\mathbb R$ such that:
$$a+b=is(a-b)\quad,\quad 
c+d=it(c-d)$$

By plugging in these values, our system of equations simplifies, as follows:
\begin{eqnarray*}
(a+b)+(c+d)&=&-1\\
|a+b|^2+|c+d|^2&=&3\\
(a+b)(\bar{c}+\bar{d})&=&-1
\end{eqnarray*}

Now observe that the last equation implies in particular that we have:
$$|a+b|^2\cdot|c+d|^2=1$$

Thus $|a+b|^2,|c+d|^2$ must be roots of the following polynomial:
$$X^2-3X+1=0$$

But this gives the following equality of sets:
$$\Big\{|a+b|\,,\,|c+d|\Big\}=\left\{\frac{\sqrt{5}+1}{2}\,,\,\frac{\sqrt{5}-1}{2}\right\}$$

This is good news, because we are now into 5-th roots of unity. To be more precise, we have 2 cases to be considered, the first one being as follows, with $z\in\mathbb T$:
$$a+b=\frac{\sqrt{5}+1}{2}\,z\quad,\quad 
c+d=-\frac{\sqrt{5}-1}{2}\,z$$

From $a+b+c+d=-1$ we obtain $z=-1$, and by using this we obtain $b=\bar{a}$, $d=\bar{c}$. Thus we have the following formulae:
$$Re(a)=\cos(2\pi/5)\quad,\quad 
Re(c)=\cos(\pi/5)$$

We conclude that we have $H\sim F_5$, as claimed. As for the second case, with $a,b$ and $c,d$ interchanged, this leads to $H\sim F_5$ as well.
\end{proof}

In view of the above, we are led to the question of finding the Hadamard matrices which are ``isolated''. Let us begin with some notations. We denote by $X_p$ an unspecified neighborhood of a point in a manifold, $p\in X$. Also, for $q\in\mathbb T_1$, meaning that $q\in\mathbb T$ is close to $1$, we define $q^r$ with $r\in\mathbb R$ by $(e^{it})^r=e^{itr}$. With these conventions, we have:

\begin{proposition}
For $H\in X_N$ and $A\in M_N(\mathbb R)$, the following are equivalent:
\begin{enumerate}
\item The following is an Hadamard matrix, for any $q\in\mathbb T_1$:
$$H_{ij}^q=H_{ij}q^{A_{ij}}$$

\item The following equations hold, for any $i\neq j$ and any $q\in\mathbb T_1$:
$$\sum_kH_{ik}\bar{H}_{jk}q^{A_{ik}-A_{jk}}=0$$

\item The following equations hold, for any $i\neq j$ and any $\varphi:\mathbb R\to\mathbb C$:
$$\sum_kH_{ik}\bar{H}_{jk}\varphi(A_{ik}-A_{jk})=0$$

\item For any $i\neq j$ and any $r\in\mathbb R$, with $E_{ij}^r=\{k|A_{ik}-A_{jk}=r\}$, we have:
$$\sum_{k\in E_{ij}^r}H_{ik}\bar{H}_{jk}=0$$
\end{enumerate}
If these conditions are satisfied, we call the matrix $H^q$ an affine deformation of $H$.
\end{proposition}

\begin{proof}
These equivalences are all elementary, and can be proved as follows:

\medskip

$(1)\iff(2)$ Indeed, the scalar products between the rows of $H^q$ are:
\begin{eqnarray*}
<H^q_i,H^q_j>
&=&\sum_kH_{ik}q^{A_{ik}}\bar{H}_{jk}\bar{q}^{A_{jk}}\\
&=&\sum_kH_{ik}\bar{H}_{jk}q^{A_{ik}-A_{jk}}
\end{eqnarray*}

$(2)\implies(4)$ This follows from the following formula, and from the fact that the power functions $\{q^r|r\in\mathbb R\}$ over the unit circle $\mathbb T$ are linearly independent:
$$\sum_kH_{ik}\bar{H}_{jk}q^{A_{ik}-A_{jk}}=\sum_{r\in\mathbb R}q^r\sum_{k\in E_{ij}^r}H_{ik}\bar{H}_{jk}$$

$(4)\implies(3)$ This follows from the following formula:
$$\sum_kH_{ik}\bar{H}_{jk}\varphi(A_{ik}-A_{jk})=\sum_{r\in\mathbb R}\varphi(r)\sum_{k\in E_{ij}^r}H_{ik}\bar{H}_{jk}$$

$(3)\implies(2)$ This simply follows by taking $\varphi(r)=q^r$.
\end{proof}

In order to understand the above deformations, which are ``affine'' in a certain sense, as suggested at the end of the statement, it is convenient to enlarge the attention to all types of deformations. We keep using the neighborhood notation $X_p$ introduced above, and we consider functions of type $f:X_p\to Y_q$, which by definition satisfy $f(p)=q$. We have the following definition, further clarifying the terminology in Proposition 11.17:

\index{affine deformation}
\index{trivial deformation}

\begin{definition}
Let $H\in M_N(\mathbb C)$ be a complex Hadamard matrix.
\begin{enumerate}
\item A deformation of $H$ is a smooth function $f:\mathbb T_1\to (X_N)_H$.

\item The deformation is called ``affine'' if $f_{ij}(q)=H_{ij}q^{A_{ij}}$, with $A\in M_N(\mathbb R)$.

\item We call ``trivial'' the deformations of type $f_{ij}(q)=H_{ij}q^{a_i+b_j}$, with $a,b\in\mathbb R^N$.
\end{enumerate}
\end{definition}

Here the adjective ``affine'', which is used in the same context as in Proposition 11.17, comes from the formula $f_{ij}(e^{it})=H_{ij}e^{iA_{ij}t}$, because the function $t\to A_{ij}t$ which produces the exponent is indeed affine. As for the adjective ``trivial'', this comes from the fact that the affine deformations of type $f(q)=(H_{ij}q^{a_i+b_j})_{ij}$ are obtained from $H$ by multiplying the rows and columns by certain numbers in $\mathbb T$, so are automatically Hadamard.

\bigskip

Getting now to the examples, and skipping some elementary theory and constructions, we have an interesting construction due to McNulty and Weigert, that we would like to explain now. This construction is based on the following simple fact:

\begin{theorem}
Assuming that $K\in M_N(\mathbb C)$ is Hadamard, so is the matrix
$$H_{ia,jb}=\frac{1}{\sqrt{Q}}K_{ij}(L_i^*R_j)_{ab}$$
provided that $\{L_1,\ldots,L_N\}\subset\sqrt{Q}U_Q$ and $\{R_1,\ldots,R_N\}\subset\sqrt{Q}U_Q$ are such that
$$\frac{1}{\sqrt{Q}}L_i^*R_j\in\sqrt{Q}U_Q$$
with $i,j=1,\ldots,N$, are complex Hadamard.
\end{theorem}

\begin{proof}
The check of the unitarity of the matrix in the statement can be done as follows, by using our various assumptions on the various matrices involved:
\begin{eqnarray*}
<H_{ia},H_{kc}>
&=&\frac{1}{Q}\sum_{jb}K_{ij}(L_i^*R_j)_{ab}\bar{K}_{kj}\overline{(L_k^*R_j)}_{cb}\\
&=&\sum_jK_{ij}\bar{K}_{kj}(L_i^*L_k)_{ac}\\
&=&N\delta_{ik}(L_i^*L_k)_{ac}\\
&=&NQ\delta_{ik}\delta_{ac}
\end{eqnarray*}

The entries of our matrix being in addition on the unit circle, we are done.
\end{proof}

As a concrete input for the above construction, we can use:

\begin{proposition}
For $q\geq3$ prime, the matrices $\{F_q,DF_q,\ldots,D^{q-1}F_q\}$, where $F_q$ is the Fourier matrix, and where
$$D=diag\left(1,1,w,w^3,w^6,w^{10},\ldots,w^{\frac{q^2-1}{8}},\ldots,w^{10},w^6,w^3,w\right)$$
with $w=e^{2\pi i/q}$, are such that $\frac{1}{\sqrt{q}}E_i^*E_j$ is complex Hadamard, for any $i\neq j$.
\end{proposition}

\begin{proof}
With by definition $0,1,\ldots,q-1$ as indices for our matrices, as usual in a Fourier analysis context, the formula of the above matrix $D$ is:
$$D_c
=w^{0+1+\ldots+(c-1)}
=w^{\frac{c(c-1)}{2}}$$

Since we have $\frac{1}{\sqrt{q}}E_i^*E_j\in\sqrt{q}U_q$, we just need to check that these matrices have entries belonging to $\mathbb T$, for any $i\neq j$. With $k=j-i$, these entries are given by:
$$\frac{1}{\sqrt{q}}(E_i^*E_j)_{ab}
=\frac{1}{\sqrt{q}}(F_q^*D^kF_q)_{ab}
=\frac{1}{\sqrt{q}}\sum_cw^{c(b-a)}D_c^k$$

Now observe that with $s=b-a$, we have the following formula:
\begin{eqnarray*}
\left|\sum_cw^{cs}D_c^k\right|^2
&=&\sum_{cd}w^{cs-ds}w^{\frac{c(c-1)}{2}\cdot k-\frac{d(d-1)}{2}\cdot k}\\
&=&\sum_{cd}w^{(c-d)\left(\frac{c+d-1}{2}\cdot k+s\right)}\\
&=&\sum_{de}w^{e\left(\frac{2d+e-1}{2}\cdot k+s\right)}\\
&=&\sum_e\left(w^{\frac{e(e-1)}{2}\cdot k+es}\sum_dw^{edk}\right)\\
&=&\sum_ew^{\frac{e(e-1)}{2}\cdot k+es}\cdot q\delta_{e0}\\
&=&q
\end{eqnarray*}

Thus the entries are on the unit circle, and we are done.
\end{proof}

Next, we have the following result, making use of Gauss sums:

\begin{proposition}
The matrices $G_k=\frac{1}{\sqrt{q}}F_q^*D^kF_q$, with $D=diag\left(w^{\frac{c(c-1)}{2}}\right)$, and with $k\neq0$ are circulant, their first row vectors $V^k$ being given by
$$V^k_i=\delta_q\left(\frac{k/2}{q}\right)w^{\frac{q^2-1}{8}\cdot k}\cdot w^{-\frac{\frac{i}{k}(\frac{i}{k}-1)}{2}}$$
where $\delta_q=1$ if $q=1(4)$ and $\delta_q=i$ if $q=3(4)$, and with all inverses being taken in $\mathbb Z_q$.
\end{proposition}

\begin{proof}
The above matrices $G_k$ are indeed circulant, their first vectors being:
$$V^k_i=\frac{1}{\sqrt{q}}\sum_cw^{\frac{c(c-1)}{2}\cdot k+ic}$$

But this is a Gauss sum, and by computing the square, we obtain:
$$(V^k_i)^2=\delta_q^2\cdot w^{\frac{q^2-1}{4}\cdot k}\cdot w^{-\frac{i}{k}(\frac{i}{k}-1)}$$

By extracting now the square root, we obtain a formula as follows:
$$V^k_i=\pm\delta_q\cdot w^{\frac{q^2-1}{8}\cdot k}\cdot w^{-\frac{\frac{i}{k}(\frac{i}{k}-1)}{2}}$$

And with Gauss computing for us the sign, this leads to the above formula.
\end{proof}

Let us combine now all the above results. We obtain the following statement:

\index{McNulty-Weigert matrix}
\index{Tao matrix}
\index{isolated matrix}

\begin{theorem}
Let $q\geq3$ be prime, consider subsets $S,T\subset\{0,1,\ldots,q-1\}$ satisfying the conditions $|S|=|T|$ and $S\cap T=\emptyset$, and write:
$$S=\{s_1,\ldots,s_N\}\quad,\quad 
T=\{t_1,\ldots,t_N\}$$
Then, with the matrix $V$ being as above, the following matrix,
$$H_{ia,jb}=K_{ij}V^{t_j-s_i}_{b-a}$$
is complex Hadamard, provided that the matrix $K\in M_N(\mathbb C)$ is complex Hadamard.
\end{theorem}

\begin{proof}
This follows indeed by using the general construction in Theorem 11.19, with input coming from Proposition 11.20 and Proposition 11.21.
\end{proof}

The above construction covers many interesting examples of Hadamard matrices, known to be isolated, such as the Tao matrix, which is as follows, with $w=e^{2\pi i/3}$:
$$T_6=\begin{pmatrix}
1&1&1&1&1&1\\ 
1&1&w&w&w^2&w^2\\ 
1&w&1&w^2&w^2&w\\
1&w&w^2&1&w&w^2\\ 
1&w^2&w^2&w&1&w\\ 
1&w^2&w&w^2&w&1
\end{pmatrix}$$

For more on Hadamard matrices, real and complex, there are many texts available.

\section*{11d. Bistochastic matrices} 

Switching again topics, but with discrete Fourier analysis being as usual around the corner, a very basic definition, that you might already know, is as follows:

\index{row-stochastic}
\index{column-stochastic}
\index{bistochastic}

\begin{definition}
A square matrix $M\in M_N(\mathbb C)$ is called bistochastic if each row and each column sum up to the same number:
$$\begin{matrix}
M_{11}&\ldots&M_{1N}&\to&\lambda\\
\vdots&&\vdots\\
M_{N1}&\ldots&M_{NN}&\to&\lambda\\
\downarrow&&\downarrow\\
\lambda&&\lambda
\end{matrix}$$
If this happens only for the rows or columns, $M$ is called row or column stochastic.
\end{definition}

As the name indicates, the above matrices are useful in statistics, with the case of the matrices having entries in $[0,1]$, summing up to $\lambda=1$, being the important one. As a basic example of a bistochastic matrix, we have the flat matrix, which is as follows:
$$\mathbb I_N=\begin{pmatrix}
1&\ldots&1\\
\vdots&&\vdots\\
1&\ldots&1
\end{pmatrix}$$

Observe that the rescaling $P_N=\mathbb I_N/N$ has the property mentioned above, namely entries in $[0,1]$, summing up to $\lambda=1$. In fact, this matrix $P_N=\mathbb I_N/N$ is a very familiar object in linear algebra, being the projection on the all-one vector, namely:
$$\xi=\begin{pmatrix}
1\\
\vdots\\
1
\end{pmatrix}$$ 

Getting back now to the general case, the various notions of stochasticity in Definition 11.23 are closely related to this vector $\xi$, due to the following simple fact:

\begin{proposition}
Let $M\in M_N(\mathbb C)$ be a square matrix.
\begin{enumerate}
\item $M$ is row stochastic, with sums $\lambda$, when $M\xi=\lambda\xi$.

\item $M$ is column stochastic, with sums $\lambda$, when $M^t\xi=\lambda\xi$.

\item $M$ is bistochastic, with sums $\lambda$, when $M\xi=M^t\xi=\lambda\xi$.
\end{enumerate}
\end{proposition}

\begin{proof}
The first assertion is clear from definitions, because when multiplying a matrix by $\xi$, we obtain the vector formed by the row sums:
$$\begin{pmatrix}
M_{11}&\ldots&M_{1N}\\
\vdots&&\vdots\\
M_{N1}&\ldots&M_{NN}
\end{pmatrix}
\begin{pmatrix}
1\\
\vdots\\
1
\end{pmatrix}
=\begin{pmatrix}
M_{11}+\ldots+M_{1N}\\
\vdots\\
M_{N1}+\ldots+M_{NN}
\end{pmatrix}$$

As for the second, and then third assertion, these are both clear from this.
\end{proof}

As an observation here, we can reformulate if we want the above statement in a purely matrix-theoretic form, by using the flat matrix $\mathbb I_N$, as follows:

\index{flat matrix}

\begin{proposition}
Let $M\in M_N(\mathbb C)$ be a square matrix.
\begin{enumerate}
\item $M$ is row stochastic, with sums $\lambda$, when $M\mathbb I_N=\lambda\mathbb I_N$.

\item $M$ is column stochastic, with sums $\lambda$, when $\mathbb I_NM=\lambda\mathbb I_N$.

\item $M$ is bistochastic, with sums $\lambda$, when $M\mathbb I_N=\mathbb I_NM=\lambda\mathbb I_N$.
\end{enumerate}
\end{proposition}

\begin{proof}
This follows from Proposition 11.24, and from the fact that both the rows and columns of the flat matrix $\mathbb I_N$ are copies of the all-one vector, namely:
$$\xi=\begin{pmatrix}
1\\
\vdots\\
1
\end{pmatrix}$$

Alternatively, we have the following formula, with $S_1,\ldots,S_N$ being the row sums of our matrix $M$, which gives the assertion (1):
$$\begin{pmatrix}
M_{11}&\ldots&M_{1N}\\
\vdots&&\vdots\\
M_{N1}&\ldots&M_{NN}
\end{pmatrix}
\begin{pmatrix}
1&\ldots&1\\
\vdots&&\vdots\\
1&\ldots&1
\end{pmatrix}
=\begin{pmatrix}
S_1&\ldots&S_1\\
\vdots&&\vdots\\
S_N&\ldots&S_N
\end{pmatrix}$$

As for the second, and then third assertion, these are both clear from this.
\end{proof}

In what follows, we will be mainly interested in the bistochastic matrices which are unitary, $M\in U_N$. As the simplest example here, which is a familiar object in quantum physics, we have the following matrix $K_N\in O_N$, obtained by suitably modifying the flat matrix $\mathbb I_N$, as to make the rows pairwise orthogonal, and of norm one:
$$K_N=\frac{1}{N}\begin{pmatrix}
2-N&2&\ldots&2\\
2&\ddots&\ddots&\vdots\\
\vdots&\ddots&\ddots&2\\
2&\ldots&2&2-N
\end{pmatrix}$$

As a first result now regarding the unitary bistochastic matrices, we have:

\begin{theorem}
For a unitary matrix $U\in U_N$, the following are equivalent:
\begin{enumerate}
\item $H$ is bistochastic, with sums $\lambda$.

\item $H$ is row stochastic, with sums $\lambda$, and $|\lambda|=1$.

\item $H$ is column stochastic, with sums $\lambda$, and $|\lambda|=1$.
\end{enumerate}
\end{theorem}

\begin{proof}
By using a symmetry argument we just need to prove $(1)\iff(2)$, and both the implications are elementary, as follows:

\medskip

$(1)\implies(2)$ If we denote by $U_1,\ldots,U_N\in\mathbb C^N$ the rows of $U$, we have indeed:
\begin{eqnarray*}
1
&=&<U_1,U_1>\\
&=&\sum_i<U_1,U_i>\\
&=&\sum_i\sum_jU_{1j}\bar{U}_{ij}\\
&=&\sum_jU_{1j}\sum_i\bar{U}_{ij}\\
&=&\sum_jU_{1j}\cdot\bar{\lambda}\\
&=&\lambda\cdot\bar{\lambda}\\
&=&|\lambda|^2
\end{eqnarray*}

$(2)\implies(1)$ Consider the all-one vector $\xi=(1)_i\in\mathbb C^N$. The fact that $U$ is row-stochastic with sums $\lambda$ reads:
\begin{eqnarray*}
\sum_jU_{ij}=\lambda,\forall i
&\iff&\sum_jU_{ij}\xi_j=\lambda\xi_i,\forall i\\
&\iff&U\xi=\lambda\xi
\end{eqnarray*}

Also, the fact that $U$ is column-stochastic with sums $\lambda$ reads:
\begin{eqnarray*}
\sum_iU_{ij}=\lambda,\forall j
&\iff&\sum_jU_{ij}\xi_i=\lambda\xi_j,\forall j\\
&\iff&U^t\xi=\lambda\xi
\end{eqnarray*}

We must prove that the first condition implies the second one, provided that the row sum $\lambda$ satisfies $|\lambda|=1$. But this follows from the following computation:
\begin{eqnarray*}
U\xi=\lambda\xi
&\implies&U^*U\xi=\lambda U^*\xi\\
&\implies&\xi=\lambda U^*\xi\\
&\implies&\xi=\bar{\lambda}U^t\xi\\
&\implies&U^t\xi=\lambda\xi
\end{eqnarray*}

Thus, we have proved both the implications, and we are done.
\end{proof}

From now on all our bistochastic matrices will be assumed to be normalized, with the sum on all rows and columns being equal to $\lambda=1$. We have the following result:

\index{bistochastic group}
\index{complex bistochastic group}
\index{real bistochastic group}

\begin{theorem}
We have groups as follows:
\begin{enumerate}
\item $B_N\subset O_N$, consisting of the orthogonal matrices which are bistochastic.

\item $C_N\subset U_N$, consisting of the unitary matrices which are bistochastic.
\end{enumerate}
\end{theorem}

\begin{proof}
We know from Theorem 11.26 that the sets of bistochastic matrices $B_N,C_N$ in the statement appear as follows, with $\xi$ being the all-one vector:
$$B_N=\left\{U\in O_N\Big|U\xi=\xi\right\}$$
$$C_N=\left\{U\in U_N\Big|U\xi=\xi\right\}$$

It is then clear that both $B_N,C_N$ are stable under the multiplication, contain the unit, and are stable by inversion. Thus, we have indeed groups, as stated.
\end{proof}

Following Idel-Wolf and related work, we would like to discuss now a non-trivial result involving the complex bistochastic group $C_N$. We will need some geometric preliminaries. The complex projective space appears by definition as follows:
$$P^{N-1}_\mathbb C=(\mathbb C^N-\{0\})\big/<x=\lambda y>$$

Inside this projective space, we have the Clifford torus, constructed as follows:
$$\mathbb T^{N-1}=\left\{(z_1,\ldots,z_N)\in P^{N-1}_\mathbb C\Big||z_1|=\ldots=|z_N|\right\}$$

With these conventions, we have the following result:

\index{complex projective space}
\index{Clifford torus}

\begin{proposition}
For a unitary matrix $U\in U_N$, the following are equivalent:
\begin{enumerate}
\item There exist $L,R\in U_N$ diagonal such that $U'=LUR$ is bistochastic.

\item The standard torus $\mathbb T^N\subset\mathbb C^N$ satisfies $\mathbb T^N\cap U\mathbb T^N\neq\emptyset$.

\item The Clifford torus $\mathbb T^{N-1}\subset P^{N-1}_\mathbb C$ satisfies $\mathbb T^{N-1}\cap U\mathbb T^{N-1}\neq\emptyset$.
\end{enumerate}
\end{proposition}

\begin{proof}
These equivalences are all elementary, as follows:

\medskip

$(1)\implies(2)$ Assuming that $U'=LUR$ is bistochastic, which in terms of the all-1 vector $\xi$ means $U'\xi=\xi$, if we set $f=R\xi\in\mathbb T^N$ we have:
$$Uf
=\bar{L}U'\bar{R}f
=\bar{L}U'\xi
=\bar{L}\xi\in\mathbb T^N$$

Thus we have $Uf\in\mathbb T^N\cap U\mathbb T^N$, which gives the conclusion.

\medskip

$(2)\implies(1)$ Given $g\in\mathbb T^N\cap U\mathbb T^N$, we can define $R,L$ as follows:
$$R=\begin{pmatrix}g_1\\&\ddots\\&&g_N\end{pmatrix}\quad,\quad 
\bar{L}=\begin{pmatrix}(Ug)_1\\&\ddots\\&&(Ug)_N\end{pmatrix}$$

With these values for $L,R$, we have then the following formulae:
$$R\xi=g\quad,\quad 
\bar{L}\xi=Ug$$

Thus the matrix $U'=LUR$ is bistochastic, because:
$$U'\xi
=LUR\xi
=LUg
=\xi$$

$(2)\implies(3)$ This is clear, because $\mathbb T^{N-1}\subset P^{N-1}_\mathbb C$ appears as the projective image of $\mathbb T^N\subset\mathbb C^N$, and so $\mathbb T^{N-1}\cap U\mathbb T^{N-1}$ appears as the projective image of $\mathbb T^N\cap U\mathbb T^N$.

\medskip

$(3)\implies(2)$ We have indeed the following equivalence:
$$\mathbb T^{N-1}\cap U\mathbb T^{N-1}\neq\emptyset
\iff\exists\lambda\neq 0,\lambda\mathbb T^N\cap U\mathbb T^N\neq\emptyset$$

But $U\in U_N$ implies $|\lambda|=1$, and this gives the result.
\end{proof}

The point now is that the condition (3) above is something familiar in symplectic geometry, and known to hold for any $U\in U_N$. We are led in this way to:

\index{bistochastic form}
\index{Sinkhorn normal form}
\index{Lagrangian submanifold}
\index{Hamiltonian isotopy}
\index{Arnold conjecture}

\begin{theorem}
Any unitary matrix $U\in U_N$ can be put in bistochastic form,
$$U'=LUR$$
with $L,R\in U_N$ being both diagonal, via a certain non-explicit method.
\end{theorem}

\begin{proof}
As already mentioned, the condition $\mathbb T^{N-1}\cap U\mathbb T^{N-1}\neq\emptyset$ in Proposition 11.28 (3) is something quite natural in symplectic geometry. To be more precise:

\medskip

(1) The Clifford torus $\mathbb T^{N-1}\subset P^{N-1}_\mathbb C$ is a Lagrangian submanifold, and the map $\mathbb T^{N-1}\to U\mathbb T^{N-1}$ is a Hamiltonian isotopy. For more on this, see Arnold \cite{ar2}.

\medskip

(2) A non-trivial result of Biran-Entov-Polterovich and Cho states that the Clifford torus $\mathbb T^{N-1}$ cannot be displaced from itself via a Hamiltonian isotopy.

\medskip

(3) Thus, the above-mentioned result tells us that  $\mathbb T^{N-1}\cap U\mathbb T^{N-1}\neq\emptyset$ holds indeed, for any $U\in U_N$. We therefore obtain the result, via Proposition 11.28. 
\end{proof}

\section*{11e. Exercises}

This was a truly exciting linear algebra chapter, and as exercises, we have:

\begin{exercise}
Learn about circulant graphs, and their adjacency matrices.
\end{exercise}

\begin{exercise}
Learn about vertex-transitive graphs, and their adjacency matrices.
\end{exercise}

\begin{exercise}
Try proving the HC or CHC, without insisting too much, tough.
\end{exercise}

\begin{exercise}
Learn about quasi-Hadamard and almost Hadamard matrices.
\end{exercise}

\begin{exercise}
Classify, with full details, the Hadamard matrices at $N=3,4$.
\end{exercise}

\begin{exercise}
Learn about the various known Hadamard matrices at $N=6$.
\end{exercise}

\begin{exercise}
Learn about the Sinkhorn algorithm, and normal form.
\end{exercise}

\begin{exercise}
Learn more about the Idel-Wolf theorem, and its ingredients.
\end{exercise}

As bonus exercise, learn more about design theory, and Hadamard matrices.

\chapter{Graph theory}

\section*{12a. Graphs, Laplacian}

We discuss here some further applications of linear algebra to discrete mathematics, and more specifically, to the finite graphs. As usual with discrete mathematics, as we got used to in this book, discrete Fourier analysis will be somewhere around the corner. 

\bigskip

Let us start with some generalities. The finite graphs $X$ are described by their adjacency matrices $d\in M_N(0,1)$, with $N$ being the number of vertices, and with this being something very useful. Consider for instance the following basic question:

\begin{question}
Given a graph $X$, with a distinguished vertex $*$:
\begin{enumerate}
\item What is the number $L_k$ of length $k$ loops on $X$, based at $*$? 

\item Equivalently, what is the measure $\mu$ having $L_k$ as moments?
\end{enumerate}
\end{question}

To be more precise, we are mainly interested in the first question, counting loops on graphs, with this being notoriously related to many applied mathematics questions, of discrete type. As for the second question, this is a useful reformulation of it.

\bigskip

In practice now, many things can be said, starting with the following basic result, featuring the adjacency matrix $d\in M_N(0,1)$ and its diagonalization, which is something quite elementary, and which can be very helpful for explicit computations:

\index{adjacency matrix}
\index{random walk}

\begin{theorem}
Given a graph $X$, with adjacency matrix $d\in M_N(0,1)$, we have:
$$L_k=(d^k)_{**}$$
When writing $d=UDU^t$ with $U\in O_N$ and $D=diag(\lambda_1,\ldots,\lambda_N)$ with $\lambda_i\in\mathbb R$, we have
$$L_k=\sum_iU_{*i}^2\lambda_i^k$$
and the real probability measure $\mu$ having these numbers as moments is given by
$$\mu=\sum_iU_{*i}^2\delta_{\lambda_i}$$
with the delta symbols standing as usual for Dirac masses.
\end{theorem}

\begin{proof}
There are several things going on here, the idea being as follows:

\medskip

(1) According to the usual rule of matrix multiplication, the formula for the powers of the adjacency matrix $d\in M_N(0,1)$ is as follows:
\begin{eqnarray*}
(d^k)_{i_0i_k}
&=&\sum_{i_1,\ldots,i_{k-1}}d_{i_0i_1}d_{i_1i_2}\ldots d_{i_{k-1}i_k}\\
&=&\sum_{i_1,\ldots,i_{k-1}}\delta_{i_0-i_1}\delta_{i_1-i_2}\ldots\delta_{i_{k-1}-i_k}\\
&=&\sum_{i_1,\ldots,i_{k-1}}\delta_{i_0-i_1-\ldots-i_{k-1}-i_k}\\
&=&\#\Big\{i_0-i_1-\ldots-i_{k-1}-i_k\Big\}
\end{eqnarray*}

In particular, with $i_0=i_k=*$, we obtain the following formula, as claimed:
$$(d^k)_{**}=\#\Big\{\!*-\,i_1-\ldots-i_{k-1}-*\Big\}=L_k$$

(2) Now since $d\in M_N(0,1)$ is symmetric, this matrix is diagonalizable, with the diagonalization being as follows, with $U\in O_N$, and $D=diag(\lambda_1,\ldots,\lambda_N)$ with $\lambda_i\in\mathbb R$:
$$d=UDU^t$$

By using this formula, we obtain the second formula in the statement:
\begin{eqnarray*}
L_k
&=&(d^k)_{**}\\
&=&(UD^kU^t)_{**}\\
&=&\sum_iU_{*i}\lambda_i^k(U^t)_{i*}\\
&=&\sum_iU_{*i}^2\lambda_i^k
\end{eqnarray*}

(3) Finally, the last assertion is clear from this, because the moments of the measure in the statement, $\mu=\sum_iU_{*i}^2\delta_{\lambda_i}$, are the following numbers:
\begin{eqnarray*}
M_k
&=&\int_\mathbb Rx^kd\mu(x)\\
&=&\sum_iU_{*i}^2\lambda_i^k\\
&=&L_k
\end{eqnarray*}

Observe also that $\mu$ is indeed of mass 1, because all rows of $U\in O_N$ must be of norm 1, and so $\sum_iU_{*i}^2=1$. Thus, we are led to the conclusions in the statement.
\end{proof}

Summarizing, foundations laid for counting loops via linear algebra, and I will leave it to you, to have some fun with all this, for some simple graphs of your choice.

\bigskip

Moving forward, in relation with this, but at a more conceptual level, we can formulate the following definition, making an interesting link with calculus and geometry:

\index{Laplacian}
\index{graph Laplacian}
\index{adjacency matrix}
\index{valence matrix}

\begin{definition}
We call Laplacian of a graph $X$ the matrix
$$L=v-d$$
with $v$ being the diagonal valence matrix, and $d$ being the adjacency matrix.
\end{definition} 

This definition is inspired by differential geometry, or just by multivariable calculus, and more specifically by the well-known Laplace operator there, given by:
$$\Delta f=\sum_{i=1}^N\frac{d^2f}{dx_i^2}$$

More on this in a moment, but as a word regarding terminology, we have:

\index{positive Laplacian}
\index{negative Laplacian}

\begin{warning}
The graph Laplacian above is in fact the negative Laplacian,
$$L=-\Delta$$
with our preference for it, negative, coming from the fact that it is positive, $L\geq0$.
\end{warning}

Which sounds like a bad joke, but this is how things are, and more on this a moment. In practice now, the graph Laplacian is given by the following formula:
$$L_{ij}=\begin{cases}
v_i&{\rm if}\ i=j\\
-1&{\rm if}\ i-j\\
0&{\rm otherwise}
\end{cases}$$

Alternatively, we have the following formula, for the entries of the Laplacian:
$$L_{ij}=\delta_{ij}v_i-\delta_{i-j}$$

With these formulae in hand, we can formulate, as our first result on the subject:

\index{harmonic function}
\index{function on graph}

\begin{proposition}
A function on a graph is harmonic, $Lf=0$, precisely when
$$f_i=\frac{1}{v_i}\sum_{i-j}f_j$$
that is, when the value at each vertex is the average over the neighbors.
\end{proposition}

\begin{proof}
We have indeed the following computation, for any function $f$:
\begin{eqnarray*}
(Lf)_i
&=&\sum_jL_{ij}f_j\\
&=&\sum_j(\delta_{ij}v_i-\delta_{i-j})f_j\\
&=&v_if_i-\sum_{i-j}f_j
\end{eqnarray*}

Thus, we are led to the conclusions in the statement.
\end{proof}

Summarizing, we have some good reasons for calling $L$ the Laplacian, because the solutions of $Lf=0$ satisfy what we would expect from a harmonic function, namely having the ``average over the neighborhood'' property. With the remark however that the harmonic functions on graphs are something trivial, due to the following fact:

\index{connected components}

\begin{proposition}
A function on a graph $X$ is harmonic in the above sense precisely when it is constant over the connected components of $X$.
\end{proposition}

\begin{proof}
This is clear from the equation that we found in Proposition 12.5, namely:
$$f_i=\frac{1}{v_i}\sum_{i-j}f_j$$

Indeed, based on this, we can say for instance that $f$ cannot have variations over a connected component, and so must be constant on these components, as stated.
\end{proof}

At a more advanced level now, let us try to understand the relation with the usual Laplacian from analysis $\Delta$, which is given by the following formula:
$$\Delta f=\sum_{i=1}^N\frac{d^2f}{dx_i^2}$$

In one dimension, $N=1$, the Laplacian is simply the second derivative, $\Delta f=f''$. Now let us recall that the first derivative of a one-variable function is given by:
$$f'(x)=\lim_{t\to 0}\frac{f(x+t)-f(x)}{t}$$

We deduce from this, or from the Taylor formula at order 2, to be fully correct, that the second derivative of a one-variable function is given by the following formula:
\begin{eqnarray*}
f''(x)
&=&\lim_{t\to 0}\frac{f'(x+t)-f'(x)}{t}\\
&=&\lim_{t\to 0}\frac{f(x+t)-2f(x)+f(x-t)}{t^2}
\end{eqnarray*}

Now since $\mathbb R$ can be thought of as appearing as the continuum limit, $t\to0$, of the graphs $t\mathbb Z\simeq\mathbb Z$, this suggests defining the Laplacian of $\mathbb Z$ by the following formula:
$$\Delta f(x)=\frac{f(x+1)-2f(x)+f(x-1)}{1^2}$$

But this is exactly what we have in Definition 12.3, up to a sign switch, the graph Laplacian of $\mathbb Z$, as constructed there, being given by the following formula:
$$Lf(x)=2f(x)-f(x+1)-f(x-1)$$

Summarizing, we have reached to the formula in Warning 12.4, namely:
$$L=-\Delta$$

In arbitrary dimensions now, everything generalizes well, and we have:

\index{lattice model}
\index{discretization}

\begin{theorem}
The Laplacian of graphs is compatible with the usual Laplacian,
$$\Delta f=\sum_{i=1}^N\frac{d^2f}{dx_i^2}$$
via the following formula, showing that our $L$ is in fact the negative Laplacian,
$$L=-\Delta$$
via regarding $\mathbb R^N$ as the continuum limit, $t\to0$, of the graphs $t\mathbb Z^N\simeq\mathbb Z^N$.
\end{theorem}

\begin{proof}
This is something that we know at $N=1$, and the proof in general is similar. Indeed, at $N=2$, to start with, the formula that we need is as follows:
\begin{eqnarray*}
\Delta f(x,y)
&=&\frac{d^2f}{dx^2}+\frac{d^2f}{dy^2}\\
&=&\lim_{t\to 0}\frac{f(x+t,y)-2f(x,y)+f(x-t,y)}{t^2}\\
&+&\lim_{t\to 0}\frac{f(x,y+t)-2f(x,y)+f(x,y-t)}{t^2}\\
&=&\lim_{t\to 0}\frac{f(x+t,y)+f(x-t,y)+f(x,y+t)+f(x,y-t)-4f(x,y)}{t^2}
\end{eqnarray*}

Now since $\mathbb R^2$ can be thought of as appearing as the continuum limit, $t\to0$, of the graphs $t\mathbb Z^2\simeq\mathbb Z^2$, this suggests defining the Laplacian of $\mathbb Z^2$ as follows:
$$\Delta f(x)
=\frac{f(x+1,y)+f(x-1,y)+f(x,y+1)+f(x,y-1)-4f(x,y)}{1^2}$$

But this is exactly what we have in Definition 12.3, up to a sign switch, the graph Laplacian of $\mathbb Z^2$, as constructed there, being given by the following formula:
$$Lf(x)=4f(x,y)-f(x+1,y)-f(x-1,y)-f(x,y+1)-f(x,y-1)$$

At higher $N\in\mathbb N$ the proof is similar, and we will leave this as an exercise.
\end{proof}

Now back to our general graph questions, and to Definition 12.3 as it is, the Laplacian of graphs as constructed there has the following basic properties:

\begin{theorem}
The graph Laplacian $L=v-d$ has the following properties:
\begin{enumerate}
\item It is symmetric, $L=L^t$.

\item It is positive definite, $L\geq 0$.

\item It is bistochastic, with row and column sums $0$.

\item It has $0$ as eigenvalue, with the other eigenvalues being positive.

\item The multiplicity of $0$ is the number of connected components.

\item In the connected case, the eigenvalues are $0<\lambda_1\leq\ldots\leq\lambda_{N-1}$.
\end{enumerate}
\end{theorem}

\begin{proof}
All this is straightforward, the idea being as follows:

\medskip

(1) This is clear from $L=v-d$, both $v,d$ being symmetric.

\medskip

(2) This follows from the following computation, for any function $f$ on the graph:
\begin{eqnarray*}
<Lf,f>
&=&\sum_{ij}L_{ij}f_if_j\\
&=&\sum_{ij}(\delta_{ij}v_i-\delta_{i-j})f_if_j\\
&=&\sum_iv_if_if_j-\sum_{i-j}f_if_j\\
&=&\sum_{i\sim j}f_i^2-\sum_{i-j}f_if_j\\
&=&\frac{1}{2}\sum_{i-j}(f_i-f_j)^2\\
&\geq&0
\end{eqnarray*}

(3) This is again clear from $L=v-d$, and from the definition of $v,d$.

\medskip

(4) Here the first assertion comes from (3), and the second one, from (2).

\medskip

(5) Given an arbitrary graph, we can label its vertices inceasingly, over the connected components, and this makes the adjacency matrix $d$, so the Laplacian $L$ as well, block diagonal. Thus, we are left with proving that for a connected graph, the multiplicity of 0 is precisely 1. But this follows from the formula from the proof of (2), namely:
$$<Lf,f>=\frac{1}{2}\sum_{i-j}(f_i-f_j)^2$$

Indeed, this formula shows in particular that we have the following equivalence:
$$Lf=0\iff f_i=f_j,\forall i-j$$

Now since our graph was assumed to be connected, as per the above beginning of proof, the condition on the right means that $f$ must be constant. Thus, the 0-eigenspace of the Laplacian follows to be 1-dimensional, spanned by the all-1 vector, as desired.

\medskip

(6) This follows indeed from (4) and (5), and with the remark that in fact we already proved this, in the proof of (5), with the formulae there being very useful in practice.
\end{proof}

\section*{12b. Kirchoff formula}

As a main application of our Laplacian technology, which will lead us deep into graph theory, let us discuss now the Kirchoff formula. This is based on the following facts:

\begin{proposition}
The following happen:
\begin{enumerate}
\item Any connected graph has a spanning tree, meaning a tree subgraph, making use of all vertices. 

\item For the complete graph $K_N$, with vertices labeled $1,\ldots,N$, the spanning trees are exactly the trees with vertices labeled $1,\ldots,N$.
\end{enumerate}
\end{proposition}

\begin{proof}
Both the assertions are trivial, the idea being as follows:

\medskip

(1) The fact that any connected graph has indeed a spanning tree is something which is very intuitive, clear on pictures, and we will leave the formal proof, which is not difficult, as an exercise. As an illustration for this, here is a picture of a quite random graph, which, after removal of some of the edges, the dotted ones, becomes indeed a tree:
$$
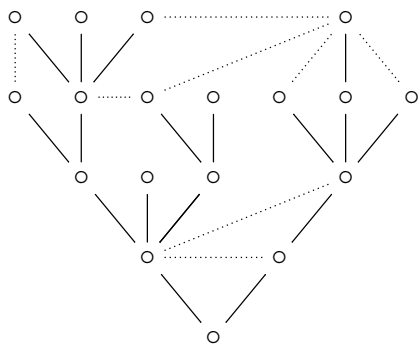
\xymatrix@R=18pt@C=12pt{
\circ\ar@{.}[d]&\circ&\circ\ar@{.}[rrr]&&&\circ\ar@{.}[dr]\ar@{.}[dlll]\ar@{.}[dl]\\
\circ&\circ\ar@{-}[ul]\ar@{-}[u]\ar@{-}[ur]\ar@{.}[r]&\circ&\circ&\circ&\circ\ar@{-}[u]&\circ\\
&\circ\ar@{-}[ul]\ar@{-}[u]&\circ&\circ\ar@{-}[ul]\ar@{-}[u]&&\circ\ar@{-}[ul]\ar@{-}[u]\ar@{-}[ur]\\
&&\circ\ar@{-}[ul]\ar@{-}[u]\ar@{-}[ur]\ar@{.}[rr]\ar@{-}[ur]\ar@{.}[urrr]&&\circ\ar@{-}[ur]\\
&&&\circ\ar@{-}[ul]\ar@{-}[ur]
}$$

(2) As for the second assertion, this is something which is clear too, and again we will leave the formal proof, which is not difficult at all, as an exercise.
\end{proof}

In view of the above, the following interesting question appears:

\begin{question}
Given a connected graph $X$, with vertices labeled $1,\ldots,N$, how to count its spanning trees? And, for the complete graph $K_N$, do we really get $N^{N-2}$ such trees, in agreement with the well-known Cayley formula, by using this method?
\end{question}

Getting to work now, following Kirchoff, the idea will be that of connecting the spanning trees of a connected graph $X$ to the combinatorics of the Laplacian of $X$, which is by definition given by the following formula, with $v$ being the valence matrix:
$$L=v-d$$

In order to get started, we will just need the fact, which is something trivial, that $L$ is bistochastic, with zero row and column sums. Indeed, this makes the link with the following basic linear algebra fact, that we can use afterwards, for our Laplacian $L$:

\index{bistochastic matrix}
\index{minors}

\begin{proposition}
For a matrix $L\in M_N(\mathbb R)$ which is bistochastic, with zero row and column sums, the signed minors
$$T_{ij}=(-1)^{i+j}\det(L^{ij})$$
do not depend on the choice of the indices $i,j$.
\end{proposition}

\begin{proof}
This is something very standard, the idea being as follows:

\medskip

(1) Before anything, let us do a quick check at $N=2$. Here the bistochastic matrices, with zero row and column sums, are as follows, with $a\in\mathbb R$:
$$L=\begin{pmatrix}a&-a\\ -a&a\end{pmatrix}$$

But this gives the result, with the number in question being $T_{ij}=a$.

\medskip

(2) Let us do as well a quick check at $N=3$. Here the bistochastic matrices, with zero row and column sums, are as follows, with $a,b,c,d\in\mathbb R$:
$$L=\begin{pmatrix}
a&b&-a-b\\
c&d&-c-d\\
-a-c&-b-d&a+b+c+d
\end{pmatrix}$$

But this gives again the result, with the number in question being $T_{ij}=ad-bc$.

\medskip

(3) In the general case now, where $N\in\mathbb N$ is arbitrary, the bistochastic matrices with zero row and column sums are as follows, with $A\in M_n(\mathbb R)$ with $n=N-1$ being an arbitary matrix, and with $R_1,\ldots,R_n$ and $C_1,\ldots,C_n$ being the row and column sums of this matrix, and $S=\sum R_i=\sum C_i$ being the total sum of this matrix:
$$L=\begin{pmatrix}
A_{11}&\ldots&A_{1n}&-R_1\\
\vdots&&\vdots&\vdots\\
A_{n1}&\ldots&A_{nn}&-R_n\\
-C_1&\ldots&-C_n&S
\end{pmatrix}$$

We want to prove that the signed minors of $L$ coincide, and by using the symmetries of the problem, it is enough to prove that the following equality holds:
$$L^{n+1,n}=-L^{n+1,n+1}$$

But, what we have on the right is $-\det A$, and what we have on the left is:
\begin{eqnarray*}
L^{n+1,n}
&=&\begin{vmatrix}
A_{11}&\ldots&A_{1,n-1}&-R_1\\
\vdots&&\vdots&\vdots\\
A_{n1}&\ldots&A_{n,n-1}&-R_n
\end{vmatrix}\\
&=&\begin{vmatrix}
A_{11}&\ldots&A_{1,n-1}&A_{11}+\ldots+A_{1,n-1}-R_1\\
\vdots&&\vdots&\vdots\\
A_{n1}&\ldots&A_{n,n-1}&A_{n1}+\ldots+A_{n,n-1}-R_n
\end{vmatrix}\\
&=&\begin{vmatrix}
A_{11}&\ldots&A_{1,n-1}&-A_{1n}\\
\vdots&&\vdots&\vdots\\
A_{n1}&\ldots&A_{n,n-1}&-A_{nn}
\end{vmatrix}\\
&=&-\det A
\end{eqnarray*}

Thus, we are led to the conclusion in the statement.
\end{proof}

We can now formulate, following Kirchoff, the following key result:

\index{Kirchoff formula}

\begin{theorem}
Given a connected graph $X$, with vertices labeled $1,\ldots,N$, the number of spanning trees inside $X$, meaning tree subgraphs using all vertices, is
$$T_X=(-1)^{i+j}\det(L^{ij})$$
with $L=v-d$ being the Laplacian, with this being independent on the chosen minor.
\end{theorem}

\begin{proof}
This is something non-trivial, the idea being as follows:

\medskip

(1) We know from Proposition 12.11 that the signed minors of $L$ coincide. In other words, we have a common formula as follows, with $T\in\mathbb Z$ being a certain number:
$$(-1)^{i+j}\det(L^{ij})=T$$

Our claim, which will prove the result, is that the number of spanning trees $T_X$ is precisely this common number $T$. That is, with $i=j=1$, our claim is that we have:
$$T_X=\det(L^{11})$$

(2) In order to prove our claim, which is non-trivial, we use a trick. We orient all the edges $e=(ij)$ of our graph as to have $i<j$, and we define the ordered incidence matrix of our graph, which is a rectangular matrix, with the vertices $i$ as row indices, and the oriented edges $e=(ij)$ as column indices, by the following formula:
$$E_{ie}=\begin{cases}
1&{\rm if}\ e=(ij)\\
-1&{\rm if}\ e=(ji)\\
0&{\rm otherwise}
\end{cases}$$

The point is that, in terms of this matrix, the Laplacian decomposes as follows:
$$L=EE^t$$

(3) Indeed, let us compute the matrix on the right. We have, by definition:
$$(EE^t)_{ij}=\sum_eE_{ie}E_{je}$$

Let us first compute the contributions of type $1\times1$, to the above sum. These come from the edges $e$ having the property $E_{ie}=E_{je}=1$. But $E_{ie}=1$ means $e=(ik)$ with $i<k$, and $E_{je}=1$ means $e=(jl)$ with $j<l$. Thus, our condition $E_{ie}=E_{je}=1$ means $i=j$, and $e=(ik)$ with $i<k$, so the contributions of type $1\times1$ are given by:
$$C_{1\times1}=\delta_{ij}\#\left\{k\Big|i<k,\ i-k\right\}$$

Similarly, the contributions of type $(-1)\times(-1)$ to our sum come from the equations $E_{ie}=E_{je}=-1$, which read $i=j$ and $e=(ki)$ with $k<i$, so are given by:
$$C_{(-1)\times(-1)}=\delta_{ij}\#\left\{k\Big|k<i,\ i-k\right\}$$

Now observe that by summing, the total $1$ contributions to our sum, be them of type $1\times1$ or $(-1)\times(-1)$, are given by the following formula, $v$ being the valence function:
\begin{eqnarray*}
C_1
&=&C_{1\times1}+C_{(-1)\times(-1)}\\
&=&\delta_{ij}\#\left\{k\Big|i<k,\ i-k\right\}+\delta_{ij}\#\left\{k\Big|k<i,\ i-k\right\}\\
&=&\delta_{ij}\#\left\{k\Big|i-k\right\}\\
&=&\delta_{ij}v_i
\end{eqnarray*}

(4) It remains to compute the total $-1$ contributions to our sum. But here, we first have the contributions of type $1\times(-1)$, coming from the equations $E_{ie}=1,E_{je}=-1$. Now since $E_{ie}=1$ means $e=(ik)$ with $i<k$, and $E_{je}=-1$ means $e=(lj)$ with $l<j$, our equations $E_{ie}=1,E_{je}=-1$ amount in saying that $e=(ij)$ with $i<j$. We conclude that the contributions of type $(-1)\times1$ to our sum are given by:
$$C_{1\times(-1)}=\delta_{i-j}\delta_{i<j}$$

Similarly, the contributions of type $(-1)\times1$ to our sum come from the equations $E_{ie}=-1,E_{je}=1$, which read $e=(ij)$ with $i<j$, so these are given by:
$$C_{(-1)\times1}=\delta_{i-j}\delta_{i>j}$$

Now by summing, the total $-1$ contributions to our sum, be them of type $1\times(-1)$ or $(-1)\times1$, are given by the following formula, $d$ being the adjacency matrix:
\begin{eqnarray*}
C_{-1}
&=&C_{1\times(-1)}+C_{(-1)\times1}\\
&=&\delta_{i-j}\delta_{i<j}+\delta_{i-j}\delta_{i>j}\\
&=&\delta_{i-j}\\
&=&d_{ij}
\end{eqnarray*}

(5) But with this, we can now finish the proof of our claim in (2), as follows:
\begin{eqnarray*}
(EE^t)_{ij}
&=&\sum_eE_{ie}E_{je}\\
&=&C_1-C_{-1}\\
&=&\delta_{ij}v_i-d_{ij}\\
&=&(v-d)_{ij}\\
&=&L_{ij}
\end{eqnarray*}

Thus, we have $EE^t=L$, and claim proved. Note in passing that our formula $EE^t=L$ gives another proof of the well-known property $L\geq0$ of the Laplacian.

\medskip

(6) Getting now towards minors, if we denote by $F$ the submatrix of $E$ obtained by deleting the first row, the one coming from the vertex 1, we have, for any $i,j>1$:
\begin{eqnarray*}
(FF^t)_{ij}
&=&\sum_eF_{ie}F_{je}\\
&=&\sum_eE_{ie}E_{je}\\
&=&(EE^t)_{ij}\\
&=&L_{ij}\\
&=&(L^{11})_{ij}
\end{eqnarray*}

We conclude that we have the following equality of matrices:
$$L^{11}=FF^t$$

(7) The point now is that, in order to compute the determinant of this latter matrix, we can use the Cauchy-Binet formula. To be more precise, the Cauchy-Binet formula says that given rectangular matrices $A,B$, of respective sizes $M\times N$ and $N\times M$, we have the following formula, with $A_S,B_S$ being both $M\times M$ matrices, obtained from $A,B$ by cutting, in the obvious way, with respect to the set of indices $S$:
$$\det(AB)=\sum_{|S|=M}\det(A_S)\det(B_S)$$

Observe that this formula tells us at $M>N$ that we have $\det(AB)=0$, as it should be, and at $M=N$ that we have $\det(AB)=\det A\det B$, again as it should be. At $M<N$, which is the interesting case, the Cauchy-Binet formula holds indeed, with the proof being a bit similar to that of the formula $\det(AB)=\det A\det B$ for the square matrices, which itself is not exactly a trivial business. We will leave clarifying all this as an exercise.

\medskip

(8) Now back to our questions, in the context of our formula $L^{11}=FF^t$ from (6), we can apply Cauchy-Binet to the matrices $A=F$ and $B=F^t$, having respective sizes $(N-1)\times N$ and $N\times(N-1)$. We are led in this way to the following formula, with $S$ ranging over the subsets of the edge set having size $N-1$, and with $F_S$ being the corresponding square submatrix of $E$, having size $(N-1)\times(N-1)$, obtained by restricting the attention to the columns indexed by the subset $S$:
\begin{eqnarray*}
\det(L^{11})
&=&\det(FF^t)\\
&=&\sum_S\det(F_S)\det(F_S^t)\\
&=&\sum_S\det(F_S)^2
\end{eqnarray*}

(9) Now comes the combinatorics. The sets $S$ appearing in the above computation specify in fact $N-1$ edges of our graph, and so specify a certain subgraph $X_S$. But, in this picture, our claim is that we have the following formula:
$$\det(F_S)=\begin{cases}
\pm1&{\rm if}\ X_S\ {\rm is\ a\ spanning\ tree}\\
0&{\rm otherwise}
\end{cases}$$

Indeed, since the subgraph $X_S$ has $N$ vertices and $N-1$ edges, it must be either a spanning tree, or have a cycle, and the study here goes as follows:

\medskip

-- In the case where $X_S$ is a spanning tree, we pick a leaf of this tree, in theory I mean, by leaving it there, on the tree. The corresponding row of $F_S$ consists then of a $\pm1$ entry, at the neighbor of the leaf, and of $0$ entries elsewhere. Thus, by developing $\det(F_S)$ over that row, we are led to a recurrence, which gives $\det(F_S)=\pm1$, as claimed above.

\medskip

-- In the other case, where $X_S$ has a cycle, the sum of the columns of $F_S$ indexed by the vertices belonging to this cycle must be $0$. We conclude that in this case we have $\det(F_S)=0$, again as claimed above, and this finishes the proof of our claim.

\medskip

(10) By putting now everything together, we obtain the following formula:
$$\det(L^{11})=T_X$$

Thus, we are led to the conclusions in the statement.
\end{proof}

As a basic application of the Kirchoff formula, let us apply it to the complete graph $K_N$. We are led in this way to another proof of the Cayley formula, as follows:

\index{complete graph}
\index{spanning tree}
\index{Cayley formula}

\begin{theorem}
The number of spanning trees of the complete graph $K_N$ is
$$T_{K_N}=N^{N-2}$$
in agreement with the Cayley formula.
\end{theorem}

\begin{proof}
This is something which is clear from the Kirchoff formula, but let us prove this slowly, as an illustration for the various computations above:

\medskip

(1) At $N=2$ the Laplacian of the segment $K_2$ is given by the following fomula:
$$L=\begin{pmatrix}1&-1\\-1&1\end{pmatrix}$$

Thus the common cofactor is 1, which equals the number of spanning trees, $2^0=1$.

\medskip

(2) At $N=3$ the Laplacian of the triangle $K_3$ is given by the following fomula:
$$L=\begin{pmatrix}
2&-1&-1\\
-1&2&-1\\
-1&-1&2
\end{pmatrix}$$

Thus the common cofactor is 3, which equals the number of spanning trees, $3^1=3$.

\medskip

(3) At $N=4$ the Laplacian of the tetrahedron $K_4$ is given by the following fomula:
$$L=\begin{pmatrix}
3&-1&-1&-1\\
-1&3&-1&-1\\
-1&-1&3&-1\\
-1&-1&-1&3
\end{pmatrix}$$

Here the cofactor is $27-11=16$, which is the number of spanning trees, $4^2=16$.

\medskip

(4) In general, for the complete graph $K_N$, the Laplacian is as follows:
$$\Delta=\begin{pmatrix}
N-1&-1&\ldots&-1&-1\\
-1&N-1&\ldots&-1&-1\\
\vdots&\vdots&&\vdots&\vdots\\
-1&-1&\ldots&N-1&-1\\
-1&-1&\ldots&-1&N-1
\end{pmatrix}$$

Thus, the common cofactor is $N^{N-2}$, in agreement with the Cayley formula.
\end{proof}

Finally, let us mention that in what regards the counting of trees having $N$ vertices, this time without labeled vertices, things here are far more complicated, and there is no formula available, for the number of such trees. We refer here to the literature.

\section*{12c. Into the waves}

Time now for some exciting physics, getting straight to the point, waves and heat. We first have the following result, regarding the waves, coming with a graph proof:

\index{wave equation}
\index{Laplace operator}
\index{lattice model}
\index{Hooke law}
\index{Newton law}

\begin{theorem}
The wave equation in $\mathbb R^N$ is
$$\ddot{\varphi}=v^2\Delta\varphi$$
where $v>0$ is the propagation speed.
\end{theorem}

\begin{proof}
Before everything, the equation in the statement is what comes out of experiments. However, allowing us a bit of imagination, and trust in this imagination, we can mathematically ``prove'' this equation, by discretizing, as follows:

\medskip

(1) Let us first consider the 1D case. In order to understand the propagation of waves, we will model $\mathbb R$ as a network of balls, with springs between them, as follows:
$$\cdots\times\!\!\!\times\!\!\!\times\bullet\times\!\!\!\times\!\!\!\times\bullet\times\!\!\!\times\!\!\!\times\bullet\times\!\!\!\times\!\!\!\times\bullet\times\!\!\!\times\!\!\!\times\bullet\times\!\!\!\times\!\!\!\times\cdots$$

Now let us send an impulse, and see how the balls will be moving. For this purpose, we zoom on one ball. The situation here is as follows, $l$ being the spring length:
$$\cdots\cdots\bullet_{\varphi(x-l)}\times\!\!\!\times\!\!\!\times\bullet_{\varphi(x)}\times\!\!\!\times\!\!\!\times\bullet_{\varphi(x+l)}\cdots\cdots$$

We have two forces acting at $x$. First is the Newton motion force, mass times acceleration, which is as follows, with $m$ being the mass of each ball:
$$F_n=m\cdot\ddot{\varphi}(x)$$

And second is the Hooke force, displacement of the spring, times spring constant. Since we have two springs at $x$, this is as follows, $k$ being the spring constant:
\begin{eqnarray*}
F_h
&=&F_h^r-F_h^l\\
&=&k(\varphi(x+l)-\varphi(x))-k(\varphi(x)-\varphi(x-l))\\
&=&k(\varphi(x+l)-2\varphi(x)+\varphi(x-l))
\end{eqnarray*}

We conclude that the equation of motion, in our model, is as follows:
$$m\cdot\ddot{\varphi}(x)=k(\varphi(x+l)-2\varphi(x)+\varphi(x-l))$$

(2) Now let us take the limit of our model, as to reach to continuum. For this purpose we will assume that our system consists of $B>>0$ balls, having a total mass $M$, and spanning a total distance $L$. Thus, our previous infinitesimal parameters are as follows, with $K$ being the spring constant of the total system, which is of course lower than $k$:
$$m=\frac{M}{B}\quad,\quad k=KB\quad,\quad l=\frac{L}{B}$$

With these changes, our equation of motion found in (1) reads:
$$\ddot{\varphi}(x)=\frac{KB^2}{M}(\varphi(x+l)-2\varphi(x)+\varphi(x-l))$$

Now observe that this equation can be written, more conveniently, as follows:
$$\ddot{\varphi}(x)=\frac{KL^2}{M}\cdot\frac{\varphi(x+l)-2\varphi(x)+\varphi(x-l)}{l^2}$$

With $N\to\infty$, and therefore $l\to0$, we obtain in this way:
$$\ddot{\varphi}(x)=\frac{KL^2}{M}\cdot\frac{d^2\varphi}{dx^2}(x)$$

We are therefore led to the wave equation in the statement, which is $\ddot{\varphi}=v^2\varphi''$ in our present $N=1$ dimensional case, the propagation speed being $v=\sqrt{K/M}\cdot L$.

\medskip

(3) In $2$ dimensions now, the same argument carries on. Indeed, we can use here a lattice model as follows, with all the edges standing for small springs:
$$
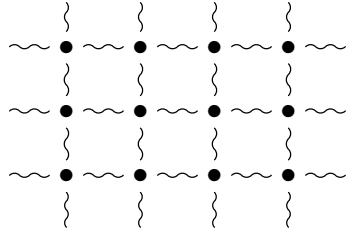
\xymatrix@R=12pt@C=15pt{
&\ar@{~}[d]&\ar@{~}[d]&\ar@{~}[d]&\ar@{~}[d]\\
\ar@{~}[r]&\bullet\ar@{~}[r]\ar@{~}[d]&\bullet\ar@{~}[r]\ar@{~}[d]&\bullet\ar@{~}[r]\ar@{~}[d]&\bullet\ar@{~}[r]\ar@{~}[d]&\\
\ar@{~}[r]&\bullet\ar@{~}[r]\ar@{~}[d]&\bullet\ar@{~}[r]\ar@{~}[d]&\bullet\ar@{~}[r]\ar@{~}[d]&\bullet\ar@{~}[r]\ar@{~}[d]&\\
\ar@{~}[r]&\bullet\ar@{~}[r]\ar@{~}[d]&\bullet\ar@{~}[r]\ar@{~}[d]&\bullet\ar@{~}[r]\ar@{~}[d]&\bullet\ar@{~}[r]\ar@{~}[d]&\\
&&&&&}$$

As before in one dimension, we send an impulse, and we zoom on one ball. The situation here is as follows, with $l$ being the spring length:
$$
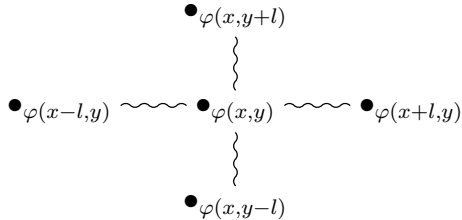
\xymatrix@R=20pt@C=20pt{
&\bullet_{\varphi(x,y+l)}\ar@{~}[d]&\\
\bullet_{\varphi(x-l,y)}\ar@{~}[r]&\bullet_{\varphi(x,y)}\ar@{~}[r]\ar@{~}[d]&\bullet_{\varphi(x+l,y)}\\
&\bullet_{\varphi(x,y-l)}}$$

We have two forces acting at $(x,y)$. First is the Newton motion force, mass times acceleration, which is as follows, with $m$ being the mass of each ball:
$$F_n=m\cdot\ddot{\varphi}(x,y)$$

And second is the Hooke force, displacement of the spring, times spring constant. Since we have four springs at $(x,y)$, this is as follows, $k$ being the spring constant:
\begin{eqnarray*}
F_h
&=&F_h^r-F_h^l+F_h^u-F_h^d\\
&=&k(\varphi(x+l,y)-\varphi(x,y))-k(\varphi(x,y)-\varphi(x-l,y))\\
&+&k(\varphi(x,y+l)-\varphi(x,y))-k(\varphi(x,y)-\varphi(x,y-l))\\
&=&k(\varphi(x+l,y)-2\varphi(x,y)+\varphi(x-l,y))\\
&+&k(\varphi(x,y+l)-2\varphi(x,y)+\varphi(x,y-l))
\end{eqnarray*}

We conclude that the equation of motion, in our model, is as follows:
\begin{eqnarray*}
m\cdot\ddot{\varphi}(x,y)
&=&k(\varphi(x+l,y)-2\varphi(x,y)+\varphi(x-l,y))\\
&+&k(\varphi(x,y+l)-2\varphi(x,y)+\varphi(x,y-l))
\end{eqnarray*}

(4) Now let us take the limit of our model, as to reach to continuum. For this purpose we will assume that our system consists of $B^2>>0$ balls, having a total mass $M$, and spanning a total area $L^2$. Thus, our previous infinitesimal parameters are as follows, with $K$ being the spring constant of the total system, taken to be equal to $k$:
$$m=\frac{M}{B^2}\quad,\quad k=K\quad,\quad l=\frac{L}{B}$$

With these changes, our equation of motion found in (3) reads:
\begin{eqnarray*}
\ddot{\varphi}(x,y)
&=&\frac{KB^2}{M}(\varphi(x+l,y)-2\varphi(x,y)+\varphi(x-l,y))\\
&+&\frac{KB^2}{M}(\varphi(x,y+l)-2\varphi(x,y)+\varphi(x,y-l))
\end{eqnarray*}

Now observe that this equation can be written, more conveniently, as follows:
\begin{eqnarray*}
\ddot{\varphi}(x,y)
&=&\frac{KL^2}{M}\times\frac{\varphi(x+l,y)-2\varphi(x,y)+\varphi(x-l,y)}{l^2}\\
&+&\frac{KL^2}{M}\times\frac{\varphi(x,y+l)-2\varphi(x,y)+\varphi(x,y-l)}{l^2}
\end{eqnarray*}

With $N\to\infty$, and therefore $l\to0$, we obtain in this way:
$$\ddot{\varphi}(x,y)=\frac{KL^2}{M}\left(\frac{d^2\varphi}{dx^2}+\frac{d^2\varphi}{dy^2}\right)(x,y)$$

Thus, we are led in this way to the following wave equation in two dimensions, with $v=\sqrt{K/M}\cdot L$ being the propagation speed of our wave:
$$\ddot{\varphi}(x,y)=v^2\left(\frac{d^2\varphi}{dx^2}+\frac{d^2\varphi}{dy^2}\right)(x,y)$$

But we recognize at right the Laplace operator, and we are done. As before in 1D, there is of course some discussion to be made here, arguing that our spring model in (3) is indeed the correct one. But do not worry, experiments confirm our findings.

\medskip

(5) In 3 dimensions now, which is the case of the main interest, corresponding to our real-life world, the same argument carries over, and the wave equation is as follows:
$$\ddot{\varphi}(x,y,z)=v^2\left(\frac{d^2\varphi}{dx^2}+\frac{d^2\varphi}{dy^2}+\frac{d^2\varphi}{dz^2}\right)(x,y,z)$$

(6) Finally, the same argument, namely a lattice model, carries on in arbitrary $N$ dimensions, and the wave equation here is as follows:
$$\ddot{\varphi}(x_1,\ldots,x_N)=v^2\sum_{i=1}^N\frac{d^2\varphi}{dx_i^2}(x_1,\ldots,x_N)$$

Thus, we are led to the conclusion in the statement.
\end{proof}

Moving on, with a bit of work, we can in fact fully solve the 1D wave equation. In order to explain this, we will need a standard calculus result, as follows:

\begin{proposition}
The derivative of a function of type
$$\varphi(x)=\int_{g(x)}^{h(x)}f(s)ds$$
is given by the formula $\varphi'(x)=f(h(x))h'(x)-f(g(x))g'(x)$.
\end{proposition}

\begin{proof}
Consider a primitive of the function that we integrate, $F'=f$. We have:
\begin{eqnarray*}
\varphi(x)
&=&\int_{g(x)}^{h(x)}f(s)ds\\
&=&\int_{g(x)}^{h(x)}F'(s)ds\\
&=&F(h(x))-F(g(x))
\end{eqnarray*}

By using now the chain rule for derivatives, we obtain from this:
\begin{eqnarray*}
\varphi'(x)
&=&F'(h(x))h'(x)-F'(g(x))g'(x)\\
&=&f(h(x))h'(x)-f(g(x))g'(x)
\end{eqnarray*}

Thus, we are led to the formula in the statement.
\end{proof}

Now back to the 1D waves, the general result here, due to d'Alembert, along with a little more, in relation with our lattice models above, is as follows:

\index{d'Alembert formula}
\index{1D wave}

\begin{theorem}
The solution of the 1D wave equation with initial value conditions $\varphi(x,0)=f(x)$ and $\dot{\varphi}(x,0)=g(x)$ is given by the d'Alembert formula, namely:
$$\varphi(x,t)=\frac{f(x-vt)+f(x+vt)}{2}+\frac{1}{2v}\int_{x-vt}^{x+vt}g(s)ds$$
In the context of our lattice model discretizations, what happens is more or less that the above d'Alembert integral gets computed via Riemann sums.
\end{theorem}

\begin{proof}
There are several things going on here, the idea being as follows:

\medskip

(1) Let us first check that the d'Alembert solution is indeed a solution of the wave equation $\ddot{\varphi}=v^2\varphi''$. The first time derivative is computed as follows:
$$\dot{\varphi}(x,t)=\frac{-vf'(x-vt)+vf'(x+vt)}{2}+\frac{1}{2v}(vg(x+vt)+vg(x-vt))$$

The second time derivative is computed as follows:
$$\ddot{\varphi}(x,t)=\frac{v^2f''(x-vt)+v^2f(x+vt)}{2}+\frac{vg'(x+vt)-vg'(x-vt)}{2}$$

Regarding now space derivatives, the first one is computed as follows:
$$\varphi'(x,t)=\frac{f'(x-vt)+f'(x+vt)}{2}+\frac{1}{2v}(g'(x+vt)-g'(x-vt))$$

As for the second space derivative, this is computed as follows:
$$\varphi''(x,t)=\frac{f''(x-vt)+f''(x+vt)}{2}+\frac{g''(x+vt)-g''(x-vt)}{2v}$$

Thus we have indeed $\ddot{\varphi}=v^2\varphi''$. As for the initial conditions, $\varphi(x,0)=f(x)$ is clear from our definition of $\varphi$, and $\dot{\varphi}(x,0)=g(x)$ is clear from our above formula of $\dot{\varphi}$.

\medskip

(2) Conversely now, we must show that our solution is unique, but instead of going here into abstract arguments, we will simply solve our equation, which among others will doublecheck out computations in (1). Let us make the following change of variables:
$$\xi=x-vt\quad,\quad\eta=x+vt$$

With this change of variables, which is quite tricky, mixing space and time variables, our wave equation $\ddot{\varphi}=v^2\varphi''$ reformulates in a very simple way, as follows:
$$\frac{d^2\varphi}{d\xi d\eta}=0$$

But this latter equation tells us that our new $\xi,\eta$ variables get separated, and we conclude from this that the solution must be of the following special form:
$$\varphi(x,t)=F(\xi)+G(\eta)=F(x-vt)+G(x+vt)$$

Now by taking into account the intial conditions $\varphi(x,0)=f(x)$ and $\dot{\varphi}(x,0)=g(x)$, and then integrating, we are led to the d'Alembert formula in the statement.

\medskip

(3) In regards now with our discretization questions, by using a 1D lattice model with balls and springs as before, what happens to all the above is more or less that the above d'Alembert integral gets computed via Riemann sums, in our model, as stated.
\end{proof}

In $N\geq2$ dimensions things are considerably more complicated, and for more on all this, we refer to any reasonably advanced theoretical physics or PDE book.

\section*{12d. Into the heat} 

Let us discuss now the heat equation, again by using a lattice model. The general equation here is quite similar to the one for the waves, as follows:

\index{heat equation}
\index{Laplace operator}
\index{thermal diffusivity}

\begin{theorem}
Heat diffusion in $\mathbb R^N$ is described by the heat equation
$$\dot{\varphi}=\alpha\Delta\varphi$$
where $\alpha>0$ is the thermal diffusivity of the medium, and $\Delta$ is the Laplace operator.
\end{theorem}

\begin{proof}
The study here is quite similar to the study of waves, as follows:

\medskip

(1) To start with, as an intuitive explanation for the equation, since the second derivative $\varphi''$ in one dimension, or the quantity $\Delta\varphi$ in general, computes the average value of a function $\varphi$ around a point, minus the value of $\varphi$ at that point, the heat equation as formulated above tells us that the rate of change $\dot{\varphi}$ of the temperature of the material at any given point must be proportional, with proportionality factor $\alpha>0$, to the average difference of temperature between that given point and the surrounding material.

\medskip

(2) The heat equation as formulated above is of course something approximative, and several improvements can be made to it, first by incorporating a term accounting for heat radiation, and then doing several fine-tunings, depending on the material involved. But more on this later, for the moment let us focus on the heat equation above.

\medskip

(3) In relation with our modelling questions, we can recover this equation a bit as we did for the wave equation before, by using a basic lattice model. Indeed, let us first assume, for simplifying, that we are in the one-dimensional case, $N=1$. Here our model looks as follows, with distance $l>0$ between neighbors:
$$\xymatrix@R=10pt@C=20pt{
\ar@{-}[r]&\circ_{x-l}\ar@{-}[r]^l&\circ_x\ar@{-}[r]^l&\circ_{x+l}\ar@{-}[r]&
}$$

In order to model heat diffusion, we have to implement the intuitive mechanism explained above, namely ``the rate of change of the temperature of the material at any given point must be proportional, with proportionality factor $\alpha>0$, to the average difference of temperature between that given point and the surrounding material''.

\medskip

(4) In practice, this leads to a condition as follows, expressing the change of the temperature $\varphi$, over a small period of time $\delta>0$:
$$\varphi(x,t+\delta)=\varphi(x,t)+\frac{\alpha\delta}{l^2}\sum_{x\sim y}\left[\varphi(y,t)-\varphi(x,t)\right]$$

To be more precise, we have made several assumptions here, as follows:

\medskip

-- General heat diffusion assumption: the change of temperature at any given point $x$ is proportional to the average over neighbors, $y\sim x$, of the differences $\varphi(y,t)-\varphi(x,t)$ between the temperatures at $x$, and at these neighbors $y$.

\medskip

-- Infinitesimal time and length conditions: in our model, the change of temperature at a given point $x$ is proportional to small period of time involved, $\delta>0$, and is inverse proportional to the square of the distance between neighbors, $l^2$.

\medskip

(5) Regarding these latter assumptions, the one regarding the proportionality with the time elapsed $\delta>0$ is something quite natural, physically speaking, and mathematically speaking too, because we can rewrite our equation as follows, making it clear that we have here an equation regarding the rate of change of temperature at $x$:
$$\frac{\varphi(x,t+\delta)-\varphi(x,t)}{\delta}=\frac{\alpha}{l^2}\sum_{x\sim y}\left[\varphi(y,t)-\varphi(x,t)\right]$$

As for the second assumption that we made above, namely inverse proportionality with $l^2$, this can be justified on physical grounds too, but again, perhaps the best is to do the math, which will show right away where this proportionality comes from. 

\medskip

(6) So, let us do the math. In the context of our 1D model the neighbors of $x$ are the points $x\pm l$, and so the equation that we wrote above takes the following form:
$$\frac{\varphi(x,t+\delta)-\varphi(x,t)}{\delta}=\frac{\alpha}{l^2}\Big[(\varphi(x+l,t)-\varphi(x,t))+(\varphi(x-l,t)-\varphi(x,t))\Big]$$

Now observe that we can write this equation as follows:
$$\frac{\varphi(x,t+\delta)-\varphi(x,t)}{\delta}
=\alpha\cdot\frac{\varphi(x+l,t)-2\varphi(x,t)+\varphi(x-l,t)}{l^2}$$

(7) As it was the case with the wave equation before, we recognize on the right the usual approximation of the second derivative, coming from calculus. Thus, when taking the continuous limit of our model, $l\to 0$, we obtain the following equation:
$$\frac{\varphi(x,t+\delta)-\varphi(x,t)}{\delta}
=\alpha\cdot\varphi''(x,t)$$

Now with $t\to0$, we are led in this way to the heat equation, namely:
$$\dot{\varphi}(x,t)=\alpha\cdot\varphi''(x,t)$$

Summarizing, we are done with the 1D case, with our proof being quite similar to the one for the wave equation, from the above. 

\medskip

(8) In practice now, there are of course still a few details to be discussed, in relation with all this, for instance at the end, in relation with the precise order of the limiting operations $l\to0$ and $\delta\to0$ to be performed, but these remain minor aspects, because our equation makes it clear, right from the beginning, that time and space are separated, and so that there is no serious issue with all this. And so, fully done with 1D.

\medskip

(9) With this done, let us discuss now 2 dimensions. Here, as before for the waves, we can use a lattice model as follows, with all lengths being $l>0$, for simplifying:
$$
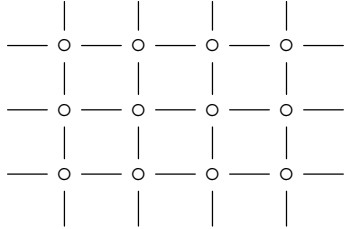
\xymatrix@R=12pt@C=15pt{
&\ar@{-}[d]&\ar@{-}[d]&\ar@{-}[d]&\ar@{-}[d]\\
\ar@{-}[r]&\circ\ar@{-}[r]\ar@{-}[d]&\circ\ar@{-}[r]\ar@{-}[d]&\circ\ar@{-}[r]\ar@{-}[d]&\circ\ar@{-}[r]\ar@{-}[d]&\\
\ar@{-}[r]&\circ\ar@{-}[r]\ar@{-}[d]&\circ\ar@{-}[r]\ar@{-}[d]&\circ\ar@{-}[r]\ar@{-}[d]&\circ\ar@{-}[r]\ar@{-}[d]&\\
\ar@{-}[r]&\circ\ar@{-}[r]\ar@{-}[d]&\circ\ar@{-}[r]\ar@{-}[d]&\circ\ar@{-}[r]\ar@{-}[d]&\circ\ar@{-}[r]\ar@{-}[d]&\\
&&&&&
}$$

(10) We have to implement now the physical heat diffusion mechanism, namely ``the rate of change of the temperature of the material at any given point must be proportional, with proportionality factor $\alpha>0$, to the average difference of temperature between that given point and the surrounding material''. In practice, this leads to a condition as follows, expressing the change of the temperature $\varphi$, over a small period of time $\delta>0$:
$$\varphi(x,y,t+\delta)=\varphi(x,y,t)+\frac{\alpha\delta}{l^2}\sum_{(x,y)\sim(u,v)}\left[\varphi(u,v,t)-\varphi(x,y,t)\right]$$

In fact, we can rewrite our equation as follows, making it clear that we have here an equation regarding the rate of change of temperature at $x$:
$$\frac{\varphi(x,y,t+\delta)-\varphi(x,y,t)}{\delta}=\frac{\alpha}{l^2}\sum_{(x,y)\sim(u,v)}\left[\varphi(u,v,t)-\varphi(x,y,t)\right]$$

(11) So, let us do the math. In the context of our 2D model the neighbors of $x$ are the points $(x\pm l,y\pm l)$, so the equation above takes the following form:
\begin{eqnarray*}
&&\frac{\varphi(x,y,t+\delta)-\varphi(x,y,t)}{\delta}\\
&=&\frac{\alpha}{l^2}\Big[(\varphi(x+l,y,t)-\varphi(x,y,t))+(\varphi(x-l,y,t)-\varphi(x,y,t))\Big]\\
&+&\frac{\alpha}{l^2}\Big[(\varphi(x,y+l,t)-\varphi(x,y,t))+(\varphi(x,y-l,t)-\varphi(x,y,t))\Big]
\end{eqnarray*}

Now observe that we can write this equation as follows:
\begin{eqnarray*}
\frac{\varphi(x,y,t+\delta)-\varphi(x,y,t)}{\delta}
&=&\alpha\cdot\frac{\varphi(x+l,y,t)-2\varphi(x,y,t)+\varphi(x-l,y,t)}{l^2}\\
&+&\alpha\cdot\frac{\varphi(x,y+l,t)-2\varphi(x,y,t)+\varphi(x,y-l,t)}{l^2}
\end{eqnarray*}

(12) As it was the case when modelling the wave equation before, we recognize on the right the usual approximation of the second derivative, coming from calculus. Thus, when taking the continuous limit of our model, $l\to 0$, we obtain the following equation:
$$\frac{\varphi(x,y,t+\delta)-\varphi(x,y,t)}{\delta}
=\alpha\left(\frac{d^2\varphi}{dx^2}+\frac{d^2\varphi}{dy^2}\right)(x,y,t)$$

Now with $t\to0$, we are led in this way to the heat equation, namely:
$$\dot{\varphi}(x,y,t)=\alpha\cdot\Delta\varphi(x,y,t)$$

Finally, in arbitrary $N$ dimensions the same argument carries over, namely a straightforward lattice model, and gives the heat equation, as formulated in the statement.
\end{proof}

Observe that we can use if we want different lenghts $l>0$ on the vertical and on the horizontal, because these will simplify anyway due to proportionality. Also, for some further mathematical fun, we can build our model on a cylinder, or a torus.

\bigskip

Also, as mentioned before, our heat equation above is something approximate, and several improvements can be made to it, first by incorporating a term accounting for heat radiation, and also by doing several fine-tunings, depending on the material involved. Some of these improvements can be implemented in the lattice model setting.

\bigskip

Regarding now the mathematics of the heat equation, many things can be said. As a first result here, often used by mathematicians, as to assume $\alpha=1$, we have:

\index{normalized heat equation}

\begin{proposition}
Up to a time rescaling, we can assume $\alpha=1$, as to deal with
$$\dot\varphi=\Delta\varphi$$
called normalized heat equation.
\end{proposition}

\begin{proof}
This is clear physically speaking, because according to our model, changing the parameter $\alpha>0$ will result in accelerating or slowing the heat diffusion, in time $t>0$. Mathematically, this follows via a change of variables, for the time variable $t$.
\end{proof}

Regarding now the resolution of the heat equation, we have here:

\index{heat kernel}

\begin{theorem}
The heat equation, normalized as $\dot\varphi=\Delta\varphi$, and with initial condition $\varphi(x,0)=f(x)$, has as solution the function
$$\varphi(x,t)=(K_t*f)(x)$$
where the function $K_t:\mathbb R^N\to\mathbb R$, called heat kernel, is given by
$$K_t(x)=(4\pi t)^{-N/2}e^{-||x||^2/4t}$$
with $||x||$ being the usual norm of vectors $x\in\mathbb R^N$.
\end{theorem}

\begin{proof}
According to the definition of the convolution operation $*$, we have to check that the following function satisfies $\dot\varphi=\Delta\varphi$, with initial condition $\varphi(x,0)=f(x)$:
$$\varphi(x,t)=(4\pi t)^{-N/2}\int_{\mathbb R^N}e^{-||x-y||^2/4t}f(y)dy$$

But both checks are elementary, coming from definitions.
\end{proof}

Regarding now discretization questions, things here are quite tricky. The idea is to use the Central Limit Theorem (CLT) from probability theory, which is as follows:

\index{CLT}
\index{central limit}
\index{normal law}
\index{Gaussian law}

\begin{theorem}[CLT]
Given random variables $f_1,f_2,f_3,\ldots\in L^\infty(X)$ which are i.i.d., centered, and with variance $t>0$, we have, with $n\to\infty$, in moments,
$$\frac{1}{\sqrt{n}}\sum_{i=1}^nf_i\sim g_t$$
where $g_t$ is the Gaussian law of parameter $t$, having as density $\frac{1}{\sqrt{2\pi t}}e^{-z^2/2t}dz$.
\end{theorem}

\begin{proof}
To start with, thanks to the Gauss formula from chapter 9, we can talk indeed about the Gaussian law $g_t$ of parameter $t>0$, having as density $\frac{1}{\sqrt{2\pi t}}e^{-z^2/2t}dz$. Also, the Fourier transform of this Gaussian law $g_t$ can be computed as follows:
\begin{eqnarray*}
F_{g_t}(x)
&=&\frac{1}{\sqrt{2\pi t}}\int_\mathbb Re^{-z^2/2t+ixz}dz\\
&=&\frac{1}{\sqrt{2\pi t}}\int_\mathbb Re^{-(z/\sqrt{2t}-\sqrt{t/2}\,iz)^2-tx^2/2}dz\\
&=&\frac{1}{\sqrt{2\pi t}}\int_\mathbb Re^{-y^2-tx^2/2}\sqrt{2t}\,dy\\
&=&\frac{1}{\sqrt{\pi}}e^{-tx^2/2}\int_\mathbb Re^{-y^2}dy\\
&=&e^{-tx^2/2}
\end{eqnarray*}

Getting now to what we want to prove, observe first that in terms of moments, the Fourier transform $F_f(x)=E(e^{ixf})$ is given by the following formula:
\begin{eqnarray*}
F_f(x)
&=&E\left(\sum_{k=0}^\infty\frac{(ixf)^k}{k!}\right)\\
&=&\sum_{k=0}^\infty\frac{(ix)^kE(f^k)}{k!}\\
&=&\sum_{k=0}^\infty\frac{i^kM_k(f)}{k!}\,x^k
\end{eqnarray*}

Thus, the Fourier transform of the variable in the statement is:
\begin{eqnarray*}
F(x)
&=&\left[F_f\left(\frac{x}{\sqrt{n}}\right)\right]^n\\
&=&\left[1-\frac{tx^2}{2n}+O(n^{-2})\right]^n\\
&\simeq&\left[1-\frac{tx^2}{2n}\right]^n\\
&\simeq&e^{-tx^2/2}
\end{eqnarray*}

But this latter function being the Fourier transform of $g_t$, we obtain the result.
\end{proof}

With the above result in hand, complemented by its higher dimensional analogues, which follow from it, we can talk afterwards about discretizing the heat kernel. We will leave some exploration and reading here as an instructive exercise.

\section*{12e. Exercises}

We had an exciting science chapter here, and as exercises on this, we have:

\begin{exercise}
Count loops on graphs, as many as you can.
\end{exercise}

\begin{exercise}
Diagonalize adjacency matrices, as many as you can.
\end{exercise}

\begin{exercise}
Learn some other proofs of the Cayley formula for graphs.
\end{exercise}

\begin{exercise}
Clarify everything in relation with the Cauchy-Binet formula.
\end{exercise}

\begin{exercise}
Learn more, from physicists, about the various types of waves.
\end{exercise}

\begin{exercise}
Learn also, from mathematicians, how to deal with waves at $N\geq2$.
\end{exercise}

\begin{exercise}
Learn more, from physicists, about the versions of the heat equation.
\end{exercise}

\begin{exercise}
Learn also, from mathematicians, how to deal with the heat equation.
\end{exercise}

As bonus exercise, now that you know about graphs, read some design theory.

\part{Matrix groups}

\ \vskip50mm

\begin{center}
{\em But here I am

Next to you

The sky is more blue

In Malibu}
\end{center}

\chapter{Finite groups}

\section*{13a. Finite groups}

We discuss in this final Part IV the various groups that the matrices can form. Let us start with some basic group theory. As a beginning for everything, we have:

\begin{definition}
A group is a set $G$ endowed with a multiplication operation 
$$(g,h)\to gh$$
which must satisfy the following conditions:
\begin{enumerate}
\item Associativity: we have, $(gh)k=g(hk)$, for any $g,h,k\in G$.

\item Unit: there is an element $1\in G$ such that $g1=1g=g$, for any $g\in G$.

\item Inverses: for any $g\in G$ there is $g^{-1}\in G$ such that $gg^{-1}=g^{-1}g=1$.
\end{enumerate}
\end{definition}

The multiplication law is not necessarily commutative. In the case where it is, in the sense that $gh=hg$, for any $g,h\in G$, we call $G$ abelian, en hommage to Abel, and we usually denote its multiplication, unit and inverse operation as follows:
$$(g,h)\to g+h\quad,\quad
0\in G\quad,\quad
g\to-g$$

However, this is not a general rule, and rather the converse is true, in the sense that if a group is denoted as above, this means that the group must be abelian.

\bigskip

Let us work out now some examples, in the finite group case. The simplest finite group is the cyclic group $\mathbb Z_N$, which is something very familiar, constructed as follows:

\index{cyclic group}
\index{roots of unity}

\begin{definition}
The cyclic group $\mathbb Z_N$ is defined as follows:
\begin{enumerate}
\item As the additive group of remainders modulo $N$.

\item As the multiplicative group of the $N$-th roots of unity.
\end{enumerate}
\end{definition}

Observe that the above two constructions are indeed equivalent, because if we set $w=e^{2\pi i/N}$, any remainder modulo $N$ defines a $N$-th root of unity, according to:
$$k\to w^k$$

We obtain in this way all the $N$-roots of unity, so our correspondence is bijective. Moreover, our correspondence transforms the sum of remainders modulo $N$ into the multiplication of the $N$-th roots of unity, due to the following formula:
$$w^kw^l=w^{k+l}$$

Thus, the groups constructed in Definition 13.2 (1) and (2) are indeed isomorphic, via $k\to w^k$, and we agree to denote by $\mathbb  Z_N$ the corresponding group. Observe that this group $\mathbb  Z_N$ is abelian. We will be back to the abelian groups later, on several occasions.

\bigskip

As a second basic example of a finite group, this time not abelian, we have the symmetric group $S_N$. This is again something very familiar, appearing as follows:

\index{symmetric group}
\index{permutation group}
\index{signature}
\index{permutation}

\begin{definition}
A permutation of $\{1,\ldots,N\}$ is a bijection, as follows:
$$\sigma:\{1,\ldots,N\}\to\{1,\ldots,N\}$$
The set of such permutations is denoted $S_N$.
\end{definition}

There are many possible notations for the permutations, the basic one consisting in writing the numbers $1,\ldots,N$, and below them, their permuted versions:
$$\sigma=\begin{pmatrix}
1&2&3&4&5\\
2&1&4&5&3
\end{pmatrix}$$

Another method, which is faster, and that I personally prefer, remember that time is money, is by denoting the permutations as diagrams, acting from top to bottom:
$$\xymatrix@R=3mm@C=3.5mm{
&\ar@{-}[ddr]&\ar@{-}[ddl]&\ar@{-}[ddrr]&\ar@{-}[ddl]&\ar@{-}[ddl]\\
\sigma=\\
&&&&&}$$

Here are some basic properties of the permutations, that you surely know about:

\begin{theorem}
The permutations have the following properties:
\begin{enumerate}
\item There are $N!$ of them.

\item They from a group.
\end{enumerate}
\end{theorem}

\begin{proof}
Indeed, in order to construct a permutation $\sigma\in S_N$, we have:

\smallskip

-- $N$ choices for the value of $\sigma(N)$.

-- $(N-1)$ choices for the value of $\sigma(N-1)$.

-- $(N-2)$ choices for the value of $\sigma(N-2)$.

$\vdots$

-- and so on, up to 1 choice for the value of $\sigma(1)$.

\smallskip

Thus, we have $N!$ choices, as claimed. As for the second assertion, this is clear.
\end{proof}

The symmetric groups $S_N$ are key objects of group theory, and they have many interesting properties. We will be back to them on many occasions, in what follows. 

\bigskip

As a third and last basic example of a finite group, for our purposes here, which is something more advanced, we have the dihedral group $D_N$, which appears as follows:

\index{dihedral group}
\index{regular polygon}
\index{symmetry group}

\begin{definition}
The dihedral group $D_N$ is the symmetry group of 
$$
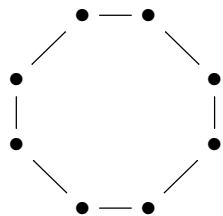
\xymatrix@R=12pt@C=12pt{
&\bullet\ar@{-}[r]\ar@{-}[dl]&\bullet\ar@{-}[dr]\\
\bullet\ar@{-}[d]&&&\bullet\ar@{-}[d]\\
\bullet\ar@{-}[dr]&&&\bullet\ar@{-}[dl]\\
&\bullet\ar@{-}[r]&\bullet}$$
that is, of the regular polygon having $N$ vertices.
\end{definition}

In order to understand how this works, here are the basic examples of regular $N$-gons, at small values of the parameter $N\in\mathbb N$, along with their symmetry groups:

\bigskip

\underline{$N=2$}. Here the $N$-gon is just a segment, and its symmetries are obviously the identity $id$, plus the symmetry $\tau$ with respect to the middle of the segment:
$$
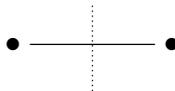
\xymatrix@R=10pt@C=20pt{
&\ar@{.}[dd]\\
\bullet\ar@{-}[rr]&&\bullet\\
&}$$

Thus we have $D_2=\{id,\tau\}$, which in group theory terms means $D_2=\mathbb Z_2$.

\bigskip

\underline{$N=3$}. Here the $N$-gon is an equilateral triangle, and we have 6 symmetries, the rotations of angles $0^\circ$, $120^\circ$, $240^\circ$, and the symmetries with respect to the altitudes: 
$$
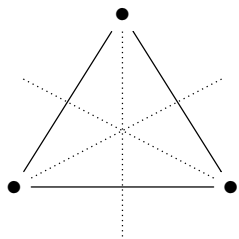
\xymatrix@R=13pt@C=28pt{
&\bullet\ar@{-}[dddr]\ar@{-}[dddl]\ar@{.}[dddd]\\
\ar@{.}[ddrr]&&\ar@{.}[ddll]\\
\\
\bullet\ar@{-}[rr]&&\bullet\\
&
}$$

Alternatively, we can say that the symmetries are all the $3!=6$ possible permutations of the vertices, and so that in group theory terms, we have $D_3=S_3$.

\bigskip

\underline{$N=4$}. Here the $N$-gon is a square, and as symmetries we have 4 rotations, of angles $0^\circ,90^\circ,180^\circ,270^\circ$, as well as 4 symmetries, with respect to the 4 symmetry axes, which are the 2 diagonals, and the 2 segments joining the midpoints of opposite sides:
$$
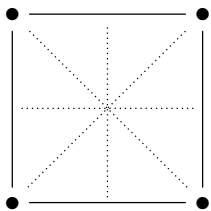
\xymatrix@R=26pt@C=26pt{
\bullet\ar@{-}[dd]\ar@{.}[ddrr]\ar@{-}[rr]&\ar@{.}[dd]&\bullet\ar@{-}[dd]\ar@{.}[ddll]\\
\ar@{.}[rr]&&\\
\bullet\ar@{-}[rr]&&\bullet
}$$

Thus, we obtain as symmetry group some sort of product between $\mathbb Z_4$ and $\mathbb Z_2$. Observe however that this product is not the usual one, our group being not abelian.

\bigskip

\underline{$N=5$}. Here the $N$-gon is a regular pentagon, and as symmetries we have 5 rotations, of angles $0^\circ,72^\circ,144^\circ,216^\circ,288^\circ$, as well as 5 symmetries, with respect to the 5 symmetry axes, which join the vertices to the midpoints of the opposite sides:
$$
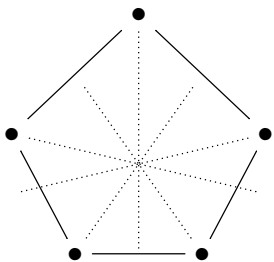
\xymatrix@R=13pt@C=11pt{
&&\bullet\ar@{-}[ddrr]\ar@{-}[ddll]\ar@{.}[dddd]\\
&&&&\\
\bullet\ar@{-}[ddr]\ar@{.}[drrrr]&&&&\bullet\ar@{-}[ddl]\ar@{.}[dllll]\\
&&&&\\
&\bullet\ar@{-}[rr]\ar@{.}[uuurr]&&\bullet\ar@{.}[uuull]&&
}$$

\underline{$N=6$}. Here the $N$-gon is a regular hexagon, and we have 6 rotations, of angles $0^\circ,60^\circ,120^\circ,180^\circ,240^\circ,300^\circ$, and 6 symmetries, with respect to the 6 symmetry axes, which are the 3 diagonals, and the 3 segments joining the midpoints of opposite sides:
$$
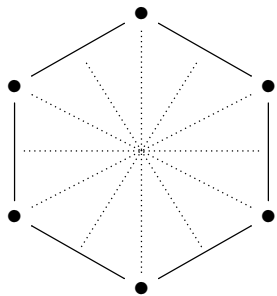
\xymatrix@R=4pt@C=14pt{
&&\bullet\ar@{-}[ddrr]\ar@{-}[ddll]\ar@{.}[dddddddd]\\
&\ar@{.}[ddddddrr]&&\ar@{.}[ddddddll]\\
\bullet\ar@{-}[dddd]\ar@{.}[ddddrrrr]&&&&\bullet\ar@{-}[dddd]\ar@{.}[ddddllll]\\
&&&&\\
\ar@{.}[rrrr]&&&&\\
&&&&\\
\bullet\ar@{-}[ddrr]&&&&\bullet\ar@{-}[ddll]\\
&&&&\\
&&\bullet
}$$

\underline{$N=7$}. Here the $N$-gon is a regular heptagon, and we have 7 rotations, of angles $kt$ with $k=0,1,\ldots,6$ and $t=360^\circ/7$, as well as 7 symmetries, with respect to the 7 symmetry axes, which join the vertices to the midpoints of the opposite sides.

\bigskip

We can see from the above that the various dihedral groups $D_N$ have many common features, and that there are some differences as well, basically coming from the parity of $N\in\mathbb N$. In general, we first have the following result, regarding the general case:

\index{rotations}
\index{symmetries}
\index{dihedral group}

\begin{proposition}
The dihedral group $D_N$ has $2N$ elements, as follows:
\begin{enumerate}
\item We have $N$ rotations $R_1,\ldots,R_N$, with $R_k$ being the rotation of angle $2k\pi/N$. When labeling the vertices of the $N$-gon $1,\ldots,N$, we have $R_k(i)=k+i$.

\item We have $N$ symmetries $S_1,\ldots,S_N$, with $S_k$ being the symmetry with respect to the $Ox$ axis rotated by $k\pi/N$. The symmetry formula is $S_k(i)=k-i$.
\end{enumerate}
\end{proposition}

\begin{proof}
This is clear from definitions. To be more precise, $D_N$ consists of:

\medskip

(1) The $N$ rotations of the $N$-gon, of angles $2k\pi/N$ with $k=1,\ldots,N$. But these are exactly the rotations $R_1,\ldots,R_N$ from the statement.

\medskip

(2) The $N$ symmetries of the $N$-gon, with respect to the $N$ medians when $N$ is odd, and with the $N/2$ diagonals plus the $N/2$ lines connecting the midpoints of opposite edges, when $N$ is even. But these are the symmetries $S_1,\ldots,S_N$ from the statement.
\end{proof}

With the above description of $D_N$ in hand, we can forget if we want about geometry and the regular $N$-gon, and talk about $D_N$ abstractly, as follows:

\index{multiplication table}

\begin{theorem}
The dihedral group $D_N$ is the group having $2N$ elements, $R_1,\ldots,R_N$ and $S_1,\ldots,S_N$, called rotations and symmetries, which multiply as follows,
$$R_kR_l=R_{k+l}\quad,\quad 
R_kS_l=S_{k+l}$$
$$S_kR_l=S_{k-l}\quad,\quad
S_kS_l=R_{k-l}$$
with all the indices being taken modulo $N$.
\end{theorem}

\begin{proof}
With notations from Proposition 13.6, the various compositions between rotations and symmetries can be computed as follows:
$$R_kR_l\ :\ i\to l+i\to k+l+i$$
$$R_kS_l\ :\ i\to l-i\to k+l-i$$
$$S_kR_l\ :\ i\to l+i\to k-l-i$$
$$S_kS_l\ :\ i\to l-i\to k-l+i$$

But these are exactly the formulae for $R_{k+l},S_{k+l},S_{k-l},R_{k-l}$, as stated. Now since a group is uniquely determined by its multiplication rules, this gives the result.
\end{proof}

Observe now that $D_N$ has the same cardinality as $E_N=\mathbb Z_N\times\mathbb Z_2$. We obviously don't have $D_N\simeq E_N$, because $D_N$ is not abelian, while $E_N$ is. So, our next goal will be that of proving that $D_N$ appears by ``twisting'' $E_N$. In order to do this, let us start with:

\begin{proposition}
The group $E_N=\mathbb Z_N\times\mathbb Z_2$ is the group having $2N$ elements, $r_1,\ldots,r_N$ and $s_1,\ldots,s_N$, which multiply according to the following rules,
$$r_kr_l=r_{k+l}\quad,\quad 
r_ks_l=s_{k+l}$$
$$s_kr_l=s_{k+l}\quad,\quad 
s_ks_l=r_{k+l}$$
with all the indices being taken modulo $N$.
\end{proposition}

\begin{proof}
With the notation $\mathbb Z_2=\{1,\tau\}$, the elements of the product group $E_N=\mathbb Z_N\times\mathbb Z_2$ can be labeled $r_1,\ldots,r_N$ and $s_1,\ldots,s_N$, as follows:
$$r_k=(k,1)\quad,\quad
s_k=(k,\tau)$$

These elements multiply then according to the formulae in the statement. Now since a group is uniquely determined by its multiplication rules, this gives the result.
\end{proof}

Let us compare now Theorem 13.7 and Proposition 13.8. In order to formally obtain $D_N$ from $E_N$, we must twist some of the multiplication rules of $E_N$, namely:
$$s_kr_l=s_{k+l}\to s_{k-l}$$
$$s_ks_l=r_{k+l}\to r_{k-l}$$

Informally, this amounts in following the rule ``$\tau$ switches the sign of what comes afterwards", and we are led in this way to the following definition:

\index{crossed product}

\begin{definition}
Given two groups $A,G$, with an action $A\curvearrowright G$, the crossed product
$$P=G\rtimes A$$
is the set $G\times A$, with multiplication $(g,a)(h,b)=(gh^a,ab)$.
\end{definition}

It is routine to check that $P$ is indeed a group. Observe that when the action is trivial, $h^a=h$ for any $a\in A$ and $h\in H$, we obtain the usual product $G\times A$. 

\bigskip

Now with this technology in hand, by getting back to the dihedral group $D_N$, we can improve Theorem 13.7, into a final result on the subject, as follows:

\index{dihedral group}
\index{crossed product decomposition}

\begin{theorem}
We have a crossed product decomposition as follows,
$$D_N=\mathbb Z_N\rtimes\mathbb Z_2$$
with $\mathbb Z_2=\{1,\tau\}$ acting on $\mathbb Z_N$ via switching signs, $k^\tau=-k$.
\end{theorem}

\begin{proof}
We have an action $\mathbb Z_2\curvearrowright\mathbb Z_N$ given by the formula in the statement, namely $k^\tau=-k$, so we can consider the corresponding crossed product group:
$$P_N=\mathbb Z_N\rtimes\mathbb Z_2$$

In order to understand the structure of $P_N$, we follow Proposition 13.8. The elements of $P_N$ can indeed be labeled $\rho_1,\ldots,\rho_N$ and $\sigma_1,\ldots,\sigma_N$, as follows:
$$\rho_k=(k,1)\quad,\quad 
\sigma_k=(k,\tau)$$

Now when computing the products of such elements, we basically obtain the formulae in Proposition 13.8, perturbed as in Definition 13.9. To be more precise, we have:
$$\rho_k\rho_l=\rho_{k+l}\quad,\quad 
\rho_k\sigma_l=\sigma_{k+l}$$
$$\sigma_k\rho_l=\sigma_{k+l}\quad,\quad
\sigma_k\sigma_l=\rho_{k+l}$$

But these are exactly the multiplication formulae for $D_N$, from Theorem 13.7. Thus, we have an isomorphism $D_N\simeq P_N$ given by $R_k\to\rho_k$ and $S_k\to\sigma_k$, as desired.
\end{proof}

\section*{13b. Symmetric groups}

We have seen some basic group theory, but you might wonder where is the linear algebra, in relation with this. In order to get to this, following Cayley, we first have:

\index{Cayley embedding}
\index{permutation group}

\begin{theorem}
Given a finite group $G$, we have an embedding as follows,
$$G\subset S_N\quad,\quad g\to(h\to gh)$$
with $N=|G|$. Thus, any finite group is a permutation group. 
\end{theorem}

\begin{proof}
Given a group element $g\in G$, we can associate to it the following map:
$$\sigma_g:G\to G\quad,\quad 
h\to gh$$

Since $gh=gh'$ implies $h=h'$, this map is bijective, and so is a permutation of $G$, viewed as a set. Thus, with $N=|G|$, we can view this map as a usual permutation, $\sigma_G\in S_N$. Summarizing, we have constructed so far a map as follows:
$$G\to S_N\quad,\quad 
g\to\sigma_g$$

Our first claim is that this is a group morphism. Indeed, this follows from:
$$\sigma_g\sigma_h(k)
=\sigma_g(hk)
=ghk
=\sigma_{gh}(k)$$

It remains to prove that this group morphism is injective. But this follows from:
\begin{eqnarray*}
g\neq h
&\implies&\sigma_g(1)\neq\sigma_h(1)\\
&\implies&\sigma_g\neq\sigma_h
\end{eqnarray*}

Thus, we are led to the conclusion in the statement.
\end{proof}

Observe that in the above statement the embedding $G\subset S_N$ that we constructed depends on a particular writing $G=\{g_1,\ldots,g_N\}$, which is needed in order to identify the permutations of $G$ with the elements of the symmetric group $S_N$. This is not very good, in practice, and as an illustration, for the basic examples of groups that we know, the Cayley theorem provides us with embeddings as follows:
$$\mathbb Z_N\subset S_N\quad,\quad 
D_N\subset S_{2N}\quad,\quad 
S_N\subset S_{N!}$$

And here the first embedding is the good one, the second one is not the best possible one, but can be useful, and the third embedding is useless. Thus, as a conclusion, the Cayley theorem remains something quite theoretical. We will be back to this later on, with a systematic study of the ``representation'' problem for the finite groups.

\bigskip

Getting back now to our main series of finite groups, $\mathbb Z_N\subset D_N\subset S_N$, these are of course permutation groups, according to the above. However, and perhaps even more interestingly, these are as well subgroups of the orthogonal group $O_N$:
$$\mathbb Z_N\subset D_N\subset S_N\subset O_N$$

In order to explain this, we first have the following key result:

\index{permutation matrix}
\index{binary matrix}

\begin{theorem}
We have a group embedding as follows, obtained by regarding $S_N$ as the permutation group of the $N$ coordinate axes of $\mathbb R^N$,
$$S_N\subset O_N$$
which makes $\sigma\in S_N$ correspond to the matrix having $1$ on row $i$ and column $\sigma(i)$, for any $i$, and having $0$ entries elsewhere.
\end{theorem}

\begin{proof}
The first assertion is clear, because the permutations of the $N$ coordinate axes of $\mathbb R^N$ are isometries. Regarding now the explicit formula, we have by definition:
$$\sigma(e_j)=e_{\sigma(j)}$$

Thus, the permutation matrix corresponding to $\sigma$ is given by:
$$\sigma_{ij}=
\begin{cases}
1&{\rm if}\ \sigma(j)=i\\
0&{\rm otherwise}
\end{cases}$$

Thus, we are led to the formula in the statement.
\end{proof}

We can combine the above result with the Cayley theorem, and we obtain:

\index{finite group}
\index{Cayley embedding}
\index{permutation group}

\begin{theorem}
Given a finite group $G$, we have an embedding as follows,
$$G\subset O_N\quad,\quad g\to(e_h\to e_{gh})$$
with $N=|G|$. Thus, any finite group is an orthogonal matrix group.
\end{theorem}

\begin{proof}
The Cayley theorem gives an embedding as follows:
$$G\subset S_N\quad,\quad g\to(h\to gh)$$

On the other hand, Theorem 13.12 provides us with an embedding as follows:
$$S_N\subset O_N\quad,\quad 
\sigma\to(e_i\to e_{\sigma(i)})$$

Thus, we are led to the conclusion in the statement.
\end{proof}

The same remarks as for the Cayley theorem apply. First, the embedding $G\subset O_N$ that we constructed depends on a particular writing $G=\{g_1,\ldots,g_N\}$. And also, for the basic examples of groups that we know, the embeddings that we obtain are as follows:
$$\mathbb Z_N\subset O_N\quad,\quad 
D_N\subset O_{2N}\quad,\quad 
S_N\subset O_{N!}$$

Summarizing, all this is not very good, and in order to advance, it is better to forget about the Cayley theorem, and build on Theorem 13.12 instead. We have here:

\begin{theorem}
We have the following finite groups of matrices:
\begin{enumerate}
\item $\mathbb Z_N\subset O_N$, the cyclic permutation matrices.

\item $D_N\subset O_N$, the dihedral permutation matrices.

\item $S_N\subset O_N$, the permutation matrices.
\end{enumerate}
\end{theorem}

\begin{proof}
This is something self-explanatory, the idea being that Theorem 13.12 provides us with embeddings as follows, given by the permutation matrices:
$$\mathbb Z_N\subset D_N\subset S_N\subset O_N$$

In practice now, the groups in the statement appear as follows:

\medskip

(1) The cyclic permutation matrices are by definition the matrices as follows, with 0 entries elsewhere, and form a group, which is isomorphic to the cyclic group $\mathbb Z_N$:
$$U=\begin{pmatrix}
&&&1&&&\\
&&&&1&&\\
&&&&&\ddots&\\
&&&&&&1\\
1&&&&&\\
&\ddots&&&&&\\
&&1&&&
\end{pmatrix}$$

(2) The dihedral matrices are the above cyclic permutation matrices, plus some suitable symmetry permutation matrices, and form a group which is isomorphic to $D_N$.

\medskip

(3) The permutation matrices, which by Theorem 13.12 form a group which is isomorphic to $S_N$, are the $0-1$ matrices having exactly one 1 on each row and column.
\end{proof}

All the above is very nice, and as an informal conclusion to this, let us record:

\begin{conclusion}
In what regards the theory of finite groups:
\begin{enumerate}
\item The central object is the symmetric group, $S_N\subset O_N$.

\item All basic finite groups appear as subgroups $G\subset O_N$.

\item Thus, the objects of study are the finite subgroups $G\subset U_N$.
\end{enumerate}
\end{conclusion}

To be more precise here, we have declared in (1) the symmetric group $S_N$ to be the most important group, and we have also suggested to look at it as a subgroup $S_N\subset O_N$, in view of the above results. Regarding (2), this is actually something which theoretically happens for any finite group, as explained above, and with the basic examples truly appearing as $G\subset O_N$, by construction. As for (3), this is a version of (2) that we have not talked about yet, but that we can expect to happen, because it is not hard to imagine that certain more complicated groups naturally appear as subgroups $G\subset U_N$.

\bigskip

Very nice all this, and good to know, but the problem is now, what to do with our finite groups, $S_N\subset O_N$, or more generally $G\subset O_N$, or even more generally $G\subset U_N$?

\bigskip

And here, you must agree with me that the answer is not very clear. A look at the group theory literature suggests doing a myriad technical things, which are all useful of course, but with none being really elementary, and inviting. So, time to ask the cat:

\begin{cat}
We cats are interested indeed in the subgroups $G\subset U_N$, which can be finite or not, and what we do with them is to compute the law of $g\to Tr(g)$.
\end{cat}

Thanks cat, this sounds quite interesting, and related indeed to linear algebra, as I was wishing for, in view of the purposes of the present book. So, let us study this problem, the linear algebra of the finite group elements, viewed as matrices, $g\in G\in U_N$. 

\bigskip

As a first observation, before talking about the trace, as cat suggests, we can have a look at the determinant, with the result here, which is good to know, being as follows:

\begin{proposition}
The determinant of the permutation matrices
$$\det:S_N\subset O_N\to\mathbb R$$
is the signature, $\varepsilon:S_N\to\mathbb Z_2$. Thus, we have $S_N\cap SO_N=A_N$, inside $O_N$.
\end{proposition}

\begin{proof}
The first assertion comes from the Sarrus type formula for the determinant from chapter 2, in terms of permutations and their signatures, which in the case of the permutation matrices simply gives $\det=\varepsilon$. As for the second assertion, this follows from this, by taking the preimage via the application $\det=\varepsilon$ of the trivial group $\{1\}$.
\end{proof}

Getting now to what cat says, study of the trace, again in the simplest case, that of $S_N\subset O_N$, things here become truly interesting, with the result being as follows:

\begin{theorem}
The trace of permutation matrices, regarded as function
$$\chi:S_N\to\mathbb N\quad,\quad \chi(\sigma)=Tr(\sigma)$$
counts the number of fixed points, according to the following formula:
$$\chi(\sigma)=\left\{i\in\{1,\ldots,N\}\Big|\sigma(i)=i\right\}$$
Moreover, the variable $\chi:S_N\to\mathbb N$ follows the Poisson law of parameter $1$,
$$p_1=\frac{1}{e}\sum_{k\in\mathbb N}\frac{\delta_k}{k!}$$
in the $N\to\infty$ limit. 
\end{theorem}

\begin{proof}
We have several things going on here, the idea being as follows:

\medskip

(1) Consider indeed the trace of permutation matrices, $\chi(\sigma)=Tr(\sigma)$. The permutation matrices being given by $\sigma_{ij}=\delta_{i\sigma(j)}$, we have the following formula, as claimed:
$$\chi(\sigma)
=\sum_i\delta_{i\sigma(i)}
=\#\left\{i\in\{1,\ldots,N\}\Big|\sigma(i)=i\right\}$$

(2) In order to prove now the second assertion, consider the following subsets of $S_N$:
$$S_N^i=\left\{\sigma\in S_N\Big|\sigma(i)=i\right\}$$

The set of permutations having no fixed points, called derangements, is then:
$$X_N=\left(\bigcup_iS_N^i\right)^c$$

Now the inclusion-exclusion principle tells us that we have:
\begin{eqnarray*}
|X_N|
&=&\left|\left(\bigcup_iS_N^i\right)^c\right|\\
&=&|S_N|-\sum_i|S_N^i|+\sum_{i<j}|S_N^i\cap S_N^j|-\ldots+(-1)^N\sum_{i_1<\ldots<i_N}|S_N^{i_1}\cap\ldots\cap S_N^{i_N}|\\
&=&N!-N(N-1)!+\binom{N}{2}(N-2)!-\ldots+(-1)^N\binom{N}{N}(N-N)!\\
&=&\sum_{k=0}^N(-1)^k\binom{N}{k}(N-k)!\\
&=&\sum_{k=0}^N(-1)^k\frac{N!}{k!}
\end{eqnarray*}

We conclude that the first probability that we are interested in, that for a random permutation $\sigma\in S_N$ to have no fixed points, is given by the following formula:
$$P(\chi=0)
=\frac{|X_N|}{N!}
=\sum_{k=0}^N\frac{(-1)^k}{k!}\simeq\frac{1}{e}$$

(3) Thus, we are on the good way for establishing the second assertion, and in order now to fully do that, we just have to fine-tune our computation. To be more precise, we would like to prove the following formula, for any $r\in\mathbb N$, in the $N\to\infty$ limit:
$$P(\chi=r)\simeq\frac{1}{r!e}$$

(4) We already know, from the above, that this formula holds at $r=0$. In the general case now, we have to count the permutations $\sigma\in S_N$ having exactly $r$ points. Now since having such a permutation amounts in choosing $r$ points among $1,\ldots,N$, and then permuting the $N-r$ points left, without fixed points allowed, we have:
\begin{eqnarray*}
\#\left\{\sigma\in S_N\Big|\chi(\sigma)=r\right\}
&=&\binom{N}{r}\#\left\{\sigma\in S_{N-r}\Big|\chi(\sigma)=0\right\}\\
&=&\frac{N!}{r!(N-r)!}\#\left\{\sigma\in S_{N-r}\Big|\chi(\sigma)=0\right\}\\
&=&N!\times\frac{1}{r!}\times\frac{\#\left\{\sigma\in S_{N-r}\Big|\chi(\sigma)=0\right\}}{(N-r)!}
\end{eqnarray*}

By dividing everything by $N!$, we obtain from this the following formula:
$$\frac{\#\left\{\sigma\in S_N\Big|\chi(\sigma)=r\right\}}{N!}=\frac{1}{r!}\times\frac{\#\left\{\sigma\in S_{N-r}\Big|\chi(\sigma)=0\right\}}{(N-r)!}$$

Now by using the computation at $r=0$, that we already have, from (2), it follows that with $N\to\infty$ we have the following estimate:
$$P(\chi=r)
\simeq\frac{1}{r!}\cdot P(\chi=0)
\simeq\frac{1}{r!}\cdot\frac{1}{e}$$

Thus, we obtain as limiting measure the Poisson law of parameter 1, as stated.
\end{proof}

The above result looks quite exciting, and as further good news, that is not the end of what can be said, because we have also the following result, which is more general:

\begin{theorem}
The truncated trace of permutation matrices, defined as
$$\chi_t:S_N\to\mathbb R\quad,\quad \chi(\sigma)=\sum_{i=1}^{[tN]}\sigma_{ii}$$
counts the truncated number of fixed points, according to the following formula:
$$\chi_t(\sigma)=\left\{i\in\{1,\ldots,[tN]\}\Big|\sigma(i)=i\right\}$$
Moreover, the variable $\chi_t:S_N\to\mathbb R$ follows the Poisson law of parameter $t$,
$$p_t=\frac{1}{e^t}\sum_{k\in\mathbb N}\frac{t^k\delta_k}{k!}$$
in the $N\to\infty$ limit. 
\end{theorem}

\begin{proof}
We have several things going on here, the idea being as follows:

\medskip

(1) Let us construct indeed the truncated trace, as in the statement. As before in the case $t=1$, we have the following computation, coming from definitions:
$$\chi_t(\sigma)
=\sum_{i=1}^{[tN]}\delta_{i\sigma(i)}
=\#\left\{i\in\{1,\ldots,[tN]\}\Big|\sigma(i)=i\right\}$$

(2) Also as before at $t=1$, we obtain by inclusion-exclusion that:
\begin{eqnarray*}
P(\chi_t=0)
&=&\frac{1}{N!}\sum_{k=0}^{[tN]}(-1)^k\sum_{i_1<\ldots<i_k<[tN]}|S_N^{i_1}\cap\ldots\cap S_N^{i_k}|\\
&=&\frac{1}{N!}\sum_{k=0}^{[tN]}(-1)^k\binom{[tN]}{k}(N-k)!\\
&=&\sum_{k=0}^{[tN]}\frac{(-1)^k}{k!}\cdot\frac{[tN]!(N-k)!}{N!([tN]-k)!}
\end{eqnarray*}

Now with $N\to\infty$, we obtain from this the following estimate:
$$P(\chi_t=0)
\simeq\sum_{k=0}^{[tN]}\frac{(-1)^k}{k!}\cdot t^k
\simeq e^{-t}$$

(3) More generally now, by counting the permutations $\sigma\in S_N$ having exactly $r$ fixed points among $1,\ldots,[tN]$, again as in the previous proof at $t=1$, we obtain:
$$P(\chi_t=r)\simeq\frac{t^r}{r!e^t}$$

Thus, we obtain in the limit a Poisson law of parameter $t$, as stated.
\end{proof}

\section*{13c. Reflection groups}

The above results regarding the symmetric group $S_N$ are quite exciting, and it is tempting to keep going this way, with similar results for $\mathbb Z_N$ and $D_N$. We first have:

\begin{proposition}
The main character of $\mathbb Z_N\subset O_N$ is given by:
$$\chi(g)=\begin{cases}
0&{\rm if}\ g\neq1\\
N&{\rm if}\ g=1
\end{cases}$$
Thus, at the probabilistic level, we have the following formula,
$$law(\chi)=\left(1-\frac{1}{N}\right)\delta_0+\frac{1}{N}\delta_N$$
telling us that the main character $\chi$ follows a Bernoulli law.
\end{proposition}

\begin{proof}
The first formula is clear, because the cyclic permutation matrices have 0 on the diagonal, and so 0 as trace, unless the matrix is the identity, having trace $N$. As for the second formula, this is a probabilistic reformulation of the first one.
\end{proof}

For the dihedral group now, which is the next one in our hierarchy, the computation is more interesting, but the final answer is no longer uniform in $N$, as follows:

\index{dihedral group}

\begin{proposition}
For the dihedral group $D_N\subset S_N$ we have
$$law(\chi)=\begin{cases}
\left(\frac{3}{4}-\frac{1}{2N}\right)\delta_0+\frac{1}{4}\delta_2+\frac{1}{2N}\delta_N&(N\ even)\\
&\\
\left(\frac{1}{2}-\frac{1}{2N}\right)\delta_0+\frac{1}{2}\delta_1+\frac{1}{2N}\delta_N&(N\ odd)
\end{cases}$$
with this law having no asymptotics, with $N\to\infty$.
\end{proposition}

\begin{proof}
This follows indeed from the fact that the dihedral group $D_N$ consists of:

\medskip

(1) $N$ symmetries, having each $1$ fixed point when $N$ is odd, and having 0 or 2 fixed points, distributed $50-50$, when $N$ is even.

\medskip

(2) $N$ rotations, each having $0$ fixed points, except for the identity, which is technically a rotation too, and which has $N$ fixed points.

\medskip

Thus, we are led to the formula in the statement, and to the final conclusion too.
\end{proof}

All the above does not look very good, so I am afraid that we are a bit stuck with our program, and that I will have to ask again the cat. And cat answers:

\begin{cat}
There are many groups $G\subset U_N$, some good, some bad. Try doing your computations for groups which really look good, geometrically speaking.
\end{cat}

Thanks cat, so forgetting now about $\mathbb Z_N$, $D_N$, and other abstract groups and mathematics that we might know, let us think deeply, and try to find a group $G\subset U_N$ which is really interesting, and that we would really like to know about. And here, we have:

\index{hyperoctahedral group}
\index{hypercube}

\begin{definition}
The hyperoctahedral group $H_N\subset O_N$ is the group of symmetries of the unit cube in $\mathbb R^N$, viewed as a graph, or equivalently, as a metric space.
\end{definition}

The hyperoctahedral group is a quite interesting group, whose definition, as a symmetry group, reminds that of the dihedral group $D_N$. So, let us start our study in the same way as we did for $D_N$, with a discussion at small values of $N\in\mathbb N$:

\medskip

\underline{$N=1$}. Here the 1-cube is the segment, whose symmetries are the identity $id$, plus the symmetry $\tau$ with respect to the middle of the segment:
$$
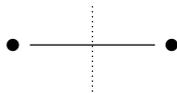
\xymatrix@R=10pt@C=20pt{
&\ar@{.}[dd]\\
\bullet\ar@{-}[rr]&&\bullet\\
&}$$

Thus, we obtain the group with 2 elements, which is a very familiar object: 
$$H_1=D_2=S_2=\mathbb Z_2$$

\underline{$N=2$}. Here the 2-cube is the square, whose symmetries are the 4 rotations, of angles $0^\circ,90^\circ,180^\circ,270^\circ$, and the 4 symmetries with respect to the 4 symmetry axes, which are the 2 diagonals, and the 2 segments joining the midpoints of opposite sides:
$$
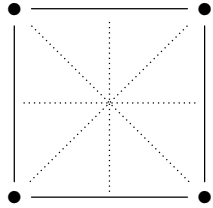
\xymatrix@R=26pt@C=26pt{
\bullet\ar@{-}[dd]\ar@{.}[ddrr]\ar@{-}[rr]&\ar@{.}[dd]&\bullet\ar@{-}[dd]\ar@{.}[ddll]\\
\ar@{.}[rr]&&\\
\bullet\ar@{-}[rr]&&\bullet
}$$

Thus, we obtain a group with 8 elements, which again is a very familiar object:
$$H_2=D_4=\mathbb Z_4\rtimes\mathbb Z_2$$

\underline{$N=3$}. Here the 3-cube is the usual cube in $\mathbb R^3$, pictured as follows:
$$
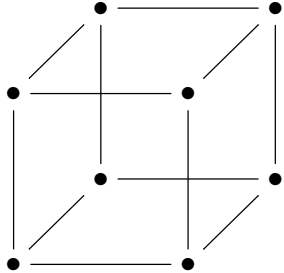
\xymatrix@R=20pt@C=20pt{
&\bullet\ar@{-}[rr]&&\bullet\\
\bullet\ar@{-}[rr]\ar@{-}[ur]&&\bullet\ar@{-}[ur]\\
&\bullet\ar@{-}[rr]\ar@{-}[uu]&&\bullet\ar@{-}[uu]\\
\bullet\ar@{-}[uu]\ar@{-}[ur]\ar@{-}[rr]&&\bullet\ar@{-}[uu]\ar@{-}[ur]
}$$

However, in relation with the symmetries, the situation now is considerably more complicated, because, thinking well, this cube has no less than 48 symmetries.

\medskip

All this looks quite complicated, but fortunately, we have the following result:

\begin{proposition}
We have the cardinality formula
$$|H_N|=2^NN!$$
coming from the fact that $H_N$ is the symmetry group of the coordinate axes of $\mathbb R^N$.
\end{proposition}

\begin{proof}
This follows from some geometric thinking, as follows:

\medskip

(1) Consider the standard cube in $\mathbb R^N$, centered at 0, and having as vertices the points having coordinates $\pm1$. With this picture in hand, it is clear that the symmetries of the cube coincide with the symmetries of the $N$ coordinate axes of $\mathbb R^N$.

\medskip

(2) In order to count now these latter symmetries, a bit as we did for the dihedral group, observe first that we have $N!$ permutations of these $N$ coordinate axes. 

\medskip

(3) But each of these permutations of the coordinate axes $\sigma\in S_N$ can be further ``decorated'' by a sign vector $e\in\{\pm1\}^N$, consisting of the possible $\pm1$ flips which can be applied to each coordinate axis, at the arrival. 

\medskip

(4) And the point is that, obviously, we obtain in this way all the elements of $H_N$. Thus, we have the following formula, for the cardinality of $H_N$:
$$|H_N|
=|S_N|\cdot|\mathbb Z_2^N|
=N!\cdot2^N$$

Thus, we are led to the conclusions in the statement. 
\end{proof}

As in the dihedral group case, it is possible to go beyond this, with a crossed product decomposition, of quite special type, called wreath product decomposition:

\index{crossed product}
\index{wreath product}
\index{hyperoctahedral group}

\begin{theorem}
We have a wreath product decomposition as follows,
$$H_N=\mathbb Z_2\wr S_N$$
which means by definition that we have a crossed product decomposition
$$H_N=\mathbb Z_2^N\rtimes S_N$$
with the permutations $\sigma\in S_N$ acting on the elements $e\in\mathbb Z_2^N$ as follows:
$$\sigma(e_1,\ldots,e_k)=(e_{\sigma(1)},\ldots,e_{\sigma(k)})$$
In particular we have, as found before, the cardinality formula $|H_N|=2^NN!$.
\end{theorem}

\begin{proof}
As explained in the proof of Proposition 13.24, the elements of $H_N$ can be identified with the pairs $g=(e,\sigma)$ consisting of a permutation $\sigma\in S_N$, and a sign vector $e\in\mathbb Z_2^N$, so that at the level of the cardinalities, we have the following formula:
$$|H_N|=|\mathbb Z_2^N\times S_N|$$

To be more precise, given an element $g\in H_N$, the element $\sigma\in S_N$ is the corresponding permutation of the $N$ coordinate axes, regarded as unoriented lines in $\mathbb R^N$, and $e\in\mathbb Z_2^N$ is the vector collecting the possible flips of these coordinate axes, at the arrival. Now observe that the product formula for two such pairs $g=(e,\sigma)$ is as follows, with the permutations $\sigma\in S_N$ acting on the elements $f\in\mathbb Z_2^N$ as in the statement:
$$(e,\sigma)(f,\tau)=(ef^\sigma,\sigma\tau)$$

Thus, we are precisely in the framework of the crossed products, as constructed in chapter 1, and we conclude that we have a crossed product decomposition, as follows:
$$H_N=\mathbb Z_2^N\rtimes S_N$$
 
Thus, we are led to the conclusion in the statement, with the formula $H_N=\mathbb Z_2\wr S_N$ being just a shorthand for the decomposition $H_N=\mathbb Z_2^N\rtimes S_N$ that we found.
\end{proof}

Regarding now the trace laws, we can compute them by using the same method as for the symmetric group $S_N$, namely inclusion-exclusion, and we have:

\index{main character}

\begin{theorem}
For the hyperoctahedral group $H_N\subset O_N$, the law of the variable
$$\chi_t=\sum_{i=1}^{[tN]}g_{ii}$$
becomes in the $N\to\infty$ limit the measure
$$b_t=e^{-t}\sum_{k=-\infty}^\infty\delta_k\sum_{p=0}^\infty \frac{(t/2)^{|k|+2p}}{(|k|+p)!p!}$$ 
called Bessel law of parameter $t>0$.
\end{theorem}

\begin{proof}
We regard $H_N$ as being the symmetry group of the graph $I_N=\{I^1,\ldots ,I^N\}$ formed by $N$ segments. The diagonal coefficients are then given by:
$$g_{ii}(g)=\begin{cases}
\ 0\ \mbox{ if $g$ moves $I^i$}\\
\ 1\ \mbox{ if $g$ fixes $I^i$}\\
-1\mbox{ if $g$ returns $I^i$}
\end{cases}$$

We denote by $\uparrow g,\downarrow g$ the number of segments among $\{I^1,\ldots ,I^s\}$ which are fixed, respectively returned by an element $g\in H_N$. With this notation, we have:
$$g_{11}+\ldots+g_{ss}=\uparrow g-\downarrow g$$

Let us denote by $P_N$ probabilities computed over the group $H_N$. The density of the law of $g_{11}+\ldots+g_{ss}$ at a point $k\geq 0$ is then given by the following formula:
\begin{eqnarray*}
D(k)
&=&P_N(\uparrow g-\downarrow g=k)\\
&=&\sum_{p=0}^\infty P_N(\uparrow g=k+p, \downarrow g=p)
\end{eqnarray*}

Assume first that we have $t=1$. We have here the following computation:
\begin{eqnarray*}
\lim_{N\to\infty}D(k)
&=&\lim_{N\to\infty}\sum_{p=0}^\infty(1/2)^{k+2p}\begin{pmatrix}k+2p\\ k+p\end{pmatrix} P_N(\uparrow g+\downarrow g=k+2p)\\ 
&=&\sum_{p=0}^\infty(1/2)^{k+2p}\begin{pmatrix}k+2p\\
k+p\end{pmatrix}\frac{1}{e(k+2p)!}\\
&=&\frac{1}{e}\sum_{p=0}^\infty \frac{(1/2)^{k+2p}}{(k+p)!p!}
\end{eqnarray*}

As for the general case $0<t\leq 1$, here the result follows by performing some modifications in the above computation. The asymptotic density is computed as follows:
\begin{eqnarray*}
\lim_{N\to\infty}D(k)
&=&\lim_{N\to\infty}\sum_{p=0}^\infty(1/2)^{k+2p}\begin{pmatrix}k+2p\\ k+p\end{pmatrix} P_N(\uparrow g+\downarrow g=k+2p)\\
&=&\sum_{p=0}^\infty(1/2)^{k+2p}\begin{pmatrix}k+2p\\
k+p\end{pmatrix}\frac{t^{k+2p}}{e^t(k+2p)!}\\
&=&e^{-t}\sum_{p=0}^\infty \frac{(t/2)^{k+2p}}{(k+p)!p!}
\end{eqnarray*}

Together with $D(-k)=D(k)$, this gives the formula in the statement.
\end{proof}

In the above result the terminology comes from the fact, up to $t\to t/2$, the density of the law is the following function, called Bessel function of the first kind:
$$f_k(t)=\sum_{p=0}^\infty \frac{t^{|k|+2p}}{(|k|+p)!p!}$$

Let us further study now these Bessel laws. A key result regarding the Poisson laws is the semigroup formula $p_s*p_t=p_{s+t}$, and in analogy with this, we have:

\index{convolution}

\begin{theorem}
The Bessel laws $b_t$ have the property
$$b_s*b_t=b_{s+t}$$
so they form a truncated one-parameter semigroup
with respect to convolution.
\end{theorem}

\begin{proof}
We use the formula that we found in Theorem 13.26, written as:
$$b_t=e^{-t}\sum_{k=-\infty}^\infty\delta_k\,f_k(t/2)$$

The Fourier transform of this measure is then given by:
$$Fb_t(y)=e^{-t}\sum_{k=-\infty}^\infty e^{ky}\,f_k(t/2)$$

We compute now the derivative with respect to $t$, as follows:
$$Fb_t(y)'=-Fb_t(y)+\frac{e^{-t}}{2}\sum_{k=-\infty}^\infty e^{ky}\,f_k'(t/2)$$

On the other hand, the derivative of $f_k$ with $k\geq 1$ is given by:
\begin{eqnarray*}
f_k'(t)
&=&\sum_{p=0}^\infty \frac{(k+2p)t^{k+2p-1}}{(k+p)!p!}\\
&=&\sum_{p=0}^\infty \frac{(k+p)t^{k+2p-1}}{(k+p)!p!}+\sum_{p=0}^\infty\frac{p\,t^{k+2p-1}}{(k+p)!p!}\\
&=&\sum_{p=0}^\infty \frac{t^{k+2p-1}}{(k+p-1)!p!}+\sum_{p=1}^\infty\frac{t^{k+2p-1}}{(k+p)!(p-1)!}\\
&=&\sum_{p=0}^\infty \frac{t^{(k-1)+2p}}{((k-1)+p)!p!}+\sum_{p=1}^\infty\frac{t^{(k+1)+2(p-1)}}{((k+1)+(p-1))!(p-1)!}\\
&=&f_{k-1}(t)+f_{k+1}(t)
\end{eqnarray*}

This computation works in fact for any $k$, so we get:
\begin{eqnarray*}
Fb_t(y)'
&=&-Fb_t(y)+\frac{e^{-t}}{2}
\sum_{k=-\infty}^\infty e^{ky} (f_{k-1}(t/2)+f_{k+1}(t/2))\\
&=&-Fb_t(y)+\frac{e^{-t}}{2} \sum_{k=-\infty}^\infty
e^{(k+1)y}f_{k}(t/2)+e^{(k-1)y}f_{k}(t/2)\\
&=&-Fb_t(y)+\frac{e^{y}+e^{-y}}{2}\,Fb_t(y)\\
&=&\left(\frac{e^{y}+e^{-y}}{2}-1\right)Fb_t(y)
\end{eqnarray*}

Thus the log of the Fourier transform is linear in $t$, and we get
the assertion.
\end{proof}

\section*{13d. Complex reflections}

In order to further discuss all this, we will need a number of probabilistic preliminaries. We recall that, conceptually speaking, the Poisson laws are the laws appearing via the Poisson Limit Theorem (PLT). In order to generalize this construction, as to cover for Bessel laws, we have the following notion, extending the Poisson limit theory:

\index{compound Poisson law}

\begin{definition}
Associated to any compactly supported positive measure $\nu$ on $\mathbb C$ is the probability measure
$$p_\nu=\lim_{n\to\infty}\left(\left(1-\frac{c}{n}\right)\delta_0+\frac{1}{n}\nu\right)^{*n}$$
where $c=mass(\nu)$, called compound Poisson law.
\end{definition}

In what follows we will be interested in the case where $\nu$ is discrete, as is for instance the case for the measure $\nu=t\delta_1$ with $t>0$, which produces via the above procedure the Poisson laws. To be more precise, we will be mainly interested in the case where $\nu$ is a multiple of the uniform measure on the $s$-th roots of unity. More on this later.

\bigskip

The following result allows us to detect the compound Poisson laws:

\index{Fourier transform}

\begin{proposition}
For $\nu=\sum_{i=1}^sc_i\delta_{z_i}$ with $c_i>0$ and $z_i\in\mathbb C$ we have
$$F_{p_\nu}(y)=\exp\left(\sum_{i=1}^sc_i(e^{iyz_i}-1)\right)$$
where $F$ denotes as usual the Fourier transform.
\end{proposition}

\begin{proof}
Let $\mu_n$ be the measure appearing in Definition 13.28, namely:
$$\mu_n=\left(1-\frac{c}{n}\right)\delta_0+\frac{1}{n}\nu$$

We have the following computation, in the context of Definition 13.28:
\begin{eqnarray*}
F_{\mu_n}(y)=\left(1-\frac{c}{n}\right)+\frac{1}{n}\sum_{i=1}^sc_ie^{iyz_i}
&\implies&F_{\mu_n^{*n}}(y)=\left(\left(1-\frac{c}{n}\right)+\frac{1}{n}\sum_{i=1}^sc_ie^{iyz_i}\right)^n\\
&\implies&F_{p_\nu}(y)=\exp\left(\sum_{i=1}^sc_i(e^{iyz_i}-1)\right)
\end{eqnarray*}

Thus, we have obtained the formula in the statement.
\end{proof}

We have as well the following result, providing an alternative to Definition 13.28, and which will be our formulation of the Compound Poisson Limit Theorem (CPLT):

\index{compound Poisson Limit theorem}
\index{CPLT}

\begin{theorem}
For $\nu=\sum_{i=1}^sc_i\delta_{z_i}$ with $c_i>0$ and $z_i\in\mathbb C$ we have
$$p_\nu={\rm law}\left(\sum_{i=1}^sz_i\alpha_i\right)$$
where the variables $\alpha_i$ are Poisson $(c_i)$, independent.
\end{theorem}

\begin{proof}
Let $\alpha$ be the sum of Poisson variables in the statement. We have:
\begin{eqnarray*}
F_{\alpha_i}(y)=\exp(c_i(e^{iy}-1)
&\implies&F_{z_i\alpha_i}(y)=\exp(c_i(e^{iyz_i}-1))\\
&\implies&F_\alpha(y)=\exp\left(\sum_{i=1}^sc_i(e^{iyz_i}-1)\right)
\end{eqnarray*}

Thus we have the same formula as in Proposition 13.29, as desired.
\end{proof}

Getting back now to the Bessel laws, we have the following result:

\index{Bessel law}

\begin{theorem}
The Bessel laws $b_t$ are compound Poisson laws, given by
$$b_t=p_{t\varepsilon}$$
where $\varepsilon=\frac{1}{2}(\delta_{-1}+\delta_1)$ is the uniform measure on $\mathbb Z_2$.
\end{theorem}

\begin{proof}
This follows indeed by comparing the formula of the Fourier transform of $b_t$, from the proof of Theorem 13.27, with the formula in Proposition 13.29.
\end{proof}

Our next task will be that of unifying and generalizing the results that we have for $S_N,H_N$. For this purpose, consider the following family of groups:

\index{complex reflection group}

\begin{definition}
The complex reflection group $H_N^s=\mathbb Z_s\wr S_N$ is given by
$$H_N^s=M_N(\mathbb Z_s\cup\{0\})\cap U_N$$
with the convention $\mathbb Z_\infty=\mathbb T$, at $s=\infty$.
\end{definition}

Here the fact that we have indeed $H_N^s=\mathbb Z_s\wr S_N$ follows as in Theorem 13.25. Observe that at $s=1,2$ we obtain the symmetric and hyperoctahedral groups:
$$H_N^1=S_N\quad,\quad 
H_N^2=H_N$$

Another important particular case is $s=\infty$, where we obtain a compact group which is actually not finite, but is of key importance, that we will denote as follows:
$$H_N^\infty=K_N$$

In order to do now the character computations for $H_N^s$, in general, we need a number of further probabilistic preliminaries. Let us start with the following definition:

\index{generalized Bessel laws}
\index{complex Bessel laws}

\begin{definition}
The Bessel law of level $s\in\mathbb N\cup\{\infty\}$ and parameter $t>0$ is
$$b_t^s=p_{t\varepsilon_s}$$
with $\varepsilon_s$ being the uniform measure on the $s$-th roots of unity.
\end{definition}

Observe that at $s=1,2$ we obtain the Poisson and real Bessel laws:
$$b^1_t=p_t\quad,\quad 
b^2_t=b_t$$

Another important particular case is $s=\infty$, where we obtain a measure which is actually not discrete, that we will denote as follows:
$$b^\infty_t=B_t$$

As a basic result on these laws, generalizing those before about $p_t,b_t$, we have:

\begin{theorem}
The generalized Bessel laws $b^s_t$ have the property
$$b^s_t*b^s_{t'}=b^s_{t+t'}$$
so they form a truncated one-parameter semigroup
with respect to convolution.
\end{theorem}

\begin{proof}
This follows indeed from the Fourier transform formula from Proposition 13.29, because for the Bessel laws, the log of this Fourier transform is linear in $t$.
\end{proof}

We can go back now to the reflection groups, and we have the following result:

\begin{theorem}
For the group $H_N^s=\mathbb Z_s\wr S_N$ we have, with $N\to\infty$,
$$\chi_t\sim b^s_t$$
where $b_t^s=p_{t\varepsilon_s}$, with $\varepsilon_s$ being the uniform measure on the $s$-th roots of unity.
\end{theorem}

\begin{proof}
In the case $t=1$, by arguing as before at $s=2$, since the limit probability for a random permutation to have exactly $k$ fixed points is $e^{-1}/k!$, we obtain:
$$\lim_{N\to\infty}law(\chi_1)=e^{-1}\sum_{k=0}^\infty \frac{1}{k!}\,\varepsilon_s^{*k}$$

On the other hand, we get from the definition of the Bessel law $b^s_1$, as desired:
\begin{eqnarray*}
b^s_1
&=&\lim_{N\to\infty}\left(\left(1-\frac{1}{N}\right)\delta_0+\frac{1}{N}\,\varepsilon_s\right)^{*N}\\
&=&\lim_{N\to\infty}\sum_{k=0}^N\begin{pmatrix}N\\ k\end{pmatrix}\left(1-\frac{1}{N}\right)^{N-k}\frac{1}{N^k}\,\varepsilon_s^{*k}\\
&=&e^{-1}\sum_{k=0}^\infty\frac{1}{k!}\,\varepsilon_s^{*k}
\end{eqnarray*}

When $t>0$ is arbitrary, we can use the same method, with some modifications where needed, again by arguing as before at $s=2$, and we obtain the result.
\end{proof}

\section*{13e. Exercises}

We had an interesting group theory chapter here, and as exercises, we have:

\begin{exercise}
Learn about the decomposition of permutations, as products of cycles.
\end{exercise}

\begin{exercise}
Review if needed the basic theory of the Poisson laws.
\end{exercise}

\begin{exercise}
Fill in all the details for the computation of $law(\chi_t)$, over $S_N$.
\end{exercise}

\begin{exercise}
Find an alternative proof for the $S_N$ result, by integrating over $S_N$.
\end{exercise}

\begin{exercise}
Learn about the Bessel functions of the first kind, appearing above.
\end{exercise}

\begin{exercise}
Learn more about the compound Poisson laws, and their properties.
\end{exercise}

\begin{exercise}
Learn about the complex reflection groups, and their classification.
\end{exercise}

\begin{exercise}
Learn about the finite abelian groups too, and their classification.
\end{exercise}

As bonus exercise, and no surprise here, read some systematic finite group theory.

\chapter{Compact groups}

\section*{14a. Lie groups}

We have seen the basic theory of the finite subgroups $G\subset U_N$. In this chapter we go for the real thing, namely basic theory of the continuous subgroups $G\subset U_N$. Indeed, these groups are the most important ones, for both mathematics and physics, and historically they came first, from the work of Felix Klein, Sophus Lie and others.

\bigskip

However, getting into this subject, continuous subgroups $G\subset U_N$, is something quite unclear, right from the beginning, because we have the following dillema:

\begin{dillema}
Shall our continuous subgroups $G\subset U_N$ be:
\begin{enumerate}
\item Continuous, or smooth.

\item Compact, or locally compact.
\end{enumerate}
\end{dillema}

Which does not look like something easy to solve, all possible combinations here seem to lead to substantial difficulties, and I am afraid that, following Klein, Lie and the others, who asked their respective cats about this, I will have to ask the cat too.

\bigskip

Unfortunately cat is gone, so in the lack of some good advice here, we will solve this problem Gordian knot style, by starting with the strongest possible axioms:

\begin{definition}
A compact Lie group is a compact group $G$ which is at the same time a smooth manifold, with the group operations being smooth.
\end{definition}

Here we are making reference to the material from chapter 10, where we discussed what a smooth manifold is. Also, we have temporarily ditched the assumption $G\subset U_N$, for keeping things simple. We can always add this condition later, coming as an assumption, or why not as a theorem, if we are lucky, say as in the case of the finite groups.

\bigskip

Now inspired by the material from chapter 10, we are led to the following questions, which should normally provide the key to the study of the compact Lie groups:

\begin{questions}
Given a compact Lie group $G$, as above:
\begin{enumerate}
\item What can we say about the tangent space at the unit, $\mathfrak g=T_1G$? 

\item Can $\mathfrak g$ be axiomatized? Do we have a correspondence $\mathfrak g\leftrightarrow G$? 

\item What about $G\subset U_N$, how does this translate in terms of $\mathfrak g$?
\end{enumerate}
\end{questions}

In answer now, inspired by what happens in the simplest cases, $G=O_N,U_N$, and by some differential geometry too, and skipping some details here, we have:

\begin{definition}
A Lie algebra is a vector space $\mathfrak g$ with an operation $(x,y)\to[x,y]$, called Lie bracket, subject to the following conditions:
\begin{enumerate}
\item $[x+y,z]=[x,z]+[y,z]$, $[x,y+z]=[x,y]+[x,z]$.

\item $[\lambda x,y]=[x,\lambda y]=\lambda[x,y]$.

\item $[x,x]=0$.

\item $[[x,y],z]+[[y,z],x]+[[z,x],y]=0$.
\end{enumerate}
\end{definition}

As a basic example here, consider a usual, associative algebra $A$. We can define then the Lie bracket on it as being the usual commutator, namely:
$$[x,y]=xy-yx$$

The above axioms (1,2,3) are then clearly satisfied, and in what regards axiom (4), called Jacobi identity, this is satisfied too, the verification being as follows:
\begin{eqnarray*}
&&[[x,y],z]+[[y,z],x]+[[z,x],y]\\
&=&[xy-yx,z]+[yz-zy,x]+[zx-xz,y]\\
&=&xyz-yxz-zxy+zyx+yzx-zyx-xyz+xzy+zxy-xzy-yzx+yxz\\
&=&0
\end{eqnarray*}

We will see in a moment that up to a certain abstract operation $\mathfrak g\to U\mathfrak g$, which is something very straightforward, called enveloping Lie algebra construction, any Lie algebra appears in this way, with its Lie bracket being formally a commutator:
$$[x,y]=xy-yx$$

In relation now with groups, we have the following fundamental result, making the connection with the theory of Lie groups, denoted as usual by $G$:

\begin{theorem}
Given a Lie group $G$, that is, a group which is a smooth manifold, with the group operations being smooth, the tangent space at the identity
$$\mathfrak g=T_1(G)$$
is a Lie algebra, with its Lie bracket being basically a usual commutator.
\end{theorem}

\begin{proof}
This is something non-trivial, the idea being as follows:

\medskip

(1) Let us first have a look at the orthogonal and unitary groups $O_N,N_N$. These are both Lie groups, and the corresponding Lie algebras $\mathfrak o_N,\mathfrak u_N$ can be computed by differentiating the equations defining $O_N,U_N$, with the conclusion being as follows:
$$\mathfrak o_N=\left\{ A\in M_N(\mathbb R)\Big|A^t=-A\right\}$$
$$\mathfrak u_N=\left\{ B\in M_N(\mathbb C)\Big|B^*=-B\right\}$$

This was for the correspondences $O_N\to\mathfrak o_N$ and $U_N\to\mathfrak u_N$. In the other sense, the correspondences $\mathfrak o_N\to O_N$ and $\mathfrak u_N\to U_N$ appear by exponentiation, the result here stating that, around 1, the orthogonal matrices can be written as $U=e^A$, with $A\in\mathfrak o_N$, and the unitary matrices can be written as $U=e^B$, with $B\in\mathfrak u_N$.

\medskip

(2) Getting now to the Lie bracket, the first observation is that both $\mathfrak o_N,\mathfrak u_N$ are stable under the usual commutator of the $N\times N$ matrices. Indeed, assuming that $A,B\in M_N(\mathbb R)$ satisfy $A^t=-A$, $B^t=-B$, their commutator satisfies $[A,B]\in M_N(\mathbb R)$, and:
\begin{eqnarray*}
[A,B]^t
&=&(AB-BA)^t\\
&=&B^tA^t-A^tB^t\\
&=&BA-AB\\
&=&-[A,B]
\end{eqnarray*}

Similarly, assuming that $A,B\in M_N(\mathbb C)$ satisfy $A^*=-A$, $B^*=-B$, their commutator $[A,B]\in M_N(\mathbb C)$ satisfies the condition $[A,B]^*=-[A,B]$.

\medskip

(3) Thus, both tangent spaces $\mathfrak o_N,\mathfrak u_N$ are Lie algebras, with the Lie bracket being the usual commutator of the $N\times N$ matrices. It remains now to see what happens to the Lie bracket when exponentiating, and the formula here is as follows:
$$e^{[A,B]}=e^{AB-BA}$$

But the term on the right can be understood in terms of the differential geometry of $O_N,U_N$, and the situation is similar when dealing with an arbitrary Lie group $G$.
\end{proof}

With this understood, let us go back to the arbitrary Lie algebras, as axiomatized in Definition 14.4.  We have the following key result, announced after Definition 14.4:

\begin{theorem}
Given a Lie algebra $\mathfrak g$, define its enveloping Lie algebra $U\mathfrak g$ as being the quotient of the tensor algebra of $\mathfrak g$, namely
$$T(\mathfrak g)=\bigoplus_{k=0}^\infty\mathfrak g^{\otimes k}$$
by the following associative algebra ideal, with $x,y$ ranging over the elements of $\mathfrak g$:
$$I=<x\otimes y-y\otimes x-[x,y]>$$
Then $U\mathfrak g$ is an associative algebra, so it is a Lie algebra too, with bracket 
$$[x,y]=xy-yx$$
and the standard embedding $\mathfrak g\subset U\mathfrak g$ is a Lie algebra embedding.\end{theorem}

\begin{proof}
This is something which is quite self-explanatory, and in what regards the examples, illustrations, and other things that can be said, for instance in relation with the Lie groups, we will leave some further reading here as an instructive exercise.
\end{proof}

Importantly, the above enveloping Lie algebra construction makes as well a link with Hopf algebra theory, and with quantum groups, via the following result:

\begin{theorem}
Given a Lie algebra $\mathfrak g$, its enveloping Lie algebra $U\mathfrak g$ is a cocommutative Hopf algebra, with comultiplication, counit and antipode given by
$$\Delta:U\mathfrak g\to U(\mathfrak g\oplus\mathfrak g)=U\mathfrak g\otimes U\mathfrak g\quad,\quad x\to x+x$$
$$\varepsilon:U\mathfrak g\to\mathbb C\quad,\quad x\to 1$$
$$S:U\mathfrak g\to U\mathfrak g^{opp}=(U\mathfrak g)^{opp}\quad,\quad x\to-x$$
via various standard identifications, for the various associative algebras involved.
\end{theorem}

\begin{proof}
Again, this is something self-explanatory, provided that you already know about Hopf algebras and quantum groups, and in what regards the examples, illustrations, and other things that can be said, we will leave some reading here as an exercise.
\end{proof}

So long for the Lie algebra basics, quicky explained in a few pages. The continuation of the story, bringing answers to Questions 14.3, is more complicated, as follows:

\begin{answers}
The following happen, under suitable assumptions:
\begin{enumerate}
\item We have indeed a correspondence $\mathfrak g\leftrightarrow G$, appearing by exponentiation.

\item The Lie algebras $\mathfrak g$ can be classified, and the compact Lie groups $G$ too.

\item In practice, we have regular cases ABCD, and exceptional cases EFG.

\item The regular compact Lie groups, of type ABCD, are $O_N,U_N,Sp_N$.
\end{enumerate}
\end{answers}

In relation with the last assertion, which is what matters the most, $O_N,U_N$ are the usual orthogonal and unitary groups, and $Sp_N\subset U_N$ with $N\in2\mathbb N$ is the symplectic group, appearing as a certain modification of the usual orthogonal group $O_N\subset U_N$.

\bigskip

Observe that, as a consequence, any compact Lie group $G$ is a unitary group, $G\subset U_N$. Finally, many further things can be said in relation with Theorem 14.7, notably with a deformation procedure for $O_N,U_N,Sp_N$, into certain ``quantum groups'', invented by Drinfeld and Jimbo. For more on all this, we refer to any good Lie algebra book.

\section*{14b. Peter-Weyl}

You might wonder at this point why not getting into details in relation with the above, which is certainly first-class mathematics, and that could easily cover the remainder of the present chapter, and perhaps even the remainder of the whole present book.

\bigskip

Well, the problem comes from the cat, who came back from his daily hunting adventures, and jumped back scared when seeing what I was typing. Here is what he says:

\begin{cat}
All this is too complicated, Lie algebras are advanced science. For an introduction, and some nice applications, go with representations of compact groups.
\end{cat}

Okay cat, I was actually sort of expecting this, a word from you, and this since I fell into Dillema 14.1. So, forgetting now about Definition 14.2, which most likely would lead us into a Nobel Prize, or a Fields Medal, but are we here for such things, let us discuss instead the representations of compact groups. We will need the following notions:

\begin{definition}
A unitary representation of a compact group $G$ is a continuous group morphism into a unitary group
$$u:G\to U_N\quad,\quad g\to u_g$$
which can be faithful or not. The character of such a representation is the function
$$\chi:G\to\mathbb C\quad,\quad g\to Tr(u_g)$$
where $Tr$ is the usual, unnormalized trace of the $N\times N$ matrices.
\end{definition}

At the level of examples, most of the compact groups that we met so far, finite or continuous, naturally appear as closed subgroups $G\subset U_N$. In this case, the embedding $G\subset U_N$ is of course a representation, called fundamental representation.

\bigskip

In general now, let us first discuss the various operations on the representations. We have here the following elementary result, coming from definitions:

\begin{proposition}
The representations of a compact group $G$ are subject to:
\begin{enumerate}
\item Making sums. Given representations $u,v$, of dimensions $N,M$, 
their sum is the $N+M$-dimensional representation $u+v=diag(u,v)$.

\item Making products. Given representations $u,v$, of dimensions $N,M$, their product is the $NM$-dimensional representation $(u\otimes v)_{ia,jb}=u_{ij}v_{ab}$.

\item Taking conjugates. Given a $N$-dimensional representation $u$, its conjugate is the $N$-dimensional representation $(\bar{u})_{ij}=\bar{u}_{ij}$.

\item Spinning by unitaries. Given a $N$-dimensional representation $u$, and a unitary $V\in U_N$, we can spin $u$ by this unitary, $u\to VuV^*$.
\end{enumerate}
\end{proposition}

\begin{proof}
The fact that the operations in the statement are indeed well-defined, among morphisms from $G$ to unitary groups, is indeed clear from definitions.
\end{proof}

In relation now with characters, we have the following result:

\begin{proposition}
We have the following formulae, regarding characters
$$\chi_{u+v}=\chi_u+\chi_v\quad,\quad 
\chi_{u\otimes v}=\chi_u\chi_v\quad,\quad 
\chi_{\bar{u}}=\bar{\chi}_u\quad,\quad
\chi_{VuV^*}=\chi_u$$
in relation with the basic operations for the representations.
\end{proposition}

\begin{proof}
All these assertions are elementary, by using the following formulae:
$$Tr(diag(U,V))=Tr(U)+Tr(V)\quad,\quad 
Tr(U\otimes V)=Tr(U)Tr(V)$$
$$Tr(\bar{U})=\overline{Tr(U)}\quad,\quad 
Tr(VUV^*)=Tr(U)$$

Thus, we are led to the character formulae in the statement.
\end{proof}

Assume now that we are given a closed subgroup $G\subset U_N$. By using the above operations, we can construct a whole family of representations of $G$, as follows:

\begin{definition}
Given a closed subgroup $G\subset U_N$, its Peter-Weyl representations are the tensor products between the fundamental representation and its conjugate:
$$u:G\subset U_N\quad,\quad 
\bar{u}:G\subset U_N$$ 
We denote these tensor products $u^{\otimes k}$, with $k=\circ\bullet\bullet\circ\ldots$ being a colored integer, with the colored tensor powers being defined according to the rules 
$$u^{\otimes\circ}=u\quad,\quad
u^{\otimes\bullet}=\bar{u}\quad,\quad
u^{\otimes kl}=u^{\otimes k}\otimes u^{\otimes l}$$
and with the convention that $u^{\otimes\emptyset}$ is the trivial representation $1:G\to U_1$.
\end{definition}

Here are a few examples of such representations, namely those coming from the colored integers of length 2, which will often appear in what follows:
$$u^{\otimes\circ\circ}=u\otimes u\quad,\quad 
u^{\otimes\circ\bullet}=u\otimes\bar{u}$$
$$u^{\otimes\bullet\circ}=\bar{u}\otimes u\quad,\quad
u^{\otimes\bullet\bullet}=\bar{u}\otimes\bar{u}$$

In order to advance, we must develop some general theory. Let us start with:

\begin{definition}
Given a compact group $G$, and two of its representations,
$$u:G\to U_N\quad,\quad 
v:G\to U_M$$
we define the space of intertwiners between these representations as being 
$$Hom(u,v)=\left\{T\in M_{M\times N}(\mathbb C)\Big|Tu_g=v_gT,\forall g\in G\right\}$$
and we use the following conventions:
\begin{enumerate}
\item We use the notations $Fix(u)=Hom(1,u)$, and $End(u)=Hom(u,u)$.

\item We write $u\sim v$ when $Hom(u,v)$ contains an invertible element.

\item We say that $u$ is irreducible, and write $u\in Irr(G)$, when $End(u)=\mathbb C1$.
\end{enumerate}
\end{definition}

The terminology here is standard, with Fix, Hom, End standing for fixed points, homomorphisms and endomorphisms. We will see later that irreducible means indecomposable, in a suitable sense. Here are now a few basic results, regarding these spaces:

\begin{theorem}
The spaces of intertwiners have the following properties:
\begin{enumerate}
\item $T\in Hom(u,v),S\in Hom(v,w)\implies ST\in Hom(u,w)$.

\item $S\in Hom(u,v),T\in Hom(w,z)\implies S\otimes T\in Hom(u\otimes w,v\otimes z)$.

\item $T\in Hom(u,v)\implies T^*\in Hom(v,u)$.
\end{enumerate}
In abstract terms, we say that the Hom spaces form a tensor $*$-category.
\end{theorem}

\begin{proof}
All the formulae in the statement are clear from definitions, via elementary computations. As for the last assertion, this is something coming from (1,2,3). We will be back to tensor categories later on, with more details on this latter fact.
\end{proof}

In order to advance, we will need the following standard linear algebra fact:

\begin{proposition}
Let $A\subset M_N(\mathbb C)$ be a $*$-algebra.
\begin{enumerate}
\item We have $1=p_1+\ldots+p_k$, with $p_i\in A$ being central minimal projections.

\item Each of the spaces $A_i=p_iAp_i$ is a non-unital $*$-subalgebra of $A$.

\item We have a non-unital $*$-algebra sum decomposition $A=A_1\oplus\ldots\oplus A_k$.

\item We have unital $*$-algebra isomorphisms $A_i\simeq M_{n_i}(\mathbb C)$, with $n_i=rank(p_i)$.

\item Thus, we have a $*$-algebra isomorphism $A\simeq M_{n_1}(\mathbb C)\oplus\ldots\oplus M_{n_k}(\mathbb C)$.
\end{enumerate}
\end{proposition}

\begin{proof}
Consider indeed an arbitrary $*$-algebra of the $N\times N$ matrices, $A\subset M_N(\mathbb C)$. Let us first look at the center of this algebra, $Z(A)=A\cap A'$. It is elementary to prove that this center, as an algebra, is of the following form:
$$Z(A)\simeq\mathbb C^k$$

Consider now the standard basis $e_1,\ldots,e_k\in\mathbb C^k$, and let  $p_1,\ldots,p_k\in Z(A)$ be the images of these vectors via the above identification. In other words, these elements $p_1,\ldots,p_k\in A$ are central minimal projections, summing up to 1:
$$p_1+\ldots+p_k=1$$

The idea is then that this partition of the unity eventually leads to the block decomposition of $A$, as in the statement, and we leave the details here as an exercise.
\end{proof}

We can now formulate our first Peter-Weyl type theorem, as follows:

\index{Peter-Weyl}

\begin{theorem}[PW1]
Let $u:G\to U_N$ be a representation, consider the algebra $A=End(u)$, and write its unit $1=p_1+\ldots+p_k$ as above. We have then 
$$u=v_1+\ldots+v_k$$
with each $v_i$ being an irreducible representation, obtained by restricting $u$ to $Im(p_i)$.
\end{theorem}

\begin{proof}
This follows indeed from Theorem 14.15 and Proposition 14.16:

\medskip

(1) We first associate to our representation $u:G\to U_N$ the corresponding action map on $\mathbb C^N$. If a linear subspace $V\subset\mathbb C^N$ is invariant, the restriction of the action map to $V$ is an action map too, which must come from a subrepresentation $v\subset u$.

\medskip

(2) Consider now a projection $p\in End(u)$. From $pu=up$ we obtain that the linear space $V=Im(p)$ is invariant under $u$, and so this space must come from a subrepresentation $v\subset u$. It is routine to check that the operation $p\to v$ maps subprojections to subrepresentations, and minimal projections to irreducible representations.

\medskip

(3) With these preliminaries in hand, let us decompose the algebra $End(u)$ as above, by using the decomposition $1=p_1+\ldots+p_k$ into central minimal projections. If we denote by $v_i\subset u$ the subrepresentation coming from the vector space $V_i=Im(p_i)$, then we obtain in this way a decomposition $u=v_1+\ldots+v_k$, as in the statement.
\end{proof}

Here is now our second Peter-Weyl theorem, complementing Theorem 14.17:

\begin{theorem}[PW2]
Given a closed subgroup $G\subset_uU_N$, any of its irreducible smooth representations 
$$v:G\to U_M$$
appears inside a tensor product of the fundamental representation $u$ and its adjoint $\bar{u}$.
\end{theorem}

\begin{proof}
This basically follows from Theorem 14.17, by reasoning as follows:

\medskip

(1) Given $v:G\to U_M$, consider its space of coefficients $C_v\subset C(G)$. The operation $w\to C_w$ is then functorial, mapping subrepresentations into linear subspaces.

\medskip

(2) A closed subgroup $G\subset_uU_N$ is a Lie group, and a representation $v:G\to U_M$ is smooth when we have an inclusion $C_v\subset<C_u>$. This is indeed well-known.

\medskip

(3) By definition of the Peter-Weyl representations, as arbitrary tensor products between the fundamental representation $u$ and its conjugate $\bar{u}$, we have:
$$<C_u>=\sum_kC_{u^{\otimes k}}$$

(4) Now by putting together the above observations (2,3) we conclude that we must have an inclusion as follows, for certain exponents $k_1,\ldots,k_p\in\mathbb N$:
$$C_v\subset C_{u^{\otimes k_1}\oplus\ldots\oplus u^{\otimes k_p}}$$

(5) By using now (1), we deduce that we have an inclusion $v\subset u^{\otimes k_1}\oplus\ldots\oplus u^{\otimes k_p}$, and by applying Theorem 14.17, this leads to the conclusion in the statement.
\end{proof}

In order to further advance, we will need the following standard fact:

\index{Haar measure}
\index{Ces\`aro limit}

\begin{theorem}
Any compact group $G$ has a unique Haar integration, which can be constructed by starting with any faithful positive unital form $\varphi\in C(G)^*$, and setting:
$$\int_G=\lim_{n\to\infty}\frac{1}{n}\sum_{k=1}^n\varphi^{*k}$$
Moreover, for any representation $v$ we have the formula
$$\left(id\otimes\int_G\right)v=P$$
where $P$ is the orthogonal projection onto $Fix(v)=\left\{\xi\in\mathbb C^n\big|v\xi=\xi\right\}$.
\end{theorem}

\begin{proof}
This is something very standard, and we will leave the proof here, first in the case where $G$ is finite, which is easier, and then in general, and an instructive analysis exercise. Of course, in case you are stuck at some point, do not hesitate to look it up.
\end{proof}

We will need as well an algebraic ingredient for our study, as follows:

\index{Frobenius isomorphism}

\begin{proposition}
We have a Frobenius type isomorphism
$$Hom(v,w)\simeq Fix(v\otimes\bar{w})$$
valid for any two representations $v,w$.
\end{proposition}

\begin{proof}
According to the definitions, we have the following equivalences:
\begin{eqnarray*}
T\in Hom(v,w)
&\iff&Tv=wT\\
&\iff&\sum_jT_{aj}v_{ji}=\sum_bw_{ab}T_{bi},\forall a,i
\end{eqnarray*}

On the other hand, we have as well the following equivalences:
\begin{eqnarray*}
T\in Fix(v\otimes\bar{w})
&\iff&(v\otimes\bar{w})T=\xi\\
&\iff&\sum_{jb}v_{ij}w_{ab}^*T_{bj}=T_{ai}\forall a,i
\end{eqnarray*}

But with this in hand, both inclusions follow from the unitarity of $v,w$.
\end{proof}

Good news, we can now formulate our third Peter-Weyl theorem, as follows:

\index{Peter-Weyl}

\begin{theorem}[PW3]
The dense subalgebra $\mathcal C(G)\subset C(G)$ generated by the coefficients of the fundamental representation decomposes as a direct sum 
$$\mathcal C(G)=\bigoplus_{v\in Irr(G)}M_{\dim(v)}(\mathbb C)$$
with the summands being pairwise orthogonal with respect to $<f,g>=\int_Gf\bar{g}$.
\end{theorem}

\begin{proof}
By combining the previous two Peter-Weyl results, we deduce that we have a linear space decomposition as follows:
$$\mathcal C(G)
=\sum_{v\in Irr(G)}C_v
=\sum_{v\in Irr(G)}M_{\dim(v)}(\mathbb C)$$

Thus, in order to conclude, it is enough to prove that for any two irreducible representations $v,w\in Irr(G)$, the corresponding spaces of coefficients are orthogonal:
$$v\not\sim w\implies C_v\perp C_w$$ 

But this follows by Frobenius duality, by integrating. Let us set indeed:
$$P_{ia,jb}=\int_Gv_{ij}\bar{w}_{ab}$$

Then $P$ is the orthogonal projection onto the following vector space:
$$Fix(v\otimes\bar{w})
\simeq Hom(v,w)
=\{0\}$$

Thus we have $P=0$, and this gives the result.
\end{proof}

Finally, we have the following result, completing the Peter-Weyl theory:

\index{Peter-Weyl}

\begin{theorem}[PW4]
The characters of irreducible representations belong to
$$\mathcal C(G)_{central}=\left\{f\in\mathcal C(G)\Big|f(gh)=f(hg),\forall g,h\in G\right\}$$
called algebra of central functions on $G$, and form an orthonormal basis of it.
\end{theorem}

\begin{proof}
Observe first that $\mathcal C(G)_{central}$ is indeed an algebra, which contains all the characters. Conversely, consider a function $f\in\mathcal C(G)$, written as follows:
$$f=\sum_{v\in Irr(G)}f_v$$

The condition $f\in\mathcal C(G)_{central}$ states then that for any $v\in Irr(G)$, we must have:
$$f_v\in\mathcal C(G)_{central}$$

But this means that $f_v$ must be a scalar multiple of $\chi_v$, so the characters form a basis of $\mathcal C(G)_{central}$, as stated. Also, the fact that we have an orthogonal basis follows from Theorem 14.21. As for the fact that the characters have norm 1, this follows from:
$$\int_G\chi_v\bar{\chi}_v
=\sum_{ij}\int_Gv_{ii}\bar{v}_{jj}
=\sum_i\frac{1}{M}
=1$$

Here we have used the fact, coming from Frobenius duality, that the various integrals $\int_Gv_{ij}\bar{v}_{kl}$ form altogether the orthogonal projection onto the following vector space:
$$Fix(v\otimes\bar{v})\simeq End(v)=\mathbb C1$$

Thus, the proof of our theorem is now complete.
\end{proof}

\section*{14c. Brauer algebras}

Getting now to more juicy material, the Lie algebra theory as developed before is not the only way of ``linearizing'' the Lie groups. As a rival principle, we have:

\begin{principle}
Any finite, or even general compact group $G$ appears as the symmetry group of its corresponding Tannakian category $C_G$,
$$G=G(C_G)$$
and by suitably delinearizing $C_G$, say via a Brauer theorem of type $C_G=span(D_G)$, we can view $G$ as symmetry group of a certain combinatorial object $D_G$.
\end{principle}

Excited about this? Does not look easy, all this material, with both Tannaka and Brauer being quite scary names, in the context of algebra. But, believe me, all this is worth learning, and it is good to have in your bag some cutting-edge technology regarding the groups, such as the results of Tannaka and Brauer. So, we will go for this.

\bigskip

Getting started now, we first have a categorical definition, as follows:

\begin{definition}
A tensor category over $H=\mathbb C^N$ is a collection $C=(C_{kl})$ of linear spaces $C_{kl}\subset\mathcal L(H^{\otimes k},H^{\otimes l})$ satisfying the following conditions:
\begin{enumerate}
\item $S,T\in C$ implies $S\otimes T\in C$.

\item If $S,T\in C$ are composable, then $ST\in C$.

\item $T\in C$ implies $T^*\in C$.

\item Each $C_{kk}$ contains the identity operator.

\item $C_{\emptyset k}$ with $k=\circ\bullet,\bullet\circ$ contain the operator $R:1\to\sum_ie_i\otimes e_i$.

\item $C_{kl,lk}$ with $k,l=\circ,\bullet$ contain the flip operator $\Sigma:a\otimes b\to b\otimes a$.
\end{enumerate}
\end{definition}

Here, as usual, the tensor powers $H^{\otimes k}$, which are Hilbert spaces depending on a colored integer $k=\circ\bullet\bullet\circ\ldots\,$, are defined by the following formulae, and multiplicativity:
$$H^{\otimes\emptyset}=\mathbb C\quad,\quad 
H^{\otimes\circ}=H\quad,\quad
H^{\otimes\bullet}=\bar{H}\simeq H$$

We have already met such categories, when dealing with the Tannakian categories of the closed subgroups $G\subset U_N$, and our knowledge can be summarized as follows:

\begin{proposition}
Given a closed subgroup $G\subset U_N$, its Tannakian category
$$C_{kl}=\left\{T\in\mathcal L(H^{\otimes k},H^{\otimes l})\Big|Tg^{\otimes k}=g^{\otimes l}T,\forall g\in G\right\}$$
is a tensor category over $H=\mathbb C^N$. Conversely, given a tensor category $C$ over $\mathbb C^N$,
$$G=\left\{g\in U_N\Big|Tg^{\otimes k}=g^{\otimes l}T,\forall k,l,\forall T\in C_{kl}\right\}$$
is a closed subgroup of $U_N$.
\end{proposition}

\begin{proof}
This is something that we basically know, the idea being as follows:

\medskip

(1) Regarding the first assertion, we have to check here the axioms (1-6) in Definition 14.24. The axioms (1-4) being all clear from definitions, let us establish (5). But this follows from the fact that each element $g\in G$ is a unitary, which can be reformulated as follows, with $R:1\to\sum_ie_i\otimes e_i$ being the map in Definition 14.24:
$$R\in Hom(1,g\otimes\bar{g})\quad,\quad 
R\in Hom(1,\bar{g}\otimes g)$$

Regarding now the condition in Definition 14.24 (6), this comes from the fact that the matrix coefficients $g\to g_{ij}$ and their conjugates $g\to\bar{g}_{ij}$ commute with each other.

\medskip

(2) Regarding the second assertion, we have to check that the subset $G\subset U_N$ constructed in the statement is a closed subgroup. But this is clear from definitions.
\end{proof}

Summarizing, we have so far precise axioms for the tensor categories $C=(C_{kl})$, given in Definition 14.24, as well as correspondences as follows:
$$G\to C_G\quad,\quad 
C\to G_C$$

We will prove in what follows that these correspondences are inverse to each other. In order to get started, we first have the following technical result:

\begin{proposition}
Consider the following conditions:
\begin{enumerate}
\item $C=C_{G_C}$, for any tensor category $C$.

\item $G=G_{C_G}$, for any closed subgroup $G\subset U_N$.
\end{enumerate}
We have then $(1)\implies(2)$. Also, $C\subset C_{G_C}$ is automatic.
\end{proposition}

\begin{proof}
Given $G\subset U_N$, we have $G\subset G_{C_G}$. On the other hand, by using (1) we have $C_G=C_{G_{C_G}}$. Thus, we have an inclusion of closed subgroups of $U_N$, which becomes an isomorphism at the level of the associated Tannakian categories, so $G=G_{C_G}$. Finally, the fact that we have an inclusion $C\subset C_{G_C}$ is clear from definitions.
\end{proof}

The point now is that it is possible to prove that we have $C_{G_C}\subset C$, by doing some abstract algebra, and we are led in this way to the following conclusion:

\begin{theorem}
The Tannakian duality constructions 
$$C\to G_C\quad,\quad 
G\to C_G$$
are inverse to each other.
\end{theorem}

\begin{proof}
This is something quite tricky, the idea being as follows:

\medskip

(1) According to Proposition 14.26, we must prove $C_{G_C}\subset C$. For this purpose, given a tensor category $C=(C_{kl})$ over a Hilbert space $H$, consider the following $*$-algebra:
$$E_C
=\bigoplus_{k,l}C_{kl}
\subset\bigoplus_{k,l}B(H^{\otimes k},H^{\otimes l})
\subset B\left(\bigoplus_kH^{\otimes k}\right)$$

Consider also, inside this $*$-algebra, the following $*$-subalgebra:
$$E_C^{(s)}
=\bigoplus_{|k|,|l|\leq s}C_{kl}
\subset\bigoplus_{|k|,|l|\leq s}B(H^{\otimes k},H^{\otimes l})
=B\left(\bigoplus_{|k|\leq s}H^{\otimes k}\right)$$

(2) It is then routine to check that we have equivalences as follows:
\begin{eqnarray*}
C_{G_C}\subset C
&\iff&E_{C_{G_C}}\subset E_C\\
&\iff&E_{C_{G_C}}^{(s)}\subset E_C^{(s)},\forall s\\
&\iff&E_{C_{G_C}}^{(s)'}\supset E_C^{(s)'},\forall s
\end{eqnarray*}

(3) Summarizing, we would like to prove that we have inclusions $E_C^{(s)'}\subset E_{C_{G_C}}^{(s)'}$. But this can be done by doing some abstract algebra, and we refer here to the standard literature on the subject. For more on all this, you have as well my book \cite{ba1}.
\end{proof}

With this piece of general theory in hand, let us go back to Principle 14.23, and develop the second idea there, namely delinearization and Brauer theorems. We have:

\begin{definition}
A category of crossing partitions is a collection $D=\bigsqcup_{k,l}D(k,l)$ of subsets $D(k,l)\subset P(k,l)$, having the following properties:
\begin{enumerate}
\item Stability under the horizontal concatenation, $(\pi,\sigma)\to[\pi\sigma]$.

\item Stability under vertical concatenation $(\pi,\sigma)\to[^\sigma_\pi]$, with matching middle symbols.

\item Stability under the upside-down turning $*$, with switching of colors, $\circ\leftrightarrow\bullet$.

\item Each set $P(k,k)$ contains the identity partition $||\ldots||$.

\item The sets $P(\emptyset,\circ\bullet)$ and $P(\emptyset,\bullet\circ)$ both contain the semicircle $\cap$.

\item The sets $P(k,\bar{k})$ with $|k|=2$ contain the crossing partition $\slash\hskip-2.0mm\backslash$.
\end{enumerate}
\end{definition} 

Observe the similarity with Definition 14.24, and more on this in a moment. In order now to construct a Tannakian category out of such a category, we will need:

\begin{proposition}
Each partition $\pi\in P(k,l)$ produces a linear map
$$T_\pi:(\mathbb C^N)^{\otimes k}\to(\mathbb C^N)^{\otimes l}$$
given by the following formula, with $e_1,\ldots,e_N$ being the standard basis of $\mathbb C^N$,
$$T_\pi(e_{i_1}\otimes\ldots\otimes e_{i_k})=\sum_{j_1\ldots j_l}\delta_\pi\begin{pmatrix}i_1&\ldots&i_k\\ j_1&\ldots&j_l\end{pmatrix}e_{j_1}\otimes\ldots\otimes e_{j_l}$$
and with the Kronecker type symbols $\delta_\pi\in\{0,1\}$ depending on whether the indices fit or not. The assignement $\pi\to T_\pi$ is categorical, in the sense that we have
$$T_\pi\otimes T_\sigma=T_{[\pi\sigma]}\quad,\quad 
T_\pi T_\sigma=N^{c(\pi,\sigma)}T_{[^\sigma_\pi]}\quad,\quad 
T_\pi^*=T_{\pi^*}$$
where $c(\pi,\sigma)$ are certain integers, coming from the erased components in the middle.
\end{proposition}

\begin{proof}
This is something elementary, the computations being as follows:

\medskip

(1) The concatenation axiom follows from the following computation:
\begin{eqnarray*}
&&(T_\pi\otimes T_\sigma)(e_{i_1}\otimes\ldots\otimes e_{i_p}\otimes e_{k_1}\otimes\ldots\otimes e_{k_r})\\
&=&\sum_{j_1\ldots j_q}\sum_{l_1\ldots l_s}\delta_\pi\begin{pmatrix}i_1&\ldots&i_p\\j_1&\ldots&j_q\end{pmatrix}\delta_\sigma\begin{pmatrix}k_1&\ldots&k_r\\l_1&\ldots&l_s\end{pmatrix}e_{j_1}\otimes\ldots\otimes e_{j_q}\otimes e_{l_1}\otimes\ldots\otimes e_{l_s}\\
&=&\sum_{j_1\ldots j_q}\sum_{l_1\ldots l_s}\delta_{[\pi\sigma]}\begin{pmatrix}i_1&\ldots&i_p&k_1&\ldots&k_r\\j_1&\ldots&j_q&l_1&\ldots&l_s\end{pmatrix}e_{j_1}\otimes\ldots\otimes e_{j_q}\otimes e_{l_1}\otimes\ldots\otimes e_{l_s}\\
&=&T_{[\pi\sigma]}(e_{i_1}\otimes\ldots\otimes e_{i_p}\otimes e_{k_1}\otimes\ldots\otimes e_{k_r})
\end{eqnarray*}

(2) The composition axiom follows from the following computation:
\begin{eqnarray*}
&&T_\pi T_\sigma(e_{i_1}\otimes\ldots\otimes e_{i_p})\\
&=&\sum_{j_1\ldots j_q}\delta_\sigma\begin{pmatrix}i_1&\ldots&i_p\\j_1&\ldots&j_q\end{pmatrix}
\sum_{k_1\ldots k_r}\delta_\pi\begin{pmatrix}j_1&\ldots&j_q\\k_1&\ldots&k_r\end{pmatrix}e_{k_1}\otimes\ldots\otimes e_{k_r}\\
&=&\sum_{k_1\ldots k_r}N^{c(\pi,\sigma)}\delta_{[^\sigma_\pi]}\begin{pmatrix}i_1&\ldots&i_p\\k_1&\ldots&k_r\end{pmatrix}e_{k_1}\otimes\ldots\otimes e_{k_r}\\
&=&N^{c(\pi,\sigma)}T_{[^\sigma_\pi]}(e_{i_1}\otimes\ldots\otimes e_{i_p})
\end{eqnarray*}

(3) Finally, the involution axiom follows from the following computation:
\begin{eqnarray*}
&&T_\pi^*(e_{j_1}\otimes\ldots\otimes e_{j_q})\\
&=&\sum_{i_1\ldots i_p}<T_\pi^*(e_{j_1}\otimes\ldots\otimes e_{j_q}),e_{i_1}\otimes\ldots\otimes e_{i_p}>e_{i_1}\otimes\ldots\otimes e_{i_p}\\
&=&\sum_{i_1\ldots i_p}\delta_\pi\begin{pmatrix}i_1&\ldots&i_p\\ j_1&\ldots& j_q\end{pmatrix}e_{i_1}\otimes\ldots\otimes e_{i_p}\\
&=&T_{\pi^*}(e_{j_1}\otimes\ldots\otimes e_{j_q})
\end{eqnarray*}

Summarizing, our correspondence is indeed categorical.
\end{proof}

We can now formulate a key theoretical result, as follows:

\begin{theorem}
Any category of crossing partitions $D\subset P$ produces a series of compact groups $G=(G_N)$, with $G_N\subset U_N$ for any $N\in\mathbb N$, via the formula
$$C_{kl}=span\left(T_\pi\Big|\pi\in D(k,l)\right)$$
for any $k,l$, and Tannakian duality. We call such groups easy.
\end{theorem}

\begin{proof}
Indeed, once we fix an integer $N\in\mathbb N$, the various axioms in Definition 14.28 show, via Proposition 14.29, that the following spaces form a Tannakian category:
$$span\left(T_\pi\Big|\pi\in D(k,l)\right)$$

Thus, Tannakian duality applies, and provides us with a closed subgroup $G_N\subset U_N$ such that the following equalities are satisfied, for any colored integers $k,l$:
$$C_{kl}=span\left(T_\pi\Big|\pi\in D(k,l)\right)$$

Thus, we are led to the conclusion in the statement.
\end{proof}

And with this, good news, done with the general theory. At the level of basic examples now, we have the following key theorem of Brauer, with the convention that a pairing is matching when it pairs $\circ-\circ$ or $\bullet-\bullet$ on the vertical, and $\circ-\bullet$ on the horizontal: 

\begin{theorem}
We have the following results:
\begin{enumerate}
\item $U_N$ is easy, coming from the category of all matching pairings $\mathcal P_2$.

\item $O_N$ is easy too, coming from the category of all pairings $P_2$.
\end{enumerate}
\end{theorem}

\begin{proof}
This can be deduced from Tannakian duality, the idea being as follows:

\medskip

(1) The unitary group $U_N$ being defined via the relations $u^*=u^{-1}$, $u^t=\bar{u}^{-1}$, the associated Tannakian category is $C=span(T_\pi|\pi\in D)$, with:
$$D
=<{\ }^{\,\cap}_{\circ\bullet}\,\,,{\ }^{\,\cap}_{\bullet\circ}>
=\mathcal P_2$$

(2) The orthogonal group $O_N\subset U_N$ being defined by imposing the relations $u_{ij}=\bar{u}_{ij}$, the associated Tannakian category is $C=span(T_\pi|\pi\in D)$, with:
$$D
=<\mathcal P_2,|^{\hskip-1.32mm\circ}_{\hskip-1.32mm\bullet},|_{\hskip-1.32mm\circ}^{\hskip-1.32mm\bullet}>
=P_2$$
  
Thus, we are led to the conclusions in the statement.
\end{proof}

Moving now towards finite groups, we first have the following result:

\begin{theorem}
The symmetric group $S_N$, regarded as group of unitary matrices,
$$S_N\subset O_N\subset U_N$$
via the permutation matrices, is easy, coming from the category of all partitions $P$.
\end{theorem}

\begin{proof}
Consider indeed the group $S_N$, regarded as a group of unitary matrices, with each permutation $\sigma\in S_N$ corresponding to the associated permutation matrix:
$$\sigma(e_i)=e_{\sigma(i)}$$

Consider as well the easy group $G\subset O_N$ coming from the category of all partitions $P$. Since $P$ is generated by the one-block ``fork'' partition $Y\in P(2,1)$, we have:
$$C(G)=C(O_N)\Big/\Big<T_Y\in Hom(u^{\otimes 2},u)\Big>$$

Now observe that we have the following formula:
$$(T_Yu^{\otimes 2})_{i,jk}
=\sum_{lm}(T_Y)_{i,lm}(u^{\otimes 2})_{lm,jk}
=u_{ij}u_{ik}$$

On the other hand, we have as well the following formula:
$$(uT_Y)_{i,jk}
=\sum_lu_{il}(T_Y)_{l,jk}
=\delta_{jk}u_{ij}$$

Thus, the relation defining $G\subset O_N$ reformulates as follows:
$$T_Y\in Hom(u^{\otimes 2},u)\iff u_{ij}u_{ik}=\delta_{jk}u_{ij},\forall i,j,k$$

In other words, the elements $u_{ij}$ must be projections, which must be pairwise orthogonal on the rows of $u=(u_{ij})$. We conclude that $G\subset O_N$ is the subgroup of matrices $g\in O_N$ having the property $g_{ij}\in\{0,1\}$. Thus we have $G=S_N$, as desired. 
\end{proof}

The hyperoctahedral group $H_N$ is easy as well, the result here being as follows:

\begin{theorem}
The hyperoctahedral group $H_N$, regarded as group of matrices,
$$S_N\subset H_N\subset O_N$$
is easy, coming from the category of partitions with even blocks $P_{even}$.
\end{theorem}

\begin{proof}
This follows as usual from Tannakian duality. To be more precise, consider the following one-block partition, which, as the name indicates, looks like a $H$ letter:
$$H\in P(2,2)$$

The linear map associated to this partition is then given by:
$$T_H(e_i\otimes e_j)=\delta_{ij}e_i\otimes e_i$$

By using this formula, we have the following computation:
\begin{eqnarray*}
(T_H\otimes id)u^{\otimes 2}(e_a\otimes e_b)
&=&(T_H\otimes id)\left(\sum_{ijkl}e_{ij}\otimes e_{kl}\otimes u_{ij}u_{kl}\right)(e_a\otimes e_b)\\
&=&(T_H\otimes id)\left(\sum_{ik}e_i\otimes e_k\otimes u_{ia}u_{kb}\right)\\
&=&\sum_ie_i\otimes e_i\otimes u_{ia}u_{ib}
\end{eqnarray*}

On the other hand, we have as well the following computation:
\begin{eqnarray*}
u^{\otimes 2}(T_H\otimes id)(e_a\otimes e_b)
&=&\delta_{ab}\left(\sum_{ijkl}e_{ij}\otimes e_{kl}\otimes u_{ij}u_{kl}\right)(e_a\otimes e_a)\\
&=&\delta_{ab}\sum_{ij}e_i\otimes e_k\otimes u_{ia}u_{ka}
\end{eqnarray*}

We conclude from this that we have the following equivalence:
$$T_H\in End(u^{\otimes 2})\iff \delta_{ik}u_{ia}u_{ib}=\delta_{ab}u_{ia}u_{ka},\forall i,k,a,b$$

We deduce from this that the corresponding closed subgroup $G\subset O_N$ consists of the matrices $g\in O_N$ which are permutation-like, with $\pm1$ nonzero entries. Thus, the corresponding group is $G=H_N$, and as a conclusion to all this, we have:
$$C(H_N)=C(O_N)\Big/\Big<T_H\in End(u^{\otimes 2})\Big>$$

But this means that the hyperoctahedral group $H_N$ is easy, coming from the category of partitions $D=<H>=P_{even}$. Thus, we are led to the conclusion in the statement.
\end{proof}

More generally now, we have in fact the following result, regarding the series of complex reflection groups $H_N^s$, which covers both the groups $S_N,H_N$:

\begin{theorem}
The complex reflection group $H_N^s=\mathbb Z_s\wr S_N$ is easy, the corresponding category $P^s$ consisting of the partitions satisfying the condition
$$\#\circ=\#\bullet(s)$$
as a weighted sum, in each block. In particular, we have the following results:
\begin{enumerate}
\item $S_N$ is easy, coming from the category $P$.

\item $H_N=\mathbb Z_2\wr S_N$ is easy, coming from the category $P_{even}$.

\item $K_N=\mathbb T\wr S_N$ is easy, coming from the category $\mathcal P_{even}$.
\end{enumerate}
\end{theorem}

\begin{proof}
This is something that we already know at $s=1,2$, from Theorems 14.32 and 14.33. In general, the proof is similar, based on Tannakian duality. To be more precise, in what regards the main assertion, the idea here is that the one-block partition $\pi\in P(s)$, which generates the category of partitions $P^s$ in the statement, implements the relations producing the subgroup $H_N^s\subset S_N$. As for the last assertions, these are all elementary:

\medskip

(1) At $s=1$ we know that we have $H_N^1=S_N$. Regarding now the corresponding category, here the condition $\#\circ=\#\bullet(1)$ is automatic, and so $P^1=P$.

\medskip

(2) At $s=2$ we know that we have $H_N^2=H_N$. Regarding now the corresponding category, here the condition $\#\circ=\#\bullet(2)$ reformulates as follows:
$$\#\circ+\,\#\bullet=0(2)$$

Thus each block must have even size, and we obtain, as claimed, $P^2=P_{even}$.

\medskip

(3) At $s=\infty$ we know that we have $H_N^\infty=K_N$. Regarding now the corresponding category, here the condition $\#\circ=\#\bullet(\infty)$ reads:
$$\#\circ=\#\bullet$$

But this is the condition defining $\mathcal P_{even}$, and so $P^\infty=\mathcal P_{even}$, as claimed.
\end{proof}

Summarizing, we have many examples. In fact, our list of easy groups has currently become quite big, and here is a selection of the main results that we have so far: 

\begin{theorem}
We have a diagram of compact groups as follows,
$$\xymatrix@R=50pt@C=50pt{
K_N\ar[r]&U_N\\
H_N\ar[u]\ar[r]&O_N\ar[u]}$$
where $H_N=\mathbb Z_2\wr S_N$ and $K_N=\mathbb T\wr S_N$, and all these groups are easy.
\end{theorem}

\begin{proof}
This follows from the above results. To be more precise, we know that the above groups are all easy, the corresponding categories of partitions being as follows:
$$
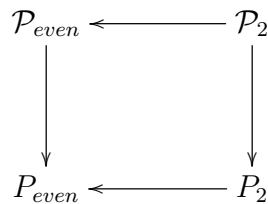
\xymatrix@R=16mm@C=18mm{
\mathcal P_{even}\ar[d]&\mathcal P_2\ar[l]\ar[d]\\
P_{even}&P_2\ar[l]}$$

Thus, we are led to the conclusion in the statement.
\end{proof}

\section*{14d. Haar integration}

In order to investigate linear independence questions for the vectors $\xi_\pi=T_\pi$, we will use the Gram matrix of these vectors. Let us begin with some standard definitions:

\begin{definition}
Let $P(k)$ be the set of partitions of $\{1,\ldots,k\}$, and let $\pi,\nu\in P(k)$.
\begin{enumerate}
\item We write $\pi\leq\nu$ if each block of $\pi$ is contained in a block of $\nu$.

\item We let $\pi\vee\nu\in P(k)$ be the partition obtained by superposing $\pi,\nu$.
\end{enumerate}
\end{definition}

As an illustration here, at $k=2$ we have $P(2)=\{||,\sqcap\}$, and the order is:
$$||\leq\sqcap$$

At $k=3$ we have $P(3)=\{|||,\sqcap|,\sqcap\hskip-3.2mm{\ }_|\,,|\sqcap,\sqcap\hskip-0.7mm\sqcap\}$, and the order relation is as follows:
$$|||\leq\sqcap|,\sqcap\hskip-3.2mm{\ }_|\,,|\sqcap\leq\sqcap\hskip-0.7mm\sqcap$$

Observe also that we have $\pi,\nu\leq\pi\vee\nu$. In fact, $\pi\vee\nu$ is the smallest partition with this property, called supremum of $\pi,\nu$. Now back to the easy groups, we have:

\begin{proposition}
The Gram matrix $G_{kN}(\pi,\nu)=<\xi_\pi,\xi_\nu>$ is given by
$$G_{kN}(\pi,\nu)=N^{|\pi\vee\nu|}$$
where $|.|$ is the number of blocks.
\end{proposition}

\begin{proof}
According to our formula of the vectors $\xi_\pi$, we have:
\begin{eqnarray*}
<\xi_\pi,\xi_\nu>
&=&\sum_{i_1\ldots i_k}\delta_\pi(i_1,\ldots,i_k)\delta_\nu(i_1,\ldots,i_k)\\
&=&\sum_{i_1\ldots i_k}\delta_{\pi\vee\nu}(i_1,\ldots,i_k)\\
&=&N^{|\pi\vee\nu|}
\end{eqnarray*}

Thus, we have obtained the formula in the statement.
\end{proof}

In order to study the Gram matrix, and more specifically to compute its determinant, we will need several standard facts about the partitions. We first have:

\begin{definition}
The M\"obius function of any lattice, and so of $P$, is given by
$$\mu(\pi,\nu)=\begin{cases}
1&{\rm if}\ \pi=\nu\\
-\sum_{\pi\leq\tau<\nu}\mu(\pi,\tau)&{\rm if}\ \pi<\nu\\
0&{\rm if}\ \pi\not\leq\nu
\end{cases}$$
with the construction being performed by recurrence.
\end{definition}

As an illustration here, let us go back to the set of 2-point partitions, $P(2)=\{||,\sqcap\}$. Here we have by definition:
$$\mu(||,||)=\mu(\sqcap,\sqcap)=1$$

Also, we know that we have $||<\sqcap$, with no intermediate partition in between, and so the above recurrence procedure gives the following formular:
$$\mu(||,\sqcap)=-\mu(||,||)=-1$$

Finally, we have $\sqcap\not\leq||$, which gives the following formula:
$$\mu(\sqcap,||)=0$$

The interest in the M\"obius function comes from the M\"obius inversion formula:
$$f(\nu)=\sum_{\pi\leq\nu}g(\pi)\implies g(\nu)=\sum_{\pi\leq\nu}\mu(\pi,\nu)f(\pi)$$

In linear algebra terms, the statement and proof of this formula are as follows:

\begin{theorem}
The inverse of the adjacency matrix of $P$, given by
$$A_{\pi\nu}=\begin{cases}
1&{\rm if}\ \pi\leq\nu\\
0&{\rm if}\ \pi\not\leq\nu
\end{cases}$$
is the M\"obius matrix of $P$, given by $M_{\pi\nu}=\mu(\pi,\nu)$.
\end{theorem}

\begin{proof}
This is well-known, coming for instance from the fact that $A$ is upper triangular. Thus, when inverting, we are led into the recurrence from Definition 14.38.
\end{proof}

As an illustration here, for $P(2)$ the formula $M=A^{-1}$ appears as follows:
$$\begin{pmatrix}1&-1\\ 0&1\end{pmatrix}=
\begin{pmatrix}1&1\\ 0&1\end{pmatrix}^{-1}$$

Now back to our Gram matrix considerations, we have the following result:

\begin{proposition}
The Gram matrix is given by $G_{kN}=AL$, where
$$L(\pi,\nu)=
\begin{cases}
N(N-1)\ldots(N-|\pi|+1)&{\rm if}\ \nu\leq\pi\\
0&{\rm otherwise}
\end{cases}$$
and where $A=M^{-1}$ is the adjacency matrix of $P(k)$.
\end{proposition}

\begin{proof}
We have the following computation:
\begin{eqnarray*}
N^{|\pi\vee\nu|}
&=&\#\left\{i_1,\ldots,i_k\in\{1,\ldots,N\}\Big|\ker i\geq\pi\vee\nu\right\}\\
&=&\sum_{\tau\geq\pi\vee\nu}\#\left\{i_1,\ldots,i_k\in\{1,\ldots,N\}\Big|\ker i=\tau\right\}\\
&=&\sum_{\tau\geq\pi\vee\nu}N(N-1)\ldots(N-|\tau|+1)
\end{eqnarray*}

According to Proposition 14.37 and to the definition of $A,L$, this formula reads:
$$(G_{kN})_{\pi\nu}
=\sum_{\tau\geq\pi}L_{\tau\nu}
=\sum_\tau A_{\pi\tau}L_{\tau\nu}
=(AL)_{\pi\nu}$$

Thus, we obtain the formula in the statement.
\end{proof}

With the above result in hand, we can now investigate the linear independence properties of the vectors $\xi_\pi$. To be more precise, we have the following result:

\begin{theorem}
The determinant of the Gram matrix $G_{kN}$ is given by
$$\det(G_{kN})=\prod_{\pi\in P(k)}\frac{N!}{(N-|\pi|)!}$$
and in particular, for $N\geq k$, the vectors $\{\xi_\pi|\pi\in P(k)\}$ are linearly independent.
\end{theorem}

\begin{proof}
According to the formula in Proposition 14.40, we have:
$$\det(G_{kN})=\det(A)\det(L)$$

Now if we order $P(k)$ as usual, with respect to the number of blocks, and then lexicographically, we see that $A$ is upper triangular, and that $L$ is lower triangular. But this shows that $\det(A)=1$, and in what concerns $\det(L)$, this can be computed as well by making the product on the diagonal, and we obtain the number in the statement. 
\end{proof}

Now back to the laws of characters, we can formulate:

\begin{proposition}
For an easy group $G=(G_N)$, coming from a category of partitions $D=(D(k,l))$, the asymptotic moments of the main character are given by
$$\lim_{N\to\infty}\int_{G_N}\chi^k=\# D(k)$$
where $D(k)=D(\emptyset,k)$, with the limiting sequence on the left consisting of certain integers, and being stationary at least starting from the $k$-th term.
\end{proposition}

\begin{proof}
This follows indeed from the Peter-Weyl theory, by using the linear independence result for the vectors $\xi_\pi$ coming from Theorem 14.41.
\end{proof}

With these preliminaries in hand, we can now state and prove:

\begin{theorem}
In the $N\to\infty$ limit, the laws of the main character for the main easy groups, real and complex, and discrete and continuous, are as follows,
$$\xymatrix@R=50pt@C=50pt{
K_N\ar[r]&U_N\\
H_N\ar[u]\ar[r]&O_N\ar[u]}\qquad
\xymatrix@R=25pt@C=50pt{\\:}
\qquad
\xymatrix@R=50pt@C=50pt{
B_1\ar[r]&G_1\\
b_1\ar[u]\ar[r]&g_1\ar[u]}$$
with these laws, namely the real and complex Gaussian and Bessel laws, being the main limiting laws in real and complex, and discrete and continuous probability.
\end{theorem}

\begin{proof}
This follows from the above results. To be more precise, we know that the above groups are all easy, the corresponding categories of partitions being as follows:
$$\xymatrix@R=16mm@C=16mm{
\mathcal P_{even}\ar[d]&\mathcal P_2\ar[l]\ar[d]\\
P_{even}&P_2\ar[l]}$$

Thus, we can use Proposition 14.42, are we are led into counting partitions, and then recovering the measures via their moments, and this leads to the result.
\end{proof}

Our aim now is to go beyond what we have, with results regarding the truncated characters. Let us start with a general formula coming from Peter-Weyl, namely:

\begin{theorem}
The Haar integration over a closed subgroup $G\subset_uU_N$ is given on the dense subalgebra of smooth functions by the Weingarten type formula
$$\int_Gg_{i_1j_1}^{e_1}\ldots g_{i_kj_k}^{e_k}\,dg=\sum_{\pi,\nu\in D(k)}\delta_\pi(i)\delta_\sigma(j)W_k(\pi,\nu)$$
valid for any colored integer $k=e_1\ldots e_k$ and any multi-indices $i,j$, where $D(k)$ is a linear basis of $Fix(u^{\otimes k})$, the associated generalized Kronecker symbols are given by
$$\delta_\pi(i)=<\pi,e_{i_1}\otimes\ldots\otimes e_{i_k}>$$
and $W_k=G_k^{-1}$ is the inverse of the Gram matrix, $G_k(\pi,\nu)=<\pi,\nu>$.
\end{theorem}

\begin{proof}
This is something very standard, coming from the fact that the above integrals form altogether the orthogonal projection $P^k$ onto the following space:
$$Fix(u^{\otimes k})=span(D(k))$$

Consider now the following linear map, with $D(k)=\{\xi_k\}$ being as in the statement:
$$E(x)=\sum_{\pi\in D(k)}<x,\xi_\pi>\xi_\pi$$

By a standard linear algebra computation, it follows that we have $P=WE$, where $W$ is the inverse of the restriction of $E$ to the following space:
$$K=span\left(T_\pi\Big|\pi\in D(k)\right)$$

But this restriction is the linear map given by the matrix $G_k$, and so $W$ is the linear map given by the inverse matrix $W_k=G_k^{-1}$, and this gives the result.
\end{proof}

In the easy case, we have the following more concrete result:

\begin{theorem}
For an easy group $G\subset U_N$, coming from a category of partitions $D=(D(k,l))$, we have the Weingarten formula
$$\int_Gg_{i_1j_1}^{e_1}\ldots g_{i_kj_k}^{e_k}\,dg=\sum_{\pi,\nu\in D(k)}\delta_\pi(i)\delta_\nu(j)W_{kN}(\pi,\nu)$$
for any $k=e_1\ldots e_k$ and any $i,j$, where $D(k)=D(\emptyset,k)$, $\delta$ are usual Kronecker type symbols, checking whether the indices match, and $W_{kN}=G_{kN}^{-1}$, with 
$$G_{kN}(\pi,\nu)=N^{|\pi\vee\nu|}$$
where $|.|$ is the number of blocks. 
\end{theorem}

\begin{proof}
We use the abstract Weingarten formula, from Theorem 14.44. Indeed, the Kronecker type symbols there are then the usual ones, as shown by:
\begin{eqnarray*}
\delta_{\xi_\pi}(i)
&=&<\xi_\pi,e_{i_1}\otimes\ldots\otimes e_{i_k}>\\
&=&\left<\sum_j\delta_\pi(j_1,\ldots,j_k)e_{j_1}\otimes\ldots\otimes e_{j_k},e_{i_1}\otimes\ldots\otimes e_{i_k}\right>\\
&=&\delta_\pi(i_1,\ldots,i_k)
\end{eqnarray*}

The Gram matrix being as well the correct one, we obtain the result.
\end{proof}

As an application of this, let us discuss the computation of the laws of characters. First, we have the following formula, in the general easy group setting:

\begin{proposition}
The moments of truncated characters are given by
$$\int_G(u_{11}+\ldots +u_{ss})^k=Tr(W_{kN}G_{ks})$$
where $G_{kN}$ and $W_{kN}=G_{kN}^{-1}$ are the associated Gram and Weingarten matrices.
\end{proposition}

\begin{proof}
We have indeed the following computation:
\begin{eqnarray*}
\int_G(u_{11}+\ldots +u_{ss})^k
&=&\sum_{i_1=1}^{s}\ldots\sum_{i_k=1}^s\int_Gu_{i_1i_1}\ldots g_{i_ki_k}\\
&=&\sum_{\pi,\sigma\in D(k)}W_{kN}(\pi,\sigma)\sum_{i_1=1}^{s}\ldots\sum_{i_k=1}^s\delta_\pi(i)\delta_\sigma(i)\\
&=&\sum_{\pi,\sigma\in D(k)}W_{kN}(\pi,\sigma)G_{ks}(\sigma,\pi)\\
&=&Tr(W_{kN}G_{ks})
\end{eqnarray*}

Thus, we have obtained the formula in the statement.
\end{proof}

The idea now is to impose a natural uniformity condition, as follows:

\begin{definition}
An easy group $G=(G_N)$, coming from a category of partitions $D\subset P$, is called uniform if it satisfies the following equivalent conditions:
\begin{enumerate}
\item $G_{N-1}=G_N\cap U_{N-1}$, via the embedding $U_{N-1}\subset U_N$ given by $u\to diag(u,1)$.

\item $G_{N-1}=G_N\cap U_{N-1}$, via the $N$ possible diagonal embeddings $U_{N-1}\subset U_N$.

\item $D$ is stable under the operation which consists in removing blocks.
\end{enumerate}
\end{definition}

Here the equivalence between the above three conditions is something standard, obtained by doing some combinatorics. Also, we have many examples of such groups, the idea here being that the most familiar easy groups $G=(G_N)$ that we know, as for instance the various real and complex rotation and reflection groups, are indeed uniform.

\bigskip

In what follows we will be mostly interested in the condition (3) above, which makes the link with our computations for truncated characters, and simplifies them. To be more precise, by imposing the uniformity condition we obtain the following result:

\begin{theorem}
For a uniform easy group $G=(G_N)$, we have the formula
$$\lim_{N\to\infty}\int_{G_N}\chi_t^k=\sum_{\pi\in D(k)}t^{|\pi|}$$
with $D\subset P$ being the associated category of partitions.
\end{theorem}

\begin{proof}
We use the general moment formula from Proposition 14.46, namely:
$$\int_G(u_{11}+\ldots +u_{ss})^k=Tr(W_{kN}G_{ks})$$

By setting $s=[tN]$, with $t>0$ being a given parameter, this formula becomes:
$$\int_{G_N}\chi_t^k=Tr(W_{kN}G_{k[tN]})$$

The point now is that in the uniform case the Gram and Weingarten matrices are asymptotically diagonal, and this leads to the formula in the statement.
\end{proof}

We can now formulate some general character results, as follows:

\begin{theorem}
With $N\to\infty$, the laws of truncated characters are as follows:
\begin{enumerate}
\item For $O_N$ we obtain the Gaussian law $g_t$.

\item For $U_N$ we obtain the complex Gaussian law $G_t$.

\item For $S_N$ we obtain the Poisson law $p_t$.

\item For $H_N$ we obtain the Bessel law $b_t$.

\item For $H_N^s$ we obtain the generalized Bessel law $b_t^s$.

\item For $K_N$ we obtain the complex Bessel law $B_t$.
\end{enumerate}
Also, for $B_N,C_N$ and for $Sp_N$ we obtain modified normal laws.
\end{theorem}

\begin{proof}
We use the formula that we found in Theorem 14.48, namely:
$$\lim_{N\to\infty}\int_{G_N}\chi_t^k=\sum_{\pi\in D(k)}t^{|\pi|}$$

By doing now some combinatorics, for instance in relation with the cumulants, this gives the results. We refer here to \cite{ba1} and various related papers.
\end{proof}

\section*{14e. Exercises}

This was a tough, modern linear algebra chapter, and as exercises, we have:

\begin{exercise}
Read more about the basic theory of Lie groups and algebras.
\end{exercise}

\begin{exercise}
Read as well about the classification of Lie groups and algebras.
\end{exercise}

\begin{exercise}
Work out all the details, for the $*$-algebras $A\subset M_N(\mathbb C)$.
\end{exercise}

\begin{exercise}
Work out all the details for the existence of the Haar measure.
\end{exercise}

\begin{exercise}
Look up and learn the full proof of Tannakian duality.
\end{exercise}

\begin{exercise}
Find some other interesting examples of easy groups.
\end{exercise}

\begin{exercise}
Work out the asymptotic law results for characters.
\end{exercise}

\begin{exercise}
Work out the asymptotic law results for truncated characters.
\end{exercise}

As bonus exercise, try unifying the theories of Lie algebras, and Brauer algebras.

\chapter{Spin matrices}

\section*{15a. Quantum physics}

Good news, done with mathematics, we have learned enough interesting things, and in the remainder of this book we will get into physics, and mathematical physics. Everything will remain of course linear algebra oriented, and we will be talking about things which are related to the most important matrix groups of them all, namely $SU_2$ and $SO_3$.

\bigskip

You have surely heard about the atomic theory, in its golden age, happening around 1900-1920. At that time, Bohr was able to put everything together, and formulate a nice conjecture regarding the functioning of hydrogen $_1{\rm H}$, and of heavier atoms as well. However, this turned difficult to prove, among others because the electrons are ``slippery'' particles, having no clear positions and speeds. Nevertheless, Heisenberg came with a clever way of solving this question, by encoding positions and speeds by certain linear operators, instead of numbers, and managed to prove the Bohr conjecture.

\bigskip

Before explaining what Heisenberg was saying, let us hear as well the point of view of Schr\"odinger, which came a few years later. His idea was to forget about exact things, and try to investigate the hydrogen atom statistically. Let us start with:

\begin{question}
In the context of the hydrogen atom, assuming that the proton is fixed, what is the probability density $\varphi_t(x)$ of the position of the electron $e$, at time $t$,
$$P_t(e\in V)=\int_V\varphi_t(x)dx$$
as function of an intial probability density $\varphi_0(x)$? Moreover, can the corresponding equation be solved, and will this prove the Bohr claims for hydrogen, statistically?
\end{question}

In order to get familiar with this question, let us first look at examples coming from classical mechanics. In the context of a particle whose position at time $t$ is given by $x_0+\gamma(t)$, the evolution of the probability density will be given by: 
$$\varphi_t(x)=\varphi_0(x)+\gamma(t)$$

However, such examples are somewhat trivial, of course not in relation with the computation of $\gamma$, usually a difficult question, but in relation with our questions, and do not apply to the electron. The point indeed is that, in what regards the electron, we have:

\begin{fact}
In respect with various simple interference experiments:
\begin{enumerate}
\item The electron is definitely not a particle in the usual sense.

\item But in most situations it behaves exactly like a wave.

\item But in other situations it behaves like a particle.
\end{enumerate}
\end{fact}

Getting back now to the Schr\"odinger question, all this suggests to use, as for the waves, an amplitude function $\psi_t(x)\in\mathbb C$, related to the density $\varphi_t(x)>0$ by the formula $\varphi_t(x)=|\psi_t(x)|^2$. Not that a big deal, you would say, because the two are related by simple formulae as follows, with $\theta_t(x)$ being an arbitrary phase function:
$$\varphi_t(x)=|\psi_t(x)|^2\quad,\quad \psi_t(x)=e^{i\theta_t(x)}\sqrt{\varphi_t(x)}$$

However, such manipulations can be crucial, raising for instance the possibility that the amplitude function satisfies some simple equation, while the density itself, maybe not. And this is what happens indeed. Schr\"odinger was led in this way to:

\begin{claim}[Schr\"odinger]
In the context of the hydrogen atom, the amplitude function of the electron $\psi=\psi_t(x)$  is subject to the Schr\"odinger equation
$$ih\dot{\psi}=-\frac{h^2}{2m}\Delta\psi+V\psi$$
$m$ being the mass, $h=h_0/2\pi$ the reduced Planck constant, and $V$ the Coulomb potential of the proton. The same holds for movements of the electron under any potential $V$.
\end{claim}

Observe the similarity with the wave equation $\ddot{\varphi}=v^2\Delta\varphi$, and with the heat equation $\dot{\varphi}=\alpha\Delta\varphi$ too. Many things can be said here. Following now Heisenberg and Schr\"odinger, and then especially Dirac, who did the axiomatization work, we have:

\begin{definition}
In quantum mechanics the states of the system are vectors of a Hilbert space $H$, and the observables of the system are linear operators 
$$T:H\to H$$
which can be densely defined, and are taken self-adjoint, $T=T^*$. The average value of such an observable $T$, evaluated on a state $\xi\in H$, is given by:
$$<T>=<T\xi,\xi>$$
In the context of the Schr\"odinger mechanics of the hydrogen atom, the Hilbert space is the space $H=L^2(\mathbb R^3)$ where the wave function $\psi$ lives, and we have
$$<T>=\int_{\mathbb R^3}T(\psi)\cdot\bar{\psi}\,dx$$
which is called ``sandwiching'' formula, with the operators 
$$x\quad,\quad-\frac{ih}{m}\nabla\quad,\quad-ih\nabla\quad,\quad -\frac{h^2\Delta}{2m}\quad,\quad -\frac{h^2\Delta}{2m}+V$$
representing the position, speed, momentum, kinetic energy, and total energy.
\end{definition}

In other words, we are doing here two things. First, we are declaring by axiom that various ``sandwiching'' formulae found before by Heisenberg hold true. And second, we are raising the possibility for other quantum mechanical systems, more complicated, to be described as well by the mathematics of operators on a certain Hilbert space $H$.

\bigskip

With this discussed, let us develop now quantum mechanics, in a mathematical way, by studying the Schr\"odinger equation from Claim 15.3. We first have:

\begin{proposition}
We have the following formula, 
$$\dot{\varphi}=\frac{ih}{2m}\left(\Delta\psi\cdot\bar{\psi}-\Delta\bar{\psi}\cdot\psi\right)$$
for the time derivative of the probability density function $\varphi=|\psi|^2$.
\end{proposition}

\begin{proof}
According to the Leibnitz product rule, we have the following formula:
$$\dot{\varphi}=\frac{d}{dt}|\psi|^2=\frac{d}{dt}(\psi\bar{\psi})=\dot{\psi}\bar{\psi}+\psi\dot{\bar{\psi}}$$

On the other hand, the Schr\"odinger equation and its conjugate read:
$$\dot{\psi}=\frac{ih}{2m}\left(\Delta\psi-\frac{2m}{h^2}V\psi\right)\quad,\quad 
\dot{\bar{\psi}}=-\frac{ih}{2m}\left(\Delta\bar{\psi}-\frac{2m}{h^2}V\bar{\psi}\right)$$

By plugging this data, we obtain the following formula:
$$\dot{\varphi}
=\frac{ih}{2m}\left[\left(\Delta\psi-\frac{2m}{h^2}V\psi\right)\bar{\psi}-\left(\Delta\bar{\psi}-\frac{2m}{h^2}V\bar{\psi}\right)\psi\right]$$

But this gives, after simplifying, the formula in the statement.
\end{proof}

As an important application of Proposition 15.5, we have:

\begin{theorem}
The Schr\"odinger equation conserves probability amplitudes,
$$\int_{\mathbb R^3}|\psi_0|^2=1\implies\int_{\mathbb R^3}|\psi_t|^2=1$$
in agreement with the basic probabilistic requirement, $P=1$ overall.
\end{theorem}

\begin{proof}
According to the formula in Proposition 15.5, we have:
\begin{eqnarray*}
\frac{d}{dt}\int_{\mathbb R^3}|\psi|^2\,dx
&=&\int_{\mathbb R^3}\frac{d}{dt}|\psi|^2\,dx\\
&=&\int_{\mathbb R^3}\dot{\varphi}\,dx\\
&=&\frac{ih}{2m}\int_{\mathbb R^3}\left(\Delta\psi\cdot\bar{\psi}-\Delta\bar{\psi}\cdot\psi\right)dx
\end{eqnarray*}

Now by remembering the definition of the Laplace operator, we have:
\begin{eqnarray*}
\frac{d}{dt}\int_{\mathbb R^3}|\psi|^2\,dx
&=&\frac{ih}{2m}\int_{\mathbb R^3}\sum_i\left(\frac{d^2\psi}{dx_i^2}\cdot\bar{\psi}-\frac{d^2\bar{\psi}}{dx_i^2}\cdot\psi\right)dx\\
&=&\frac{ih}{2m}\sum_i\int_{\mathbb R^3}\frac{d}{dx_i}\left(\frac{d\psi}{dx_i}\cdot\bar{\psi}-\frac{d\bar{\psi}}{dx_i}\cdot\psi\right)dx\\
&=&\frac{ih}{2m}\sum_i\int_{\mathbb R^2}\left[\frac{d\psi}{dx}\cdot\bar{\psi}-\frac{d\bar{\psi}}{dx}\cdot\psi\right]_{-\infty}^\infty\frac{dx}{dx_i}\\
&=&0
\end{eqnarray*}

Thus, we are led to the conclusion in the statement.
\end{proof}

Moving now towards hydrogen, we have here the following result:

\begin{theorem}
In the case of time-independent potentials $V$, including the Coulomb potential of the proton, the separated solutions of the Schr\"odinger equation
$$\psi_t(x)=w_t\phi(x)$$
are given by the following formulae, with $E$ being a certain constant,
$$w=e^{-iEt/h}w_0\quad,\quad E\phi=-\frac{h^2}{2m}\Delta\phi+V\phi$$
with the equation for $\phi$ being called time-independent Schr\"odinger equation.
\end{theorem}

\begin{proof}
By dividing by $\psi$, the Schr\"odinger equation becomes:
$$ih\cdot\frac{\dot{w}}{w}=-\frac{h^2}{2m}\cdot\frac{\Delta\phi}{\phi}+V$$

Now since the left-hand side depends only on time, and the right-hand side depends only on space, both quantities must equal a constant $E$, and this gives the result.
\end{proof}

Moving ahead with theory, we can further build on Theorem 15.7, as follows:

\begin{theorem}
In the case of time-independent potentials $V$, the Schr\"odinger equation and its time-independent version have the following properties:
\begin{enumerate}
\item For solutions of type $\psi=w_t\phi(x)$, the density $\varphi=|\psi|$ is time-independent, and more generally, all quantities of type $<T>$ are time-independent.

\item The time-independent Schr\"odinger equation can be written as $\widehat{H}\phi=E\phi$, with $H=T+V$ being the total energy, of Hamiltonian.

\item For solutions of type $\psi=w_t\phi(x)$ we have $<H^k>=E^k$ for any $k$. In particular we have $<H>=E$, and the variance is $<H^2>-<H>^2=0$.
\end{enumerate}
\end{theorem}

\begin{proof}
All the formulae are clear indeed from the fact that, when using the sandwiching formula for computing averages, the phases will cancel:
$$<T>=\int_{\mathbb R^3}\bar{\psi}\cdot T\cdot\psi\,dx
=\int_{\mathbb R^3}\bar{\phi}\cdot T\cdot\phi\,dx$$

Thus, we are led to the various conclusions in the statement.
\end{proof}

We have as well the following key result, mathematical this time:

\begin{theorem}
The solutions of the Schr\"odinger equation with time-independent potential $V$ appear as linear combinations of separated solutions
$$\psi=\sum_nc_ne^{-iE_nt/h}\phi_h$$
with the absolute values of the coefficients being given by
$$<H>=\sum_n|c_n|^2E_n$$
$|c_n|$ being the probability for a measurement to return the energy value $E_n$.
\end{theorem}

\begin{proof}
This is something standard, which follows from Fourier analysis, which allows the decomposition of $\psi$ as in the statement, and that we will not really need, in what follows next. As before, for a physical discussion here, we refer to Griffiths \cite{gr2}.
\end{proof}

In order to solve now the hydrogen atom, by using the Schr\"odinger equation, the idea will be that of reformulating this equation in spherical coordinates. We have:

\begin{theorem}
The time-independent Schr\"odinger equation in spherical coordinates separates, for solutions of type $\phi=\rho(r)\alpha(s,t)$, into two equations, as follows,
$$\frac{d}{dr}\left(r^2\cdot\frac{d\rho}{dr}\right)
-\frac{2mr^2}{h^2}(V-E)\rho=K\rho$$
$$\sin s\cdot\frac{d}{ds}\left(\sin s\cdot\frac{d\alpha}{ds}\right)
+\frac{d^2\alpha}{dt^2}=-K\sin^2s\cdot\alpha$$
with $K$ being a constant, called radial equation, and angular equation.
\end{theorem}

\begin{proof}
We use the following well-known formula for the Laplace operator in spherical coordinates, whose proof can be found in any advanced calculus book:
$$\Delta=\frac{1}{r^2}\cdot\frac{d}{dr}\left(r^2\cdot\frac{d}{dr}\right)
+\frac{1}{r^2\sin s}\cdot\frac{d}{ds}\left(\sin s\cdot\frac{d}{ds}\right)
+\frac{1}{r^2\sin^2s}\cdot\frac{d^2}{dt^2}$$

By using this formula, the time-independent Schr\"odinger equation reformulates as:
$$(V-E)\phi=\frac{h^2}{2m}\left[\frac{1}{r^2}\cdot\frac{d}{dr}\left(r^2\cdot\frac{d\phi}{dr}\right)
+\frac{1}{r^2\sin s}\cdot\frac{d}{ds}\left(\sin s\cdot\frac{d\phi}{ds}\right)
+\frac{1}{r^2\sin^2s}\cdot\frac{d^2\phi}{dt^2}\right]$$

Let us look now for separable solutions for this latter equation, consisting of a radial part and an angular part, as in the statement, namely:
$$\phi(r,s,t)=\rho(r)\alpha(s,t)$$

By plugging this function into our equation, we obtain:
$$(V-E)\rho\alpha=\frac{h^2}{2m}\left[\frac{\alpha}{r^2}\cdot\frac{d}{dr}\left(r^2\cdot\frac{d\rho}{dr}\right)
+\frac{\rho}{r^2\sin s}\cdot\frac{d}{ds}\left(\sin s\cdot\frac{d\alpha}{ds}\right)
+\frac{\rho}{r^2\sin^2s}\cdot\frac{d^2\alpha}{dt^2}\right]$$

By multiplying everything by $2mr^2/(h^2\rho\alpha)$, and then moving the radial terms to the left, and the angular terms to the right, this latter equation can be written as follows:
$$\frac{2mr^2}{h^2}(V-E)-\frac{1}{\rho}\cdot\frac{d}{dr}\left(r^2\cdot\frac{d\rho}{dr}\right)
=\frac{1}{\alpha\sin^2s}\left[\sin s\cdot\frac{d}{ds}\left(\sin s\cdot\frac{d\alpha}{ds}\right)
+\frac{d^2\alpha}{dt^2}\right]$$

Since this latter equation is now separated between radial and angular variables, both sides must be equal to a certain constant $-K$, and this gives the result.
\end{proof}

Let us first study the angular equation. The result here is as follows:

\begin{theorem}
The separated solutions $\alpha=\sigma(s)\theta(t)$ of the angular equation,
$$\sin s\cdot\frac{d}{ds}\left(\sin s\cdot\frac{d\alpha}{ds}\right)
+\frac{d^2\alpha}{dt^2}=-K\sin^2s\cdot\alpha$$
are given by the following formulae, where $l\in\mathbb N$ is such that $K=l(l+1)$,
$$\sigma(s)=P_l^m(\cos s)\quad,\quad\theta(t)=e^{imt}$$
and where $m\in\mathbb Z$ is a constant, and with $P_l^m$ being the Legendre function,
$$P_l^m(x)=(-1)^m(1-x^2)^{m/2}\left(\frac{d}{dx}\right)^mP_l(x)$$
where $P_l$ are the Legendre polynomials, given by the following formula:
$$P_l(x)=\frac{1}{2^ll!}\left(\frac{d}{dx}\right)^l(x^2-1)^l$$
These solutions $\alpha=\sigma(s)\theta(t)$ are called spherical harmonics.
\end{theorem}

\begin{proof}
This follows indeed from a routine study, and with the comment that everything is taken up to linear combinations. We will normalize the wave function later.
\end{proof}

In order to finish our study, it remains to solve the radial equation, for the Coulomb potential $V$ of the proton. As a first manipulation on the radial equation, we have:

\begin{proposition}
The radial equation, written with $K=l(l+1)$, 
$$(r^2\rho')'-\frac{2mr^2}{h^2}(V-E)\rho=l(l+1)\rho$$
takes with $\rho=u/r$ the following form, called modified radial equation,
$$Eu=-\frac{h^2}{2m}\cdot u''+\left(V+\frac{h^2l(l+1)}{2mr^2}\right)u$$
which is a time-independent 1D Schr\"odinger equation.
\end{proposition}

\begin{proof}
With $\rho=u/r$ as in the statement, we have:
$$\rho=\frac{u}{r}\quad,\quad \rho'=\frac{u'r-u}{r^2}\quad,\quad (r^2\rho')'=u''r$$

By plugging this data into the radial equation, this becomes:
$$u''r-\frac{2mr}{h^2}(V-E)u=\frac{l(l+1)}{r}\cdot u$$

By multiplying everything by $h^2/(2mr)$, this latter equation becomes:
$$\frac{h^2}{2m}\cdot u''-(V-E)u=\frac{h^2l(l+1)}{2mr^2}\cdot u$$

But this gives the formula in the statement. As for the interpretation, as time-independent 1D Schr\"odinger equation, this is clear as well, and with the comment here that the term added to the potential $V$ is some sort of centrifugal term.
\end{proof}

It remains to solve the above equation, for the Coulomb potential of the proton. And we have here the following result, which proves the original claims by Bohr:

\begin{theorem}[Schr\"odinger]
In the case of the hydrogen atom, where $V$ is the Coulomb potential of the proton, the modified radial equation, which reads
$$Eu=-\frac{h^2}{2m}\cdot u''+\left(-\frac{Ke^2}{r}+\frac{h^2l(l+1)}{2mr^2}\right)u$$
leads to the Bohr formula for allowed energies,
$$E_n=-\frac{m}{2}\left(\frac{Ke^2}{h}\right)^2\cdot\frac{1}{n^2}$$
with $n\in\mathbb N$, the binding energy being 
$$E_1\simeq-2.177\times 10^{-18}$$
with means $E_1\simeq-13.591\ {\rm eV}$.
\end{theorem}

\begin{proof}
This is again something non-trivial, the idea being as follows:

\medskip

(1) By dividing our modified radial equation by $E$, this becomes:
$$-\frac{h^2}{2mE}\cdot u''=\left(1+\frac{Ke^2}{Er}-\frac{h^2l(l+1)}{2mEr^2}\right)u$$

In terms of $\gamma=\sqrt{-2mE}/h$, this equation takes the following form:
$$\frac{u''}{\gamma^2}=\left(1+\frac{Ke^2}{Er}+\frac{l(l+1)}{(\gamma r)^2}\right)u$$

In terms of the new variable $p=\gamma r$, this latter equation reads:
$$u''=\left(1+\frac{\gamma Ke^2}{Ep}+\frac{l(l+1)}{p^2}\right)u$$

Now let us introduce a new constant $S$ for our problem, as follows:
$$S=-\frac{\gamma Ke^2}{E}$$

In terms of this new constant, our equation reads:
$$u''=\left(1-\frac{S}{p}+\frac{l(l+1)}{p^2}\right)u$$

(2) The idea will be that of looking for a solution written as a power series, but before that, we must ``peel off'' the asymptotic behavior. Which is something that can be done, of course, heuristically. With $p\to\infty$ we are led to $u''=u$, and ignoring the solution $u=e^p$ which blows up, our approximate asymptotic solution is:
$$u\sim e^{-p}$$

Similarly, with $p\to0$ we are led to $u''=l(l+1)u/p^2$, and ignoring the solution $u=p^{-l}$ which blows up, our approximate asymptotic solution is:
$$u\sim p^{l+1}$$

(3) The above heuristic considerations suggest writing our function $u$ as follows:
$$u=p^{l+1}e^{-p}v$$

So, let us do this. In terms of $v$, we have the following formula:
$$u'=p^le^{-p}\left[(l+1-p)v+pv'\right]$$

Differentiating a second time gives the following formula:
$$u''=p^le^{-p}\left[\left(\frac{l(l+1)}{p}-2l-2+p\right)v+2(l+1-p)v'+pv''\right]$$

Thus the radial equation, as modified in (1) above, reads:
$$pv''+2(l+1-p)v'+(S-2(l+1))v=0$$

(4) We will be looking for a solution $v$ appearing as a power series:
$$v=\sum_{j=0}^\infty c_jp^j$$

But our equation leads to the following recurrence formula for the coefficients:
$$c_{j+1}=\frac{2(j+l+1)-S}{(j+1)(j+2l+2)}\cdot c_j$$

(5) We are in principle done, but we still must check that, with this choice for the coefficients $c_j$, our solution $v$, or rather our solution $u$, does not blow up. And the whole point is here. Indeed, at $j>>0$ our recurrence formula reads, approximately:
$$c_{j+1}\simeq\frac{2c_j}{j}$$

But, surprisingly, this leads to $v\simeq c_0e^{2p}$, and so to $u\simeq c_0p^{l+1}e^p$, which blows up.

\medskip

(6) As a conclusion, the only possibility for $u$ not to blow up is that where the series defining $v$ terminates at some point. Thus, we must have for a certain index $j$:
$$2(j+l+1)=S$$

In other words, we must have, for a certain integer $n>l$:
$$S=2n$$

(7) We are almost there. Recall from (1) above that $S$ was defined as follows:
$$S=-\frac{\gamma Ke^2}{E}\quad:\quad\gamma=\frac{\sqrt{-2mE}}{h}$$

Thus, we have the following formula for the square of $S$:
$$S^2=\frac{\gamma^2K^2e^4}{E^2}=-\frac{2mE}{h^2}\cdot\frac{K^2e^4}{E^2}=-\frac{2mK^2e^4}{h^2E}$$

Now by using the formula $S=2n$ from (6), the energy $E$ must be of the form:
$$E=-\frac{2mK^2e^4}{h^2S^2}=-\frac{mK^2e^4}{2h^2n^2}$$

Calling this energy $E_n$, depending on $n\in\mathbb N$, we have, as claimed:
$$E_n=-\frac{m}{2}\left(\frac{Ke^2}{h}\right)^2\cdot\frac{1}{n^2}$$

(8) Thus, we proved the Bohr formula. Regarding numerics, the data is as follows:
$$K=8.988\times10^9\quad,\quad
e=1.602\times10^{-19}$$
$$h=1.055\times10^{-34}\quad,\quad 
m=9.109\times10^{-31}$$

But this gives the formula of $E_1$ in the statement.
\end{proof}

In order to further advance, let us formulate our conclusions so far as follows:

\begin{theorem}
The wave functions of the hydrogen atom are the following functions, labelled by three quantum numbers, $n,l,m$,
$$\phi_{nlm}(r,s,t)=\rho_{nl}(r)\alpha_l^m(s,t)$$
where $\rho_{nl}(r)=p^{l+1}e^{-p}v(p)/r$ with $p=\alpha r$ as before, with the coefficients of $v$ subject to
$$c_{j+1}=\frac{2(j+l+1-n)}{(j+1)(j+2l+2)}\cdot c_j$$
and $\alpha_l^m(s,t)$ being the spherical harmonics found before.
\end{theorem}

\begin{proof}
This follows indeed by putting together all the results obtained so far, and with the remark that everything is up to the normalization of the wave function.
\end{proof}

In order to improve our results, we will need the following standard fact:

\begin{proposition}
The polynomials $v(p)$ are given by the formula
$$v(p)=L_{n-l-1}^{2l+1}(p)$$
where the polynomials on the right, called associated Laguerre polynomials, are given by
$$L_q^p(x)=(-1)^p\left(\frac{d}{dx}\right)^pL_{p+q}(x)$$
with $L_{p+q}$ being the Laguerre polynomials, given by the following formula,
$$L_q(x)=\frac{e^x}{q!}\left(\frac{d}{dx}\right)^q(e^{-x}x^q)$$
called Rodrigues formula for the Laguerre polynomials.
\end{proposition}

\begin{proof}
The story here is very similar to that of the Legendre polynomials. Consider the Hilbert space $H=L^2[0,\infty)$, with the following scalar product on it:
$$<f,g>=\int_0^\infty f(x)g(x)e^{-x}\,dx$$

The orthogonal basis obtained by applying Gram-Schmidt to the Weierstrass basis $\{x^q\}$ is then formed by the Laguerre polynomials $\{L_q\}$, and this gives the results.
\end{proof}

With the above result in hand, we can now improve our main results, as follows:

\begin{theorem}
The wave functions of the hydrogen atom are given by
$$\phi_{nlm}(r,s,t)=\sqrt{\left(\frac{2}{na}\right)^3\frac{(n-l-1)!}{2n(n+l)!}}e^{-r/na}\left(\frac{2r}{na}\right)^lL_{n-l-1}^{2l+1}\left(\frac{2r}{na}\right)\alpha_l^m(s,t)$$
with $\alpha_l^m(s,t)$ being the spherical harmonics found before.
\end{theorem}

\begin{proof}
This follows indeed by putting together what we have, and then doing some remaining work, concerning the normalization of the wave function.
\end{proof}

\section*{15b. Angular momentum}

What is next? All sorts of corrections to the solution of the hydrogen atom discussed above, coming from relativity theory, electron spin, and other phenomena. Indeed, and with experiments confirming this, what we have above is just the tip of the iceberg.

\bigskip

However, all this is non-trivial and long business, and as a main objective for us, we would like to talk about electron spin. Following Uhlenbeck, Goudsmit, Pauli and others, we will first talk angular momentum, and then we will axiomatize spin as being the quantity which naturally ``complements'' the angular momentum. We first have:

\begin{proposition}
The components of the position operator $x=(x_1,x_2,x_3)$ and momentum operator $p=-ih\nabla$ satisfy the following relations,
$$[x_i,x_j]=[p_i,p_j]=0$$
$$[x_i,p_j]=ih\delta_{ij}$$
where $[a,b]=ab-ba$, called canonical commutation relations.
\end{proposition}

\begin{proof}
All the above formulae are elementary, as follows:

\medskip

(1) The components of the position operator $x=(x_1,x_2,x_3)$ obviously commute with each other, $x_ix_j=x_jx_i$, which makes their commutators vanish, $[x_i,x_j]=0$. 

\medskip

(2) Regarding the momentum operator $p=-ih\nabla$, its components are as follows:
$$p_1=-ih\cdot\frac{d}{dx_1}\quad,\quad 
p_2=-ih\cdot\frac{d}{dx_2}\quad,\quad
p_3=-ih\cdot\frac{d}{dx_3}$$

Since partial derivatives commute with each other, we obtain $[p_i,p_j]=0$.

\medskip

(3) It remains to prove the last formula, and we have here:
\begin{eqnarray*}
[x_i,p_j]f
&=&(x_ip_j-p_jx_i)f\\
&=&-ih\left(x_i\cdot\frac{df}{dx_j}-\frac{d}{dx_j}(x_if)\right)\\
&=&-ih\left(x_i\cdot\frac{df}{dx_j}-\frac{dx_i}{dx_j}\cdot f-
x_i\cdot\frac{df}{dx_j}\right)\\
&=&ih\cdot\frac{dx_i}{dx_j}\cdot f\\
&=&ih\delta_{ij}\cdot f
\end{eqnarray*}

Thus, we are led to the conclusion in the statement.
\end{proof}

The above might look a bit complicated, and the simplest way to remember it is that ``everyhing commutes'', that is, $ab=ba$, except for the coordinates and momenta coordinates taken in the same direction, which are subject to the following rule:
$$x_ip_i=p_ix_i+ih$$

Getting now to angular momentum, it is convenient to change notation, with $(x,y,z)$ instead of $(x_1,x_2,x_3)$. We have the following result, to start with:

\begin{theorem}
The components of the angular momentum operator
$$L=x\times(-ih\nabla)$$
satisfy the following equations,
$$[L_x,L_y]=ihL_z\quad,\quad 
[L_y,L_z]=ihL_x\quad,\quad 
[L_z,L_x]=ihL_y$$
called commutation relations for the angular momentum.
\end{theorem}

\begin{proof}
With the more familiar notation $p=-ih\nabla$ for the momentum, or rather for the associated operator, the components of the angular momentum operator are:
$$L_x=yp_z-zp_y\quad,\quad 
L_y=zp_x-xp_z\quad,\quad
L_z=xp_y-yp_x$$

Let us prove the first commutation relation. We have:
\begin{eqnarray*}
[L_x,L_y]
&=&[yp_z-zp_y,zp_x-xp_z]\\
&=&[yp_z,zp_x]-[yp_z,xp_z]-[zp_y,zp_x]+[zp_y,xp_z]
\end{eqnarray*}

By heavily using the commutation relations from Proposition 15.17, we have:
$$[yp_z,zp_x]=yp_zzp_x-zp_xyp_z=y(zp_z-ih)p_x-zyp_xp_z=-ihyp_x$$
$$[yp_z,xp_z]=yp_zxp_z-xp_zyp_z=0$$
$$[zp_y,zp_x]=zp_yzp_x-zp_xzp_y=0$$
$$[zp_y,xp_z]=zp_yxp_z-xp_zzp_y=zxp_yp_z-x(zp_z-ih)p_y=ihxp_y$$

We conclude that the commutator that we were computing is given by the following formula, which is precisely the one in the statement:
\begin{eqnarray*}
[L_x,L_y]
&=&-ihyp_x+ihxp_y\\
&=&ih(xp_y-yp_x)\\
&=&ihL_z
\end{eqnarray*}

The proof of the other two commutation relations is similar, or can be simply obtained by invoking the cyclic invariance $x\to y\to z\to x$ of our problem, which cyclic invariance is not broken by the vector product $\times$ used, and so can indeed be invoked.
\end{proof}

As an interesting consequence of Theorem 15.18, we have:

\begin{proposition}
The following operator, called square of angular momentum
$$L^2=L_x^2+L_y^2+L_z^2$$
commutes with all $3$ operators $L_x,L_y,L_z$.
\end{proposition}

\begin{proof}
We have the following computation, to start with:
\begin{eqnarray*}
[L^2,L_x]
&=&(L_x^2+L_y^2+L_z^2)L_x-L_x(L_x^2+L_y^2+L_z^2)\\
&=&L_y^2L_x+L_z^2L_x-L_xL_y^2-L_xL_z^2\\
&=&[L_y^2,L_x]+[L_z^2,L_x]
\end{eqnarray*}

The first commutator can be computed with a trick, as follows:
\begin{eqnarray*}
[L_y^2,L_x]
&=&L_yL_yL_x-L_xL_yL_y\\
&=&L_yL_yL_x-L_yL_xL_y+L_yL_xL_y-L_xL_yL_y\\
&=&L_y[L_y,L_x]+[L_y,L_x]L_y\\
&=&L_y(-ihL_z)+(-ihL_z)L_y\\
&=&-ih(L_yL_z+L_zL_y)
\end{eqnarray*}

The second commutator can be computed with the same trick, as follows:
\begin{eqnarray*}
[L_z^2,L_x]
&=&L_zL_zL_x-L_xL_zL_z\\
&=&L_zL_zL_x-L_zL_xL_z+L_zL_xL_z-L_xL_zL_z\\
&=&L_z[L_z,L_x]+[L_z,L_x]L_z\\
&=&L_z(ihL_y)+(ihL_y)L_z\\
&=&ih(L_zL_y+L_yL_z)
\end{eqnarray*}

Now by summing we obtain the following commutation relation, as desired:
$$[L^2,L_x]=0$$

The proof of the other two commutation relations is similar, or we can simply invoke here the cyclic symmetry argument from the end of the proof of Theorem 15.18.
\end{proof}

Let us discuss now the diagonalization of $L_x,L_y,L_z$. Since these operators do not commute, we cannot hope for a joint diagonalization. Thus, we must choose one of them, and for reasons that will become clear later, when writing things in spherical coordinates, we will choose $L_x$. In view of Proposition 15.19, this operator $L_x$ does commute with $L^2$, and so we can hope for a joint diagonalization of $L^2,L_x$. And, so is what happens:

\begin{theorem}
The operators $L^2,L_x$ diagonalize as
$$L^2f_l^m=h^2l(l+1)f_l^m$$
$$L_xf_l^m=hmf_l^m$$
where $l\in\mathbb N/2$ and $m=-l,-l+1,\ldots,l-1,l$.
\end{theorem} 

\begin{proof}
This is something quite long, the idea being as follows:

\medskip

(1) For reasons that will become clear later on, let us introduce two operators as follows, called raising and lowering operators:
$$L_+=L_y+iL_z\quad,\quad 
L_-=L_y-iL_z$$

We will often deal with these operators at the same time, using the following notation:
$$L_\pm=L_y\pm iL_z$$

(2) In order to get started, we first have the following computation:
\begin{eqnarray*}
[L_x,L_\pm]
&=&[L_x,L_y]\pm i[L_x,L_z]\\
&=&ihL_z\pm i(-ihL_y)\\
&=&h(iL_z\pm L_y)\\
&=&\pm h(\pm iL_z+L_y)\\
&=&\pm hL_\pm
\end{eqnarray*}

(3) Our claim now is that the conditions $L^2f=\lambda f$, $L_xf=\mu f$ imply:
$$L^2(L_\pm f)=\lambda(L_\pm f)$$
$$L_x(L_\pm f)=(\mu\pm h)(L_\pm f)$$

Indeed, the first formula follows from the following computation:
\begin{eqnarray*}
L^2(L_\pm f)
&=&L_\pm(L^2 f)\\
&=&L_\pm(\lambda f)\\
&=&\lambda(L_\pm f)
\end{eqnarray*}

As for the second formula, this follows from the following computation:
\begin{eqnarray*}
L_x(L_\pm f)
&=&L_xL_\pm f\\
&=&(L_xL_\pm-L_\pm L_x)f+L_\pm L_xf\\
&=&\pm hL_\pm f+L_\pm(\mu f)\\
&=&(\mu\pm h)(L_\pm f)
\end{eqnarray*}

(4) Now in view of the formulae found in (3), the raising and lowering operators act on the joint eigenfunctions of $L^2,L_x$, by leaving the $L^2$ eigenvalue unchanged, and by raising and lowering the eigenvalue of $L_x$. But both this raising process and lowering process for the eigenvalue of $L_x$ cannot go on forever, because of the following estimate:
\begin{eqnarray*}
\lambda
&=&<L^2>\\
&=&<L_x^2>+<L_y^2>+<L_z^2>\\
&=&\mu^2+<L_y^2>+<L_z^2>\\
&\geq&\mu^2
\end{eqnarray*}

(5) In order to see exactly how the raising and lowering processes terminate, we will need some more computations. We first have the following computation:
\begin{eqnarray*}
L_\pm L_\mp
&=&(L_y\pm iL_z)(L_y\mp iL_z)\\
&=&L_y^2+L_z^2\mp i(L_yL_z-L_zL_y)\\
&=&L_y^2+L_z^2\mp i(ihL_x)\\
&=&L_y^2+L_z^2\pm hL_x\\
&=&L^2-L_x^2\pm hL_x
\end{eqnarray*}

We conclude from this that we have the following formula:
$$L^2=L_\pm L_\mp+L_x^2\mp hL_x$$

Now assuming $L_xf=hlf$, at termination of the raising process, we have:
\begin{eqnarray*}
L^2(f)
&=&(L_-L_++L_x^2+hL_x)f\\
&=&(0+h^2l^2+h^2l)f\\
&=&h^2l(l+1)f
\end{eqnarray*}

Similarly, assuming $L_xf=hl'f$, at termination of the lowering process, we have:
\begin{eqnarray*}
L^2(f)
&=&(L_+-L_-+L_x^2-hL_x)f\\
&=&(0+h^2l'^2-h^2l')f\\
&=&h^2l'(l'-1)f
\end{eqnarray*}

Thus $l(l+1)=l'(l'-1)$, and since $l'=l+1$ is impossible, due to raising vs lowering, we must have $l'=-l$, and this leads to the conclusion in the statement.
\end{proof}

Moving ahead now, the above is all we need. The idea in what follows will be that of writing everything in spherical coordinates, and then finding the eigenfunctions. 

\bigskip

And, what happens is that we have here the following remarkable result:

\index{spherical harmonics}
\index{joint eigenfunctions}

\begin{theorem}
In spherical coordinates $r,s,t$ we have
$$L_x=-\frac{ih}{dt}$$
$$L_y=ih\left(\frac{\sin t}{ds}+\frac{\cos s\cos t}{\sin s}\cdot\frac{1}{dt}\right)$$
$$L_z=-ih\left(\frac{\cos t}{ds}-\frac{\cos s\sin t}{\sin s}\cdot\frac{1}{dt}\right)$$
and the spherical harmonics are joint eigenfunctions of $L^2,L_x$.
\end{theorem}

\begin{proof}
We recall that, according to our usual, $N$-dimensional looking conventions, the spherical coordinates that we use are as follows, with $r\in[0,\infty)$ being the radius, $s\in[0,\pi]$ being the polar angle, and $t\in[0,2\pi]$ being the azimuthal angle:
$$\begin{cases}
x\!\!\!&=\ r\cos s\\
y\!\!\!&=\ r\sin s\cos t\\
z\!\!\!&=\ r\sin s\sin t
\end{cases}$$

(1) We know that we have $L=-ihx\times\nabla$, so let us first compute $\nabla$ in spherical coordinates. We have here, according to the chain rule for derivatives:
\begin{eqnarray*}
\nabla
&=&\begin{pmatrix}
dr/dx&ds/dx&dt/dx\\
dr/dy&ds/dy&dt/dy\\
dr/dz&ds/dz&dt/dz
\end{pmatrix}
\begin{pmatrix}
d/dr\\
d/ds\\
d/dt
\end{pmatrix}\\
&=&\begin{pmatrix}
dx/dr&dy/dr&dz/dr\\
dx/ds&dy/ds&dz/ds\\
dx/dt&dy/dt&dz/dt
\end{pmatrix}^{-1}
\begin{pmatrix}
d/dr\\
d/ds\\
d/dt
\end{pmatrix}
\end{eqnarray*}

(2) On the other hand, it is routine to check that we have:
$$\begin{pmatrix}
dx/dr&dx/ds&dx/dt\\
dy/dr&dy/ds&dy/dt\\
dz/dr&dz/ds&dz/dt
\end{pmatrix}
=\begin{pmatrix}
\cos s&-r\sin s&0\\
\sin s\cos t&r\cos s\cos t&-r\sin s\sin t\\
\sin s\sin t&r\cos s\sin t&r\sin s\cos t
\end{pmatrix}$$

It is also routine to see that this latter matrix, say $A$, satisfies:
$$A^tA=\begin{pmatrix}1&0&0\\ 0&r^2&0\\ 0&0&r^2\sin^2s\end{pmatrix}$$

Now if we call $D$ the diagonal matrix on the right, we conclude that the matrix, say $B$, appearing in the above formula of $\nabla$ is given by:
\begin{eqnarray*}
B
&=&(A^t)^{-1}\\
&=&AD^{-1}\\
&=&\begin{pmatrix}
\cos s&-r\sin s&0\\
\sin s\cos t&r\cos s\cos t&-r\sin s\sin t\\
\sin s\sin t&r\cos s\sin t&r\sin s\cos t
\end{pmatrix}
\begin{pmatrix}1&0&0\\ 0&1/r^2&0\\ 0&0&1/(r^2\sin^2s)\end{pmatrix}\\
&=&\begin{pmatrix}
\cos s&-\sin s/r&0\\
\sin s\cos t&\cos s\cos t/r&-\sin t/(r\sin s)\\
\sin s\sin t&\cos s\sin t/r&\cos t/(r\sin s)
\end{pmatrix}
\end{eqnarray*}

(3) Thus, the angular momentum operator that we are looking for, $L=-ihx\times\nabla$, written more conveniently as $L=-ihx/r\times r\nabla$, is given by:
$$L
=-ih
\begin{pmatrix}
\cos s\\
\sin s\cos t\\
\sin s\sin t
\end{pmatrix}\times
\begin{pmatrix}
r\cos s&-\sin s&0\\
r\sin s\cos t&\cos s\cos t&-\sin t/\sin s\\
r\sin s\sin t&\cos s\sin t&\cos t/\sin s
\end{pmatrix}
\begin{pmatrix}
d/dr\\
d/ds\\
d/dt
\end{pmatrix}$$

And computing now the vector product gives the formula for $L$ in the statement.

\medskip

(4) Now with our explicit formula for $L$ in hand, we next find that the raising and lowering operators are given by:
$$L_\pm=\pm he^{\pm it}\left(\frac{d}{ds}\pm i\,\frac{\cos s}{\sin s}\cdot\frac{1}{dt}\right)$$

Next, we find that these two operators satisfy the following formula:
$$L_+L_-=-h^2\left(\frac{d^2}{ds^2}+\frac{\cos s}{\sin s}\cdot\frac{d}{ds}+\frac{\cos^2s}{\sin^2s}\cdot\frac{d^2}{dt^2}+i\frac{d}{dt}\right)$$

And finally, by using this latter formula, we find that $L^2$ is given by:
$$L^2=-h^2\left(\frac{1}{\sin s}\cdot\frac{d}{ds}\left(\sin s\cdot\frac{d}{ds}\right)+\frac{1}{\sin^2s}\cdot\frac{d^2}{dt^2}\right)$$

(5) With all these formulae in hand, we can now finish. The eigenfunction equation for the above operator $L^2$, with eigenvalue $h^2l(l+1)$, is as follows:
$$-h^2\left(\frac{1}{\sin s}\cdot\frac{d}{ds}\left(\sin s\cdot\frac{d}{ds}\right)+\frac{1}{\sin^2s}\cdot\frac{d^2}{dt^2}\right)f=h^2l(l+1)f$$

But this is precisely the angular equation found before. As for the eigenfunction equation for the operator $L_x$, with eigenvalue $hm$, this is as follows:
$$-\frac{ih}{dt}\,f=hmf$$

But this is equivalent to the azimuthal equation, and this gives the result.
\end{proof}

\section*{15c. Pauli matrices}

In order to talk now about spin, we will regard, a bit as in the classical mechanics case, the spin and the angular momentum as being similar quantities. Thus, in analogy with the basic equations for the angular momentum, we should have:

\begin{definition}
The components of the spin operator are subject to
$$[S_x,S_y]=ihS_z$$
$$[S_y,S_z]=ihS_x$$
$$[S_z,S_x]=ihS_y$$
called commutation relations for the spin operator.
\end{definition}

The point now is that, with the above relations in hand, which are identical to the commutation relations for the angular momentum, all the general results from the previous  section, based on that commutation relations, extend to our present setting, simply by changing $L$ into $S$ everywhere. And in particular, we are led in this way to:

\begin{theorem}
We have the following diagonalization formulae
$$S^2f_s^m=h^2s(s+1)f_s^m$$
$$S_xf_s^m=hmf_s^m$$
$$S_\pm f_s^m=h\sqrt{s(s+1)-m(m\pm1)}\,f_s^{m\pm1}$$
involving the operators $S^2=S_x^2+S_y^2+S_z^2$, $S_x$ and $S_\pm=S_y\pm iS_z$.
\end{theorem}

\begin{proof}
Here the first two formulae are something that we already know, from the previous section, with $L,j$ being replaced by $S,s$. As for the last formula, this is something that we did not need, in the $L,j$ context, but that we will need now. We want to compute the constants $C_{s,\pm}^m$ making work the raising and lowering formula, namely:
$$S_\pm f_s^m=C_{s,\pm}^mf_s^{m\pm1}$$

But this can be done by using $S^2=S_\pm S_\mp+S_x^2\mp hS_x$ and $S_\pm^*=S_\mp$, and we get:
$$C_{s,+}^m=h\sqrt{s(s+1)-m(m+1)}$$
$$C_{s,-}^m=h\sqrt{s(s+1)-m(m-1)}$$

Thus, we are led to the last formula in the statement, and we are done.
\end{proof}

In practice now, let us look for the simplest realization of spin. We are led, for fixed particles, to a quantum mechanics over $H=\mathbb C^2$, with spin up and down being represented by the basis vectors.It remains to see the equations in Theorem 15.23 reformulate, in this $H=\mathbb C^2$ setting. But here, not many choices, and we are led in this way to:

\begin{definition}
In the quantum mechanics of the spin, over $H=\mathbb C^2$, with
$$e_1=\binom{1}{0}\quad,\quad e_2=\binom{0}{1}$$
being spin up and down, the spin is subject to the following equations, for $f=e_1,e_2$, 
$$S^2f=h^2s(s+1)f$$
$$S_xf=hm_ff$$
$$S_\pm f=h\sqrt{s(s+1)-m_f(m_f\pm1)}\,\check{f}$$
with parameters $s=1/2$, $m_{e_1}=1/2$, $m_{e_2}=-1/2$, and with $\{e_1,e_2\}=\{f,\check{f}\}$.
\end{definition}

Here all the choices, and notably $s=1/2$, are very natural in view of Theorem 15.23, because these are the choices providing a ``minimal'' realization of the equations in Theorem 15.23, in the smallest possible number of dimensions, namely $N=2$. However, all this comes with a shade of mystery, or at least is not rock-solid enough as to be called theorem, and it is probably safer to use the term ``definition'', as we did above.

\bigskip

The point now is that the above questions can be solved, the result being:

\begin{theorem}
In the above $H=\mathbb C^2$ context, of the mechanics of a single, fixed electron, the components of the normalized spin $\sigma=2S/h$ are as follows,
$$\sigma_x=\begin{pmatrix}1&0\\0&-1\end{pmatrix}
\quad,\quad \sigma_y=\begin{pmatrix}0&1\\1&0\end{pmatrix}
\quad,\quad \sigma_z=\begin{pmatrix}0&-i\\i&0\end{pmatrix}$$
called Pauli matrices. In the general, dynamic context, where we already have a Hilbert space $H$ for the wave function, spin can be introduced by using the space
$$H'=H\otimes\mathbb C^2$$
and using the above Pauli matrices for it, acting on the $\mathbb C^2$ part.
\end{theorem}

\begin{proof}
As a first observation, we recognize in the above the Pauli matrices from chapter 3, which appeared there mathematically, in relation with $SU_2$, slightly modified. In what follows we will use the above new convention for the Pauli matrices. Regarding now the proof, the equations in Definition 15.24, written in full detail, are as follows:
$$S^2\binom{1}{0}=\frac{3h^2}{4}\binom{1}{0}
\quad,\quad
S^2\binom{0}{1}=\frac{3h^2}{4}\binom{0}{1}$$
$$S_x\binom{1}{0}=\frac{h}{2}\binom{1}{0}
\quad,\quad
S_x\binom{0}{1}=-\frac{h}{2}\binom{0}{1}$$
$$S_+\binom{1}{0}=\binom{0}{0}\quad,\quad S_+\binom{0}{1}=h\binom{1}{0}$$
$$S_-\binom{1}{0}=h\binom{0}{1}\quad,\quad S_-\binom{0}{1}=\binom{0}{0}$$

Thus, we have the following formulae, for the various matrices involved:
$$S^2=\frac{3h^2}{4}\begin{pmatrix}1&0\\0&1\end{pmatrix}
\quad,\quad S_x=\frac{h}{2}\begin{pmatrix}1&0\\0&-1\end{pmatrix}$$
$$S_+=h\begin{pmatrix}0&1\\0&0\end{pmatrix}
\quad,\quad S_-=h\begin{pmatrix}0&0\\1&0\end{pmatrix}$$

In relation with what we want to prove, we have obtained the formula of $S_x$. Regarding now the formulae of $S_y,S_z$, these follow by solving the following system:
$$S_+=S_y+iS_z\quad,\quad 
S_-=S_y-iS_z$$

To be more precise, the computation for $S_y$ goes as follows:
$$S_y=\frac{S_++S_-}{2}=\frac{h}{2}\begin{pmatrix}0&1\\1&0\end{pmatrix}$$

As for the computation for $S_z$, this goes as follows:
$$S_z=\frac{S_+-S_-}{2i}=\frac{h}{2i}\begin{pmatrix}0&1\\-1&0\end{pmatrix}=
\frac{h}{2}\begin{pmatrix}0&-i\\i&0\end{pmatrix}$$

Thus, we are led to the conclusions in the statement.
\end{proof}

As a first consequence of the above, looking quite good, we have:

\begin{fact}
Electrons have spin $1/2$.
\end{fact}

This comes indeed from the formula $s=1/2$ in Definition 15.24, and some further speculations are certainly possible. For instance the Pauli matrices all square up to one, $\sigma_i^2=1$, so we can say that ``it takes $720^\circ$ instead of the usual $360^\circ$ to turn an electron back in place'', leading to the conclusion that the electron spin is $360/720=1/2$.

\section*{15d. Dirac matrices} 

Getting now to more particle physics, as a continuation of the above, things are quite tricky, and as a guiding principle here, we know that the electron cannot live without the photon. Indeed, in the context of the basic physics of atoms, electrons can jump between energy levels, emitting or absorbing photons, and with this being known to happen even in the absence of external stimuli. Thus, the true ``brother'' of the electron is not the proton or the neutron, but rather the photon. And so, the minimal extension of quantum mechanics, that we should try to build now, should deal with electrons and photons.

\bigskip

In view of this, let us first look into the photon, try to understand how to make it fit into our theory, and leave the electron for later. As a starting point, we have:

\begin{fact}
The master equation for free electromagnetic radiation, that is, for free photons, is the wave equation at speed $v=c$, namely:
$$\ddot{\varphi}=c^2\Delta\varphi$$
This equation can be reformulated in the more symmetric form
$$\left(\frac{1}{c^2}\cdot\frac{d^2}{dt^2}-\Delta\right)\varphi=0$$
with the operator on the left being called the d'Alembertian.
\end{fact}

In relation now with the electron, there is an obvious similarity here with the free Schr\"odinger equation, without potential $V$, which reads:
$$\left(i\frac{d}{dt}+\frac{h}{2m}\,\Delta\right)\psi=0$$

This similarity suggests looking for a relativistic version of the Schr\"odinger equation, which is compatible with the wave equation at $v=c$. And coming up with such an equation is not very complicated, the straightforward answer being as follows:

\begin{definition}
The following abstract mathematical equation,
$$\left(-\frac{1}{c^2}\cdot\frac{d^2}{dt^2}+\Delta\right)\psi=\frac{m^2c^2}{h^2}\,\psi$$
on a function $\psi=\psi_t(x)$, is called the Klein-Gordon equation.
\end{definition}

To be more precise, what we have here is some sort of a speculatory equation, formally obtained from the Schr\"odinger equation, via a few simple manipulations, as to make it relativistic. And with the relation with photons being something very simple, the thing being that at zero mass, $m=0$, we obtain precisely the wave equation at $v=c$.

\bigskip

All this is very nice, looks like we have a beginning of theory here, both making the electrons relativistic, and unifying them with photons. And isn't this too beautiful to be true. Going ahead now with physics, the following question appears:

\begin{question}
What does the Klein-Gordon equation really describe?
\end{question}

And here, unfortunately, bad news all the way. A closer look at the Klein-Gordon equation reveals all sorts of bugs, making it unusable for anything reasonable. And with the main bug, which is enough for disqualifying it, being that, unlike the Schr\"odinger equation which preserves probability amplitudes $|\psi|^2$, the Klein-Gordon equation does not have this property. Thus, even before trying to understand what the Klein-Gordon equation really describes, we are left with the conclusion that this equation cannot really describe anything reasonable, due to the formal nature of the function $\psi$ involved.

\bigskip

So, this was for the story of the Klein-Gordon equation. Actually this equation was first discovered by Schr\"odinger himself, in the context of his original work on the Schr\"odinger equation. But noticing the above bugs with it, Schr\"odinger dismissed it right away, and then downgraded his objectives, looking for something non-relativistic instead, and then found the Schr\"odinger equation, leading to the story that we know.

\bigskip

This being said, the Klein-Gordon equation found later a number of interesting applications, the continuation of the story being as follows:

\bigskip

(1) Dirac found a clever way of extracting the ``square root'' of the Klein-Gordon equation. And this square root equation, called Dirac equation, turned out to be the correct one, making exactly what the Klein-Gordon equation was supposed to do.

\bigskip

(2) Technically speaking, the Klein-Gordon equation is very useful for investigating the Dirac equation, because the components of the solutions of the Dirac equation satisfy the Klein-Gordon equation. More on this later, when discussing the Dirac equation. 

\bigskip

(3) Finally, the Klein-Gordon equation was later recognized to describe well the spin 0 particles. But with these particles being something specialized, including the unstable and sowewhat fringe ``pions'', and the Higgs boson, which is something complicated. 

\bigskip

We will briefly discuss all this, in what follows. Getting to work now, following Dirac, the idea is that of extracting the square root of the Klein-Gordon operator, as follows:

\begin{proposition}
We can extract the square root of the Klein-Gordon operator, via a formula as follows,
$$-\frac{1}{c^2}\cdot\frac{d^2}{dt^2}+\Delta=\left(\frac{i}{c}\cdot\frac{Pd}{dt}+\frac{Qd}{dx}+\frac{Rd}{dy}+\frac{Sd}{dz}\right)^2$$
by using matrices $P,Q,R,S$ which anticommute, $AB=-BA$, and whose squares equal one, $A^2=1$.
\end{proposition}

\begin{proof}
We have the following computation, valid for any matrices $P,Q,R,S$, with the notation $\{A,B\}=AB+BA$:
\begin{eqnarray*}
\left(\frac{i}{c}\cdot\frac{Pd}{dt}+\frac{Qd}{dx}+\frac{Rd}{dy}+\frac{Sd}{dz}\right)^2
&=&-\frac{1}{c^2}\cdot\frac{P^2d^2}{dt^2}+\frac{Q^2d^2}{dx^2}+\frac{R^2d^2}{dy^2}+\frac{S^2d^2}{dz^2}\\
&+&\frac{i}{c}\left(\frac{\{P,Q\}d^2}{dtdx}+\frac{\{P,R\}d^2}{dtdy}+\frac{\{P,S\}d^2}{dtdz}\right)\\
&+&\frac{\{Q,R\}d^2}{dxdy}+\frac{\{Q,S\}d^2}{dxdz}+\frac{\{R,S\}d^2}{dydz}
\end{eqnarray*}

Thus, in order to obtain in this way the Klein-Gordon operator, the conditions in the statement must be satisfied.
\end{proof}

As a technical comment here, normally when extracting a square root, we should look for a self-adjoint operator. In view of this, observe that we have:
$$\left(\frac{i}{c}\cdot\frac{Pd}{dt}+\frac{Qd}{dx}+\frac{Rd}{dy}+\frac{Sd}{dz}\right)^*
=-\frac{i}{c}\cdot\frac{P^*d}{dt}+\frac{Q^*d}{dx}+\frac{R^*d}{dy}+\frac{S^*d}{dz}$$

Thus, we should normally add the conditions $P^*=-P$ and $Q^*=Q$, $R^*=R$, $S^*=S$ to those above. But, the thing is that due to some subtle reasons, the natural square root of the Klein-Gordon operator is not self-adjoint. More on this later.

\bigskip

Looking for matrices $P,Q,R,S$ as above is not exactly trivial, and the simplest solutions appear in $M_4(\mathbb C)$, in connection with the Pauli matrices, as follows:

\begin{proposition}
The simplest matrices $P,Q,R,S$ as above appear as
$$P=\gamma_0\quad,\quad Q=i\gamma_1\quad,\quad R=i\gamma_2\quad,\quad S=i\gamma_3$$
with $\gamma_0,\gamma_1,\gamma_2,\gamma_3$ being the Dirac matrices, given by
$$\gamma_0=\begin{pmatrix}1&0\\0&-1\end{pmatrix}
\quad,\quad\gamma_i=\begin{pmatrix}0&\sigma_i\\-\sigma_i&0\end{pmatrix}$$
where $\sigma_1,\sigma_2,\sigma_3$ are the Pauli spin matrices, given by
$$\sigma_1=\begin{pmatrix}1&0\\0&-1\end{pmatrix}
\quad,\quad \sigma_2=\begin{pmatrix}0&1\\1&0\end{pmatrix}
\quad,\quad \sigma_3=\begin{pmatrix}0&-i\\i&0\end{pmatrix}$$
that we met before, in the context of the electron spin.
\end{proposition}

\begin{proof}
We have $\gamma_0^2=1$, and by using $\sigma_i^2=1$ for any $i=1,2,3$, we have as well the following formula, which shows that we have $(i\gamma_i)^2=1$, as needed:
$$\gamma_i^2=\begin{pmatrix}0&\sigma_i\\-\sigma_i&0\end{pmatrix}\begin{pmatrix}0&\sigma_i\\-\sigma_i&0\end{pmatrix}=\begin{pmatrix}-1&0\\0&-1\end{pmatrix}$$

As in what regards the commutators, we first have, for any $i=1,2,3$, the following equalities, which show that $\gamma_0$ anticommutes indeed with $\gamma_i$:
$$\gamma_0\gamma_i=\begin{pmatrix}1&0\\0&-1\end{pmatrix}\begin{pmatrix}0&\sigma_i\\-\sigma_i&0\end{pmatrix}=\begin{pmatrix}0&\sigma_i\\\sigma_i&0\end{pmatrix}$$
$$\gamma_i\gamma_0=\begin{pmatrix}0&\sigma_i\\-\sigma_i&0\end{pmatrix}\begin{pmatrix}1&0\\0&-1\end{pmatrix}=\begin{pmatrix}0&-\sigma_i\\-\sigma_i&0\end{pmatrix}$$

Regarding now the remaining commutators, observe here that we have:
$$\gamma_i\gamma_j=\begin{pmatrix}0&\sigma_i\\-\sigma_i&0\end{pmatrix}
\begin{pmatrix}0&\sigma_j\\-\sigma_j&0\end{pmatrix}
=\begin{pmatrix}-\sigma_i\sigma_j&0\\0&-\sigma_i\sigma_j\end{pmatrix}$$

Now since the Pauli matrices anticommute, we obtain $\gamma_i\gamma_j=-\gamma_j\gamma_i$, as desired.
\end{proof}

We can now put everything together, and we obtain:

\begin{theorem}
The following operator, called Dirac operator,
$$D=i\left(\frac{\gamma_0d}{cdt}+\frac{\gamma_1d}{dx}+\frac{\gamma_2d}{dy}+\frac{\gamma_3d}{dz}\right)$$
has the property that its square is the Klein-Gordon operator.
\end{theorem}

\begin{proof}
With notations from Proposition 15.30 and Proposition 15.31, and by making the choices in Proposition 15.26, we have:
\begin{eqnarray*}
\frac{i}{c}\cdot\frac{Pd}{dt}+\frac{Qd}{dx}+\frac{Rd}{dy}+\frac{Sd}{dz}
&=&\frac{i}{c}\cdot\frac{\gamma_0d}{dt}+\frac{i\gamma_1d}{dx}+\frac{i\gamma_2d}{dy}+\frac{i\gamma_3d}{dz}\\
&=&i\left(\frac{\gamma_0d}{cdt}+\frac{\gamma_1d}{dx}+\frac{\gamma_2d}{dy}+\frac{\gamma_3d}{dz}\right)
\end{eqnarray*}

Thus, we have here a square root of the Klein-Gordon operator, as desired.
\end{proof}

We can now extract the square root of the Klein-Gordon equation, as follows:

\begin{theorem}
We have the following equation, called Dirac equation,
$$ih\left(\frac{\gamma_0d}{cdt}+\frac{\gamma_1d}{dx}+\frac{\gamma_2d}{dy}+\frac{\gamma_3d}{dz}\right)\psi=mc\psi$$
obtained by extracting the square root of the Klein-Gordon equation.
\end{theorem}

\begin{proof}
This is more of a definition, based on the above, but we have called it Theorem, in view of its importance, and for finishing this chapter in beauty, with a theorem.
\end{proof}

In practice now, as usual with such theoretical physics equations, extreme caution is recommended, at least to start with. However, with a bit of patience, the Dirac equation can be systematically studied, and the good news is that, passed a few difficulties, this is indeed a true, magic equation, basically solving the problems mentioned in the beginning of this section. For more on this, you can check any particle physics book.

\section*{15e. Exercises}

Tough physics chapter that we had here, and as useful exercises, we have:

\begin{exercise}
Learn if needed the basics of electrostatics.
\end{exercise}

\begin{exercise}
Learn as well, if needed, the basics of electrodynamics.
\end{exercise}

\begin{exercise}
Read about radiation, light, optics and spectroscopy.
\end{exercise}

\begin{exercise}
Learn about spectral lines, and Lyman, Balmer, Paschen.
\end{exercise}

\begin{exercise}
Read about the Ritz-Rydberg combination principle.
\end{exercise}

\begin{exercise}
Read also about Max Planck, quanta, and Bohr's claims.
\end{exercise}

\begin{exercise}
Read about the Heisenberg matrix mechanics.
\end{exercise}

\begin{exercise}
With all the above learned, read again the present chapter.
\end{exercise}

As bonus exercise, for more, read some quantum electrodynamics. All good stuff.

\chapter{Random matrices}

\section*{16a. Random matrices}

With quantum physics discussed you would say, end of the story, with math and physics and everything, and in particular, end of the present book. However, no advanced linear algebra book would be complete without a word on the random matrices, which are a natural, far-reaching generalization of the usual matrices. These matrices, which appear in a wide array of questions in mathematics and physics, and even in more bizarre contexts, such as economics and finance, are something very simple, as follows:

\begin{definition}
A random matrix is a matrix with random variables as entries,
$$Z\in M_N(L^\infty(X))$$
with $X$ being a probability space.
\end{definition}

Regarding now the mathematics of such matrices, we will be mainly interested in computing their law. Recall indeed from chapter 3 that any scalar matrix $A\in M_N(\mathbb C)$ has a certain abstract law, which in the normal case, $AA^*=A^*A$, is a usual probability measure on $\mathbb C$, and in the self-adjoint case, $A=A^*$, is a usual probability measure on $\mathbb R$. We will see in a moment that the same happens for the random matrices, axiomatized as above, and in view of this, the following question makes sense:

\begin{question}
What are the laws of the various basic types of random matrices, as for instance those having i.i.d. entries, subject to simple constraints? Also, what happens to these laws in the $N>>0$ regime, do we have some interesting asymptotics?
\end{question}

So, this was for the definition and main question regarding the random matrices, and with the motivations for all this coming from a remarkable mixture of first class mathematics and physics, featuring all sorts of interesting questions in probability theory, operator algebras, statistical mechanics, quantum mechanics, and many more.

\bigskip

Excited about this? So am I, and before starting, a piece of advertisement too:

\begin{answer}
The above random matrix questions can all be solved, with very interesting answers, involving among others our favorite groups, $SU_2$ and $SO_3$.
\end{answer}

Getting to work now, as a first task, mentioned above, we must extend the spectral measure material from chapter 3, from the case of the matrix algebras $M_N(\mathbb C)$ to the case of the random matrix algebras $M_N(L^\infty(X))$. Now since these latter algebras are infinite dimensional, the best is to start our study with a discussion of the operator algebras, coming as a continuation of the operator theory from chapter 8. We first have:

\begin{definition}
An abstract operator algebra, or $C^*$-algebra, is a complex algebra $A$ having a norm $||.||$ and an involution $*$, subject to the following conditions:
\begin{enumerate}
\item $A$ is closed with respect to the norm.

\item We have $||aa^*||=||a||^2$, for any $a\in A$.
\end{enumerate}
\end{definition}

In other words, what we did here is to axiomatize the abstract properties of the operator algebras $A\subset B(H)$, coming from the various general results about linear operators from chapter 8, without any reference to the ambient Hilbert space $H$.

\bigskip

As basic examples now, we have the usual matrix algebras $M_N(\mathbb C)$, with the norm and the involution being the usual matrix norm and involution, given by:
$$||A||=\sup_{||x||=1}||Ax||\quad,\quad 
(A^*)_{ij}=\overline{A}_{ji}$$

Some other basic examples are the algebras $L^\infty(X)$ of essentially bounded functions $f:X\to\mathbb C$ on a measured space $X$, with the usual norm and involution, namely:
$$||f||=\sup_{x\in X}|f(x)|\quad,\quad 
f^*(x)=\overline{f(x)}$$

We can put these two basic classes of examples together, as follows:

\begin{proposition}
The random matrix algebras $A=M_N(L^\infty(X))$ are $C^*$-algebras, with their usual norm and involution, given by:
$$||Z||=\sup_{x\in X}||Z_x||\quad,\quad 
(Z^*)_{ij}=\overline{Z}_{ij}$$
These algebras generalize both the algebras $M_N(\mathbb C)$, and the algebras $L^\infty(X)$.
\end{proposition}

\begin{proof}
The fact that the $C^*$-algebra axioms are satisfied is clear from definitions. As for the last assertion, this follows by taking $X=\{.\}$ and $N=1$, respectively.
\end{proof}

We can in fact say more about the above algebras, as follows:

\begin{theorem}
Any algebra of type $L^\infty(X)$ is an operator algebra, as follows:
$$L^\infty(X)\subset B(L^2(X))\quad,\quad 
f\to(g\to fg)$$
More generally, any random matrix algebra is an operator algebra, as follows,
$$M_N(L^\infty(X))\subset B\left(\mathbb C^N\otimes L^2(X)\right)$$
with the embedding being the above one, tensored with the identity.
\end{theorem}

\begin{proof}
We have two assertions to be proved, the idea being as follows:

\medskip

(1) Given $f\in L^\infty(X)$, consider the following operator, acting on $H=L^2(X)$:
$$T_f(g)=fg$$

Observe that $T_f$ is indeed well-defined, and bounded as well, because:
$$||fg||_2
=\sqrt{\int_X|f(x)|^2|g(x)|^2d\mu(x)}
\leq||f||_\infty||g||_2$$

The application $f\to T_f$ being linear, involutive, continuous, and injective as well, we obtain in this way a $C^*$-algebra embedding $L^\infty(X)\subset B(H)$, as desired.

\medskip

(2) Regarding the second assertion, this is best viewed in the following way:
\begin{eqnarray*}
M_N(L^\infty(X))
&=&M_N(\mathbb C)\otimes L^\infty(X)\\
&\subset&M_N(\mathbb C)\otimes B(L^2(X))\\
&=&B\left(\mathbb C^N\otimes L^2(X)\right)
\end{eqnarray*}

Here we have used (1), and some standard tensor product identifications.
\end{proof}

Our purpose in what follows is to develop the spectral theory of the $C^*$-algebras, and in particular that of the random matrix algebras $A=M_N(L^\infty(X))$ that we are interested in, one of our objectives being that of talking about spectral measures, in the normal case, in analogy with what we know about the usual matrices. Let us start with:

\begin{theorem}
Given an element $a\in A$ of a $C^*$-algebra, define its spectrum as:
$$\sigma(a)=\left\{\lambda\in\mathbb C\Big|a-\lambda\notin A^{-1}\right\}$$
The following spectral theory results hold, exactly as in the $A=B(H)$ case:
\begin{enumerate}
\item We have $\sigma(ab)\cup\{0\}=\sigma(ba)\cup\{0\}$.

\item We have $\sigma(f(a))=f(\sigma(a))$, for any $f\in\mathbb C(X)$ having poles outside $\sigma(a)$.

\item The spectrum $\sigma(a)$ is compact, non-empty, and contained in $D_0(||a||)$.

\item The spectra of unitaries $(u^*=u^{-1})$ and self-adjoints $(a=a^*)$ are on $\mathbb T,\mathbb R$.

\item The spectral radius of normal elements $(aa^*=a^*a)$ is given by $\rho(a)=||a||$.
\end{enumerate}
In addition, assuming $a\in A\subset B$, the spectra of $a$ with respect to $A$ and to $B$ coincide.
\end{theorem}

\begin{proof}
Here the assertions (1-5), which are of course formulated a bit informally, are well-known for the full operator algebra $A=B(H)$, and the proof in general is similar:

\medskip

(1) Assuming that $1-ab$ is invertible, with inverse $c$, we have $abc=cab=c-1$, and it follows that $1-ba$ is invertible too, with inverse $1+bca$. Thus $\sigma(ab),\sigma(ba)$ agree on $1\in\mathbb C$, and by linearity, it follows that $\sigma(ab),\sigma(ba)$ agree on any point $\lambda\in\mathbb C^*$.

\medskip

(2) The formula $\sigma(f(a))=f(\sigma(a))$ is clear for polynomials, $f\in\mathbb C[X]$, by factorizing $f-\lambda$, with $\lambda\in\mathbb C$. Then, the extension to the rational functions is straightforward, because $P(a)/Q(a)-\lambda$ is invertible precisely when $P(a)-\lambda Q(a)$ is.

\medskip

(3) By using $1/(1-b)=1+b+b^2+\ldots$ for $||b||<1$ we obtain that $a-\lambda$ is invertible for $|\lambda|>||a||$, and so $\sigma(a)\subset D_0(||a||)$. It is also clear that $\sigma(a)$ is closed, so what we have is a compact set. Finally, assuming $\sigma(a)=\emptyset$ the function $f(\lambda)=\varphi((a-\lambda)^{-1})$ is well-defined, for any $\varphi\in A^*$, and by Liouville we get $f=0$, contradiction.

\medskip

(4) Assuming $u^*=u^{-1}$ we have $||u||=1$, and so $\sigma(u)\subset D_0(1)$. But with $f(z)=z^{-1}$ we obtain via (2) that we have as well $\sigma(u)\subset f(D_0(1))$, and this gives $\sigma(u)\subset\mathbb T$. As for the result regarding the self-adjoints, this can be obtained from the result for the unitaries, by using (2) with functions of type $f(z)=(z+it)/(z-it)$, with $t\in\mathbb R$.

\medskip

(5) It is routine to check, by integrating quantities of type $z^n/(z-a)$ over circles centered at the origin, and estimating, that the spectral radius is given by $\rho(a)=\lim||a^n||^{1/n}$. But in the self-adjoint case, $a=a^*$, this gives $\rho(a)=||a||$, by using exponents of type $n=2^k$, and then the extension to the general normal case is straightforward.

\medskip 

(6) Regarding now the last assertion, the inclusion $\sigma_B(a)\subset\sigma_A(a)$ is clear. For the converse, assume $a-\lambda\in B^{-1}$, and set $b=(a-\lambda )^*(a-\lambda )$. We have then:
$$\sigma_A(b)-\sigma_B(b)=\left\{\mu\in\mathbb C-\sigma_B(b)\Big|(b-\mu)^{-1}\in B-A\right\}$$

Thus this difference in an open subset of $\mathbb C$. On the other hand $b$ being self-adjoint, its two spectra are both real, and so is their difference. Thus the two spectra of $b$ are equal, and in particular $b$ is invertible in $A$, and so $a-\lambda\in A^{-1}$, as desired.
\end{proof}

We can now a prove a key result about operator algebras, as follows:

\begin{theorem}[Gelfand]
If $X$ is a compact space,  the algebra $C(X)$ of continuous functions on it $f:X\to\mathbb C$ is a $C^*$-algebra, with usual norm and involution, namely:
$$||f||=\sup_{x\in X}|f(x)|\quad,\quad 
f^*(x)=\overline{f(x)}$$
Conversely, any commutative $C^*$-algebra is of this form, $A=C(X)$, with 
$$X=\Big\{\chi:A\to\mathbb C\ ,\ {\rm normed\ algebra\ character}\Big\}$$
with topology making continuous the evaluation maps $ev_a:\chi\to\chi(a)$.
\end{theorem}

\begin{proof}
There are several things going on here, the idea being as follows:

\medskip

(1) The first assertion is clear from definitions. Observe that we have indeed:
$$||ff^*||
=\sup_{x\in X}|f(x)|^2
=||f||^2$$

Observe also that the algebra $C(X)$ is commutative, because $fg=gf$.

\medskip

(2) Conversely, given a commutative $C^*$-algebra $A$, let us define $X$ as in the statement. Then $X$ is compact, and $a\to ev_a$ is a morphism of algebras, as follows:
$$ev:A\to C(X)$$

(3) We first prove that $ev$ is involutive. We use the following formula, which is similar to the $z=Re(z)+iIm(z)$ decomposition formula for usual complex numbers:
$$a=\frac{a+a^*}{2}+i\cdot\frac{a-a^*}{2i}$$

Thus it is enough to prove $ev_{a^*}=ev_a^*$ for the self-adjoint elements $a$. But this is the same as proving that $a=a^*$ implies that $ev_a$ is a real function, which is in turn true, by Theorem 16.7, because $ev_a(\chi)=\chi(a)$ is an element of $\sigma(a)$, contained in $\mathbb R$.

\medskip

(4) Since $A$ is commutative, each element is normal, so $ev$ is isometric:
$$||ev_a||
=\rho(a)
=||a||$$

It remains to prove that $ev$ is surjective. But this follows from the Stone-Weierstrass theorem, because $ev(A)$ is a closed subalgebra of $C(X)$, which separates the points.
\end{proof}

As a main consequence of the Gelfand theorem, we have:

\begin{theorem}
For any normal element $a\in A$ we have an identification as follows:
$$<a>=C(\sigma(a))$$
In addition, given a function $f\in C(\sigma(a))$, we can apply it to $a$, and we have
$$\sigma(f(a))=f(\sigma(a))$$
which generalizes the previous rational calculus formula, in the normal case.
\end{theorem}

\begin{proof}
Since $a$ is normal, the $C^*$-algebra $<a>$ that is generates is commutative, so if we denote by $X$ the space of the characters $\chi:<a>\to\mathbb C$, we have:
$$<a>=C(X)$$

Now since the map $X\to\sigma(a)$ given by evaluation at $a$ is bijective, we obtain:
$$<a>=C(\sigma(a))$$

Thus, we are dealing here with usual functions, and this gives all the assertions.
\end{proof}

In order to get now towards noncommutative probability, we first have to develop the theory of positive elements, and linear forms. First, we have the following result:

\begin{proposition}
For an element $a\in A$, the following are equivalent:
\begin{enumerate}
\item $a$ is positive, in the sense that $\sigma(a)\subset[0,\infty)$.

\item $a=b^2$, for some $b\in A$ satisfying $b=b^*$.

\item $a=cc^*$, for some $c\in A$.
\end{enumerate}
\end{proposition}

\begin{proof}
This is something very standard, as follows:

\medskip

$(1)\implies(2)$ Observe first that $\sigma(a)\subset\mathbb R$ implies $a=a^*$. Thus the algebra $<a>$ is commutative, and by using Theorem 16.9, we can set $b=\sqrt{a}$.

\medskip

$(2)\implies(3)$ This is trivial, because we can simply set $c=b$. 

\medskip

$(2)\implies(1)$ This is clear too, because we have:
$$\sigma(a)
=\sigma(b^2)
=\sigma(b)^2\subset\mathbb R^2
=[0,\infty)$$

$(3)\implies(1)$ We can proceed here by contradiction. Indeed, by multiplying $c$ by a suitable element of the algebra $<cc^*>$, we are led to the existence of an element $d\neq0$ satisfying $-dd^*\geq0$. By writing now $d=x+iy$ with $x=x^*,y=y^*$ we have:
$$dd^*+d^*d
=2(x^2+y^2)
\geq0$$

Thus $d^*d\geq0$, which is easily seen to contradict the condition $-dd^*\geq0$.
\end{proof}

We can talk as well about positive linear forms, as follows:

\begin{definition}
Consider a linear map $\varphi:A\to\mathbb C$.
\begin{enumerate}
\item $\varphi$ is called positive when $a\geq0\implies\varphi(a)\geq0$.

\item $\varphi$ is called faithful and positive when $a\geq0,a\neq0\implies\varphi(a)>0$.
\end{enumerate}
\end{definition}

In the commutative case, $A=C(X)$, the positive linear forms appear as follows, with $\mu$ being positive, and strictly positive if we want $\varphi$ to be faithful and positive:
$$\varphi(f)=\int_Xf(x)d\mu(x)$$

In general, the positive linear forms can be thought of as being integration functionals with respect to some underlying ``positive measures''. Based on this, we can formulate:

\begin{definition}
Let $A$ be a $C^*$-algebra, given with a positive trace $tr:A\to\mathbb C$.
\begin{enumerate}
\item The elements $a\in A$ are called random variables.

\item The moments of such a variable are the numbers $M_k(a)=tr(a^k)$.

\item The law of such a variable is the functional $\mu_a:P\to tr(P(a))$.
\end{enumerate}
\end{definition}

Here the exponent $k=\circ\bullet\bullet\circ\ldots$ is by definition a colored integer, and the powers $a^k$ are defined by the following formulae, and multiplicativity: 
$$a^\emptyset=1\quad,\quad
a^\circ=a\quad,\quad
a^\bullet=a^*$$ 

As for the polynomial $P$, this is a noncommuting $*$-polynomial in one variable: 
$$P\in\mathbb C<X,X^*>$$

Observe that the law is uniquely determined by the moments, because we have:
$$P(X)=\sum_k\lambda_kX^k
\implies\mu_a(P)=\sum_k\lambda_kM_k(a)$$

At the level of the general theory, we have the following key result, extending the various results that we have, regarding the self-adjoint and normal matrices:

\index{normal element}
\index{spectral measure}

\begin{theorem}
Let $A$ be a $C^*$-algebra, with a trace $tr$, and consider an element $a\in A$ which is normal, in the sense that $aa^*=a^*a$.
\begin{enumerate}
\item $\mu_a$ is a complex probability measure, satisfying $supp(\mu_a)\subset\sigma(a)$.

\item In the self-adjoint case, $a=a^*$, this measure $\mu_a$ is real.

\item Assuming that $tr$ is faithful, we have $supp(\mu_a)=\sigma(a)$.
\end{enumerate}
\end{theorem}

\begin{proof}
In the normal case, $aa^*=a^*a$, the Gelfand theorem, or rather the subsequent continuous functional calculus theorem, tells us that we have: 
$$<a>=C(\sigma(a))$$

Thus the functional $f(a)\to tr(f(a))$ can be regarded as an integration functional on the algebra $C(\sigma(a))$, and by the Riesz theorem, this latter functional must come from a probability measure $\mu$ on the spectrum $\sigma(a)$, in the sense that we must have:
$$tr(f(a))=\int_{\sigma(a)}f(z)d\mu(z)$$

We are therefore led to the conclusions in the statement, with the uniqueness assertion coming from the fact that the elements $a^k$, taken as usual with respect to colored integer exponents, $k=\circ\bullet\bullet\circ\ldots$\,, generate the whole $C^*$-algebra $C(\sigma(a))$.
\end{proof}

As a first concrete application now, by getting back to the random matrices, and to the various questions raised in the beginning of this chapter, we have:

\begin{theorem}
Given a random matrix $Z\in M_N(L^\infty(X))$ which is normal, 
$$ZZ^*=Z^*Z$$
its law, which is by definition the following abstract functional,
$$\mu:\mathbb C<X,X^*>\to\mathbb C\quad,\quad 
P\to\frac{1}{N}\int_Xtr(P(Z))$$
when restricted to the usual polynomials in two variables,
$$\mu:\mathbb C[X,X^*]\to\mathbb C\quad,\quad 
P\to\frac{1}{N}\int_Xtr(P(Z))$$
must come from a probability measure on the spectrum $\sigma(Z)\subset\mathbb C$, as follows:
$$\mu(P)=\int_{\sigma(T)}P(x)d\mu(x)$$
We agree to use the symbol $\mu$ for all these notions.
\end{theorem}

\begin{proof}
This follows indeed from what we know from Theorem 16.13, applied to the normal element $a=Z$, belonging to the $C^*$-algebra $A=M_N(L^\infty(X))$. 
\end{proof}

At the level of the basic examples, the situation is as follows:

\begin{theorem}
The basic random matrices $Z\in M_N(L^\infty(X))$ are as follows:
\begin{enumerate}
\item In the case $N=1$ the random matrix is a usual random variable, $f\in L^\infty(X)$, automatically normal, and its law as defined above is the usual law.

\item In the case $X=\{.\}$ the random matrix is a usual scalar matrix, $A\in M_N(\mathbb C)$, and in the diagonalizable case, the law is $\mu=\frac{1}{N}\left(\delta_{\lambda_1}+\ldots+\delta_{\lambda_N}\right)$.
\end{enumerate}
\end{theorem}

\begin{proof}
This is clear indeed, with the first assertion coming from definitions, and the second assertion coming by diagonalizing the matrix, as explained in chapter 3.
\end{proof}

At a more advanced level now, the main problem regarding the random matrices is that of computing the law of various classes of such matrices, coming in series:

\begin{question}
What is the law of random matrices coming in series
$$Z_N\in M_N(L^\infty(X))$$
in the $N>>0$ regime?
\end{question}

The general strategy here, coming from physicists, is that of computing first the asymptotic law $\mu^0$, in the $N\to\infty$ limit, and then looking for the higher order terms as well, as to finally reach to a series in $N^{-1}$ giving the law of $Z_N$, as follows: 
$$\mu_N=\mu^0+N^{-1}\mu^1+N^{-2}\mu^2+\ldots$$

As a basic example here, of particular interest are the matrices having i.i.d. complex normal entries, under the constraint $Z=Z^*$. Here the asymptotic law $\mu^0$ is the Wigner semicircle law on $[-2,2]$. We will discuss this in a moment, after some preliminaries.

\section*{16b. Gaussian matrices}

We recall that a random matrix algebra is an algebra of type $A=M_N(L^\infty(X))$, and that we are interested in the computation of the laws of the operators $Z\in A$, called random matrices. Regarding the precise classes of random matrices that we are interested in, first we have the complex Gaussian matrices, which are constructed as follows:

\begin{definition}
A complex Gaussian matrix is a random matrix of type
$$Z\in M_N(L^\infty(X))$$
which has i.i.d. complex normal entries.
\end{definition}

We will see that the above matrices have an interesting, and ``central'' combinatorics, among all kinds of random matrices, with the study of the other random matrices being usually obtained as a modification of the study of the Gaussian matrices.

\bigskip

As a somewhat surprising remark, using real normal variables in Definition 16.17, instead of the complex ones appearing there, leads nowhere. The correct real versions of the Gaussian matrices are the Wigner random matrices, constructed as follows: 

\begin{definition}
A Wigner matrix is a random matrix of type
$$Z\in M_N(L^\infty(X))$$
which has i.i.d. complex normal entries, up to the constraint $Z=Z^*$.
\end{definition}

In other words, a Wigner matrix must be as follows, with the diagonal entries being real normal variables, $a_i\sim g_t$, for some $t>0$, the upper diagonal entries being complex normal variables, $b_{ij}\sim G_t$, the lower diagonal entries being the conjugates of the upper diagonal entries, as indicated, and with all the variables $a_i,b_{ij}$ being independent: 
$$Z=\begin{pmatrix}
a_1&b_{12}&\ldots&\ldots&b_{1N}\\
\bar{b}_{12}&a_2&\ddots&&\vdots\\
\vdots&\ddots&\ddots&\ddots&\vdots\\
\vdots&&\ddots&a_{N-1}&b_{N-1,N}\\
\bar{b}_{1N}&\ldots&\ldots&\bar{b}_{N-1,N}&a_N
\end{pmatrix}$$

As a comment here, for many concrete applications the Wigner matrices are in fact the central objects in random matrix theory, and in particular, they are often more important than the Gaussian matrices. In fact, these are the random matrices which were first considered and investigated, a long time ago, by Wigner himself.

\bigskip

Finally, we will be interested as well in the complex Wishart matrices, which are the positive versions of the above random matrices, constructed as follows: 

\begin{definition}
A complex Wishart matrix is a random matrix of type
$$Z=YY^*\in M_N(L^\infty(X))$$
with $Y$ being a complex Gaussian matrix.
\end{definition}

As before with the Gaussian and Wigner matrices, there are many possible comments that can be made here, of technical or historical nature. First, using in the above real Gaussian variables instead of complex ones leads to a less interesting combinatorics. Also, these matrices were introduced and studied by Marchenko-Pastur not long after Wigner, and so historically came second. Finally, in what regards their combinatorics and applications, these matrices quite often come first, before both the Gaussian and the Wigner ones, with all this being of course a matter of knowledge and taste.

\bigskip

Summarizing, we have three main types of random matrices, which can be somehow designated as ``complex'', ``real'' and ``positive'', and that we will study in what follows. Let us also mention that there are many other interesting classes of random matrices, usually appearing as modifications of the above. More on these later.

\bigskip

Getting to work, let us fix some definitions, for the various types of normal variables appearing above. In the real case, the definition that we will need is as follows:

\begin{definition}
The normal law of parameter $1$ is the following measure:
$$g_1=\frac{1}{\sqrt{2\pi}}e^{-x^2/2}dx$$
More generally, the normal law of parameter $t>0$ is the following measure:
$$g_t=\frac{1}{\sqrt{2\pi t}}e^{-x^2/2t}dx$$
These are also called Gaussian distributions, with ``g'' standing for Gauss.
\end{definition}

The above laws are usually denoted $\mathcal N(0,1)$ and $\mathcal N(0,t)$, but since we will be doing here all kinds of probability, we will use simplified notations for all our measures. Let us also mention that the normal laws traditionally have 2 parameters, the mean and variance, but here we will not need the mean, all our theory using centered laws. Finally, observe that the above laws have indeed mass 1, as they should, due to the Gauss formula:
$$\int_\mathbb R e^{-x^2/2t}dx
=\int_\mathbb R e^{-y^2}\sqrt{2t}\,dy
=\sqrt{2\pi t}$$

Many things can be said about the normal laws, notably with the following result, that will play a key role in what follows, in our various moment computations:

\begin{theorem}
The moments of the normal law are the numbers
$$M_k(g_t)=\sum_{\pi\in P_2(k)}t^{|\pi|}$$
where $P_2(k)$ is the set of pairings of $\{1,\ldots,k\}$, and $|.|$ is the number of blocks.
\end{theorem}

\begin{proof}
The moments of the normal law are subject to the following formula:
\begin{eqnarray*}
M_k(g_t)
&=&\frac{1}{\sqrt{2\pi t}}\int_\mathbb Rx^ke^{-x^2/2t}dx\\
&=&\frac{1}{\sqrt{2\pi t}}\int_\mathbb R(tx^{k-1})\left(-e^{-x^2/2t}\right)'dx\\
&=&\frac{1}{\sqrt{2\pi t}}\int_\mathbb Rt(k-1)x^{k-2}e^{-x^2/2t}dx\\
&=&t(k-1)\times\frac{1}{\sqrt{2\pi t}}\int_\mathbb Rx^{k-2}e^{-x^2/2t}dx\\
&=&t(k-1)M_{k-2}(g_t)
\end{eqnarray*}

On the other hand, let us count the pairings of $\{1,\ldots,k\}$. In order to have such a pairing, we must pair $1$ with one of the numbers $2,\ldots,k$, and then use a pairing of the remaining $k-2$ numbers. Thus, we have the following recurrence formula:
$$|P_2(k)|=(k-1)|P_2(k-2)|$$

Thus we obtain the result at $t=1$, and the general case $t>0$ follows too.
\end{proof}

In the complex case now, the definition that we will need is as follows:

\begin{definition}
The complex Gaussian law of parameter $t>0$ is
$$G_t=law\left(\frac{1}{\sqrt{2}}(a+ib)\right)$$
where $a,b$ are independent, each following the law $g_t$.
\end{definition}

As before with the real Gaussian laws, many things can be said here, notably with the following result, that will play a key role in what follows, for our computations: 

\begin{theorem}
The moments of the complex normal law are the numbers
$$M_k(G_t)=\sum_{\pi\in\mathcal P_2(k)}t^{|\pi|}$$
where $\mathcal P_2(k)$ are the matching pairings of $\{1,\ldots,k\}$, and $|.|$ is the number of blocks.
\end{theorem}

\begin{proof}
We can assume that we are in the case $t=1$, and here a straightforward computation shows that the moments are given by the following formula:
$$M_k=\begin{cases}
(|k|/2)!&(k\ {\rm uniform})\\
0&(k\ {\rm not\ uniform})
\end{cases}$$

On the other hand, the numbers $|\mathcal P_2(k)|$ are given by the same formula. Indeed, in order to have a matching pairing of $k$, our exponent $k=\circ\bullet\bullet\circ\ldots$ must be uniform, consisting of $p$ copies of $\circ$ and $p$ copies of $\bullet$, with $p=|k|/2$. But then the matching pairings of $k$ correspond to the permutations of the $\bullet$ symbols, as to be matched with $\circ$ symbols, and so we have $p!$ such pairings. Thus, we are led to the result.
\end{proof}

In practice, we also need to know how to compute joint moments of independent normal variables. We have here the following result, to be heavily used later on:

\begin{theorem}
Given independent variables $X_i$, each following the complex normal law $G_t$, with $t>0$ being a fixed parameter, we have the Wick formula
$$E\left(X_{i_1}^{k_1}\ldots X_{i_s}^{k_s}\right)=t^{s/2}\#\left\{\pi\in\mathcal P_2(k)\Big|\pi\leq\ker i\right\}$$
where $k=k_1\ldots k_s$ and $i=i_1\ldots i_s$, for the joint moments of these variables.
\end{theorem}

\begin{proof}
This is something well-known, and the basis for all possible computations with complex normal variables, which can be proved in two steps, as follows:

\medskip

(1) Let us first discuss the case where we have a single variable $X$, which amounts in taking $X_i=X$ for any $i$ in the formula in the statement. What we have to compute here are the moments of $X$, with respect to colored integer exponents $k=\circ\bullet\bullet\circ\ldots\,$, and the formula in the statement is equivalent to the one from Theorem 16.23, namely:
$$E(X^k)=t^{|k|/2}|\mathcal P_2(k)|$$

(2) In general now, the point is that we obtain the formula in the statement. Indeed, when expanding the product $X_{i_1}^{k_1}\ldots X_{i_s}^{k_s}$ and rearranging the terms, we are left with doing a number of computations as in (1), and then making the product of the expectations that we found. But this amounts precisely in counting the partitions in the statement, with the condition $\pi\leq\ker i$ there standing precisely for the fact that we are doing the various type (1) computations independently, and then making the product.
\end{proof}

Now by getting back to the Gaussian matrices, we have the following result, with $\mathcal{NC}_2(k)=\mathcal P_2(k)\cap NC(k)$ standing for the noncrossing pairings of a colored integer $k$:

\begin{theorem}
Given a sequence of Gaussian random matrices
$$Z_N\in M_N(L^\infty(X))$$
having independent $G_t$ variables as entries, for some fixed $t>0$, we have
$$M_k\left(\frac{Z_N}{\sqrt{N}}\right)\simeq t^{|k|/2}|\mathcal{NC}_2(k)|$$
for any colored integer $k=\circ\bullet\bullet\circ\ldots\,$, in the $N\to\infty$ limit.
\end{theorem}

\begin{proof}
This is something standard, which can be done as follows:

\medskip

(1) We fix $N\in\mathbb N$, and we let $Z=Z_N$. Let us first compute the trace of $Z^k$. With $k=k_1\ldots k_s$, and with the convention $(ij)^\circ=ij,(ij)^\bullet=ji$, we have:
\begin{eqnarray*}
Tr(Z^k)
&=&Tr(Z^{k_1}\ldots Z^{k_s})\\
&=&\sum_{i_1=1}^N\ldots\sum_{i_s=1}^N(Z^{k_1})_{i_1i_2}(Z^{k_2})_{i_2i_3}\ldots(Z^{k_s})_{i_si_1}\\
&=&\sum_{i_1=1}^N\ldots\sum_{i_s=1}^N(Z_{(i_1i_2)^{k_1}})^{k_1}(Z_{(i_2i_3)^{k_2}})^{k_2}\ldots(Z_{(i_si_1)^{k_s}})^{k_s}
\end{eqnarray*}

(2) Next, we rescale our variable $Z$ by a $\sqrt{N}$ factor, as in the statement, and we also replace the usual trace by its normalized version, $tr=Tr/N$. Our formula becomes:
$$tr\left(\left(\frac{Z}{\sqrt{N}}\right)^k\right)=\frac{1}{N^{s/2+1}}\sum_{i_1=1}^N\ldots\sum_{i_s=1}^N(Z_{(i_1i_2)^{k_1}})^{k_1}(Z_{(i_2i_3)^{k_2}})^{k_2}\ldots(Z_{(i_si_1)^{k_s}})^{k_s}$$

Thus, the moment that we are interested in is given by:
$$M_k\left(\frac{Z}{\sqrt{N}}\right)=\frac{1}{N^{s/2+1}}\sum_{i_1=1}^N\ldots\sum_{i_s=1}^N\int_X(Z_{(i_1i_2)^{k_1}})^{k_1}(Z_{(i_2i_3)^{k_2}})^{k_2}\ldots(Z_{(i_si_1)^{k_s}})^{k_s}$$

(3) Let us apply now the Wick formula, from Theorem 16.24. We conclude that the moment that we are interested in is given by the following formula:
\begin{eqnarray*}
&&M_k\left(\frac{Z}{\sqrt{N}}\right)\\
&=&\frac{t^{s/2}}{N^{s/2+1}}\sum_{i_1=1}^N\ldots\sum_{i_s=1}^N\#\left\{\pi\in\mathcal P_2(k)\Big|\pi\leq\ker\left((i_1i_2)^{k_1},(i_2i_3)^{k_2},\ldots,(i_si_1)^{k_s}\right)\right\}\\
&=&t^{s/2}\sum_{\pi\in\mathcal P_2(k)}\frac{1}{N^{s/2+1}}\#\left\{i\in\{1,\ldots,N\}^s\Big|\pi\leq\ker\left((i_1i_2)^{k_1},(i_2i_3)^{k_2},\ldots,(i_si_1)^{k_s}\right)\right\}
\end{eqnarray*}

(4) Our claim now is that in the $N\to\infty$ limit the combinatorics of the above sum simplifies, with only the noncrossing partitions contributing to the sum, and with each of them contributing precisely with a 1 factor, so that we will have, as desired:
\begin{eqnarray*}
M_k\left(\frac{Z}{\sqrt{N}}\right)
&=&t^{s/2}\sum_{\pi\in\mathcal P_2(k)}\Big(\delta_{\pi\in NC_2(k)}+O(N^{-1})\Big)\\
&\simeq&t^{s/2}\sum_{\pi\in\mathcal P_2(k)}\delta_{\pi\in NC_2(k)}\\
&=&t^{s/2}|\mathcal{NC}_2(k)|
\end{eqnarray*}

(5) In order to prove this, the first observation is that when $k$ is not uniform, in the sense that it contains a different number of $\circ$, $\bullet$ symbols, we have $\mathcal P_2(k)=\emptyset$, and so:
$$M_k\left(\frac{Z}{\sqrt{N}}\right)=t^{s/2}|\mathcal{NC}_2(k)|=0$$

(6) Thus, we are left with the case where $k$ is uniform. Let us examine first the case where $k$ consists of an alternating sequence of $\circ$ and $\bullet$ symbols, as follows:
$$k=\underbrace{\circ\bullet\circ\bullet\ldots\ldots\circ\bullet}_{2p}$$

In this case it is convenient to relabel our multi-index $i=(i_1,\ldots,i_s)$, with $s=2p$, in the form $(j_1,l_1,j_2,l_2,\ldots,j_p,l_p)$. With this done, our moment formula becomes:
$$M_k\left(\frac{Z}{\sqrt{N}}\right)
=t^p\sum_{\pi\in\mathcal P_2(k)}\frac{1}{N^{p+1}}\#\left\{j,l\in\{1,\ldots,N\}^p\Big|\pi\leq\ker\left(j_1l_1,j_2l_1,j_2l_2,\ldots,j_1l_p\right)\right\}$$

Now observe that, with $k$ being as above, we have an identification $\mathcal P_2(k)\simeq S_p$, obtained in the obvious way. With this done too, our moment formula becomes:
$$M_k\left(\frac{Z}{\sqrt{N}}\right)
=t^p\sum_{\pi\in S_p}\frac{1}{N^{p+1}}\#\left\{j,l\in\{1,\ldots,N\}^p\Big|j_r=j_{\pi(r)+1},l_r=l_{\pi(r)},\forall r\right\}$$

(7) We are now ready to do our asymptotic study, and prove the claim in (4). Let indeed $\gamma\in S_p$ be the full cycle, which is by definition the following permutation:
$$\gamma=(1 \, 2 \, \ldots \, p)$$

In terms of $\gamma$, the conditions $j_r=j_{\pi(r)+1}$ and $l_r=l_{\pi(r)}$ found above read:
$$\gamma\pi\leq\ker j\quad,\quad 
\pi\leq\ker l$$

Counting the number of free parameters in our moment formula, we obtain:
$$M_k\left(\frac{Z}{\sqrt{N}}\right)
=\frac{t^p}{N^{p+1}}\sum_{\pi\in S_p}N^{|\pi|+|\gamma\pi|}
=t^p\sum_{\pi\in S_p}N^{|\pi|+|\gamma\pi|-p-1}$$

(8) The point now is that the last exponent is well-known to be $\leq 0$, with equality precisely when the permutation $\pi\in S_p$ is geodesic, which in practice means that $\pi$ must come from a noncrossing partition. Thus we obtain, in the $N\to\infty$ limit, as desired:
$$M_k\left(\frac{Z}{\sqrt{N}}\right)\simeq t^p|\mathcal{NC}_2(k)|$$

This finishes the proof in the case of the exponents $k$ which are alternating, and the case where $k$ is an arbitrary uniform exponent is similar, by permuting everything.
\end{proof}

As a conclusion to this, we have obtained as asymptotic law for the Gaussian matrices a certain mysterious distribution, having as moments some  numbers which are similar to the moments of the usual normal laws, but with the ``underlying matching pairings being now replaced by underlying matching noncrossing pairings''. More on this later.

\section*{16c. Wigner and Wishart}
 
Regarding now the Wigner matrices, we have here the following result, coming as a consequence of Theorem 16.25, via some simple algebraic manipulations:

\begin{theorem}
Given a sequence of Wigner random matrices
$$Z_N\in M_N(L^\infty(X))$$
having independent $G_t$ variables as entries, with $t>0$, up to $Z_N=Z_N^*$, we have
$$M_k\left(\frac{Z_N}{\sqrt{N}}\right)\simeq t^{k/2}|NC_2(k)|$$
for any integer $k\in\mathbb N$, in the $N\to\infty$ limit.
\end{theorem}

\begin{proof}
This can be deduced from a direct computation based on the Wick formula, similar to that from the proof of Theorem 16.25, but the best is to deduce this result from Theorem 16.25 itself. Indeed, we know from there that for Gaussian matrices $Y_N\in M_N(L^\infty(X))$ we have the following formula, valid for any colored integer $K=\circ\bullet\bullet\circ\ldots\,$, in the $N\to\infty$ limit, with $\mathcal{NC}_2$ standing for noncrossing matching pairings:
$$M_K\left(\frac{Y_N}{\sqrt{N}}\right)\simeq t^{|K|/2}|\mathcal{NC}_2(K)|$$

By doing some combinatorics, we deduce from this that we have the following formula for the moments of the matrices $Re(Y_N)$, with respect to usual exponents, $k\in\mathbb N$:
\begin{eqnarray*}
M_k\left(\frac{Re(Y_N)}{\sqrt{N}}\right)
&=&2^{-k}\cdot M_k\left(\frac{Y_N}{\sqrt{N}}+\frac{Y_N^*}{\sqrt{N}}\right)\\
&=&2^{-k}\sum_{|K|=k}M_K\left(\frac{Y_N}{\sqrt{N}}\right)\\
&\simeq&2^{-k}\sum_{|K|=k}t^{k/2}|\mathcal{NC}_2(K)|\\
&=&2^{-k}\cdot t^{k/2}\cdot 2^{k/2}|\mathcal{NC}_2(k)|\\
&=&2^{-k/2}\cdot t^{k/2}|NC_2(k)|
\end{eqnarray*}

Now since the matrices $Z_N=\sqrt{2}Re(Y_N)$ are of Wigner type, this gives the result.
\end{proof}

Summarizing, all this brings us into counting noncrossing pairings. So, let us start with some preliminaries here. We first have the following well-known result:

\begin{theorem}
The Catalan numbers, which are by definition given by
$$C_k=|NC_2(2k)|$$
satisfy the following recurrence formula, with initial data $C_0=C_1=1$,
$$C_{k+1}=\sum_{a+b=k}C_aC_b$$ 
their generating series $f(z)=\sum_{k\geq0}C_kz^k$ satisfies the equation
$$zf^2-f+1=0$$
and is given by the following explicit formula,
$$f(z)=\frac{1-\sqrt{1-4z}}{2z}$$ 
and we have the following explicit formula for these numbers:
$$C_k=\frac{1}{k+1}\binom{2k}{k}$$
Numerically, these numbers are $1,1,2,5,14,42,132,429,1430,4862,16796,\ldots$
\end{theorem}

\begin{proof}
We must count the noncrossing pairings of $\{1,\ldots,2k\}$. Now observe that such a pairing appears by pairing 1 to an odd number, $2a+1$, and then inserting a noncrossing pairing of $\{2,\ldots,2a\}$, and a noncrossing pairing of $\{2a+2,\ldots,2l\}$. We conclude that we have the following recurrence formula for the Catalan numbers:
$$C_k=\sum_{a+b=k-1}C_aC_b$$ 

In terms of the generating series $f(z)=\sum_{k\geq0}C_kz^k$, this recurrence formula reads:
\begin{eqnarray*}
zf^2
&=&\sum_{a,b\geq0}C_aC_bz^{a+b+1}\\
&=&\sum_{k\geq1}\sum_{a+b=k-1}C_aC_bz^k\\
&=&\sum_{k\geq1}C_kz^k\\
&=&f-1
\end{eqnarray*}

Thus $f$ satisfies $zf^2-f+1=0$, and by solving this equation, and choosing the solution which is bounded at $z=0$, we obtain the following formula:
$$f(z)=\frac{1-\sqrt{1-4z}}{2z}$$ 

In order to finish, we use the generalized binomial formula, which gives:
$$\sqrt{1+t}=1-2\sum_{k=1}^\infty\frac{1}{k}\binom{2k-2}{k-1}\left(\frac{-t}{4}\right)^k$$

Now back to our series $f$, we obtain the following formula for it:
\begin{eqnarray*}
f(z)
&=&\frac{1-\sqrt{1-4z}}{2z}\\
&=&\sum_{k=1}^\infty\frac{1}{k}\binom{2k-2}{k-1}z^{k-1}\\
&=&\sum_{k=0}^\infty\frac{1}{k+1}\binom{2k}{k}z^k
\end{eqnarray*}

It follows that the Catalan numbers are given by:
$$C_k=\frac{1}{k+1}\binom{2k}{k}$$

Thus, we are led to the conclusion in the statement.
\end{proof}

In order to recapture now the Wigner measure from its moments, we can use:

\begin{proposition}
The Catalan numbers are the even moments of 
$$\gamma_1=\frac{1}{2\pi}\sqrt{4-x^2}dx$$
called standard semicircle law. As for the odd moments of $\gamma_1$, these all vanish. 
\end{proposition}

\begin{proof}
The even moments of $\gamma_1$ can be computed with the change of variable $x=2\cos t$, and some trigonometric know-how, and we are led to the following formula:
\begin{eqnarray*}
M_{2k}
&=&\frac{1}{\pi}\int_0^2\sqrt{4-x^2}x^{2k}dx\\
&=&\frac{4^{k+1}}{\pi}\int_0^{\pi/2}\cos^{2k}t\sin^2t\,dt\\
&=&C_k
\end{eqnarray*}

As for the odd moments, these all vanish, because the density of $\gamma_1$ is an even function. Thus, we are led to the conclusion in the statement.
\end{proof}

More generally, we have the following result, involving a parameter $t>0$:

\begin{proposition}
Given $t>0$, the real measure having as even moments the numbers $M_{2k}=t^kC_k$ and having all odd moments $0$ is the measure
$$\gamma_t=\frac{1}{2\pi t}\sqrt{4t-x^2}dx$$
called Wigner semicircle law on $[-2\sqrt{t},2\sqrt{t}]$.
\end{proposition}

\begin{proof}
This follows indeed from Proposition 16.28, via a change of variables.
\end{proof}

Now by putting everything together, we obtain the Wigner theorem, as follows:

\begin{theorem}
Given a sequence of Wigner random matrices
$$Z_N\in M_N(L^\infty(X))$$
which by definition have i.i.d. complex normal entries, up to $Z_N=Z_N^*$, we have
$$Z_N\sim\gamma_t$$
in the $N\to\infty$ limit, where $\gamma_t=\frac{1}{2\pi t}\sqrt{4t-x^2}dx$ is the Wigner semicircle law. 
\end{theorem}

\begin{proof}
This follows indeed from all the above, and more specifically, by combining Theorem 16.26, Theorem 16.27 and Proposition 16.29.
\end{proof}

Let us discuss now the Wishart matrices, which are the positive analogues of the Wigner matrices. Quite surprisingly, the computation here leads to the Catalan numbers, but not in the same way as for the Wigner matrices, the result being as follows:

\begin{theorem}
Given a sequence of complex Wishart matrices
$$W_N=Y_NY_N^*\in M_N(L^\infty(X))$$
with $Y_N$ being $N\times N$ complex Gaussian of parameter $t>0$, we have
$$M_k\left(\frac{W_N}{N}\right)\simeq t^kC_k$$
for any exponent $k\in\mathbb N$, in the $N\to\infty$ limit.
\end{theorem}

\begin{proof}
There are several possible proofs for this result, as follows:

\medskip

(1) A first method is by using the formula that we have in Theorem 16.25, for the Gaussian matrices $Y_N$. Indeed, we know from there that we have the following formula, valid for any colored integer $K=\circ\bullet\bullet\circ\ldots\,$, in the $N\to\infty$ limit:
$$M_K\left(\frac{Y_N}{\sqrt{N}}\right)\simeq t^{|K|/2}|\mathcal{NC}_2(K)|$$

With $K=\circ\bullet\circ\bullet\ldots\,$, alternating word of length $2k$, with $k\in\mathbb N$, this gives:
$$M_k\left(\frac{Y_NY_N^*}{N}\right)\simeq t^k|\mathcal{NC}_2(K)|$$

Thus, in terms of the Wishart matrix $W_N=Y_NY_N^*$ we have, for any $k\in\mathbb N$:
$$M_k\left(\frac{W_N}{N}\right)\simeq t^k|\mathcal{NC}_2(K)|$$

The point now is that, by doing some combinatorics, we have:
$$|\mathcal{NC}_2(K)|=|NC_2(2k)|=C_k$$

Thus, we are led to the formula in the statement.

\medskip

(2) A second method, that we will explain now as well, is by proving the result directly, starting from definitions. The matrix entries of our matrix $W=W_N$ are given by:
$$W_{ij}=\sum_{r=1}^NY_{ir}\bar{Y}_{jr}$$

Thus, the normalized traces of powers of $W$ are given by the following formula:
\begin{eqnarray*}
tr(W^k)
&=&\frac{1}{N}\sum_{i_1=1}^N\ldots\sum_{i_k=1}^NW_{i_1i_2}W_{i_2i_3}\ldots W_{i_ki_1}\\
&=&\frac{1}{N}\sum_{i_1=1}^N\ldots\sum_{i_k=1}^N\sum_{r_1=1}^N\ldots\sum_{r_k=1}^NY_{i_1r_1}\bar{Y}_{i_2r_1}Y_{i_2r_2}\bar{Y}_{i_3r_2}\ldots Y_{i_kr_k}\bar{Y}_{i_1r_k}
\end{eqnarray*}

By rescaling now $W$ by a $1/N$ factor, as in the statement, we obtain:
$$tr\left(\left(\frac{W}{N}\right)^k\right)=\frac{1}{N^{k+1}}\sum_{i_1=1}^N\ldots\sum_{i_k=1}^N\sum_{r_1=1}^N\ldots\sum_{r_k=1}^NY_{i_1r_1}\bar{Y}_{i_2r_1}Y_{i_2r_2}\bar{Y}_{i_3r_2}\ldots Y_{i_kr_k}\bar{Y}_{i_1r_k}$$

By using now the Wick rule, we obtain the following formula for the moments, with $K=\circ\bullet\circ\bullet\ldots\,$, alternating word of lenght $2k$, and with $I=(i_1r_1,i_2r_1,\ldots,i_kr_k,i_1r_k)$:
\begin{eqnarray*}
M_k\left(\frac{W}{N}\right)
&=&\frac{t^k}{N^{k+1}}\sum_{i_1=1}^N\ldots\sum_{i_k=1}^N\sum_{r_1=1}^N\ldots\sum_{r_k=1}^N\#\left\{\pi\in\mathcal P_2(K)\Big|\pi\leq\ker(I)\right\}\\
&=&\frac{t^k}{N^{k+1}}\sum_{\pi\in\mathcal P_2(K)}\#\left\{i,r\in\{1,\ldots,N\}^k\Big|\pi\leq\ker(I)\right\}
\end{eqnarray*}

In order to compute this quantity, we use the standard bijection $\mathcal P_2(K)\simeq S_k$. By identifying the pairings $\pi\in\mathcal P_2(K)$ with their counterparts $\pi\in S_k$, we obtain:
\begin{eqnarray*}
M_k\left(\frac{W}{N}\right)
&=&\frac{t^k}{N^{k+1}}\sum_{\pi\in S_k}\#\left\{i,r\in\{1,\ldots,N\}^k\Big|i_s=i_{\pi(s)+1},r_s=r_{\pi(s)},\forall s\right\}
\end{eqnarray*}

Now let $\gamma\in S_k$ be the full cycle, which is by definition the following permutation:
$$\gamma=(1 \, 2 \, \ldots \, k)$$

The general factor in the product computed above is then 1 precisely when following two conditions are simultaneously satisfied:
$$\gamma\pi\leq\ker i\quad,\quad 
\pi\leq\ker r$$

Counting the number of free parameters in our moment formula, we obtain:
$$M_k\left(\frac{W}{N}\right)
=t^k\sum_{\pi\in S_k}N^{|\pi|+|\gamma\pi|-k-1}$$

The point now is that the last exponent is well-known to be $\leq 0$, with equality precisely when the permutation $\pi\in S_k$ is geodesic, which in practice means that $\pi$ must come from a noncrossing partition. Thus we obtain, in the $N\to\infty$ limit:
$$M_k\left(\frac{W}{N}\right)\simeq t^kC_k$$

Thus, we are led to the conclusion in the statement.
\end{proof}

As a consequence of the above result, we have a new look on the Catalan numbers, which is more adapted to our present Wishart matrix considerations, as follows:

\begin{proposition}
The Catalan numbers $C_k=|NC_2(2k)|$ appear as well as
$$C_k=|NC(k)|$$
where $NC(k)$ is the set of all noncrossing partitions of $\{1,\ldots,k\}$.
\end{proposition}

\begin{proof}
This follows indeed from the proof of Theorem 16.31. Observe that we obtain as well a formula in terms of matching pairings of alternating colored integers.
\end{proof}

The direct explanation for the above formula, relating noncrossing partitions and pairings, comes form the following result, which is very useful, and good to know:

\begin{proposition}
We have a bijection between noncrossing partitions and pairings
$$NC(k)\simeq NC_2(2k)$$
which is constructed as follows:
\begin{enumerate}
\item The application $NC(k)\to NC_2(2k)$ is the ``fattening'' one, obtained by doubling all the legs, and doubling all the strings as well.

\item Its inverse $NC_2(2k)\to NC(k)$ is the ``shrinking'' application, obtained by collapsing pairs of consecutive neighbors.
\end{enumerate}
\end{proposition}

\begin{proof}
The fact that the two operations in the statement are indeed inverse to each other is clear, by computing the corresponding two compositions, with the remark that the construction of the fattening operation requires the partitions to be noncrossing.
\end{proof}

Getting back now to probability, we are led to the question of finding the law having the Catalan numbers as moments, in the above way. The result here is as follows:

\begin{proposition}
The real measure having the Catalan numbers as moments is
$$\pi_1=\frac{1}{2\pi}\sqrt{4x^{-1}-1}\,dx$$
called Marchenko-Pastur law of parameter $1$.
\end{proposition}

\begin{proof}
The moments of the law $\pi_1$ in the statement can be computed with the change of variable $x=4\cos^2t$, and some trigonometric know-how, as follows:
\begin{eqnarray*}
M_k
&=&\frac{1}{2\pi}\int_0^4\sqrt{4x^{-1}-1}\,x^kdx\\
&=&\frac{4^{k+1}}{\pi}\int_0^{\pi/2}\cos^{2k}t\sin^2t\,dt\\
&=&C_k
\end{eqnarray*}

Thus, we are led to the conclusion in the statement.
\end{proof}

Now back to the Wishart matrices, we are led to the following result:

\begin{theorem}
Given a sequence of complex Wishart matrices
$$W_N=Y_NY_N^*\in M_N(L^\infty(X))$$
with $Y_N$ being $N\times N$ complex Gaussian of parameter $t>0$, we have
$$\frac{W_N}{tN}\sim\frac{1}{2\pi}\sqrt{4x^{-1}-1}\,dx$$
with $N\to\infty$, with the limiting measure being the Marchenko-Pastur law $\pi_1$.
\end{theorem}

\begin{proof}
This follows indeed from Theorem 16.31 and Proposition 16.34.
\end{proof}

As a comment now, while the above result is definitely something interesting at $t=1$, at general $t>0$ this looks more like a ``fake'' generalization of the $t=1$ result, because the law $\pi_1$ stays the same, modulo a trivial rescaling. The reasons behind this phenomenon are quite subtle, and skipping some discussion, the point is that Theorem 16.35 is indeed something ``fake'' at general $t>0$, and the correct generalization of the $t=1$ computation, involving more general classes of complex Wishart matrices, is as follows:

\begin{theorem}
Given a sequence of general complex Wishart matrices
$$W_N=Y_NY_N^*\in M_N(L^\infty(X))$$
with $Y_N$ being $N\times M$ complex Gaussian of parameter $1$, we have
$$\frac{W_N}{N}\sim\max(1-t,0)\delta_0+\frac{\sqrt{4t-(x-1-t)^2}}{2\pi x}\,dx$$
with $M=tN\to\infty$, with the limiting measure being the Marchenko-Pastur law $\pi_t$.
\end{theorem}

\begin{proof}
This follows once again by using the moment method, the limiting moments in the $M=tN\to\infty$ regime being as follows, after doing the combinatorics:
$$M_k\left(\frac{W_N}{N}\right)\simeq\sum_{\pi\in NC(k)}t^{|\pi|}$$

But these numbers are the moments of the Marchenko-Pastur law $\pi_t$, which in addition has the density given by the formula in the statement, and this gives the result.
\end{proof}

\section*{16d. Back to groups}

Many things can be said about the limiting laws found by Wigner and Marchenko-Pastur, and in what concerns us, we will present an interpretation of them related to our favorite matrix groups, $SU_2$ and $SO_3$. In order to discuss this, we will need the following formula, coming as a continuation of some previous formulae from chapter 9:

\begin{theorem}
We have the following formula,
$$\int_0^{\pi/2}\cos^rt\sin^st\,dt=\left(\frac{\pi}{2}\right)^{\varepsilon(r)\varepsilon(s)}\frac{r!!s!!}{(r+s+1)!!}$$
where $\varepsilon(r)=1$ if $r$ is even, and $\varepsilon(r)=0$ if $r$ is odd.
\end{theorem}

\begin{proof}
Let us call $I_{rs}$ the integral in the statement. In order to do the partial integration, observe that we have the following formula:
\begin{eqnarray*}
(\cos^rt\sin^st)'
&=&r\cos^{r-1}t(-\sin t)\sin^st+\cos^rt\cdot s\sin^{s-1}t\cos t\\
&=&-r\cos^{r-1}t\sin^{s+1}t+s\cos^{r+1}t\sin^{s-1}t
\end{eqnarray*}

By integrating between $0$ and $\pi/2$, we obtain, for $r,s>0$:
$$rI_{r-1,s+1}=sI_{r+1,s-1}$$

Thus, we can compute $I_{rs}$ by recurrence, and when $s$ is even we obtain:
\begin{eqnarray*}
I_{rs}
&=&\frac{r!!s!!}{(r+s)!!}\,I_{r+s}\\
&=&\frac{r!!s!!}{(r+s)!!}\left(\frac{\pi}{2}\right)^{\varepsilon(r+s)}\frac{(r+s)!!}{(r+s+1)!!}\\
&=&\left(\frac{\pi}{2}\right)^{\varepsilon(r)\varepsilon(s)}\frac{r!!s!!}{(r+s+1)!!}
\end{eqnarray*}

Observe that this gives the result for $r$ even as well, by symmetry. In the remaining case now, where both the exponents $r,s$ are odd, we can use once again the equality $rI_{r-1,s+1}=sI_{r+1,s-1}$ found above, and the recurrence gives the following formula:
\begin{eqnarray*}
I_{rs}
&=&\frac{r!!s!!}{(r+s-1)!!}\,I_{r+s-1,1}\\
&=&\frac{r!!s!!}{(r+s-1)!!}\cdot\frac{1}{r+s}\\
&=&\left(\frac{\pi}{2}\right)^{\varepsilon(r)\varepsilon(s)}\frac{r!!s!!}{(r+s+1)!!}
\end{eqnarray*}

Thus, we obtain indeed the formula in the statement.
\end{proof}

As an application of the above trigonometric formula, we have:

\begin{theorem}
The moments of the hyperspherical variables are
$$\int_{S^{N-1}_\mathbb R}x_i^pdx=\frac{(N-1)!!p!!}{(N+p-1)!!}$$
and the rescaled variables $y_i=\sqrt{N}x_i$ become normal and independent with $N\to\infty$.
\end{theorem}

\begin{proof}
By using spherical coordinates, with input from Theorem 16.37, we are led to the folllowing formula, for the joint moments of the hyperspherical variables:
$$\int_{S^{N-1}_\mathbb R}x_1^{k_1}\ldots x_N^{k_N}\,dx
=\frac{(N-1)!!k_1!!\ldots k_N!!}{(N+\Sigma k_i-1)!!}$$

But this gives the formula in the statement, and the last assertion too.
\end{proof}

Getting back now to our random matrix problematics, here is how the quite mysterious semicircle Wigner law $\gamma_1$ appears, geometrically, in relation with the group $SU_2$: 

\begin{theorem}
The main character of $SU_2$ follows the following law,
$$\gamma_1=\frac{1}{2\pi}\sqrt{4-x^2}dx$$
which is the Wigner law of parameter $1$.
\end{theorem}

\begin{proof}
We know that $SU_2$ is the group of unitary rotations $U\in U_2$ of determinant one, and as explained in chapter 3, by solving the equation $U^*=U^{-1}$, we are led to:
$$SU_2=\left\{\begin{pmatrix}a+ib&c+id\\ -c+id&a-ib\end{pmatrix}\ \Big|\ a^2+b^2+c^2+d^2=1\right\}$$

In this picture, the main character is given by the following formula:
$$\chi\begin{pmatrix}a+ib&c+id\\ -c+id&a-ib\end{pmatrix}=2a$$

We are therefore left with computing the law of the following variable:
$$a\in C(S^3_\mathbb R)$$

But this is something very familiar, namely a hyperspherical variable at $N=4$, so we can use here Theorem 16.38. We obtain the following moment formula:
\begin{eqnarray*}
\int_{S^3_\mathbb R}a^{2k}
&=&\frac{3!!(2k)!!}{(2k+3)!!}\\
&=&2\cdot\frac{(2k)!}{2^kk!2^{k+1}(k+1)!}\\
&=&\frac{1}{4^k}\cdot\frac{1}{k+1}\binom{2k}{k}\\
&=&\frac{C_k}{4^k}
\end{eqnarray*}

Thus the variable $2a\in C(S^3_\mathbb R)$ follows the Wigner semicircle law $\gamma_1$, as claimed.
\end{proof}

Quite nice all this, and things are not over here. We have as well a similar result regarding the Marchenko-Pastur law $\pi_1$, involving this time the group $SO_3$, as follows:

\begin{theorem}
The main character of $SO_3$, modified by adding $1$ to it, given in standard Euler-Rodrigues coordinates by
$$\chi=4a^2$$
follows a squared semicircle law, or Marchenko-Pastur law $\pi_1$.
\end{theorem}

\begin{proof}
This follows by using the quotient map $SU_2\to SO_3$, and the result for $SU_2$. Indeed, by using the Euler-Rodrigues formula, in the context of Theorem 16.39 and its proof, the main character of $SO_3$, modified by adding $1$ to it, is given by:
$$\chi=(3a^2-b^2-c^2-d^2)+1=4a^2$$

Now recall from the proof of Theorem 16.39 that we have:
$$2a\sim\gamma_1$$

On the other hand, a quick comparison between the moment formulae for the Wigner and Marchenko-Pastur laws, which are very similar, shows that we have:
$$f\sim\gamma_1\implies f^2\sim\pi_1$$

Thus, with $f=2p$, we obtain the result in the statement.
\end{proof}

And with this, done with random matrices. Of course, all the above was just the basic theory, and for more, have a look at any random matrix or free probability book.

\section*{16e. Exercises}

Congratulations for having read this book, and no exercises for this final chapter. However, there is a lot of further advanced linear algebra to be discovered, quite often in relation with operators and infinite dimensions. In the hope that you will go this way.

\baselineskip=14pt

\printindex

\end{document}